\documentclass[preprint]{aastex61}

\newcommand\aastex{AAS\TeX}

\received{\today}
\submitjournal{AAS}

\shorttitle{\aastex\ ALMA Spatially-resolved dense molecular gas observations 
of nearby ULIRGs}
\shortauthors{Imanishi et al.}

\begin{document}

\title{ALMA Spatially-resolved Dense Molecular Gas Survey of 
Nearby Ultraluminous Infrared Galaxies}

\correspondingauthor{Masatoshi Imanishi}
\email{masa.imanishi@nao.ac.jp}

\author[0000-0001-6186-8792]{Masatoshi Imanishi}
\affil{National Astronomical Observatory of Japan, National Institutes 
of Natural Sciences (NINS), 2-21-1 Osawa, Mitaka, Tokyo 181-8588, Japan}
\affil{Department of Astronomy, School of Science, Graduate
University for Advanced Studies (SOKENDAI), Mitaka, Tokyo 181-8588}

\author[0000-0002-6939-0372]{Kouichiro Nakanishi}
\affil{National Astronomical Observatory of Japan, National Institutes 
of Natural Sciences (NINS), 2-21-1 Osawa, Mitaka, Tokyo 181-8588, Japan}
\affil{Department of Astronomy, School of Science, Graduate
University for Advanced Studies (SOKENDAI), Mitaka, Tokyo 181-8588}

\author[0000-0001-9452-0813]{Takuma Izumi}
\affil{National Astronomical Observatory of Japan, National Institutes 
of Natural Sciences (NINS), 2-21-1 Osawa, Mitaka, Tokyo 181-8588, Japan}



\begin{abstract}

We present the results of our ALMA HCN J=3--2 and HCO$^{+}$ J=3--2 line 
observations of a uniformly selected sample ($>$25) of nearby 
ultraluminous infrared galaxies (ULIRGs) at $z <$ 0.15.
The emission of these dense molecular gas tracers and continuum are
spatially resolved in the majority of observed ULIRGs for the first time 
with achieved synthesized beam sizes of $\lesssim$0$\farcs$2 
or $\lesssim$500 pc. 
In most ULIRGs, the HCN-to-HCO$^{+}$ J=3--2 flux ratios in the nuclear 
regions within the beam size are systematically higher than those 
in the spatially extended regions. 
The elevated nuclear HCN J=3--2 emission could be related to 
(a) luminous buried active galactic nuclei, 
(b) the high molecular gas density and temperature in 
ULIRG's nuclei, and/or (c) mechanical heating by spatially 
compact nuclear outflows.
A small fraction of the observed ULIRGs display higher HCN-to-HCO$^{+}$ 
J=3--2 flux ratios in localized off-nuclear regions than those of the nuclei, 
which may be due to mechanical heating by spatially extended outflows.
The observed nearby ULIRGs are generally rich in dense 
($>$10$^{5}$ cm$^{-3}$) molecular gas, with an estimated mass of 
$>$10$^{9}$M$_{\odot}$ within the nuclear (a few kpc) regions, and 
dense gas can dominate the total molecular mass there.
We find a low detection rate ($<$20\%) regarding the possible 
signature of a vibrationally excited (v$_{2}$=1f) HCN J=3--2 emission 
line in the vicinity of the bright HCO$^{+}$ J=3--2 line that may 
be due, in part, to the large molecular line widths of ULIRGs. 

\end{abstract}




\section{Introduction} \label{sec:intro}

Ultraluminous infrared galaxies (ULIRGs) emit 
very strong infrared radiation, with infrared (8--1000 $\mu$m) luminosity 
of L$_{\rm IR}$ $>$ 10$^{12}$L$_{\odot}$. 
ULIRGs are mostly found in gas-rich major galaxy mergers in the 
nearby universe at $z <$ 0.15 (e.g.) \citep{san88,cle96,mur96,duc97}.
The large observed infrared luminosities indicate the presence of luminous 
energy sources; however, these sources may be hidden 
and/or surrounded by dust.
Numerical simulations of gas-rich major galaxy mergers predict 
not only that active star formation (a starburst) is triggered but also that 
existing supermassive black holes (SMBHs) can achieve high mass accretion rates 
and become luminous active galactic nuclei (AGNs); both are likely to occur in 
heavily dust-obscured regions \citep{hop06}.
It is important to observationally test this widely accepted 
scenario, if we are to understand the nature of the ULIRG population 
in the local universe \citep{san88,ale12}.

For this purpose, it is necessary to conduct observations using wavelengths 
of low dust extinction, given that the ULIRG energy sources, 
particularly AGNs, are buried deep in the nuclear gas and dust. 
Dust extinction effects are much reduced in the infrared and hard X-ray range, 
compared to the optical range. 
Spectroscopic investigations of such buried AGNs 
in nearby ULIRGs have been undertaken using an infrared range of 2.5--35 $\mu$m 
(e.g.) 
\citep{gen98,imd00,soi02,ima06a,ima06,arm07,ima07a,nar08,ima08,nar09,vei09,ima09b,ima10a,ima10b,nar10,lee12,ich14,alo16} 
and hard X-ray range of $>$10 keV (e.g.) \citep{ten15,oda17,ric17,iwa17}.
This is because AGNs and starbursts display different relative strengths of 
polycyclic aromatic hydrocarbon (PAH) emission features and dust 
continuum emission in the infrared, and also have different hard X-ray to 
bolometric luminosity ratios. 
In fact, these observations have revealed optically elusive, but 
intrinsically luminous, buried AGNs in many nearby ULIRGs. 
However, there remains a large fraction of nearby ULIRGs for which no 
clear AGN signature has been detected in infrared and hard X-ray surveys. 
It remains unclear whether the non-detection of AGN signatures in these 
sources is due to the fact that (a) AGNs are not sufficiently luminous 
(i.e., mass accretion rates onto SMBHs are not so high) to be 
detectable relative to strong starbursts, 
or (b) luminous AGNs are present but are too highly obscured to be 
detectable even in the infrared and X-ray range.
Observations at wavelengths of even stronger dust/gas penetration 
power may reveal extremely deeply buried AGNs ($\tau$ $>$ 1 at 
infrared 2.5--35 $\mu$m and X-ray $\sim$10 keV) in nearby ULIRGs.

(Sub)millimeter wavelengths at 0.3--3.6 mm (84--950 GHz) are of particular 
interest in this regard. 
If the Galactic extinction curve by gas and dust is assumed, the extinction 
is a factor of $>$30 smaller than those at infrared $\sim$20 $\mu$m and 
X-ray $\sim$10 keV \citep{hil83}. 
With the Atacama Large Millimeter Array (ALMA), 
it is now possible to conduct high-sensitivity and high-spatial-resolution
observations of nearby ULIRGs in this potentially powerful (sub)millimeter 
wavelength range.
As the energy-generation mechanisms are different between a starburst 
and AGN (nuclear fusion versus a mass-accreting SMBH, respectively), 
the physical/chemical effects to the surrounding molecular gas are expected 
to differ between the two energy sources.
Rotational J-transition lines of many molecules reside in the 
(sub)millimeter wavelength range, and multiple molecular line flux 
ratios can differ depending on the primary energy sources. 
In fact, (sub)millimeter molecular rotational J-transition line flux 
ratios have been used to separate AGNs from starbursts in nearby 
optically identified luminous AGNs and starburst galaxies 
(e.g.) \citep{koh05,kri08}.
(Sub)millimeter molecular line observations have been applied 
also to a limited number of nearby ULIRGs, using pre-ALMA observing 
facilities and the early phase of ALMA. 
Some fraction of ULIRGs had been diagnosed to contain optically elusive, 
but infrared-detectable luminous buried AGNs.
A trend was found that such ULIRGs displayed different molecular 
rotational J-transition line flux ratios from starburst-dominated galaxies, 
although a small fraction of infrared-classified starbursts showed 
similar molecular line flux ratios to AGN-important galaxies (e.g.) 
\citep{ima04,ima06b,ima07b,ima09a,cos11,ima14b,pri15,izu16,ima16c,ima18b}.
Compared to old single dish telescope data, the trend has been seen 
more clearly in recent interferometric data, most likely because 
(a) only nuclear molecular gas significantly affected by luminous AGNs is 
selectively probed (with reduced contaminations from 
spatially extended starburst-dominated emission) and 
(b) multiple molecular line data were taken simultaneously 
(which minimized possible systematic uncertainties in molecular 
line flux ratios). 

Based on sensitive ALMA studies, a few candidates of 
optically/infrared/X-ray-elusive but (sub)millimeter-detectable 
extremely deeply buried AGNs have been found \citep{ima16c,ima18b}. 
This has been accomplished (a) by finding molecular rotational 
J-transition line flux ratios (at the vibrational ground level; v=0) 
usually seen in AGN-important galaxies and (b) by detecting 
molecular J-transition lines at a vibrationally excited 
(v$_{2}$=1, l=1f; hereafter v$_{2}$=1f) level.
The latter detection can be the signature of a luminous buried AGN, 
particularly if the vibrationally excited (v$_{2}$=1f) molecular line 
luminosity, relative to vibrational ground (v=0) molecular line and/or 
infrared luminosity, is significantly higher than those seen in active 
star-forming regions \citep{aal15a,aal15b,ima16c,ima18b}. 
This is because (a) molecular vibrational excitation is most likely 
to be caused by mid-infrared (5--25 $\mu$m) radiative pumping 
\citep{sak10} and (b) the mid-infrared to bolometric luminosity ratio 
in an AGN can be much higher than a starburst, due to AGN-heated hot 
dust emission. 
However, these molecular line observations of nearby ULIRGs have only been 
conducted for arbitrarily selected interesting sources with 
known luminous AGN signatures at other wavelengths and/or some other 
unique observational properties.
Target selection has been non-uniform, hampering any statistical 
discussion about the nature of the nearby ULIRG population in general. 
Specifically, it is still unclear (a) whether ULIRGs with 
infrared-identified luminous buried AGNs consistently show molecular line 
flux ratios expected for AGN-important galaxies and (b) what fraction 
of nearby ULIRGs contain infrared/X-ray-elusive, but 
(sub)millimeter-detectable, extremely deeply buried AGNs 
($\tau$ $>$ 1 at 2.5--35 $\mu$m and $\sim$10 keV).
It is a natural next step to apply the powerful (sub)millimeter 
molecular line method to an extended number of carefully and uniformly 
selected nearby ULIRG samples, using the highly sensitive ALMA. 

In this paper, we present the results of ALMA molecular line 
observations of an increasing number of nearby ULIRGs.
Molecular line and continuum emission were spatially resolved 
in the majority of observed ULIRGs, allowing for the first 
time detailed and systematic investigations of the internal spatial 
variation of molecular line flux ratios in a statistically significant 
number of the nearby ULIRG population.
Throughout this paper, we adopt H$_{0}$ $=$ 71 km s$^{-1}$ Mpc$^{-1}$, 
$\Omega_{\rm M}$ = 0.27, and $\Omega_{\rm \Lambda}$ = 0.73.
Unless otherwise stated, ``molecular line flux ratio'' refers to 
the ``rotational J-transition line flux ratio at the vibrational 
ground level (v=0)''.

\section{Selected Targets and Molecular Lines} 

Our targets are nearby ULIRGs in the IRAS 1 Jy sample \citep{kim98}.
Optical spectroscopic classifications \citep{vei99,yua10}, 
as well as infrared spectroscopic energy diagnostics (a) at 5--35 $\mu$m 
using Spitzer IRS 
(e.g.) \citep{vei09,ima07a,ima09b,ima10a,nar08,nar09,nar10} 
and (b) at 2.5--5 $\mu$m using AKARI IRC 
(e.g.) \citep{ima08,ima10b}, have extensively been conducted for this sample.
We selected ULIRGs classified optically as LINERs or HII-regions 
(i.e., no obvious optical AGN signature) \citep{vei99} at $z <$ 0.15 and 
at declinations $<$ $+$20$^{\circ}$ to be best observable from the ALMA site 
in Chile. 
Our primary goal was to unveil optically and/or 
infrared/X-ray-elusive, but intrinsically luminous, buried AGNs in 
nearby ULIRGs, based on the proposed powerful (sub)millimeter approach. 

In nearby ULIRGs, which are mostly gas-rich major galaxy mergers, molecular 
gas is concentrated in the nuclear regions (e.g.) 
\citep{dow98,eva02,per18}, and gas density becomes high there \citep{sco15}. 
Observations using dense gas tracers are crucial to understanding the 
physical properties of molecular gas in nearby ULIRGs 
\citep{gao04a,gao04b}.
HCN and HCO$^{+}$ rotational J-transition lines are better 
probes of such dense molecular gas, due to larger dipole moments, than 
the widely used bright CO low-J (J=1--0 or J=2--1) transition lines 
\citep{shi15}.
Furthermore, the trend of higher HCN-to-HCO$^{+}$ line flux ratios 
in AGNs than in starbursts have been argued 
mostly in recent high-quality interferometric data (where 
HCN and HCO$^{+}$ lines were observed simultaneously),
at J=1--0 \citep{koh05,kri08,ima06b,ima07b,ima09a}, 
J=3--2 \citep{kri08,ima16a,ima16c}, 
and J=4--3 \citep{izu13,ima14b,gar14,izu15,izu16,ima18b}.
Thus, observing both HCN and HCO$^{+}$ lines can be an effective way 
to disentangle the energetic role of AGNs and starbursts.

Our next step was to select ULIRGs with modestly bright dense molecular 
emission lines from the nuclear regions, where the effects of the 
putative AGNs to molecular line flux ratios are expected to be strong.
Regarding the nuclear buried energy sources of ULIRGs, it is expected 
that dense molecular emission line luminosity correlates with infrared 
dust emission luminosity \citep{gao04a,gao04b}, regardless of whether 
the energy 
sources are stars and/or AGNs, because both dense molecular gas and 
spatially coexisting dust are heated and produce dense molecular line 
and infrared dust continuum emission, respectively. 
Although the infrared luminosities of ULIRGs are usually derived 
from IRAS's large-aperture ($>$60$''$) measurements (Table 1), 
the bulk of the infrared luminosities in nearby ULIRGs 
originate in nuclear ($<$a few kpc) regions, with small contributions 
from spatially extended ($>$3 kpc) star formation in the host galaxies 
\citep{soi00,dia10,ima11}.
Namely, ULIRGs with high IRAS-based infrared fluxes strongly suggest 
high {\it nuclear} infrared dust continuum emission fluxes and thereby
high {\it nuclear} dense molecular emission line fluxes 
\citep{ima14b}.

Next, we determined which rotational J-transition lines should be observed. 
Historically, HCN and HCO$^{+}$ J=1--0 lines have long been used to diagnose 
energy sources (e.g.) 
\citep{koh05,kri08,ima04,ima06b,ima07b,ima09a,cos11,pri15}.
However, using ALMA, the J=1--0 lines can be observed only for nearby 
sources at $z <$ 0.05; above this redshift, the lines 
are redshifted beyond the frequency coverage of band 3 (84--116 GHz), 
the longest wavelength band of the current ALMA.
Many interesting nearby ULIRGs in the IRAS 1 Jy sample are at $z >$ 0.05 
\citep{kim98}.

Higher J-transition lines at higher frequencies (shorter wavelengths) 
can be covered with ALMA for higher redshift sources. 
More importantly, higher J-transition lines probe denser and warmer 
molecular gas, due to higher critical density and excitation energy level, 
than lower J-transition lines \citep{shi15}.
Given that dense, warm molecular gas locates preferentially in the nuclear 
regions of galaxies, molecular emission line flux ratios 
at higher-J should be more sensitive to the properties of molecular gas 
in close vicinity to the ULIRG's nuclear primary energy sources 
($<$a few kpc), with small contaminations from low dense and cool 
molecular gas in the spatially extended ($>$3 kpc) host galaxies.
Given that the J=2--1 lines of HCN and HCO$^{+}$ for nearby sources at 
$z <$ 0.15 were not observable before ALMA Cycle 4  
(due to the unavailability of band 5 [163--211 GHz] at that time), 
we started our dense molecular line observations of nearby ULIRGs at J=4--3 
\citep{ima14b,izu16,ima18b} and J=3--2 \citep{ima16c}. 
We decided to expand our HCN and HCO$^{+}$ line survey of nearby 
ULIRGs at J=3--2 in band 6 (211--275 GHz) for the following
reasons.
First, due to less requirements for weather conditions in band 6, 
the chance of program execution is higher than that in 
band 7 (275--373 GHz) where J=4--3 lines of HCN and HCO$^{+}$ are 
observable for nearby sources. 
Second, at J=3--2, in addition to the widely used HCN and HCO$^{+}$ 
rotational J-transition lines at the vibrational ground level (v=0), 
we can simultaneously cover the vibrationally excited (v$_{2}$=1f) 
emission lines of both HCN and HCO$^{+}$ in ALMA band 6, which can 
provide additional signatures of luminous buried AGNs surrounded by a large 
column density of dense molecular gas and dust ($\S$1).

Assuming a fixed HCN J=3--2 to J=1--0 flux ratio, we ordered ULIRGs 
in the IRAS 1 Jy sample at $z <$ 0.15 and declinations $<$ +20$^{\circ}$, 
from the highest infrared flux.
We selected 26 sources with infrared fluxes above a certain threshold, 
which are summarized in Table 1.
Of these, HCN J=3--2 and HCO$^{+}$ J=3--2 emission line data of 
three ULIRGs (IRAS 12112$+$0305, IRAS 20414$-$1651, and IRAS 22491$-$1808)
had been obtained during ALMA Cycle 2 or 3 \citep{ima16c}.
We thus observed the remaining 23 ULIRGs in Cycle 5.
These 26 ULIRGs were selected solely based on the infrared flux 
(or expected nuclear HCN J=3--2 emission line flux). 
This flux limited sample was complete and unbiased with respect 
to the energetic importance of optically elusive, but infrared-detected, 
buried AGNs. 

\section{Observations and Data Analyses} 

Our HCN J=3--2 and HCO$^{+}$ J=3--2 observations of 23 ULIRGs in band 6 
(211--275 GHz) were conducted using our ALMA Cycle 5 program 
2017.1.00057.S (PI = M. Imanishi). 
The observation details are summarized in Table 2.
Observations of the majority of the ULIRGs were conducted in 2017 December 
with the longest baseline of $>$2500 m, excluding 
IRAS 21329$-$2346 and IRAS 22206$-$2715 whose observations were 
made in 2018 April with the $\sim$500 m longest baseline. 
We adopted the widest 1.875 GHz width mode and three spectral windows 
to simultaneously cover HCN J=3--2 ($\nu_{\rm rest}$ = 265.886 GHz), 
HCO$^{+}$ J=3--2 ($\nu_{\rm rest}$ = 267.558 GHz) and 
HCN v$_{2}$=1f J=3--2 ($\nu_{\rm rest}$ = 267.199 GHz), and 
HCO$^{+}$ v$_{2}$=1f J=3--2 ($\nu_{\rm rest}$ = 268.689 GHz) lines in each of 
the windows.

We started our data reduction from calibrated data provided by ALMA, 
using CASA (https://casa.nrao.edu).
Based on channels that did not show obvious emission and/or absorption 
lines, we determined the continuum level, and subtracted it using the 
CASA task ``uvcontsub''.
Then we applied the ``clean'' task 
(Briggs-weighting; robust $=$ 0.5 and gain $=$ 0.1)
for the continuum-only and 
continuum-subtracted molecular line data to create maps. 
The final velocity resolution was $\sim$20 km s$^{-1}$.
The pixel scales were 0$\farcs$02 pixel$^{-1}$ for the majority of the 
observed ULIRGs with a final synthesized beam size of 
0$\farcs$1--0$\farcs$2 (150--530 pc), except IRAS 21329$-$2346 
and IRAS 22206$-$2715, for which 0$\farcs$04 pixel$^{-1}$ was adopted 
due to the larger (0$\farcs$6--0$\farcs$8 or 1.3--1.7 kpc) 
synthesized beam sizes. 
According to the ALMA Cycle 5 Proposer's Guide, the maximum recoverable 
scale (MRS) is 1--3$''$ (1.4--7.5 kpc) for sources with 
0$\farcs$1--0$\farcs$2 synthesized beam sizes and 7--9$''$ 
(15--21 kpc) for the two sources with 0$\farcs$6--0$\farcs$8 
synthesized beam sizes.
Diffuse continuum and molecular line emission with a spatial 
scale larger than the MRS can be missed in our ALMA data.
The absolute flux calibration uncertainty of band 6 data is expected 
to be $<$10\% in ALMA Cycle 5. 

\section{Results} 

Figure 1 and Table 3 show the continuum maps and detailed continuum 
emission properties, respectively.
Significant continuum emission was detected from all sources, except 
IRAS 10485$-$1447 and IRAS 02411$+$0353. 
Multiple continuum emission components are clearly seen in the five ULIRGs 
IRAS 04103$-$2838 (Fig. 1c), IRAS 09039$+$0503 (Fig. 1d), 
IRAS 11095$-$0238 (Fig. 1f), IRAS 13335$-$2612 (Fig. 1g), 
and IRAS 14348$-$1447 (Fig. 1h).
Other specific comments on the continuum emission for several 
selected sources are provided in the Appendices A and D.

Figure 2 (left) displays the continuum-subtracted 
full frequency coverage spectra at the 
continuum peak position within the synthesized beam size.
Gaussian fits for the detected HCN J=3--2 and HCO$^{+}$ J=3--2 emission 
lines were applied, after removing data points possibly affected by 
other emission and/or absorption features, and fitting results are listed 
in Tables 4 and 5, respectively. 
The best Gaussian fits are shown in the Appendix B.
Figure 3 shows integrated intensity (moment 0) maps of the HCN J=3--2 
and HCO$^{+}$ J=3--2 emission lines, by summing velocity channels with 
significant line signals.
The peak flux values of the HCN J=3--2 and HCO$^{+}$ J=3--2 
emission lines in the moment 0 maps are summarized in Tables 4 and 5, 
respectively. 
These values roughly agree with the Gaussian fit estimates 
(columns 2 and 8 in Tables 4 and 5).
For the HCN J=3--2 and HCO$^{+}$ J=3--2 emission line fluxes, 
Gaussian fit estimates will be used in our discussion. 
For the HCO$^{+}$ J=3--2 emission line flux estimates, possible 
contamination from the nearby vibrationally excited (v$_{2}$=1f) HCN 
J=3--2 emission lines cannot be completely removed. However, their 
contributions should be negligible given much smaller fluxes than the 
HCO$^{+}$ J=3--2 emission line fluxes (see $\S$5.2).

Intensity-weighted mean velocity (moment 1) and intensity-weighted
velocity dispersion (moment 2) maps for HCN J=3--2 and 
HCO$^{+}$ J=3--2 emission lines are shown in the Appendix C for selected 
ULIRGs with bright molecular lines.

Table 6 tabulates the estimated intrinsic emission sizes of the 
ALMA-detected continuum, HCN J=3--2, and HCO$^{+}$ J=3--2 lines 
after deconvolution, using the CASA task ``imfit''.
These values will be used for the comparison of the spatial extents 
of molecular line and continuum emission, with respect to beam sizes.
For most of the sources observed with small beam sizes 
(0$\farcs$1--0$\farcs$2 or 150--530 pc in physical scale), 
the deconvolved intrinsic emission sizes are significantly 
larger than the beam sizes, suggesting that the emission is spatially 
resolved.
We extracted spectra within 1$''$ (1.4--2.5 kpc) diameter circular 
apertures around the continuum peak positions, which are displayed 
in Figure 2 (right).
This aperture size was adopted, as ULIRG's continuum and dense 
molecular line emission are usually concentrated in compact nuclear 
($<$a few kpc) regions 
(e.g.)
\citep{wil08,ima14b,sco15,tun15,mar16,ima16c,sak17,man17,sli17,bar17,bar18,ima18b}.
IRAS 09039$+$0503 (Figs. 3g and 3h) and IRAS 11095$-$0238 
(Figs. 3k and 3l) show double nuclear molecular line emission with 
the separation of $\sim$0$\farcs$5, for which we created spatially 
integrated spectra with 1$\farcs$5 diameter circular apertures centered 
at the middle point of the two nuclei, to include emission from both nuclei 
(Figs. 2d' and 2h').
In general, the molecular line peak fluxes are higher in these 
spatially integrated spectra than the peak-position beam-sized 
spectra (Fig. 2, left), 
due to the contribution from spatially resolved (a few 100 pc to 
a few kpc) nuclear emission components. 
 
For IRAS 21329$-$2346, IRAS 22206$-$2715, IRAS 12112$+$0305, 
IRAS 20414$-$1651, and IRAS 22491$-$1808 which were observed with larger beam 
sizes (0$\farcs$5--0$\farcs$9 or 0.9--1.7 kpc), we created spatially integrated 
spectra with 1$\farcs$5 or 2$''$ diameter circular apertures 
(Figs. 2o', 2aa', 2ab', 2ac', 2ad', and 2ae').
For these five ULIRGs, the flux increase from the peak position within 
the beam size (Tables 4 and 5), to the integrated area (Table 7), 
is relatively modest ($<$20--30\%) compared to other ULIRGs observed 
with small 0$\farcs$1--0$\farcs$2 synthesized beam sizes, as expected, 
because the bulk of molecular line fluxes in these five ULIRGs 
have already been covered in the peak-position beam-sized 
spectra.

Figure 4 compares the intrinsic deconvolved size of continuum versus 
HCN J=3--2 and HCO$^{+}$ J=3--2 emission lines (left) and 
HCN J=3--2 vs. HCO$^{+}$ J=3--2 (right).
While the HCN J=3--2 line emission size is comparable to the continuum 
emission size (Fig. 4, left), the emission size of the 
HCO$^{+}$ J=3--2 line is larger overall than those of the continuum and 
HCN J=3--2 line (Fig. 4, left and right).
The more spatially extended emission for HCO$^{+}$ J=3--2 than for 
HCN J=3--2 is reasonable, given that the critical density of 
HCO$^{+}$ J=3--2 is a factor of $\sim$5 smaller than that of HCN J=3--2 
\citep{shi15}. Thus, HCO$^{+}$ can be collisionally excited to J=3 
to a greater extent than 
HCN by spatially extended low-density and low-temperature molecular gas 
(compared to nuclear dense, warm molecular gas).

\section{Discussion} 

\subsection{HCN-to-HCO$^{+}$ J=3--2 Flux Ratios}

Table 8 tabulates the HCN-to-HCO$^{+}$ J=3--2 flux ratios 
based on Gaussian fit measurements in the nuclear beam-sized 
spectra and spatially integrated spectra for the 24 ULIRGs with 
significant dense molecular line detection in both spectra, 
excluding IRAS 10485$-$1447 and IRAS 02411$-$0353. 
These measurements are compared in Figure 5.
The ratios in the beam-sized spectra at the nuclear peak position  
tend to be higher than those in the spatially integrated spectra, 
because most objects are located below the thick solid line 
in Fig. 5.
This suggests that the HCN-to-HCO$^{+}$ J=3--2 flux ratios are 
generally higher at the compact nuclear cores than in spatially 
resolved regions.
  
To better investigate the enhanced HCN J=3--2 emission at the  
compact nuclear cores, we created maps of the ratios  
of the HCN J=3--2 to HCO$^{+}$ J=3--2 flux (in Jy km s$^{-1}$) for ULIRGs 
observed with small (0$\farcs$1--0$\farcs$2) beam sizes (Fig. 6). 
ULIRGs observed with larger beam sizes (0$\farcs$5--0$\farcs$9) are 
not plotted, as we were unable to obtain meaningful information regarding 
the spatial variation in the HCN-to-HCO$^{+}$ J=3--2 flux ratios. 
If an AGN produces stronger HCN emission than star formation ($\S$1), 
it is expected that the HCN-to-HCO$^{+}$ J=3--2 flux ratios are 
higher at the compact nuclear cores where AGNs reside than in 
spatially resolved off-nuclear regions where star formation is usually 
energetically dominant. 
This trend is clearly seen in the nearby well-studied 
optically identified luminous AGN, NGC 7469 (z = 0.016), in which the 
HCN-to-HCO$^{+}$ J=3--2 flux ratio is significantly higher at the 
AGN-dominated nuclear core than off-nuclear bright star-forming knots 
(Fig. 6a). 
In Figure 6, HCN-to-HCO$^{+}$ J=3--2 flux ratios in nearby ULIRGs are 
also generally higher in the compact nuclear cores than the 
spatially resolved regions.
AGN effects are a plausible scenario.
In fact, an enhanced HCN abundance, relative to HCO$^{+}$, 
in molecular gas in the close vicinity 
of luminous AGNs is observationally suggested \citep{nak18,ima18b,sai18},
which can explain the elevated HCN emission of compact nuclear cores 
\footnote{
A decreased HCO$^{+}$ abundance is another possibility and 
is argued in highly turbulent regions \citep{pap07}.
However, we regard that this could happen in wide areas of merging 
ULIRGs, rather than only at the compact nuclear cores. 
}. 
Alternatively, the volume number density and temperature of molecular 
gas are expected to be very high in the nuclei of ULIRGs \citep{sco15}. 
More HCN collisional excitation by nuclear dense and warm 
molecular gas than in the spatially resolved regions \citep{ima18b} 
can explain the observed HCN J=3--2 flux enhancement in ULIRG's nuclei 
as well.
ALMA multiple J-transition HCN and/or HCO$^{+}$ observations (e.g., 
J=2--1 and J=4--3, in addition to J=3--2) with similar beam sizes 
(0$\farcs$1--0$\farcs$2) are needed to constrain the spatial variation 
in molecular gas density and temperature, and thereby excitation 
conditions, in the observed ULIRGs. 
It is also possible that mechanical heating by shocks caused by 
spatially compact nuclear outflows \citep{aal12,aal17,pri17,bar18} 
contributes a lot for the observed nuclear molecular line flux ratios.
An AGN is a plausible origin for such a compact outflow 
\citep{aal12,aal17}, but a very compact starburst remains an 
alternative origin.

Several ULIRGs show higher HCN-to-HCO$^{+}$ J=3--2 flux 
ratios in off-nuclear local regions than the nuclear cores.
Examples include IRAS 23234$+$0946 and IRAS 01004$-$2237, in which  
excesses in the HCN-to-HCO$^{+}$ J=3--2 flux ratios are found both  
at the southeastern and northwestern sides of nuclei in 
Figures 6l and 6o. 
Such off-nuclear local regions with enhanced HCN-to-HCO$^{+}$ J=3--2 flux  
ratios are also seen in the western part of the nuclei for 
IRAS 00188$-$0856 and IRAS 01298$-$0744 (Figs. 6b and 6q). 
Although molecular line emission is generally fainter in off-nuclear  
edge regions than in nuclei, we display in Figure 6 only regions with 
$\gtrsim$3$\sigma$ HCO$^{+}$ J=3--2 emission line detection in the 
moment 0 maps in Figure 3. 
Mechanical heating by spatially extended outflow-related shocks 
may be responsible for these high HCN-to-HCO$^{+}$ J=3--2 flux ratios 
at off-nuclear regions \citep{izu13}. 
It is also possible that molecular gas becomes locally dense and warm  
in off-nuclear regions through merger-induced processes. 
 
Figure 7 is a plot of the nuclear HCN-to-HCO$^{+}$ J=3--2 flux ratios 
as a function of the infrared spectroscopically estimated AGN bolometric 
contribution \citep{nar10}.
ULIRGs with significant infrared-estimated AGN contributions tend to show 
HCN-to-HCO$^{+}$ J=3--2 flux ratios larger than unity, as expected 
for AGN-important galaxies \citep{kri08,ima16c}. 
Not only the infrared-identified AGN-significant ULIRGs but also 
the majority of ULIRGs with no significant infrared-detected AGNs 
show HCN J=3--2 emission stronger than HCO$^{+}$ J=3--2 emission.
The HCN-to-HCO$^{+}$ J=3--2 flux ratios in known-starburst-dominated regions 
(mostly less infrared luminous than ULIRGs) are usually less than unity 
\citep{kri08,ima16c}.
Plausible scenarios include (a) the presence of infrared-elusive but 
(sub)millimeter-detectable extremely deeply buried ($\tau$ $>$ 1 at 
infrared 2.5--35 $\mu$m) luminous AGNs and their effects to nuclear 
molecular line fluxes (including both AGN radiative effects and 
mechanical heating by AGN-related compact outflow shocks) and/or 
(b) more HCN excitation to J=3 than in less infrared luminous starburst 
galaxies due to dense and warm molecular gas in ULIRG's nuclei.

\subsection{Vibrationally Excited HCN J=3--2 Emission Lines}

Vibrationally excited (v$_{2}$=1f) HCN J=3--2 emission lines 
at $\nu_{\rm rest}$ = 267.199 GHz were 
detected in several buried-AGN-hosting infrared luminous galaxies and 
such candidates \citep{sak10,aal15a,aal15b,mar16,ima16b}.
This emission is believed to be vibrationally excited by 
infrared radiative pumping, by absorbing $\sim$14 $\mu$m infrared photons, 
rather than collisionally excited \citep{sak10,aal15b}.
Because an AGN can emit $\sim$14 $\mu$m infrared photons more efficiently 
than star formation for given bolometric luminosity, 
due to AGN-heated hot dust thermal radiation, 
detection of the strong HCN v$_{2}$=1f J=3--2 
emission line can be used to argue for the presence of a luminous AGN, 
particularly if the HCN v$_{2}$=1f J=3--2 to HCN J=3--2 
(v=0) and/or infrared luminosity ratio is significantly larger than those
found in pure star-forming regions \citep{aal15b,ima16c,ima17}.
We thus searched for signatures of the HCN v$_{2}$=1f J=3--2 
emission line. 
 
The detection of the HCN v$_{2}$=1f J=3--2 emission line is 
straightforward if the molecular line widths are small, say $<$200 km s$^{-1}$ 
full width at half maximum (FWHM), because we can clearly separate 
the contribution 
from the nearby strong HCO$^{+}$ J=3--2 (v=0) emission line at $\nu_{\rm rest}$ = 267.558 GHz 
($\sim$400 km s$^{-1}$ separation) \citep{sak10,aal15b,mar16,ima16b}. 
However, the detection is not easy for the majority of ULIRGs whose 
molecular line widths are usually large.
The nuclear beam-sized spectra of IRAS 10378$+$1108, IRAS 11095$-$0238 NE, 
IRAS 14348$-$1447 SW, and IRAS 16090$-$0139 show possible signatures 
of the excess emission on the lower-frequency side of the HCO$^{+}$ J=3--2 
line emission (Figs. 2f, 2h, 2l, and 2n), which may be attributable 
to the contribution from the HCN v$_{2}$=1f J=3--2 emission line.
We compared the emission line profiles of HCN J=3--2 and HCO$^{+}$ J=3--2 
in detail, after normalizing peak flux values, and looked for sources that 
showed significant flux excess at the lower-frequency (high-velocity) side 
of the HCO$^{+}$ J=3--2 line, compared to the HCN J=3--2 line. 
IRAS 10378$+$1108 and IRAS 11095$-$0238 NE may correspond to such 
cases (Fig. 8).
If the excess is due to the vibrationally excited HCN v$_{2}$=1f J=3--2 
emission line, then its estimated fluxes are roughly 
$\sim$0.1 (Jy km s$^{-1}$) for both sources.
The HCN v$_{2}$=1f to v=0 flux ratios at J=3--2 are $\sim$0.05, 
which is as high as an AGN-hosting ULIRG with a clearly detected 
HCN v$_{2}$=1f J=3--2 emission line (i.e., IRAS 20551$-$4250) 
\citep{ima16b,ima17}, but smaller than other such ULIRGs 
\citep{aal15b,mar16}. 

The detection rate of the HCN v$_{2}$=1f J=3--2 emission line in ULIRGs 
observed in ALMA Cycle 5 is at most $\sim$9 \% (=2/23).
Because IRAS 12112$+$0305 NE and IRAS 22491$-$1808, and possibly
IRAS 20414$-$1651, observed in ALMA Cycle 2 or 3, display some signs 
of the HCN v$_{2}$=1f J=3--2 emission lines \citep{ima16c}, the detection 
rate in the complete sample could be as high as $\sim$19 \% (=5/26), 
which is still low.
However, wide molecular emission lines in ULIRGs preclude the 
identification of the faint HCN v$_{2}$=1f J=3--2 emission line inside 
the emission tail of the bright HCO$^{+}$ J=3--2 (v=0) emission line.
The limited S/N ratios can also hamper the recognition of the HCN 
v$_{2}$=1f J=3--2 emission line.
The actual fraction of ULIRGs showing the HCN v$_{2}$=1f J=3--2 
emission lines could be higher.
It is also possible that emission from the vicinity of a buried AGN 
is opaque even to the (sub)millimeter continuum \citep{aal15b}, in which 
case the equivalent width of the HCN v$_{2}$=1f J=3--2 emission line 
could be low. 
If this were the case, it would mean that luminous buried AGNs can be 
elusive even through (sub)millimeter observations.

The signatures of the HCO$^{+}$ v$_{2}$=1f J=3--2 ($\nu_{\rm rest}$ = 
268.689 GHz) emission lines were searched for in the obtained ALMA spectra, 
but were not detected clearly in any of the observed ULIRGs. 

\subsection{Molecular Inflow and Outflow}
 
IRAS 00091$-$0738 (Fig. 2q) shows double-peaked emission line profiles 
for HCN J=3--2 and HCO$^{+}$ J=3--2.
For HCO$^{+}$ J=3--2, because the flux at the central dip is below the 
continuum level, the double-peaked profile of the HCO$^{+}$ J=3--2 emission 
line cannot be solely due to rotation \citep{sco17} or self-absorption 
\citep{aal15b}.
Foreground absorption in front of the molecular line-emitting region is 
required at least for HCO$^{+}$ J=3--2.
We created an integrated intensity (moment 0) map of this HCO$^{+}$ J=3--2 
absorption component only (below the continuum level) and confirmed that 
the peak position of this absorption component spatially 
agrees with the continuum emission peak position in Table 3, supporting 
the foreground absorption scenario.
IRAS 01004$-$2237 (Fig. 2s) also displays an absorption feature 
at the lower-frequency side of the HCO$^{+}$ J=3--2 emission line. 
This absorption peak position is also spatially coincident with the 
continuum emission peak position in Table 3 within 1 pixel (0$\farcs$02).
These absorption features may be related to inflowing and/or outflowing 
molecular gas. 

Molecular outflow signatures in nearby ULIRGs are usually seen as 
broad emission wings for bright molecular lines, mostly in CO 
J-transition lines (e.g.) \citep{mai12,cic14,gar15,vei17,ima17,per18}.
The detection of the broad emission line wings in the HCN and HCO$^{+}$ 
lines in nearby ULIRGs has been limited \citep{aal15a,pri17,bar18}, simply due 
to the fainter emission line fluxes of these dense gas tracers compared to 
the brightest CO lines. 
The broad emission wings in HCN J=3--2 and HCO$^{+}$ J=3--2 lines 
were not clearly seen in any of the observed ULIRGs in this paper 
(Fig. 2). 

\subsection{Nuclear Dense Molecular Mass}

The luminosities of the HCN J=3--2 and HCO$^{+}$ J=3--2 emission lines 
are summarized in Table 9; equations (1) and (3) 
of \citet{sol05} were used for the calculations from the estimated fluxes 
in spatially integrated (1--2$''$) spectra (Table 7).
These luminosities can be converted into the mass of dense 
($>$10$^{5}$ cm$^{-3}$) molecular gas within the nuclear (a few kpc) regions 
of the observed ULIRGs.
When molecular gas emission lines are optically thick and 
thermally excited, i.e., where the excitation temperature T$_{\rm ex}$ equals 
the gas kinetic temperature T$_{\rm kin}$ (T$_{\rm ex}$ $=$ T$_{\rm kin}$), 
the molecular line luminosities in units of [K km s$^{-1}$ pc$^{2}$] 
should be comparable with respect to different J-transition lines.
The conversion factor from molecular line luminosity to dense 
molecular gas mass is estimated to be $\alpha_{\rm HCN}$=6--40 for 
HCN J=1--0 in various starbursts and AGNs 
(M$_{\rm dense}$ = 6--40 $\times$ HCN J=1--0 luminosity 
[M$_{\odot}$ (K km s$^{-1}$ pc$^{2}$)$^{-1}$]) \citep{gao04a,kri08,gre09,ler15}.
Although the $\alpha_{\rm HCN}$ value has some uncertainty, assuming 
the lower range of $\alpha_{\rm HCN}$ = 10, 
the derived dense molecular gas mass from the HCN J=3--2 emission line 
luminosity is as high as $>$10$^{9}$M$_{\odot}$ for the observed nearby 
ULIRGs, except IRAS 10485$-$1447, IRAS 02411$+$0353, and 
IRAS 12112$+$0305 SW. 
Because the conversion factor ($\alpha$) is estimated to be comparable for 
HCN J=1--0 and HCO$^{+}$ J=1--0 \citep{ler15,ler17}, we obtain similarly 
a high dense molecular gas mass of $>$10$^{9}$M$_{\odot}$ from the 
HCO$^{+}$ J=3--2 emission line luminosity. 
When HCN and HCO$^{+}$ are only sub-thermally excited at J=3--2
(T$_{\rm ex}$ $<$ T$_{\rm kin}$), the luminosity in units of 
[K km s$^{-1}$ pc$^{2}$] is expected to be higher at J=1--0 than at J=3--2, 
so that the dense molecular gas mass will be even higher.
The observed nearby ULIRGs are dense molecular gas-rich 
in the nuclear few-kpc regions.

The {\it total} molecular mass within a certain volume is often 
estimated from CO J=1--0 and/or CO J=2--1 emission line luminosities 
\citep{bol13}.
For IRAS 14348$-$1447 SW and NE and IRAS 12112$+$0305 NE and SW, 
the {\it total} molecular mass in the nuclear regions 
(using the adopted cosmological parameters in Table 1)
are estimated to be 12 $\times$ 10$^{9}$M$_{\odot}$, 
7.4 $\times$ 10$^{9}$M$_{\odot}$, 11 $\times$ 10$^{9}$M$_{\odot}$, 
and 3.6 $\times$ 10$^{9}$M$_{\odot}$, 
respectively, from pre-ALMA interferometric CO J=1--0 observations 
\citep{eva00,eva02}. 
The dense molecular masses estimated from HCN J=3--2 data 
(Table 9, column 4) are 
$>$5.2 $\times$ 10$^{9}$M$_{\odot}$, $>$3.0 $\times$ 10$^{9}$M$_{\odot}$, 
$>$4.8 $\times$ 10$^{9}$M$_{\odot}$, and $>$0.3 $\times$ 10$^{9}$M$_{\odot}$, 
respectively, where $\alpha_{\rm HCN}$=10 is assumed conservatively and 
the lower limit means that HCN could be sub-thermally excited at J=3--2.
With the exception of the faint secondary nucleus, 
IRAS 12112$+$0305 SW, dense molecular gas masses can account for 
the bulk of the nuclear total molecular masses.
For IRAS 22491$-$1808, we can estimate the nuclear total molecular 
gas mass from the ALMA-measured CO J=2--1 flux of 57.4$\pm$0.1 (Jy km s$^{-1}$) 
\citep{per18} to be $\sim$6 $\times$ 10$^{9}$M$_{\odot}$, where 
the conversion factor from CO luminosity in (K km s$^{-1}$ pc$^{2}$) 
to molecular gas mass with $\alpha_{\rm CO}$=1.5 \citep{eva02} 
and optically thick thermal excitation for CO J=2--1 
(T$_{\rm ex}$ $\sim$ T$_{\rm kin}$) are assumed.
The dense molecular mass based on HCN J=3--2 data (Table 9, column 4) 
is $>$4.3 $\times$ 10$^{9}$M$_{\odot}$, which is as high as 
the total molecular mass in the IRAS 22491$-$1808 nucleus.
Comparisons of the nuclei of these selected ULIRGs suggest 
that dense gas is the dominant molecular phase in the nuclei of 
nearby ULIRGs in general.

\subsection{Relation between Dense Molecular Line and Infrared Luminosity}

For energy sources surrounded by dense molecular gas and dust as in 
the nuclear regions of ULIRGs, whatever the energy sources are 
(either buried AGNs and/or nuclear starbursts), HCN and HCO$^{+}$ 
emission line luminosities (=mass tracers of dense molecular gas 
excited by energy sources) are expected to roughly correlate with 
infrared luminosity (=mass tracer of infrared continuum emitting 
dust heated by energy sources) at a first approximation, because 
these luminosities become higher for more luminous energy sources ($\S$2).
Their correlation has been found in Galactic star-forming regions and 
various types of external galaxies 
\citep{gao04a,wu05,eva06,gra08,ma13,zha14,liu16,tan18,ima18b}.
Figure 9 shows a comparison of the HCN J=3--2 and HCO$^{+}$ J=3--2 
emission line luminosities measured with 1--2$''$ diameter circular 
apertures, with IRAS-measured infrared luminosity.
As the energetically essential nuclear regions (a few kpc) of 
nearby ULIRGs are covered ($\S$2), the bulk of dense molecular gas 
emission in individual sources is expected to be included in our 
ALMA measurements. 

The best-fits for HCN J=4--3, HCO$^{+}$ J=4--3, and HCN J=1--0 
\citep{gao04b,zha14,tan18} are plotted together.
For sub-thermal excitation at J=4--3, the best-fit lines at J=4--3 
should be at the left side of those at J=1--0.
This is the case both for HCN and HCO$^{+}$.
For HCN, our data at J=3--2 are located to the right of the best-fit 
lines for J=4--3 (solid and dotted straight lines in Fig. 9a), but 
left of that for J=1--0 (dashed straight line in Fig. 9a).
These trends can be explained if HCN is sub-thermally excited at J=3--2 
and J=4--3, and the deviation below the thermal excitation is larger 
at J=4--3 than at J=3--2. 
For HCO$^{+}$, the two best-fit lines for J=4--3 \citep{zha14,tan18} 
are largely different, possibly because of different individual 
sources used for the fitting; however, our HCO$^{+}$ J=3--2 data are 
located on the left side of the best-fit line for J=1--0 (dashed 
straight line in Fig. 9b). 
Sub-thermal HCO$^{+}$ excitation at J=3--2 is indicated.

HCN J=3--2 emission in these ULIRGs is estimated to be sub-thermally 
excited, and yet shows higher HCN-to-HCO$^{+}$ J=3--2 flux ratios 
than known less-infrared luminous starburst-dominated galaxies 
($\S$5.1).
This suggests that in the comparison starburst sample, the ratio of 
HCN J=3--2 excitation temperature to molecular gas kinetic temperature 
(T$_{\rm ex}$/T$_{\rm kin}$) is even lower (more sub-thermal further below 
thermal excitation) than the observed ULIRGs. 

\subsection{Spatially Integrated Continuum Emission and Spectral Energy 
Distribution}

Table 10 summarizes spatially integrated continuum fluxes for the observed 
ULIRGs. 
The same aperture size as employed for molecular line flux measurement 
(Table 7) is used for individual sources. 
Namely, a circular aperture with a diameter of 1--2$''$ is adopted, 
depending on the synthesized beam size and emission morphology.
The spatially integrated continuum fluxes shown in Table 10 are substantially 
(a factor of 1.5--6) larger than the continuum fluxes at the nuclear 
peak positions within the beam size listed in Table 3, particularly for 
sources observed with small (0$\farcs$1--0$\farcs$2 or 150--530 pc) 
beam sizes, suggesting the presence of a strong, spatially resolved 
(a few 100 pc to a few kpc) continuum emission component at $\sim$240 GHz. 
While the continuum emission from a compact nuclear core should 
come from an AGN and/or very compact ($<$500 pc) starbursts, the 
spatially resolved continuum emission is most likely to come from 
a few kpc-scale starburst activity. 
The following three physical mechanisms in starbursts can contribute 
to the observed spatially resolved continuum fluxes: 
(1) dust thermal radiation, (2) free-free emission from 
HII-regions, and (3) synchrotron radiation.
Although the third synchrotron radiation component is important 
at lower frequencies ($<$20 GHz), $\sim$240 GHz continuum emission 
from starbursts is usually dominated by dust thermal radiation and 
free-free emission. 
We thus estimated these two components.

For a small fraction of the observed ULIRGs, there are 
photometric data by Herschel at 250 $\mu$m, 350 $\mu$m, and 500 $\mu$m 
\citep{cle18}, which are usually dominated by dust thermal radiation.
These data, together with IRAS 60 $\mu$m and 100 $\mu$m fluxes and 
our ALMA spatially integrated continuum fluxes at 
$\sim$240 GHz ($\sim$1250 $\mu$m), are plotted in Figure 10. 
Although the aperture sizes of the Herschel and IRAS data are much 
larger, our ALMA measurements roughly agree with the extrapolation from 
the shorter-wavelength Herschel and IRAS data \citep{cle18}, 
supporting the scenario that nearby ULIRGs are energetically 
dominated by compact nuclear ($<$a few kpc) regions \citep{soi00,dia10,ima11}, 
and that our ALMA data recover the bulk of the nuclear continuum emission.
The ALMA data point in IRAS 11095$-$0238 (Figure 10c) is significantly 
higher than the extrapolation, which may indicate a significant 
contribution from additional free-free emission.
For IRAS 22491$-$1808 (Figure 10g), the ALMA data point is slightly below the 
extrapolation, which may be due to the fit because (1) the longest 
wavelength Herschel data point is also below the fit and 
(2) the maximum recoverable scale is large (i.e., ALMA missing flux 
is expected to be small) among the observed ULIRGs ($\S$3).
For the remaining ULIRGs, such Herschel photometric measurements 
are not reported, which hampers the reliable estimate of the contribution 
from the dust thermal radiation at $\sim$240 GHz.

On the other hand, the contributions from the free-free emission in 
HII-regions can be estimated for all sources, from the 
far-infrared (40--500 $\mu$m) luminosities obtained from IRAS data 
(Table 1).
The flux of the free-free emission from HII-regions (in mJy) is 
expressed as 
\begin{eqnarray}
{\rm Flux (mJy)} &=& 2.28 \times 10^{-7} \times (\frac{\rm D}{\rm Mpc})^{-2} 
\times (\frac{\rm T_e}{\rm 10^{4} K})^{0.59} \times (\frac{\nu}{\rm GHz})^{-0.1} 
\times \frac{\rm L_{FIR}}{\rm L_{\odot}} 
\end{eqnarray}
\citep{nak05}, 
where D is the luminosity distance (in Mpc), 
T$_{\rm e}$ is the electron temperature (in K), $\nu$ is the 
frequency (in GHz), and L$_{\rm FIR}$ is the far-infrared luminosity 
(in L$_{\odot}$).
In fact, an AGN can also contribute to the far-infrared luminosity, 
so that the actual free-free continuum flux from HII-regions in 
starbursts is even smaller. 
These estimates of the free-free emission are shown in Table 10 
(column 3), where T$_{\rm e}$ = 10$^{4}$ K is assumed.
In the majority of the observed ULIRGs, the observed continuum flux is 
significantly higher than the upper limit of the free-free emission 
from HII-regions in starbursts, suggesting that dust thermal radiation 
has an important contribution at $\sim$240 GHz.

No clear continuum emission was detected at the nuclear peak positions 
of IRAS 10485$-$1447 and IRAS 02411$+$0353. 
For both sources, we obtained continuum fluxes within a 1$''$ diameter 
circular aperture of $<$2 mJy at $\sim$240 GHz, 
whereas the estimated free-free continuum fluxes at 
$\sim$240 GHz from the far-infrared luminosity measured with 
IRAS's large aperture ($>$60$''$) are $<$0.5 mJy.
The faint continuum fluxes in IRAS 10485$-$1447 and IRAS 02411$+$0353, 
as measured with our ALMA data, do not pose any serious inconsistencies.

For IRAS 10485$-$1447, radio synchrotron emission is detected with fluxes 
of 6.45 mJy and 3.37 mJy at 1.4 GHz and 4.8 GHz, 
respectively \citep{baa06}.
By fitting these radio data with a power law function and extrapolating 
to our ALMA frequency, the synchrotron emission component 
at $\sim$240 GHz is estimated to be $\sim$0.4 mJy, which is again 
sufficiently small compared to the observed flux upper limit.
It is possible that the observed 1.4 GHz flux is significantly 
attenuated by free-free absorption. If we 
make a power law fit using an intrinsic 1.4 GHz flux, the extrapolated 
synchrotron emission flux at $\sim$240 GHz will be even smaller.

\section{Summary} 

We presented our ALMA Cycle 5 observational results of the complete 
sample of 26 nearby ($z <$ 0.15) ULIRGs with no obvious signatures of 
optically detectable luminous AGNs, in the HCN J=3--2 and HCO$^{+}$ J=3--2 
lines.
The sample was selected from the IRAS 1 Jy sample, based on the 
declination of $<$+20$^{\circ}$ and infrared flux above a certain threshold 
to secure sufficiently high nuclear dense molecular emission line peak flux.
The majority of the ULIRGs were observed with 
a 0$\farcs$1--0$\farcs$2 (150--530 pc) synthesized beam size; 
0$\farcs$5--0$\farcs$9 (0.9--1.7 kpc) was used for the remaining 
small fraction of ULIRGs. 
We found the following main results.

\begin{enumerate}

\item The continuum, HCN J=3--2, and HCO$^{+}$ J=3--2 emission lines were 
spatially resolved in the majority of the observed ULIRGs. 
The HCN J=3--2 and continuum emission sizes are roughly 
comparable; 
however, HCO$^{+}$ J=3--2 emission is spatially more extended 
than the HCN J=3--2 emission.
This can naturally be explained by the smaller critical density of 
HCO$^{+}$ J=3--2 than that of HCN J=3--2, where low density 
and temperature molecular gas in spatially resolved regions outside 
compact nuclear cores can collisionally excite HCO$^{+}$ to J=3 
more than HCN, and produce stronger HCO$^{+}$ J=3--2 emission there.

\item The HCN-to-HCO$^{+}$ J=3--2 flux ratios at the compact nuclear cores 
of the ULIRGs, within the beam size, are generally higher than
(a) unity, (b) those seen in less-infrared luminous 
starburst-dominated galaxies, and (c) those in spatially integrated 
areas of the same ULIRGs measured with 1--2$''$ diameter circular 
apertures. 
For ULIRGs observed with small synthesized beam sizes with 
0$\farcs$1--0$\farcs$2, we created the maps of the HCN-to-HCO$^{+}$ J=3--2 
flux ratios and investigated their spatial variation. 
In the majority of the ULIRGs, the regions of elevated HCN-to-HCO$^{+}$ 
J=3--2 flux ratios spatially coincide with nuclear cores. 
It is likely that the putative buried AGNs and/or high HCN collisional 
excitation by dense and warm molecular gas at the nuclear cores are 
responsible for the enhanced HCN emission there. 
Mechanical heating by shocks originated in spatially compact nuclear 
outflows can also contribute.

\item 
Several ULIRGs show higher HCN-to-HCO$^{+}$ J=3--2 flux ratios at 
off-nuclear local regions than the nuclear cores, which is difficult 
to explain with the above scenario.
Mechanical heating by spatially extended outflow-origin shock 
activity may be responsible.

\item 
We estimated that in the majority of the observed ULIRGs, 
dense ($>$10$^{5}$ cm$^{-3}$) molecular gas mass within the nuclear 
few kpc regions is as high as $>$10$^{9}$M$_{\odot}$, based on the 
observed HCN J=3--2 and HCO$^{+}$ J=3--2 emission line luminosities, 
suggesting that nearby ULIRG's nuclei are dense molecular gas rich.
For selected ULIRGs with available nuclear total molecular masses 
based on interferometric CO J-transition line data, dense molecular 
masses can be as high as the nuclear total molecular masses in 
most cases, suggesting that dense gas is a dominant molecular phase in 
nearby ULIRG's nuclei.

\item 
A comparison of the HCN J=3--2 and HCO$^{+}$ J=3--2 emission line 
luminosities with infrared luminosity suggests that HCN and HCO$^{+}$ 
are only sub-thermally excited at J=3--2.
Even with this sub-thermal HCN J=3--2 excitation in ULIRGs, 
their HCN-to-HCO$^{+}$ J=3--2 flux ratios are higher than starburst 
galaxies.
The ratios of HCN J=3--2 excitation temperature to molecular gas 
kinetic temperature (T$_{\rm ex}$/T$_{\rm kin}$) in the comparison 
starburst sample are suggested to be even smaller than the 
ULIRG sample. 

\item 
Signatures of the excess emission at the lower-frequency 
side of the HCO$^{+}$ J=3--2 (v=0) emission lines are recognizable 
in IRAS 10378$+$1108 and IRAS 11095$-$0238, in addition to 
three ULIRGs reported in \citet{ima16c} (IRAS 12112$+$0305, 
IRAS 20414$-$1651, and IRAS 22491$-$1808), out of the 26 observed ULIRGs. 
The excess emission can be due to contributions from
vibrationally excited (v$_{2}$=1f) HCN J=3--2 emission lines.
The HCN v$_{2}$=1f to v=0 flux ratios at J=3--2 could be as high as 
an AGN-hosting ULIRG with a clearly detected HCN v$_{2}$=1f J=3--2 emission 
line.
If the HCN vibrational excitation to v$_{2}$=1f is due to infrared 
radiative pumping, by absorbing infrared $\sim$14 $\mu$m photons 
coming from AGN-heated hot dust grains, these five ULIRGs may 
contain (sub)millimeter-detectable deeply buried AGNs. 
The low detection rate (5/26 $\sim$ 19$\%$) of the signatures of the 
HCN v$_{2}$=1f J=3--2 emission lines may be due to the combination 
of limited signal to noise ratios and large molecular line widths in ULIRGs 
that hamper the identification of faint HCN v$_{2}$=1f J=3--2 emission lines, 
by clearly separating from the nearby much brighter 
HCO$^{+}$ J=3--2 (v=0) emission lines. 
Signatures of the HCO$^{+}$ v$_{2}$=1f J=3--2 emission lines were not seen 
in any of the observed ULIRGs.

\item 
High HCN-to-HCO$^{+}$ J=3--2 flux ratios are found in (a) 
all ULIRGs with significant ($>$20$\%$) energetic contributions 
from infrared-spectroscopically identified luminous buried AGNs, 
and (b)
some fraction of ULIRGs without such infrared-identified luminous 
buried AGNs. 
Although the HCN v$_{2}$=1f J=3--2 emission lines (=another potential 
AGN indicator) were not clearly detected in the majority of these ULIRGs, 
if the observed high HCN-to-HCO$^{+}$ J=3--2 flux ratios are due to 
AGN effects (including both AGN radiation feedback and mechanical 
feedback by AGN-origin compact nuclear outflows), then our (sub)millimeter 
molecular line method (a) provides a consistent picture with the 
infrared spectroscopic method about the presence of energetically 
significant luminous buried AGNs for the former ULIRGs, and 
(b) may suggest the presence of infrared-elusive but 
(sub)millimeter-detectable extremely deeply buried luminous AGNs 
for the latter ULIRGs. 

\item 
Molecular outflows have usually been discussed based on the detection 
of broad components in the bright CO emission lines, but no such broad 
components were clearly seen in the HCN J=3--2 and HCO$^{+}$ J=3--2 emission 
lines in the observed ULIRGs.

\item 
Spatially integrated continuum fluxes at $\sim$240 GHz are significantly 
higher than the estimated free-free emission fluxes from HII-regions 
in spatially resolved (a few kpc) starbursts, suggesting that thermal 
radiation from dust contributes significantly to the continuum flux 
in band 6 ($\sim$240 GHz).

\end{enumerate}

\acknowledgments

We thank Dr. K. Saigo and F. Egusa for their supports
regarding ALMA data analysis, and the anonymous referee for 
his/her valuable comments which helped improve the clarity 
of this manuscript. 
M.I. is supported by JSPS KAKENHI Grant Number 15K05030.
This paper made use of the following ALMA data:
ADS/JAO.ALMA\#2017.1.00057.S, 2013.1.00032.S, and 2015.1.00027.S.
ALMA is a partnership of ESO (representing its member states), NSF (USA) 
and NINS (Japan), together with NRC (Canada), NSC and ASIAA
(Taiwan), and KASI (Republic of Korea), in cooperation with the Republic
of Chile. The Joint ALMA Observatory is operated by ESO, AUI/NRAO, and
NAOJ. 
Data analysis was in part carried out on the open use data analysis
computer system at the Astronomy Data Center, ADC, of the National
Astronomical Observatory of Japan. 
This research has made use of NASA's Astrophysics Data System and the
NASA/IPAC Extragalactic Database (NED) which is operated by the Jet
Propulsion Laboratory, California Institute of Technology, under
contract with the National Aeronautics and Space Administration. 

%

\vspace{5mm}
\facilities{ALMA}






\appendix

\section{Comments on Continuum Emission for Selected Galaxies}

For IRAS 04103$-$2838, the HCN J=3--2 and HCO$^{+}$ J=3--2 emission line 
peaks are significantly displaced from the continuum peak 
(04 12 19.445, $-$28 30 25.00)ICRS (Fig. 1c).
The spectrum in Figure 2c was extracted at the peak position of the 
HCN J=3--2 and HCO$^{+}$ J=3--2 emission lines.

IRAS 09039$+$0503 is a single-nucleus ULIRG in seeing-sized 
optical and near-infrared K-band (2.2 $\mu$m) images \citep{kim02}. 
However, our high-spatial-resolution ALMA molecular line and continuum 
data reveal double nuclear morphology (SW and NE), with $\sim$0$\farcs$5 
separation in Figures 1d, 3g, and 3h. 
The spectra shown in Figures 2d and 2e were extracted for both nuclei 
separately. 

IRAS 11095$-$0238 is also a single-nucleus ULIRG in the optical and 
near-infrared \citep{kim02}; however, our ALMA data show a faint secondary 
emission component in the $\sim$0$\farcs$5 southwestern part of the main 
nucleus in the continuum and HCO$^{+}$ J=3--2 line maps 
(Figs. 1f and 3l).
The HCO$^{+}$ J=3--2 emission line shows a significant feature in-between 
these two continuum-emitting regions in IRAS 11095$-$0238 (Fig. 1f).

For IRAS 01166$-$0844, continuum and molecular line emission were 
detected at the SE nucleus, but not at the NW nucleus \citep{kim02}.

For IRAS 10485$-$1447, neither the continuum nor HCN J=3--2 and HCO$^{+}$ 
J=3--2 emission lines were detected.
No detectable CO J=1--0 emission line was seen in pre-ALMA 
interferometric observations \citep{tru01}.
Thus, IRAS 10485$-$1447 may be deficient in molecular gas.

\section{Comments on Selected Galaxies}

In Figure 11, the best Gaussian fits for the detected molecular emission 
lines are overplotted on the actual data.

\section{Intensity-weighted Mean Velocity and Intensity-weighted Velocity Dispersion Maps}

Figures 12 and 13 display intensity-weighted mean velocity (moment 1) and 
intensity-weighted velocity dispersion (moment 2) maps of HCN J=3--2 
and HCO$^{+}$ J=3--2 emission lines of selected ULIRGs, for which we can 
obtain meaningful dynamical information of these dense gas tracers 
(mostly $>$10$\sigma$ detection in their integrated intensity [moment 0] 
maps).
For most of the ULIRGs with spatially resolved molecular line emission, 
a rotation pattern with blueshifted and redshifted components are visible 
(Fig. 12).

\section{A Continuum-emitting Source within the IRAS 13509$+$0442 Field}

A bright continuum-emitting source was detected at the $\sim$10$''$ 
northern side of IRAS 13509$+$0442 at $\sim$235 GHz and $\sim$310 GHz 
\citep{ima16c,ima18b}.
This source is considered a distant submillimeter galaxy (SMG) 
candidate 
\citep{ima16c,ima18b}.
We made a clean map with a wider field of view 
(1024 $\times$ 1024 pixels of 0$\farcs$02 pixel$^{-1}$) to investigate this 
source in our ALMA Cycle 5 high-spatial-resolution data.

A continuum map created using all channels (because there are no 
obvious 
emission lines in this source) is displayed in Figure 14 (left).
The continuum emission is clearly detected with the peak flux of 0.84 
(mJy beam$^{-1}$) (16$\sigma$) and is spatially resolved with the beam size 
of 220 (mas) $\times$ 160 (mas) (position angle = 29$^{\circ}$ east of north). 
The estimated intrinsic deconvolved emission size is 
421$\pm$43 (mas) $\times$ 323$\pm$35 (mas) (position angle is 
56$\pm$23$^{\circ}$ east of north). 
The spatially integrated continuum flux within a 1$''$ diameter circular 
aperture is 3.6 (mJy).
Figure 14 (right) shows a spectrum without continuum subtraction 
at the peak position within the beam size.
No significant emission line is recognizable.

\clearpage


\begin{deluxetable}{lcccrrrrcccc}
\tabletypesize{\scriptsize}
\tablecaption{Basic Properties of the Observed Ultraluminous Infrared Galaxies 
\label{tbl-1}} 
\tablewidth{0pt}
\tablehead{
\colhead{Object} & \colhead{Redshift $\tablenotemark{A}$} & 
\colhead{d$_{\rm L}$} & \colhead{Scale} & 
\colhead{f$_{\rm 12}$} & 
\colhead{f$_{\rm 25}$} & 
\colhead{f$_{\rm 60}$} & 
\colhead{f$_{\rm 100}$} & 
\colhead{log L$_{\rm IR}$} & 
\colhead{log L$_{\rm FIR}$} &
\colhead{Optical} & \colhead{IR-AGN} 
\\
\colhead{} & \colhead{} & \colhead{[Mpc]} & \colhead{[kpc/$''$]}  
& \colhead{[Jy]}
& \colhead{[Jy]} & \colhead{[Jy]} & \colhead{[Jy]}  &
\colhead{[L$_{\odot}$]} & \colhead{[L$_{\odot}$]} & \colhead{Class} 
& \colhead{Fraction [\%]} \\  
\colhead{(1)} & \colhead{(2)} & \colhead{(3)} & \colhead{(4)} & 
\colhead{(5)} & \colhead{(6)} & \colhead{(7)} & \colhead{(8)} & 
\colhead{(9)} & \colhead{(10)} & \colhead{(11)} & \colhead{(12)} 
}
\startdata
IRAS 00188$-$0856 & 0.1285 (0.128) & 596 & 2.3 & $<$0.12 & 0.37 & 2.59 & 3.40 
& 12.4 & 12.3 & LINER & 35$\pm$4 \\    
IRAS 03250$+$1606 & 0.1286 (0.129) & 596 & 2.3 & $<$0.10 & $<$0.15 & 1.38 
& 1.77 & 12.1 & 12.0 & LINER & $<$0.2 \\   
IRAS 04103$-$2838 & 0.118 & 543 & 2.1 & 0.08 & 0.54 & 1.82 & 1.71 & 12.2 &
12.0 & LINER & 5.4$\pm$0.8 \\ 
IRAS 09039$+$0503 & 0.1257 $\tablenotemark{B}$ & 578 & 2.2 & 0.07 & 0.12 
& 1.48 & 2.06 & 12.1 & 12.0 & LINER & 5.8$^{+1.1}_{-0.9}$ \\  
IRAS 10378$+$1108 & 0.1365 (0.136) & 636 & 2.4 & $<$0.11 & 0.24 & 2.28 
& 1.82 & 12.3 & 12.2 & LINER & 14$\pm$2 \\   
IRAS 10485$-$1447 & 0.133 & 619 & 2.3 & $<$0.11 & 0.25 & 1.73 & 1.66 
& 12.2 & 12.1 & LINER & 5.7$\pm$1.0 \\   
IRAS 11095$-$0238 & 0.1066 $\tablenotemark{C}$ & 484 & 1.9 & 0.06 & 0.42 
& 3.25 & 2.53 & 12.3 & 12.1 & LINER & 60$\pm$4 \\   
IRAS 13335$-$2612 & 0.125 $\tablenotemark{D}$ & 578 & 2.2 & $<$0.13 
& $<$0.14 & 1.40 & 2.10 & 12.1 & 12.0 & LINER & $<$0.03 \\
\enddata
\end{deluxetable}

\begin{deluxetable}{llccc|ccc}
\tabletypesize{\scriptsize}
\tablecaption{Log of Our ALMA Cycle 5 Observations \label{tbl-2}} 
\tablewidth{0pt}
\tablehead{
\colhead{Object} & \colhead{Date} & \colhead{Antenna} & 
\colhead{Baseline} & \colhead{Integration} & \multicolumn{3}{c}{Calibrator} \\ 
\colhead{} & \colhead{[UT]} & \colhead{Number} & \colhead{[m]} &
\colhead{[min]} & \colhead{Bandpass} & \colhead{Flux} & \colhead{Phase}  \\
\colhead{(1)} & \colhead{(2)} & \colhead{(3)} & \colhead{(4)} &
\colhead{(5)} & \colhead{(6)} & \colhead{(7)}  & \colhead{(8)} 
}
\startdata 
IRAS 00188$-$0856 & 2017 December 17 & 45 & 15--2517 & 36 & J0006$-$0623
& J0006$-$0623 & J0017$-$0512 \\
IRAS 03250$+$1606 & 2017 December 10 & 46 & 15--3321 & 39 & J0238$+$1636 &
J0238$+$1636 & J0309$+$1029 \\
IRAS 04103$-$2838 & 2017 December 9 & 44 & 15--3083 & 15 & J0522$-$3627
& J0522$-$3627 & J0348$-$2749 \\
IRAS 09039$+$0503 & 2017 December 23 & 48 & 15--2517 & 22 & J0854$+$2006 & 
J0854$+$2006 & J0914$+$0245 \\
IRAS 10378$+$1108 & 2017 December 17 & 45 & 15--2517 & 28 & J1058$+$0133 &
J1058$+$0133 & J1025$+$1253 \\
IRAS 10485$-$1447 & 2017 December 11 & 44 & 15--3321 & 18 & J1037$-$2934
& J1037$-$2934 & J1048$-$1909 \\
IRAS 11095$-$0238 & 2017 December 11 & 44 & 15--3321 & 12 & J1229$+$0203
& J1229$+$0203 & J1058$+$0133 \\
IRAS 13335$-$2612 & 2017 December 23 & 42 & 15--2517 & 21 & J1337$-$1257
& J1337$-$1257 & J1351$-$2912 \\
\enddata

\tablenotetext{A}{Observed in ALMA Cycle 2 or 3.}

\tablecomments{ 
Col.(1): Object name. 
Col.(2): Observation date in UT. 
Col.(3): Number of antennas used for observations. 
Col.(4): Baseline length in meters. Minimum and maximum baseline lengths are 
shown.  
Col.(5): Net on source integration time in minutes.
Cols.(6), (7), and (8): Bandpass, flux, and phase calibrator for the 
target source, respectively.
}

\end{deluxetable}

\begin{deluxetable}{llrlcl}
\tabletypesize{\scriptsize}
\tablecaption{Continuum Emission \label{tbl-3}}
\tablewidth{0pt}
\tablehead{
\colhead{Object} & \colhead{Frequency} & \colhead{Peak flux} & 
\colhead{Peak coordinate} & \colhead{rms} & \colhead{Synthesized beam} \\
\colhead{} & \colhead{[GHz]} & \colhead{[mJy/beam]} & 
\colhead{(RA,DEC)ICRS} & \colhead{[mJy/beam]} & 
\colhead{[$''$ $\times$ $''$] ($^{\circ}$)} \\  
\colhead{(1)} & \colhead{(2)} & \colhead{(3)}  & \colhead{(4)}  &
\colhead{(5)} & \colhead{(6)} 
}
\startdata 
IRAS 00188$-$0856 & 234.7--239.7 & 1.4 (35$\sigma$) & (00 21 26.514,
$-$08 39 26.01) & 0.040 & 0.18$\times$0.13 ($-$79$^{\circ}$) \\
IRAS 03250$+$1606 & 234.6--239.6 & 0.41 (14$\sigma$) & (03 27 49.812,
$+$16 16 59.31) & 0.029 & 0.13$\times$0.10 ($-$28$^{\circ}$) \\
IRAS 04103$-$2838 N & 236.9--241.8 & 0.21 (4.4$\sigma$) & (04 12 19.445,
$-$28 30 25.00) & 0.047 & 0.11$\times$0.10 (73$^{\circ}$) \\
IRAS 04103$-$2838 S & 236.9--241.8 & 0.18 (3.9$\sigma$) & (04 12 19.445,
$-$28 30 25.14) & 0.047 & 0.11$\times$0.10 (73$^{\circ}$) \\
IRAS 04103$-$2838 NE & 236.9--241.8 & 0.18 (3.9$\sigma$) & (04 12
19.440, $-$28 30 24.90) & 0.047 & 0.11$\times$0.10 (73$^{\circ}$) \\
IRAS 09039$+$0503 SW & 235.4--240.3 & 0.83 (19$\sigma)$ & (09 06 34.027,
$+$04 51 27.43) & 0.044 & 0.18$\times$0.14 (77$^{\circ}$) \\
IRAS 09039$+$0503 NE & 235.4--240.3 & 0.46 (10$\sigma$) & (09 06 34.045,
$+$04 51 27.87) & 0.044 & 0.18$\times$0.14 (77$^{\circ}$) \\
IRAS 10378$+$1108 & 233.0--238.0 & 1.5 (33$\sigma$) & (10 40 29.170, $+$10
53 18.15) & 0.045 & 0.17$\times$0.15 ($-$81$^{\circ}$) \\
\enddata

\tablenotetext{A}{No emission with $>$3$\sigma$ is detected within
0$\farcs$2 around the near-infrared K-band (2.2 $\mu$m) peak position of 
IRAS 10485$-$1447 E at (10 51 03.67, $-$15 03 19.8) in J2000 \citep{kim02}.}

\tablenotetext{B}{HCO$^{+}$ J=3--2 emission line peak.}

\tablenotetext{C}{Taken from \citet{ima16c}.}

\tablecomments{Col.(1): Object name.
Col.(2): Frequency range (in GHz) used for continuum extraction.
Frequencies of obvious emission lines and absorption features are removed.
Col.(3): Flux (in mJy beam$^{-1}$) at the emission peak.
Value at the highest flux pixel (0$\farcs$02--0$\farcs$04 pixel$^{-1}$)
is extracted. 
The detection significance relative to the root mean square (rms) noise is shown in parentheses. 
Possible systematic uncertainties, coming from the absolute flux
calibration ambiguity in individual ALMA observation and 
choice of frequency range for the continuum level determination, 
are not included. 
Col.(4): Coordinate of the continuum emission peak in ICRS.
Col.(5): The rms noise level (1$\sigma$) (in mJy beam$^{-1}$), derived
from the standard deviation of sky signals in each continuum map.
Col.(6): Synthesized beam (in arcsec $\times$ arcsec) and position angle
(in degrees). 
The position angle is 0$^{\circ}$ along the north--south direction
and increases in the counterclockwise direction.}

\end{deluxetable}

\begin{deluxetable}{l|lcl|cccc}
\tabletypesize{\scriptsize}
\tablecaption{Observed Properties of the HCN J=3--2 Emission Line 
at the Nuclear Peak Position 
\label{tbl-4}} 
\tablewidth{0pt}
\tablehead{
\colhead{Object} & \multicolumn{3}{c}{Integrated intensity (moment 0) map} & 
\multicolumn{4}{c}{Gaussian fit} \\  
\colhead{} & \colhead{Peak} &
\colhead{rms} & \colhead{Beam} & \colhead{Velocity}
& \colhead{Peak} & \colhead{FWHM} & \colhead{Flux} \\ 
\colhead{} & \multicolumn{2}{c}{[Jy beam$^{-1}$ km s$^{-1}$]} & 
\colhead{[$''$ $\times$ $''$] ($^{\circ}$)} &
\colhead{[km s$^{-1}$]} & \colhead{[mJy]} & \colhead{[km s$^{-1}$]} &
\colhead{[Jy km s$^{-1}$]} \\  
\colhead{(1)} & \colhead{(2)} & \colhead{(3)} & \colhead{(4)} & 
\colhead{(5)} & \colhead{(6)} & \colhead{(7)} & \colhead{(8)} 
}
\startdata 
IRAS 00188$-$0856 & 1.6 (25$\sigma$) & 0.065 & 0.18$\times$0.13
($-$79$^{\circ}$) & 38544$\pm$6 & 5.5$\pm$0.2 & 355$\pm$13 & 1.9$\pm$0.1
\\ 
IRAS 03250$+$1606 & 0.56 (11$\sigma$) & 0.049 & 0.13$\times$0.10
($-$28$^{\circ}$) & 38572$\pm$26 & 1.3$\pm$0.1 & 526$\pm$60 &
0.63$\pm$0.09 \\    
IRAS 04103$-$2838 \tablenotemark{A} & 0.44 (7.3$\sigma)$ & 0.060 &
0.11$\times$0.10 (73$^{\circ}$) & 35339$\pm$26 & 1.4$\pm$0.2 &
407$\pm$69 & 0.53$\pm$0.12 \\
IRAS 09039$+$0503 SW & 0.86 (13$\sigma$) & 0.067 & 0.19$\times$0.14
(78$^{\circ}$) & 37685$\pm$16 & 2.4$\pm$0.2 & 419$\pm$41 & 0.96$\pm$0.12 \\ 
IRAS 09039$+$0503 NE & 0.67 (10$\sigma$) & 0.067 & 0.19$\times$0.14
(78$^{\circ}$) & 37528$\pm$12 & 2.6$\pm$0.3 & 236$\pm$45 & 0.57$\pm$0.12 \\ 
IRAS 10378$+$1108 & 1.9 (26$\sigma$) & 0.074 & 0.17$\times$0.15
($-$83$^{\circ}$) & 40946$\pm$6 & 5.8$\pm$0.2 & 411$\pm$12 & 2.2$\pm$0.1 \\
IRAS 10485$-$1447 & $<$0.24 ($<$3$\sigma$) \tablenotemark{B} & 0.078 &
0.13$\times$0.10 ($-$78$^{\circ}$) & --- & --- & --- & --- \\
IRAS 11095$-$0238 NE & 1.7 (14$\sigma$) & 0.11 & 0.13$\times$0.12
($-$60$^{\circ}$) & 31956$\pm$10 & 4.1$\pm$0.2 & 437$\pm$29 & 1.7$\pm$0.1 \\
\enddata

\tablenotetext{A}{HCN J=3--2 emission line peak 
at (04 12 19.448, $-$28 30 24.92)ICRS.}

\tablenotetext{B}{No emission with $>$3$\sigma$ is detected within
0$\farcs$2 around the near-infrared K-band peak position 
at (10 51 03.67, $-$15 03 19.8) in J2000 \citep{kim02}.
} 

\tablenotetext{C}{Taken from \citet{ima16c}.} 

\tablecomments{Col.(1): Object name.
Col.(2): Integrated intensity of the HCN J=3--2 emission line 
($\nu_{\rm rest}$=265.886 GHz) (in Jy beam$^{-1}$ km s$^{-1}$) at the 
emission peak. 
Detection significance relative to the rms noise (1$\sigma$) in the 
moment 0 map is shown in parentheses. 
Possible systematic uncertainty is not included. 
Col.(3): The rms noise (1$\sigma$) level in the moment 0 map (in 
Jy beam$^{-1}$ km s$^{-1}$), derived from the standard deviation 
of sky signals in each moment 0 map. 
Col.(4): Beam size (in arcsec $\times$ arcsec) and position angle (in
degrees). Position angle is 0$^{\circ}$ along the north-south direction, 
and increases counterclockwise. 
Cols.(5)--(8): Gaussian fit of the HCN J=3--2 emission line in the 
spectrum at the continuum peak position, within the beam size. 
For faint emission lines, we fixed some parameters to the best fit 
or fiducial values.
Col.(5): Optical local standard of rest (LSR) velocity (v$_{\rm opt}$) 
of emission line peak (in km s$^{-1}$). 
Col.(6): Peak flux (in mJy). 
Col.(7): Observed full width at half maximum (FWHM) (in km s$^{-1}$).
Col.(8): Flux (in Jy km s$^{-1}$). 
For the flux estimate, the observed FWHM (in 
km s$^{-1}$) in column 7 is divided by ($1+z$) to obtain the
intrinsic FWHM (in km s$^{-1}$).}

\end{deluxetable}

\clearpage

\begin{deluxetable}{l|lcl|cccc}
\tabletypesize{\scriptsize}
\tablecaption{Observed Properties of the HCO$^{+}$ J=3--2 Emission Line 
at the Nuclear Peak Position \label{tbl-5}} 
\tablewidth{0pt}
\tablehead{
\colhead{Object} & \multicolumn{3}{c}{Integrated intensity (moment 0) map} & 
\multicolumn{4}{c}{Gaussian fit} \\  
\colhead{} & \colhead{Peak} &
\colhead{rms} & \colhead{Beam} & \colhead{Velocity}
& \colhead{Peak} & \colhead{FWHM} & \colhead{Flux} \\ 
\colhead{} & \multicolumn{2}{c}{[Jy beam$^{-1}$ km s$^{-1}$]} & 
\colhead{[$''$ $\times$ $''$] ($^{\circ}$)} &
\colhead{[km s$^{-1}$]} & \colhead{[mJy]} & \colhead{[km s$^{-1}$]} &
\colhead{[Jy km s$^{-1}$]} \\  
\colhead{(1)} & \colhead{(2)} & \colhead{(3)} & \colhead{(4)} & 
\colhead{(5)} & \colhead{(6)} & \colhead{(7)} & \colhead{(8)} 
}
\startdata 
IRAS 00188$-$0856 & 0.86 (15$\sigma$) & 0.056 & 0.18$\times$0.13 
($-$79$^{\circ}$) & 38538$\pm$10 & 3.1$\pm$0.2 & 345$\pm$23 &
1.0$\pm$0.1 \\
IRAS 03250$+$1606 & 0.40 (8.5$\sigma$) & 0.047 & 0.13$\times$0.10
($-$28$^{\circ}$) & 38558$\pm$29 & 0.98$\pm$0.11 & 535$\pm$59 &
0.49$\pm$0.08 \\  
IRAS 04103$-$2838 \tablenotemark{A} & 0.40 (6.4$\sigma$) & 0.062 & 
0.11$\times$0.10 (74$^{\circ}$) & 35346$\pm$18 & 1.6$\pm$0.2 &
300$\pm$48 & 0.46$\pm$0.09 \\
IRAS 09039$+$0503 SW & 0.46 (7.4$\sigma$) & 0.062 & 0.18$\times$0.14
(80$^{\circ}$) & 37721$\pm$23 & 1.4$\pm$0.2 & 413$\pm$58 & 0.56$\pm$0.10
\\ 
IRAS 09039$+$0503 NE & 0.55 (8.9$\sigma$) & 0.062 & 0.18$\times$0.14
(80$^{\circ}$) & 37555$\pm$12 & 2.1$\pm$0.2 & 257$\pm$32 & 0.51$\pm$0.08 \\ 
IRAS 10378$+$1108 & 1.8 (24$\sigma$) & 0.075 & 0.17$\times$0.15
($-$81$^{\circ}$) & 40948$\pm$7 & 5.3$\pm$0.2 & 399$\pm$17 & 2.0$\pm$0.1
\\
IRAS 10485$-$1447 & $<$0.23 ($<$3$\sigma$) \tablenotemark{B} & 0.076 &
0.13$\times$0.10 ($-$76$^{\circ}$) & --- & --- & --- & --- \\
IRAS 11095$-$0238 NE & 0.89 (10$\sigma$) & 0.087 & 0.13$\times$0.12
($-$63$^{\circ}$) & 32017$\pm$26 & 2.1$\pm$0.2 & 467$\pm$68 &
0.92$\pm$0.16  \\
\enddata

\tablenotetext{A}{HCO$^{+}$ J=3--2 emission line peak 
at (04 12 19.448, $-$28 30 24.92)ICRS.}

\tablenotetext{B}{No emission with $>$3$\sigma$ is detected within
0$\farcs$2 around the near-infrared K-band peak position 
at (10 51 03.67, $-$15 03 19.8) in J2000 \citep{kim02}.} 

\tablenotetext{C}{Taken from \citet{ima16c}.} 

\tablecomments{ 
Col.(1): Object name.
Col.(2): Integrated intensity of the HCO$^{+}$ J=3--2 emission line 
($\nu_{\rm rest}$=267.558 GHz) (in Jy beam$^{-1}$ km s$^{-1}$) at the 
emission peak. 
Detection significance relative to the rms noise (1$\sigma$) in the 
moment 0 map is shown in parentheses. 
Possible systematic uncertainty is not included. 
Col.(3): The rms noise (1$\sigma$) level in the moment 0 map (in 
Jy beam$^{-1}$ km s$^{-1}$), derived from the standard deviation 
of sky signals in each moment 0 map. 
Col.(4): Beam size (in arcsec $\times$ arcsec) and position angle 
(in degree). Position angle is 0$^{\circ}$ along the north-south direction, 
and increases counterclockwise. 
Cols.(5)--(8): Gaussian fit of the HCO$^{+}$ J=3--2 emission line 
in the spectrum at the continuum peak position, within the beam size. 
For faint emission lines, we fixed some parameters to the best fit 
or fiducial values.
Col.(5): Optical LSR velocity (v$_{\rm opt}$) of emission line peak 
(in km s$^{-1}$).  
Col.(6): Peak flux (in mJy). 
Col.(7): Observed FWHM (in km s$^{-1}$).
Col.(8): Flux (in Jy km s$^{-1}$). 
}

\end{deluxetable}

\clearpage

\begin{deluxetable}{lcccc}
\tabletypesize{\scriptsize}
\tablecaption{Emission Size after Deconvolution \label{tbl-6}} 
\tablewidth{0pt}
\tablehead{
\colhead{Object} & \colhead{continuum}  & \colhead{HCN J=3--2}  &
\colhead{HCO$^{+}$ J=3--2}  & \colhead{Beam size} \\
\colhead{} & \colhead{[mas $\times$ mas ($^{\circ}$)]} & 
\colhead{[mas $\times$ mas ($^{\circ}$)]} & 
\colhead{[mas $\times$ mas ($^{\circ}$)]} & 
\colhead{[mas $\times$ mas ($^{\circ}$)]}  \\ 
\colhead{(1)} & \colhead{(2)} & \colhead{(3)} & \colhead{(4)} & \colhead{(5)} 
}
\startdata
IRAS 00188$-$0856 & 248$\pm$19 $\times$ 150$\pm$11 (93$\pm$6) & 
332$\pm$29 $\times$  177$\pm$16 (91$\pm$5) & 
352$\pm$38 $\times$ 218$\pm$24 (87$\pm$10) & 180 $\times$ 130 ($-$79) \\
IRAS 03250$+$1606 & 314$\pm$40 $\times$ 284$\pm$39 (4$\pm$52) & 
226$\pm$32 $\times$ 186$\pm$28 (136$\pm$76) & 
440$\pm$86 $\times$ 254$\pm$53 (2$\pm$16) & 130 $\times$ 100 ($-$28) \\
IRAS 04103$-$2838 & 380$\pm$112 $\times$ 105$\pm$54 (161$\pm$7) & 
165$\pm$52 $\times$ 120$\pm$50 (52$\pm$40) &
284$\pm$71 $\times$ 123$\pm$38 (88$\pm$11) & 110 $\times$100 (73) \\
IRAS 09039$+$0503 SW & 254$\pm$54 $\times$ 156$\pm$53 (29$\pm$29) & 
401$\pm$85 $\times$ 305$\pm$71 (57$\pm$45) & 
705$\pm$117 $\times$ 467$\pm$83 (17$\pm$18) & 180 $\times$ 140 (77) \\
IRAS 09039$+$0503 NE & 587$\pm$143 $\times$ 398$\pm$104 (40$\pm$29) & 
387$\pm$86 $\times$ 311$\pm$77 (52$\pm$41) & 
541$\pm$90 $\times$ 400$\pm$71 (42$\pm$55) & 180 $\times$140 (77) \\
IRAS 10378$+$1108 & 137$\pm$22 $\times$ 130$\pm$25 (49$\pm$84) & 
220$\pm$22 $\times$ 162$\pm$22 (32$\pm$17) & 
263$\pm$24 $\times$ 163$\pm$21 (42$\pm$9) & 170 $\times$150 ($-$81) \\
IRAS 11095$-$0238 NW & 208$\pm$43 $\times$ 159$\pm$35 (2$\pm$33) & 
224$\pm$39 $\times$ 165$\pm$33 (12$\pm$25) & 
495$\pm$86 $\times$ 205$\pm$41 (52$\pm$7) & 130 $\times$ 120 ($-$83) \\
IRAS 13335$-$2612 N & 137$\pm$17 $\times$ 44$\pm$18 (89$\pm$6) & 
278$\pm$56 $\times$ 178$\pm$46 (58$\pm$31) & 
242$\pm$60 $\times$ 224$\pm$64 (69$\pm$83) & 160 $\times$ 120 ($-$87) \\
\enddata

\tablenotetext{A}{Taken from \citet{ima16c}.}

\tablecomments{Col.(1): Object name. 
Cols.(2),(3),(4): Intrinsic emission size (in mas) along the major and 
minor axis, after deconvolution, using the CASA task ``imfit'', for the 
continuum, HCN J=3--2, and HCO$^{+}$ J=3--2, respectively. 
The position angle (in degrees) is shown in parentheses.
Col.(5): Synthesized beam size for the continuum data shown for comparison, 
because it is difficult to constrain the intrinsic emission 
size much smaller than the beam size.
The size (in mas) along the major and minor axis, and the position angle 
(in degrees) (in parentheses) are shown.}

\end{deluxetable}

\clearpage

\begin{deluxetable}{l|cccc|cccc}
\rotate
\tabletypesize{\scriptsize}
\tablecaption{Gaussian-fit Fluxes of HCN J=3--2 and 
HCO$^{+}$ J=3--2 Emission Lines in Spatially Integrated Spectra \label{tbl-7}} 
\tablewidth{0pt}
\tablehead{
\colhead{Object} & \multicolumn{4}{c}{Gaussian fit (HCN J=3--2)}
& \multicolumn{4}{c}{Gaussian fit (HCO$^{+}$ J=3--2)} \\  
\colhead{} & \colhead{Velocity} & \colhead{Peak} & \colhead{FWHM} & 
\colhead{Flux} & \colhead{Velocity} & \colhead{Peak} & \colhead{FWHM} & 
\colhead{Flux} \\ 
\colhead{} & \colhead{[km s$^{-1}$]} & \colhead{[mJy]} & 
\colhead{[km s$^{-1}$]} & \colhead{[Jy km s$^{-1}$]} & 
\colhead{[km s$^{-1}$]} & \colhead{[mJy]} & \colhead{[km s$^{-1}$]} 
& \colhead{[Jy km s$^{-1}$]} \\  
\colhead{(1)} & \colhead{(2)} & \colhead{(3)} & \colhead{(4)} & 
\colhead{(5)} & \colhead{(6)} & \colhead{(7)} & \colhead{(8)} & 
\colhead{(9)} 
}
\startdata 
IRAS 00188$-$0856 & 38548$\pm$6 & 19$\pm$1 & 339$\pm$14 & 6.1$\pm$0.3
& 38514$\pm$10 & 12$\pm$1 & 319$\pm$20 & 3.5$\pm$0.3 \\
IRAS 03250$+$1606 & 38620$\pm$28 & 5.8$\pm$0.6 & 540$\pm$54 & 2.9$\pm$0.4 &
38652$\pm$26 & 6.8$\pm$0.6 & 520$\pm$58 & 3.3$\pm$0.5 \\
IRAS 04103$-$2838 & 35327$\pm$86 & 4.9$\pm$1.9 & 382 (fix) & 1.8$\pm$0.7
& 35267$\pm$57 & 7.0$\pm$1.8 & 315 (fix) & 2.1$\pm$0.5 \\
IRAS 09039$+$0503 (1$\farcs$5) & 37583$\pm$12 & 19$\pm$1 & 376$\pm$29 & 
6.8$\pm$0.7 & 37569$\pm$11 & 19$\pm$1 & 384$\pm$26 & 7.0$\pm$1.2 \\ 
IRAS 10378$+$1108 & 40911$\pm$9 & 16$\pm$1 & 384$\pm$24 & 5.8$\pm$0.5 &
40895$\pm$9 & 16$\pm$1 & 378$\pm$23 & 5.7$\pm$0.4 \\
IRAS 11095$-$0238 (1$\farcs$5) & 31949$\pm$40 & 14$\pm$2 & 458$\pm$114 & 
6.1$\pm$1.9 & 31903$\pm$31 & 14$\pm$2 & 397$\pm$86 & 5.5$\pm$1.5 \\
IRAS 13335$-$2612 N & 37486$\pm$52 & 4.3$\pm$0.6 & 785$\pm$136 & 3.2$\pm$0.7 
& 37471$\pm$64 & 3.3$\pm$0.7 & 646$\pm$122 & 2.0$\pm$0.6 \\ 
IRAS 13335$-$2612 S & 37718$\pm$62 & 2.8$\pm$1.0 & 338 (fix) & 0.89$\pm$0.33
& 37681$\pm$79 & 1.9$\pm$0.5 & 460 (fix) & 0.81$\pm$0.22 \\
\enddata

\tablecomments{ 
Col.(1): Object name.
Cols.(2)--(5): Gaussian fit of the HCN J=3--2 emission line 
($\nu_{\rm rest}$=265.886 GHz) (in Jy km s$^{-1}$) in the spectrum 
within a 1--2$''$ diameter circular aperture around the continuum 
peak position.
Col.(2): Optical LSR velocity (v$_{\rm opt}$) of the emission line peak 
(in km s$^{-1}$).  
Col.(3): Peak flux (in mJy). 
Col.(4): Observed FWHM (in km s$^{-1}$).
Col.(5): Flux (in Jy km s$^{-1}$). 
Cols.(6)--(9): Gaussian fit of the HCO$^{+}$ J=3--2 emission line 
($\nu_{\rm rest}$=267.558 GHz) (in Jy km s$^{-1}$) in the spectrum 
within a 1--2$''$ diameter circular aperture around the continuum 
peak position.
Col.(6): Optical LSR velocity (v$_{\rm opt}$) of the emission line peak 
(in km s$^{-1}$).  
Col.(7): Peak flux (in mJy). 
Col.(8): Observed FWHM (in km s$^{-1}$).
Col.(9): Flux (in Jy km s$^{-1}$). 
}
\end{deluxetable}

\clearpage

\begin{deluxetable}{l|cc}
\tabletypesize{\scriptsize}
\tablecaption{HCN-to-HCO$^{+}$ Flux Ratio at J=3--2 \label{tbl-8}} 
\tablewidth{0pt}
\tablehead{
\colhead{Object} & \colhead{Peak (beam size)} & \colhead{Spatially integrated} \\
\colhead{(1)} & \colhead{(2)} & \colhead{(3)} 
}
\startdata 
IRAS 00188$-$0856 & 1.9$\pm$0.2 & 1.7$\pm$0.2 \\
IRAS 03250$+$1606 & 1.3$\pm$0.3 &  0.9$\pm$0.2 \\
IRAS 04103$-$2838 & 1.2$\pm$0.3 & 0.8$\pm$0.4 \\
IRAS 09039$+$0503 SW & 1.7$\pm$0.4 & 1.0$\pm$0.2 \tablenotemark{A} \\
IRAS 09039$+$0503 NE & 1.1$\pm$0.3 &  1.0$\pm$0.2 \tablenotemark{A} \\
IRAS 10378$+$1108 & 1.1$\pm$0.1 & 1.0$\pm$0.1 \\
IRAS 11095$-$0238 NE & 1.9$\pm$0.4 & 1.1$\pm$0.5 \tablenotemark{A} \\
IRAS 11095$-$0238 SW & 0.6$\pm$0.5 & 1.1$\pm$0.5 \tablenotemark{A} \\
\enddata

\tablenotetext{A}{A 1$\farcs$5 diameter circular aperture measurement. 
Fluxes from both nuclei are included.}

\tablenotetext{B}{
A 1$\farcs$5 diameter circular aperture measurement. 
}

\tablenotetext{C}{
Adopted from \citet{ima16c}.
}

\tablenotetext{D}{
A 2$''$ diameter circular aperture measurement.
}

\tablecomments{Col.(1): Object name.
Col.(2): The ratio of HCN J=3--2 to HCO$^{+}$ J=3--2 flux (in Jy km s$^{-1}$) 
at the nuclear peak position within the beam size. 
Col.(3): The ratio of HCN J=3--2 to HCO$^{+}$ J=3--2 flux (in Jy km s$^{-1}$) 
within a 1$''$ diameter circular aperture (unless otherwise noted).
}

\end{deluxetable}

\clearpage

\begin{deluxetable}{lcccc}
\tabletypesize{\scriptsize}
\tablecaption{Luminosity of the HCN and HCO$^{+}$ J=3--2 
Emission Lines \label{tbl-9}} 
\tablewidth{0pt}
\tablehead{
\colhead{Object} & \colhead{HCN J=3--2} & \colhead{HCO$^{+}$ J=3--2} &
\colhead{HCN J=3--2} & \colhead{HCO$^{+}$ J=3--2} \\
\colhead{} & \colhead{10$^{5}$ [L$_{\odot}$]} & \colhead{10$^{5}$ [L$_{\odot}$]} & 
\colhead{10$^{7}$ [K km s$^{-1}$ pc$^{2}$]} & \colhead{10$^{7}$ [K km s$^{-1}$ pc$^{2}$]} \\       
\colhead{(1)} & \colhead{(2)} & \colhead{(3)} & \colhead{(4)} & \colhead{(5)} 
}
\startdata
IRAS 00188$-$0856 & 5.3$\pm$0.3 & 3.0$\pm$0.3 & 88$\pm$5 & 50$\pm$4 \\
IRAS 03250$+$1606 & 2.6$\pm$0.4 & 2.9$\pm$0.4 & 42$\pm$6 & 48$\pm$7 \\
IRAS 04103$-$2838 & 1.3$\pm$0.5 & 1.5$\pm$0.4 & 22$\pm$8 & 25$\pm$6 \\
IRAS 09039$+$0503 (1$\farcs$5) & 5.6$\pm$0.6 & 5.8$\pm$1.0 & 93$\pm$9 & 
94$\pm$16 \\
IRAS 10378$+$1108 & 5.7$\pm$0.5 & 5.7$\pm$0.4 & 95$\pm$8 & 93$\pm$7 \\
IRAS 10485$-$1447 & $<$0.23 \tablenotemark{A} & 
$<$0.22 \tablenotemark{A} & $<$3.8 \tablenotemark{A} & 
$<$3.6 \tablenotemark{A} \\
IRAS 11095$-$0238 (1$\farcs$5) & 3.6$\pm$1.1 & 3.2$\pm$0.9 & 60$\pm$18 
& 53$\pm$14 \\
IRAS 13335$-$2612 N & 2.6$\pm$0.6 & 1.7$\pm$0.5 & 44$\pm$9 & 27$\pm$7 \\
\enddata

\tablenotetext{A}{Based on the peak value in the integrated intensity 
(moment 0) map.}

\tablecomments{Col.(1): Object name. 
Cols.(2) and (3): Luminosity of the HCN J=3--2 and HCO$^{+}$ J=3--2 emission
lines (in units of L$_{\odot}$), respectively, derived from 
the spatially integrated (1--2$''$) emission line flux (Table 7). 
Cols.(4) and (5): Luminosity of the HCN J=3--2 and HCO$^{+}$ J=3--2 emission
lines (in units of K km s$^{-1}$ pc$^{2}$), respectively, derived from 
the spatially integrated (1--2$''$) emission line flux (Table 7).
}

\end{deluxetable}

\clearpage

\begin{deluxetable}{lcc}
\tabletypesize{\scriptsize}
\tablecaption{Spatially Integrated Continuum Flux \label{tbl-10}} 
\tablewidth{0pt}
\tablehead{
\colhead{Object} & \colhead{Observed flux} & \colhead{Free-free emission} \\
\colhead{} & \colhead{[mJy]} & \colhead{[mJy]}  \\
\colhead{(1)} & \colhead{(2)}  & \colhead{(3)}  
}
\startdata
IRAS 00188$-$0856 & 4.0 & $\lesssim$0.8 \\
IRAS 03250$+$1606 & 2.6 & $\lesssim$0.5 \\
IRAS 04103$-$2838 & 0.71 & $\lesssim$0.5 \\
IRAS 09039$+$0503 (1$\farcs$5) & 3.8 & $\lesssim$0.5 \\
IRAS 10378$+$1108 & 3.2 & $\lesssim$0.6 \\
IRAS 10485$-$1447 & $<$2 & $\lesssim$0.5 \\
IRAS 11095$-$0238 (1$\farcs$5) & 4.2 & $\lesssim$0.8 \\
IRAS 13335$-$2612 N & 2.3 & $\lesssim$0.5 \\
\enddata

\tablecomments{ 
Col.(1): Object name. 
Col.(2): Spatially integrated continuum flux at $\sim$240 GHz (in mJy). 
A 1$''$ diameter circular aperture is basically employed.
A 1$\farcs$5 or 2$''$ diameter circular aperture is used for a small 
fraction of sources, which are indicated in column 1.
Col.(3): Estimated free-free emission flux from HII-regions in starbursts 
at $\sim$240 GHz (in mJy). 
Upper limits mean possible AGN contribution to the observed far-infrared 
luminosity ($\S$5.6).
}

\end{deluxetable}

\clearpage

\begin{figure}
\begin{center}
\includegraphics[angle=0,scale=.424]{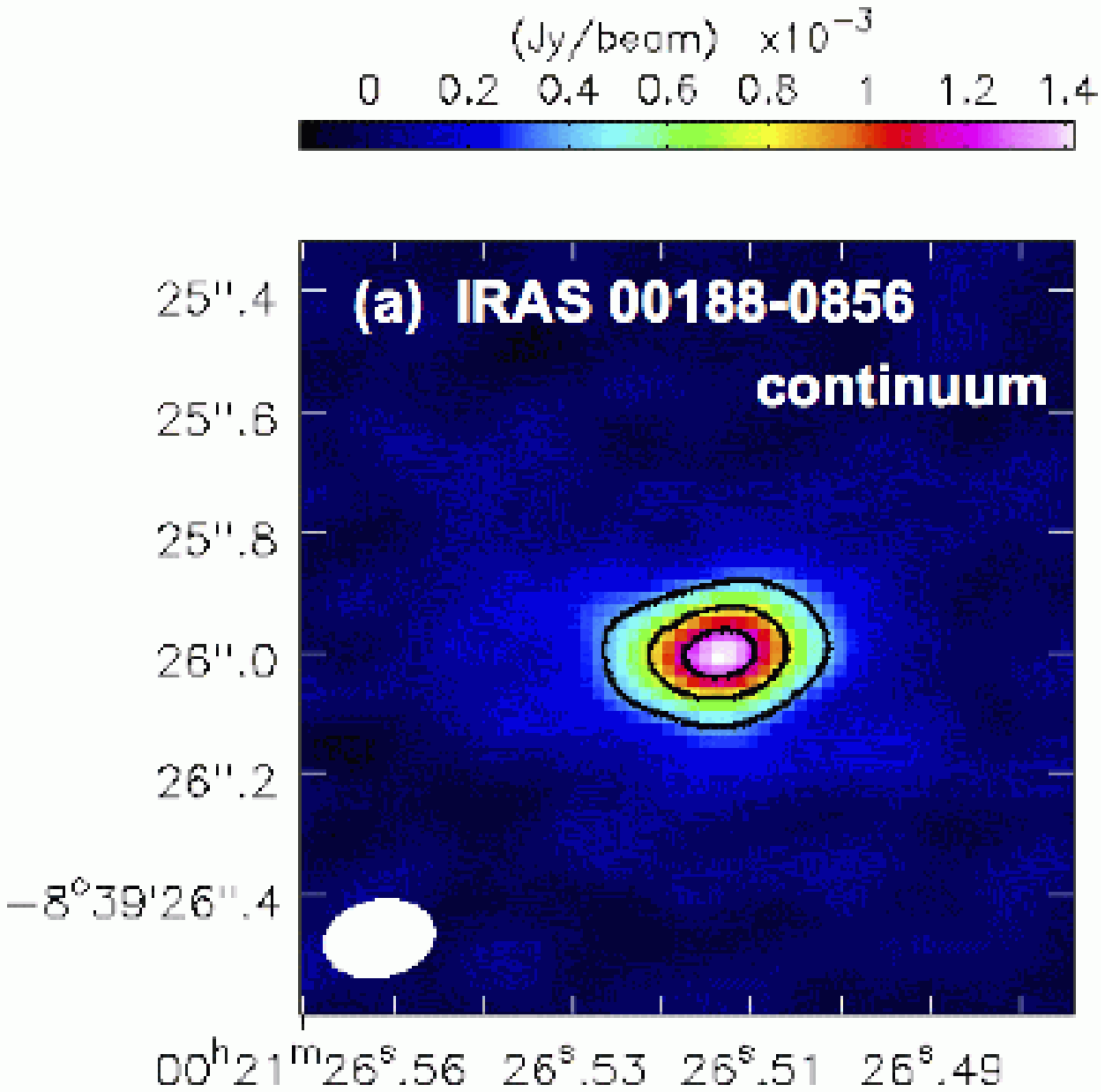} 
\includegraphics[angle=0,scale=.424]{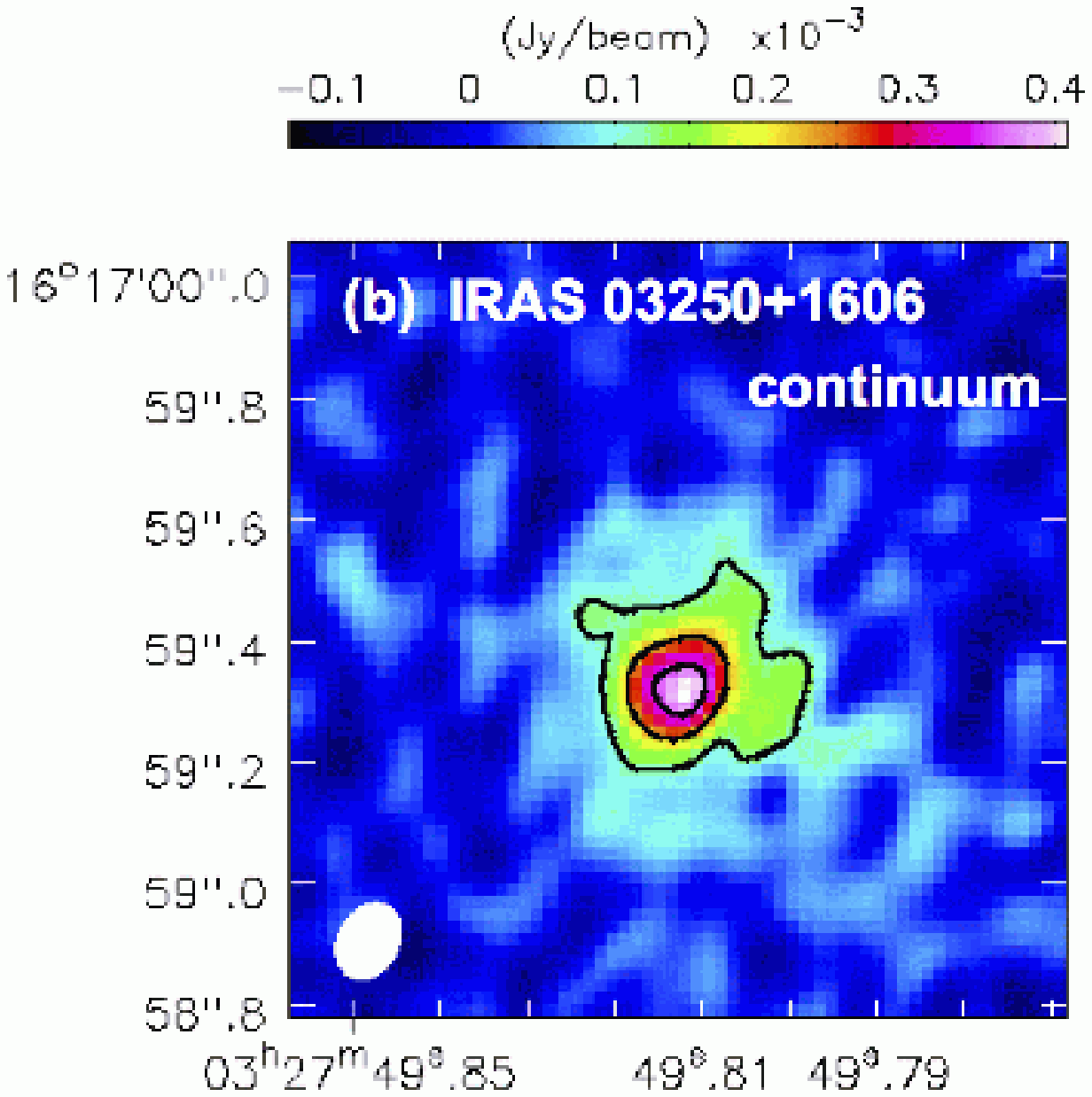} 
\includegraphics[angle=0,scale=.424]{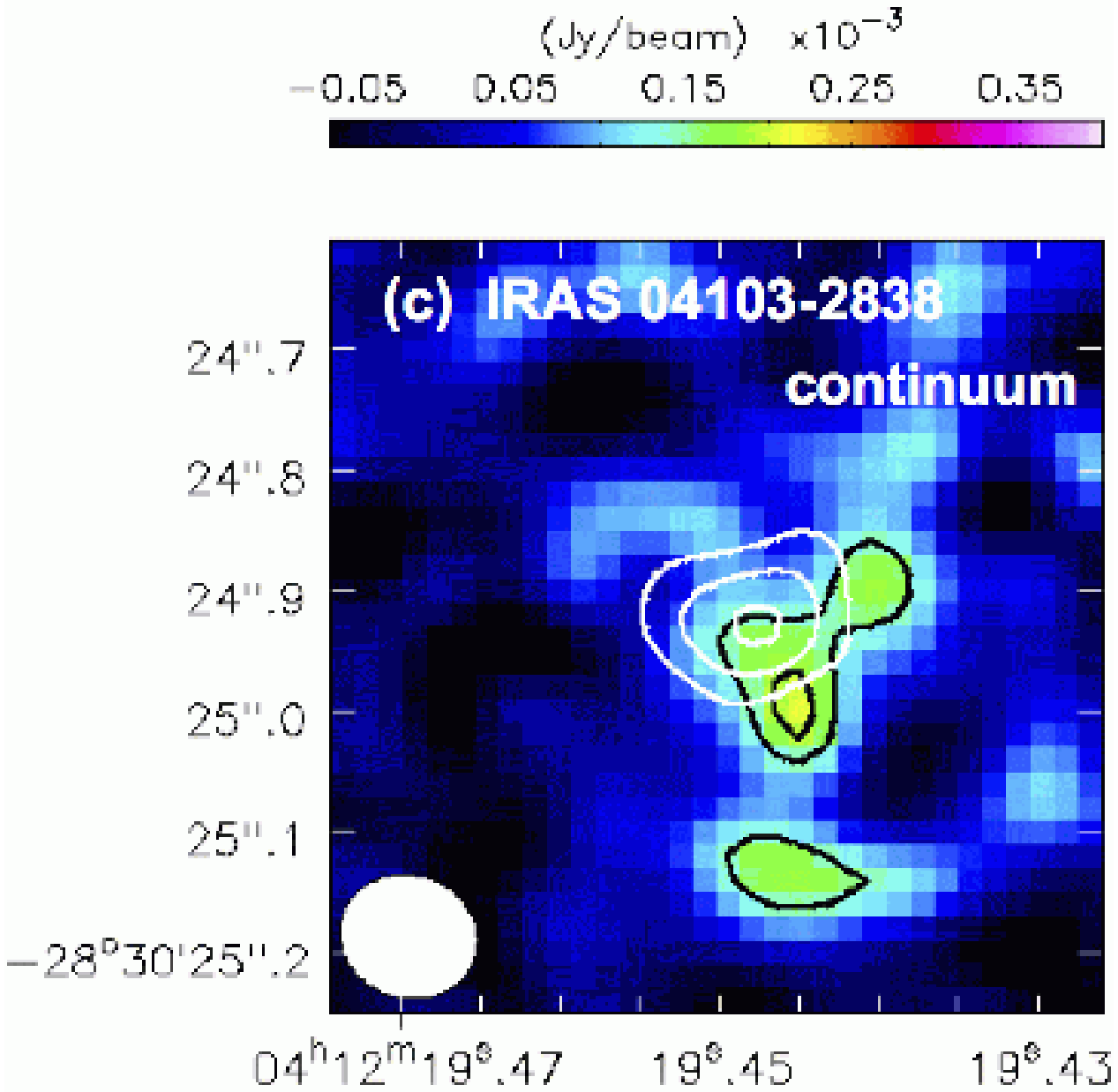}\\ 
\includegraphics[angle=0,scale=.424]{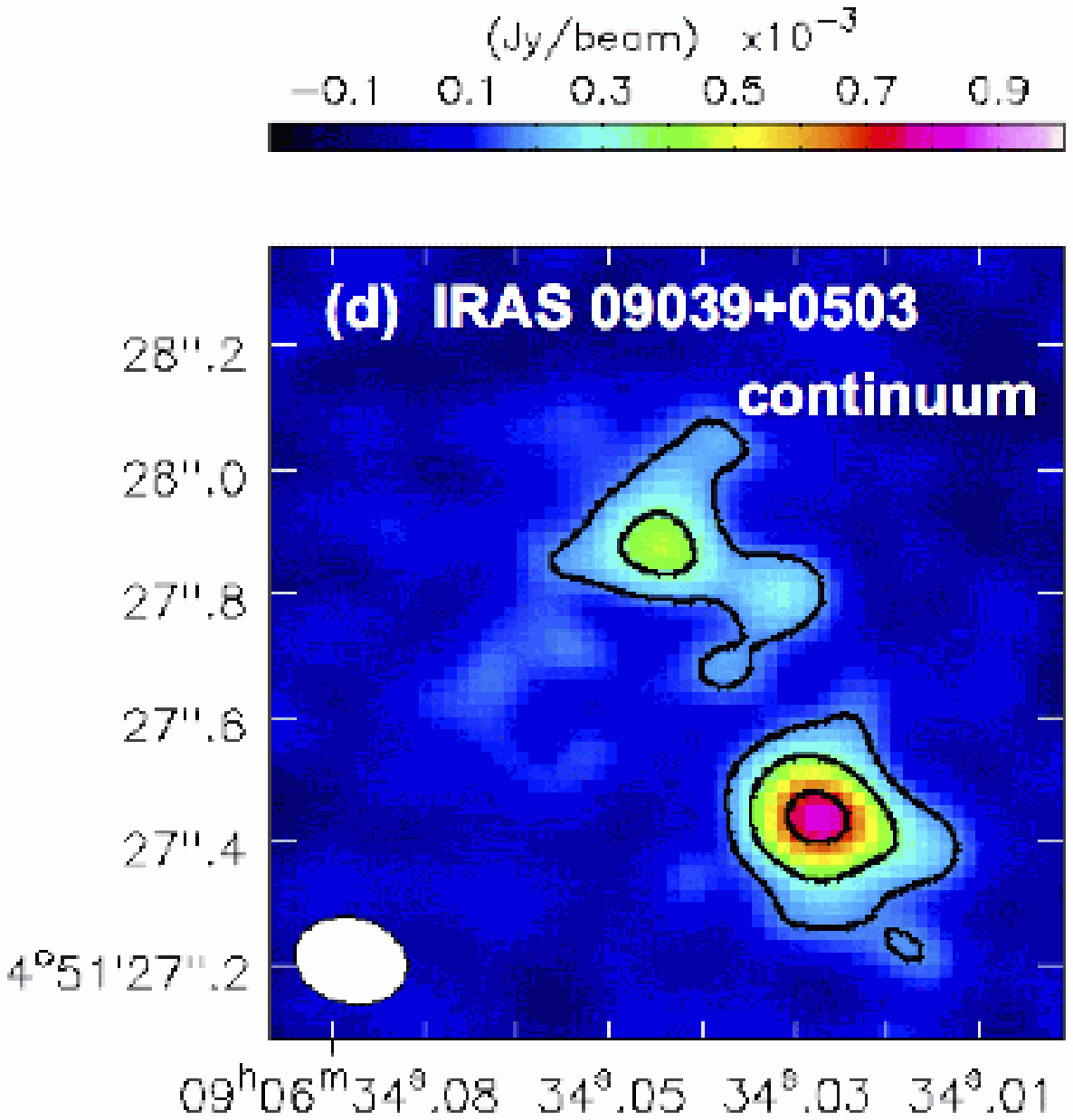} 
\includegraphics[angle=0,scale=.424]{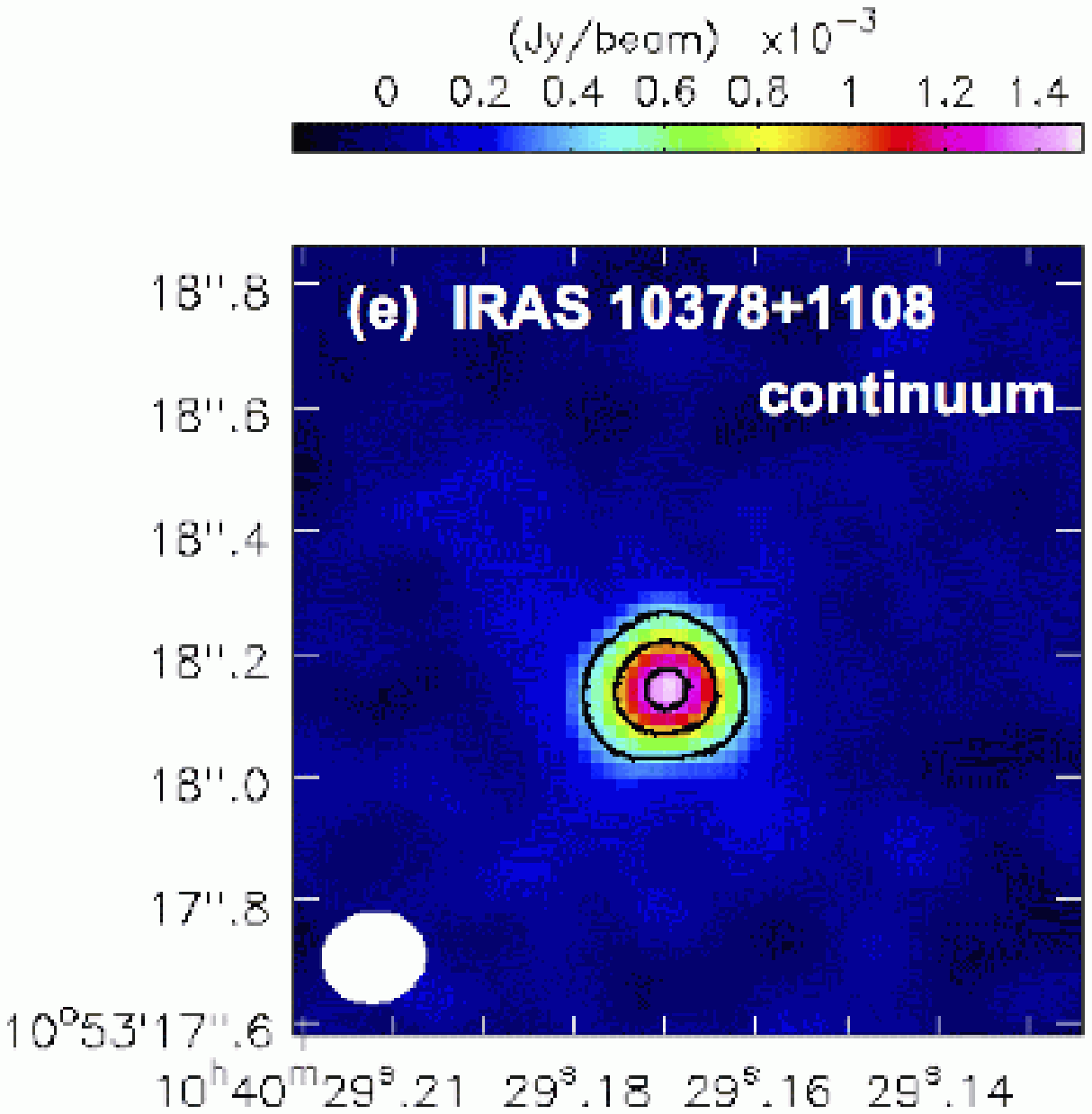} 
\includegraphics[angle=0,scale=.424]{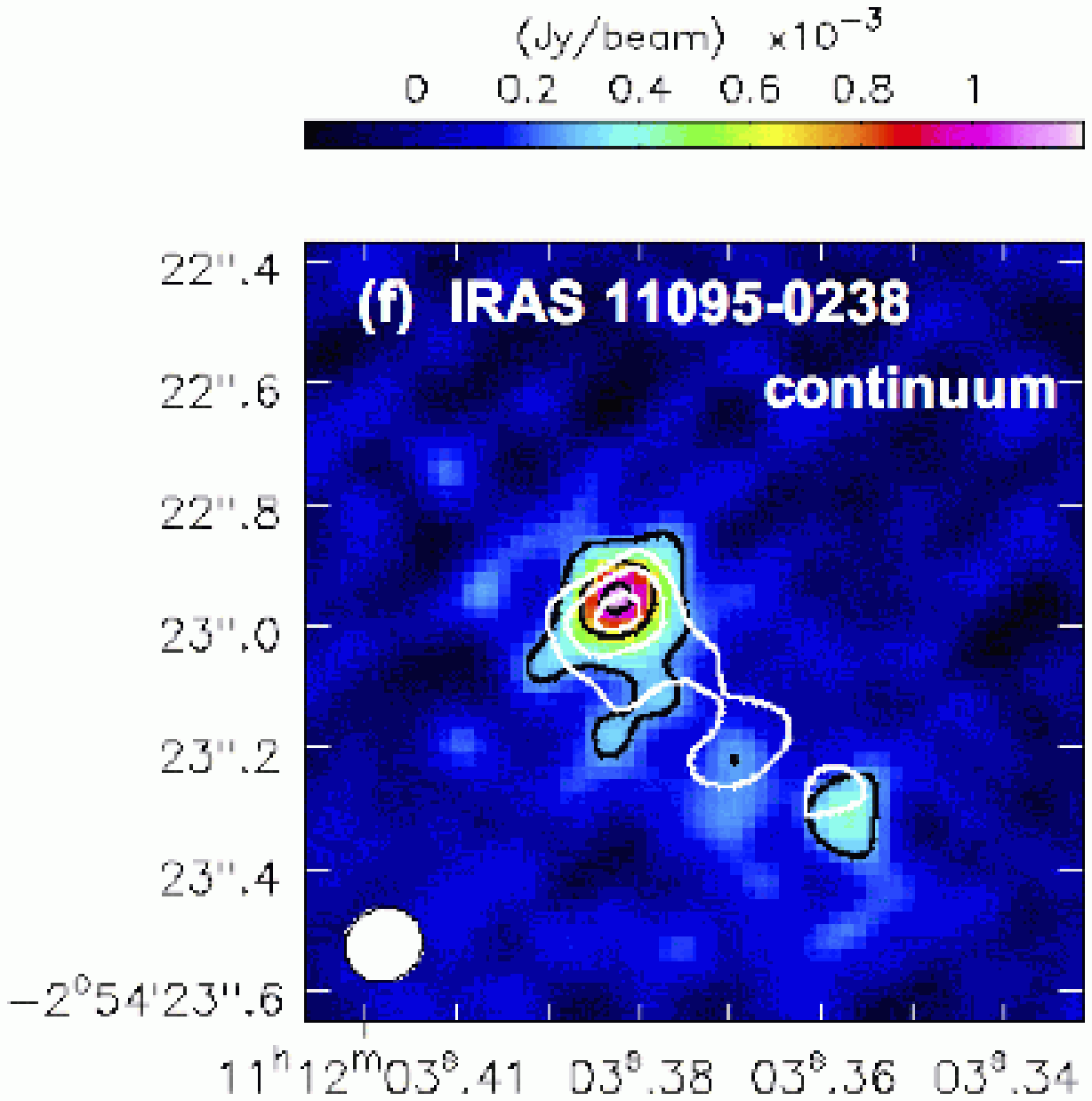}\\ 
\includegraphics[angle=0,scale=.424]{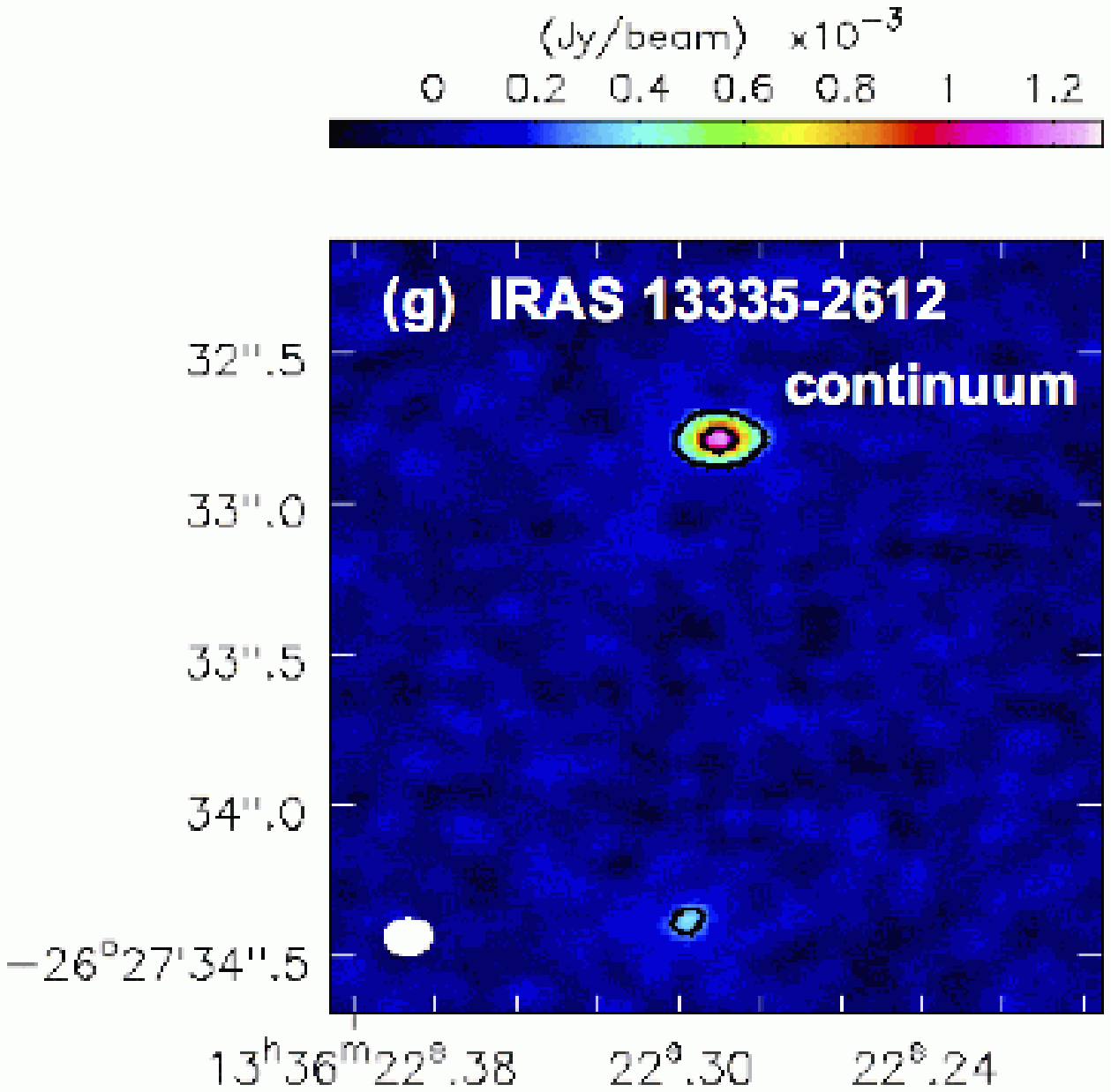}
\includegraphics[angle=0,scale=.424]{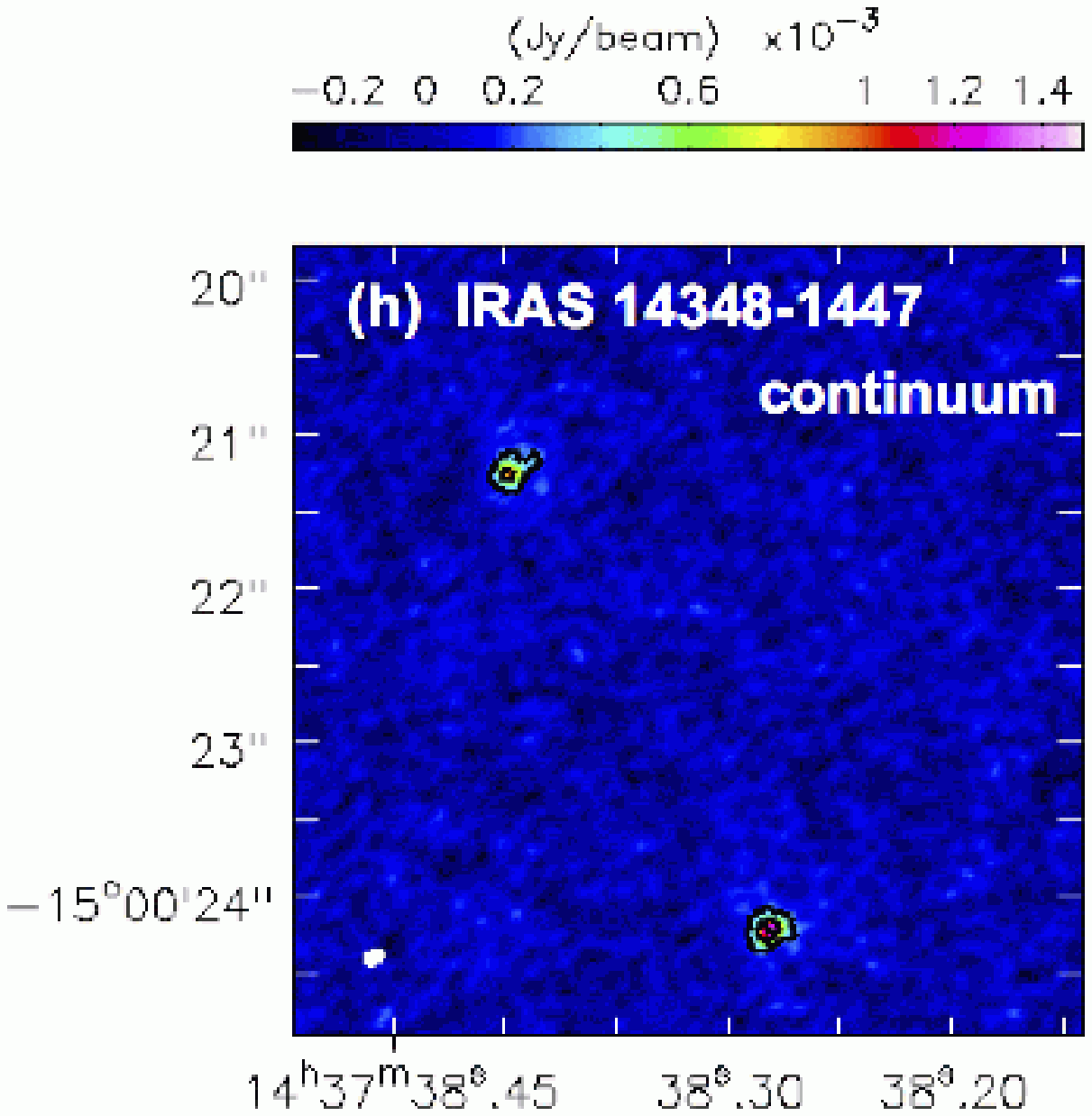}
\includegraphics[angle=0,scale=.424]{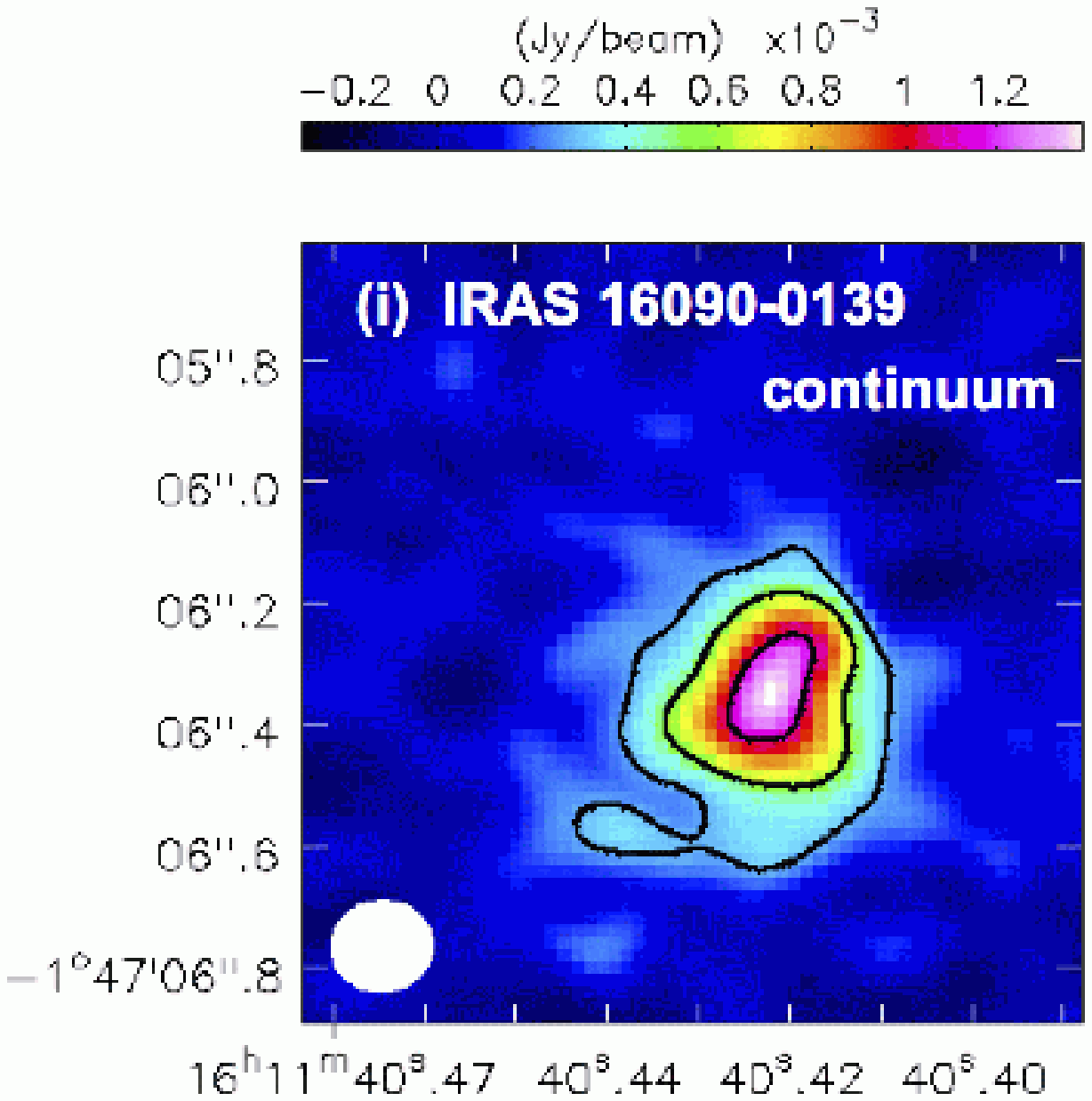}\\ 
\includegraphics[angle=0,scale=.424]{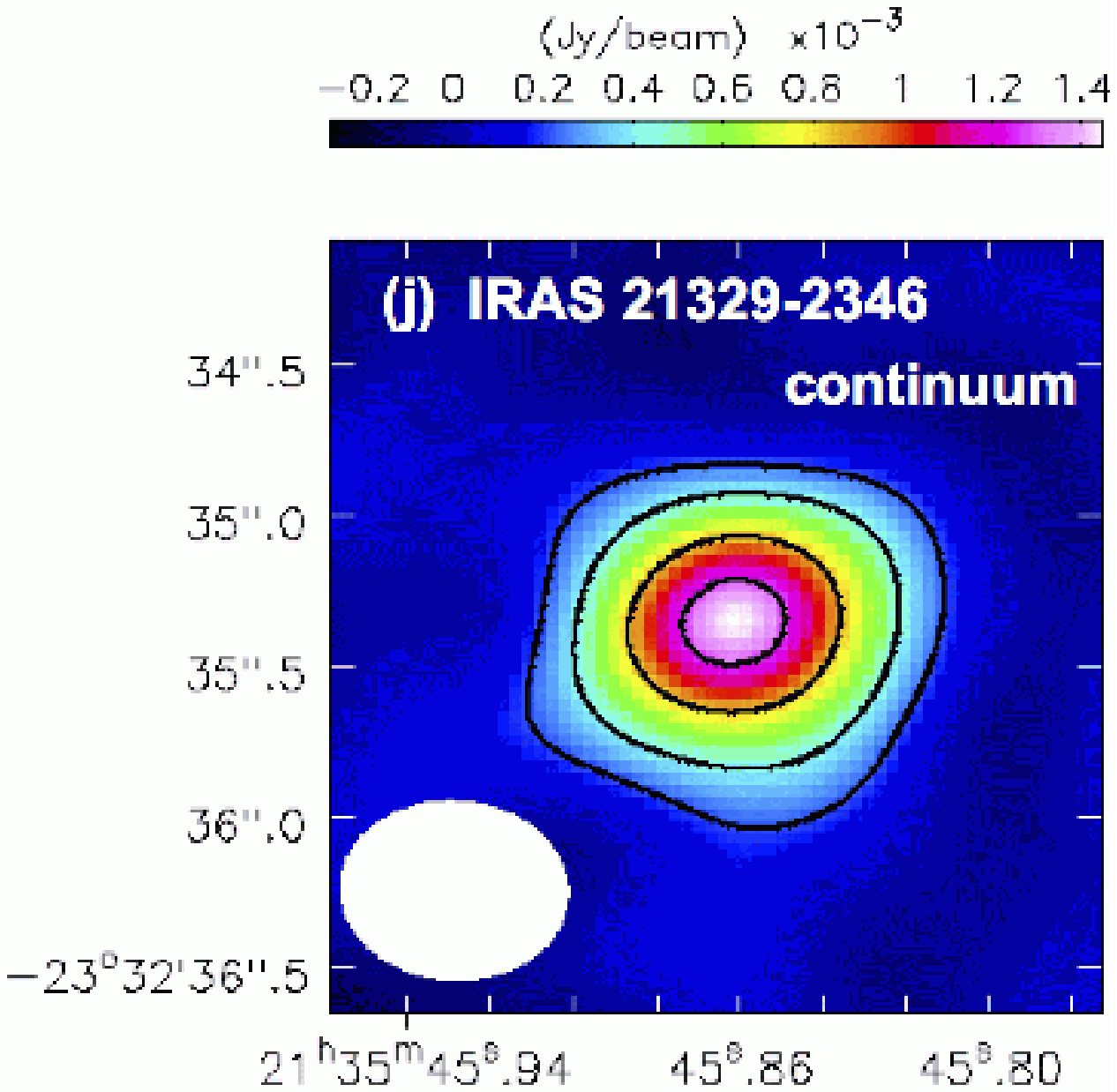} 
\includegraphics[angle=0,scale=.424]{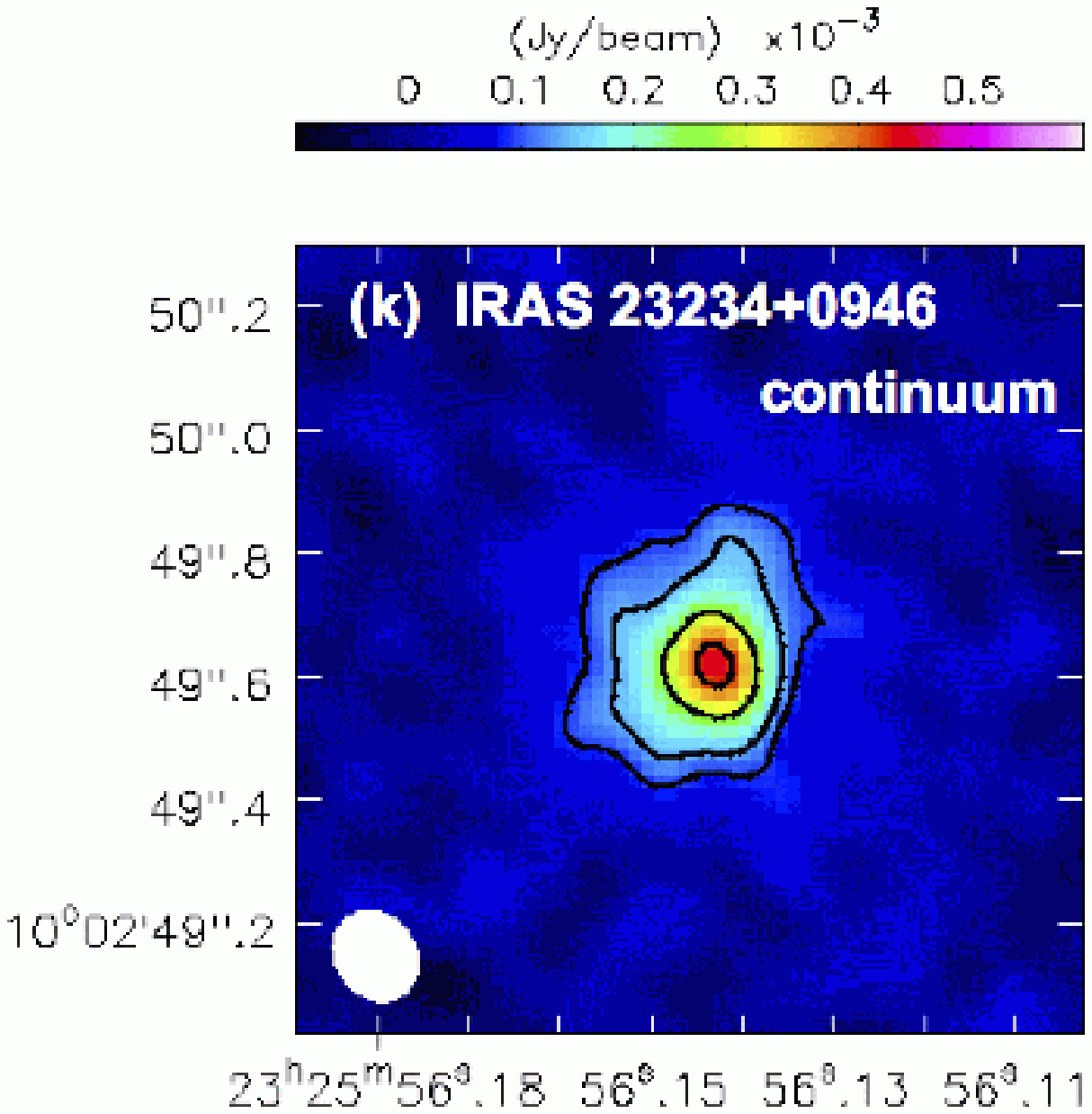} 
\includegraphics[angle=0,scale=.424]{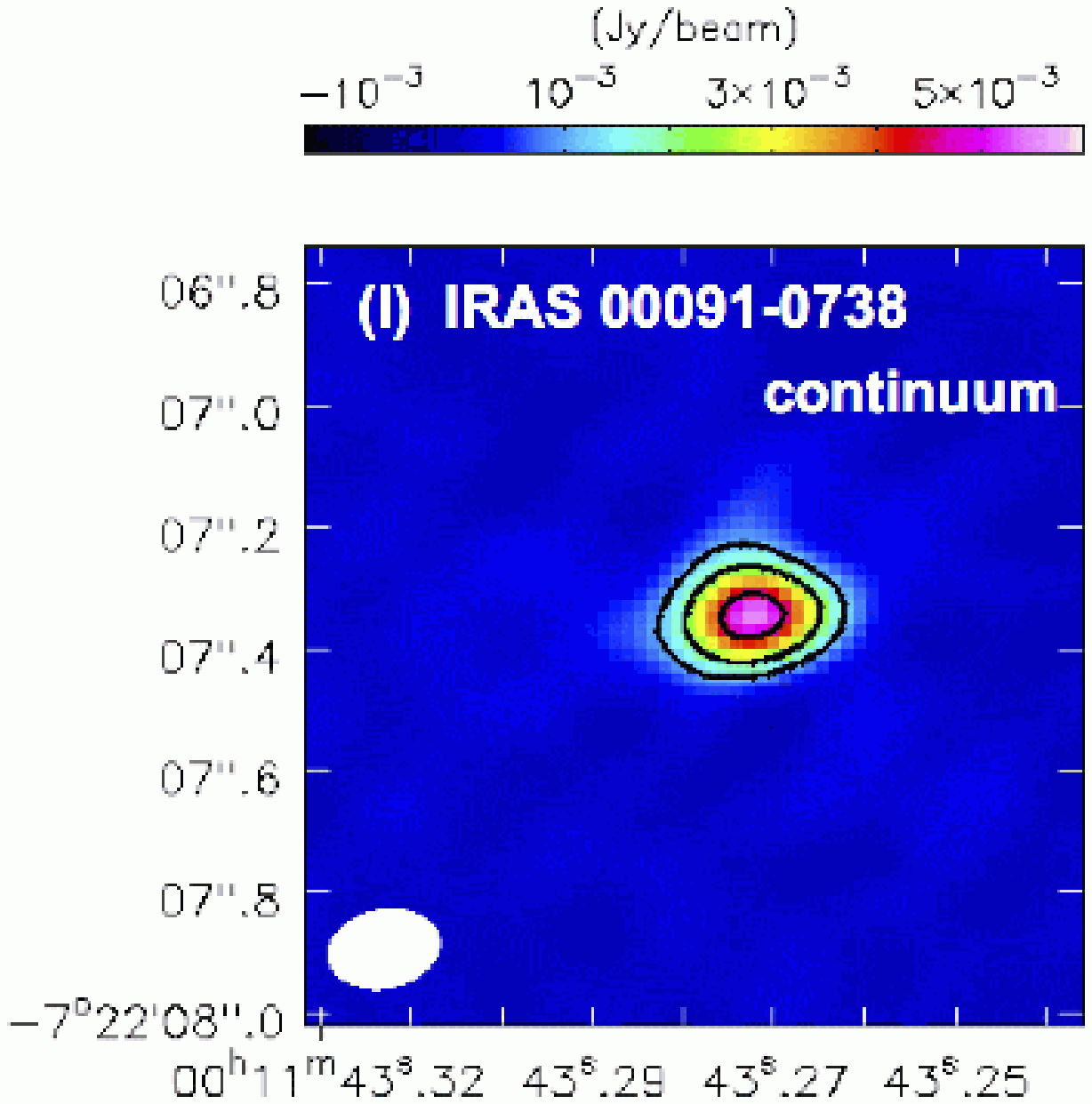} \\
\end{center}
\end{figure}

\clearpage

\begin{figure}
\begin{center}
\includegraphics[angle=0,scale=.424]{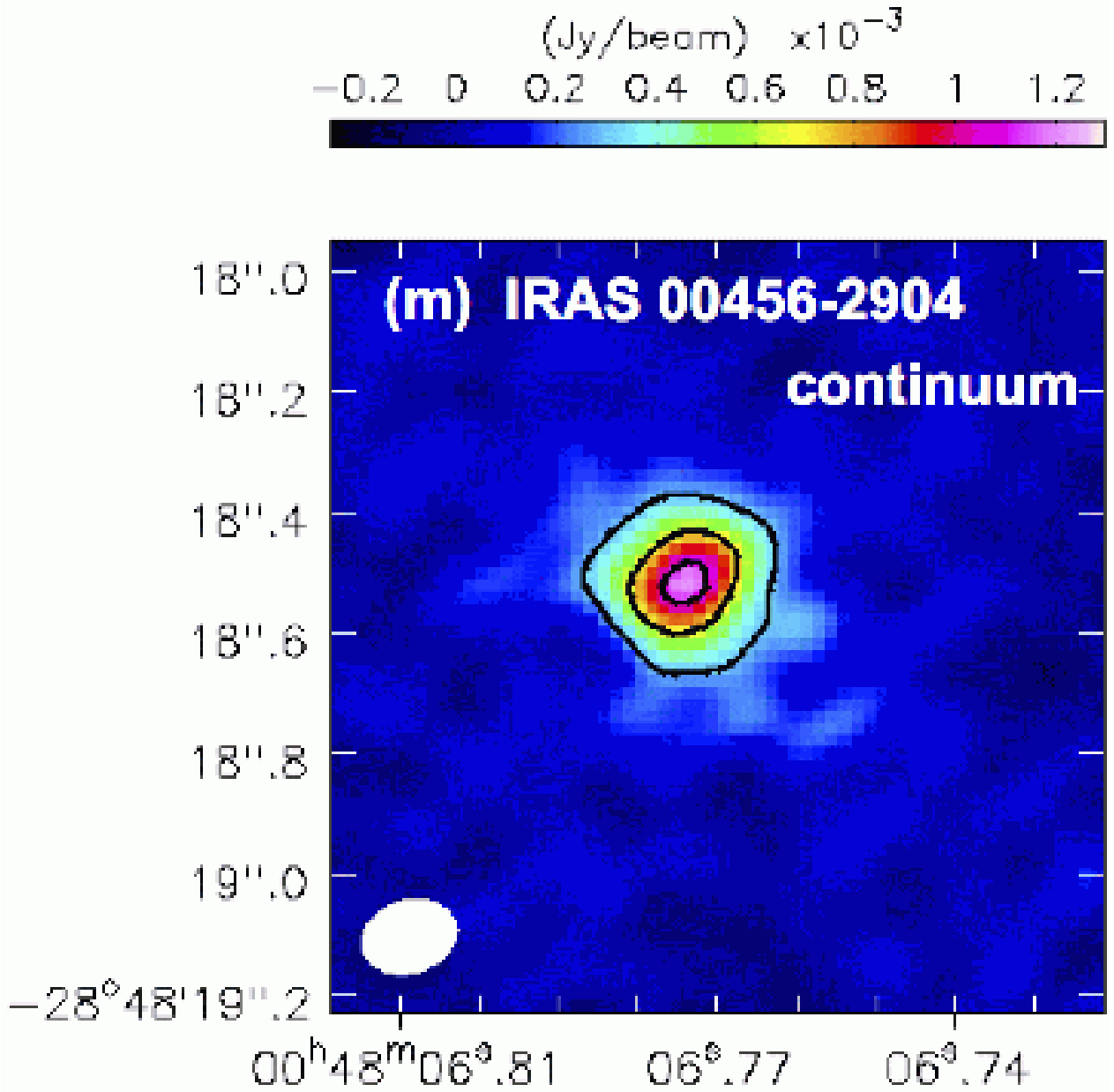}
\includegraphics[angle=0,scale=.424]{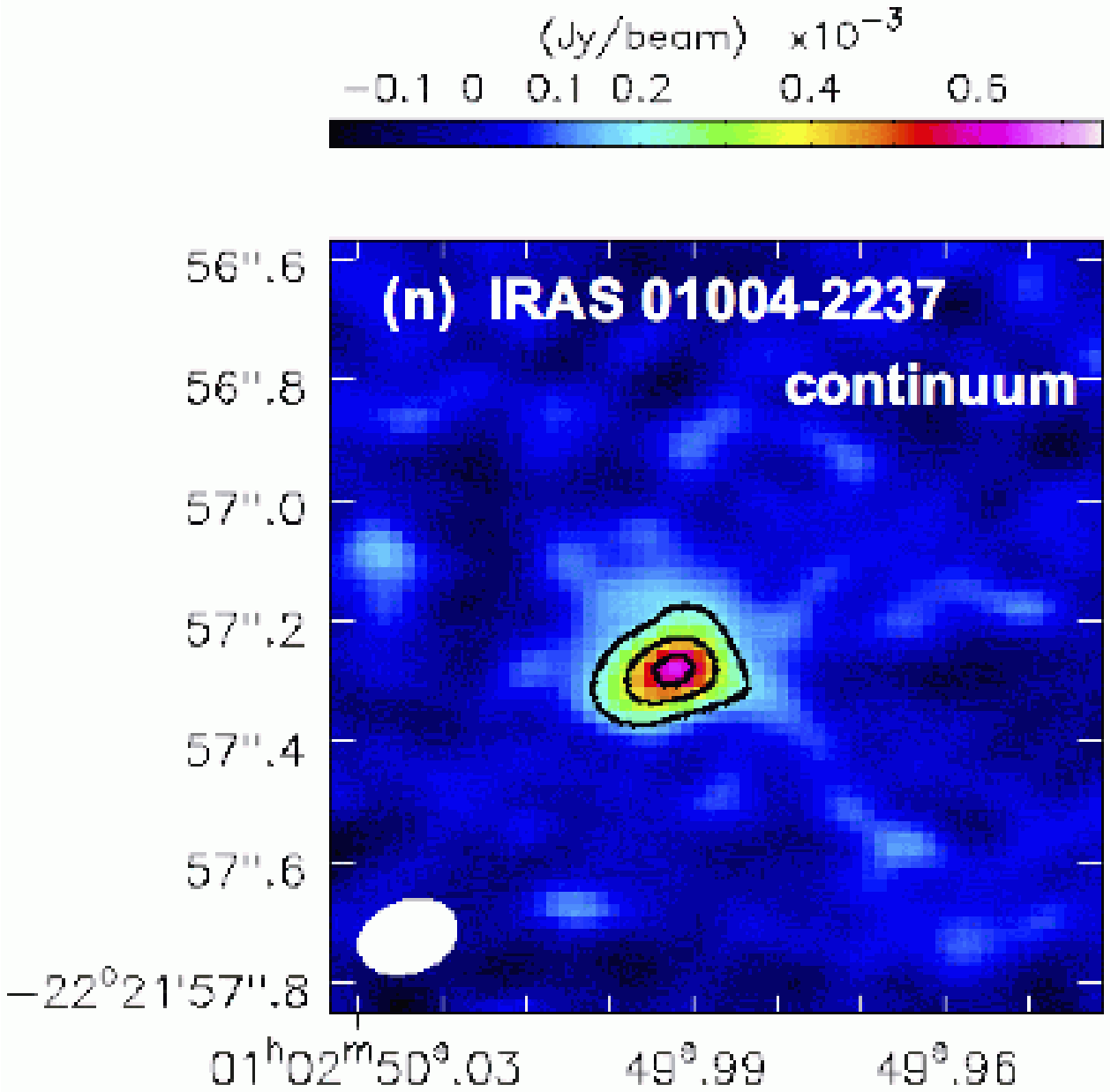} 
\includegraphics[angle=0,scale=.424]{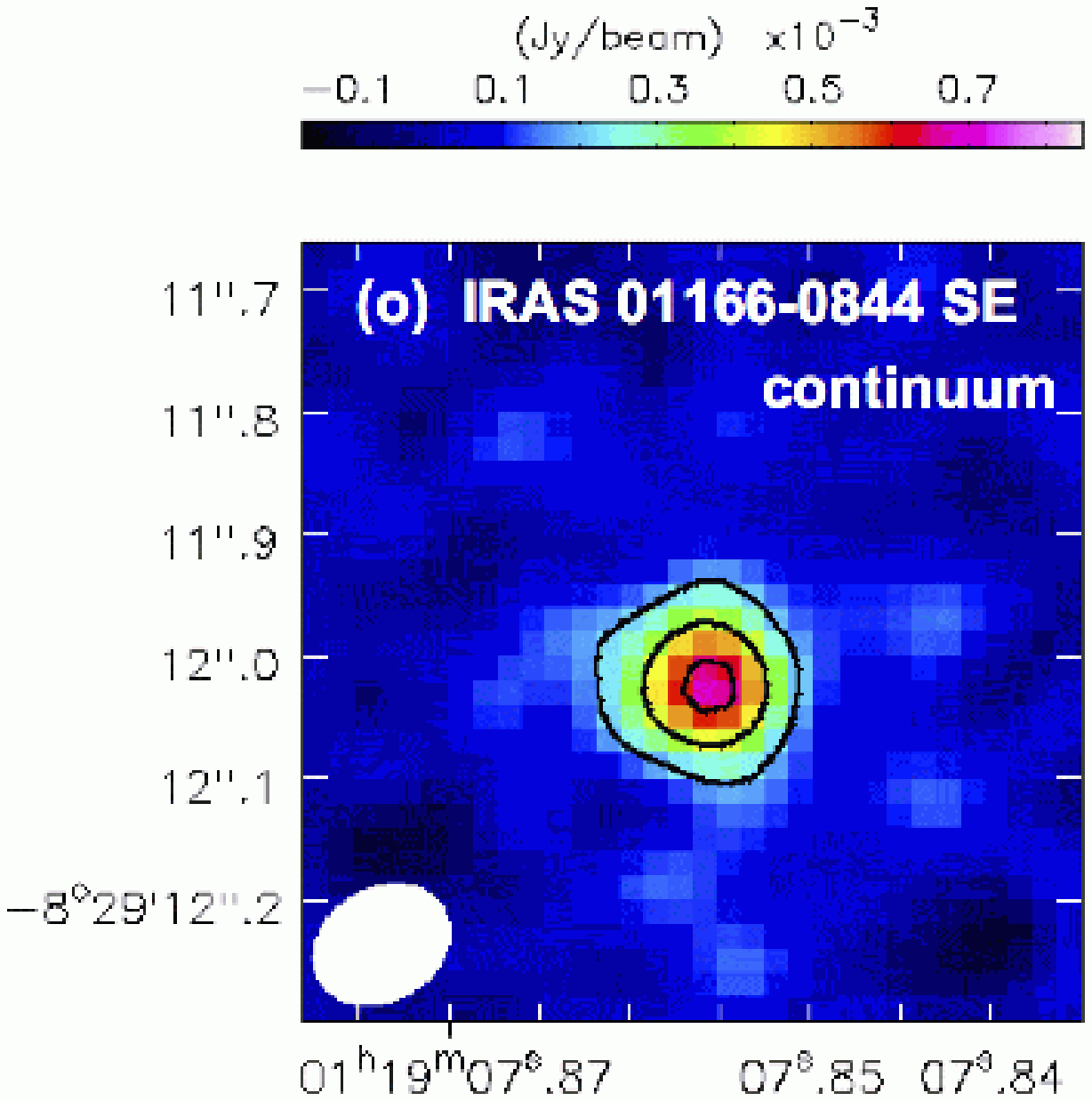}\\ 
\includegraphics[angle=0,scale=.424]{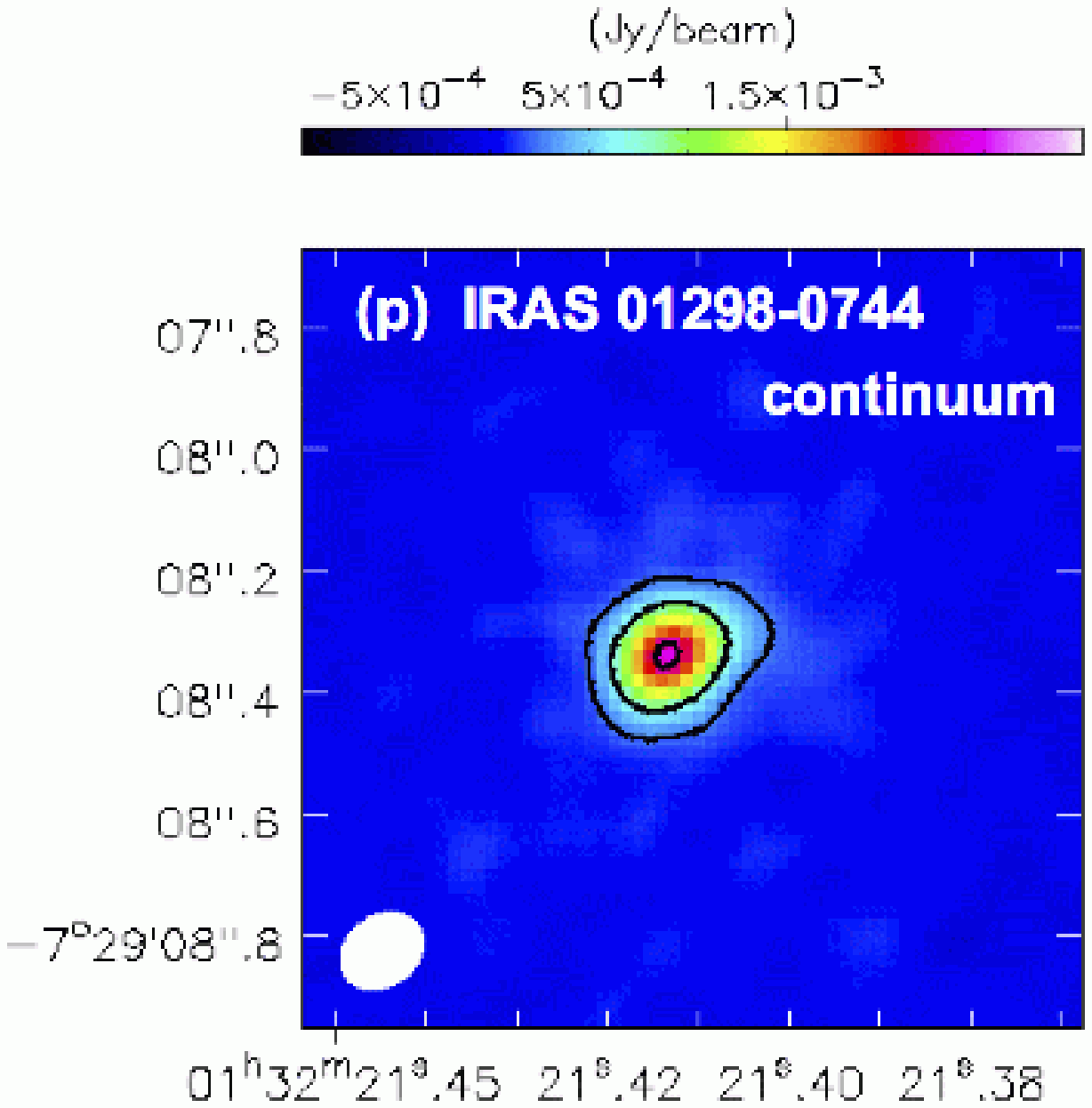} 
\includegraphics[angle=0,scale=.424]{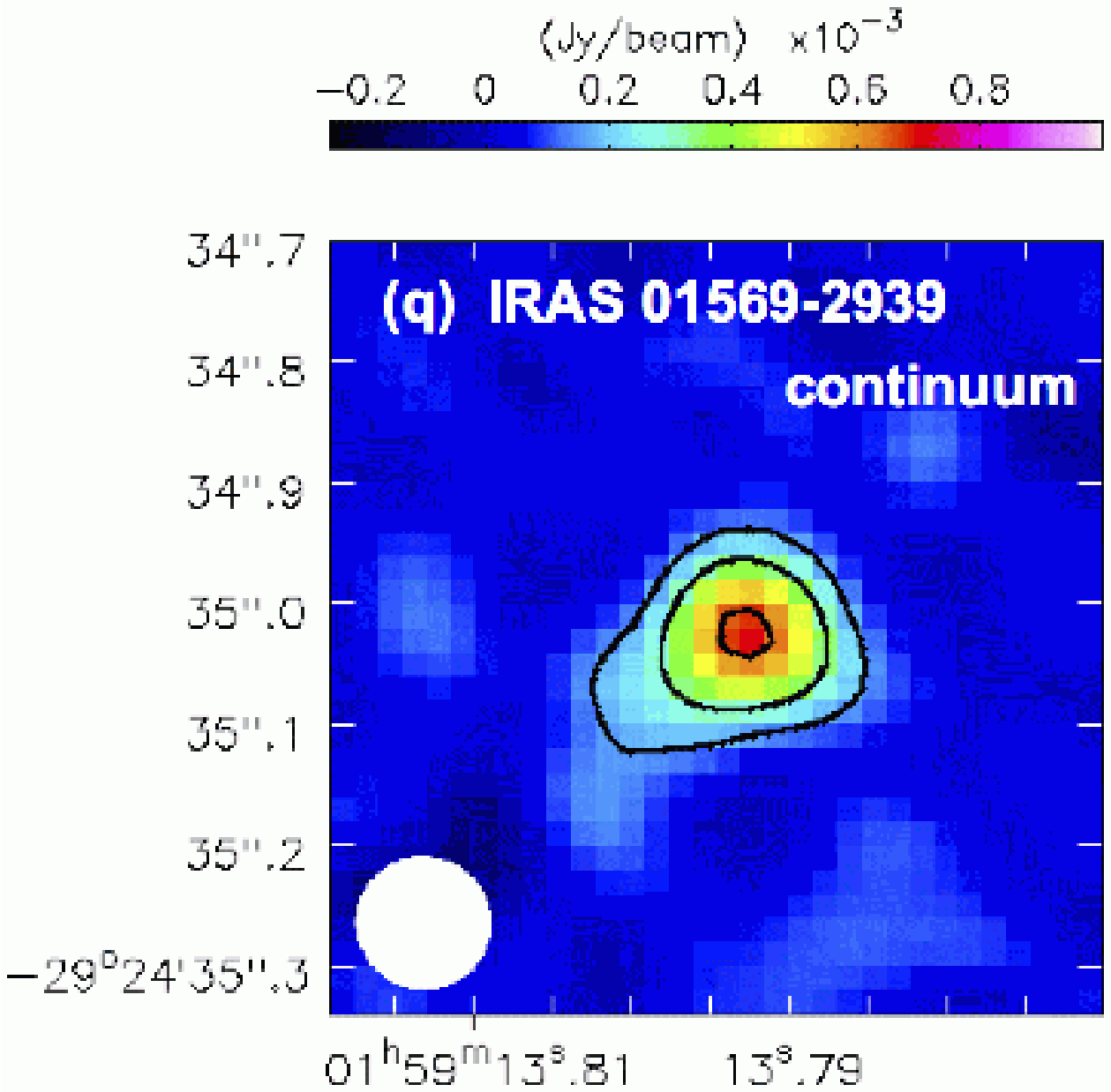} 
\includegraphics[angle=0,scale=.424]{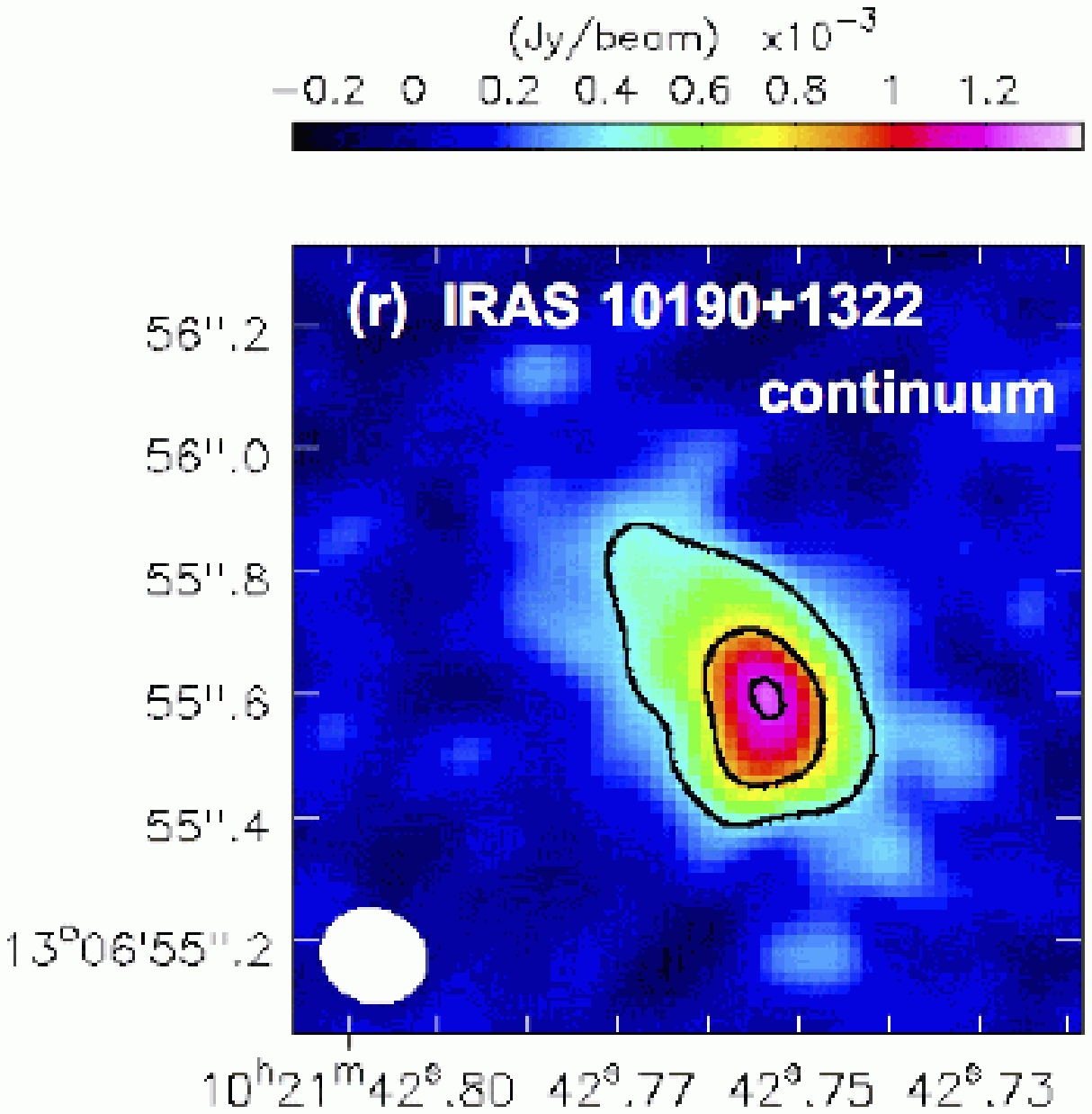}\\ 
\includegraphics[angle=0,scale=.424]{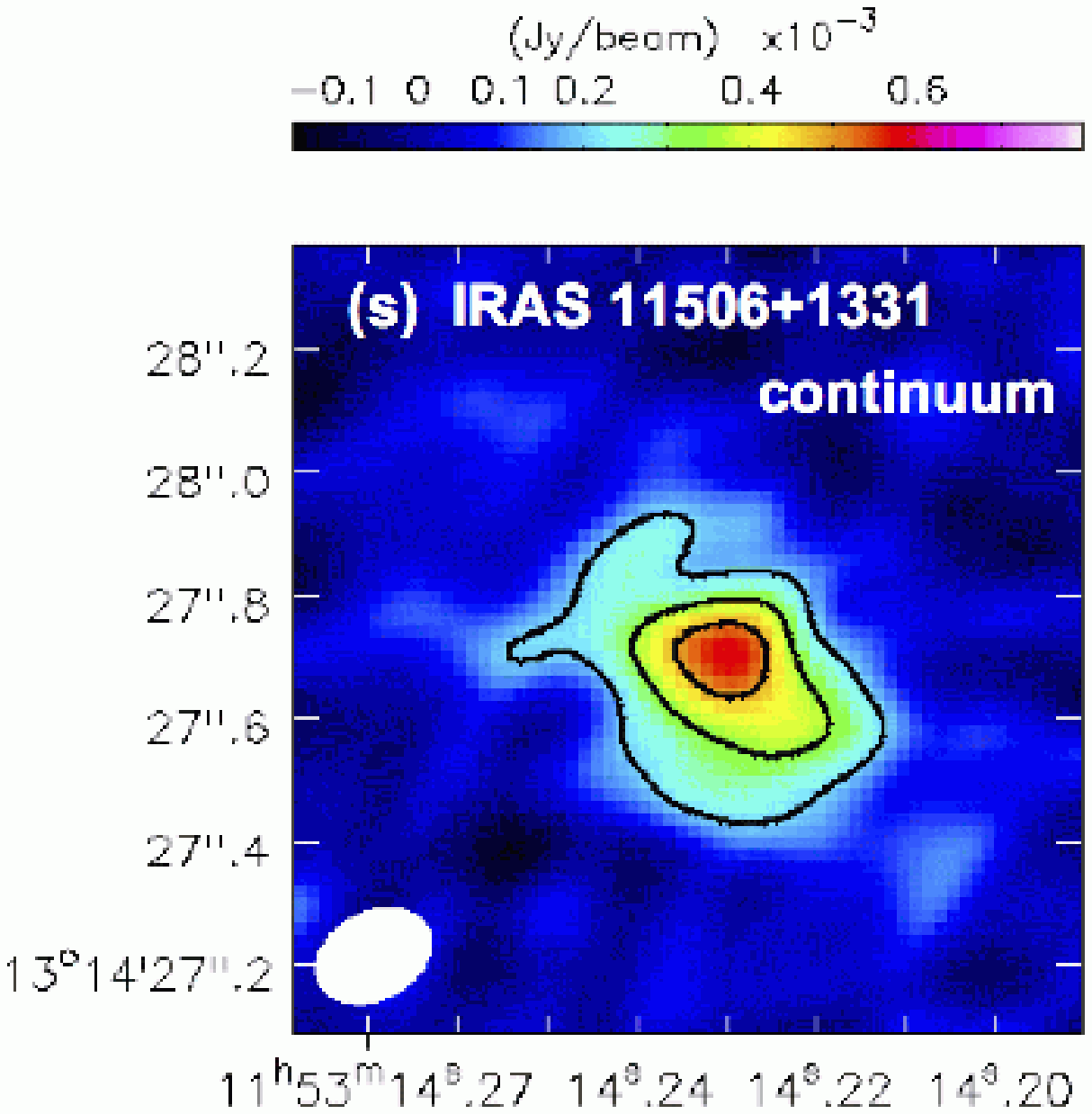} 
\includegraphics[angle=0,scale=.424]{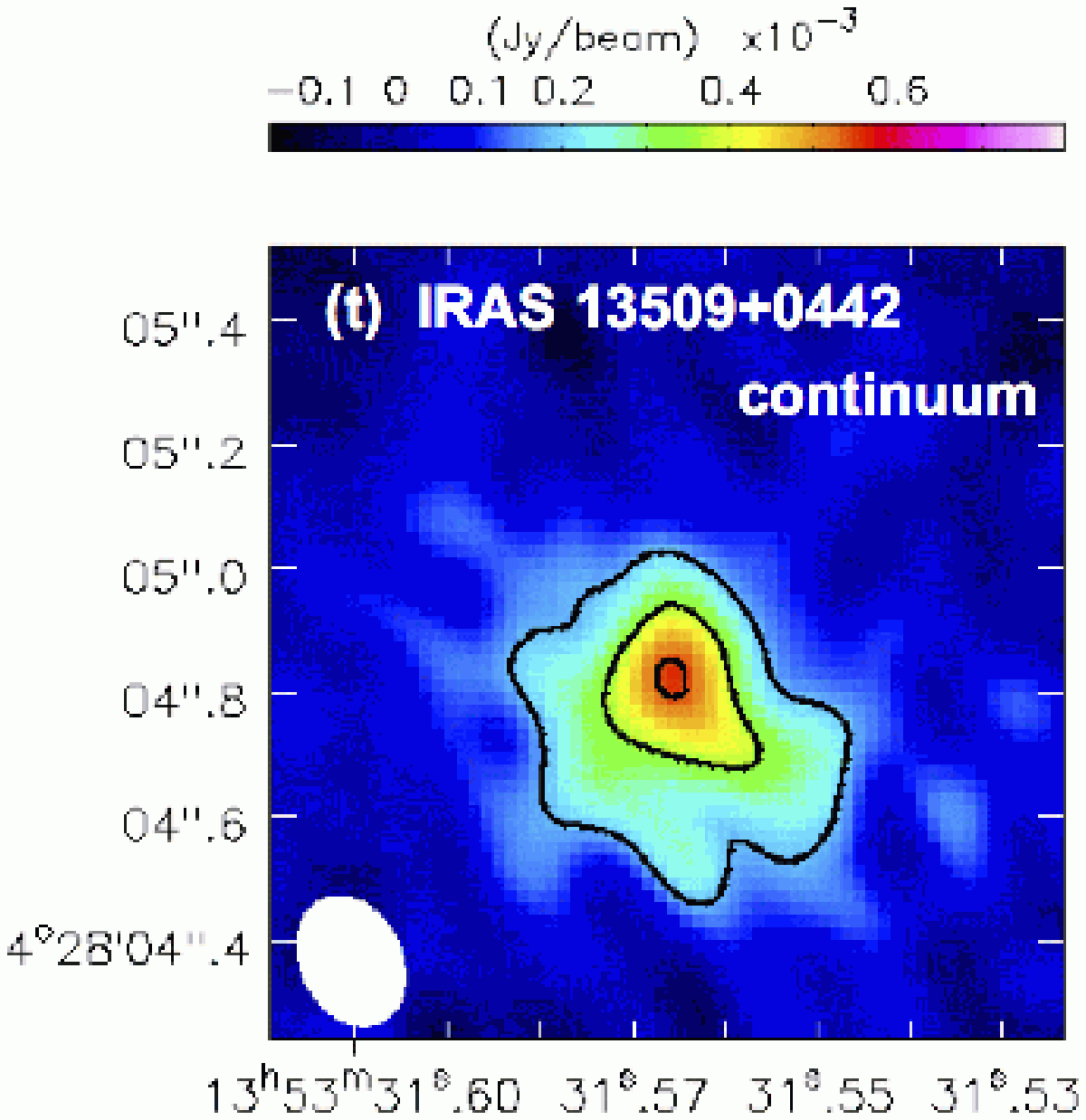} 
\includegraphics[angle=0,scale=.424]{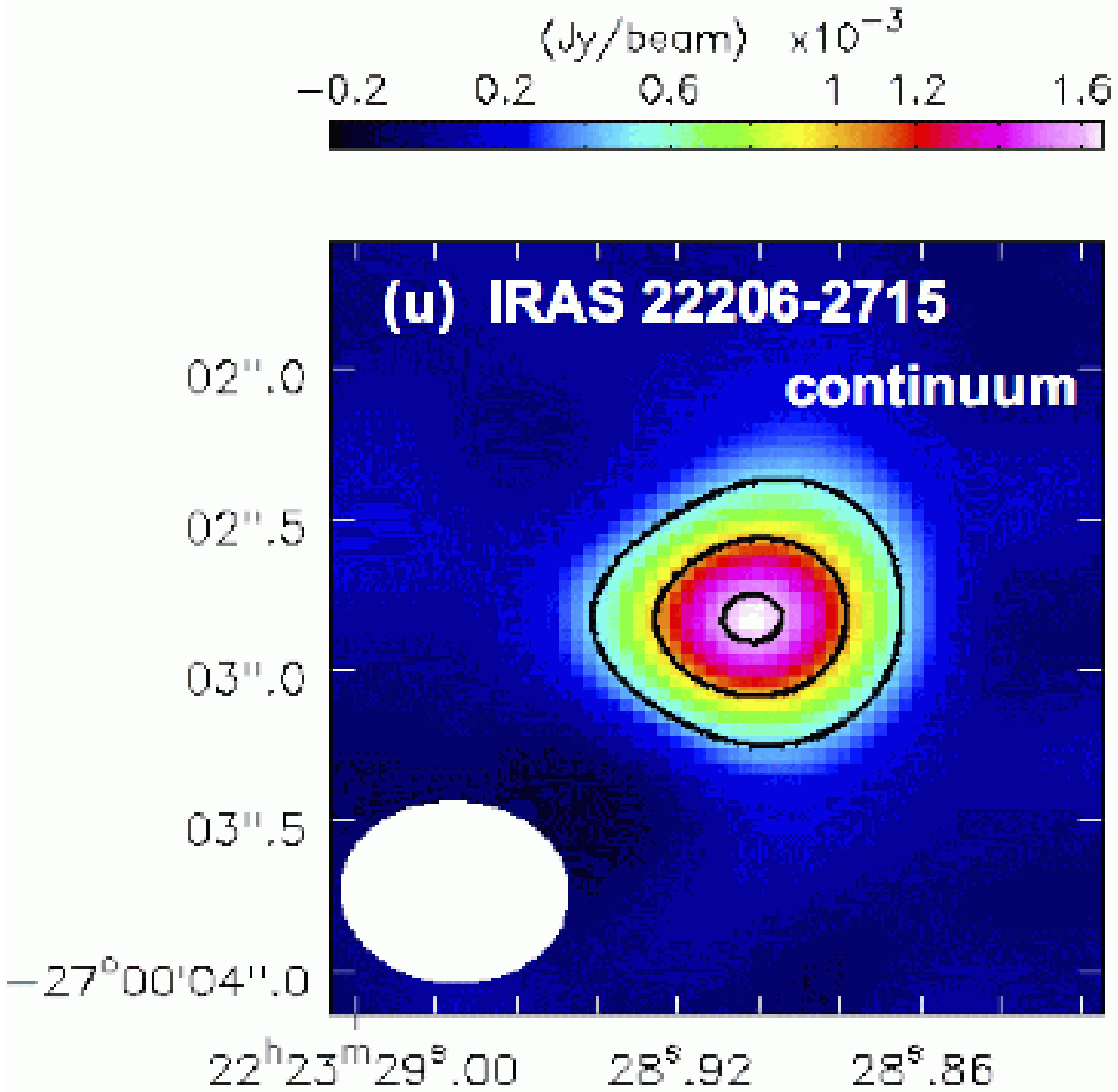} \\
\vspace*{0.2cm}
\hspace*{0.3cm}
\includegraphics[angle=0,scale=.424]{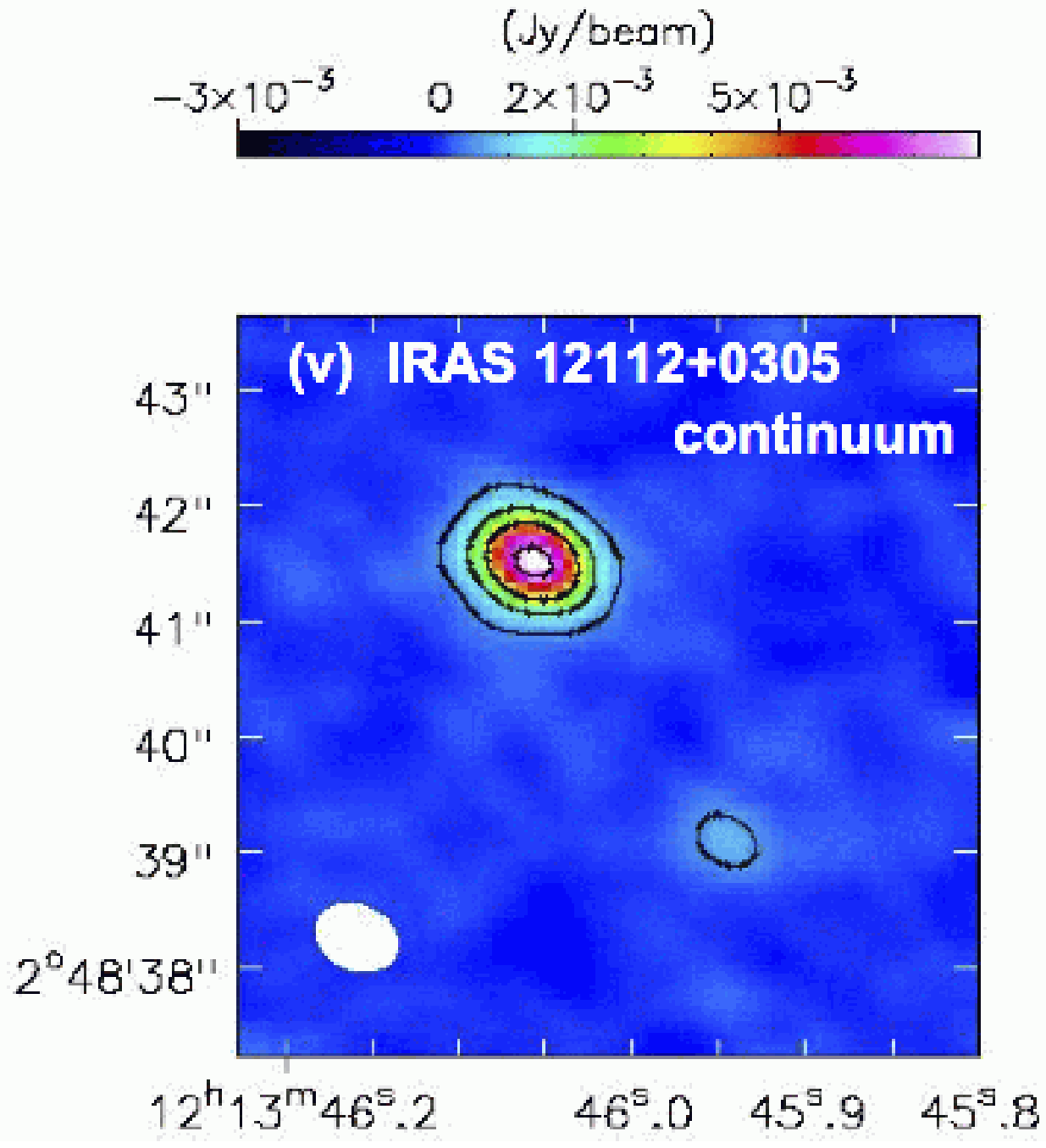} 
\includegraphics[angle=0,scale=.424]{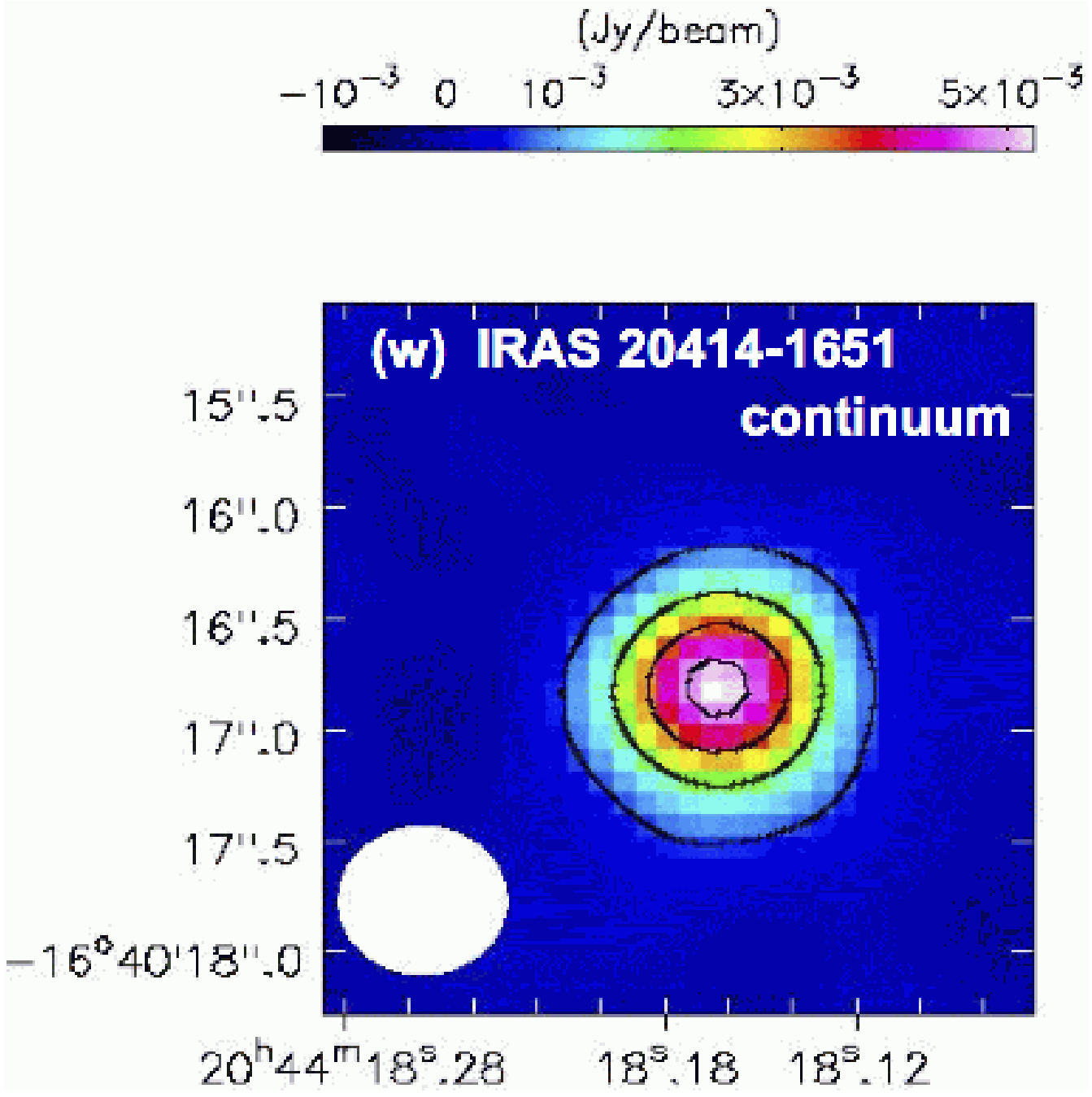} 
\includegraphics[angle=0,scale=.424]{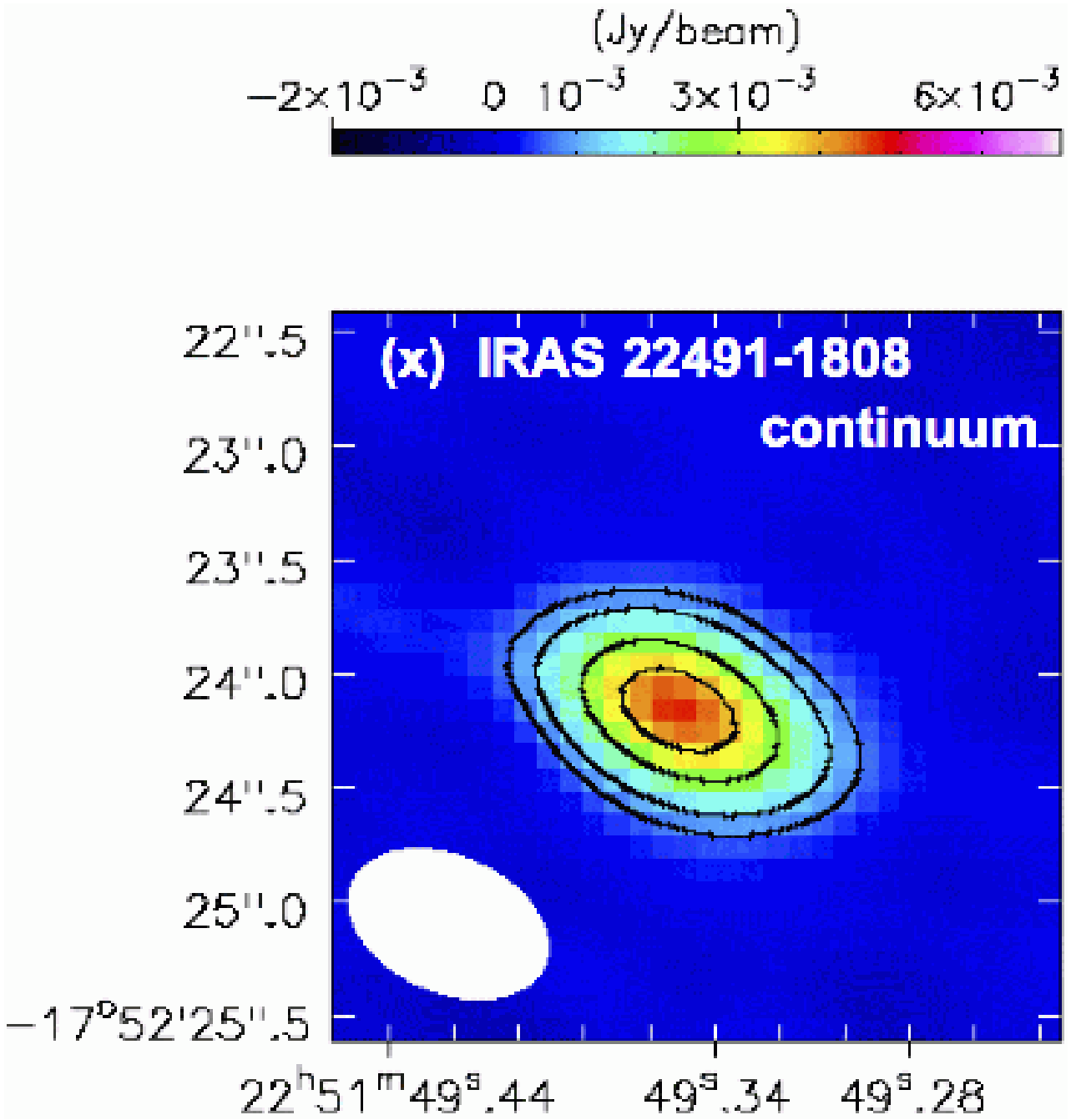} \\
\end{center}
\end{figure}

\clearpage

\begin{figure}
\caption{
Continuum emission maps of ULIRGs with $>$3$\sigma$ detection.
The abscissa and ordinate are right ascension and declination in ICRS,
respectively.
The black contours are 
10$\sigma$, 20$\sigma$, 30$\sigma$ for IRAS 00188$-$0856, 
4$\sigma$, 8$\sigma$, 12$\sigma$ for IRAS 03250+1606, 
3$\sigma$, 4$\sigma$ for IRAS 04103$-$2838, 
4$\sigma$, 8$\sigma$, 16$\sigma$ for IRAS 09039+0503, 
10$\sigma$, 20$\sigma$, 30$\sigma$ for IRAS 10378+1108, 
4$\sigma$, 10$\sigma$, 16$\sigma$ for IRAS 11095$-$0238, 
6$\sigma$, 20$\sigma$ for IRAS 13335$-$2612, 
4$\sigma$, 12$\sigma$, 20$\sigma$ for IRAS 14348$-$1447, 
4$\sigma$, 8$\sigma$, 16$\sigma$ for IRAS 16090$-$0139, 
3$\sigma$, 6$\sigma$, 12$\sigma$, 18$\sigma$ for IRAS 21329$-$2346, 
4$\sigma$, 6$\sigma$, 12$\sigma$, 18$\sigma$ for IRAS 23234+0946, 
12$\sigma$, 24$\sigma$, 48$\sigma$ for IRAS 00091$-$0738, 
6$\sigma$, 14$\sigma$, 22$\sigma$ for IRAS 00456$-$2904, 
4$\sigma$, 8$\sigma$, 12$\sigma$ for IRAS 01004$-$2237, 
5$\sigma$, 10$\sigma$, 15$\sigma$ for IRAS 01166$-$0844 SE, 
6$\sigma$, 18$\sigma$, 54$\sigma$ for IRAS 01298$-$0744, 
4$\sigma$, 8$\sigma$, 16$\sigma$ for IRAS 01569$-$2939, 
5$\sigma$, 10$\sigma$, 15$\sigma$ for IRAS 10190+1322, 
4$\sigma$, 7$\sigma$, 10$\sigma$ for IRAS 11506+1331, 
4$\sigma$, 8$\sigma$, 12$\sigma$ for IRAS 13509+0442, 
7$\sigma$, 15$\sigma$, 23$\sigma$ for IRAS 22206$-$2715,
3$\sigma$, 10$\sigma$, 20$\sigma$, 40$\sigma$ for IRAS 12112$+$0305,
5$\sigma$, 15$\sigma$, 25$\sigma$, 35$\sigma$ for IRAS 20414$-$1651, and
3$\sigma$, 5$\sigma$, 10$\sigma$, 15$\sigma$ for IRAS 22491$-$1808. 
For IRAS 04103$-$2838 and IRAS 11095$-$0238, the white contours of the 
HCN J=3--2 emission line (3$\sigma$, 5$\sigma$, 7$\sigma$ in the 
moment 0 map in Fig. 3e) and HCO$^{+}$ J=3--2 emission line 
(4$\sigma$, 6$\sigma$, 9$\sigma$ in the moment 0 map in Fig. 3l) 
are plotted together, respectively, to highlight the 
significant positional offset between the continuum and dense 
molecular line emission.
IRAS 10485$-$1447 and IRAS 02411$+$0353 are not shown, as there was 
no significant ($>$3$\sigma$) continuum emission. 
The 1$\sigma$ rms noise level in each source is tabulated 
in Table 3 (column 5).
Beam sizes are shown as filled circles in the lower-left region.
}
\end{figure}

\clearpage

\begin{figure}
\begin{center}
\includegraphics[angle=0,scale=.4]{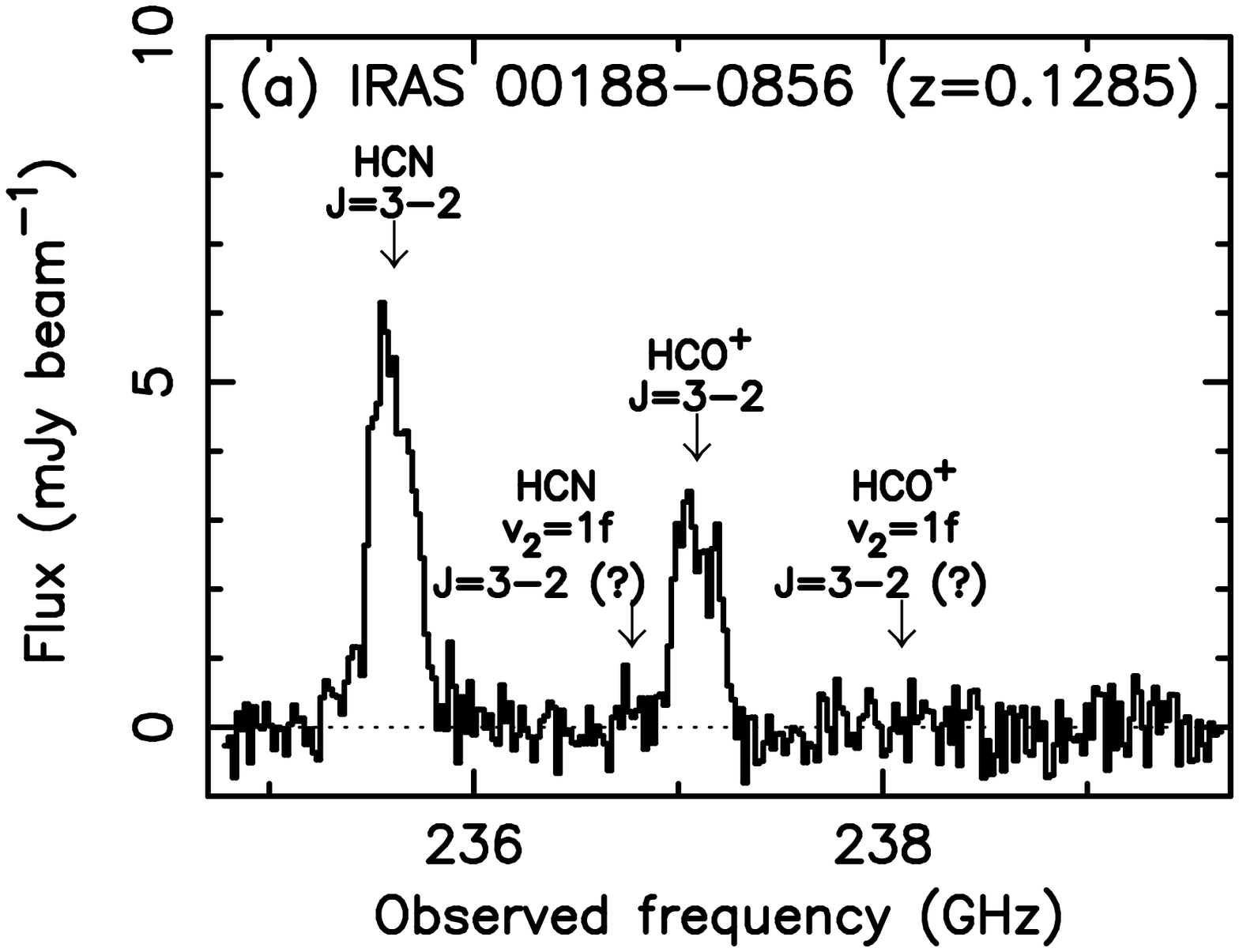} 
\includegraphics[angle=0,scale=.4]{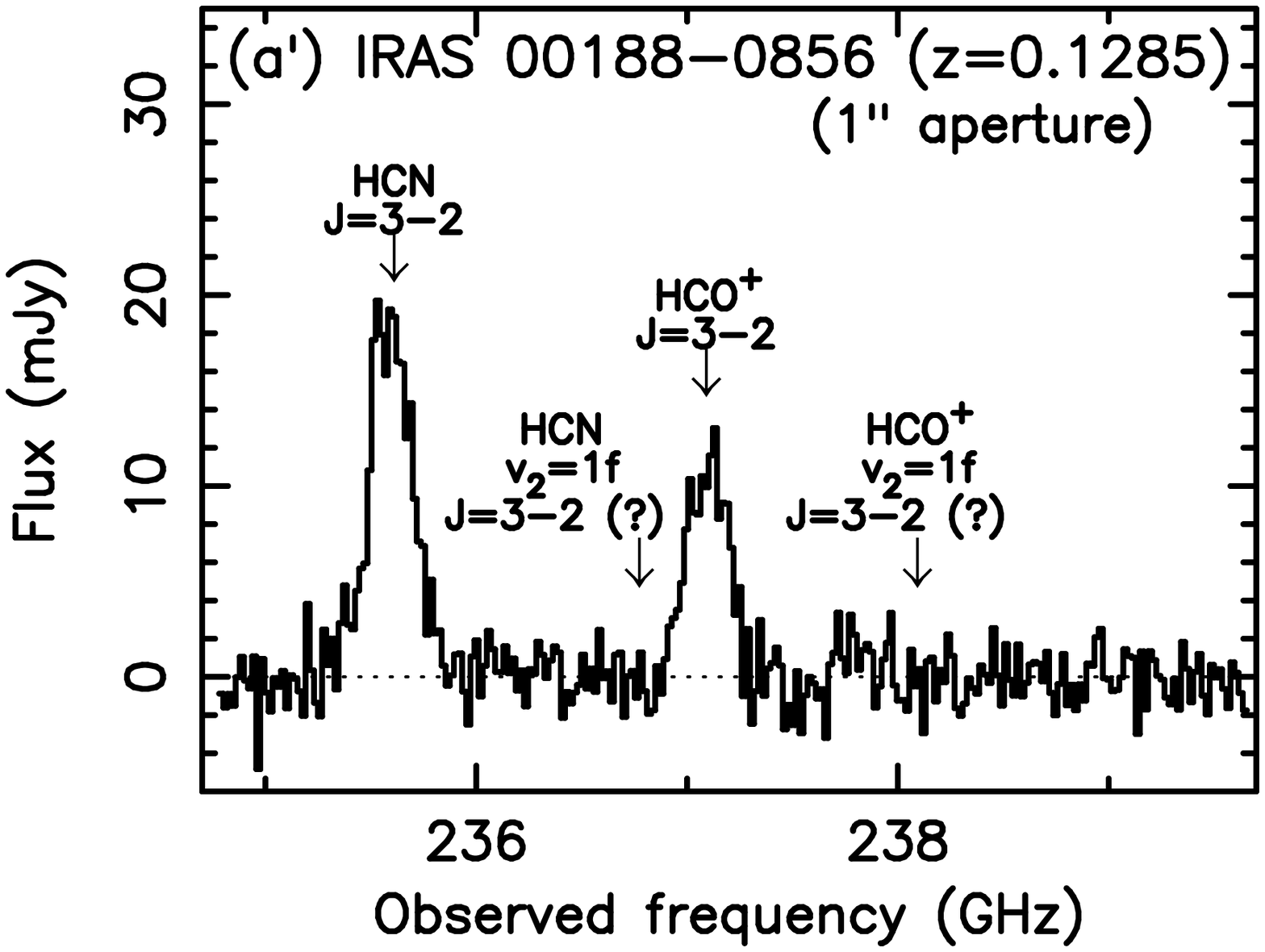} \\
\includegraphics[angle=0,scale=.4]{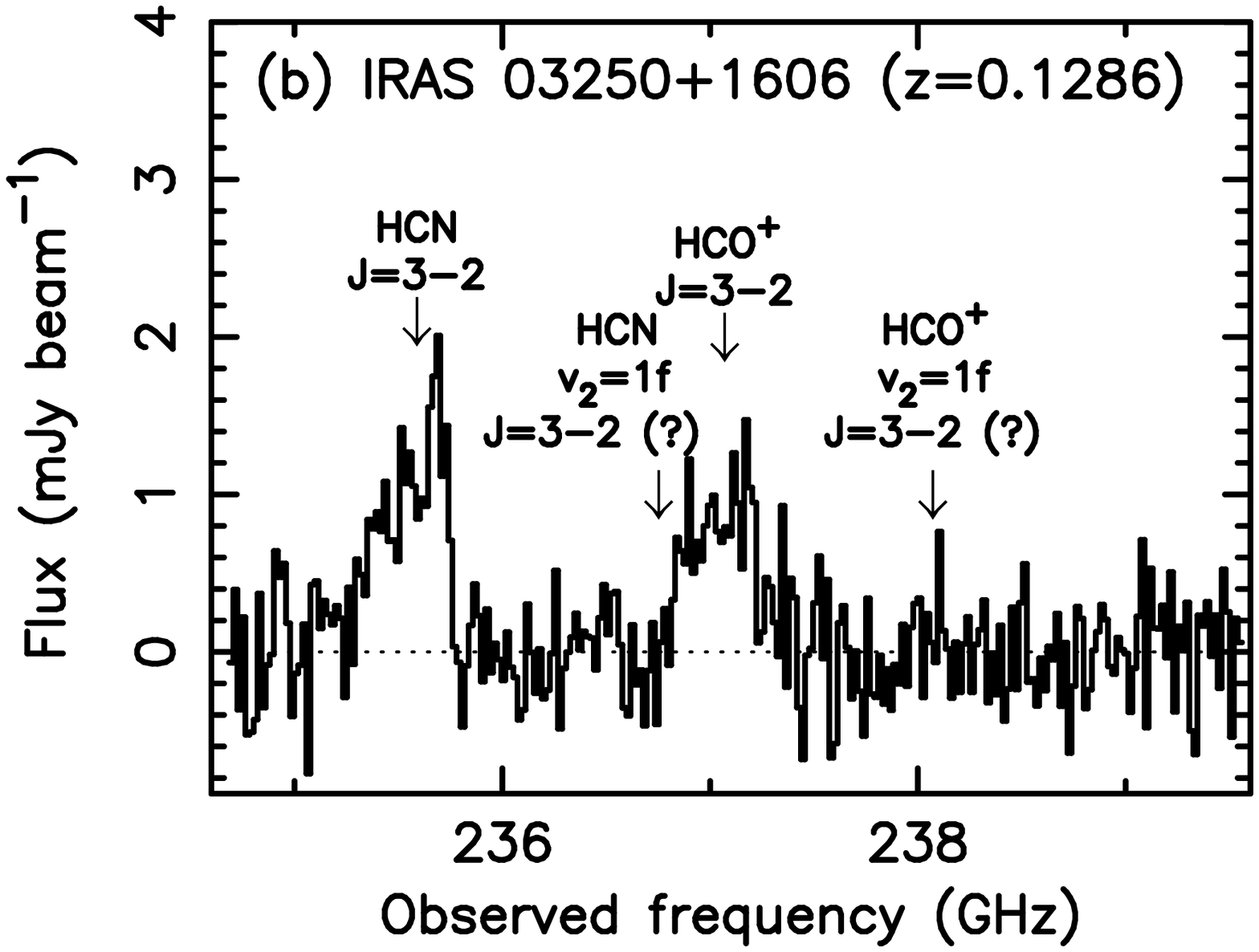} 
\includegraphics[angle=0,scale=.4]{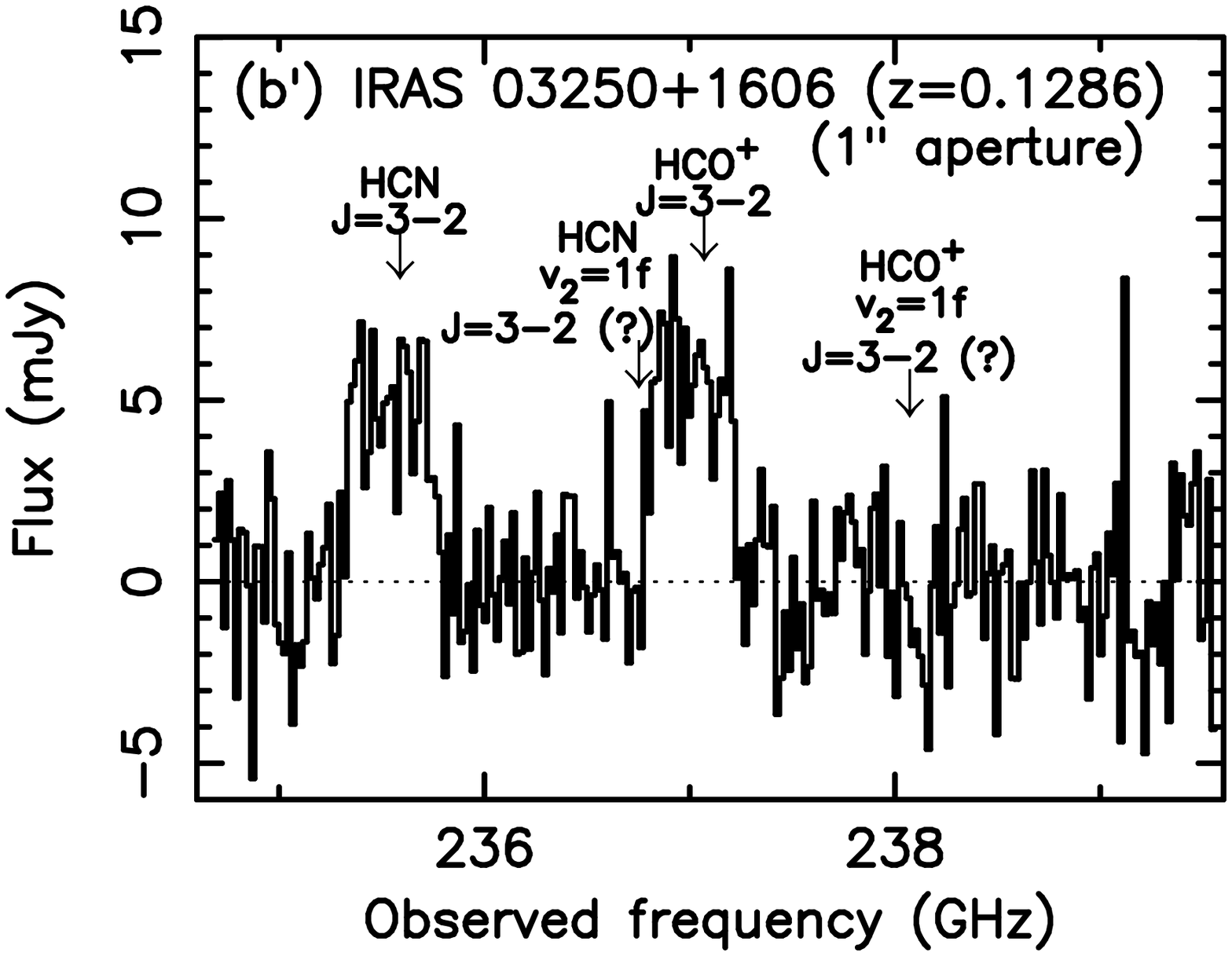} \\
\includegraphics[angle=0,scale=.4]{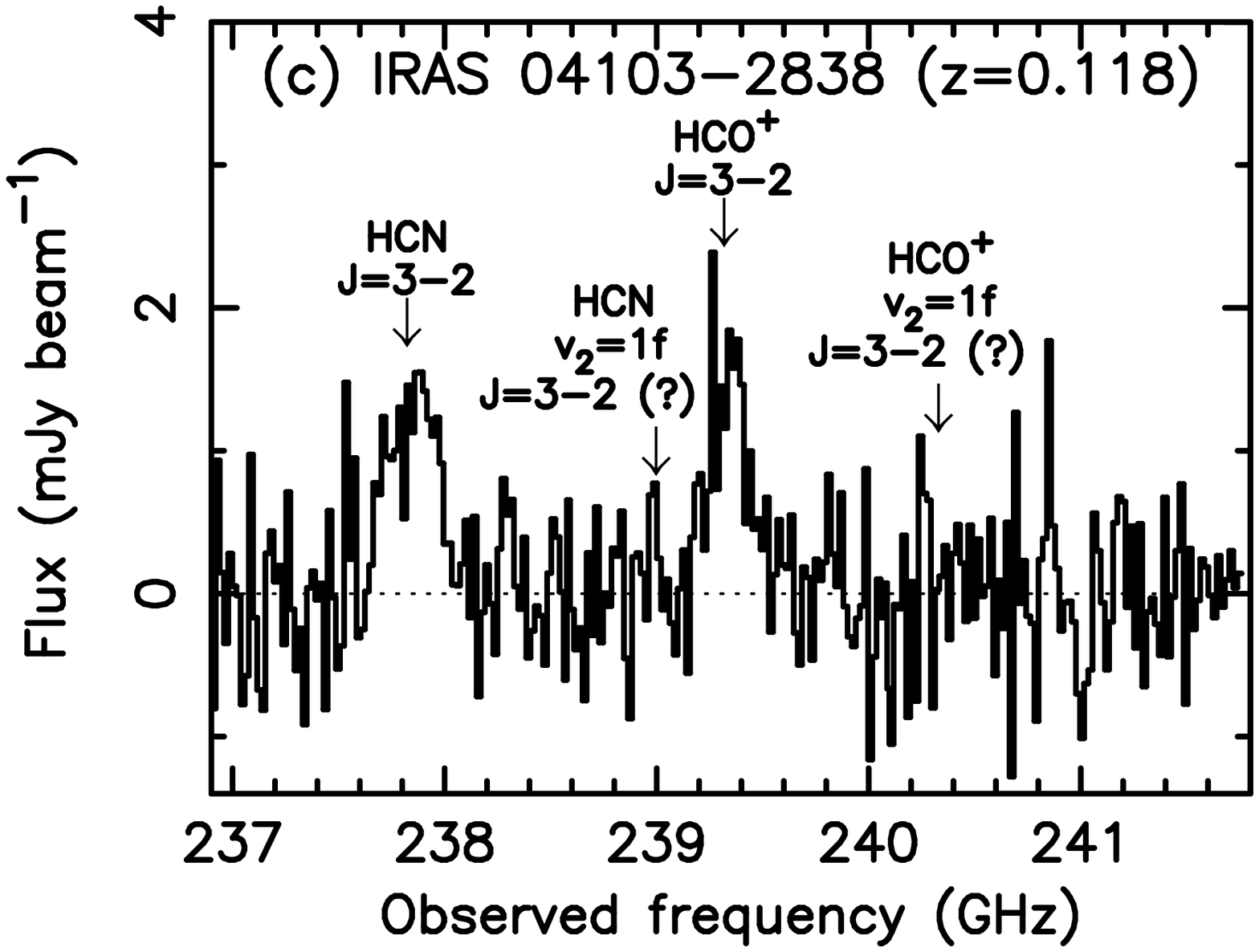} 
\includegraphics[angle=0,scale=.4]{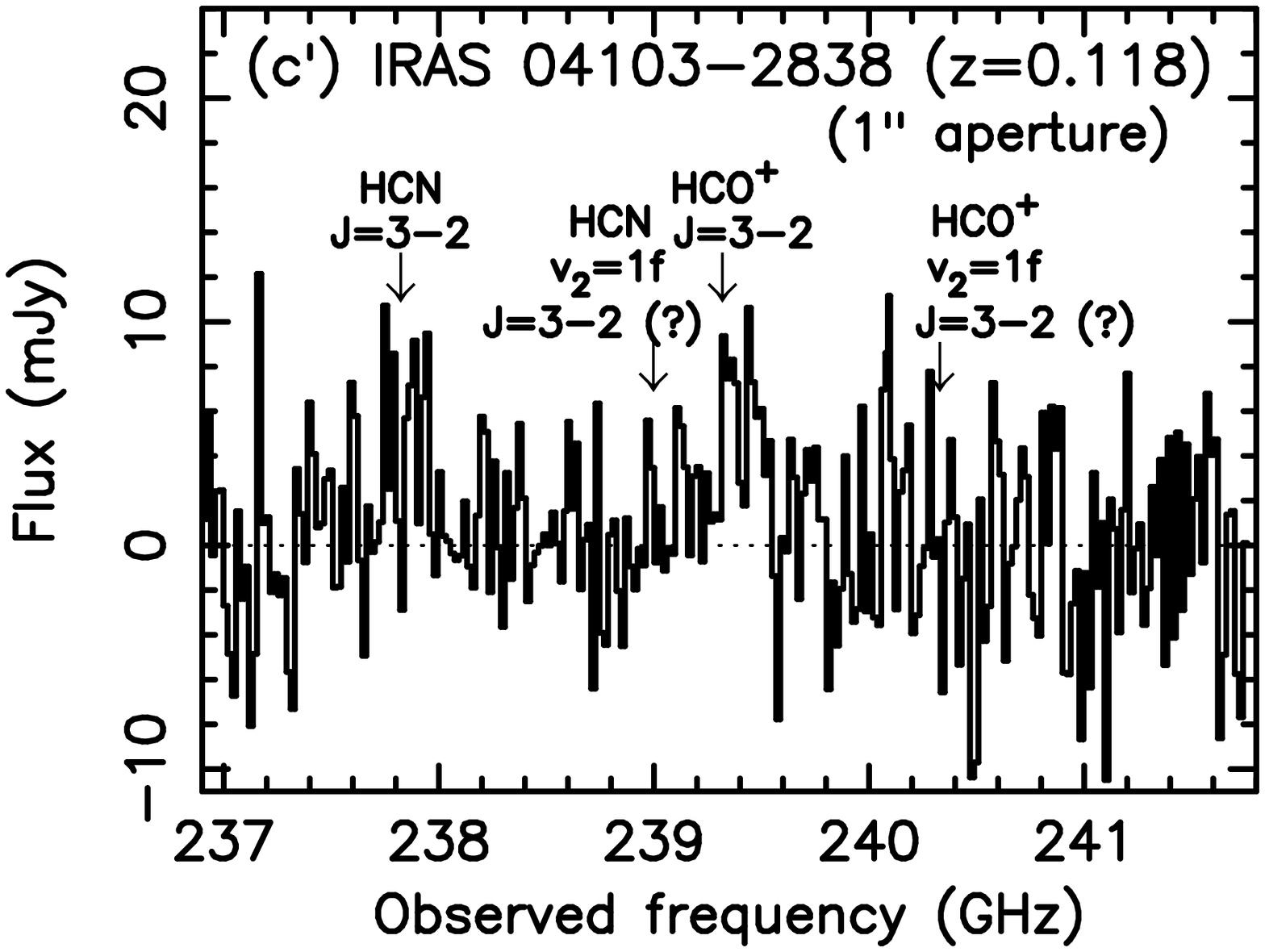} \\ 
\includegraphics[angle=0,scale=.4]{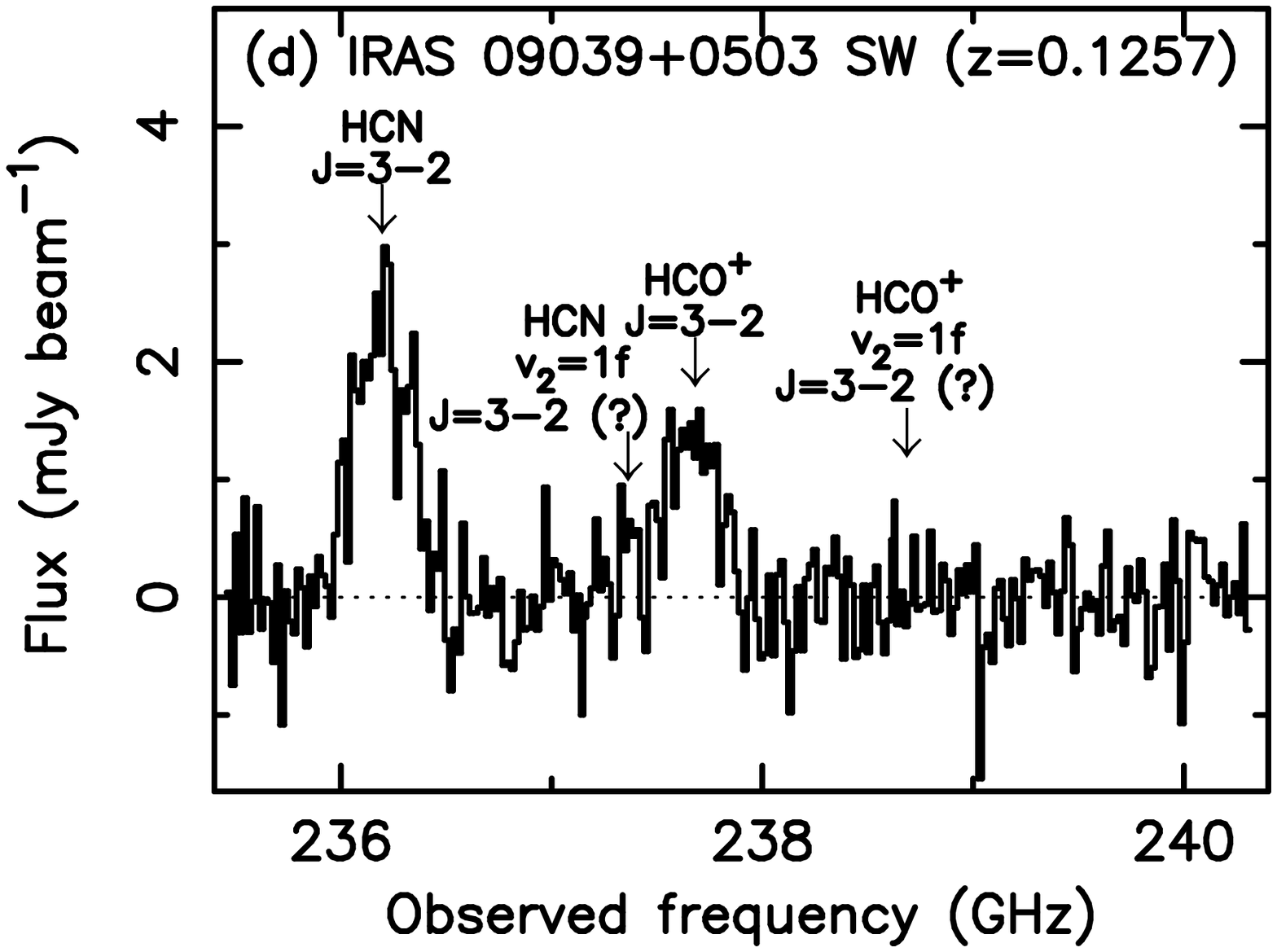}
\includegraphics[angle=0,scale=.4]{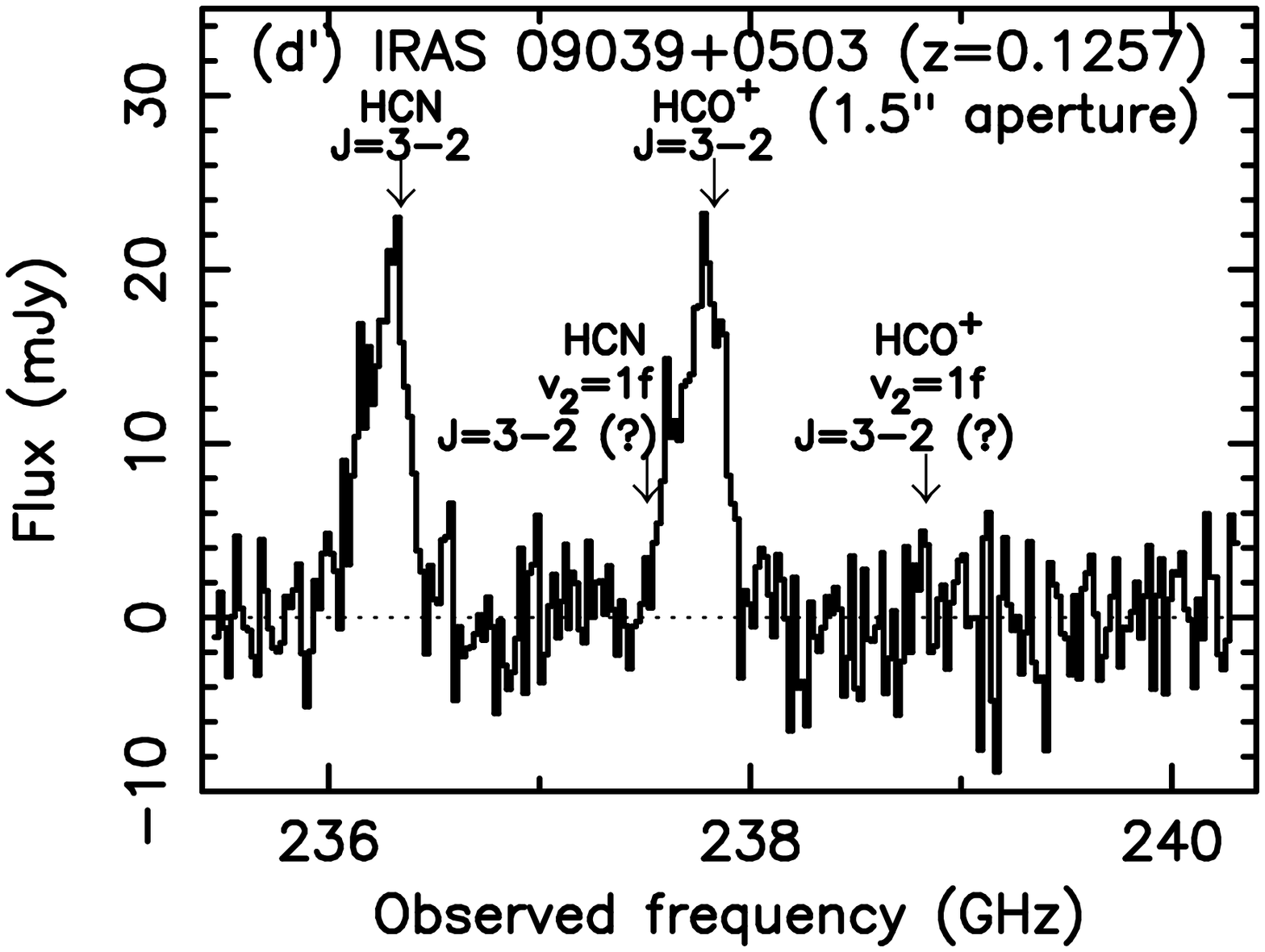} \\
\end{center}
\end{figure}

\clearpage

\begin{figure}
\begin{center}
\hspace*{-8.2cm} \includegraphics[angle=0,scale=.4]{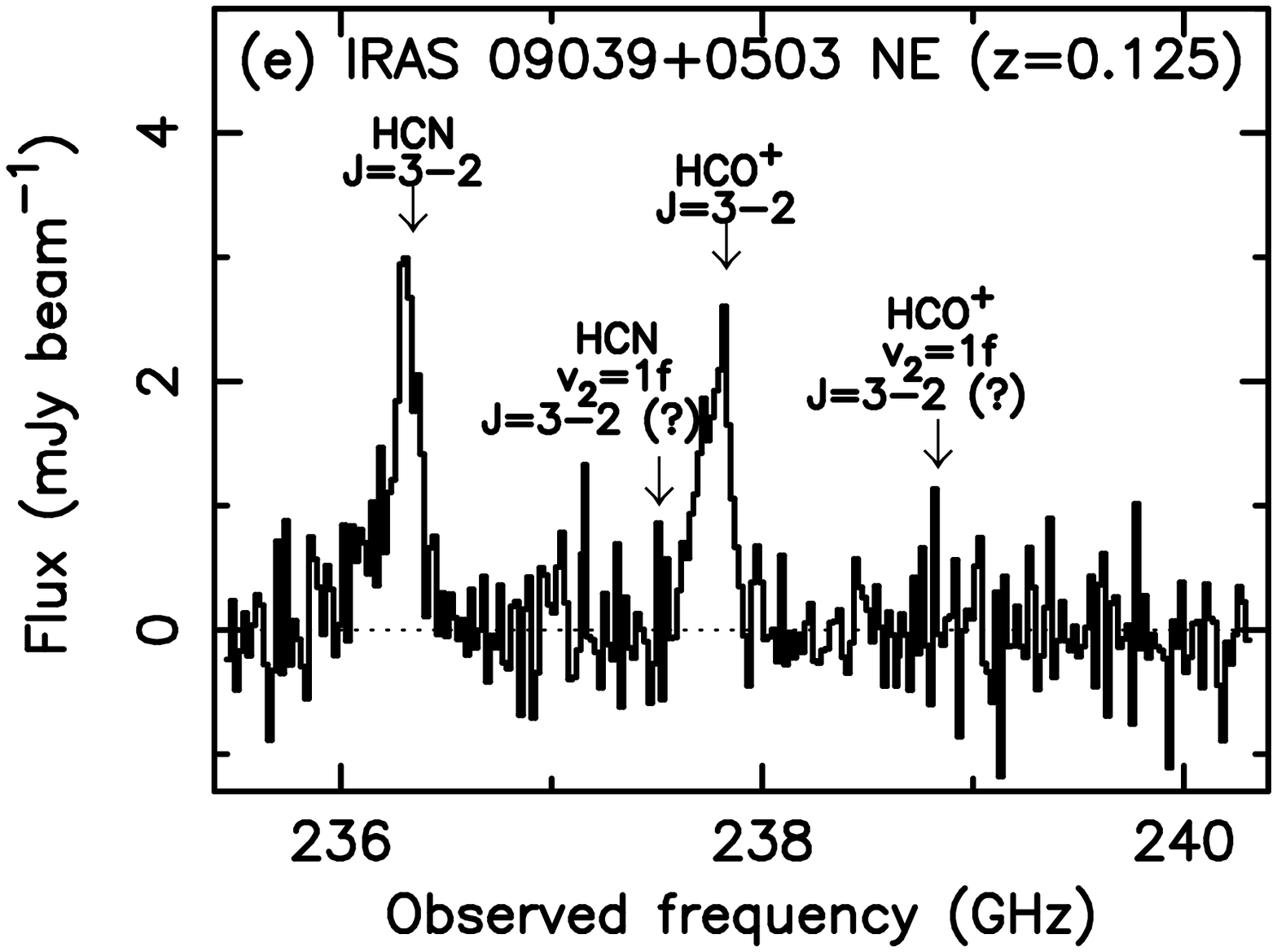} \\
\includegraphics[angle=0,scale=.4]{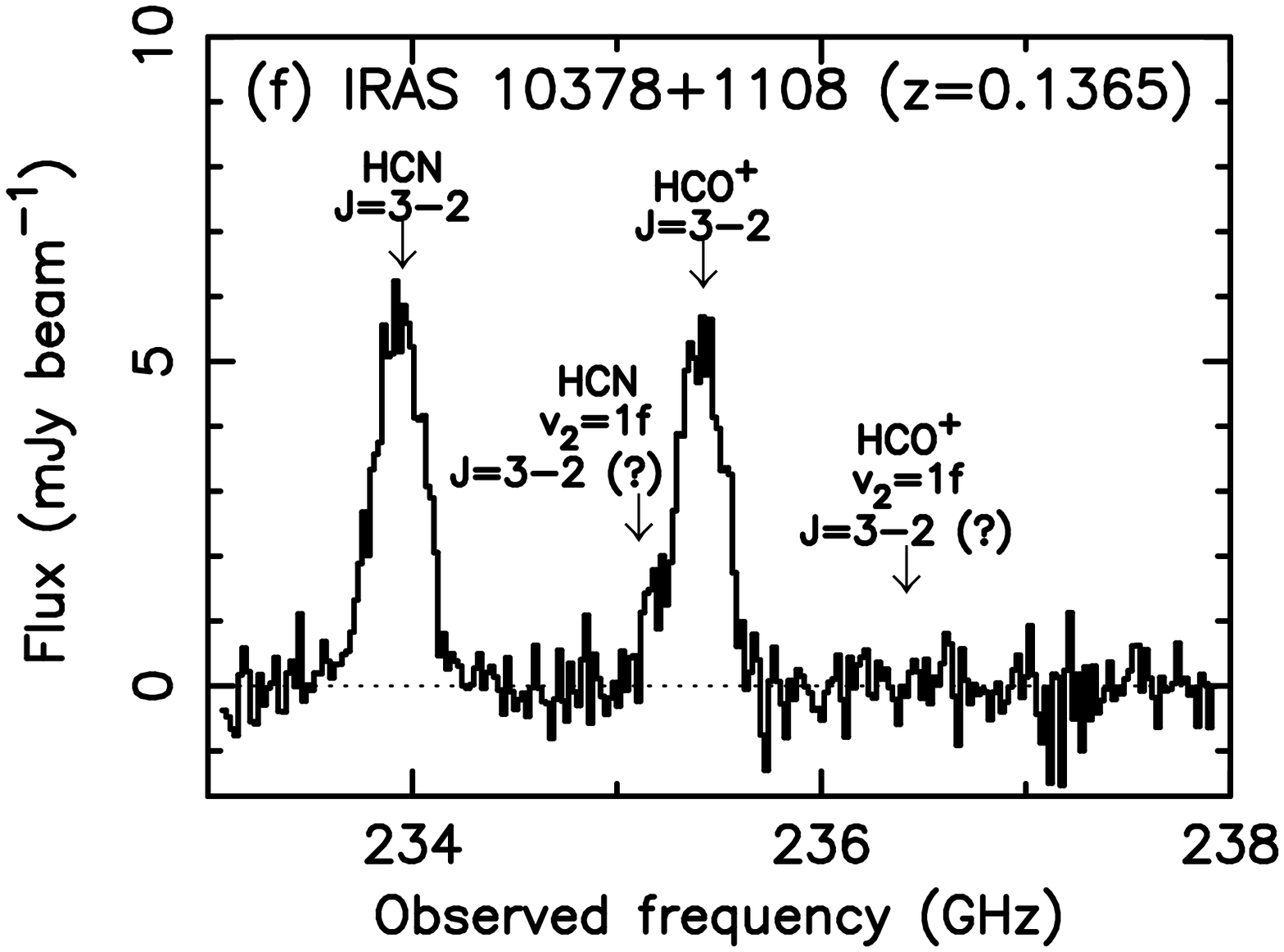} 
\includegraphics[angle=0,scale=.4]{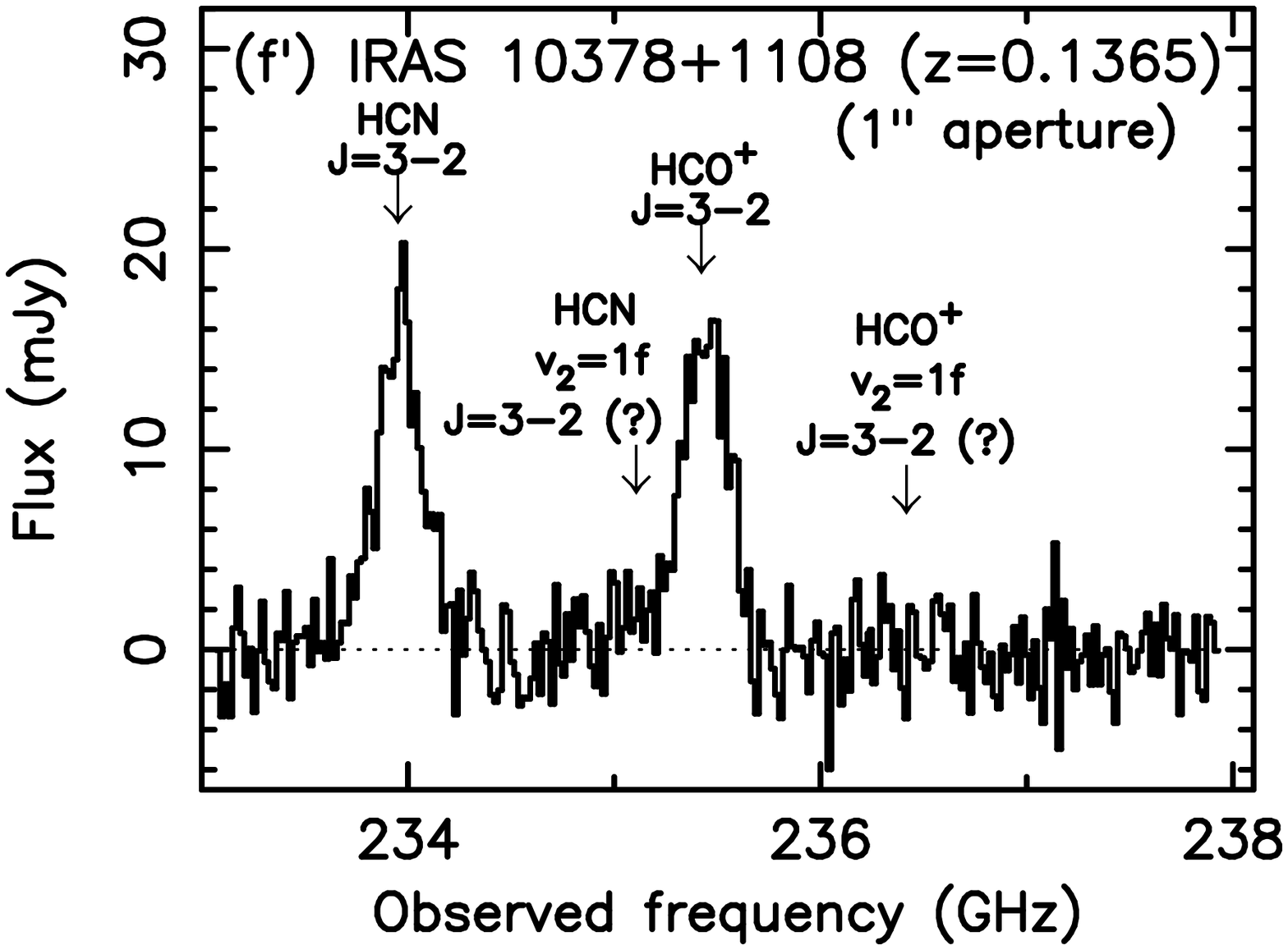} \\ 
\hspace*{-8.2cm} \includegraphics[angle=0,scale=.4]{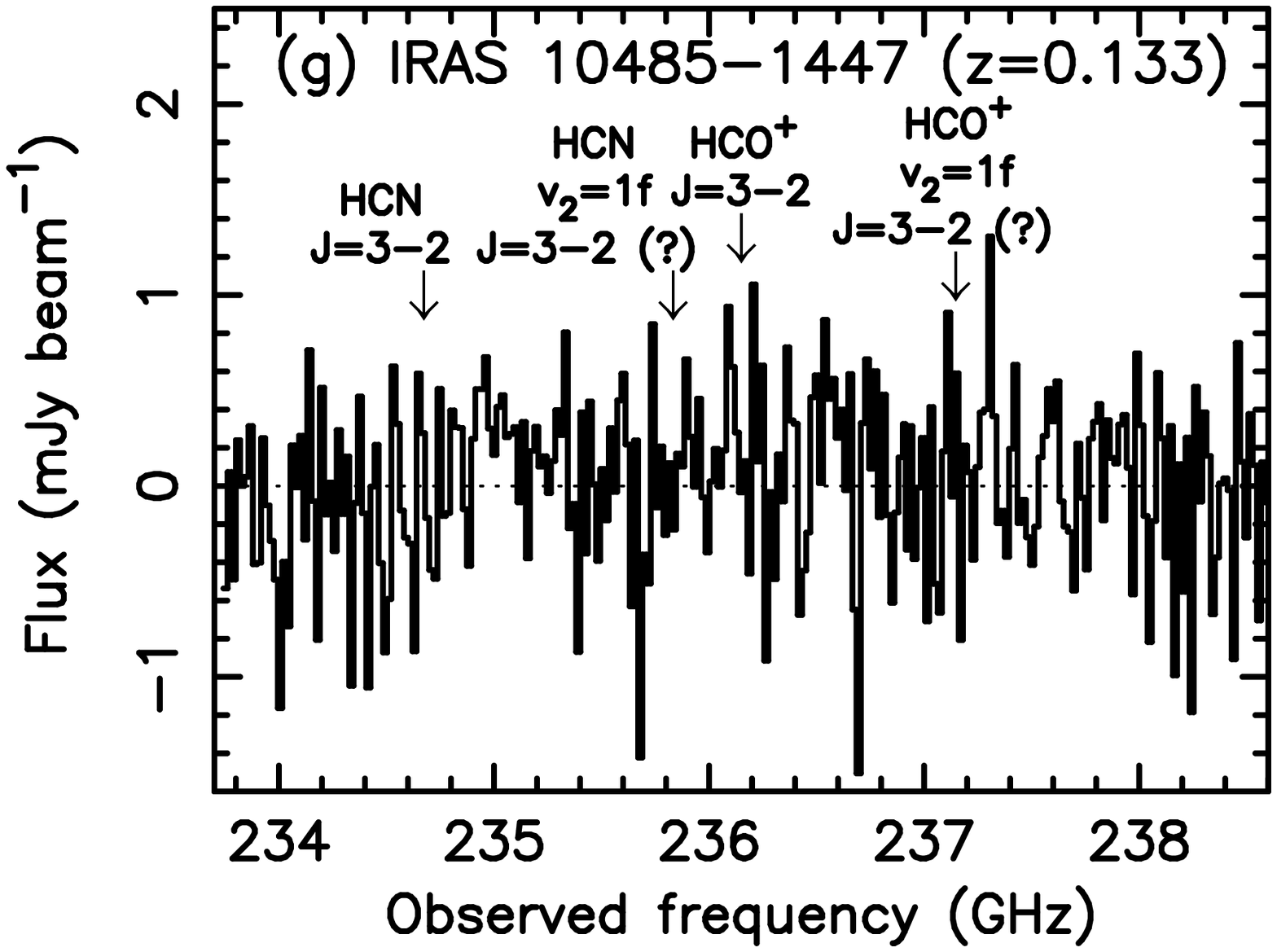} \\
\includegraphics[angle=0,scale=.4]{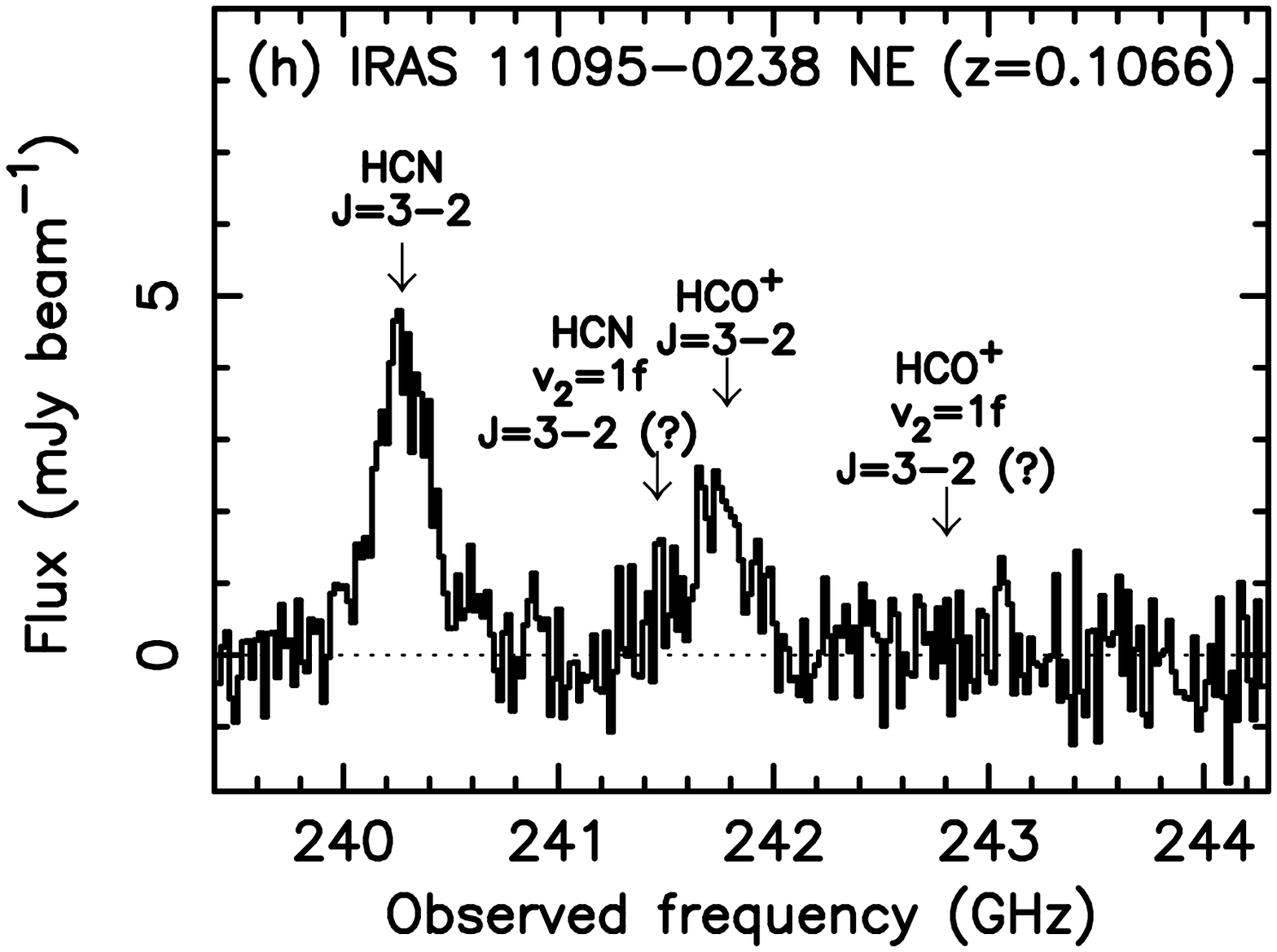} 
\includegraphics[angle=0,scale=.4]{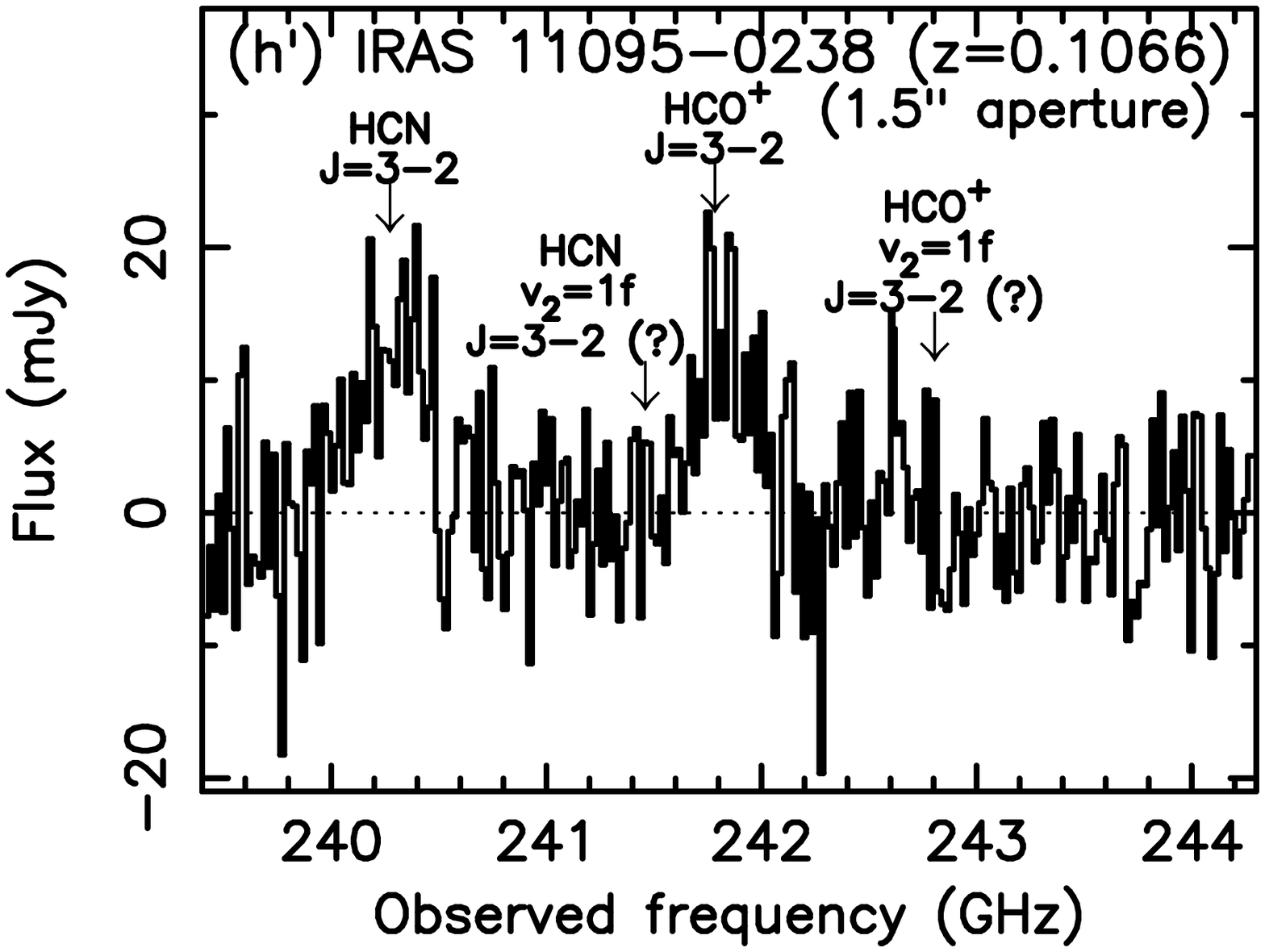} \\ 
\end{center}
\end{figure}

\clearpage

\begin{figure}
\begin{center}
\hspace*{-8.2cm}  \includegraphics[angle=0,scale=.4]{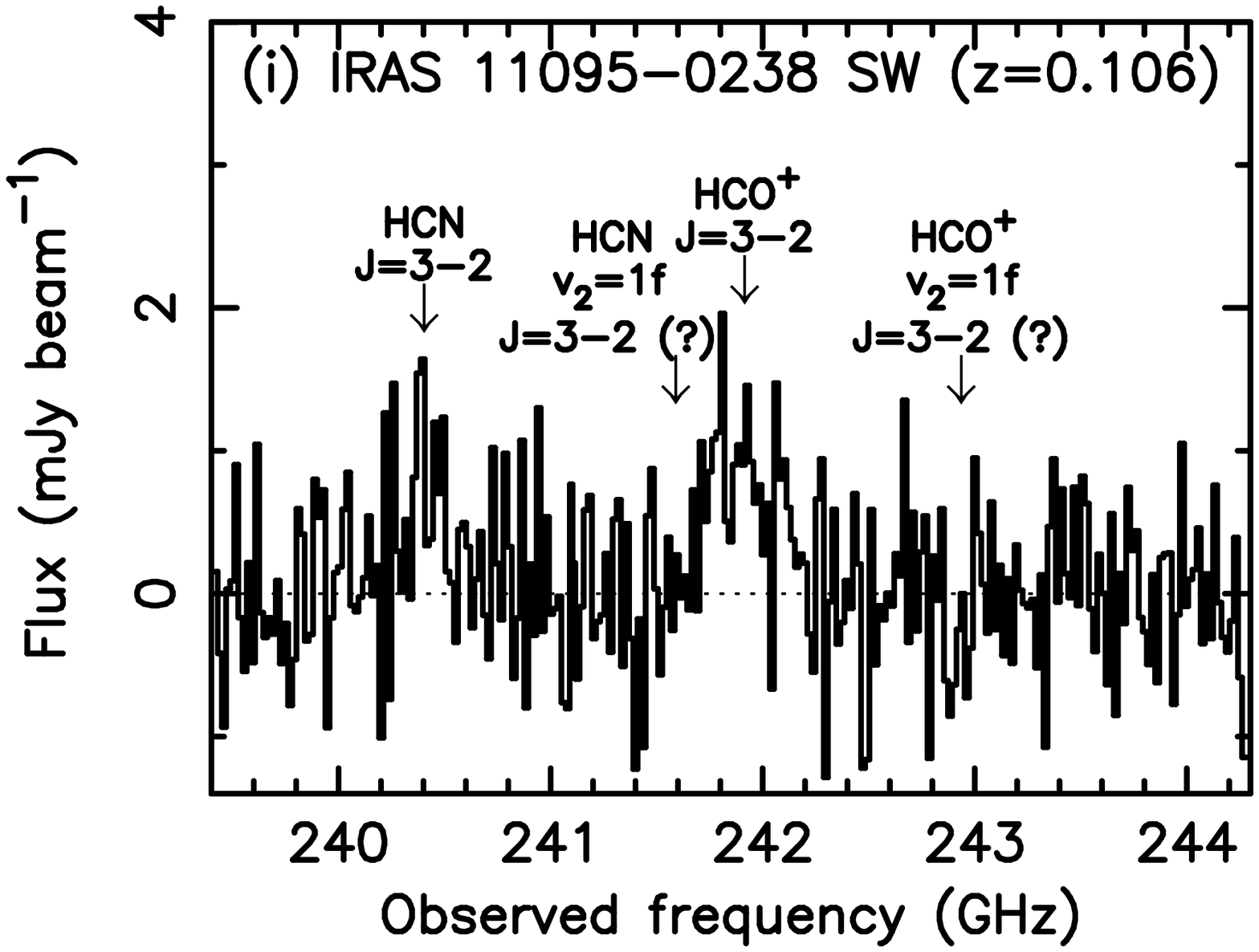} \\
\includegraphics[angle=0,scale=.4]{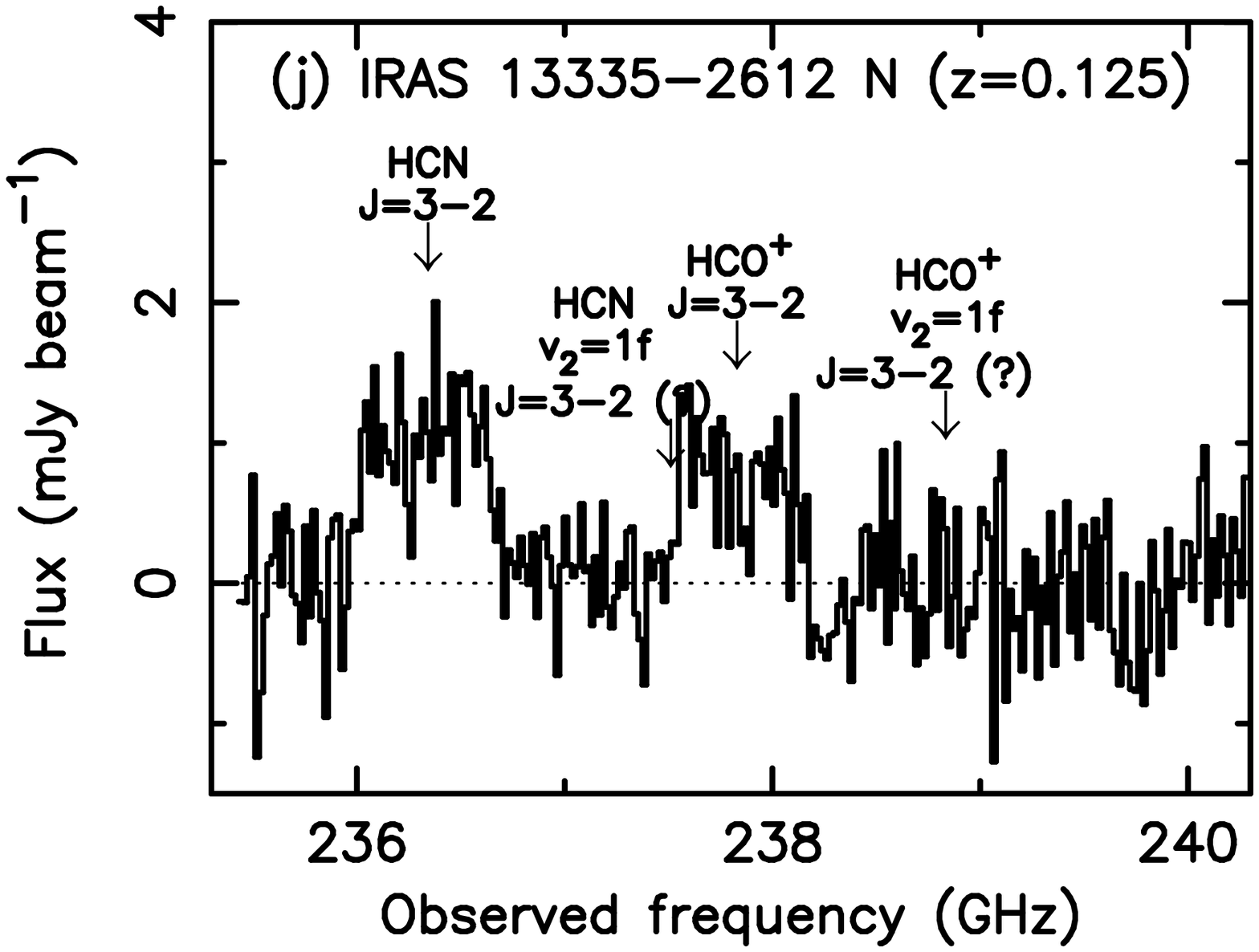} 
\includegraphics[angle=0,scale=.4]{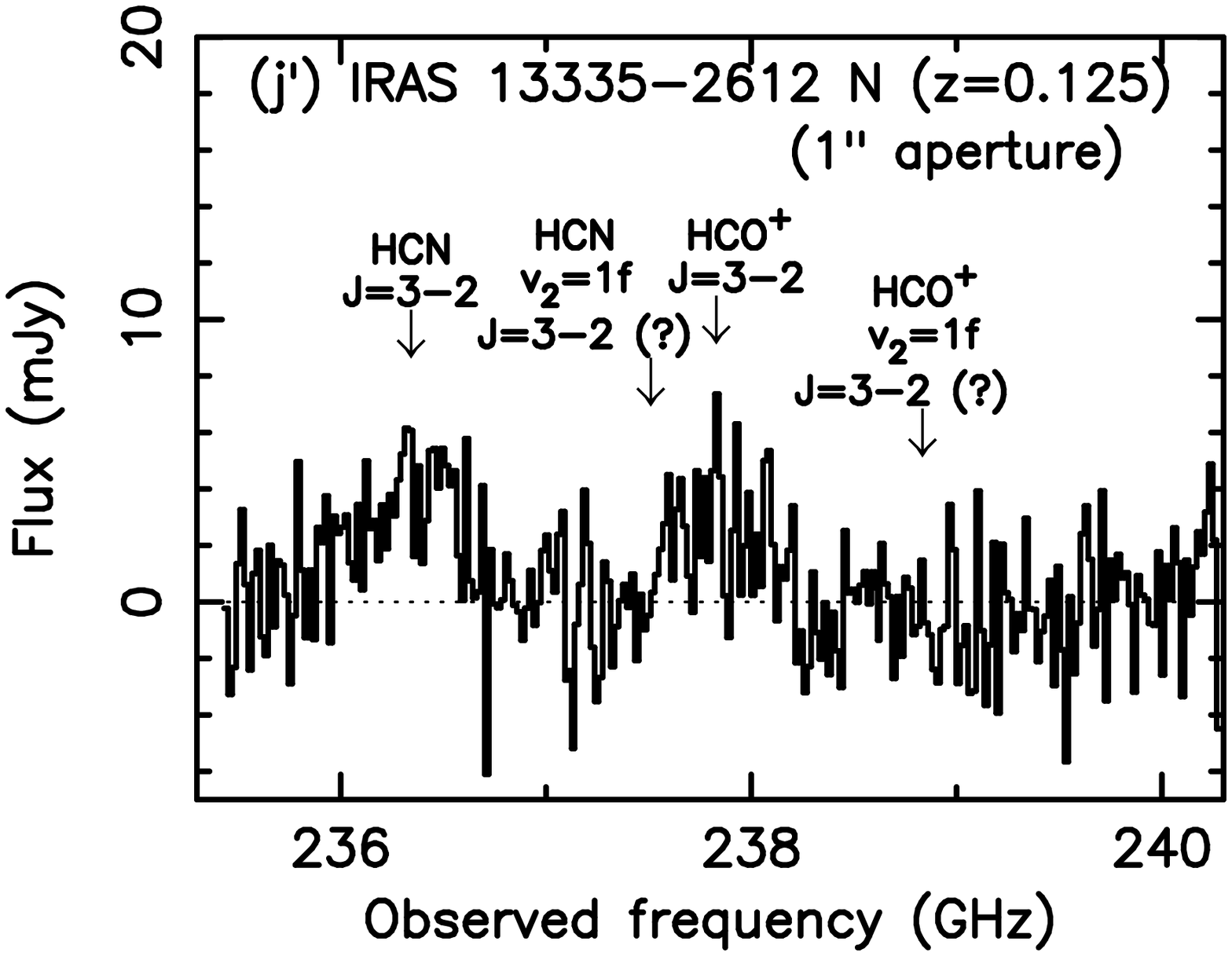} \\ 
\includegraphics[angle=0,scale=.4]{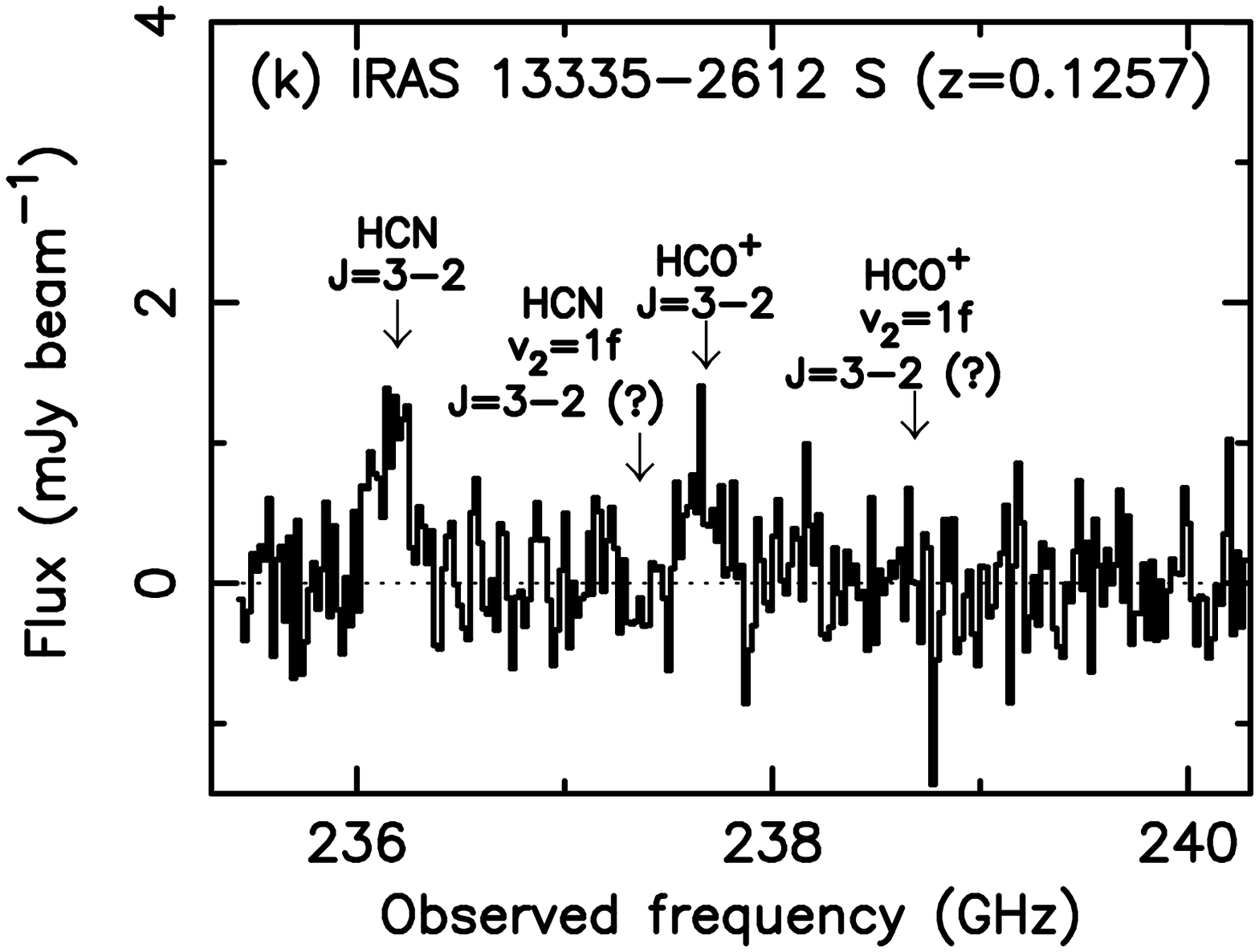} 
\includegraphics[angle=0,scale=.4]{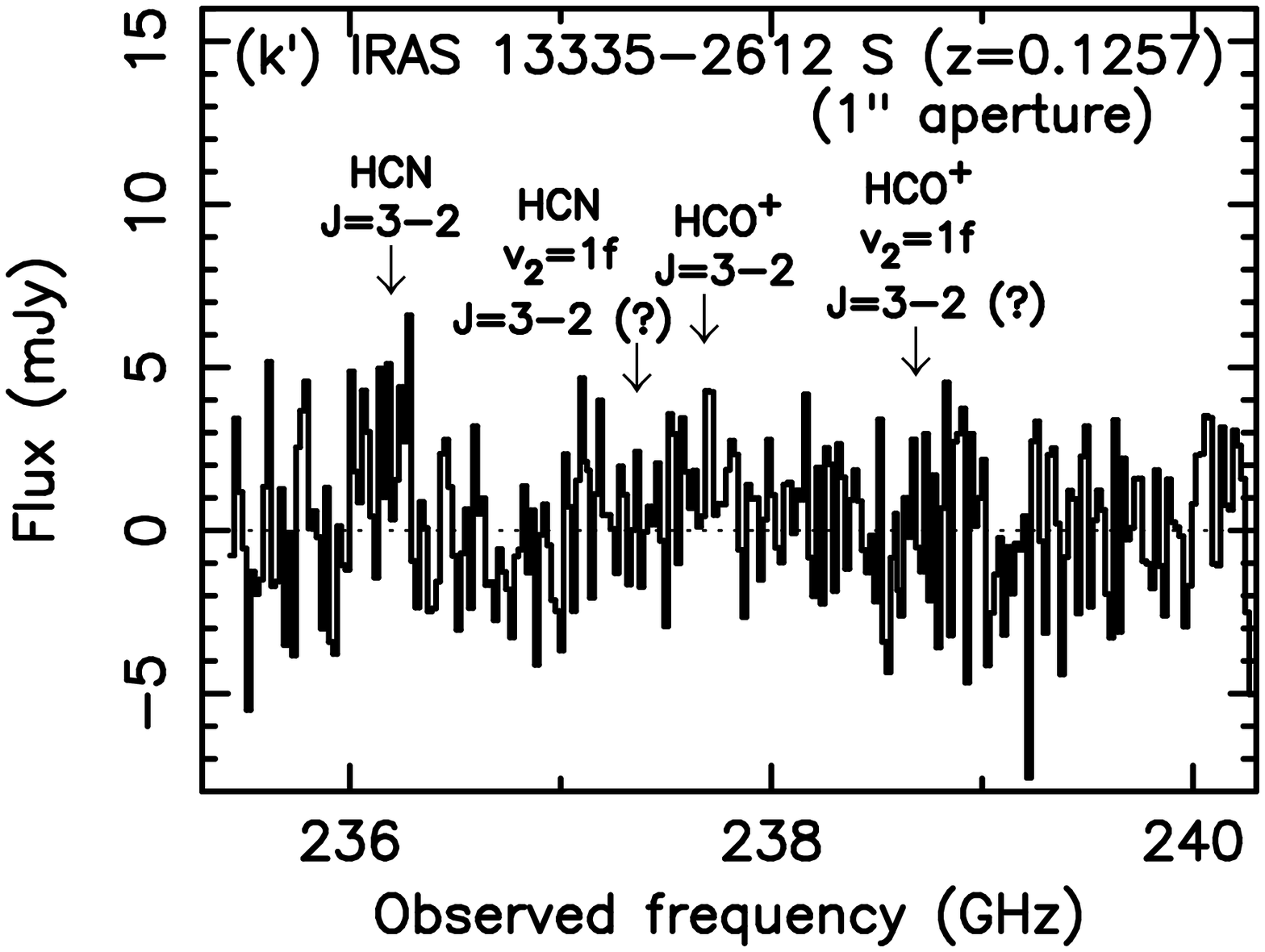} \\
\includegraphics[angle=0,scale=.4]{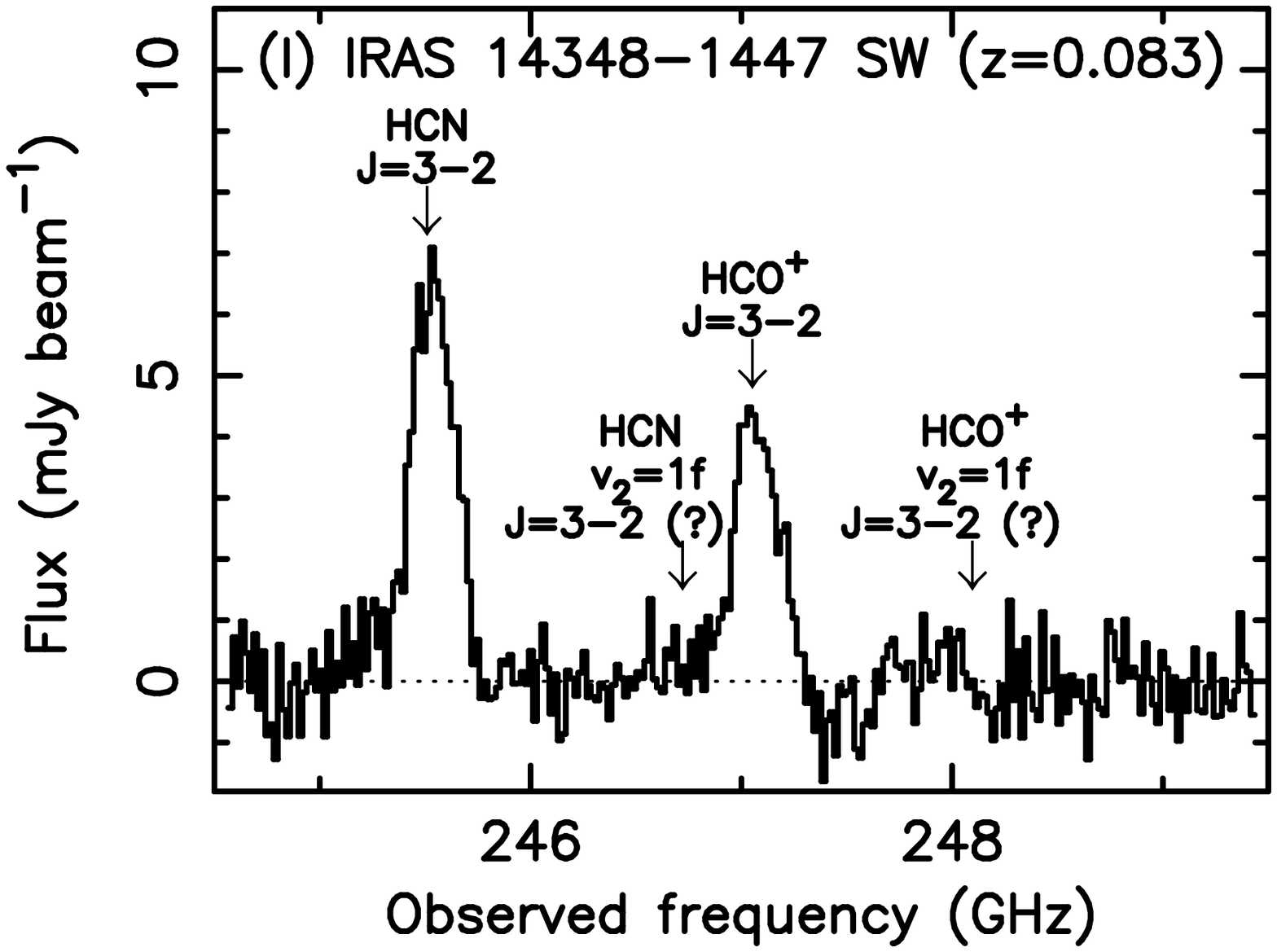} 
\includegraphics[angle=0,scale=.4]{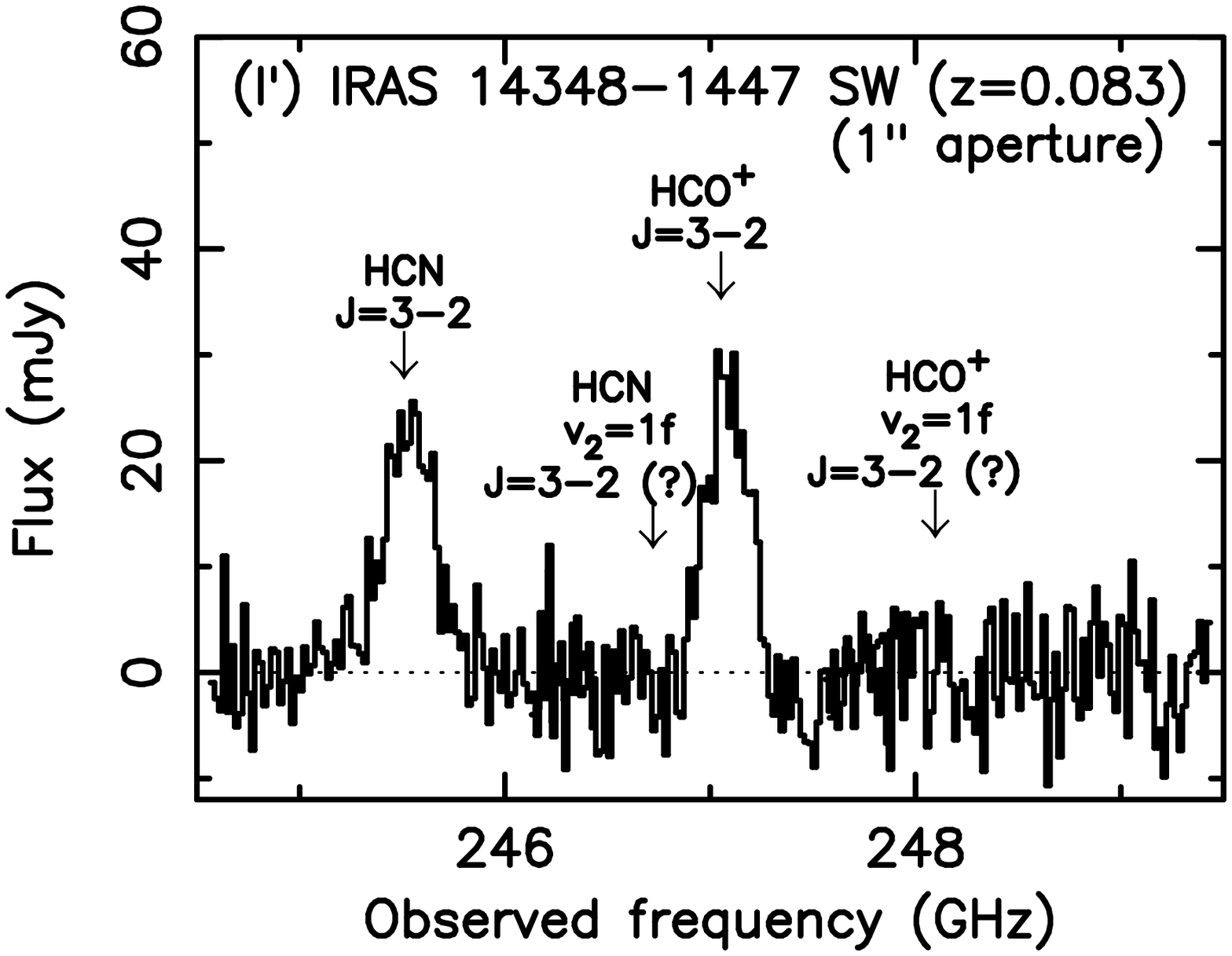} \\
\end{center}
\end{figure}

\clearpage

\begin{figure}
\begin{center}
\includegraphics[angle=0,scale=.4]{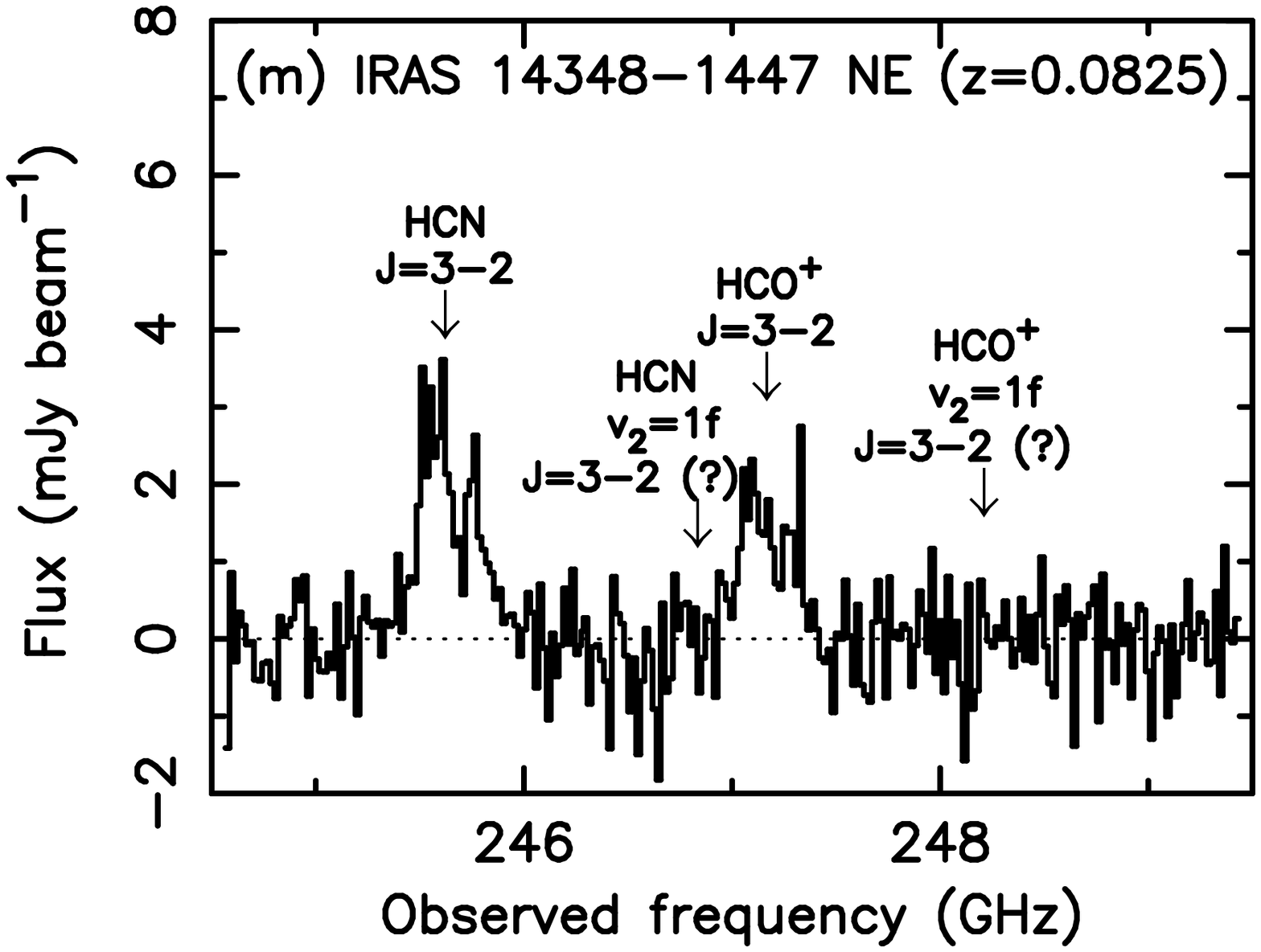} 
\includegraphics[angle=0,scale=.4]{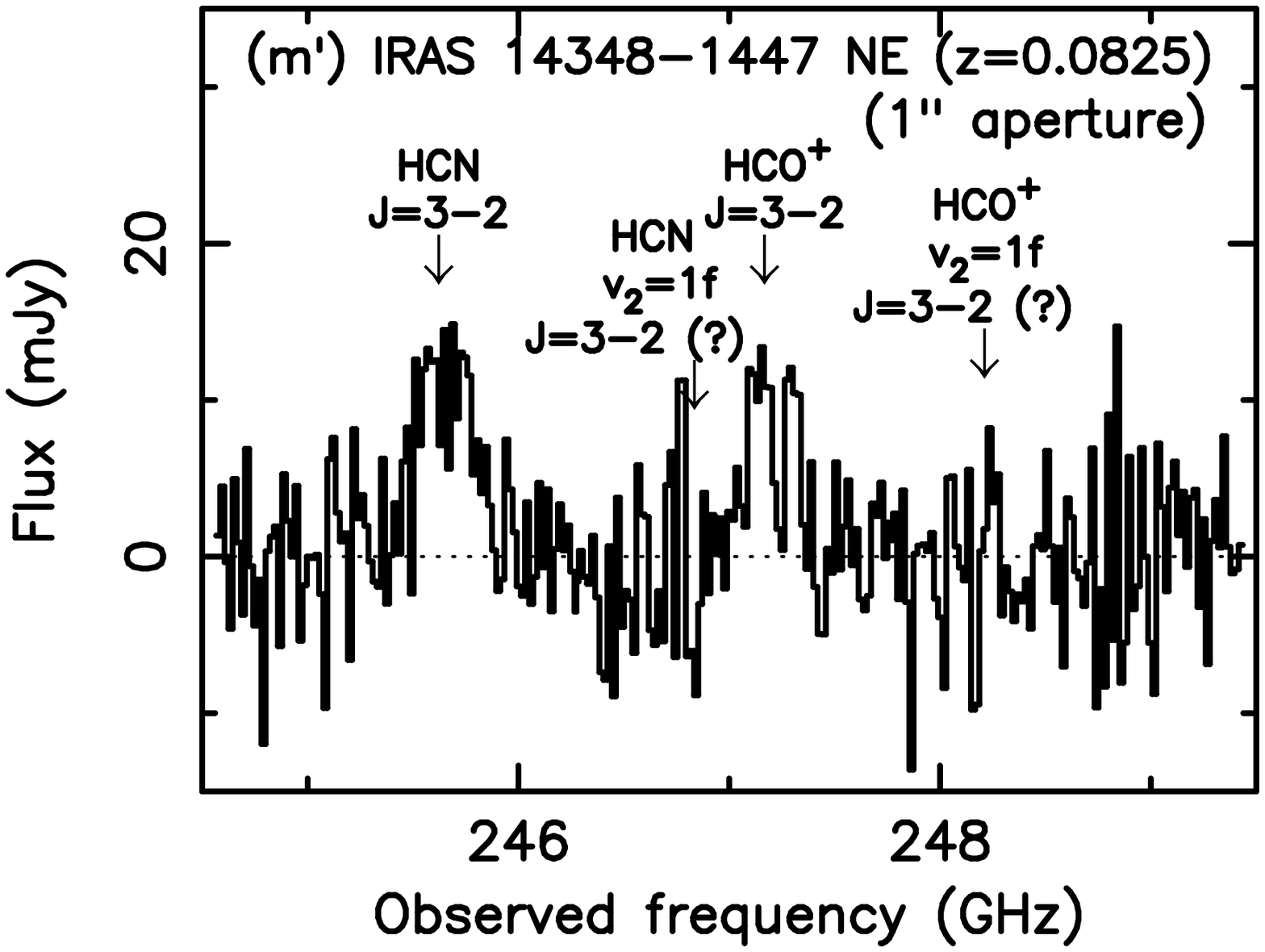} \\
\includegraphics[angle=0,scale=.4]{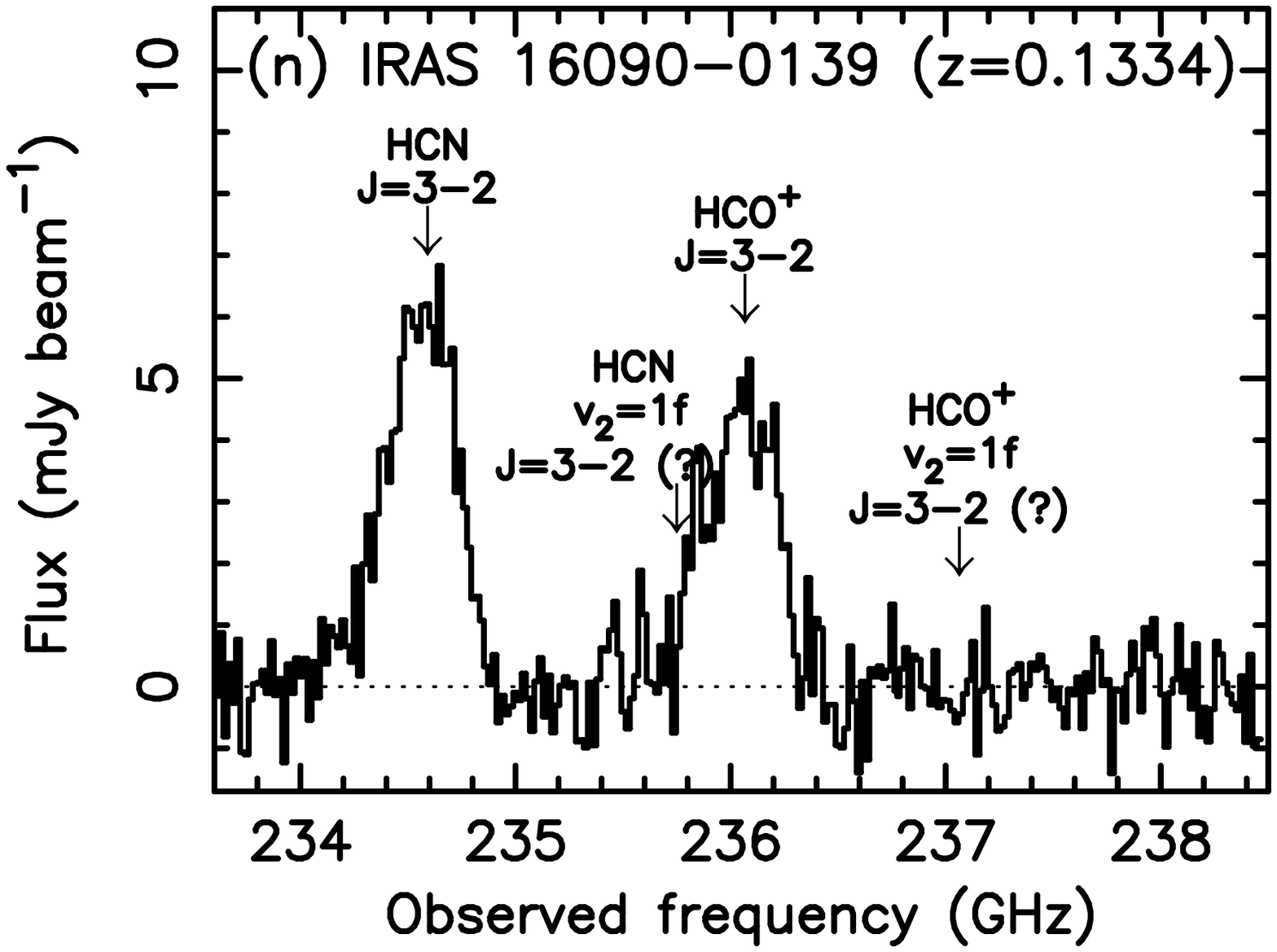} 
\includegraphics[angle=0,scale=.4]{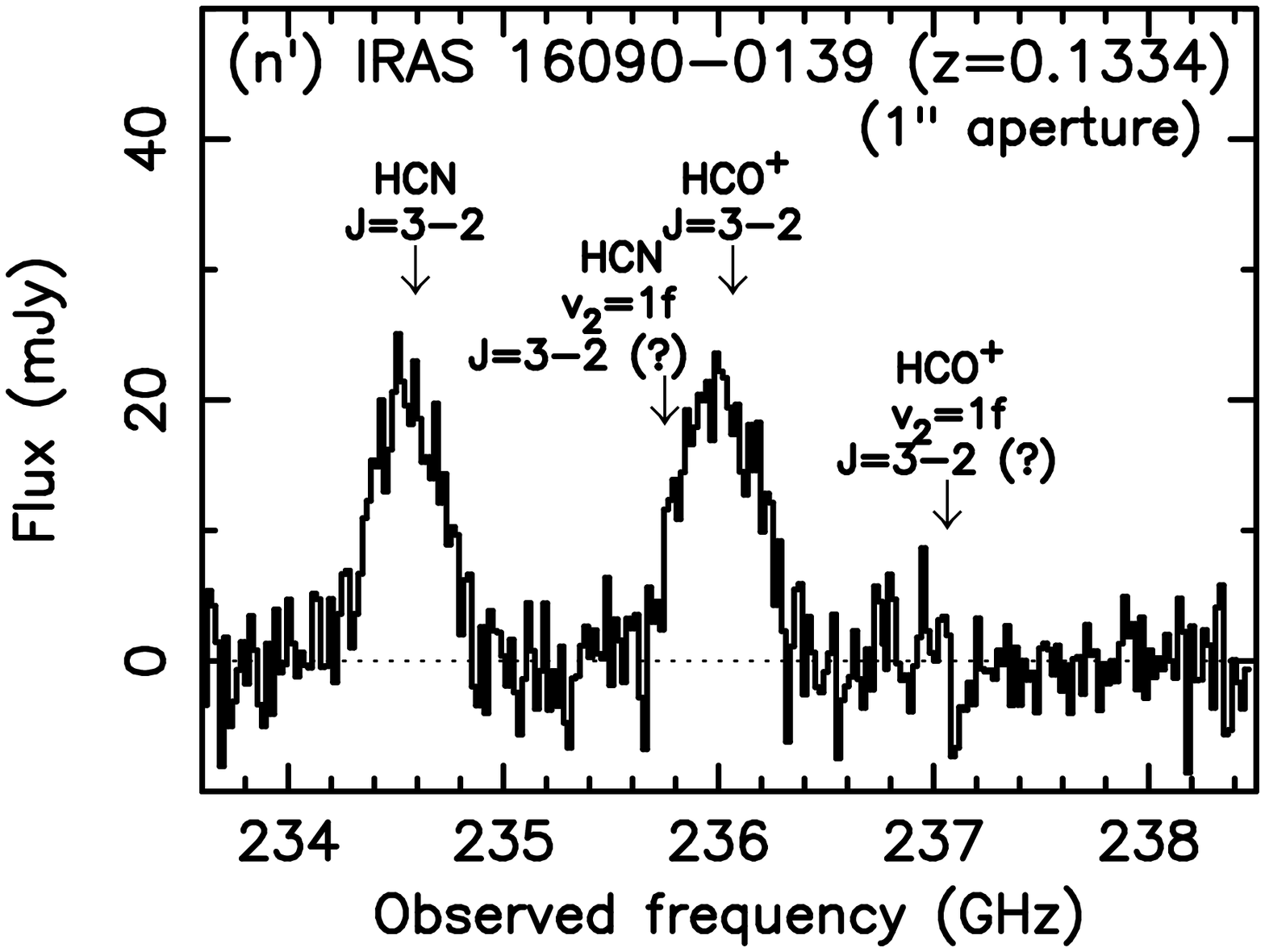} \\ 
\includegraphics[angle=0,scale=.4]{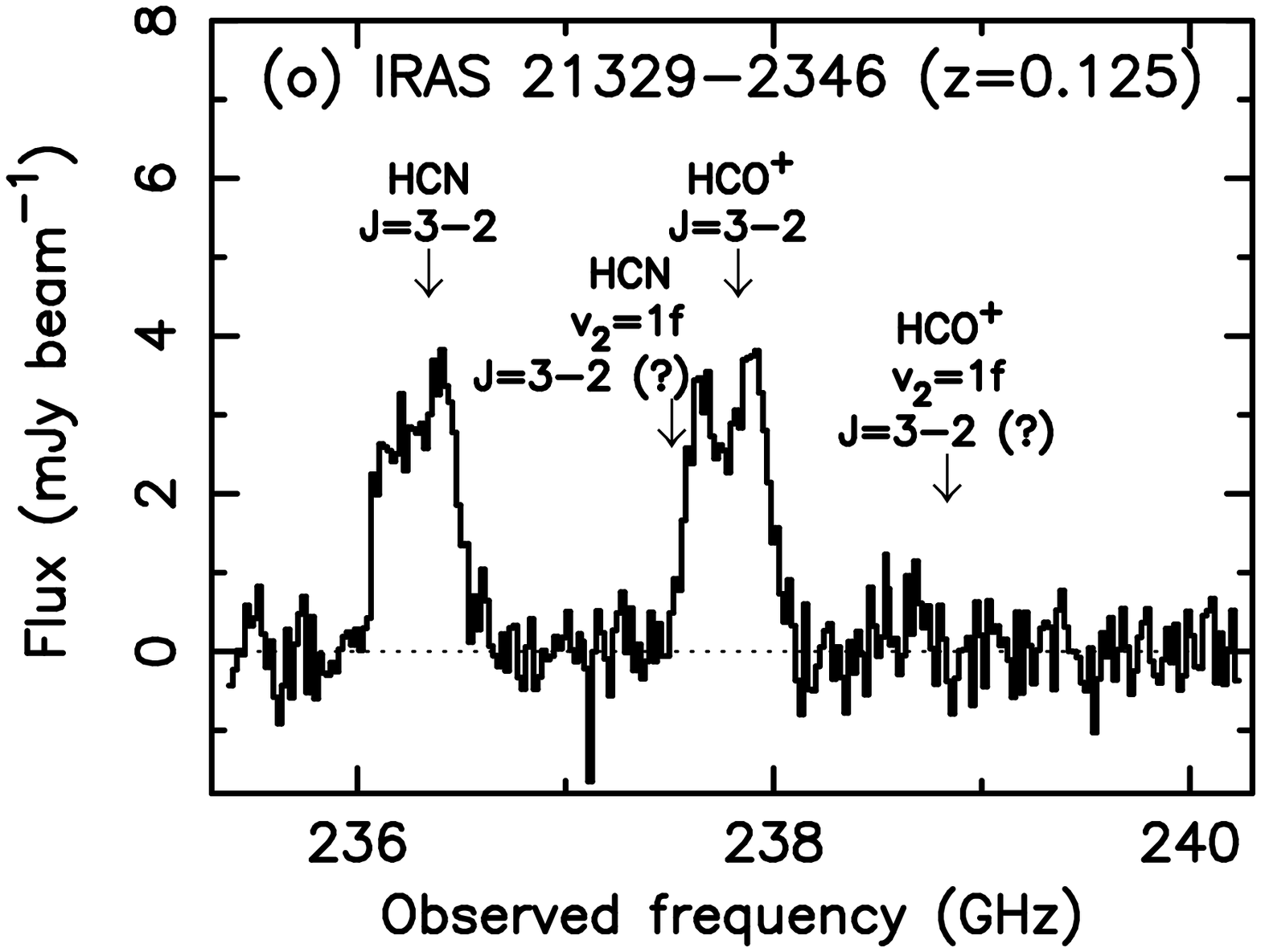} 
\includegraphics[angle=0,scale=.4]{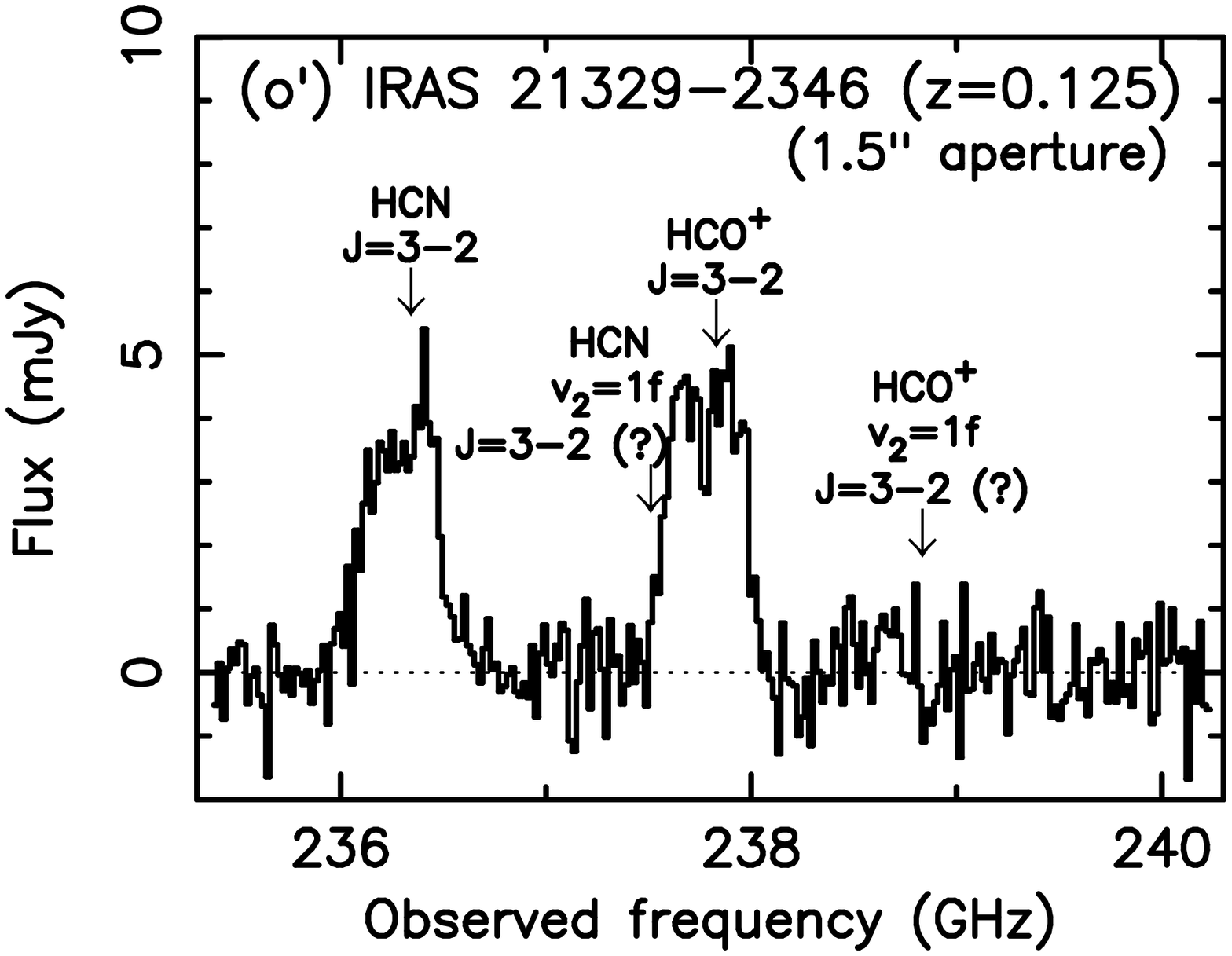} \\
\includegraphics[angle=0,scale=.4]{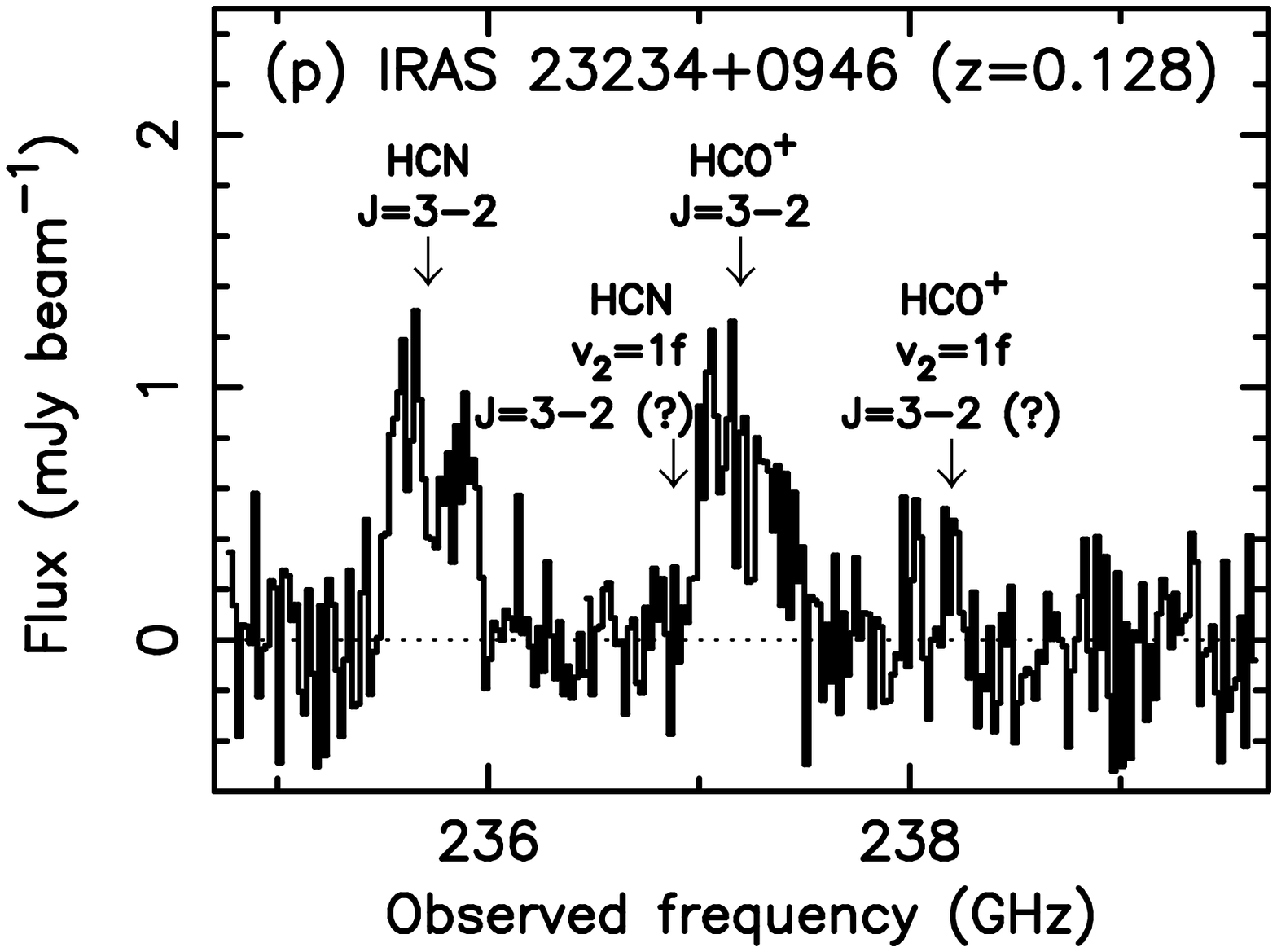} 
\includegraphics[angle=0,scale=.4]{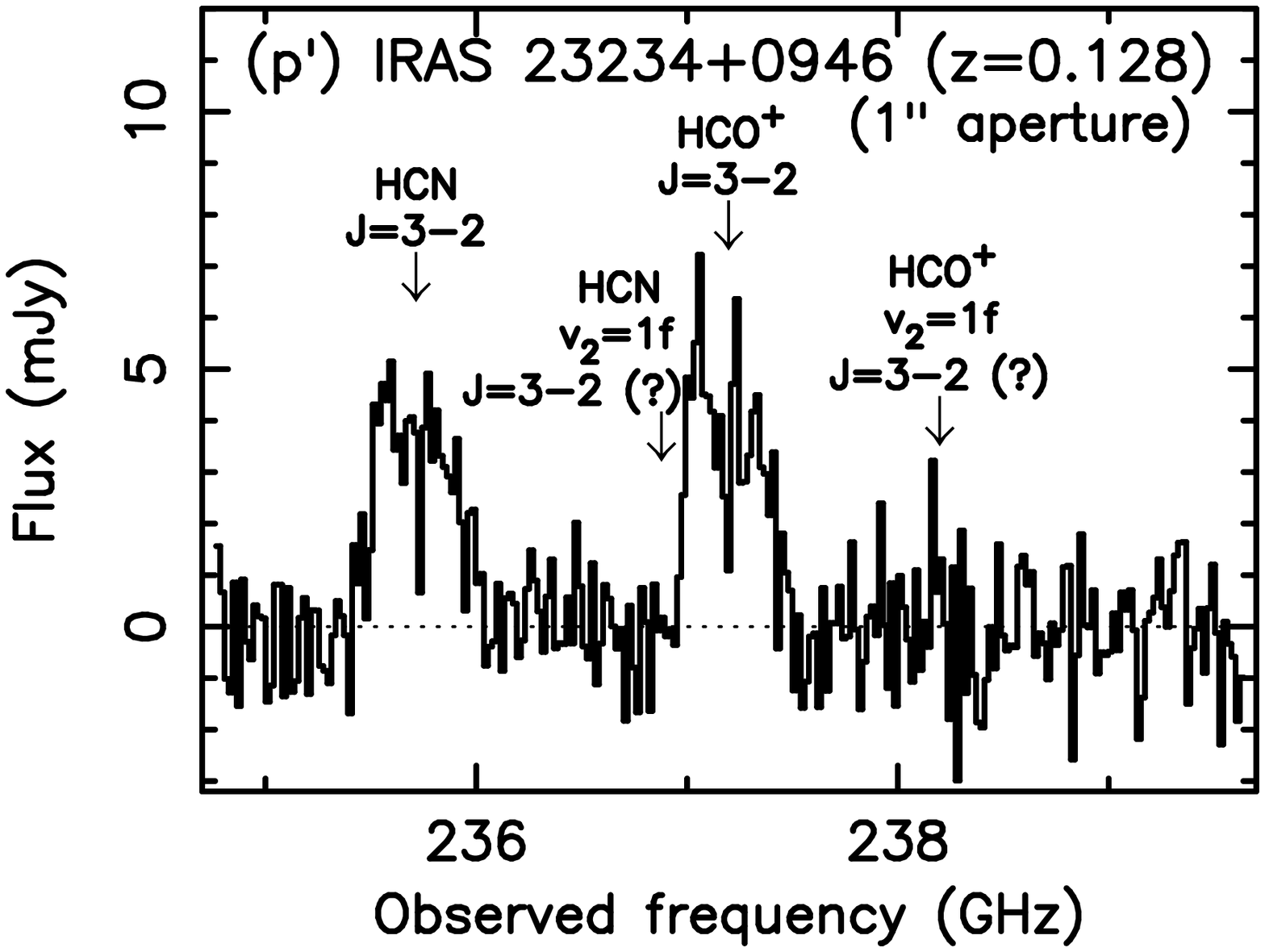} \\
\end{center}
\end{figure}

\clearpage

\begin{figure}
\begin{center}
\includegraphics[angle=0,scale=.4]{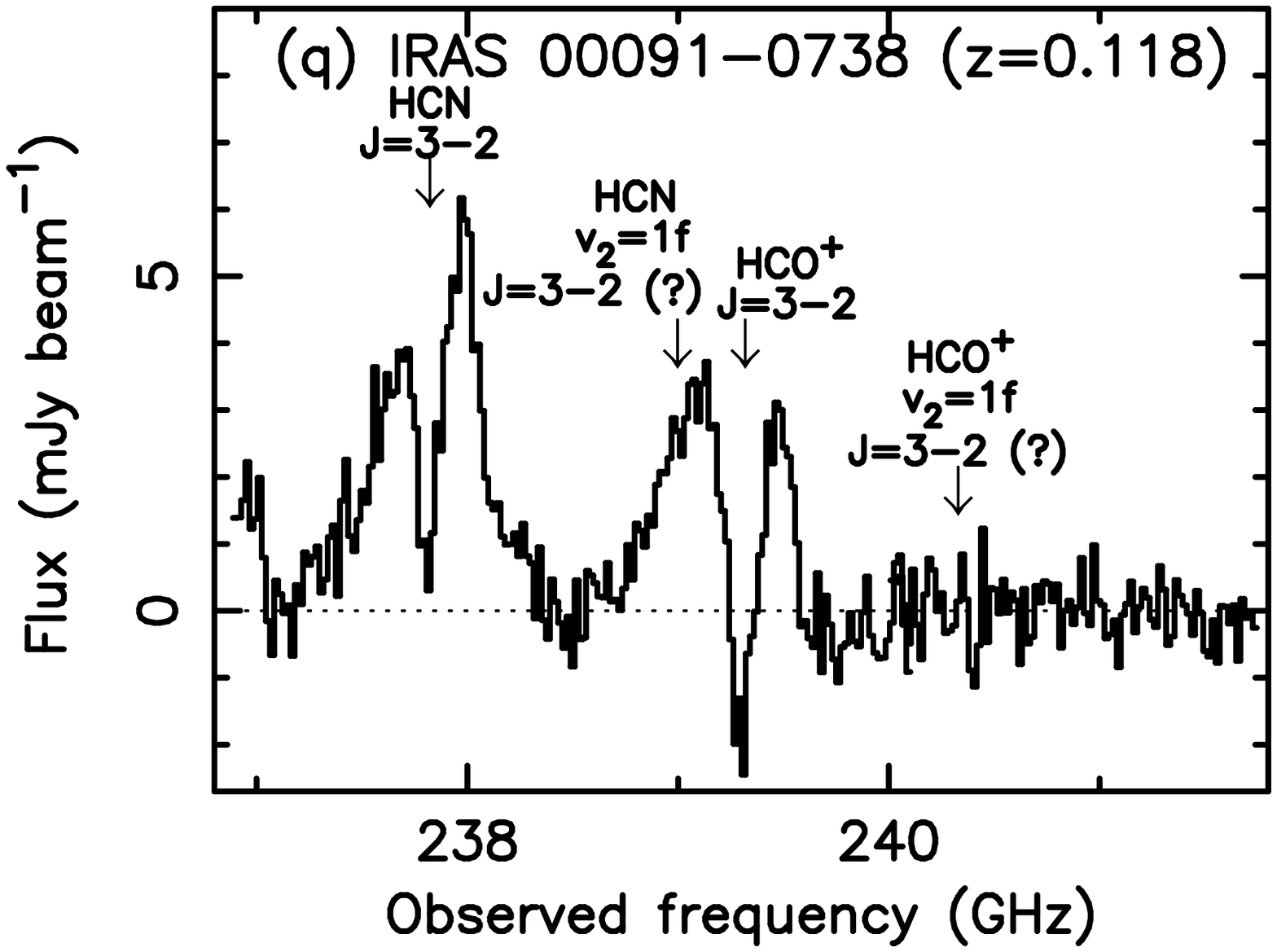} 
\includegraphics[angle=0,scale=.4]{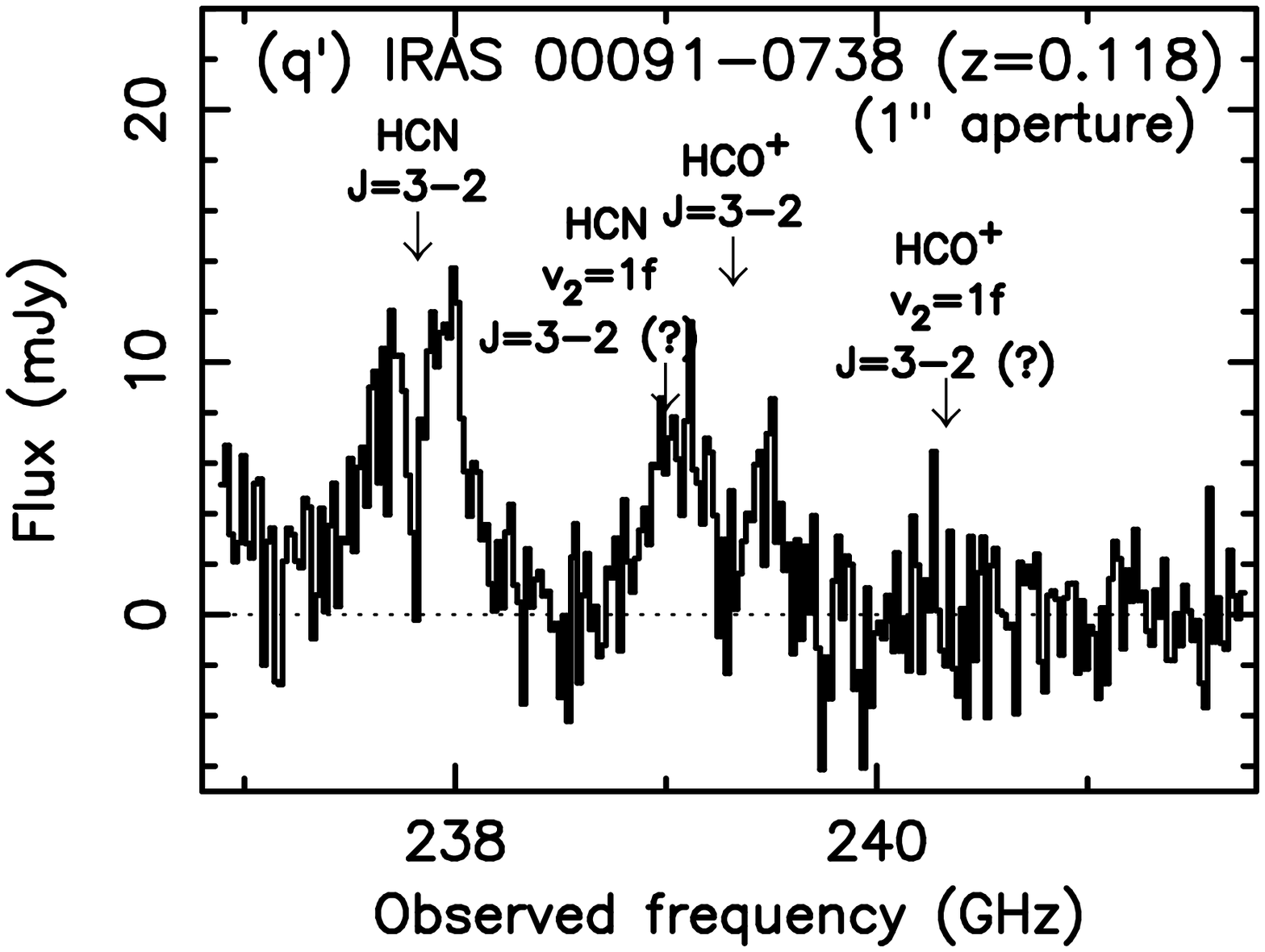} \\ 
\includegraphics[angle=0,scale=.4]{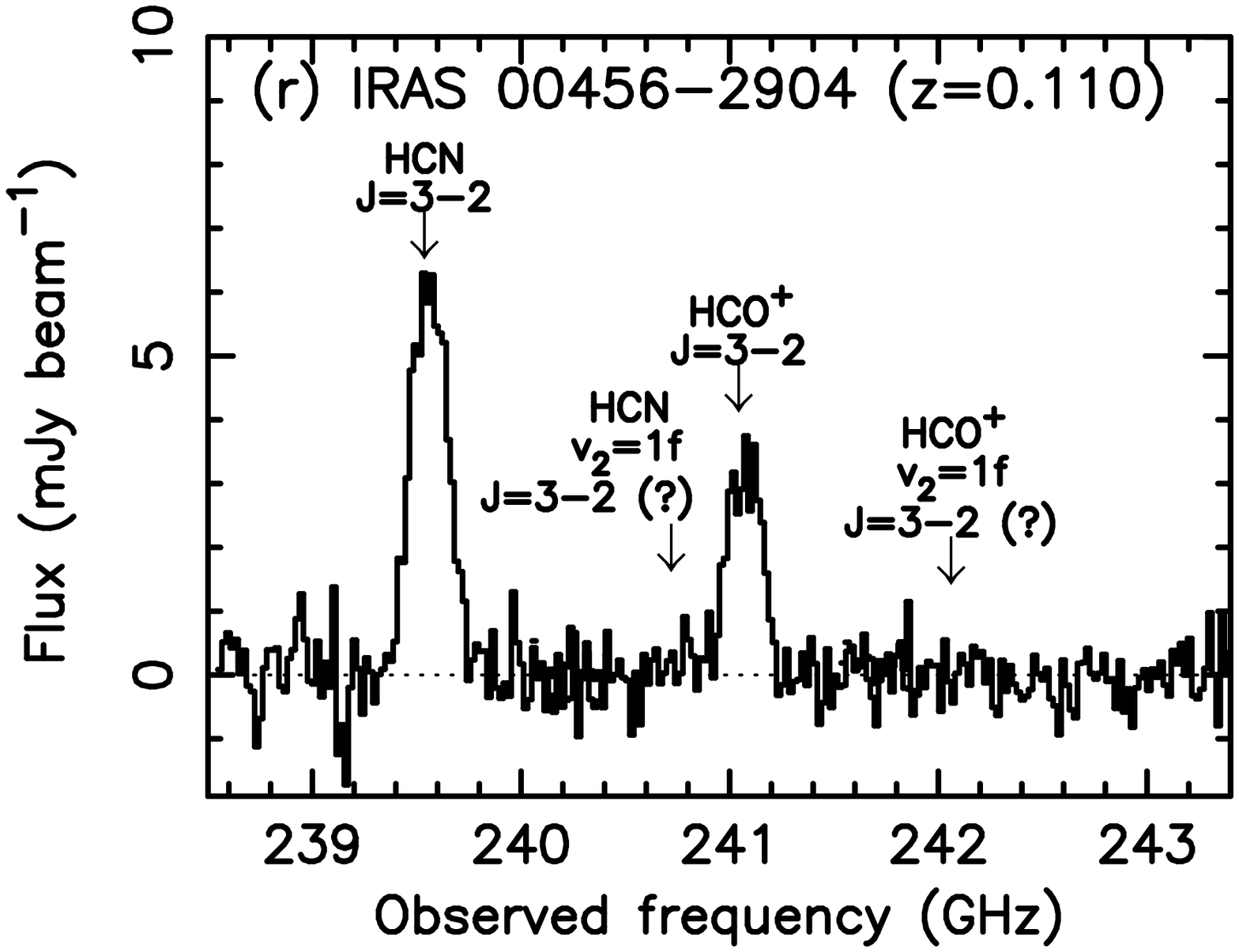} 
\includegraphics[angle=0,scale=.4]{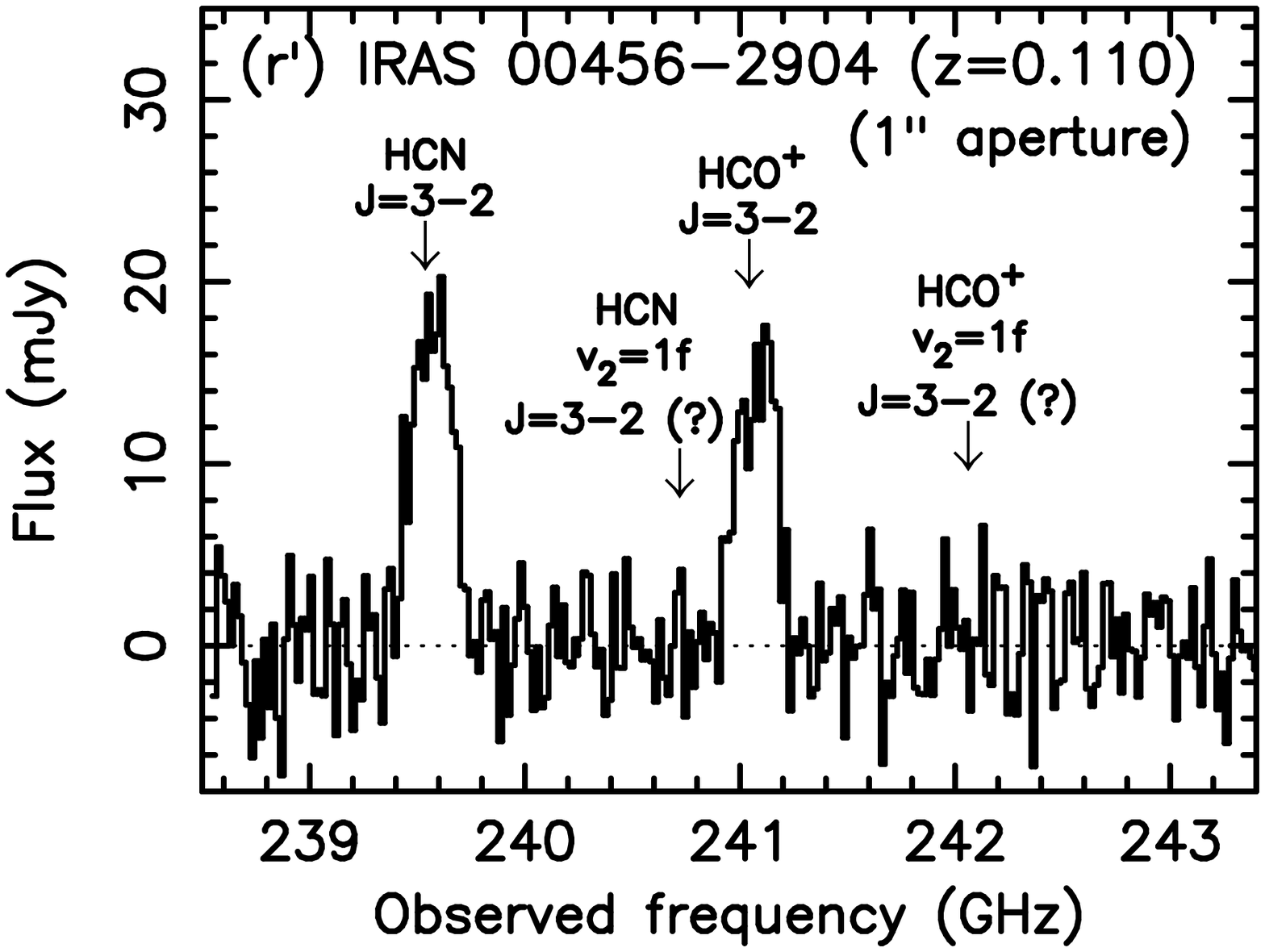} \\
\includegraphics[angle=0,scale=.4]{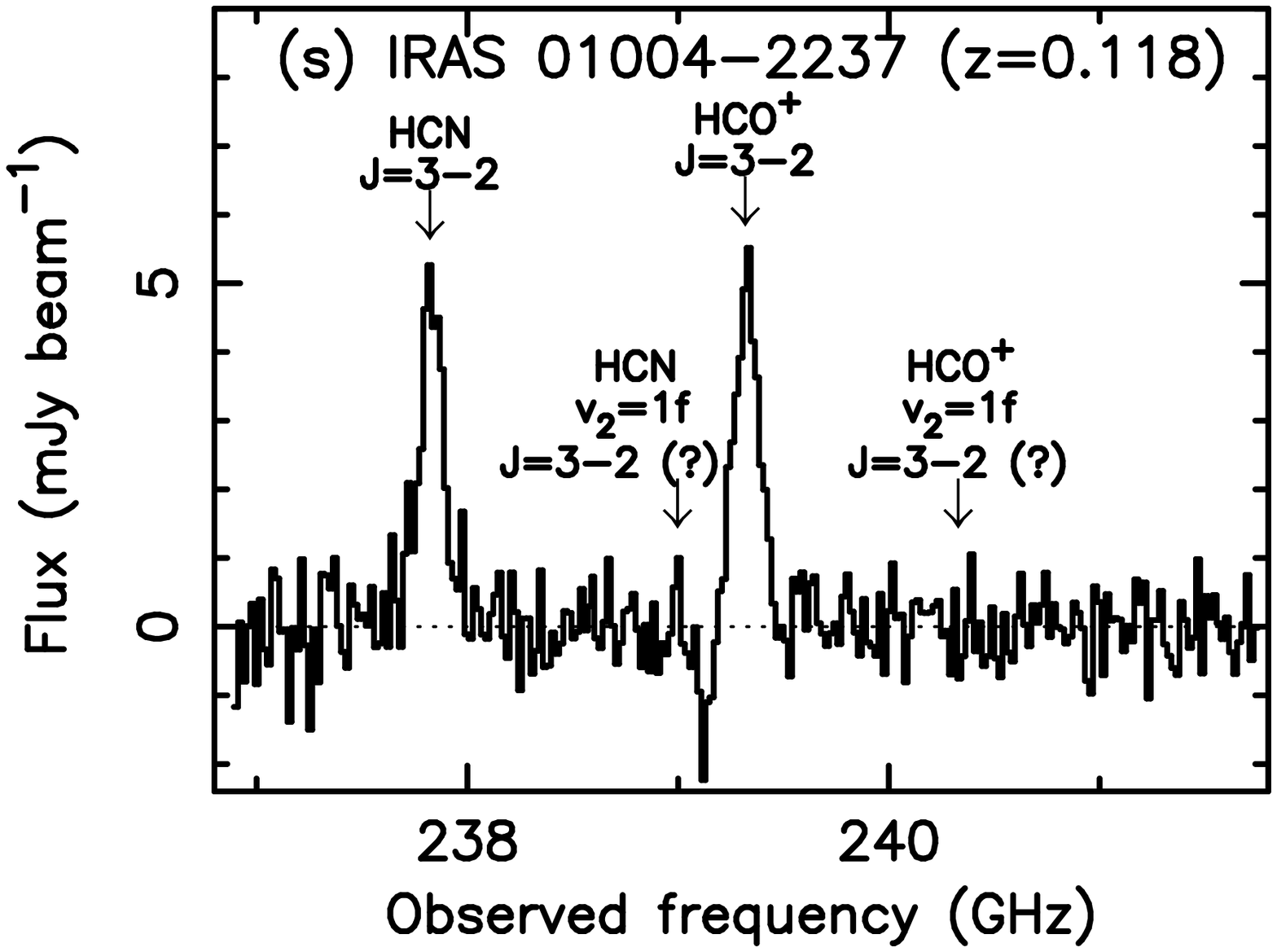} 
\includegraphics[angle=0,scale=.4]{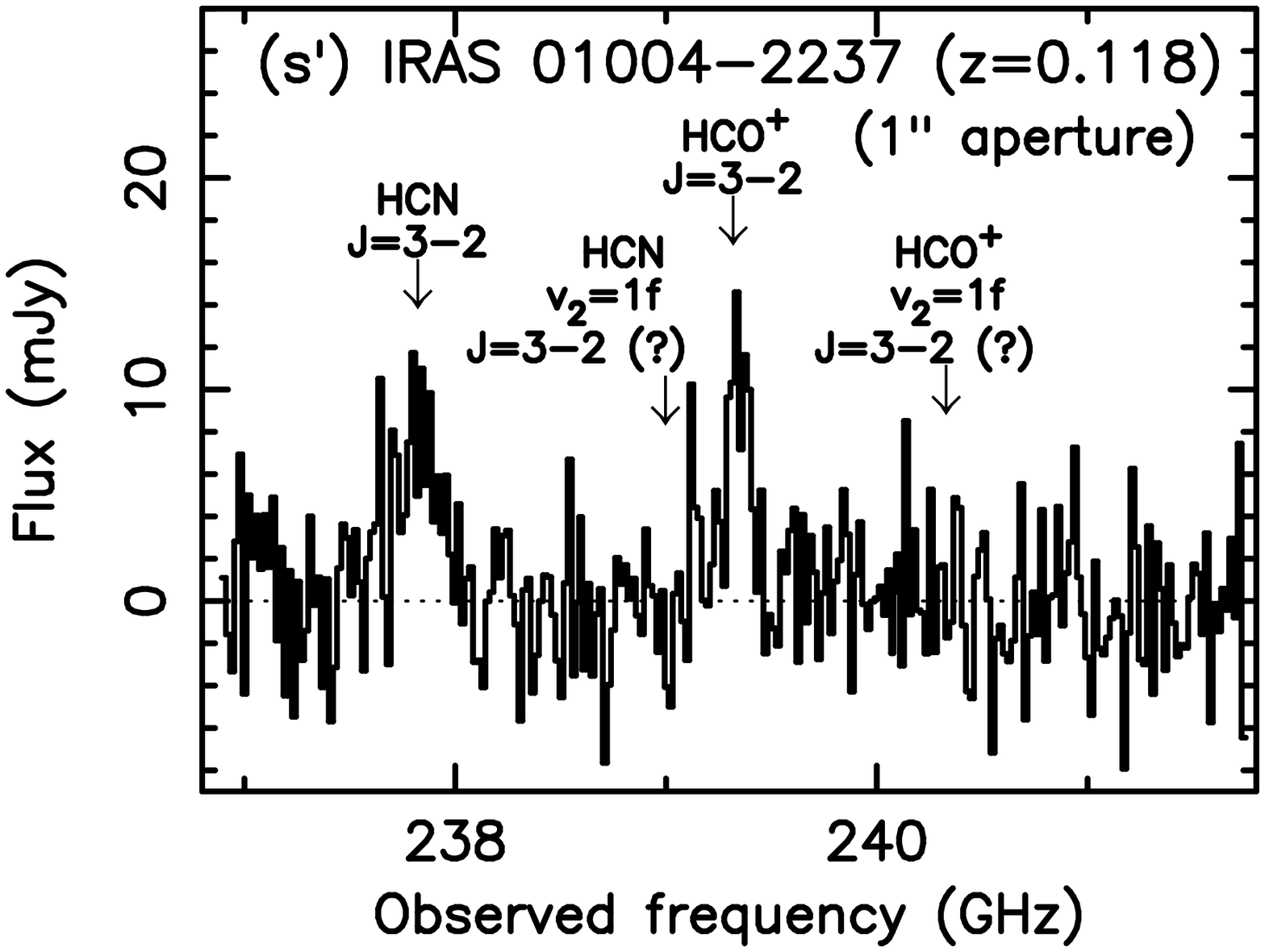} \\
\includegraphics[angle=0,scale=.4]{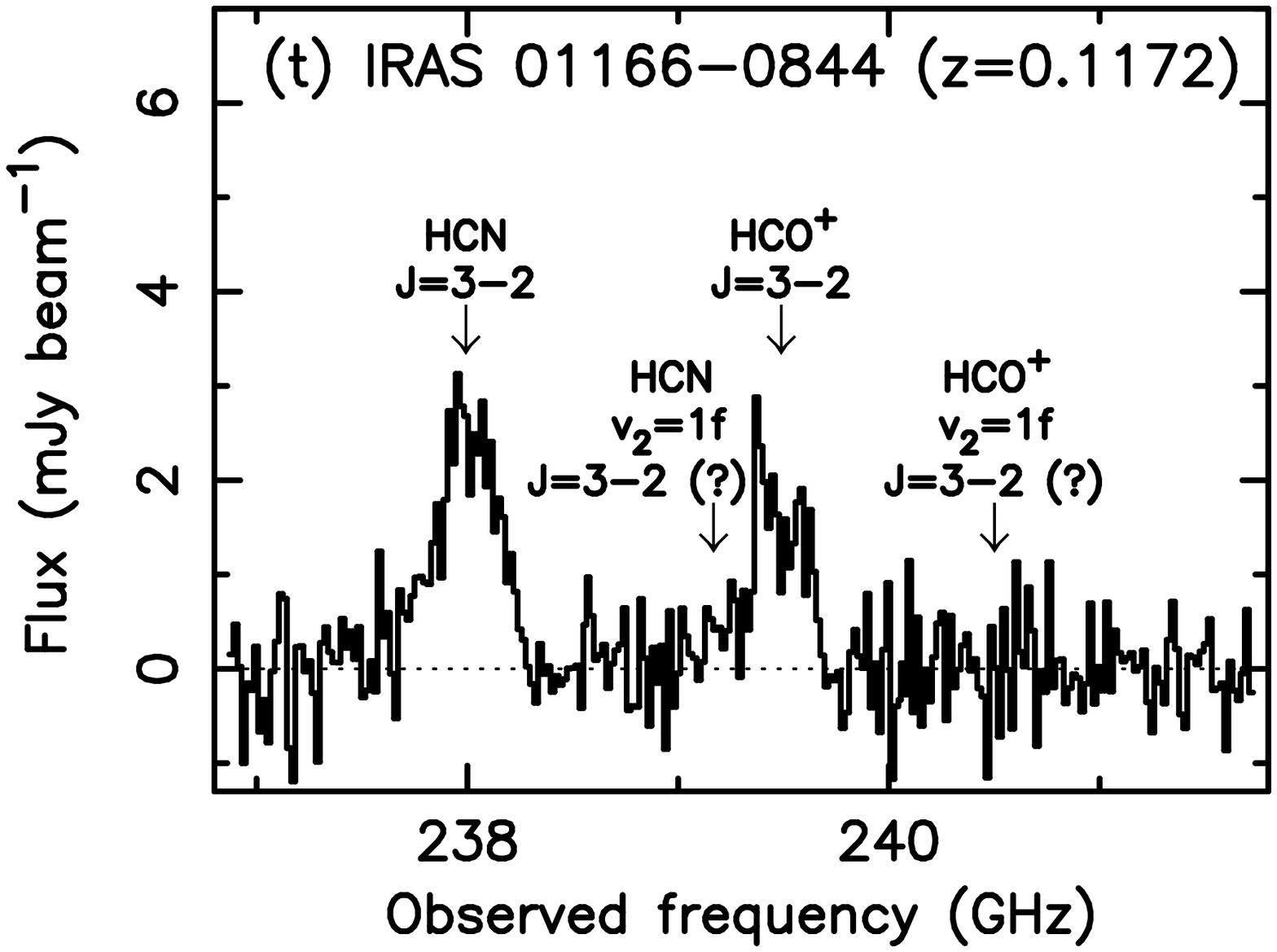}
\includegraphics[angle=0,scale=.4]{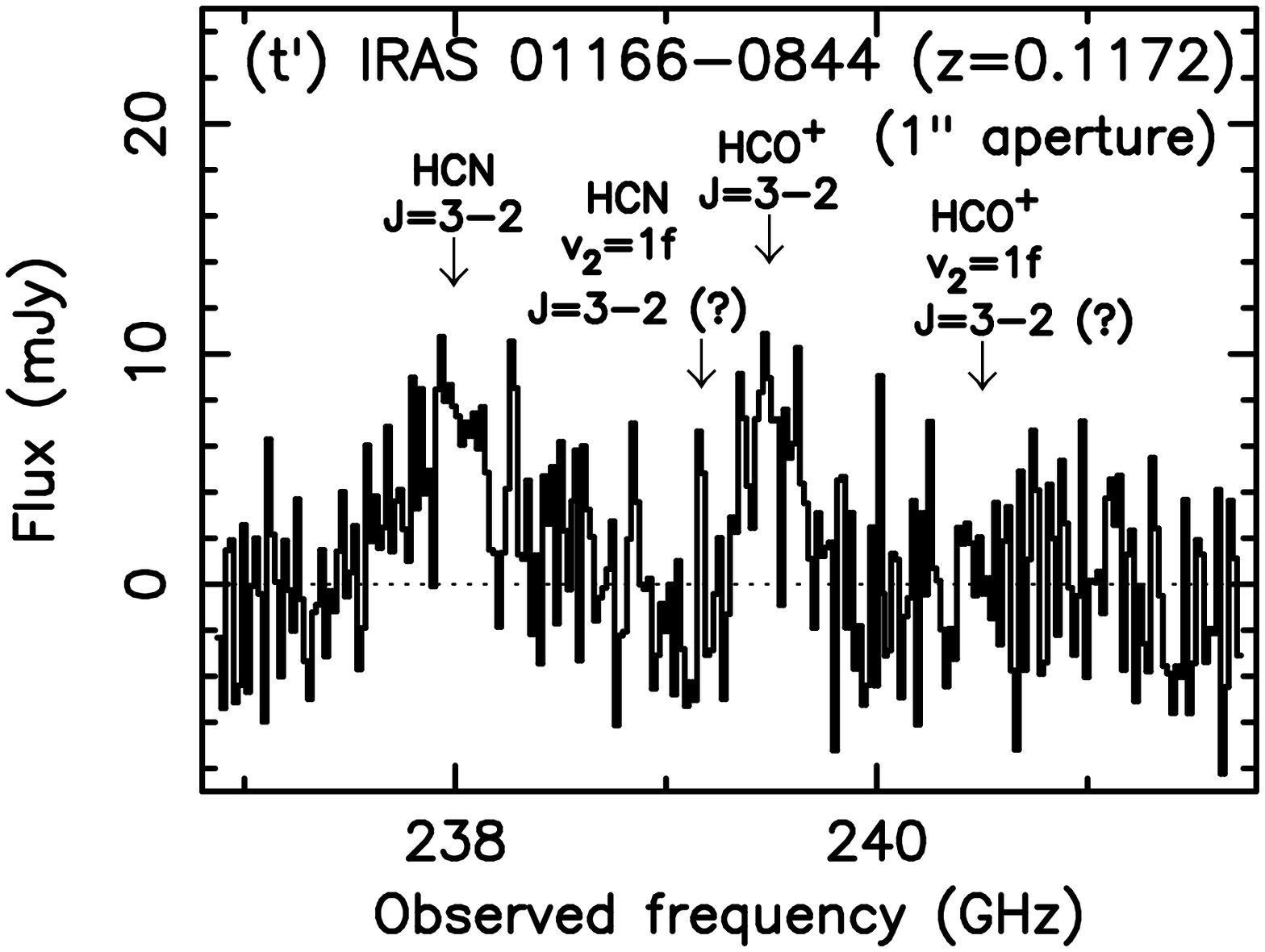} \\
\end{center}
\end{figure}

\clearpage

\begin{figure}
\begin{center}
\includegraphics[angle=0,scale=.4]{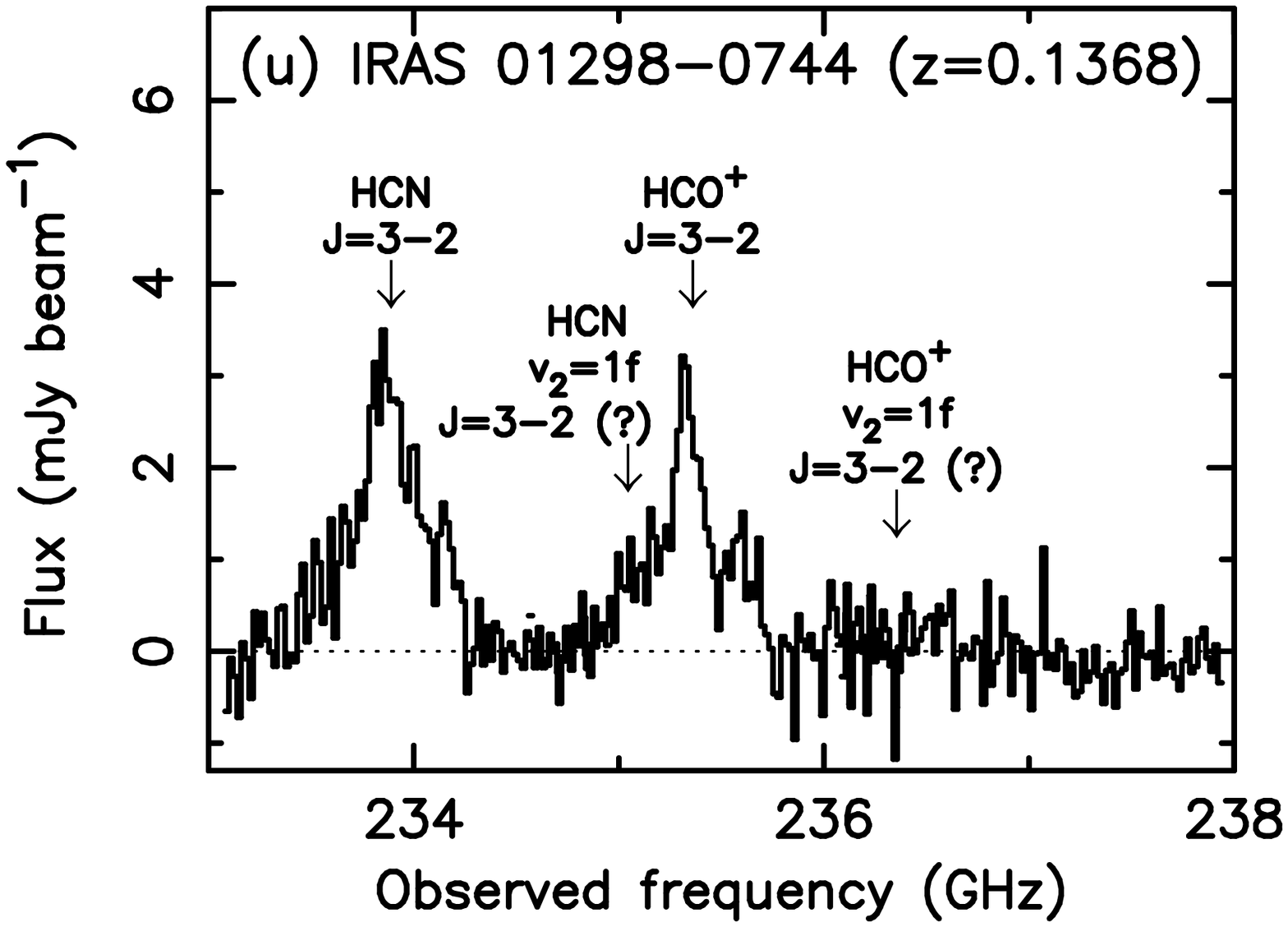} 
\includegraphics[angle=0,scale=.4]{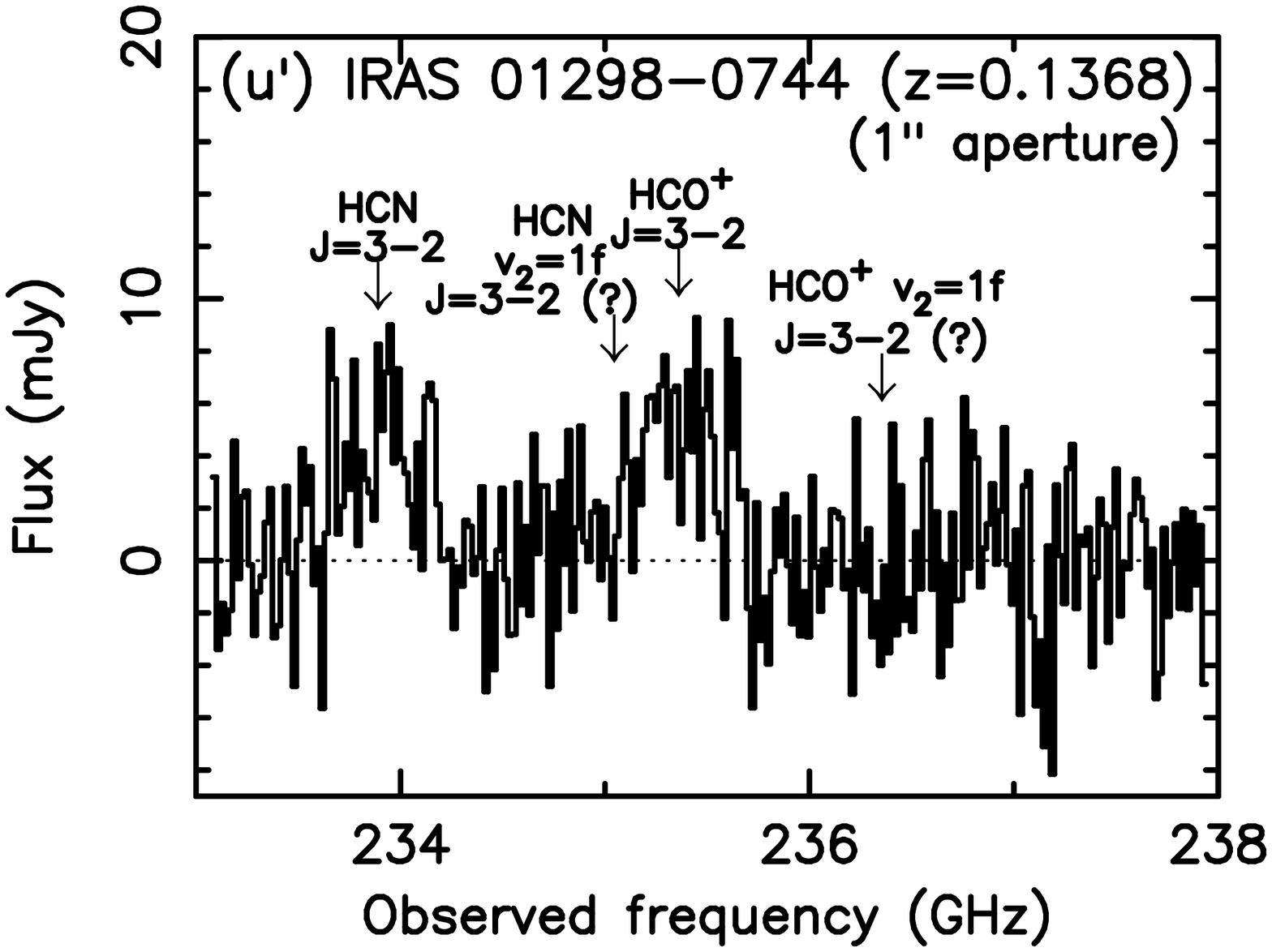} \\
\includegraphics[angle=0,scale=.4]{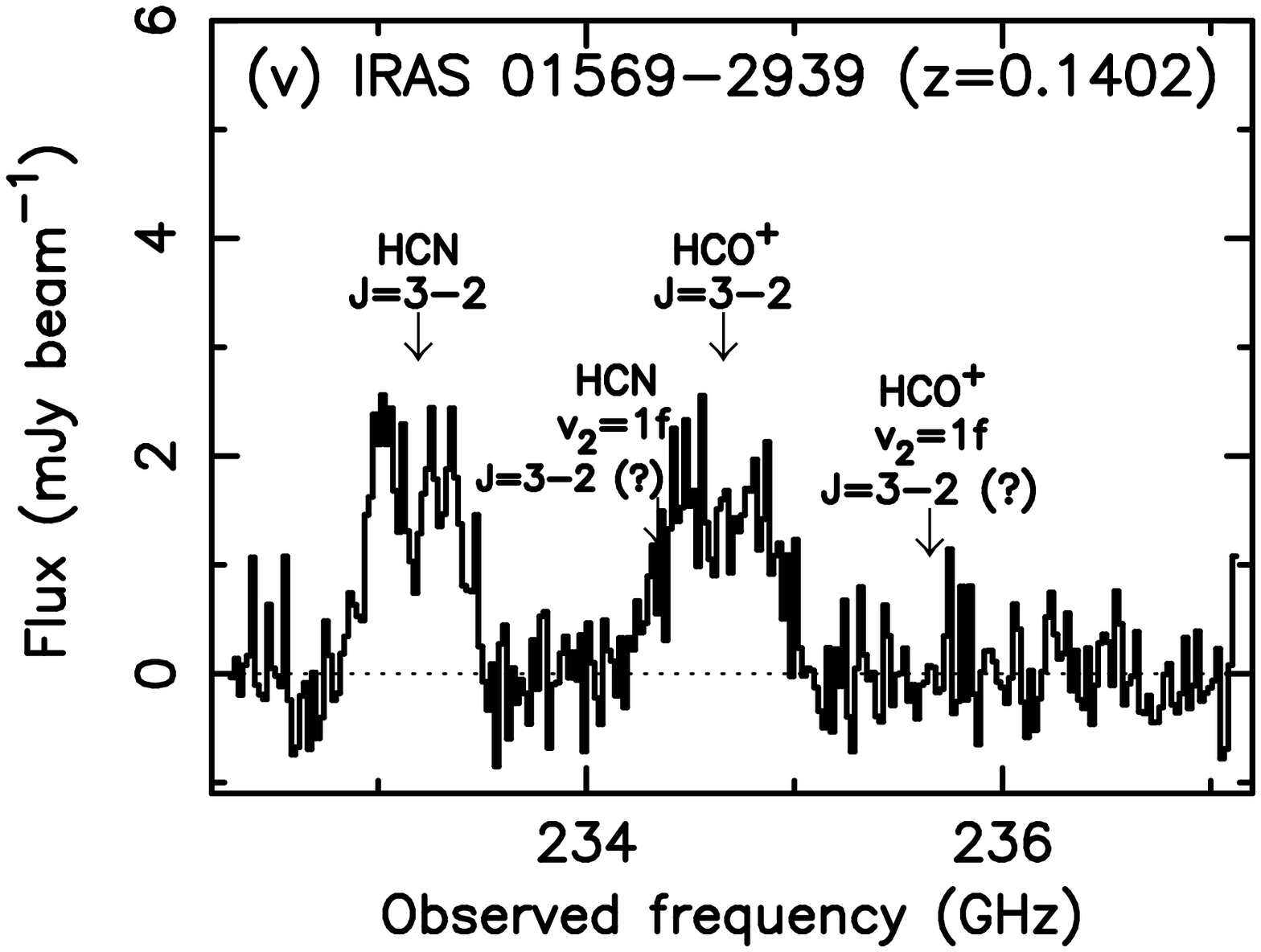} 
\includegraphics[angle=0,scale=.4]{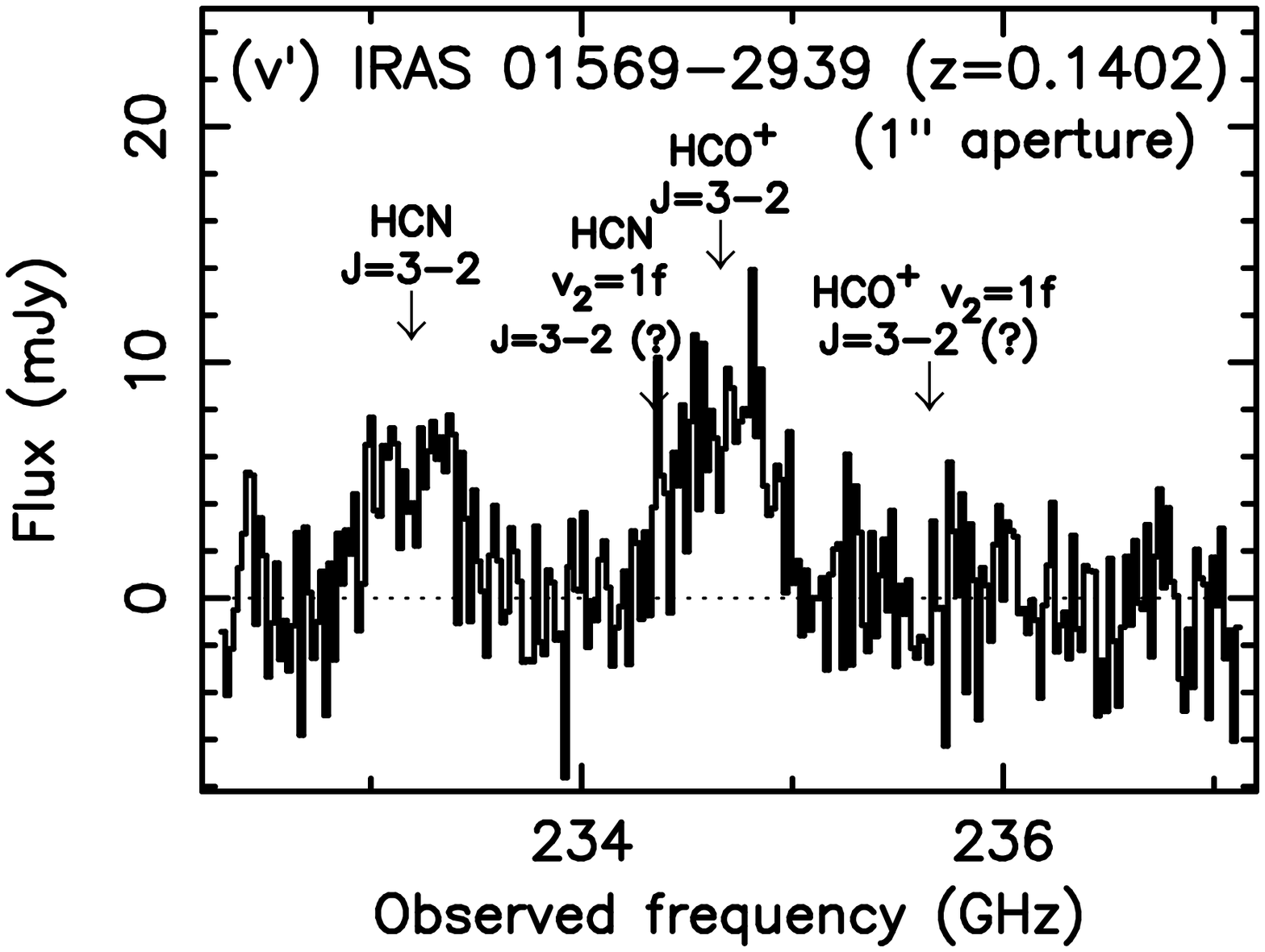} \\
\hspace*{-8.2cm} 
\includegraphics[angle=0,scale=.4]{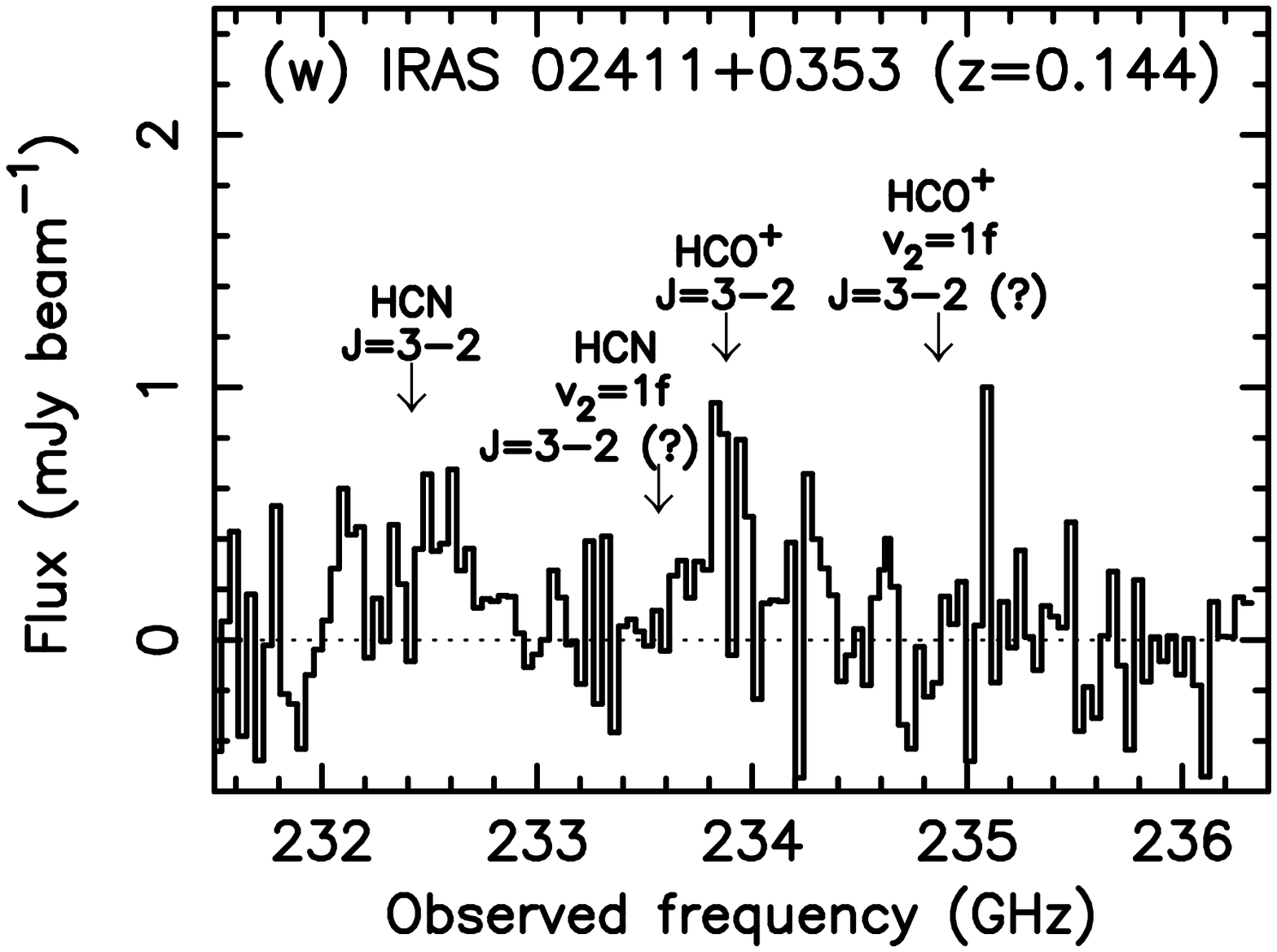} \\
\includegraphics[angle=0,scale=.4]{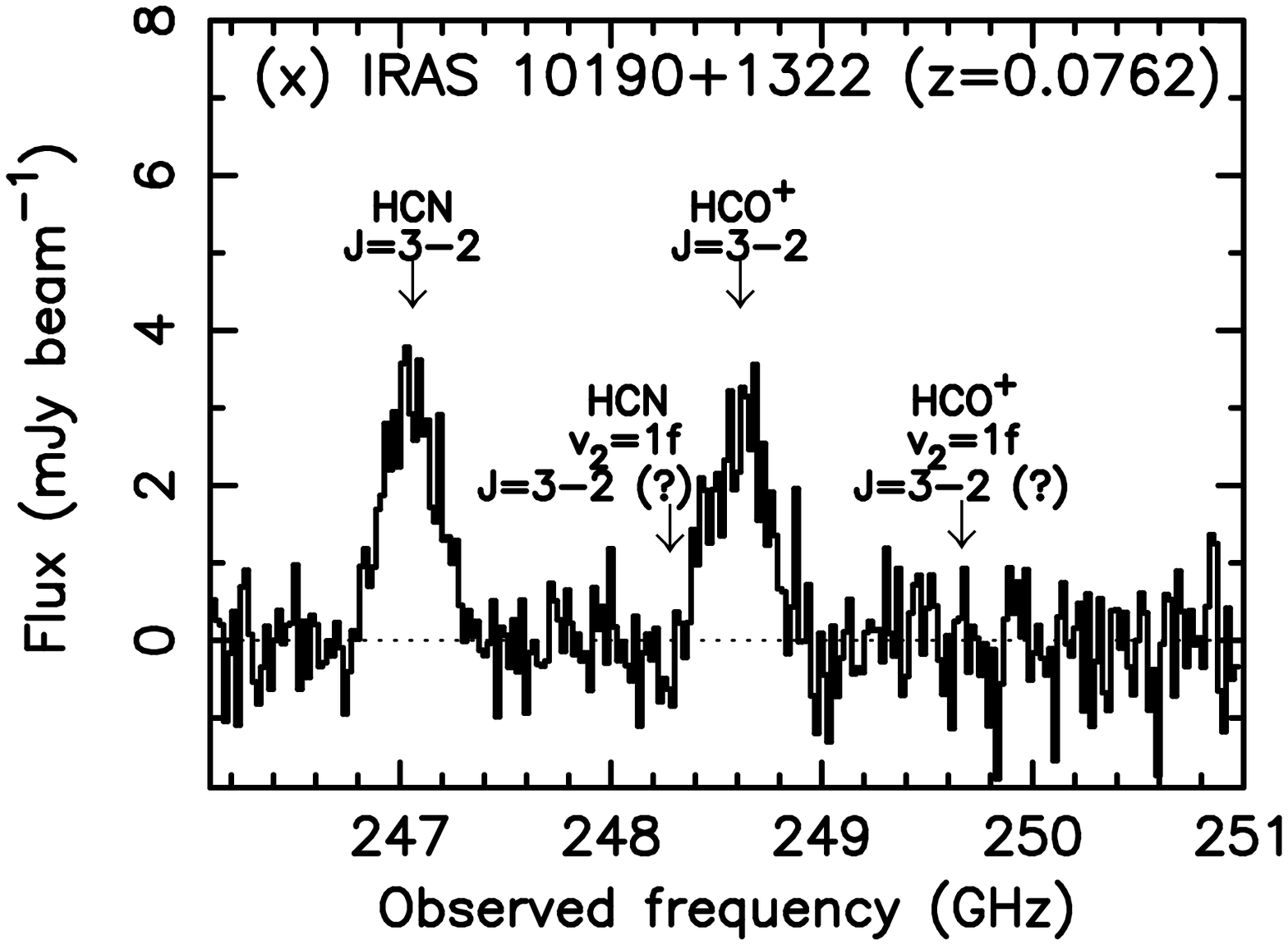} 
\includegraphics[angle=0,scale=.4]{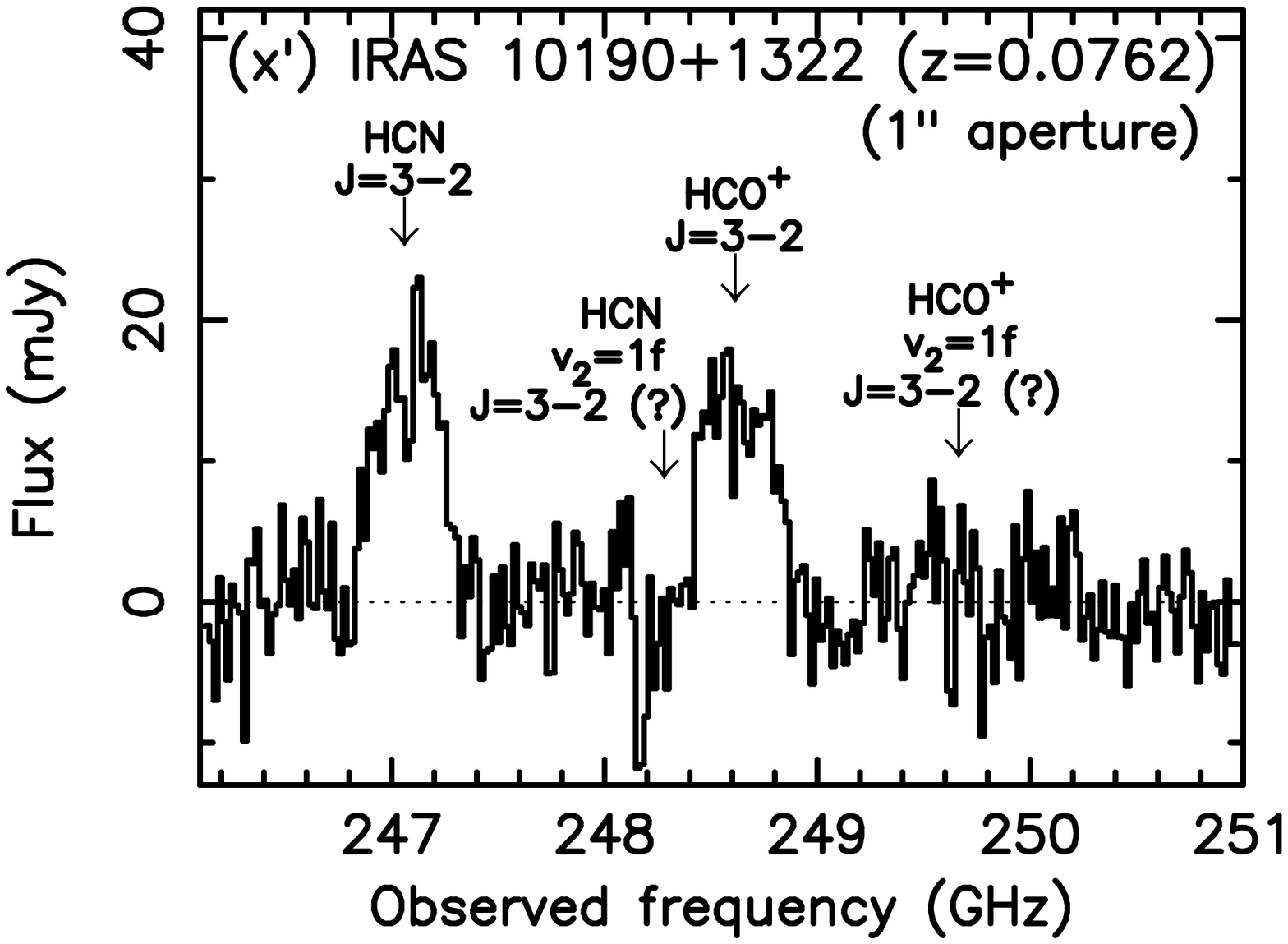} \\
\end{center}
\end{figure}

\clearpage

\begin{figure}
\begin{center}
\includegraphics[angle=0,scale=.4]{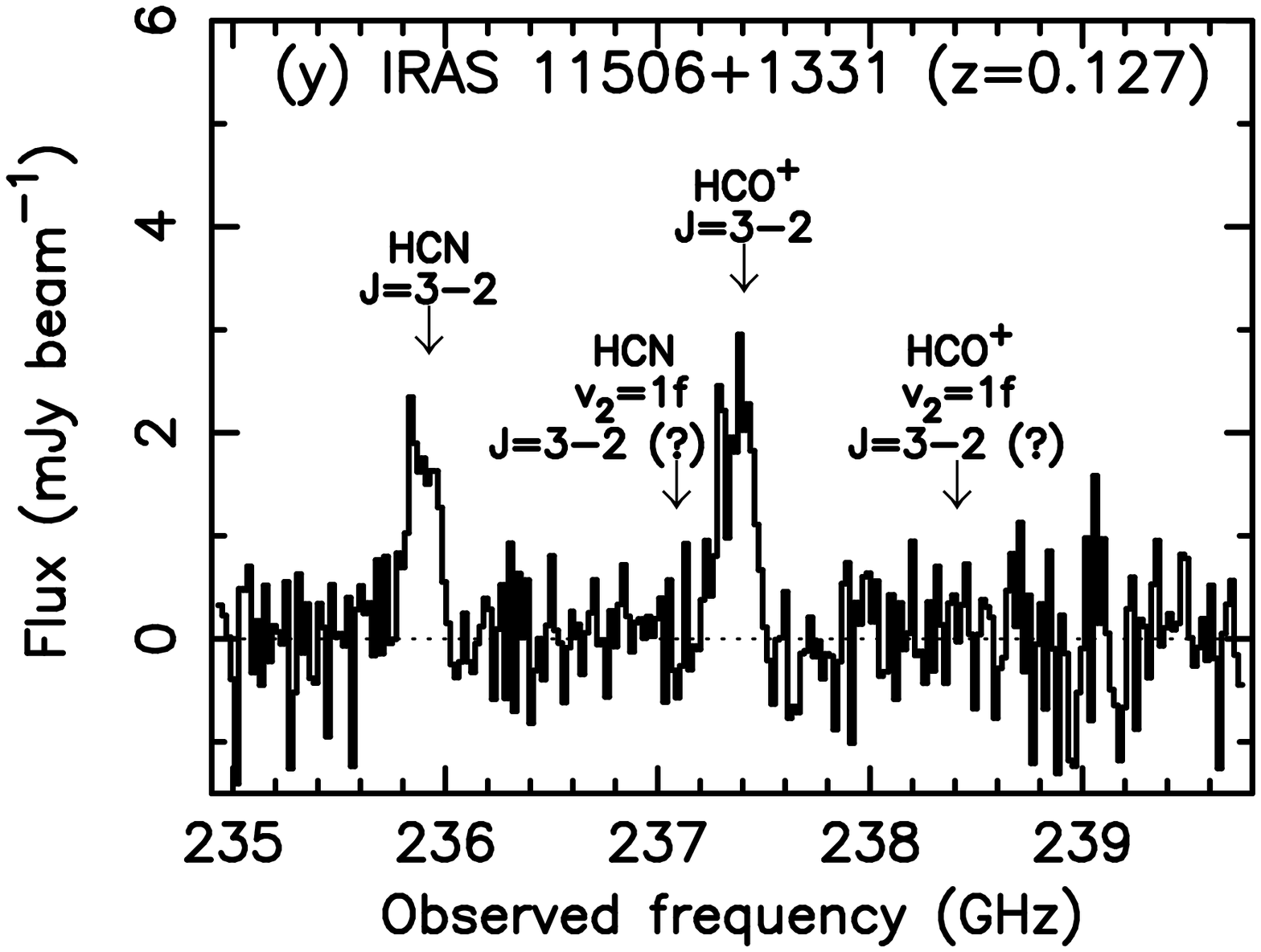}  
\includegraphics[angle=0,scale=.4]{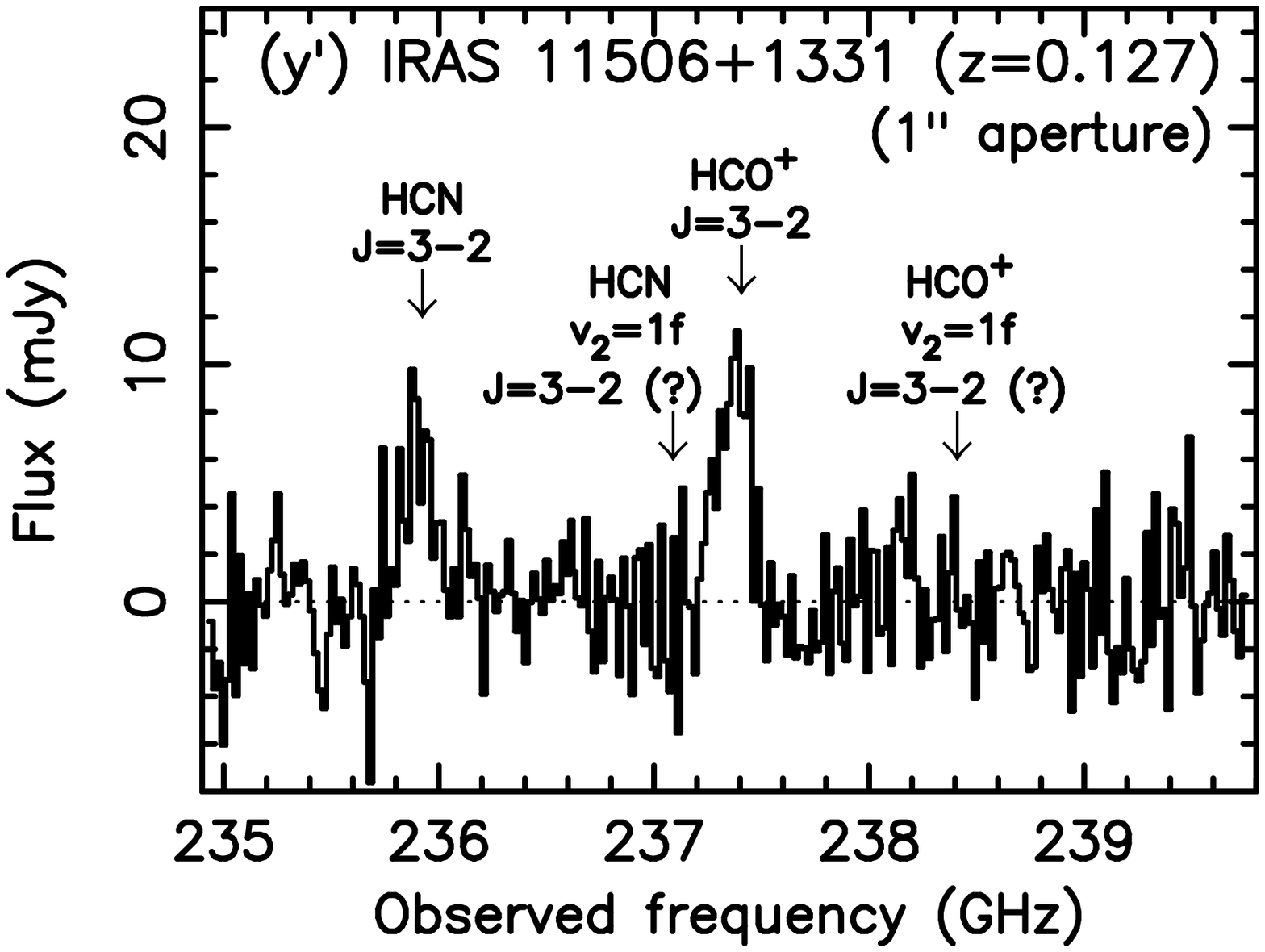} \\  
\includegraphics[angle=0,scale=.4]{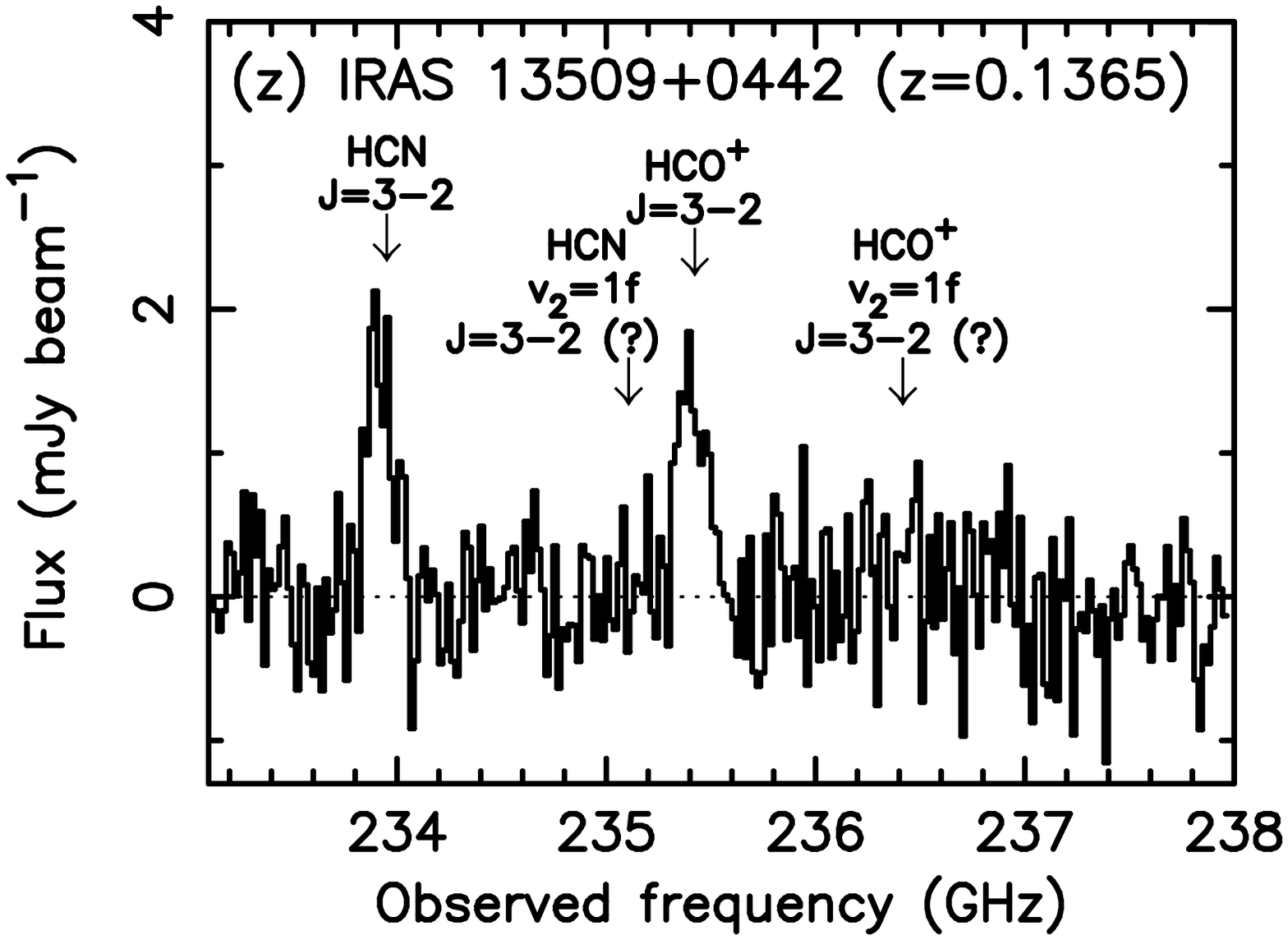} 
\includegraphics[angle=0,scale=.4]{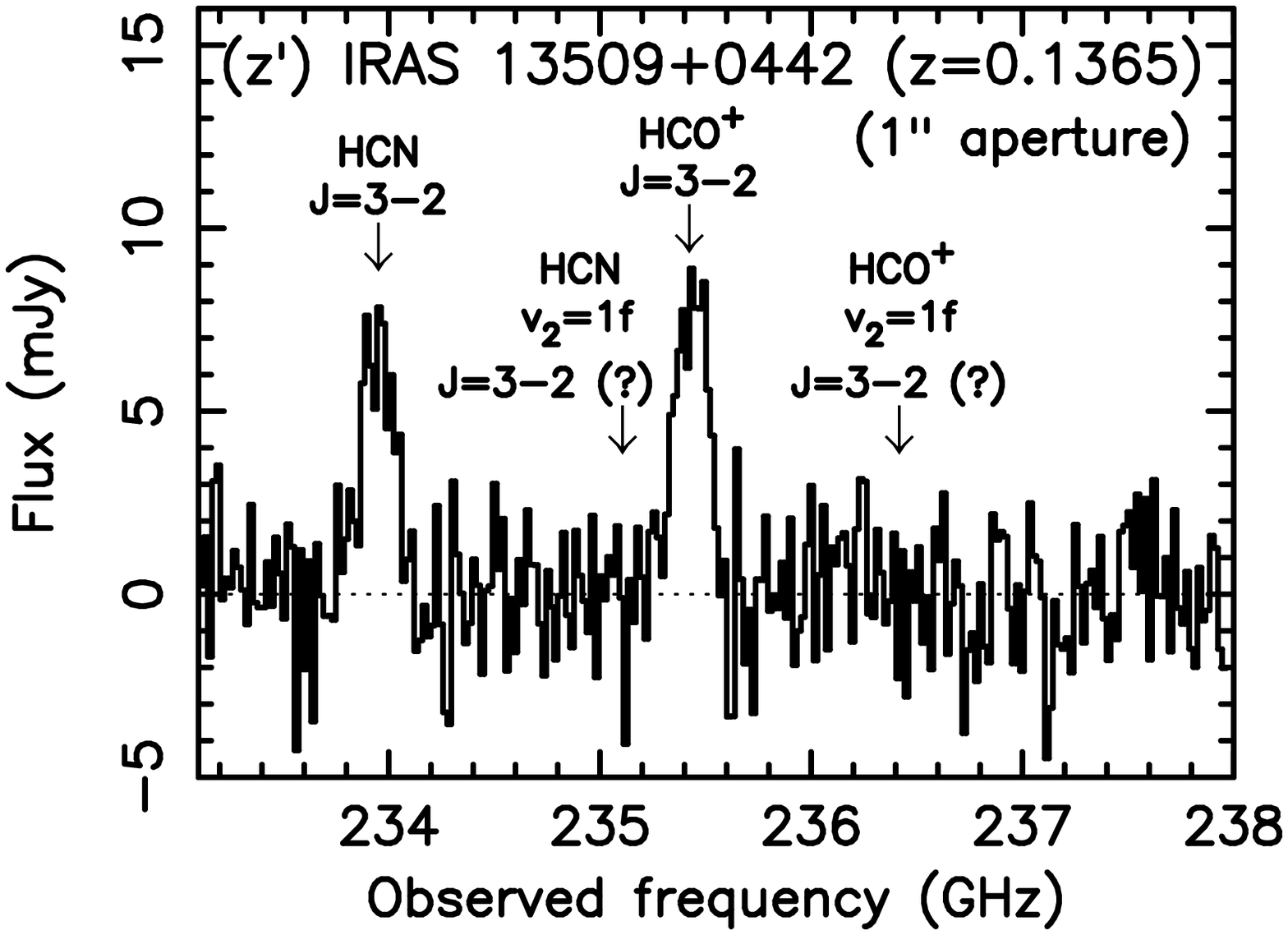} \\
\includegraphics[angle=0,scale=.4]{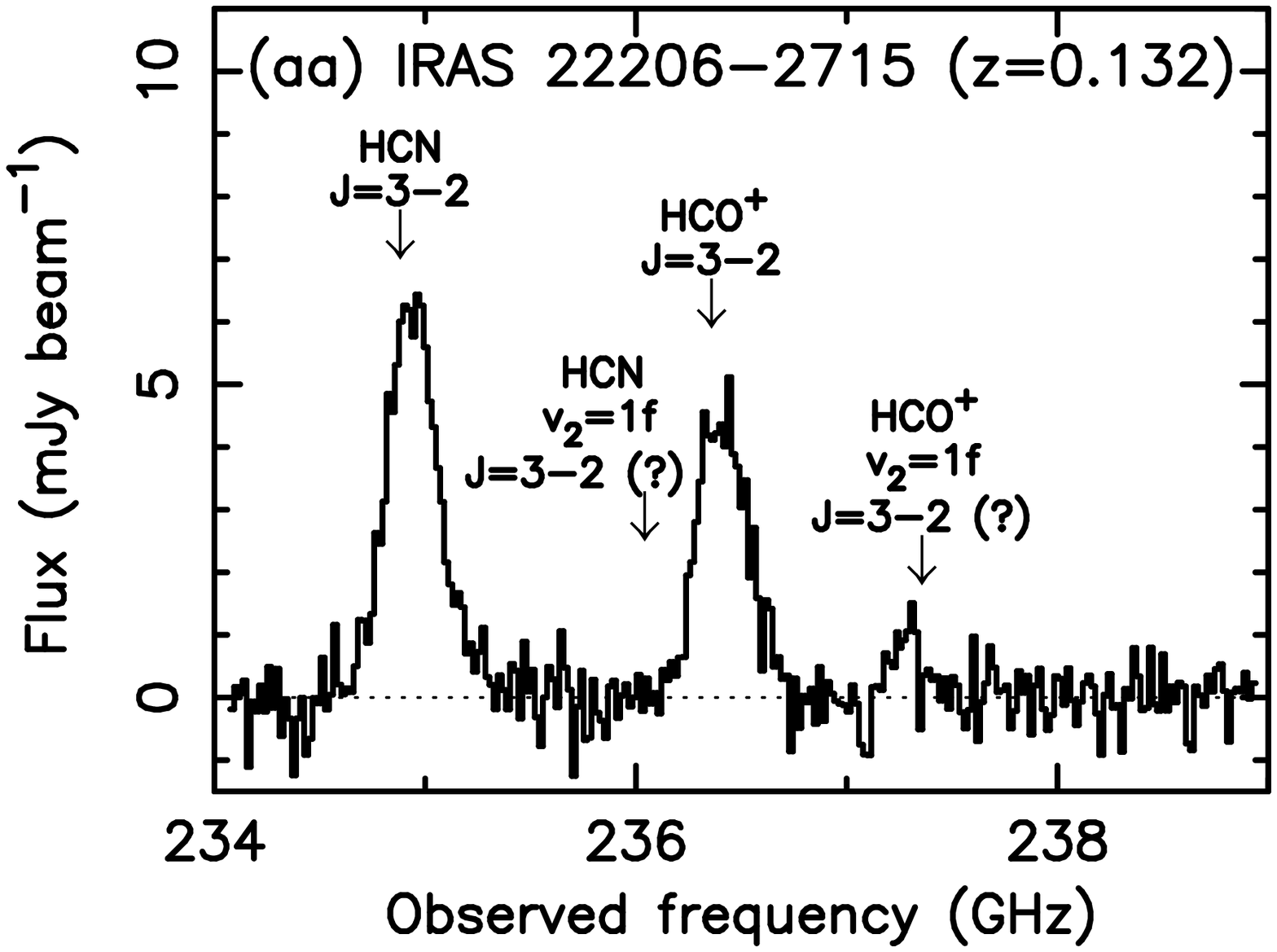} 
\includegraphics[angle=0,scale=.4]{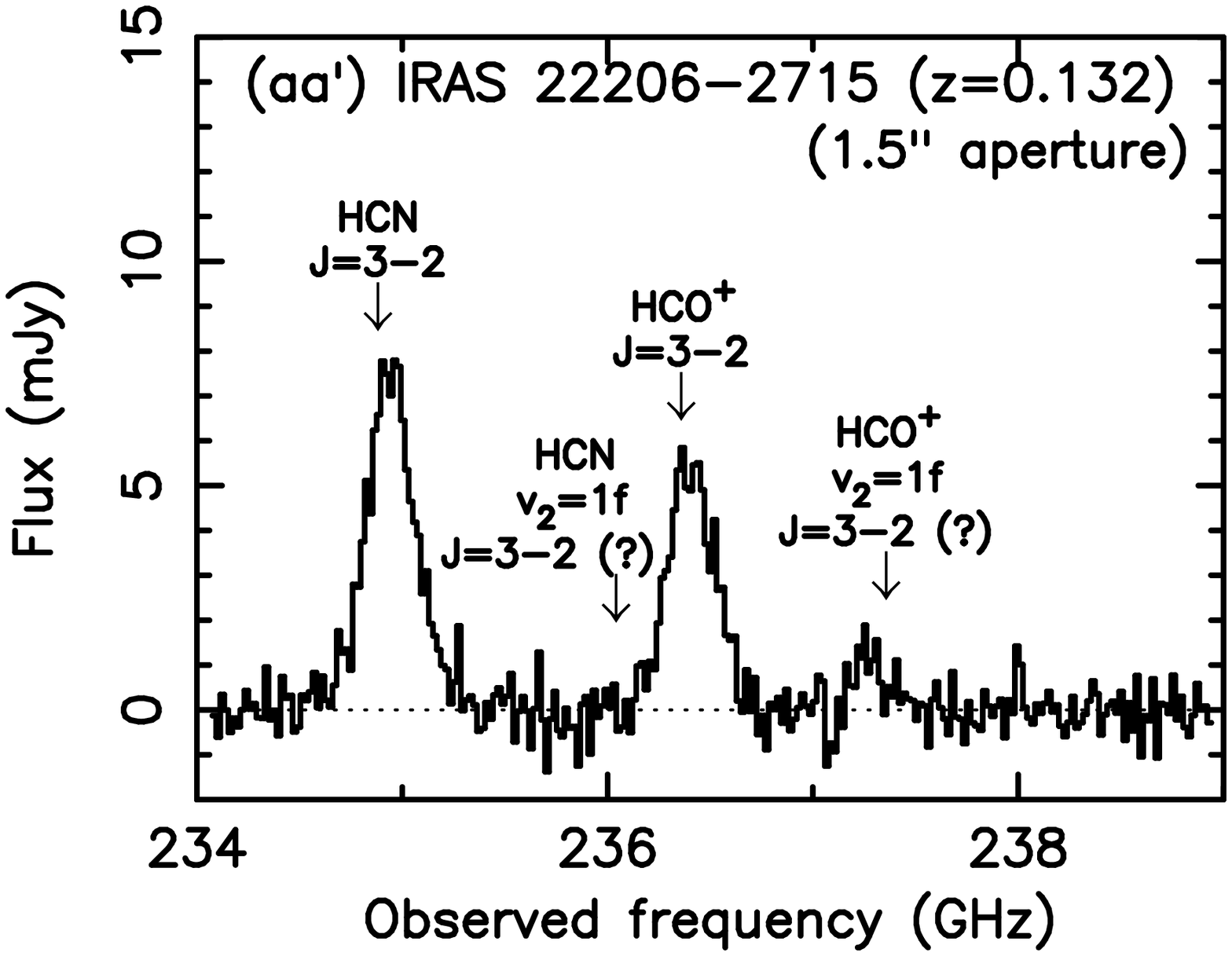} \\ 
\includegraphics[angle=0,scale=.4]{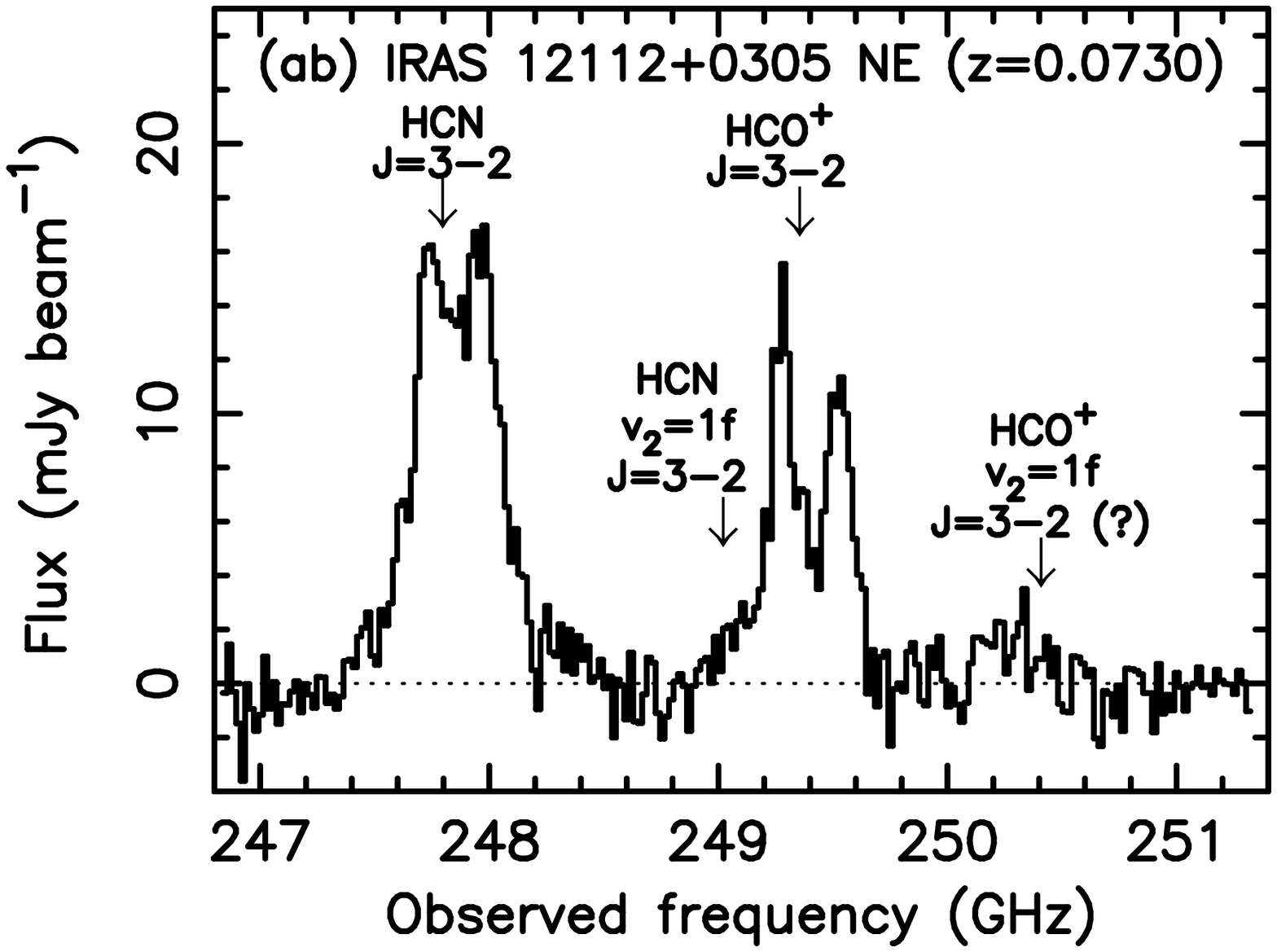} 
\includegraphics[angle=0,scale=.4]{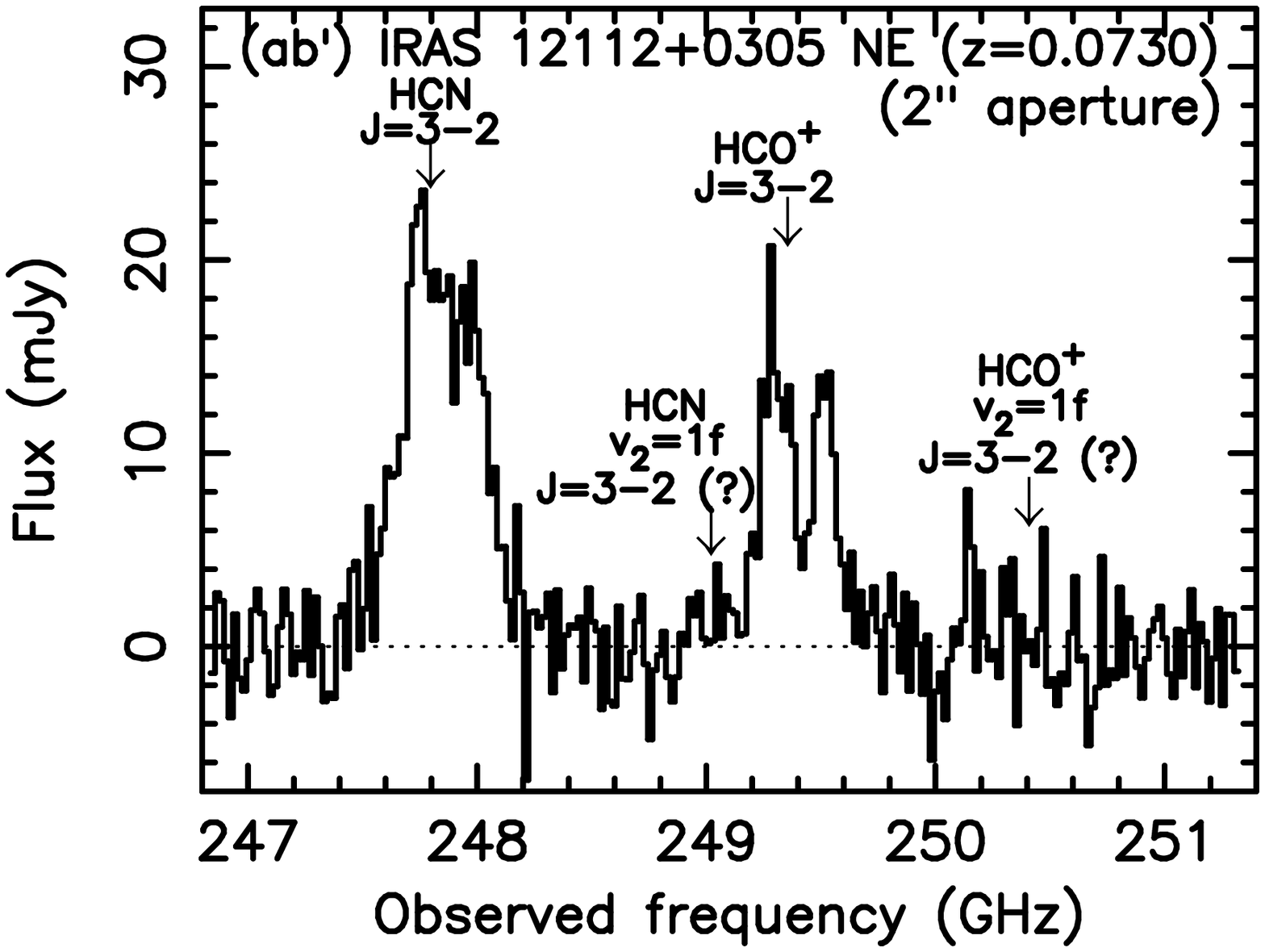} \\ 
\end{center}
\end{figure}

\clearpage

\begin{figure}
\begin{center}
\includegraphics[angle=0,scale=.4]{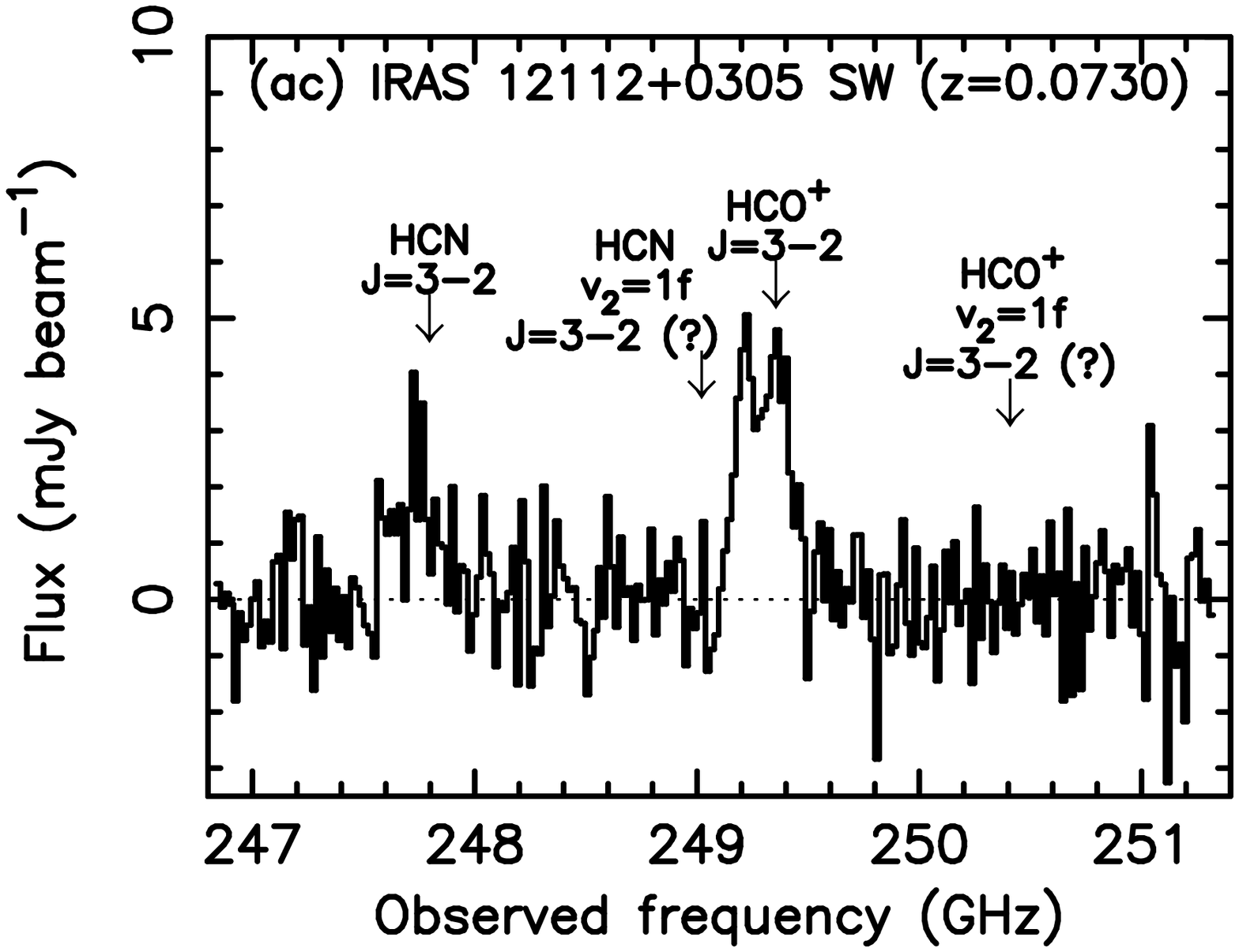} 
\includegraphics[angle=0,scale=.4]{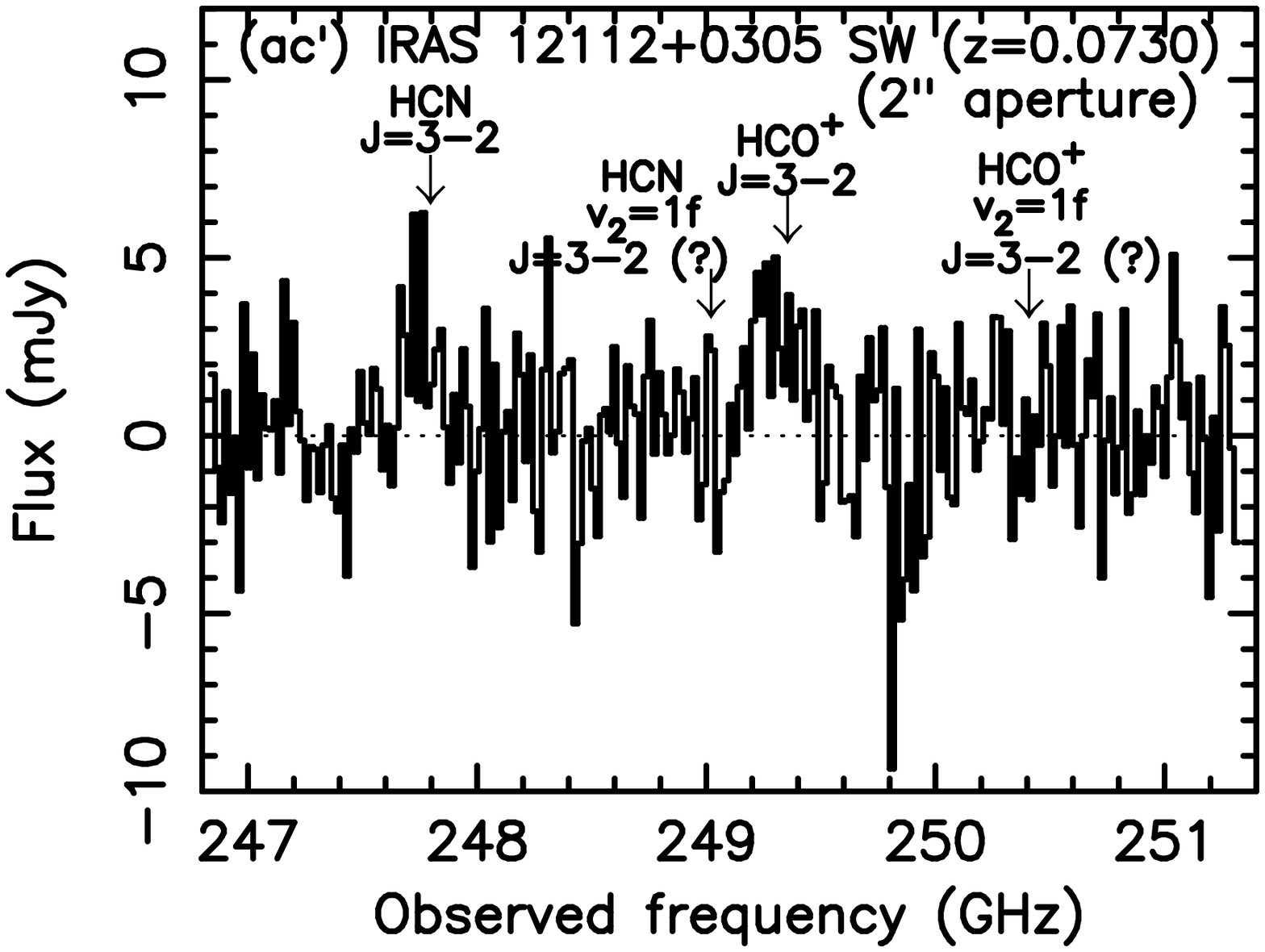} \\ 
\includegraphics[angle=0,scale=.4]{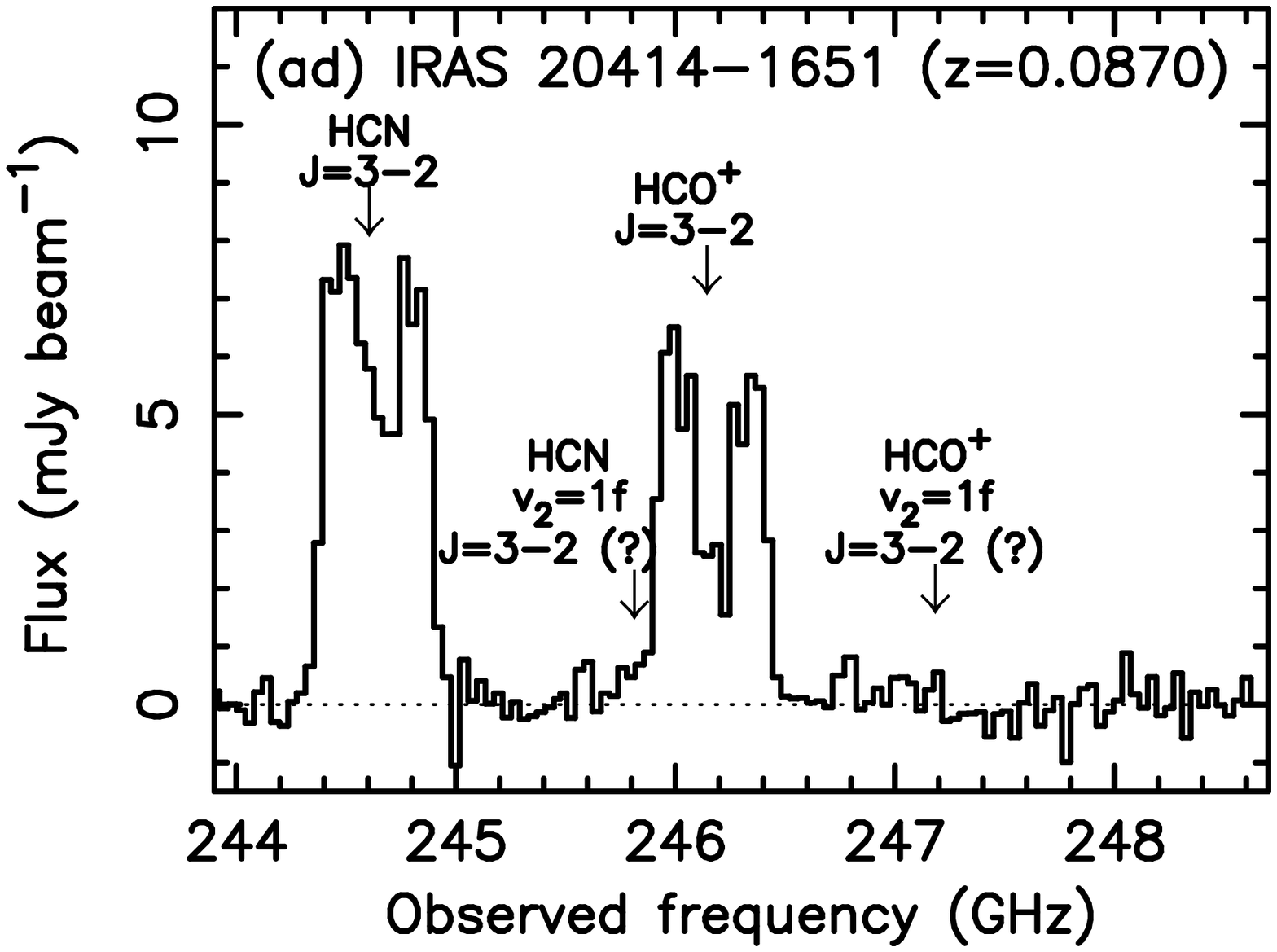} 
\includegraphics[angle=0,scale=.4]{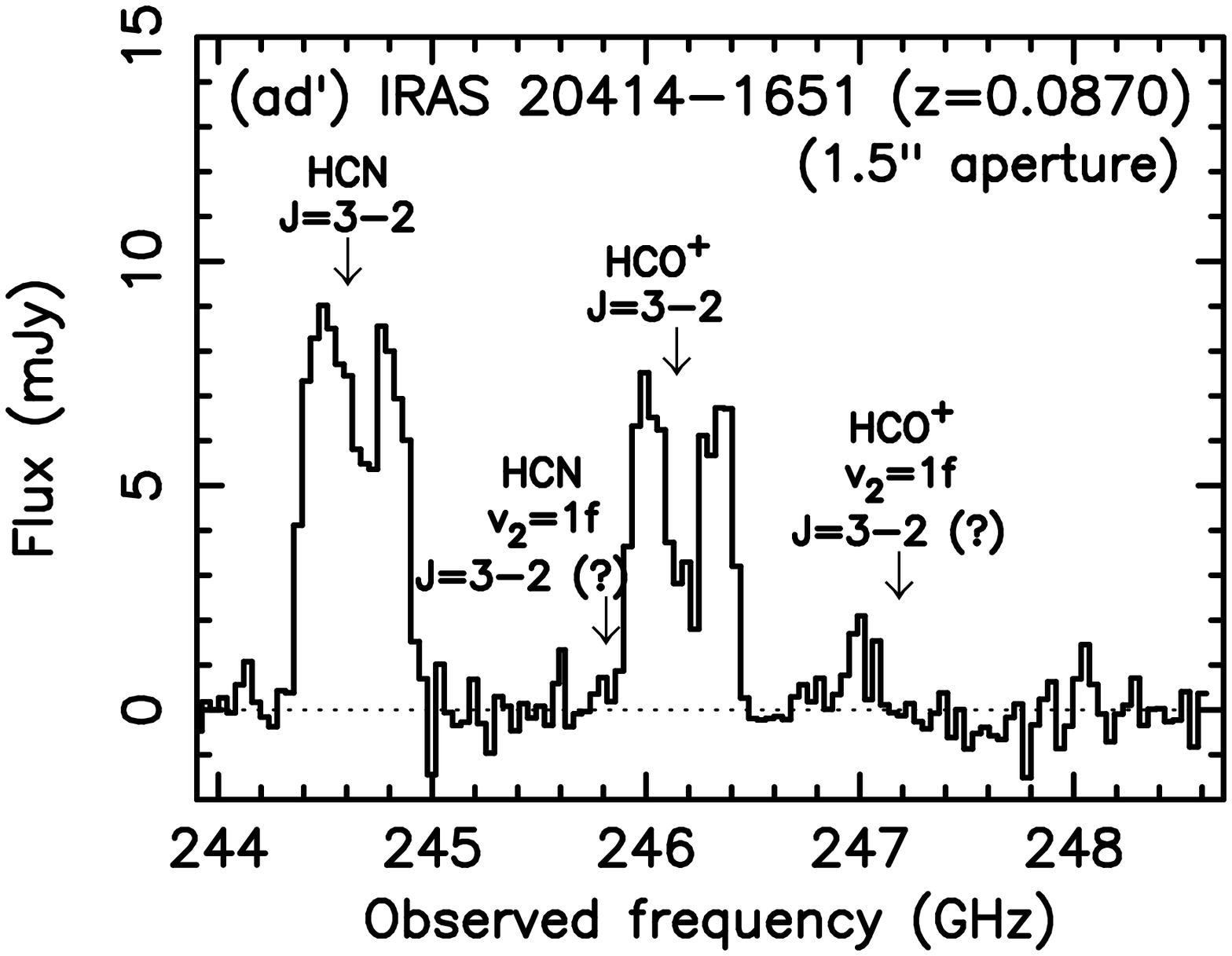} \\ 
\includegraphics[angle=0,scale=.4]{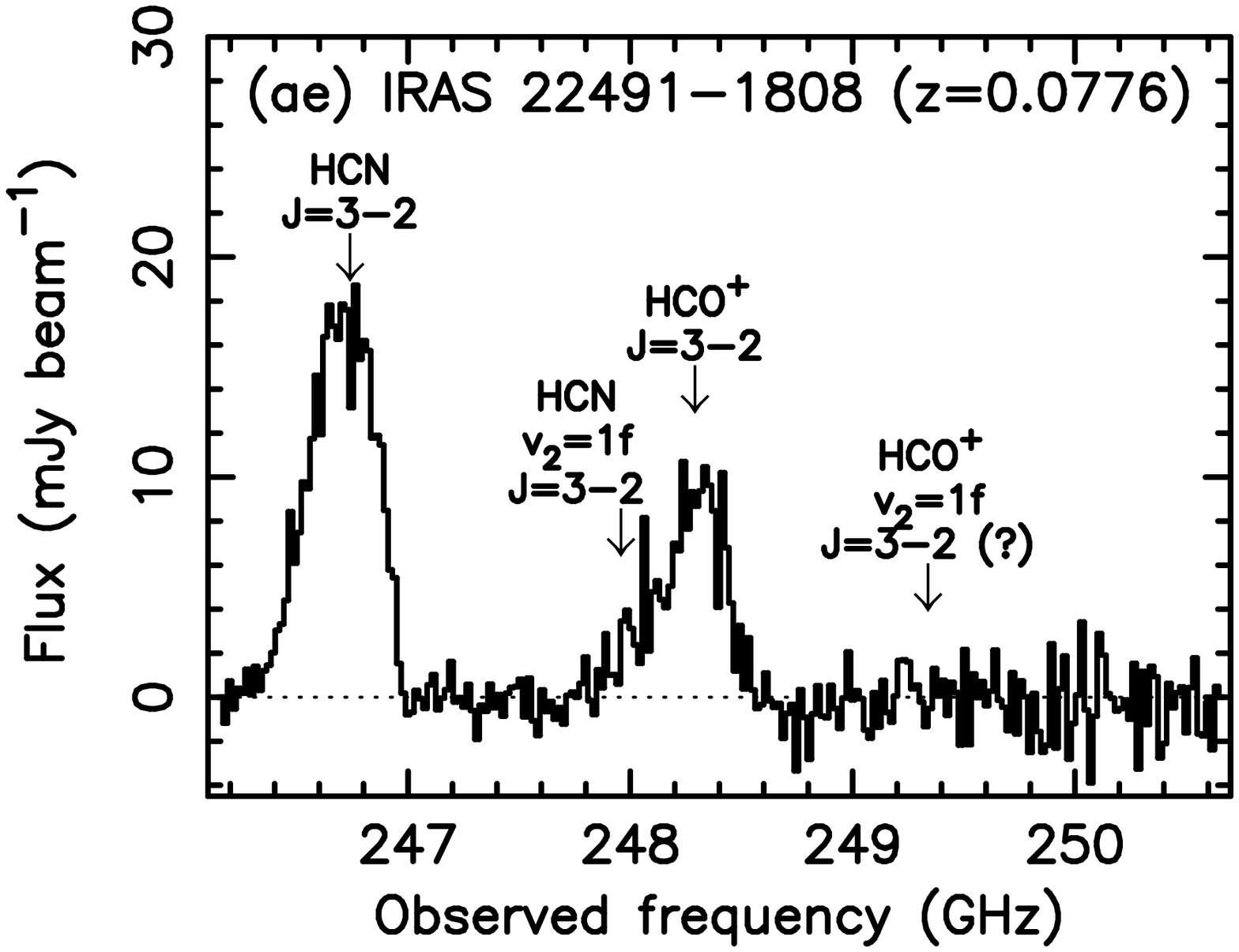} 
\includegraphics[angle=0,scale=.4]{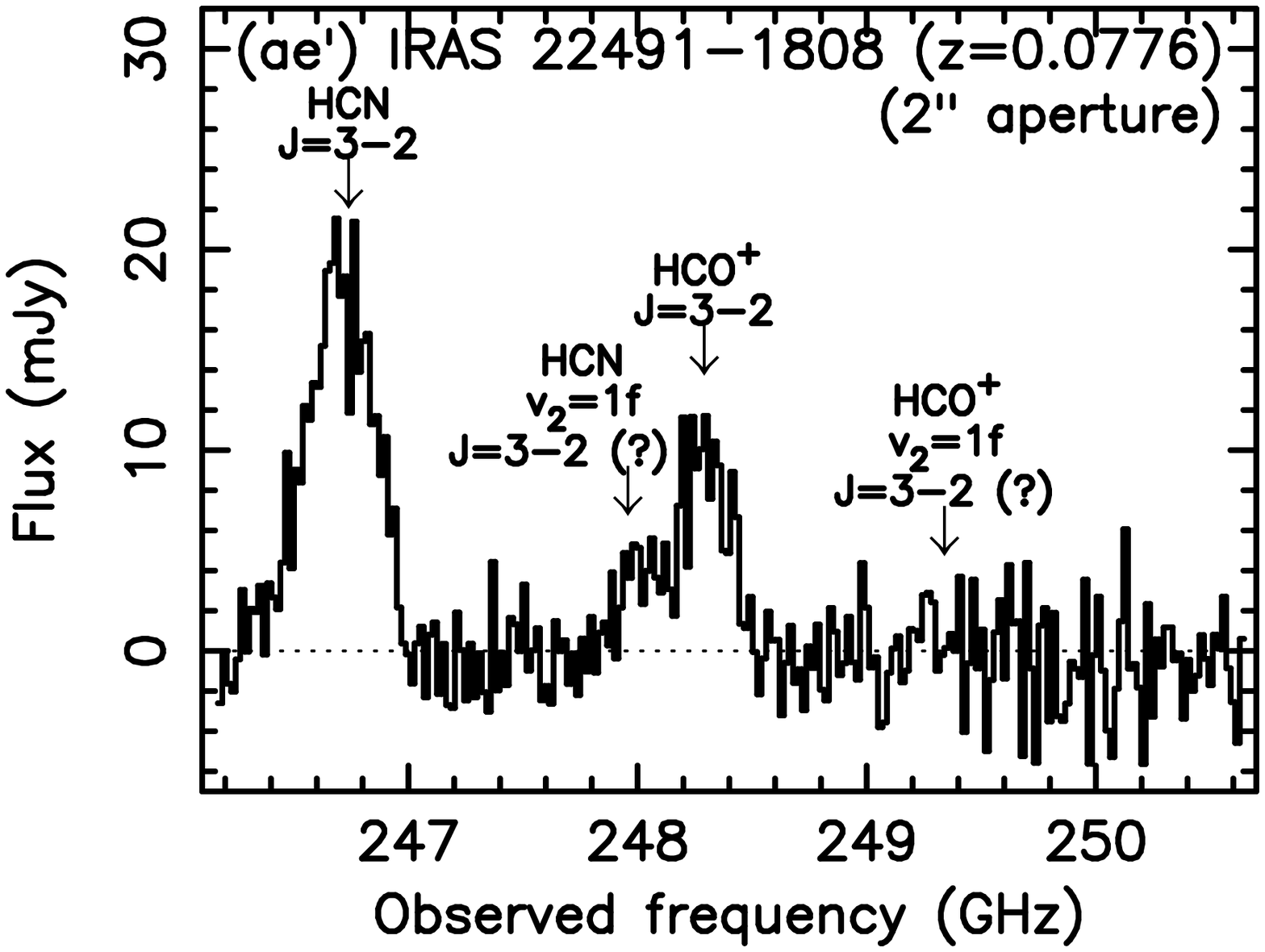} \\ 
\end{center}
\caption{
ALMA full frequency coverage spectra of the observed ULIRGs. 
{\it (Left)}: Spectra at the continuum peak position within the beam size. 
For IRAS 04103$-$2838 (Fig. 2c), the spectrum is extracted at the 
peak position of both the HCN J=3--2 and HCO$^{+}$ J=3--2 emission lines, 
which is significantly displaced from the continuum peak position 
(Fig. 1c).
{\it (Right)}: Spectra within a 1$''$ diameter circular aperture around the 
continuum peak position.
For IRAS 09039$+$0503 (Fig. 2d') and IRAS 11095$-$0238 
(Fig. 2h'), spatially integrated spectra 
with a 1$\farcs$5 diameter circular aperture are shown, which include 
emission from multiple components.
A 1$\farcs$5 (IRAS 21329$-$2346 and IRAS 22206$-$2715, and IRAS 20414$-$1651) 
or 2$''$ (IRAS 12112$+$0305 NE and SW, and IRAS 22491$-$1808) diameter circular 
aperture is used for sources with large synthesized beam sizes 
(0$\farcs$5--0$\farcs$9).
Spatially integrated spectra are not shown for IRAS 10485$-$1447 and 
IRAS 02411$-$0353, because molecular emission lines are too faint 
to be detectable, due to increased noise in wider aperture spectra.
}
\end{figure}

\clearpage

\begin{figure}
\begin{center}
\includegraphics[angle=0,scale=.314]{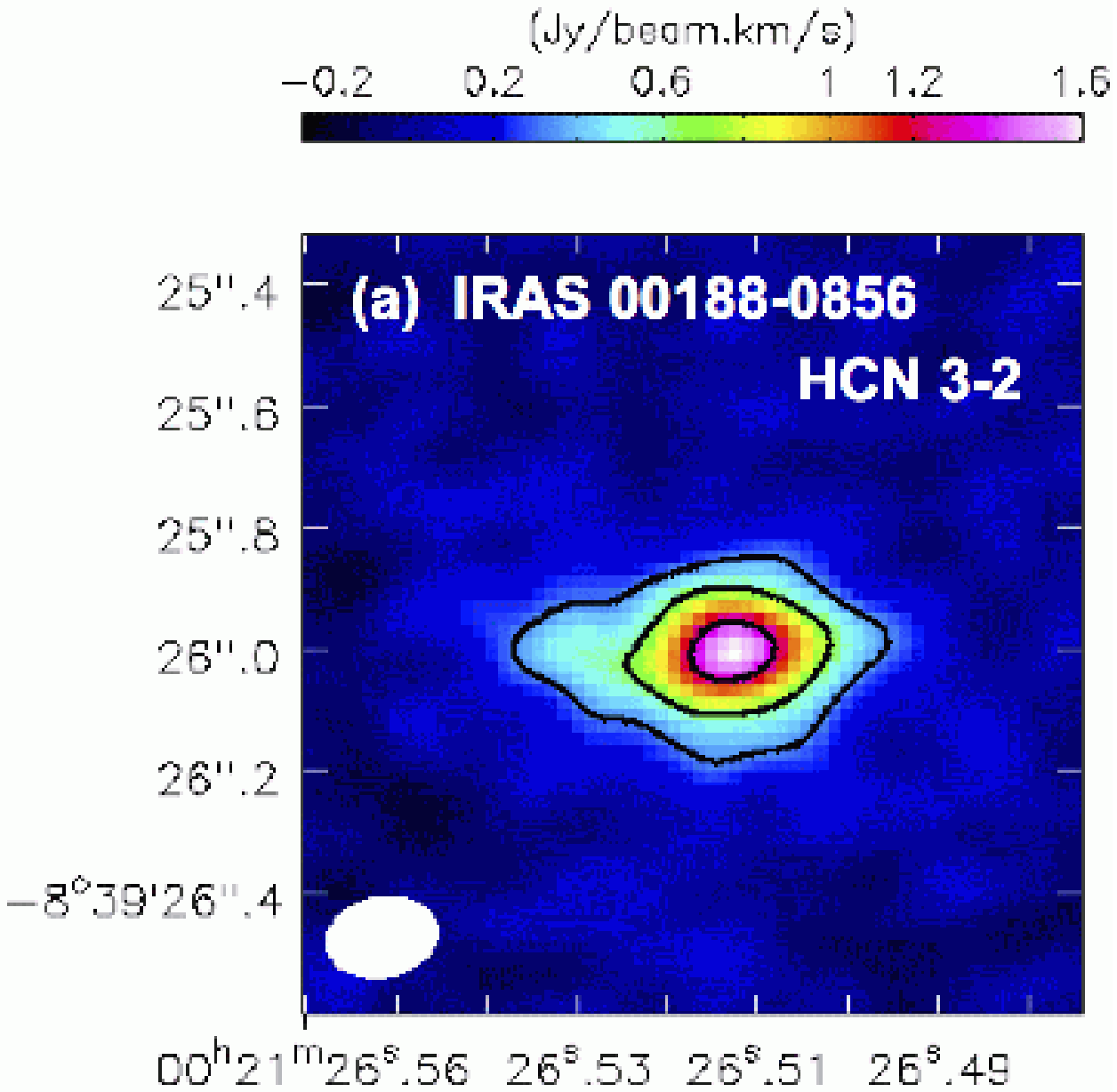} 
\includegraphics[angle=0,scale=.314]{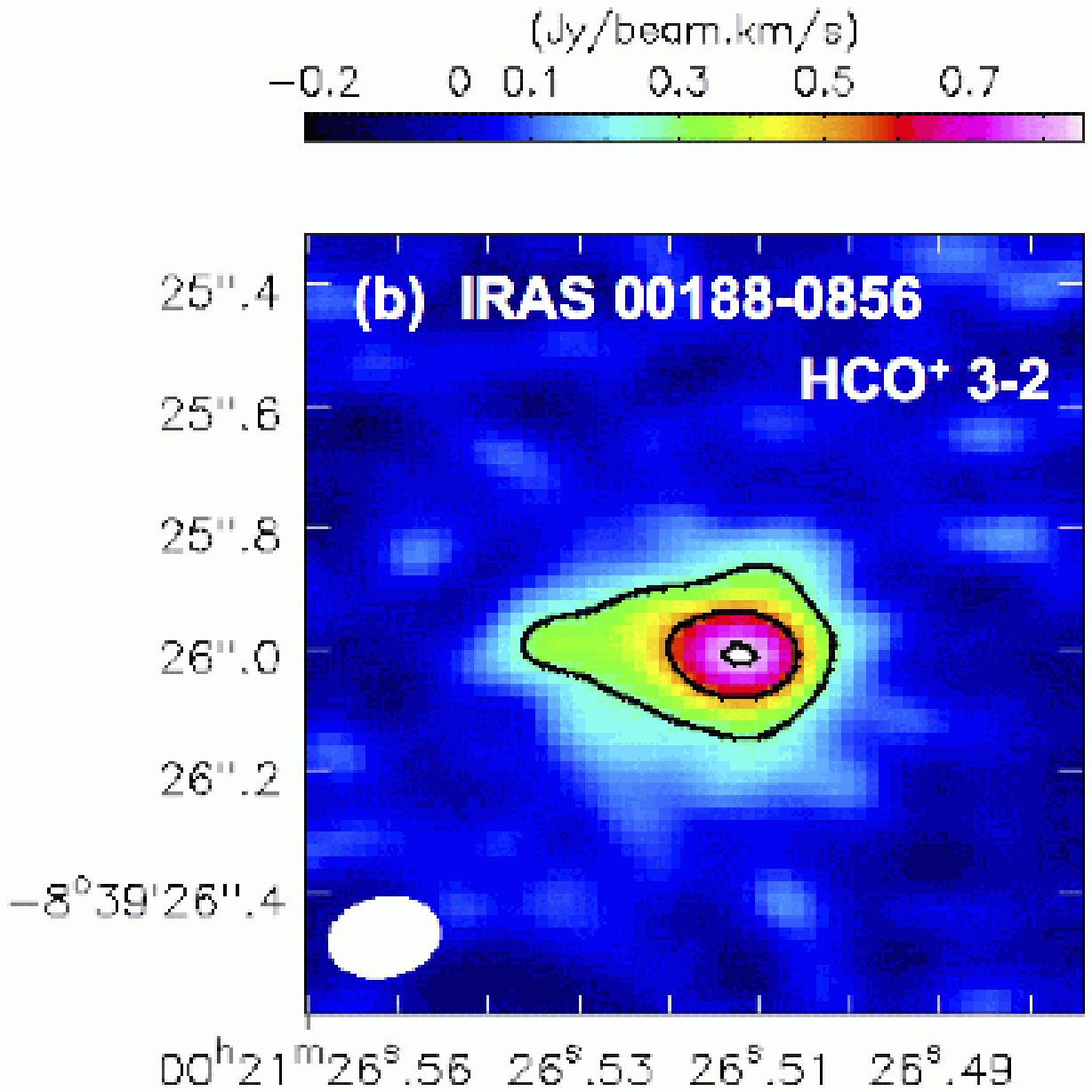}
\includegraphics[angle=0,scale=.314]{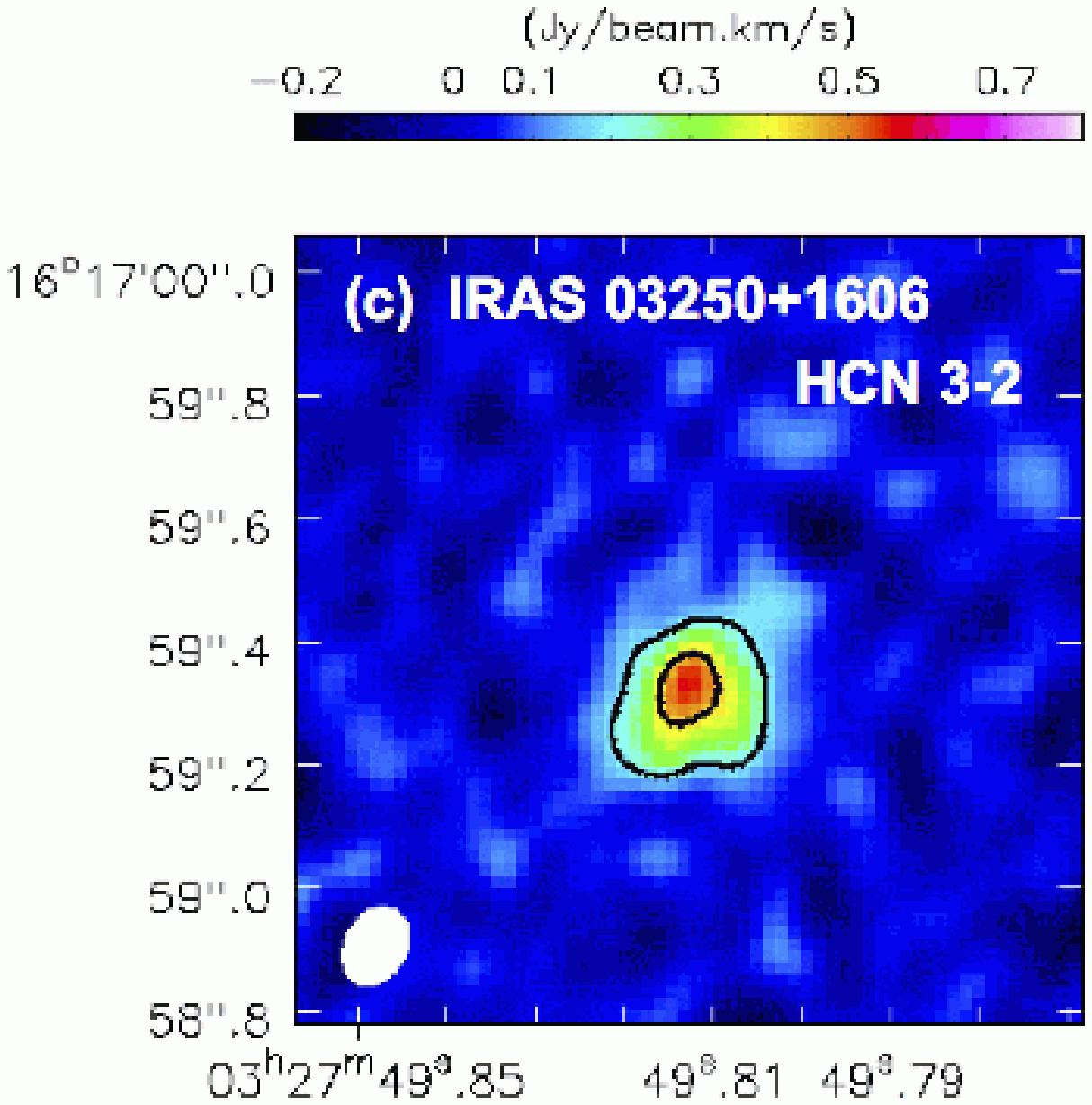} 
\includegraphics[angle=0,scale=.314]{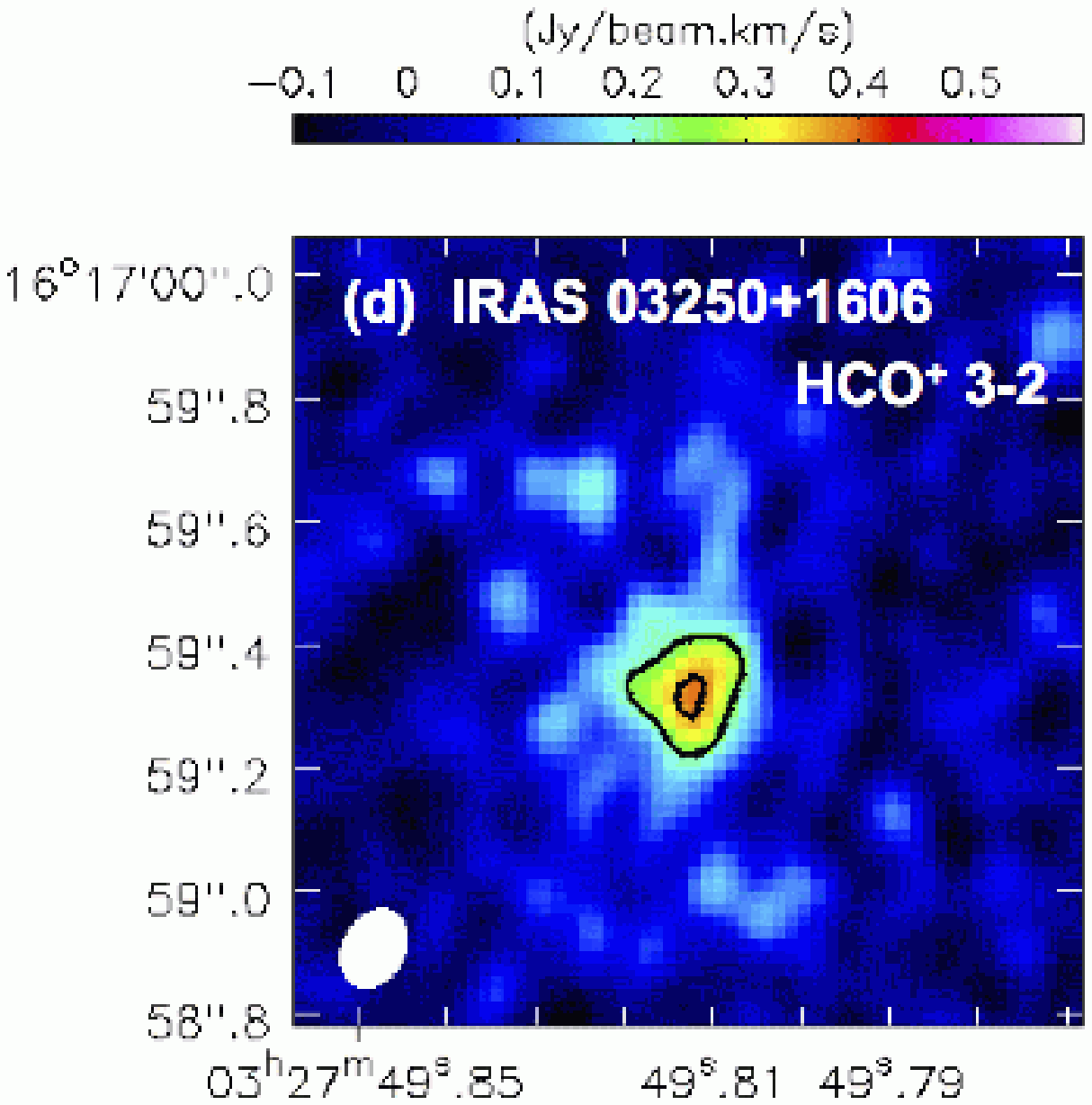}\\ 
\includegraphics[angle=0,scale=.314]{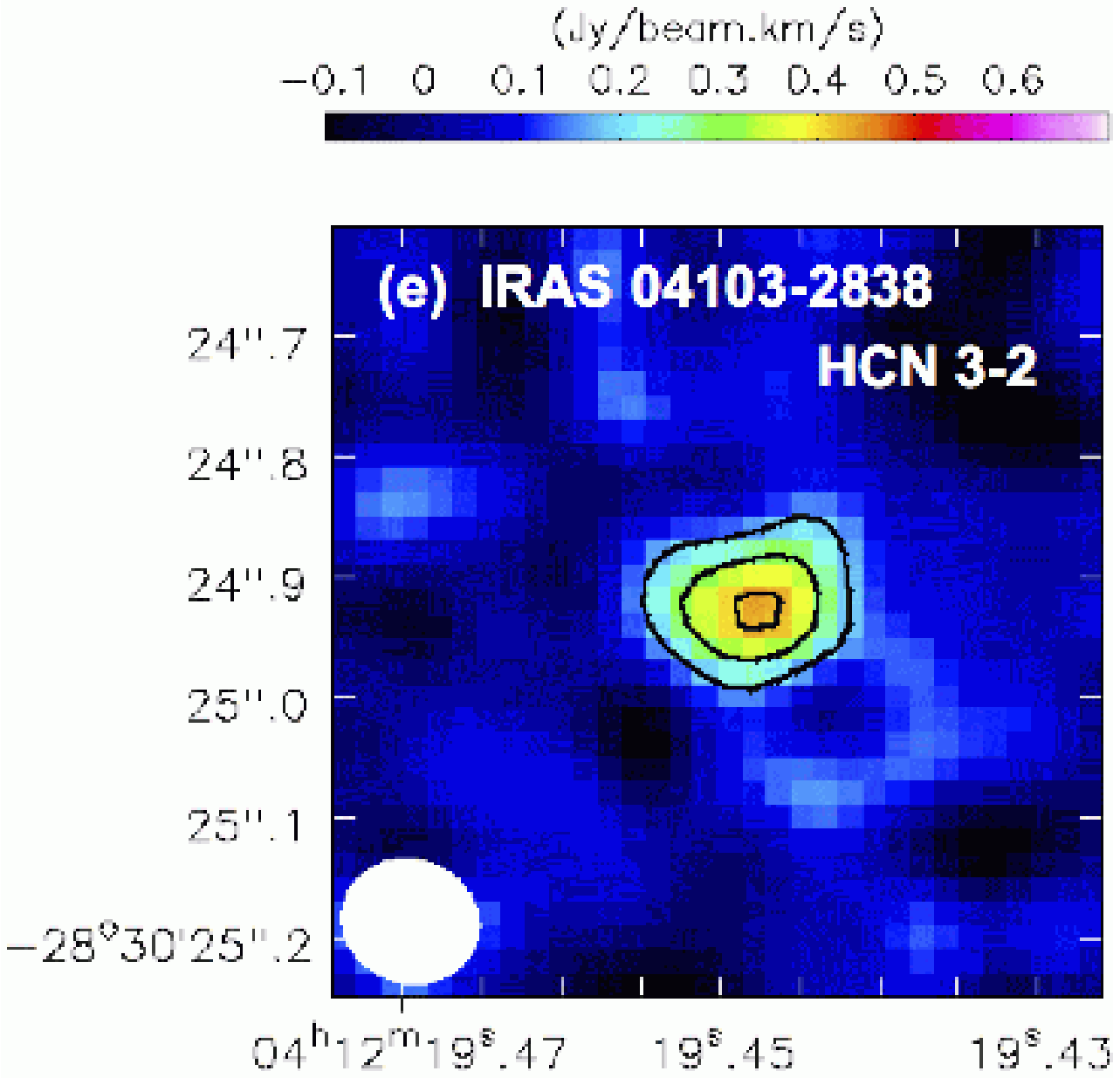} 
\includegraphics[angle=0,scale=.314]{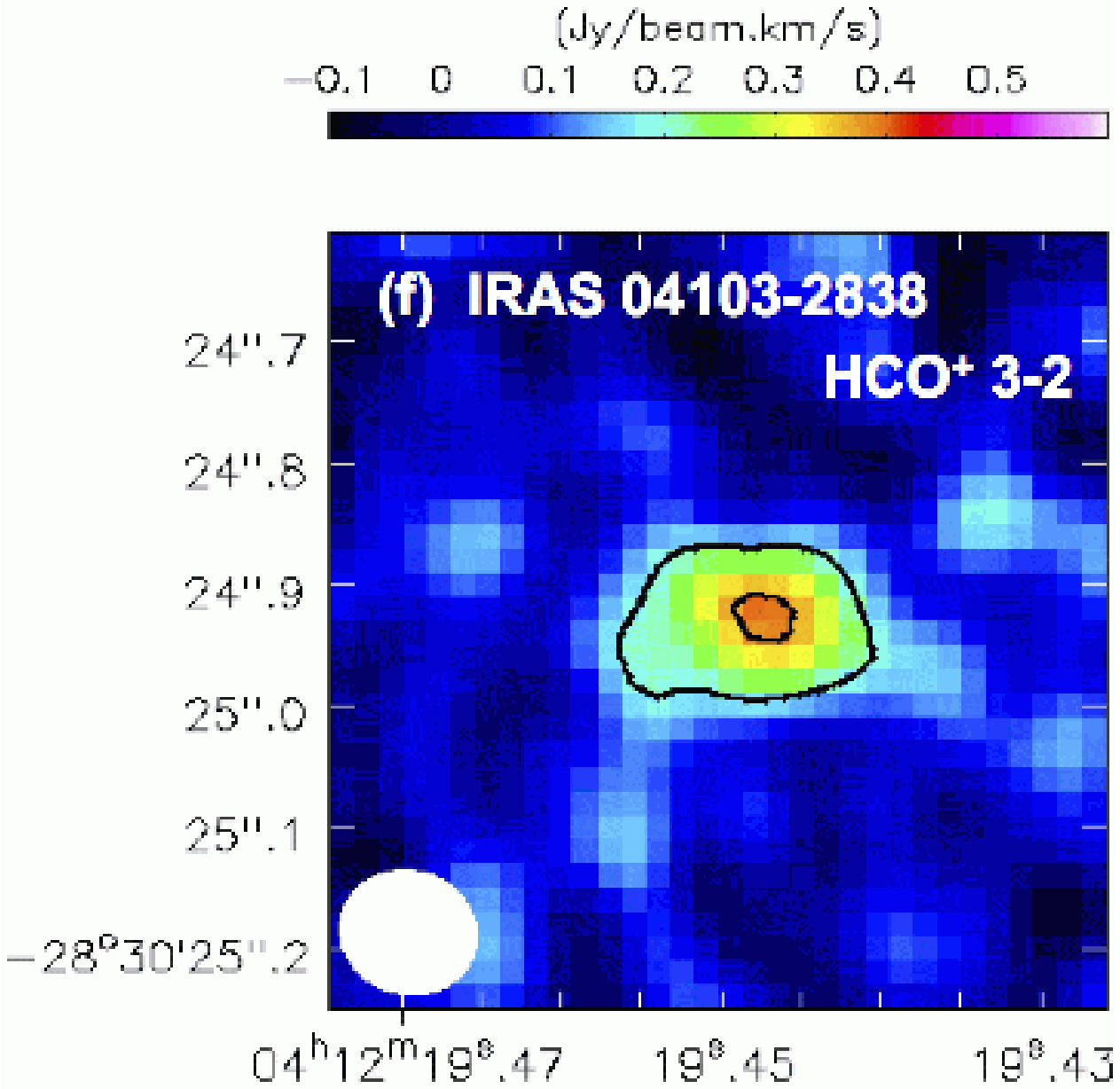} 
\includegraphics[angle=0,scale=.314]{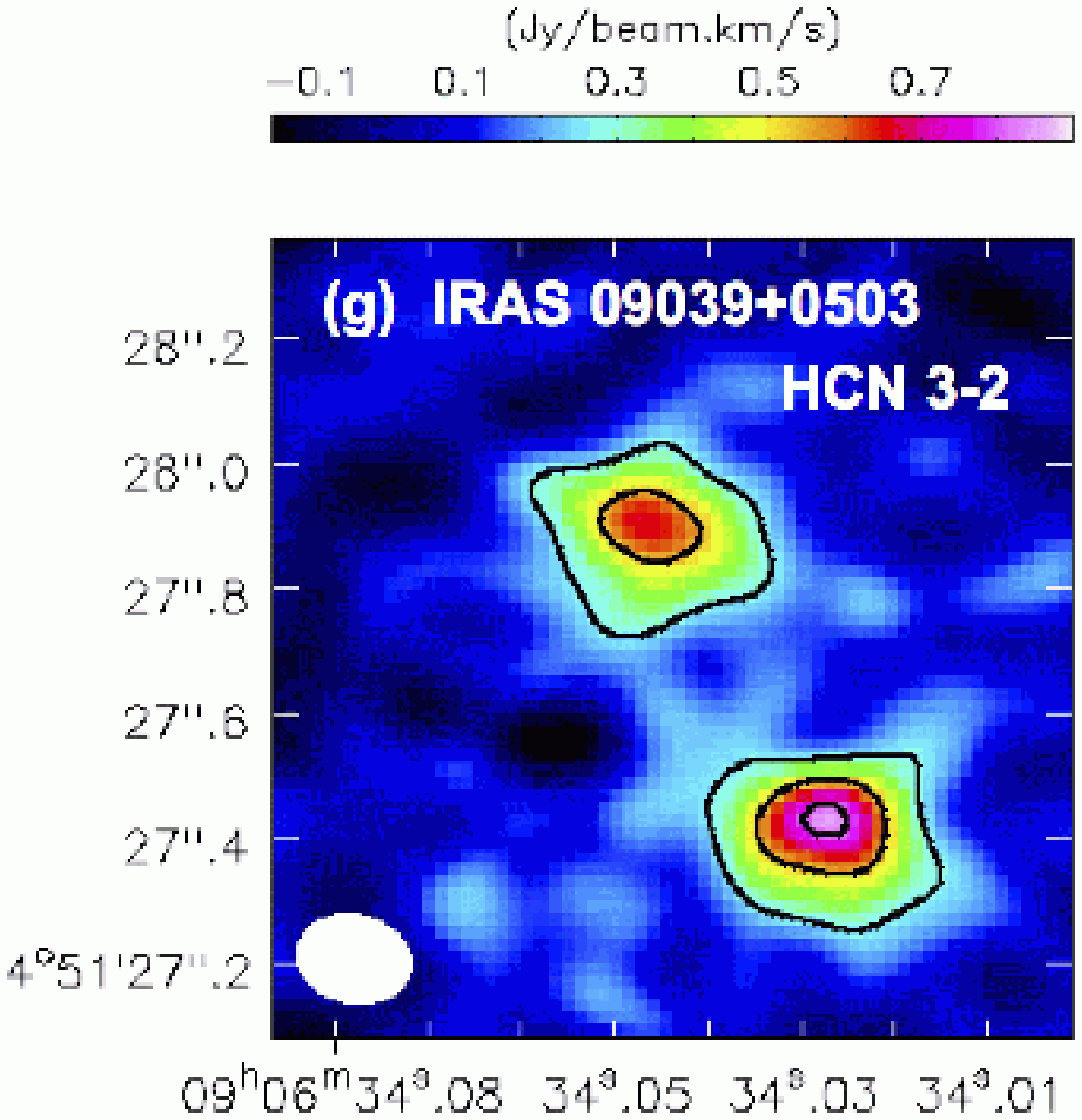} 
\includegraphics[angle=0,scale=.314]{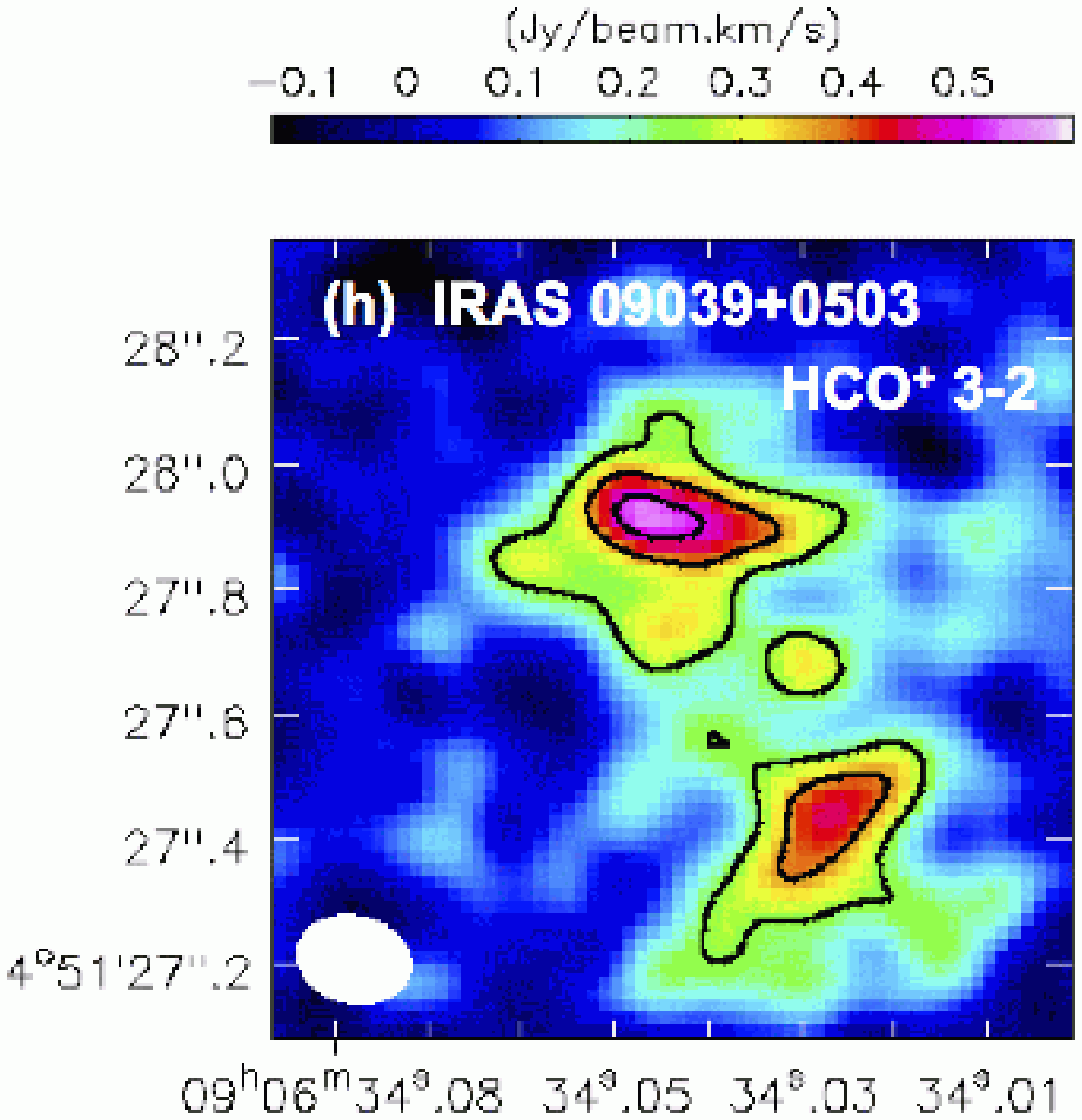}\\ 
\includegraphics[angle=0,scale=.314]{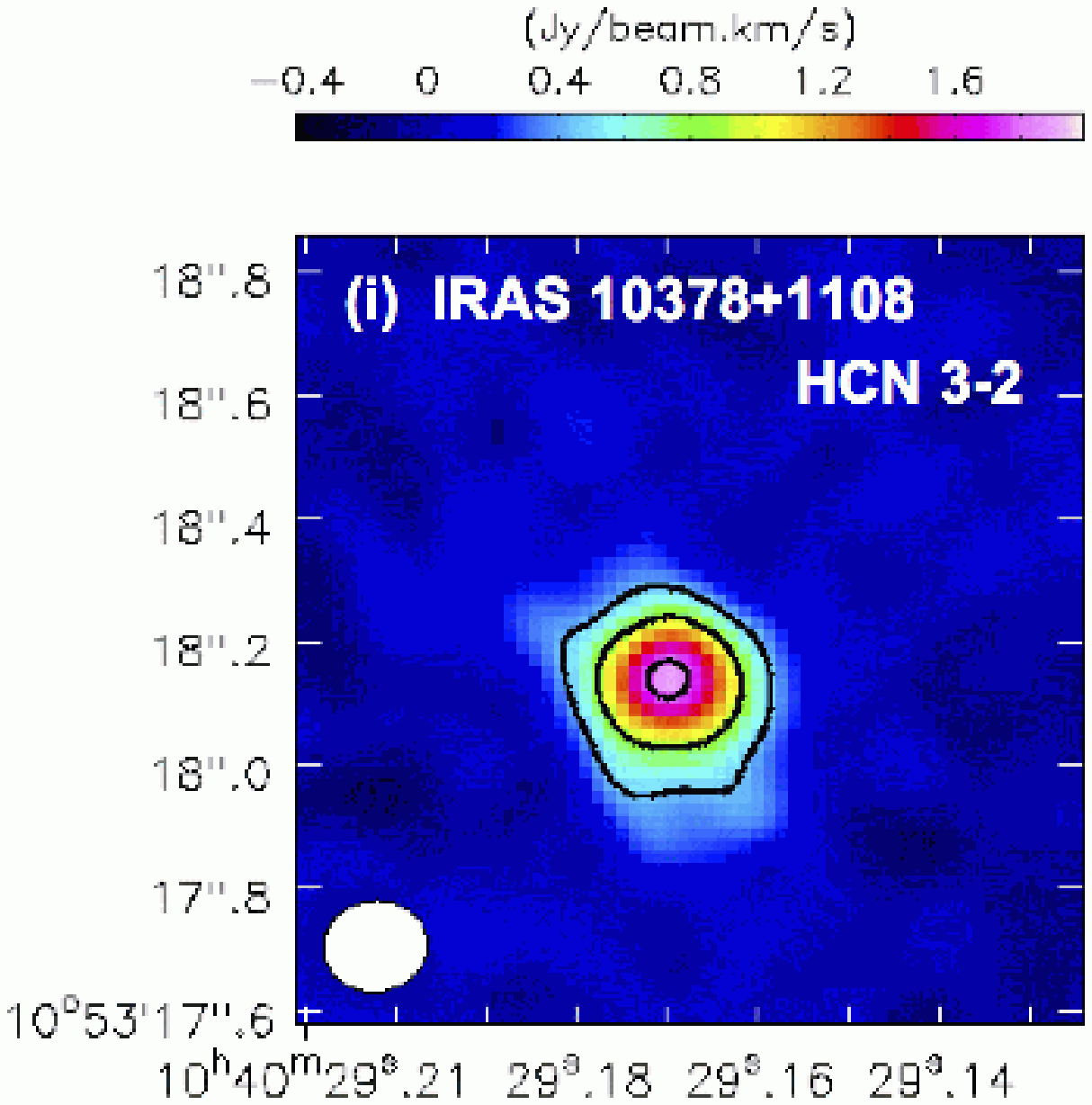} 
\includegraphics[angle=0,scale=.314]{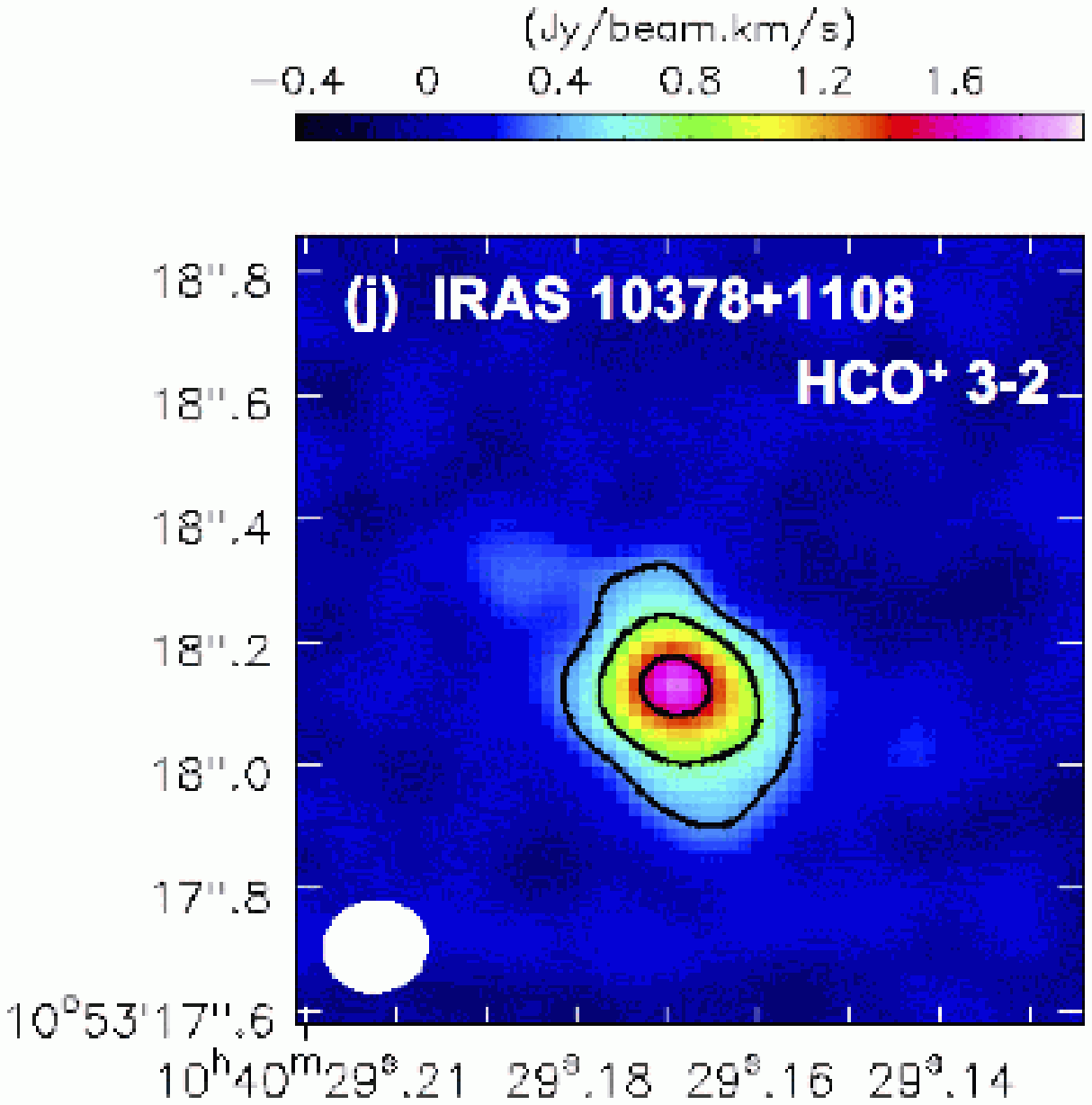} 
\includegraphics[angle=0,scale=.314]{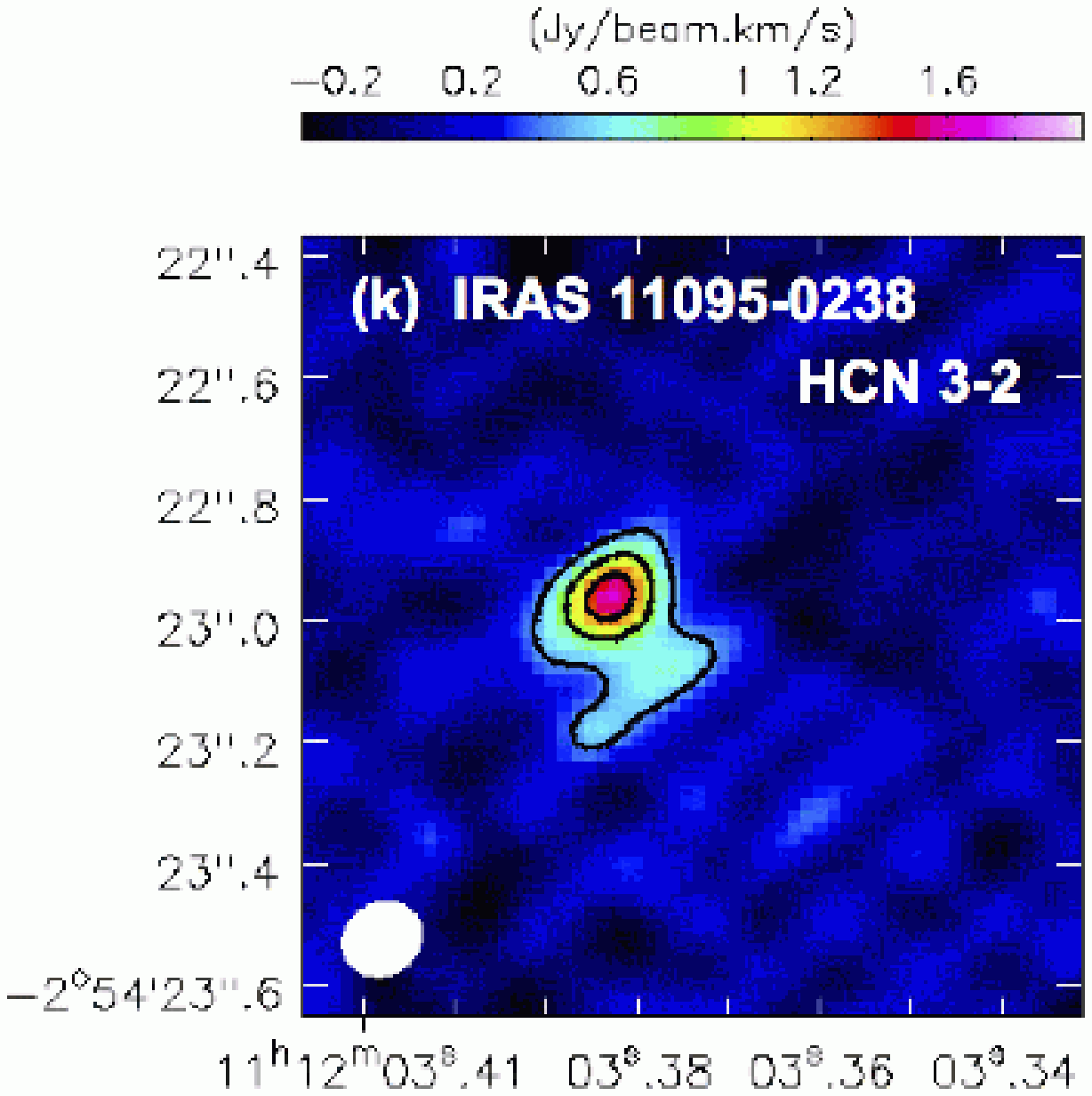} 
\includegraphics[angle=0,scale=.314]{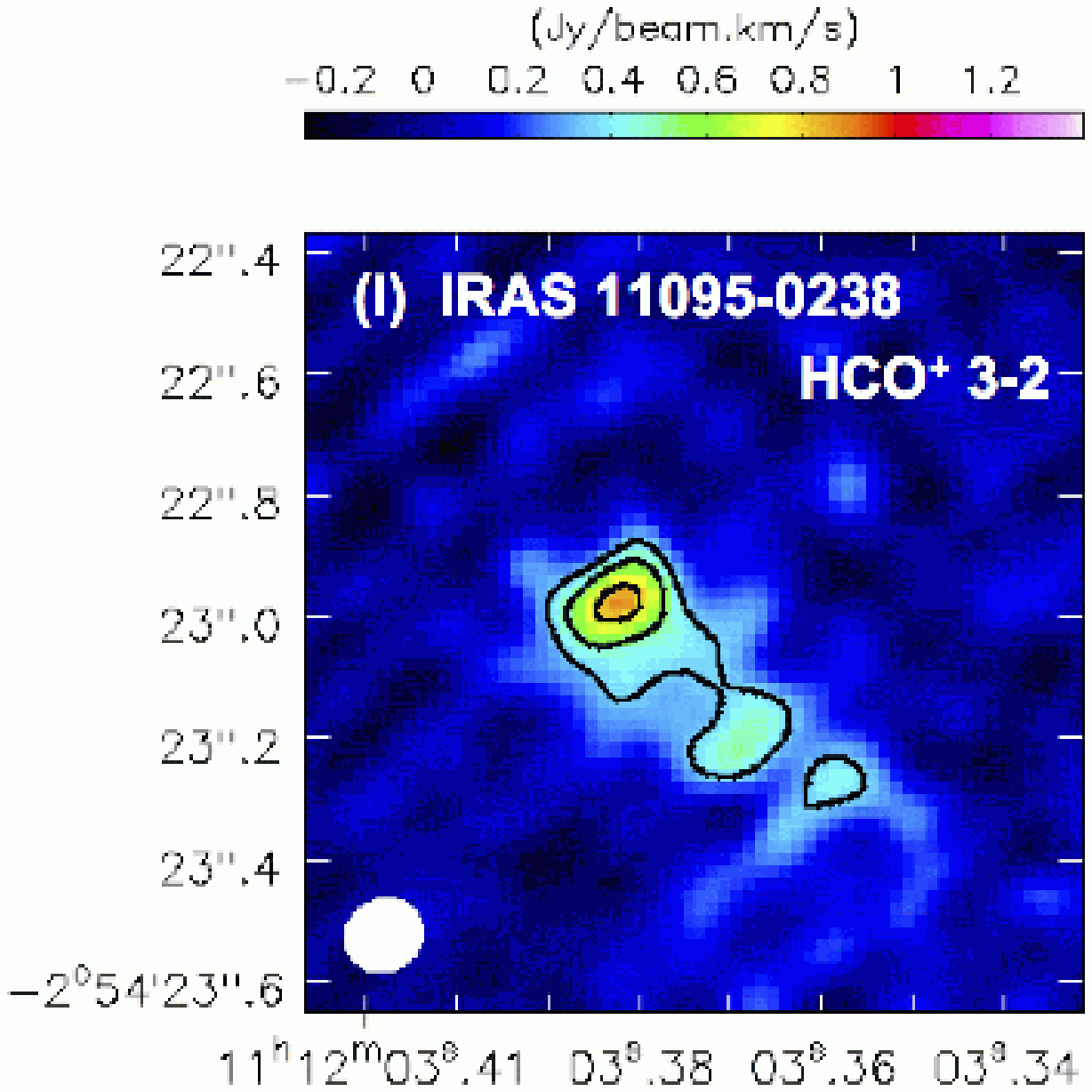}\\ 
\includegraphics[angle=0,scale=.314]{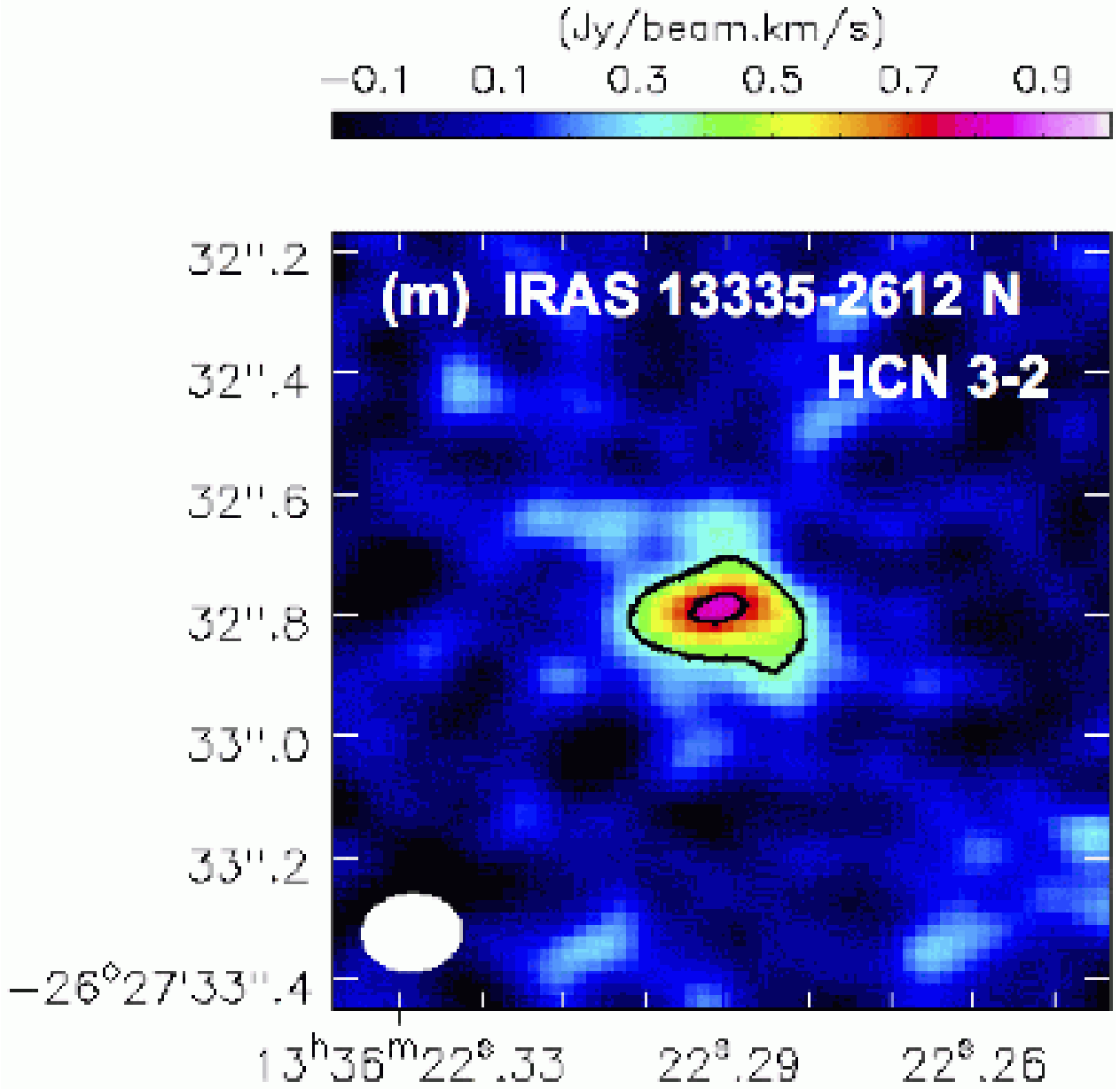} 
\includegraphics[angle=0,scale=.314]{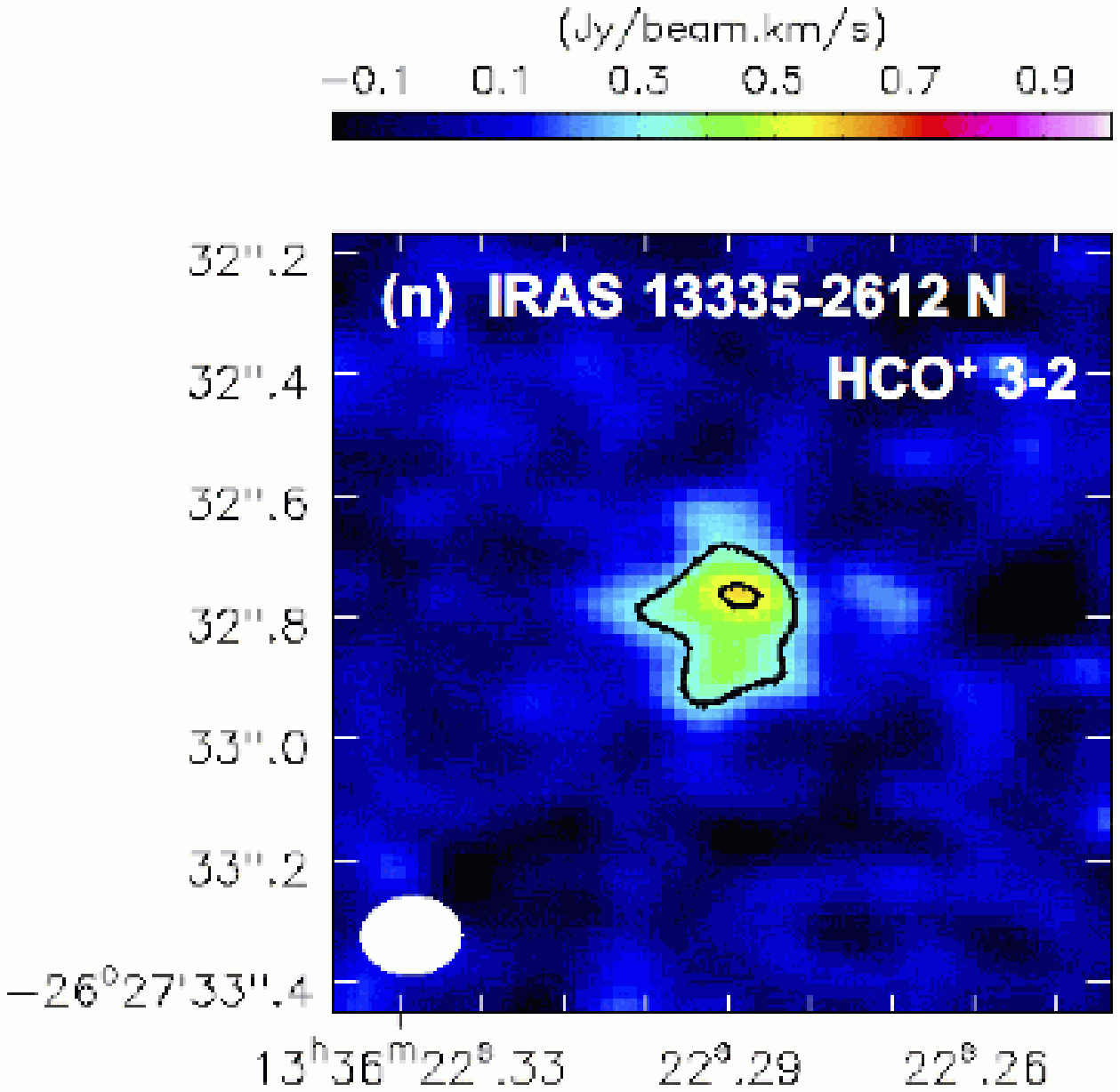} 
\includegraphics[angle=0,scale=.314]{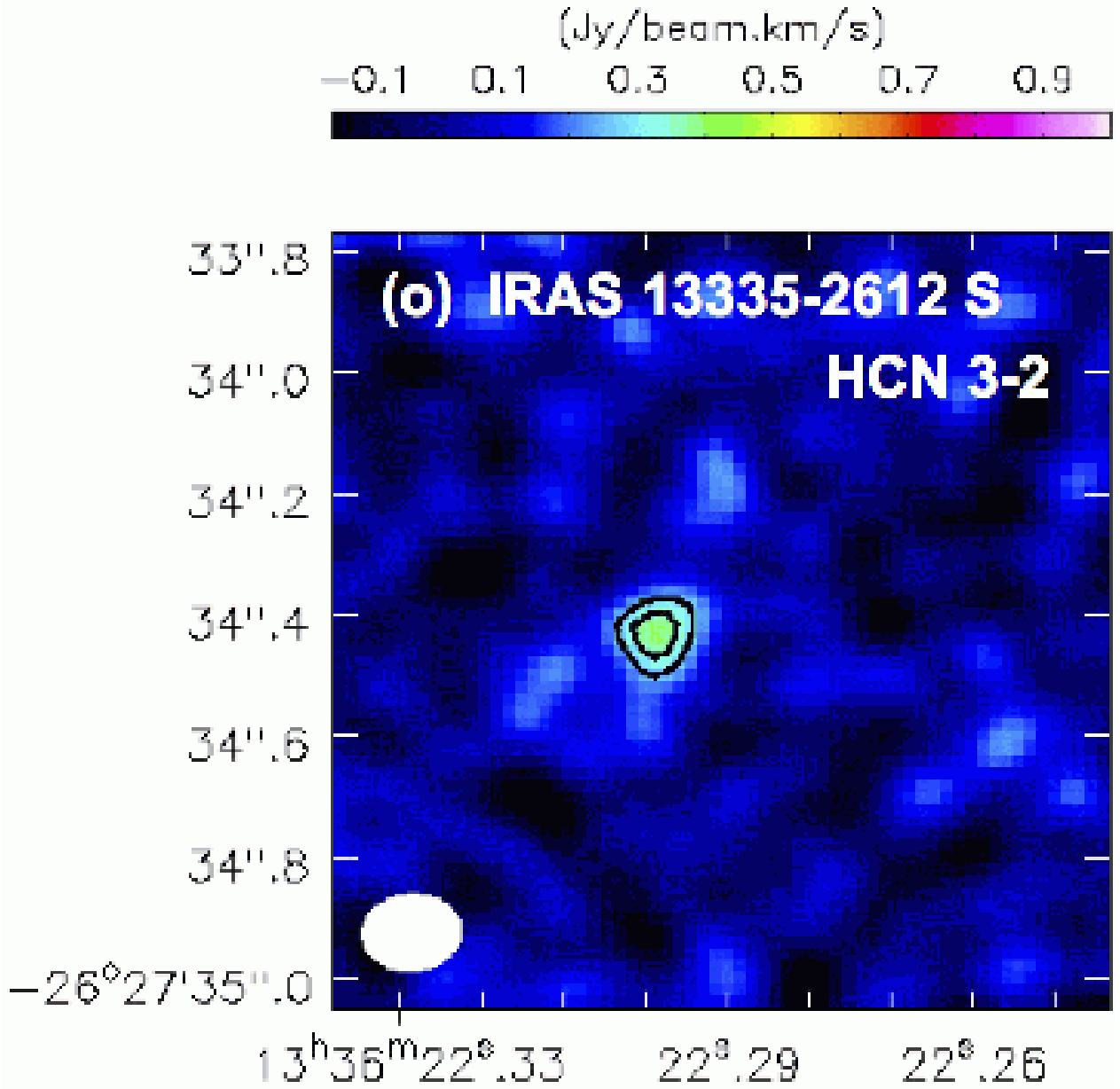} 
\includegraphics[angle=0,scale=.314]{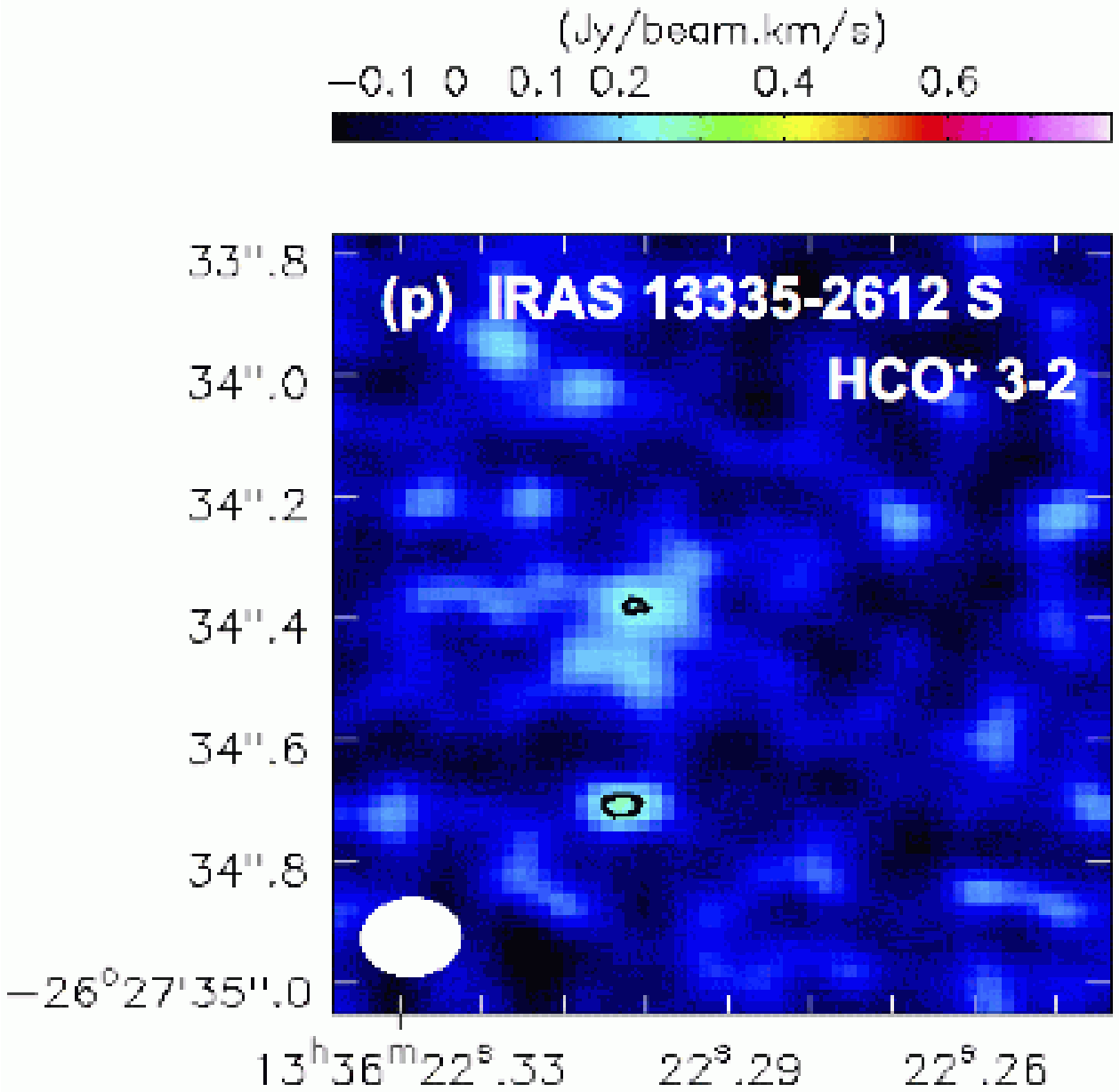} \\
\includegraphics[angle=0,scale=.314]{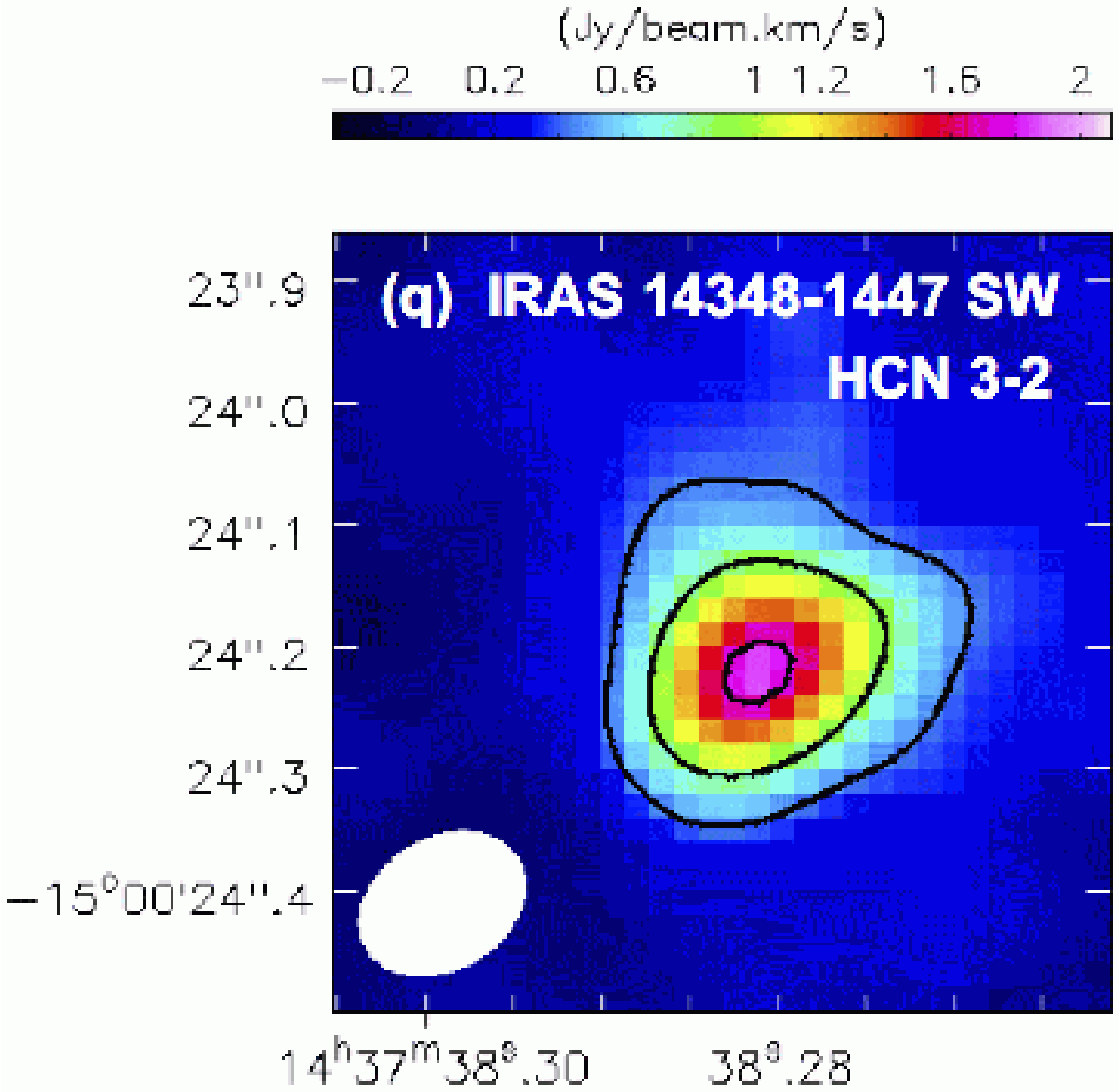} 
\includegraphics[angle=0,scale=.314]{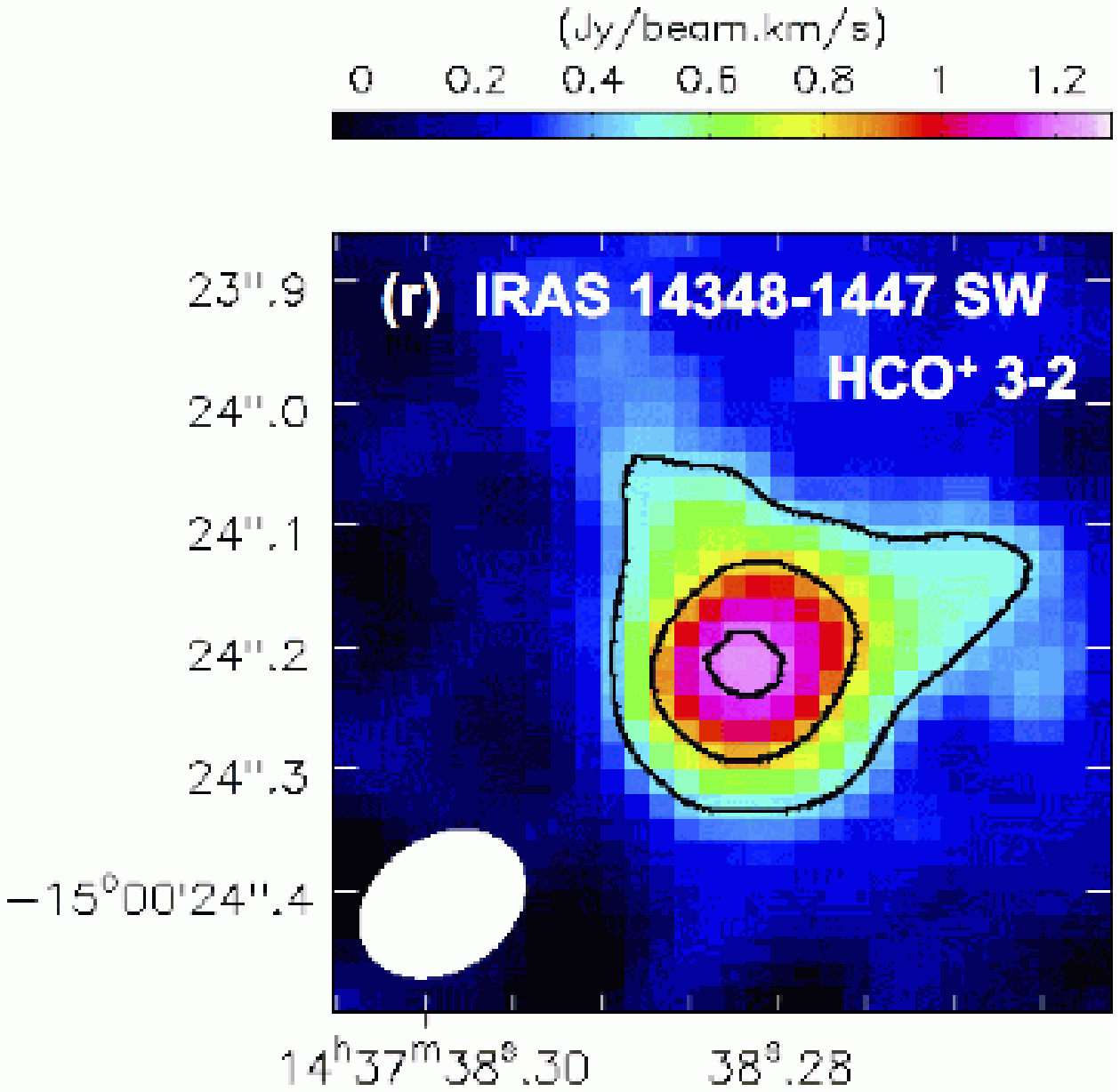} 
\includegraphics[angle=0,scale=.3104]{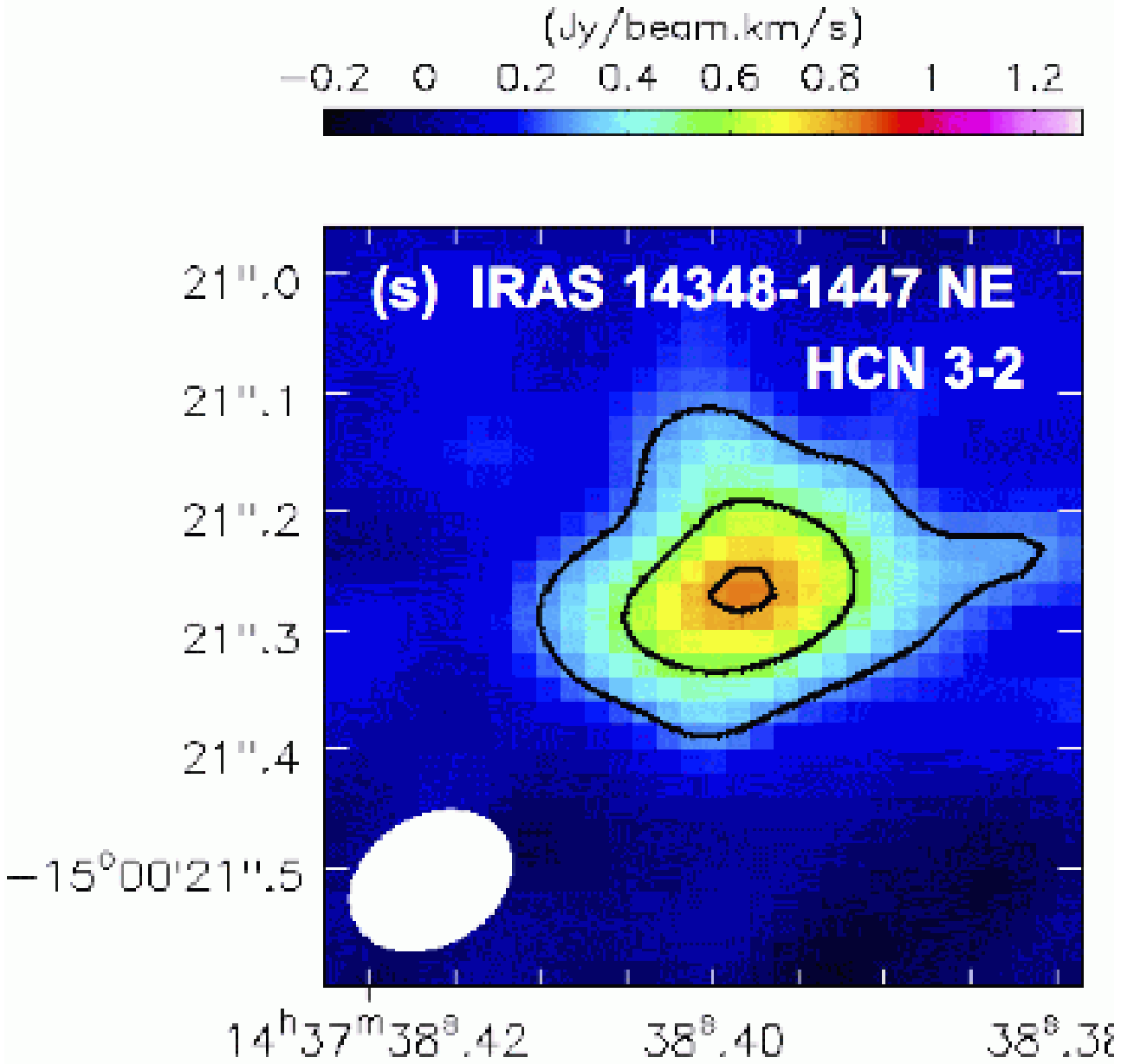} 
\includegraphics[angle=0,scale=.3104]{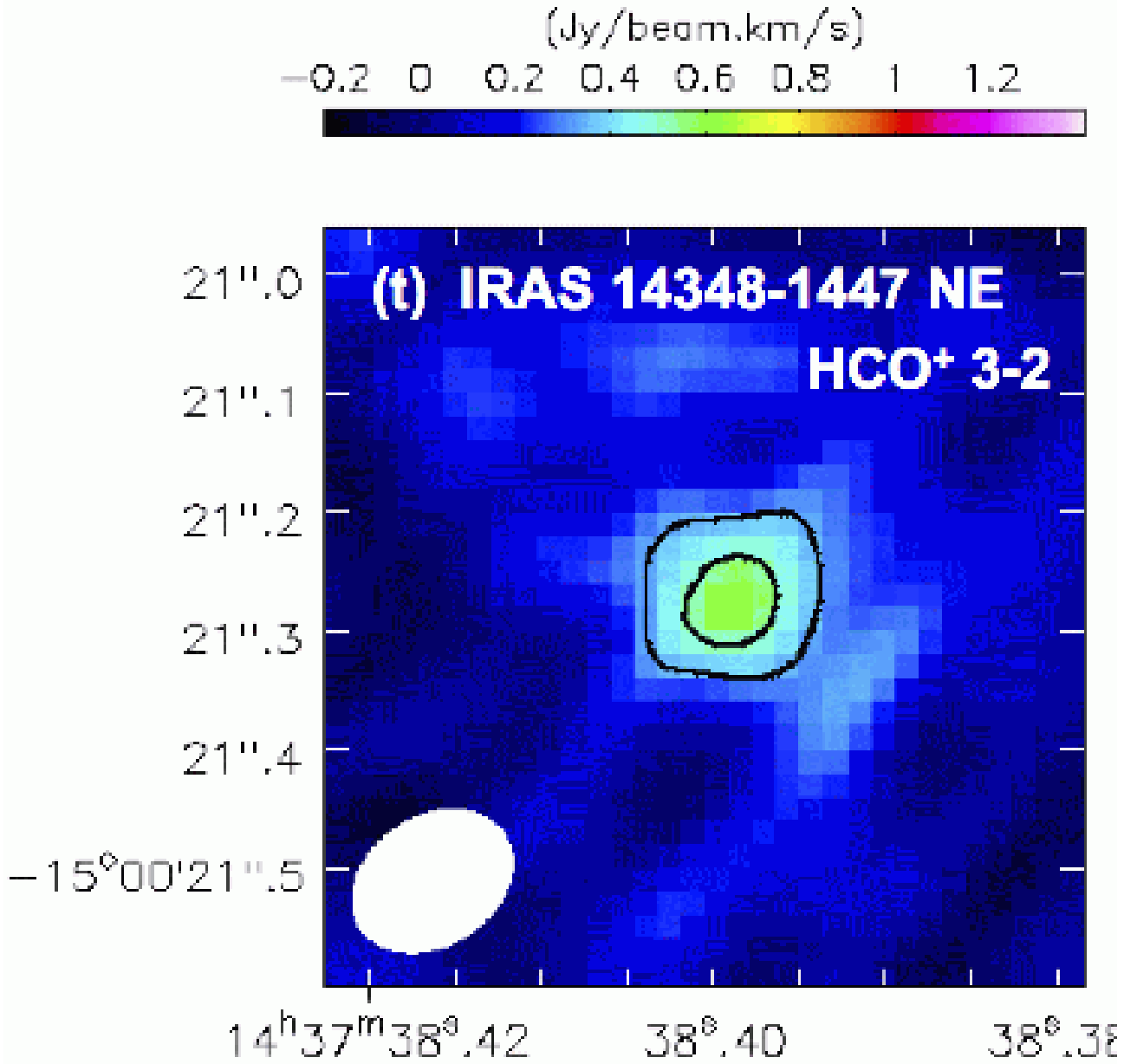} \\
\includegraphics[angle=0,scale=.314]{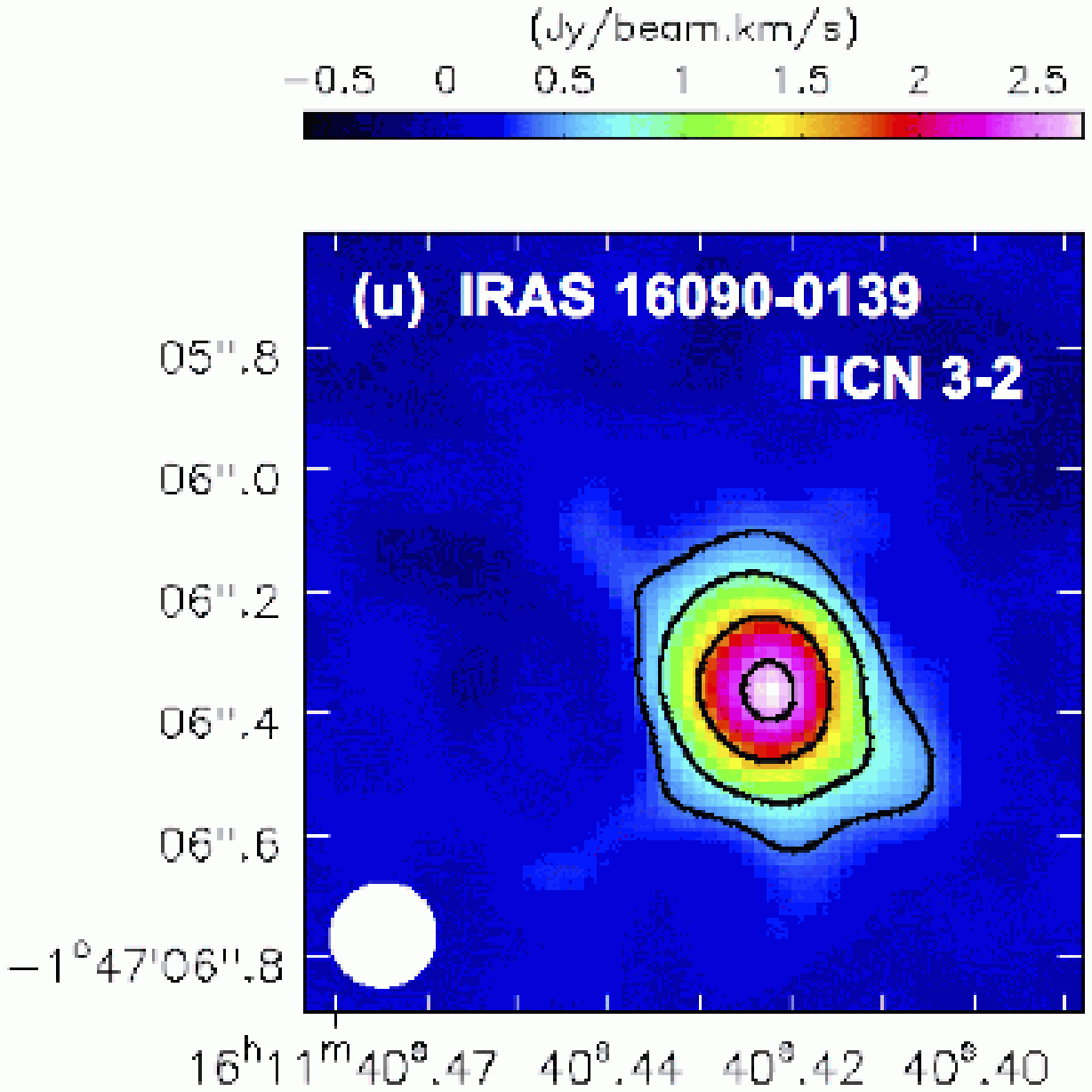} 
\includegraphics[angle=0,scale=.314]{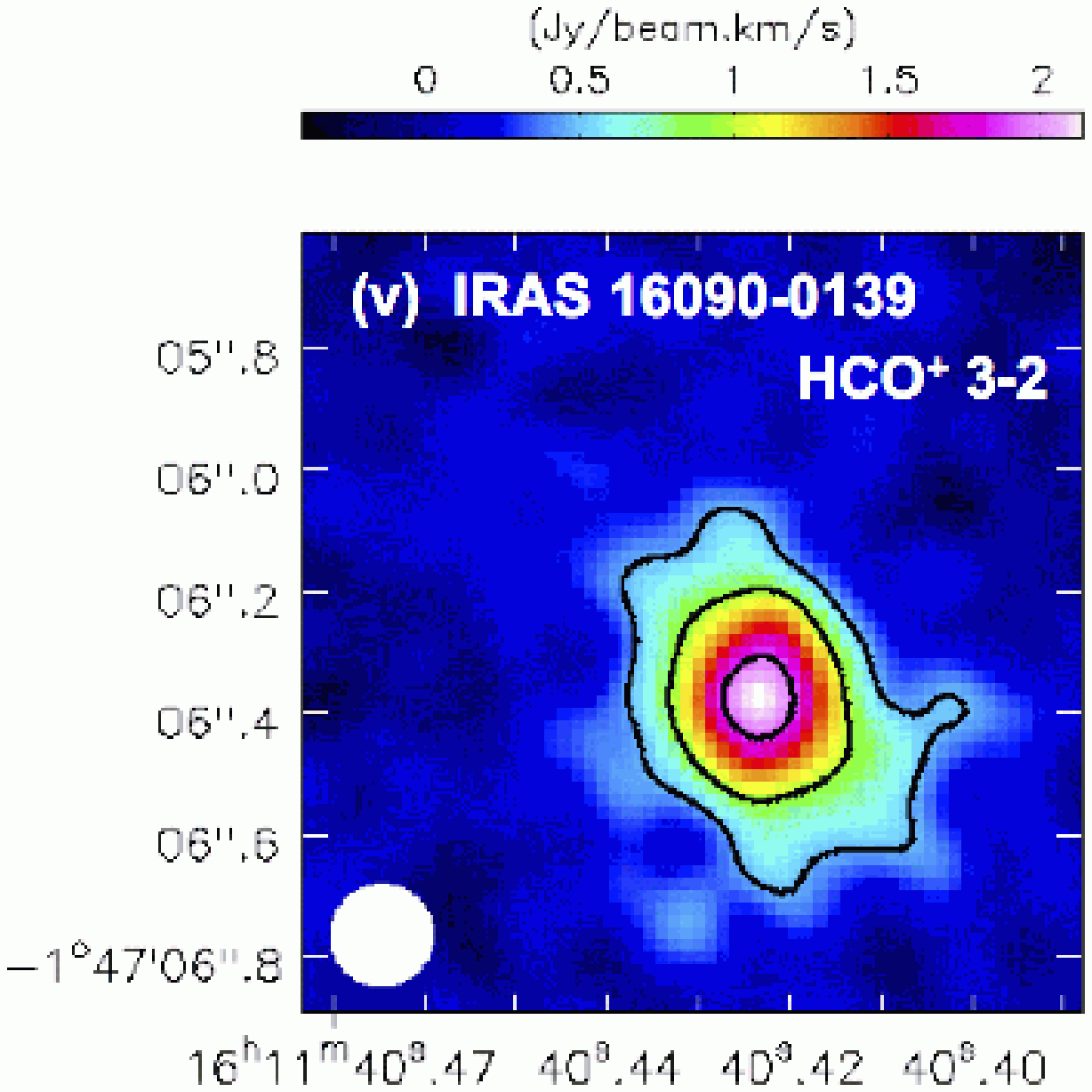} 
\includegraphics[angle=0,scale=.314]{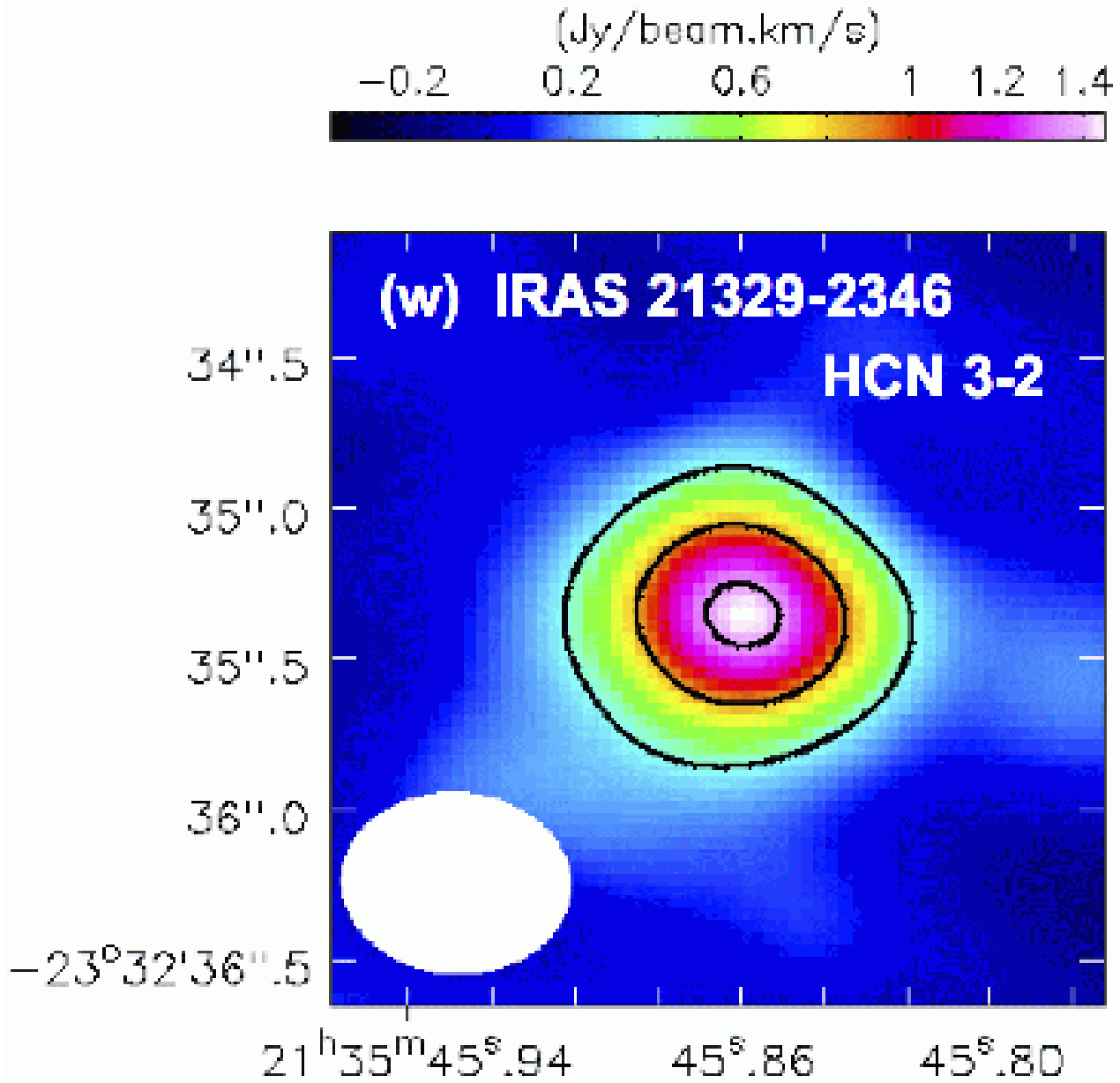} 
\includegraphics[angle=0,scale=.314]{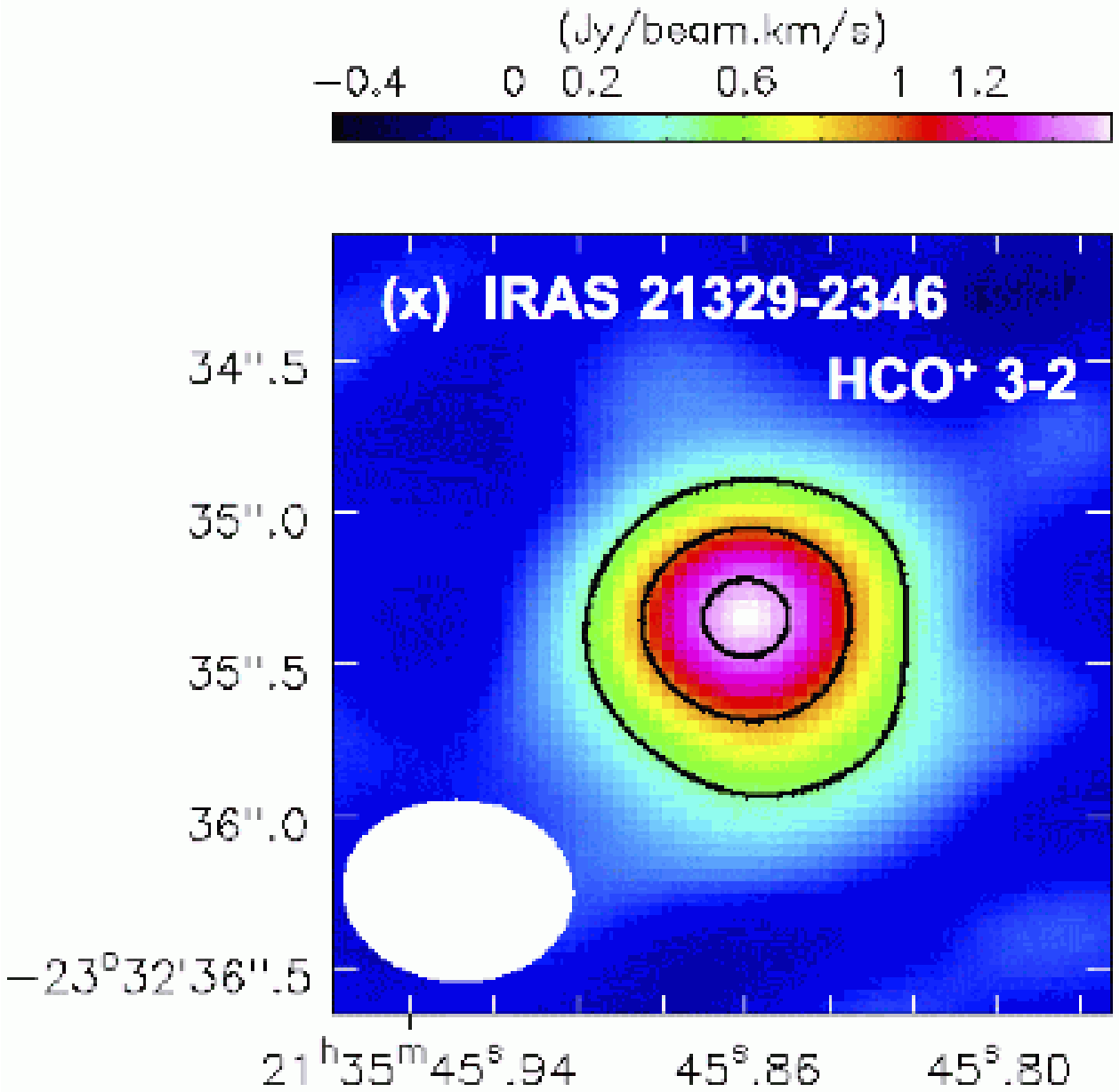} \\
\end{center}
\end{figure}

\clearpage

\begin{figure}
\begin{center}
\includegraphics[angle=0,scale=.314]{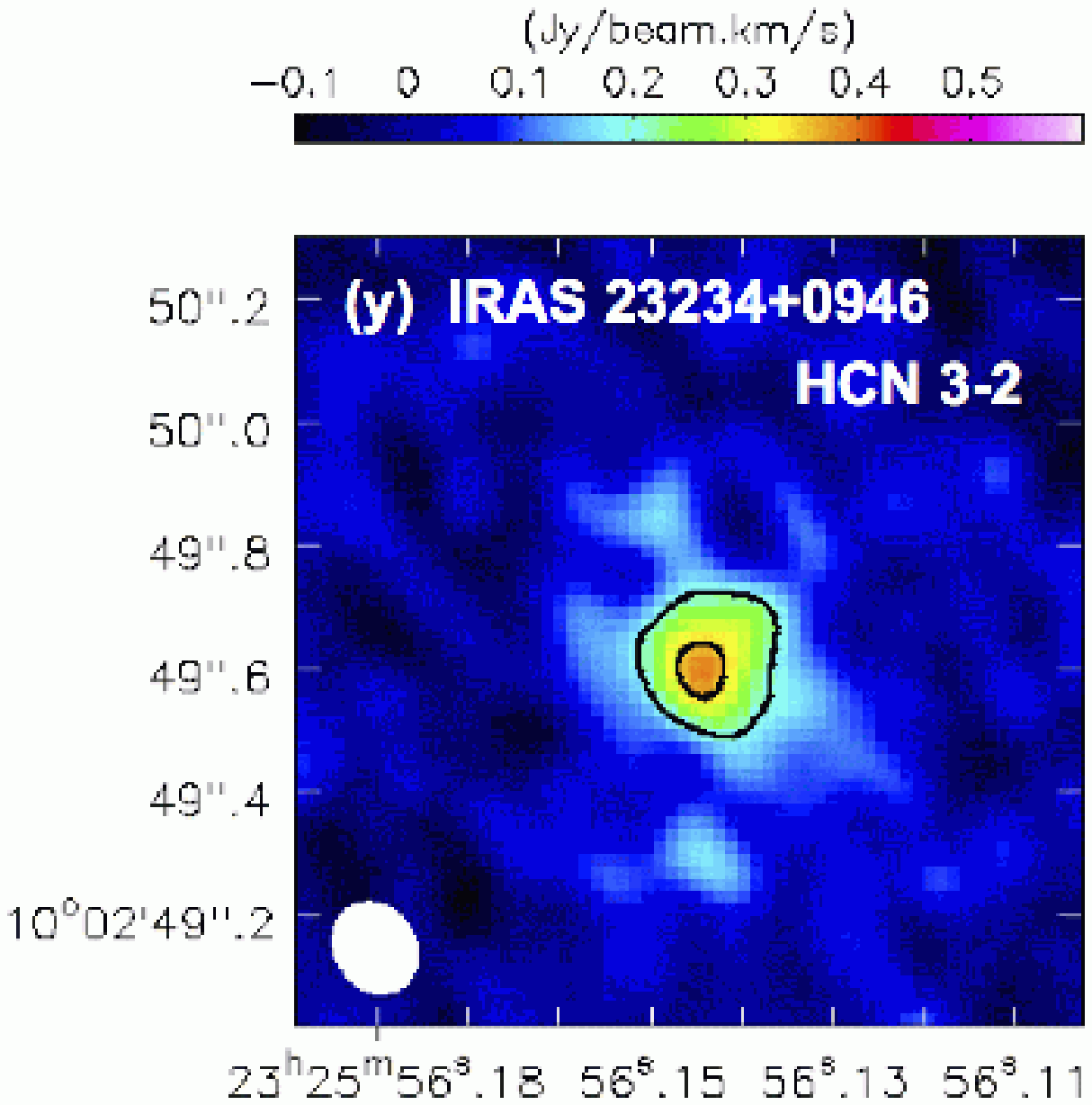} 
\includegraphics[angle=0,scale=.314]{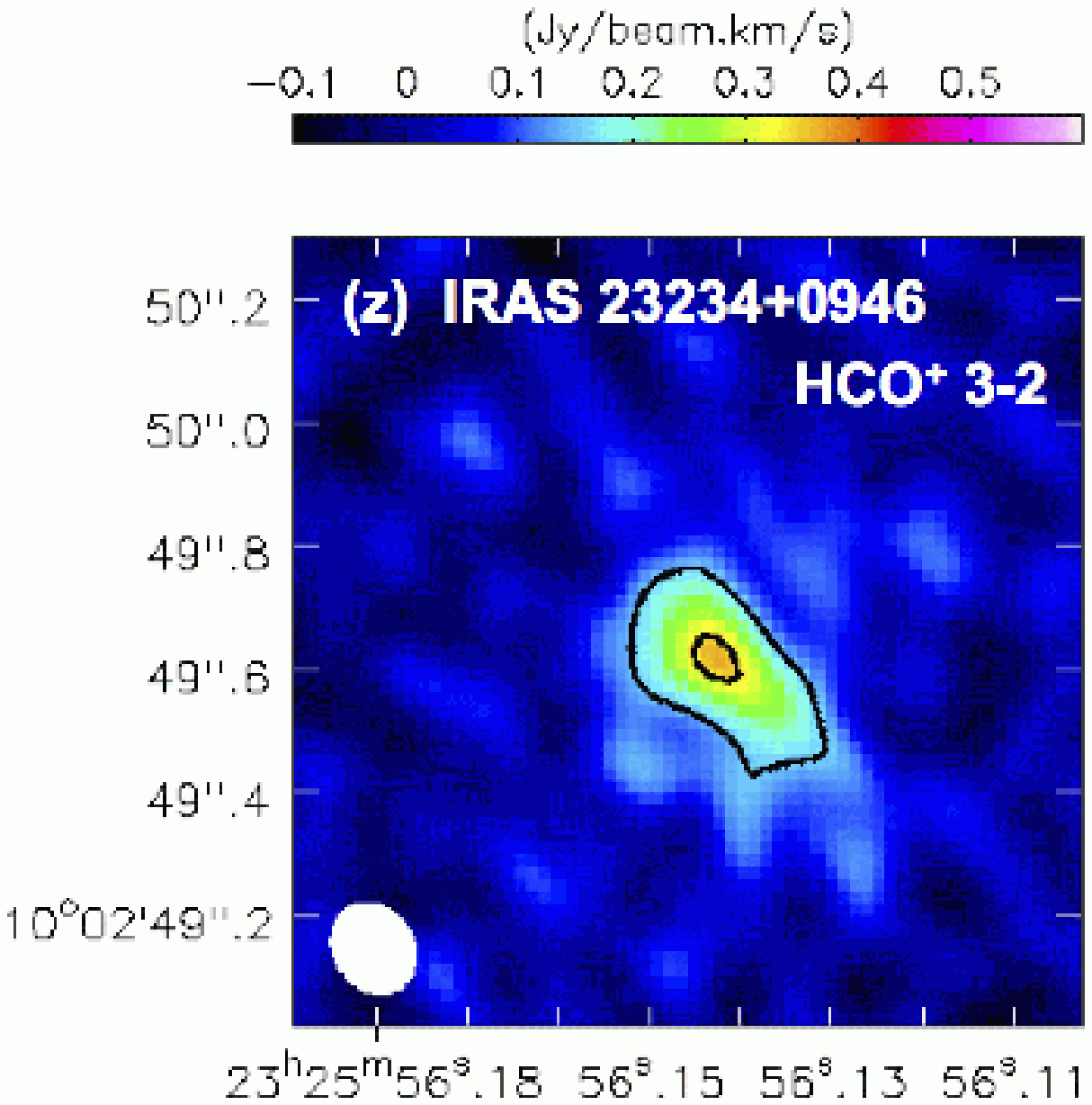} 
\includegraphics[angle=0,scale=.314]{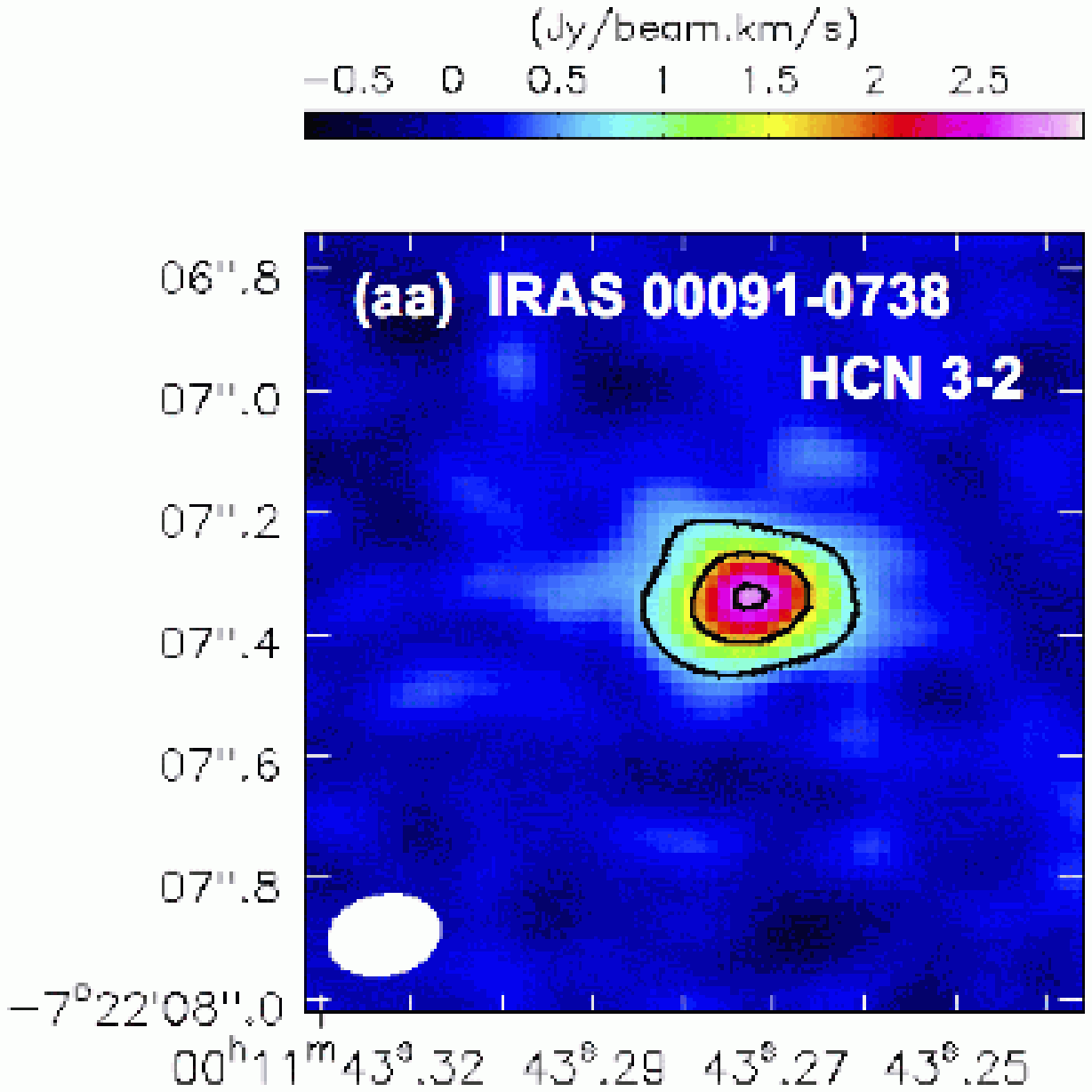} 
\includegraphics[angle=0,scale=.314]{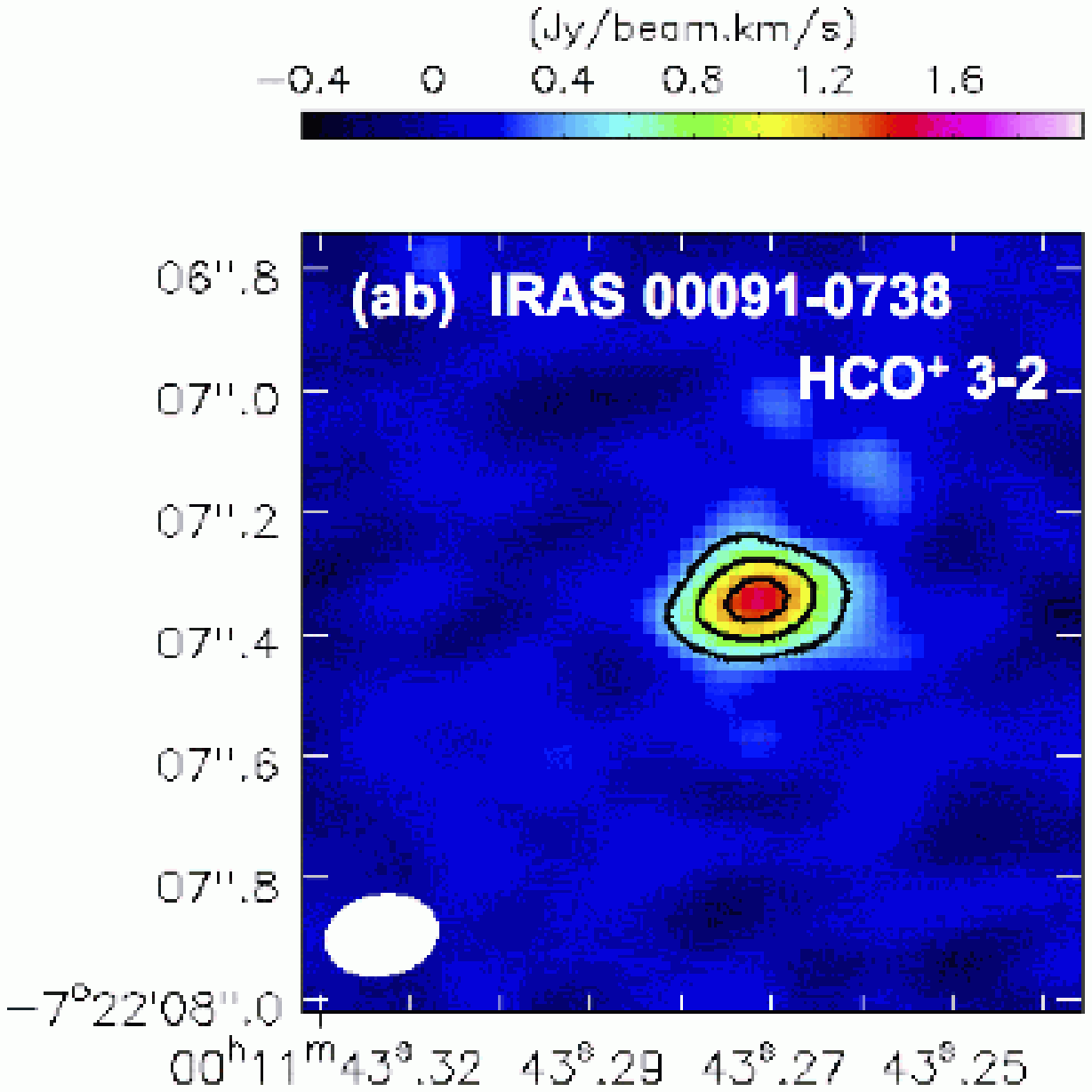} \\
\includegraphics[angle=0,scale=.314]{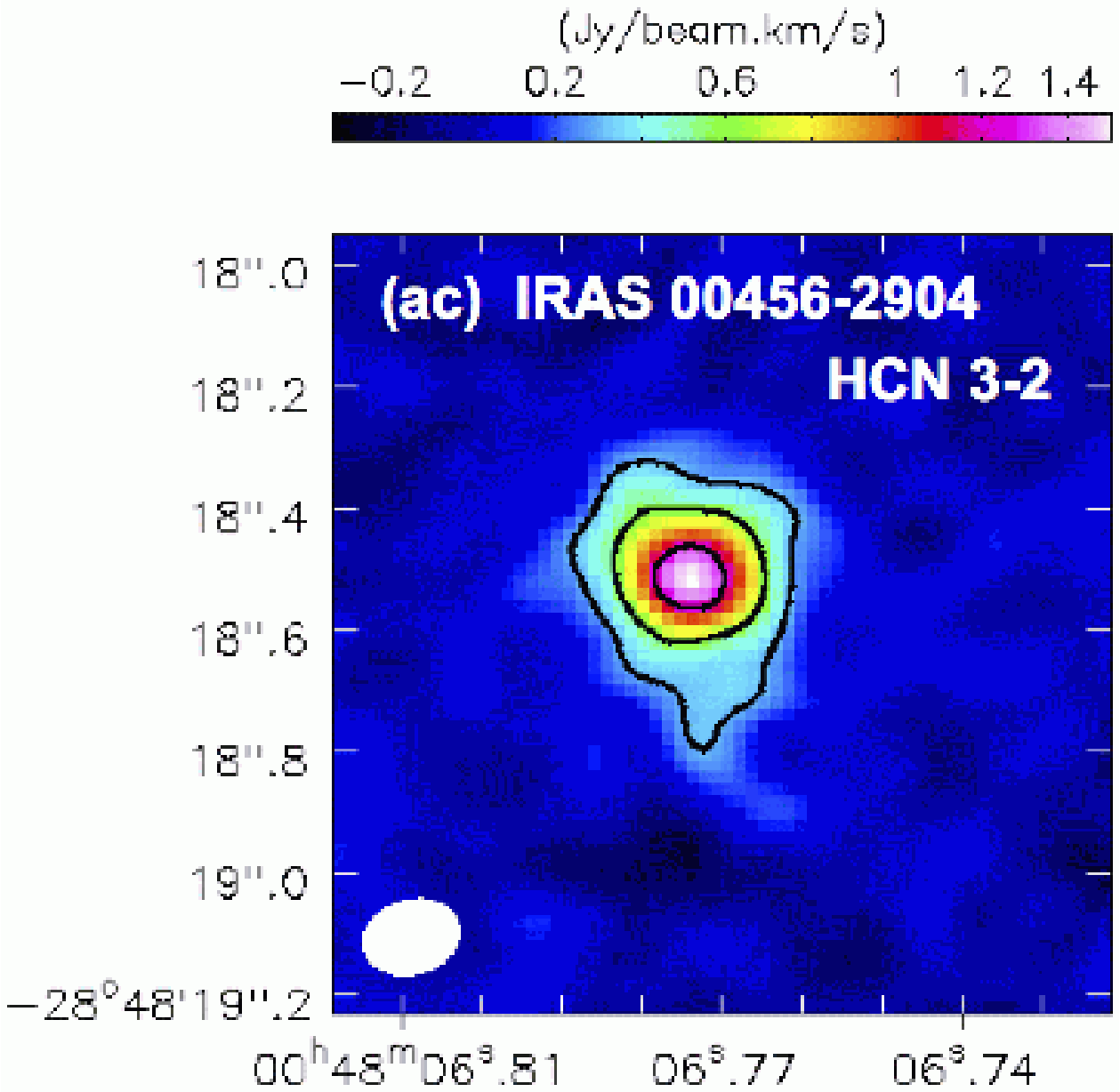} 
\includegraphics[angle=0,scale=.314]{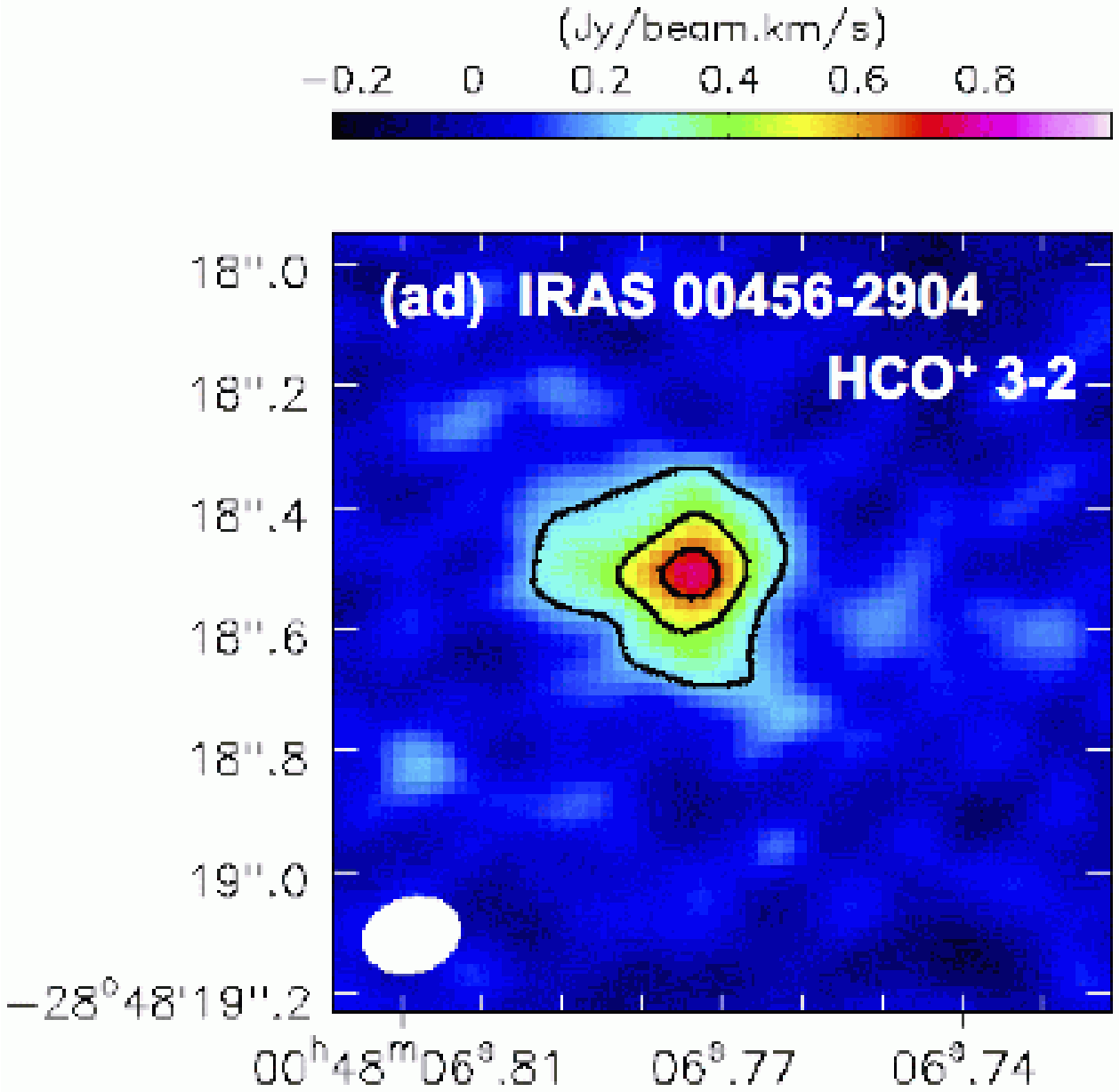} 
\includegraphics[angle=0,scale=.314]{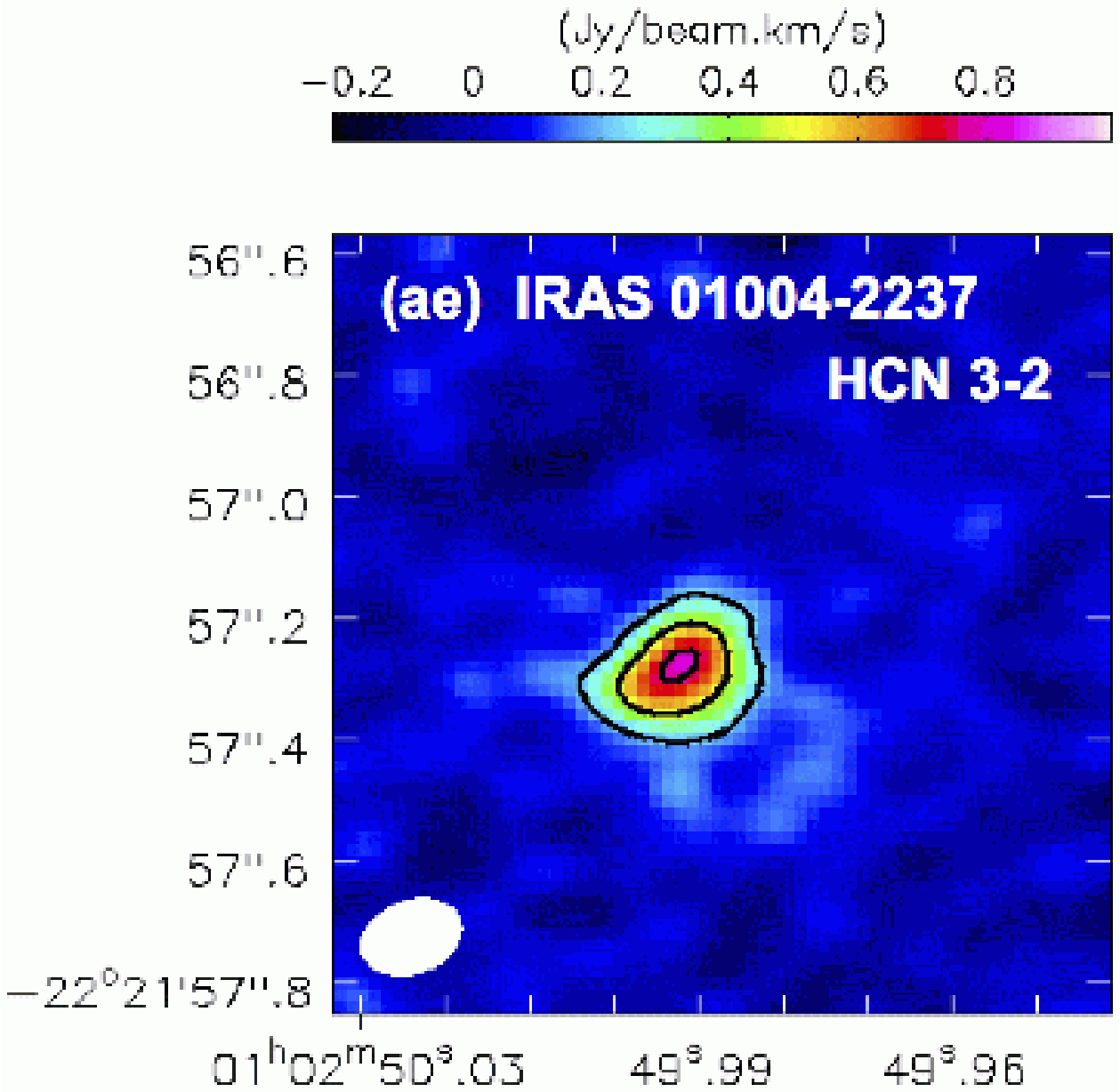} 
\includegraphics[angle=0,scale=.314]{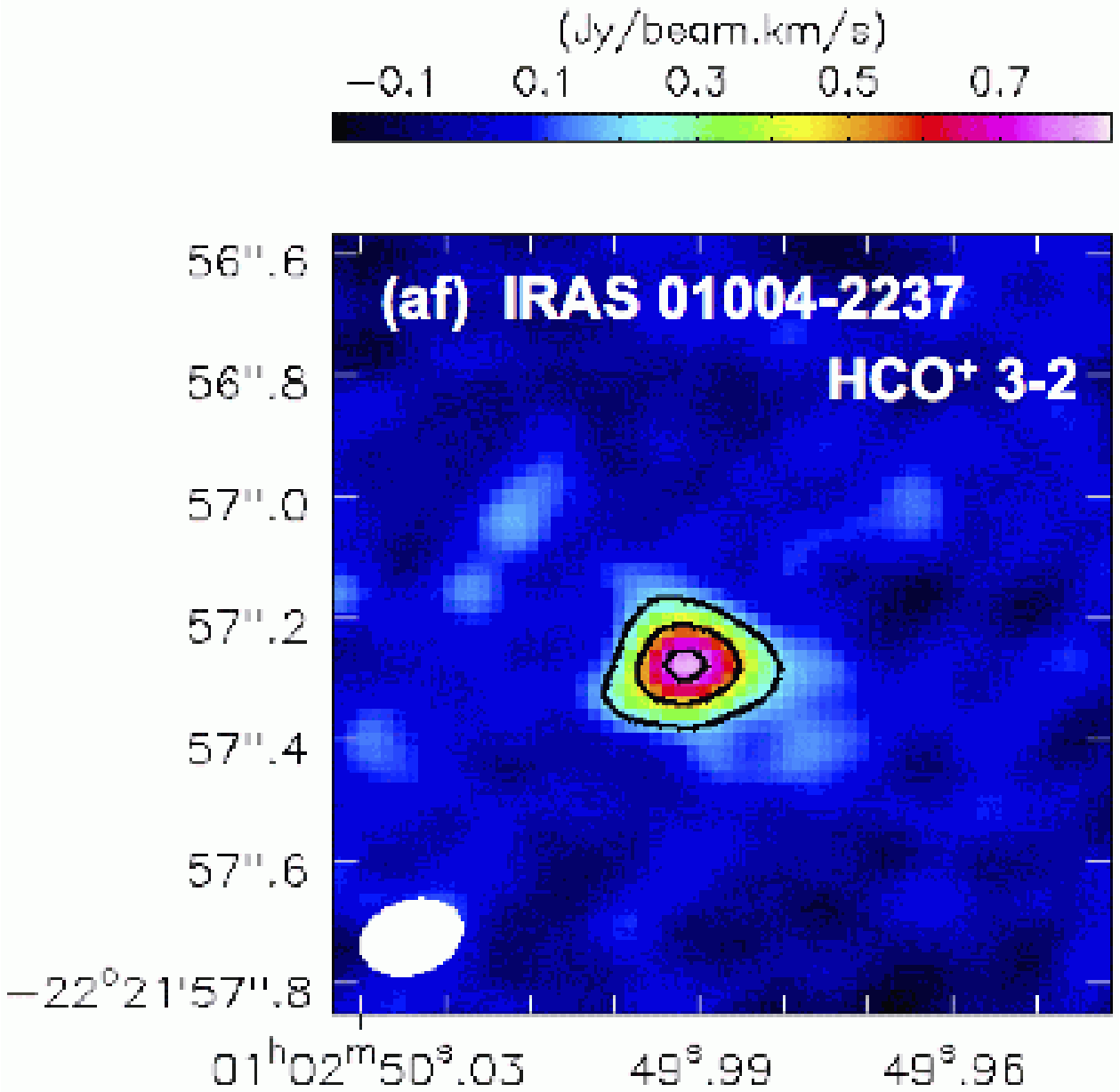} \\ 
\includegraphics[angle=0,scale=.314]{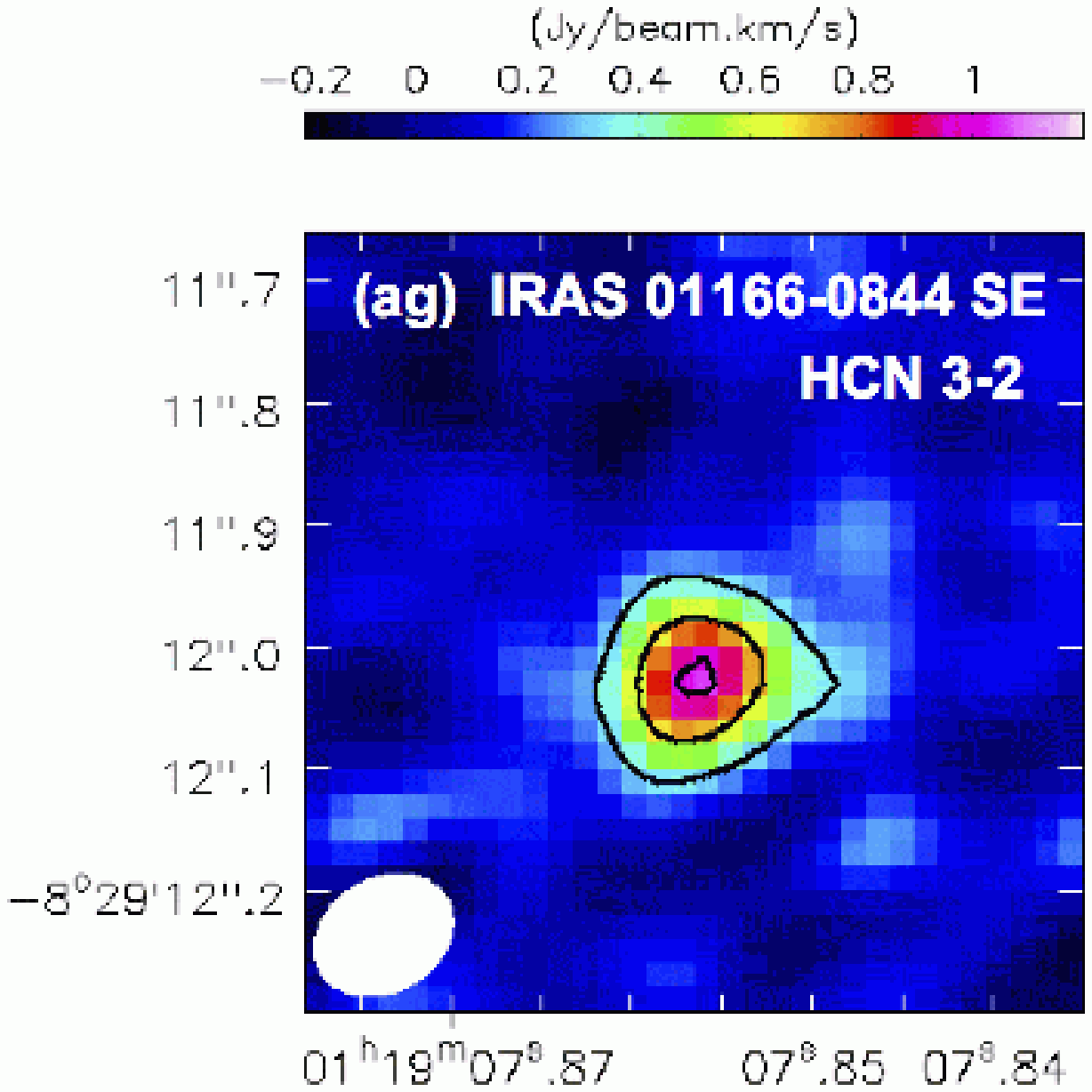} 
\includegraphics[angle=0,scale=.314]{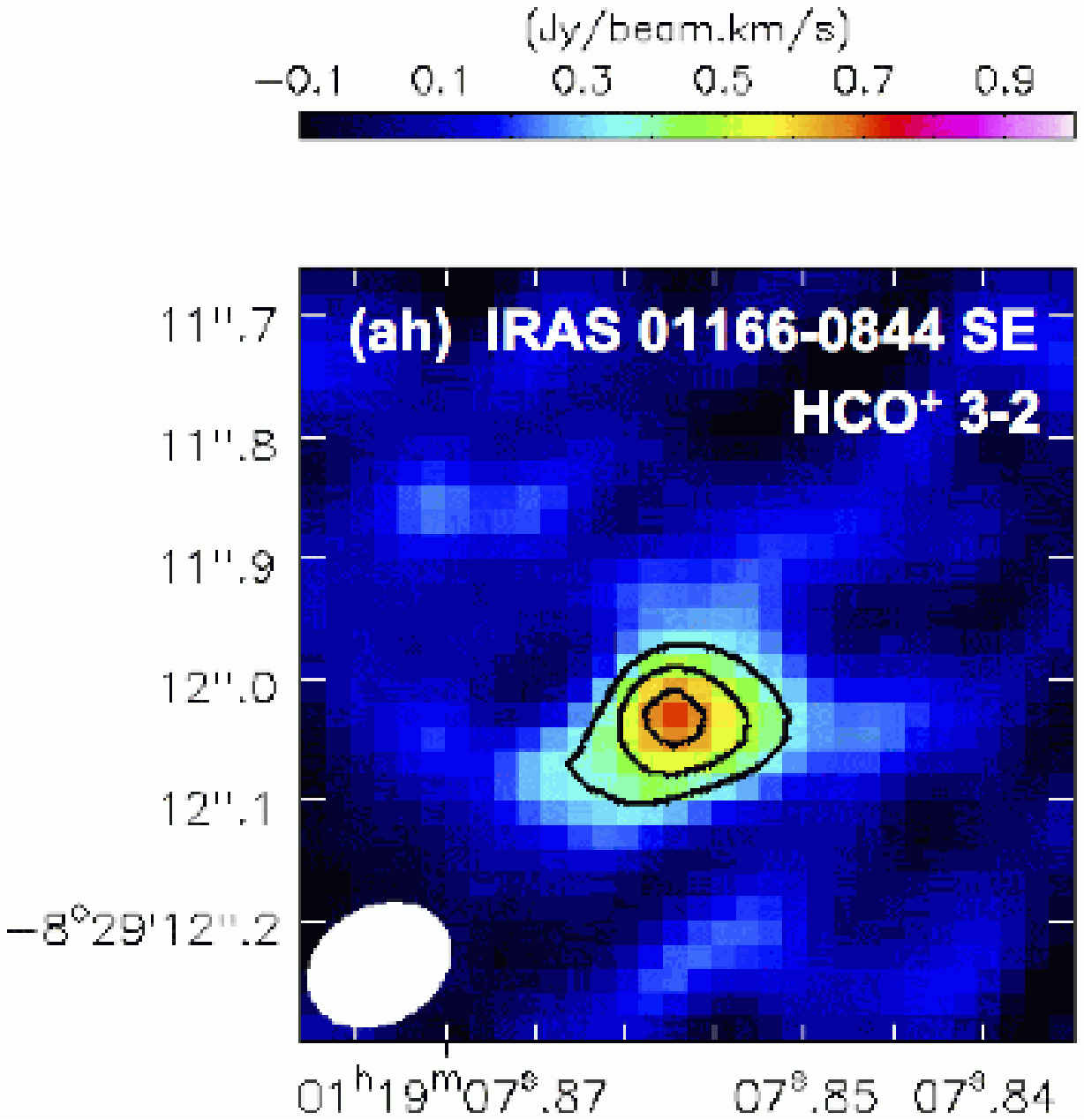} 
\includegraphics[angle=0,scale=.314]{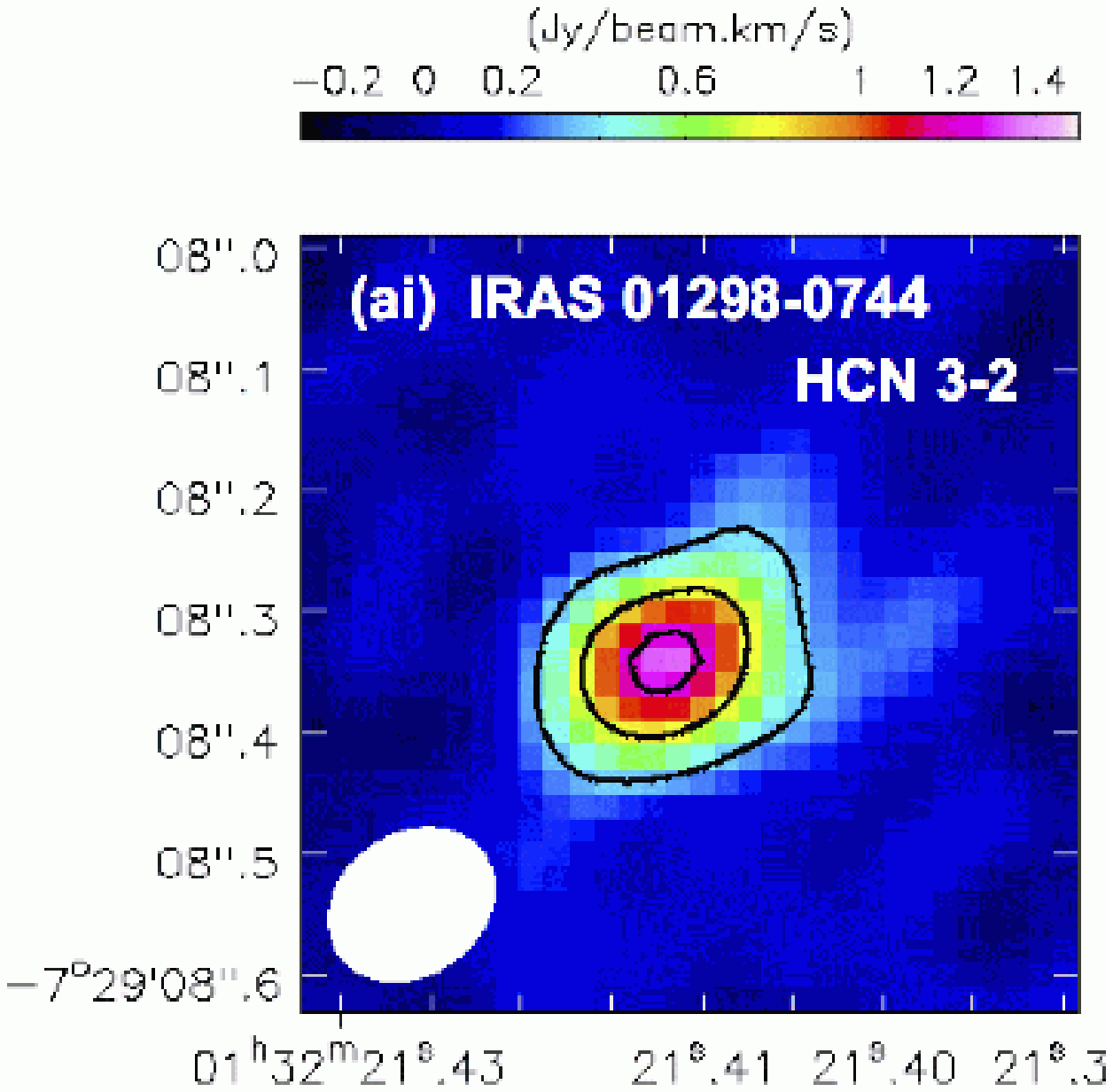} 
\includegraphics[angle=0,scale=.314]{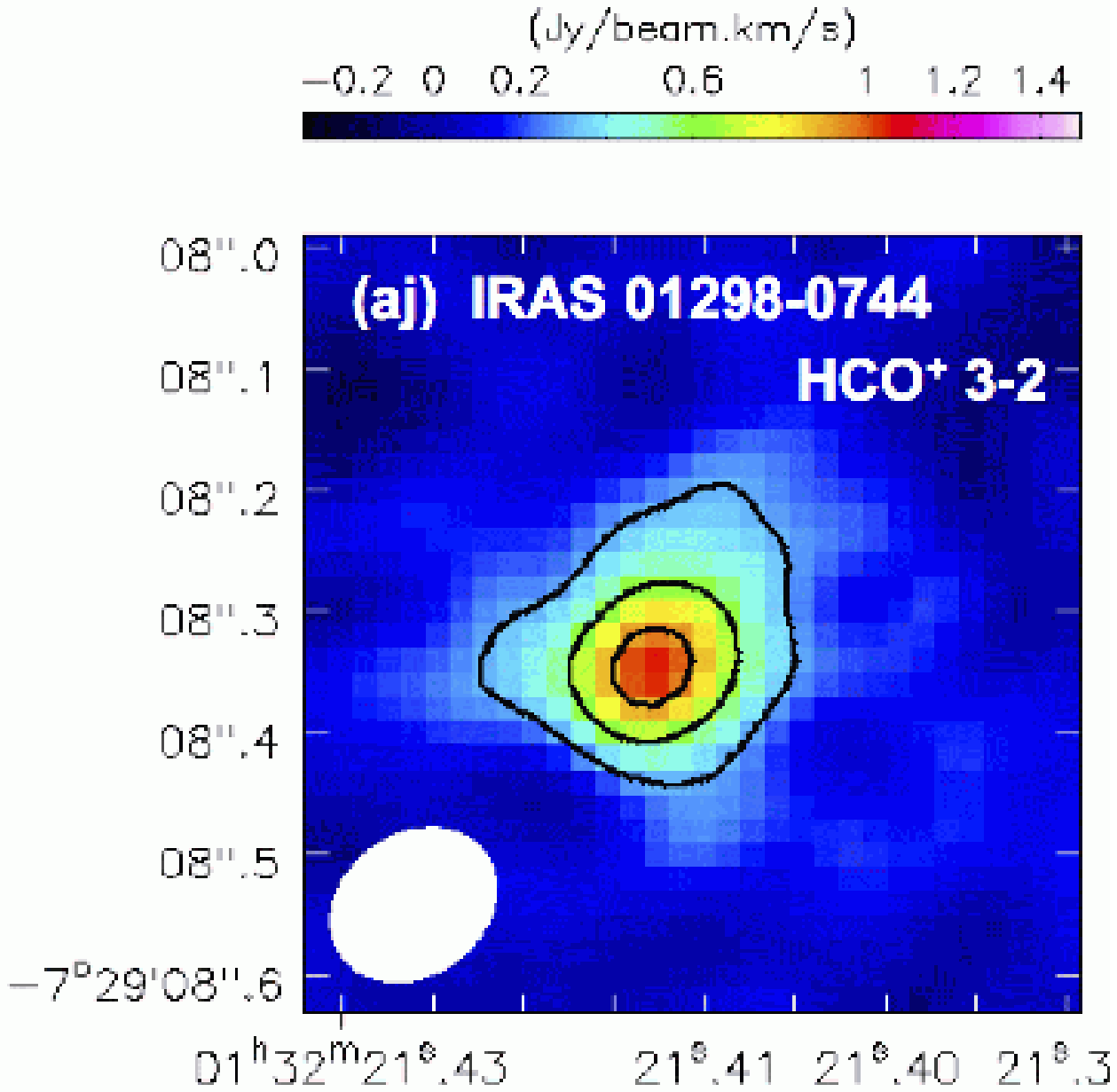} \\ 
\includegraphics[angle=0,scale=.314]{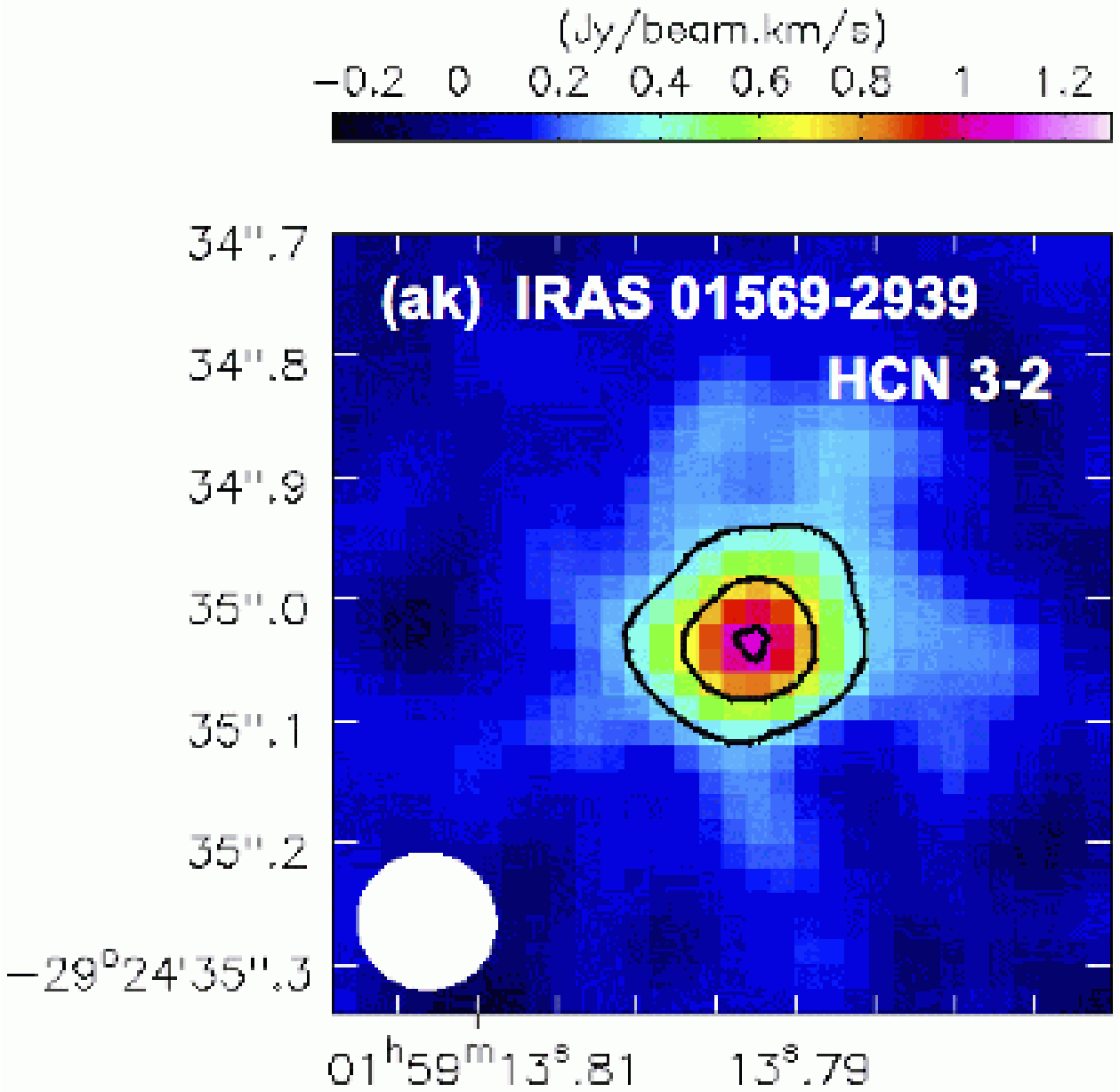} 
\includegraphics[angle=0,scale=.314]{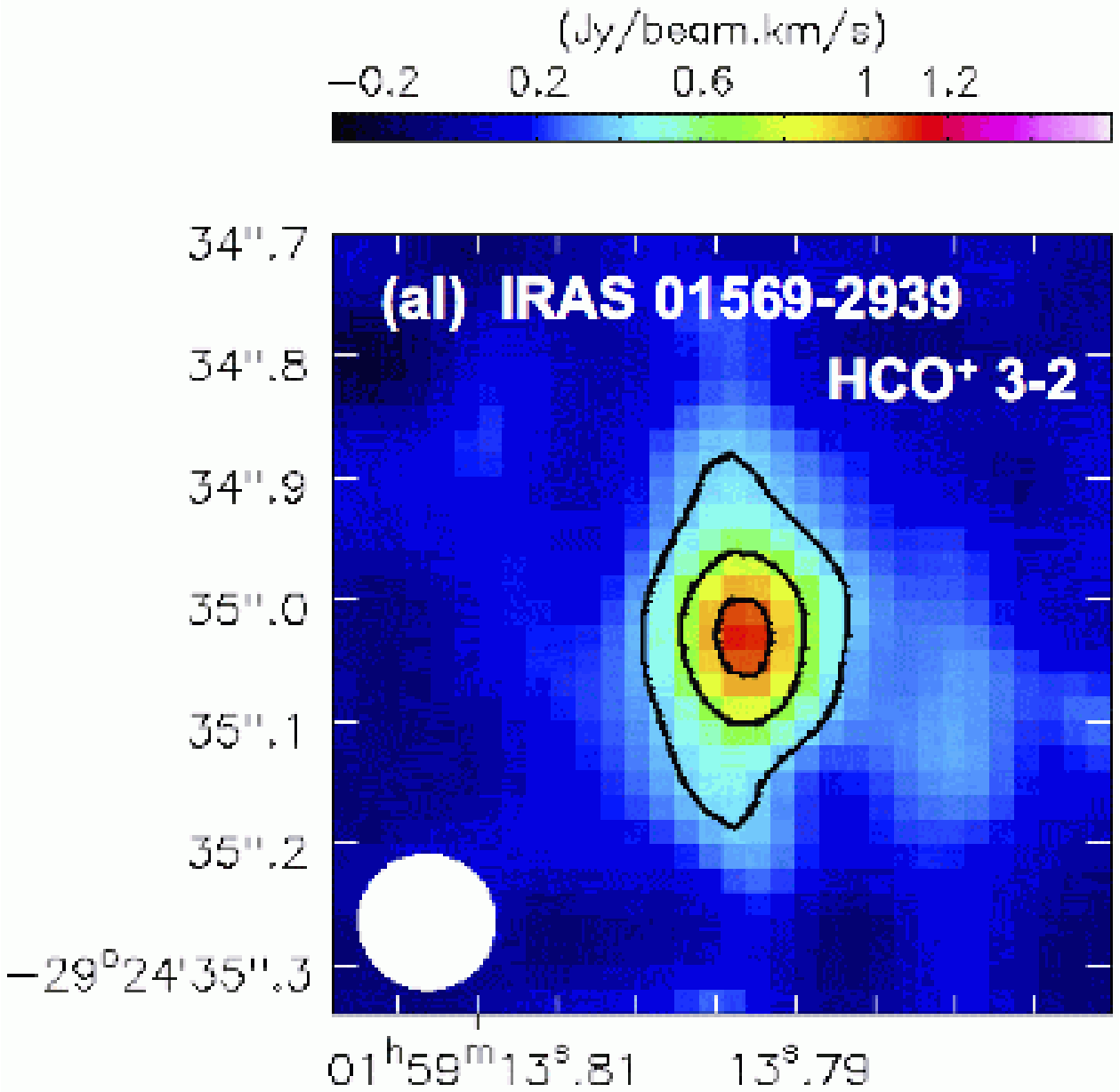} 
\includegraphics[angle=0,scale=.314]{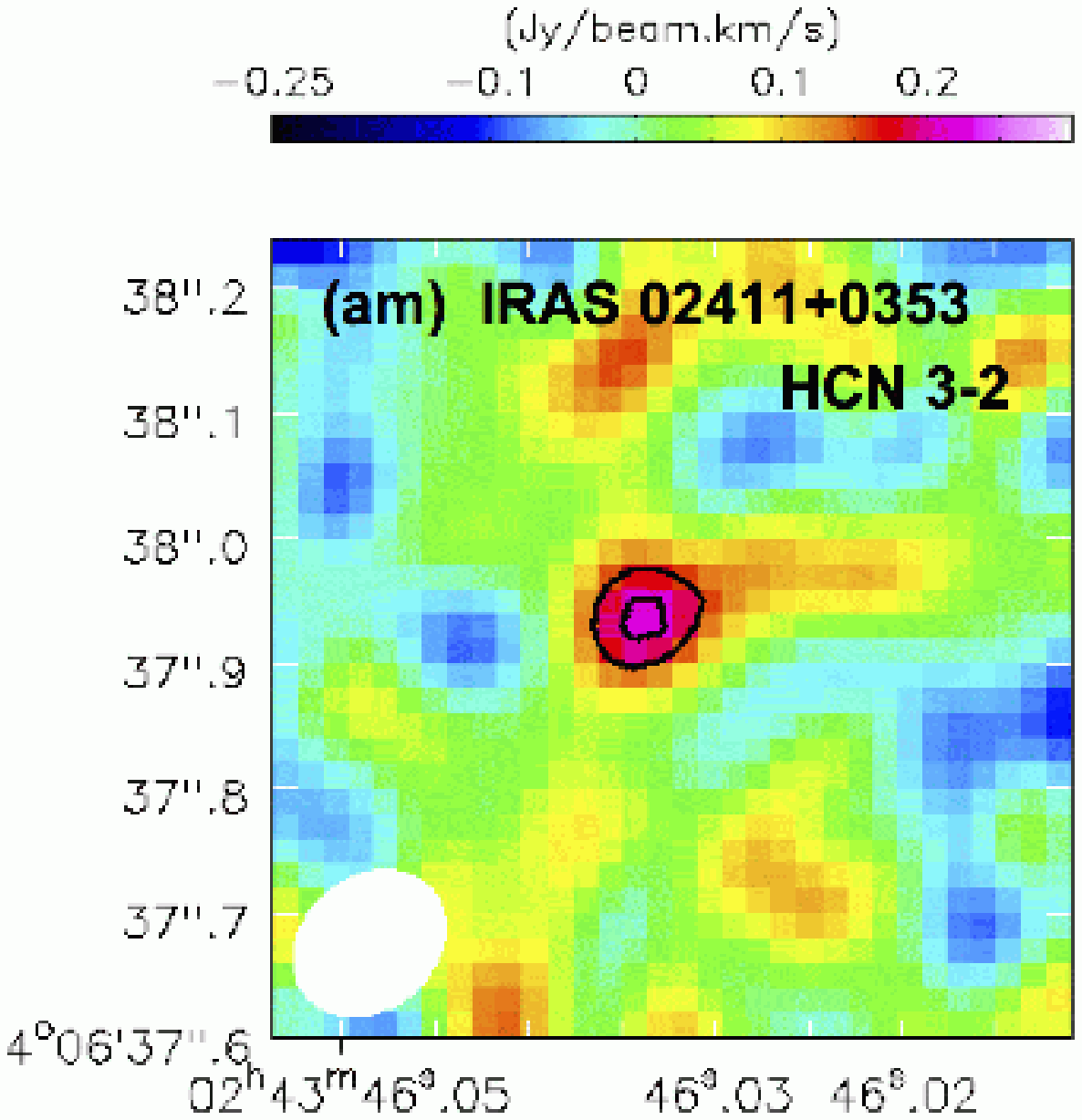} 
\includegraphics[angle=0,scale=.314]{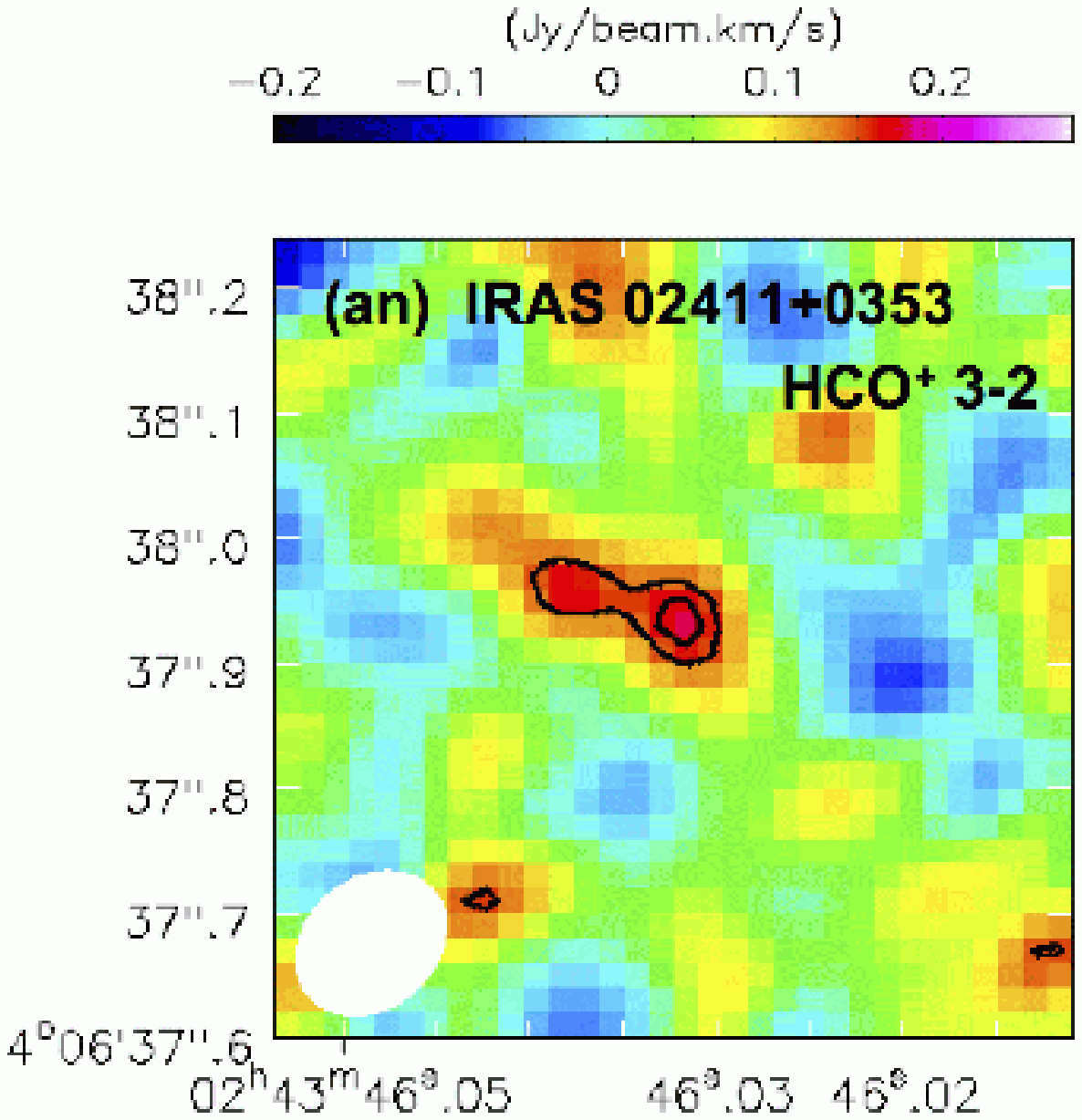} \\   
\includegraphics[angle=0,scale=.314]{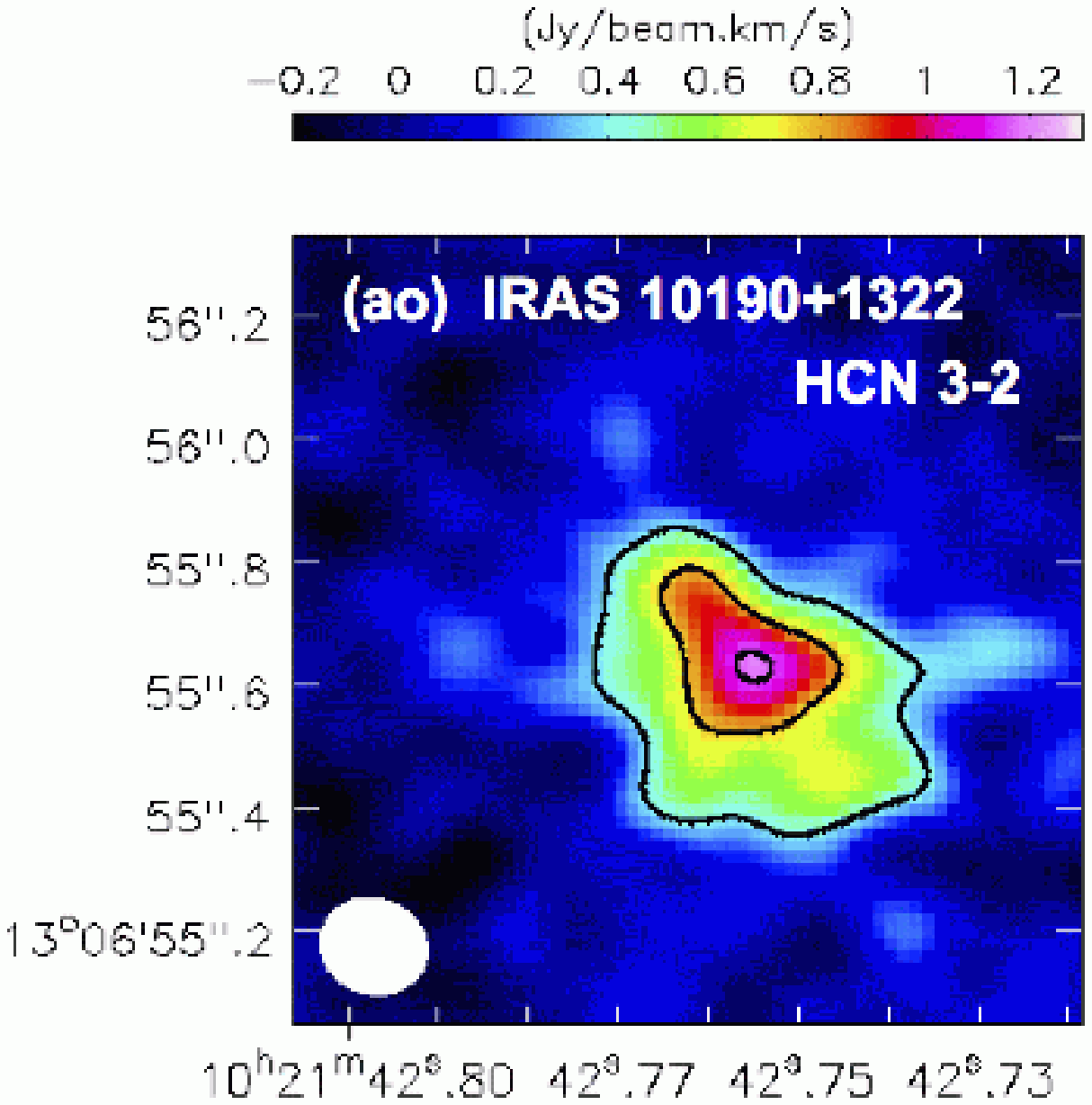} 
\includegraphics[angle=0,scale=.314]{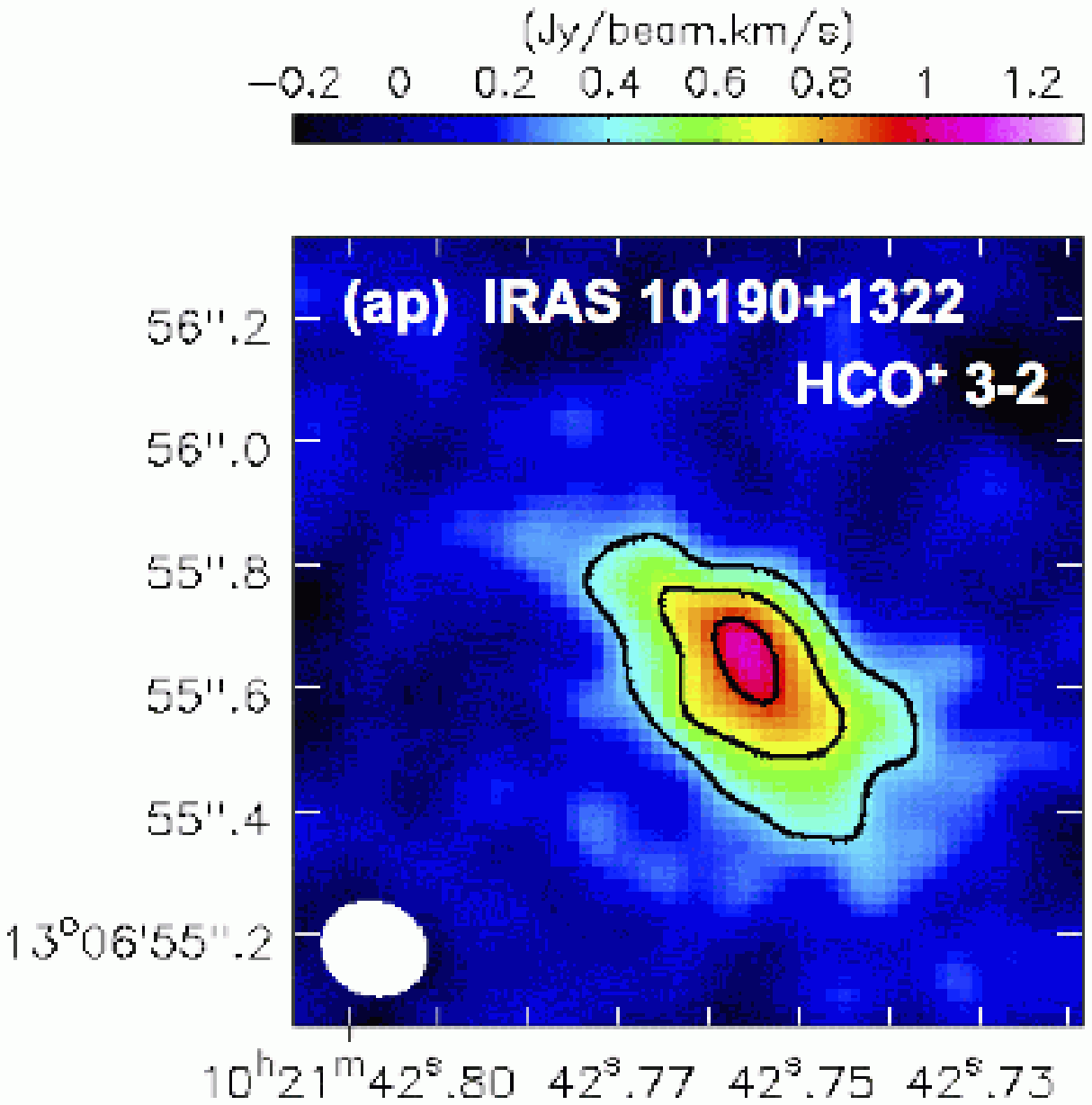} 
\includegraphics[angle=0,scale=.314]{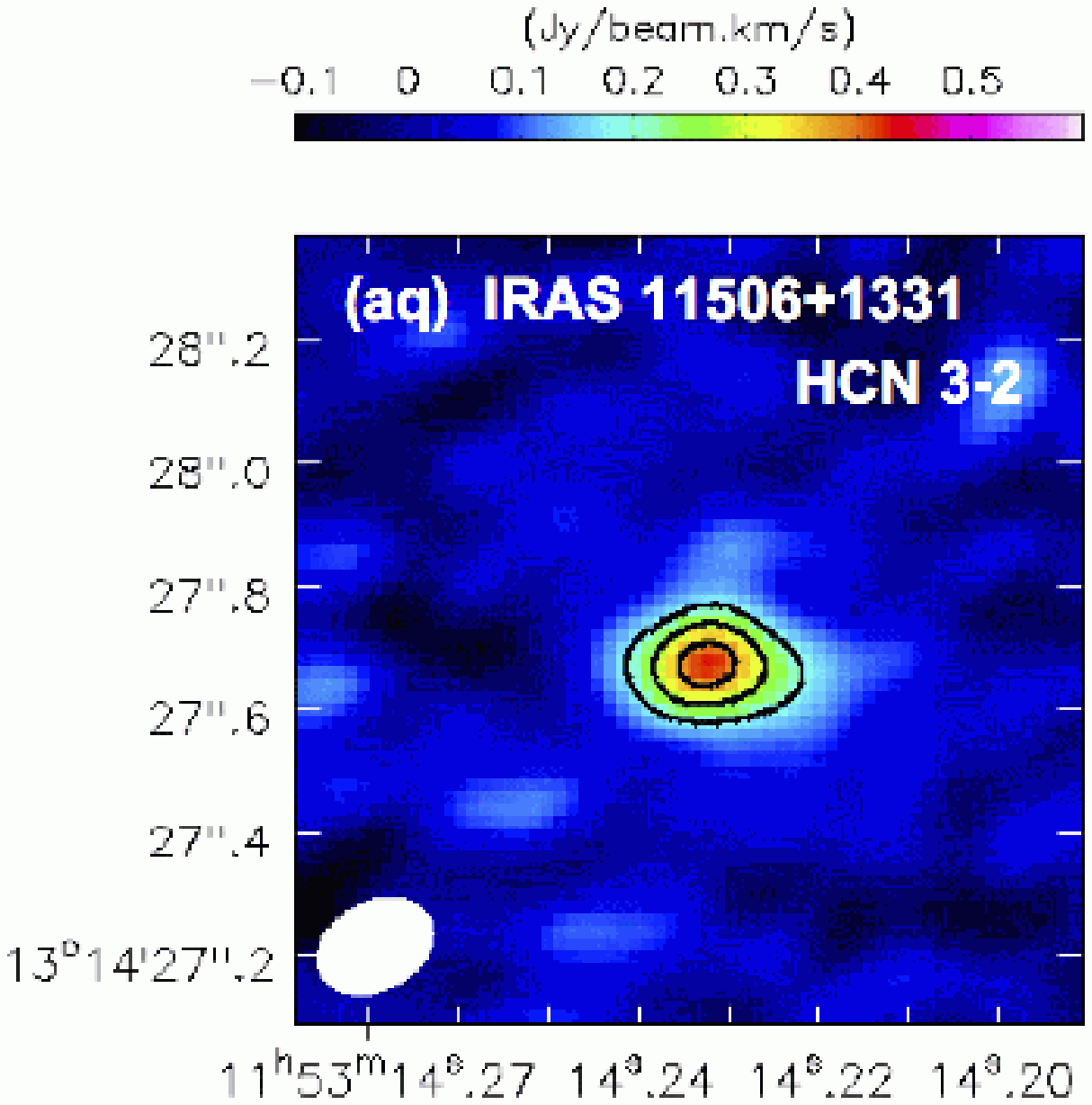} 
\includegraphics[angle=0,scale=.314]{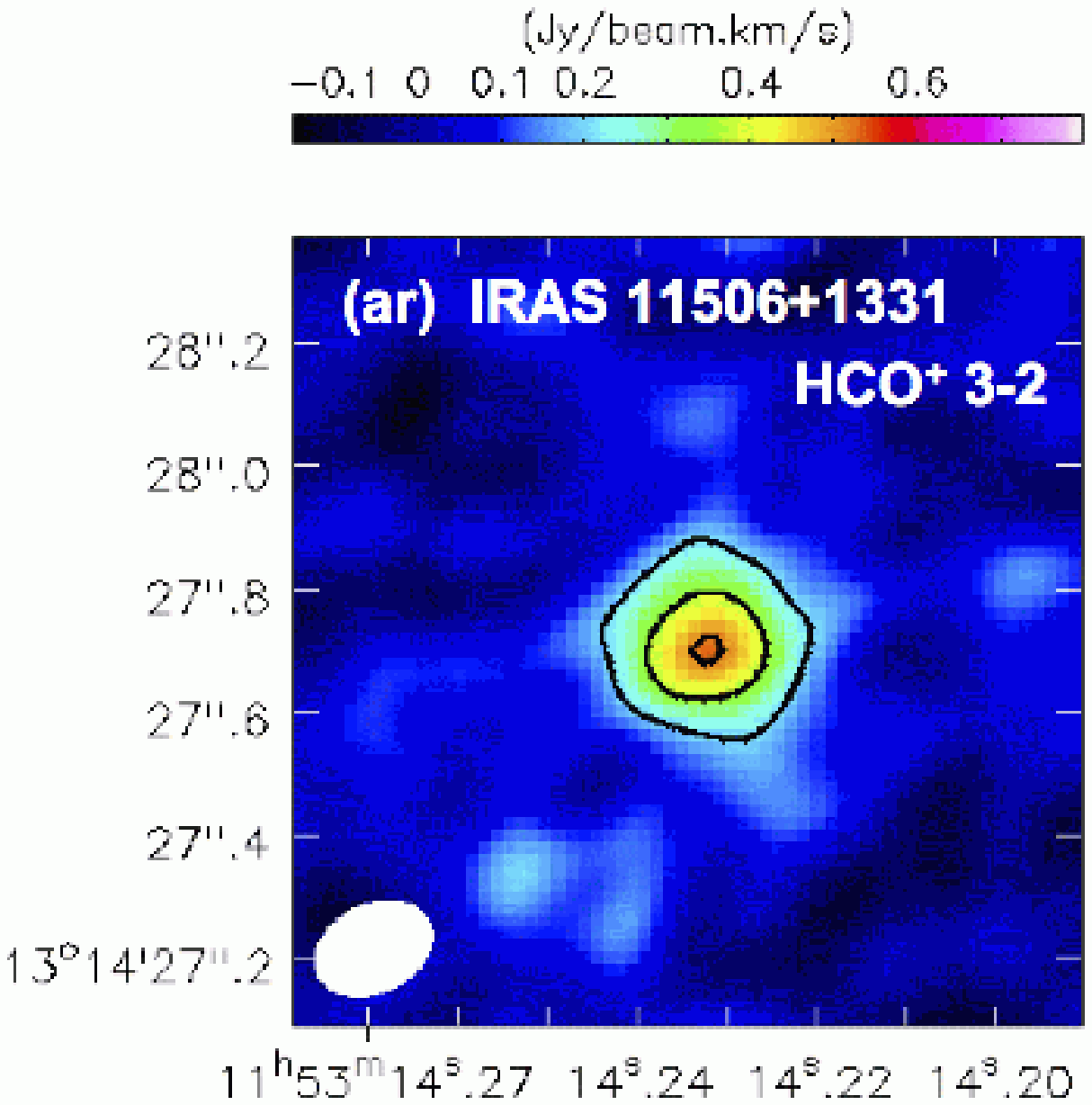} \\   
\includegraphics[angle=0,scale=.314]{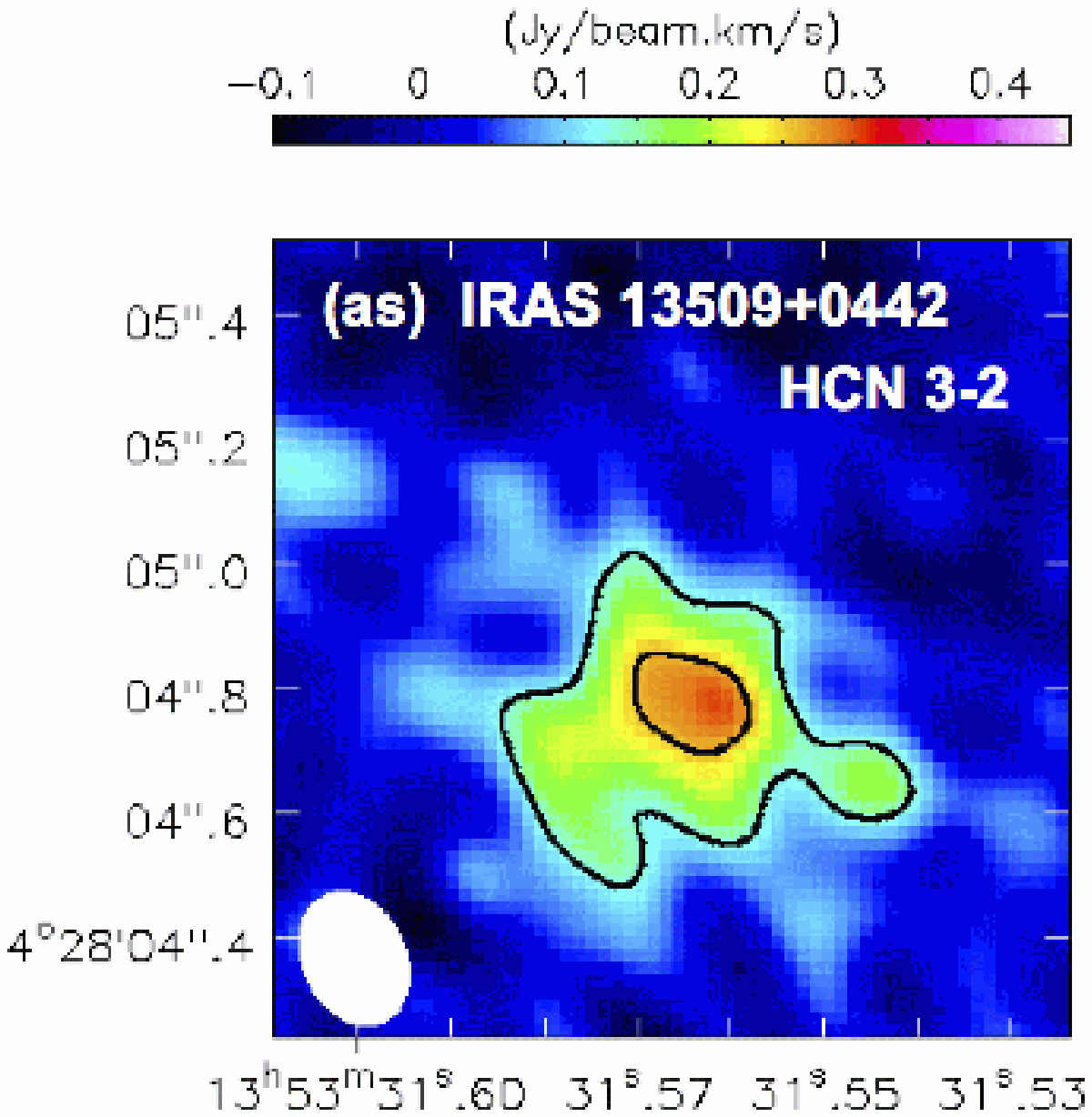} 
\includegraphics[angle=0,scale=.314]{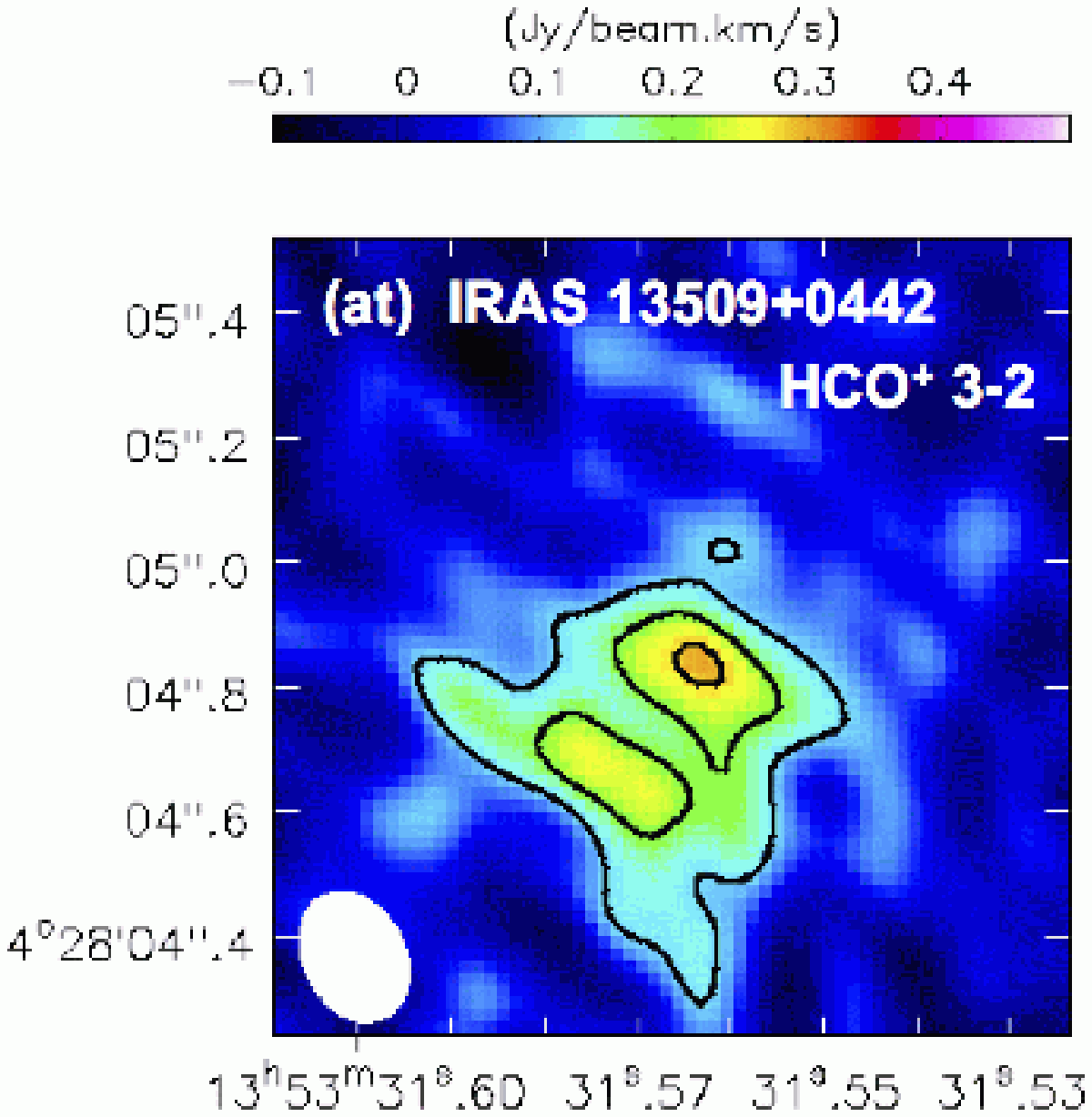}
\includegraphics[angle=0,scale=.314]{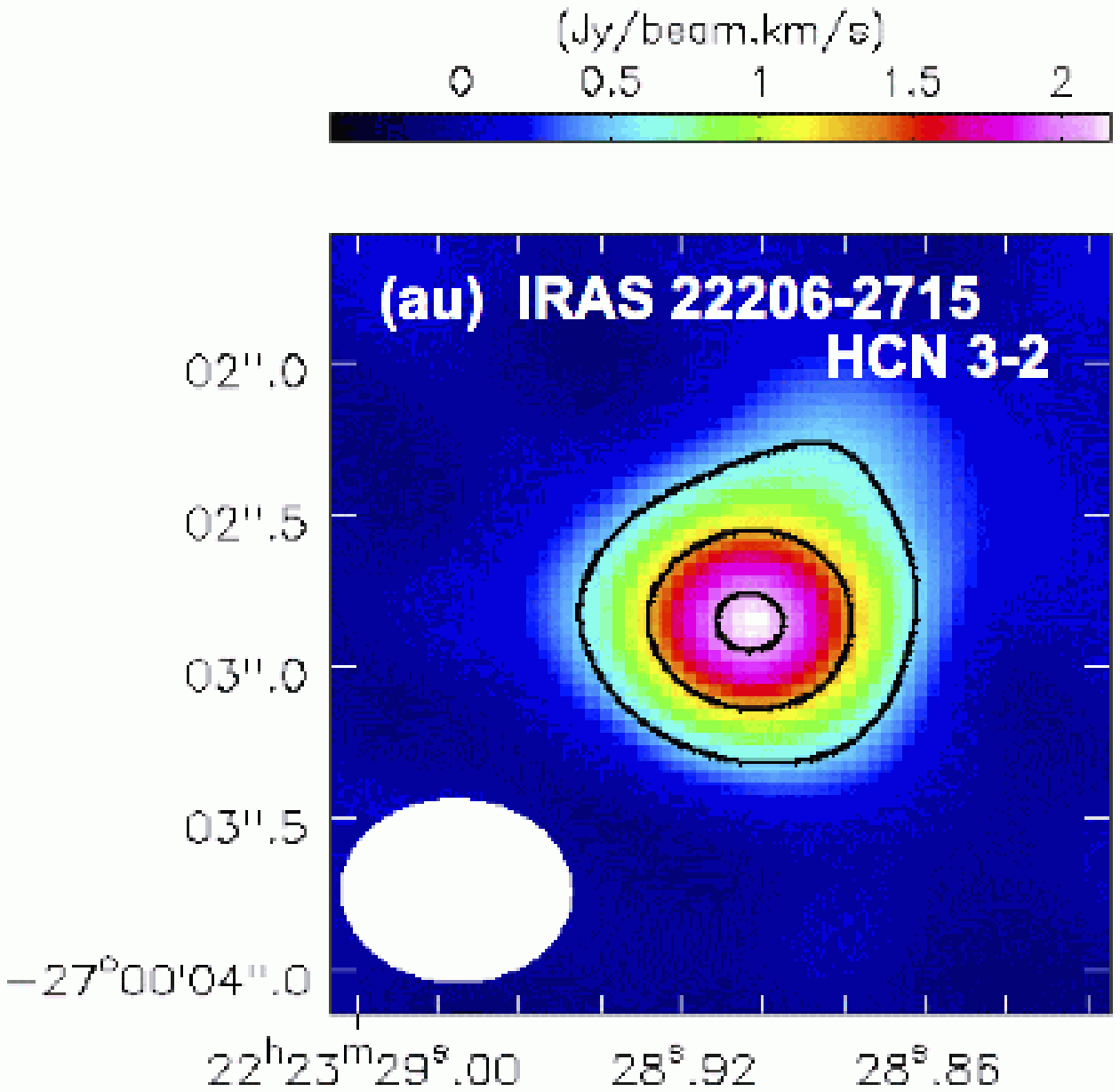} 
\includegraphics[angle=0,scale=.314]{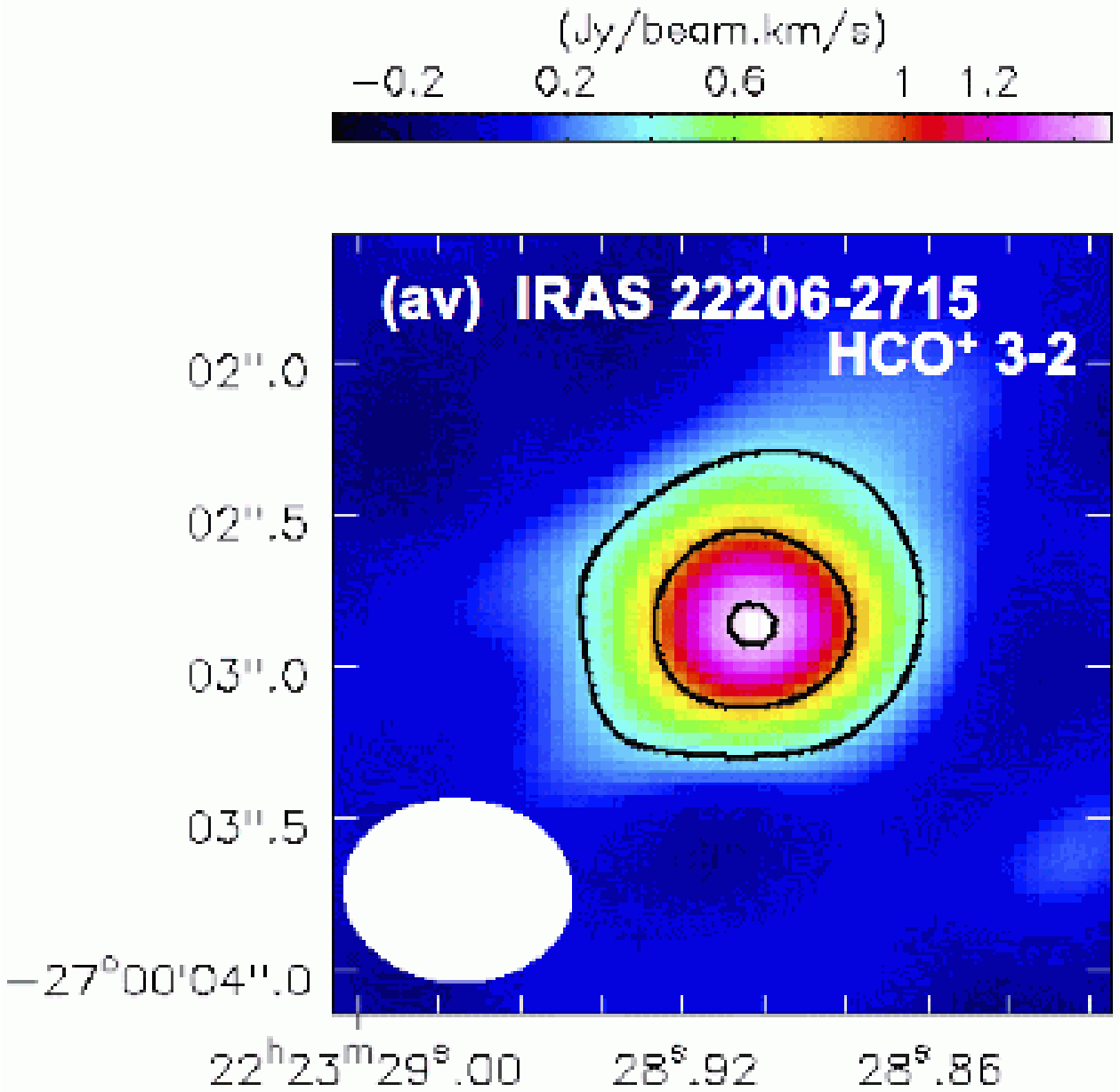} \\
\end{center}
\end{figure}

\clearpage

\begin{figure}
\begin{center}
\includegraphics[angle=0,scale=.314]{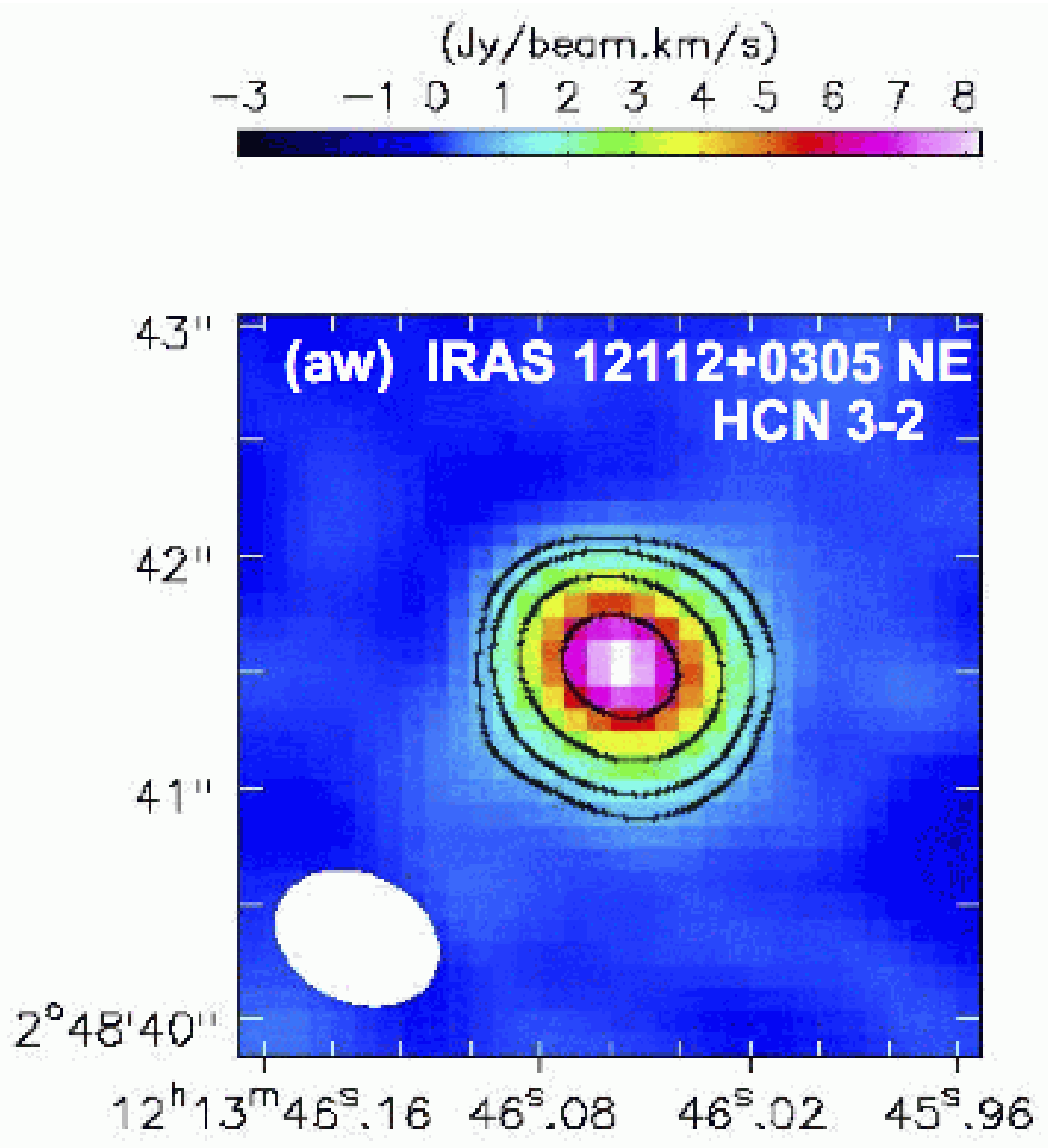} 
\includegraphics[angle=0,scale=.314]{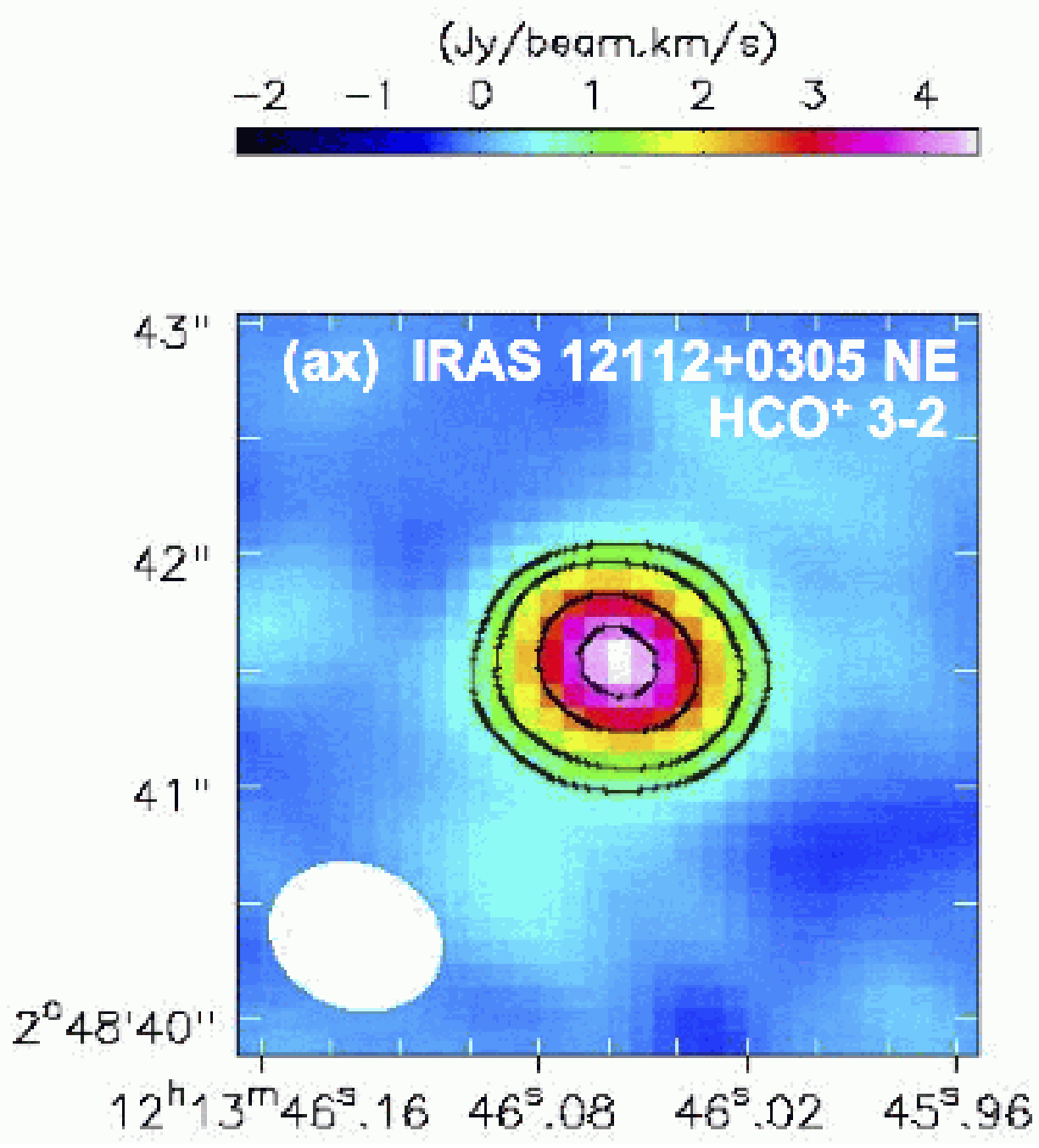}
\includegraphics[angle=0,scale=.314]{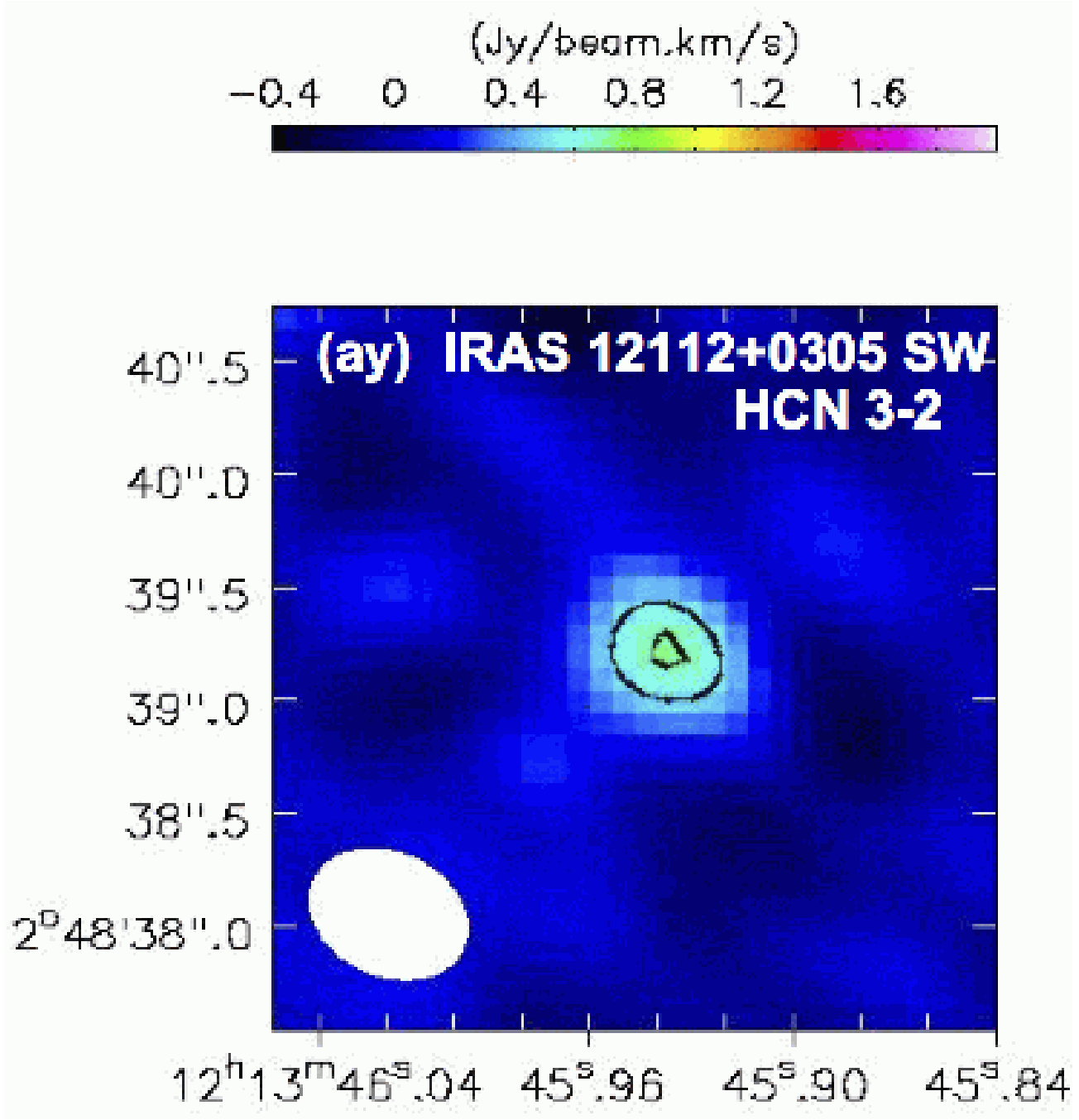} 
\includegraphics[angle=0,scale=.314]{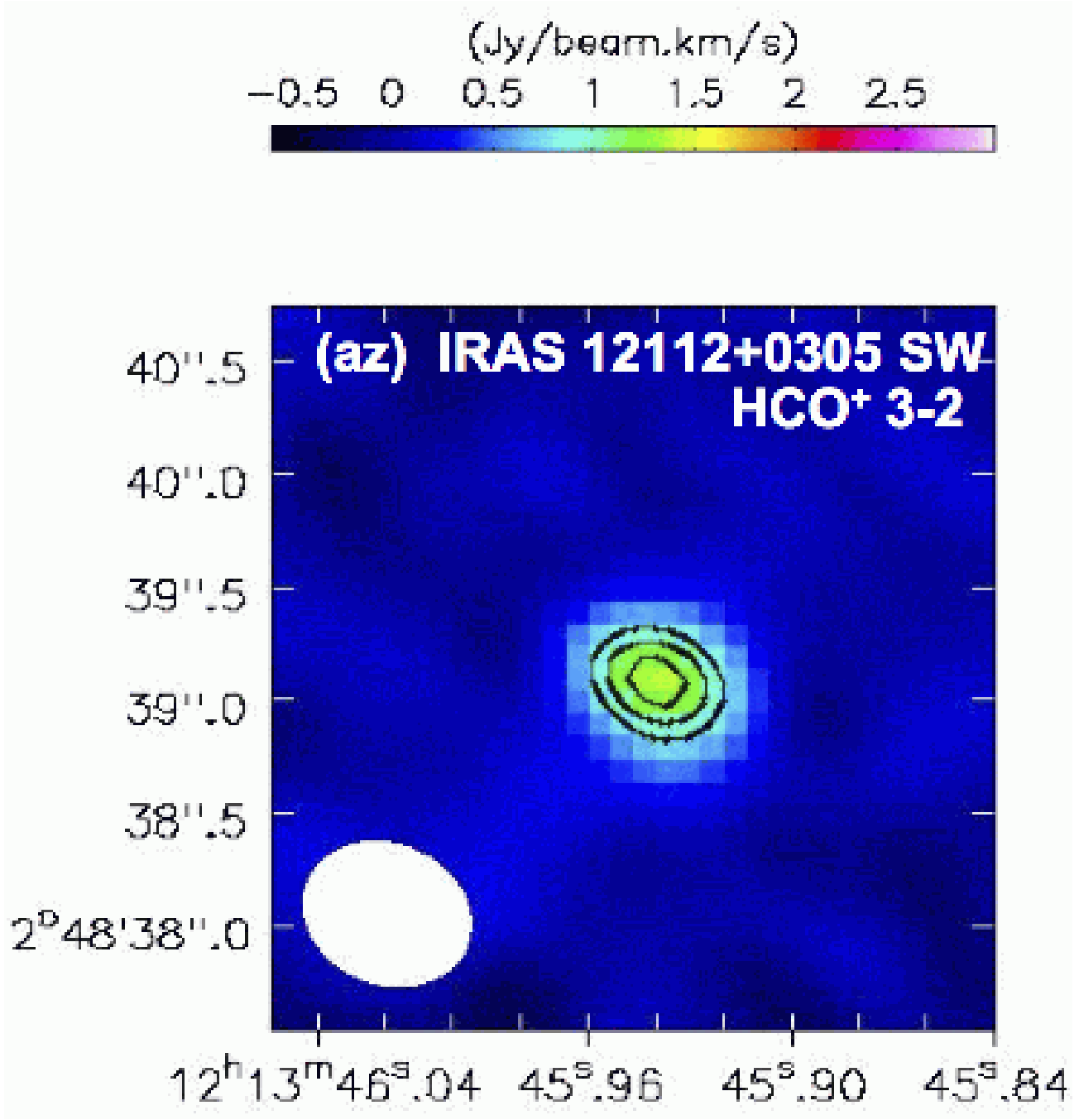} \\
\vspace*{0.2cm}
\hspace*{-0.8cm}
\includegraphics[angle=0,scale=.314]{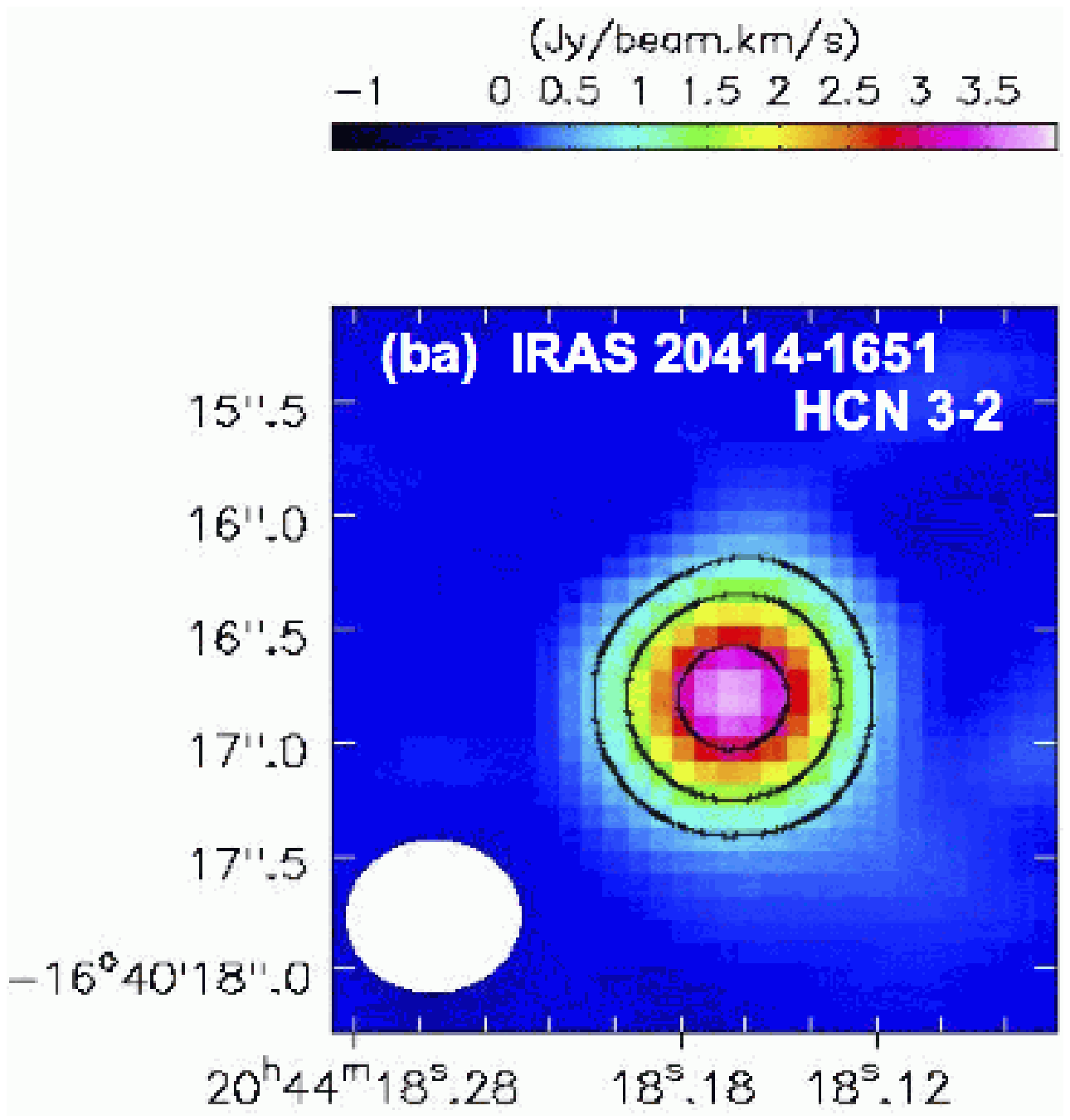} 
\includegraphics[angle=0,scale=.314]{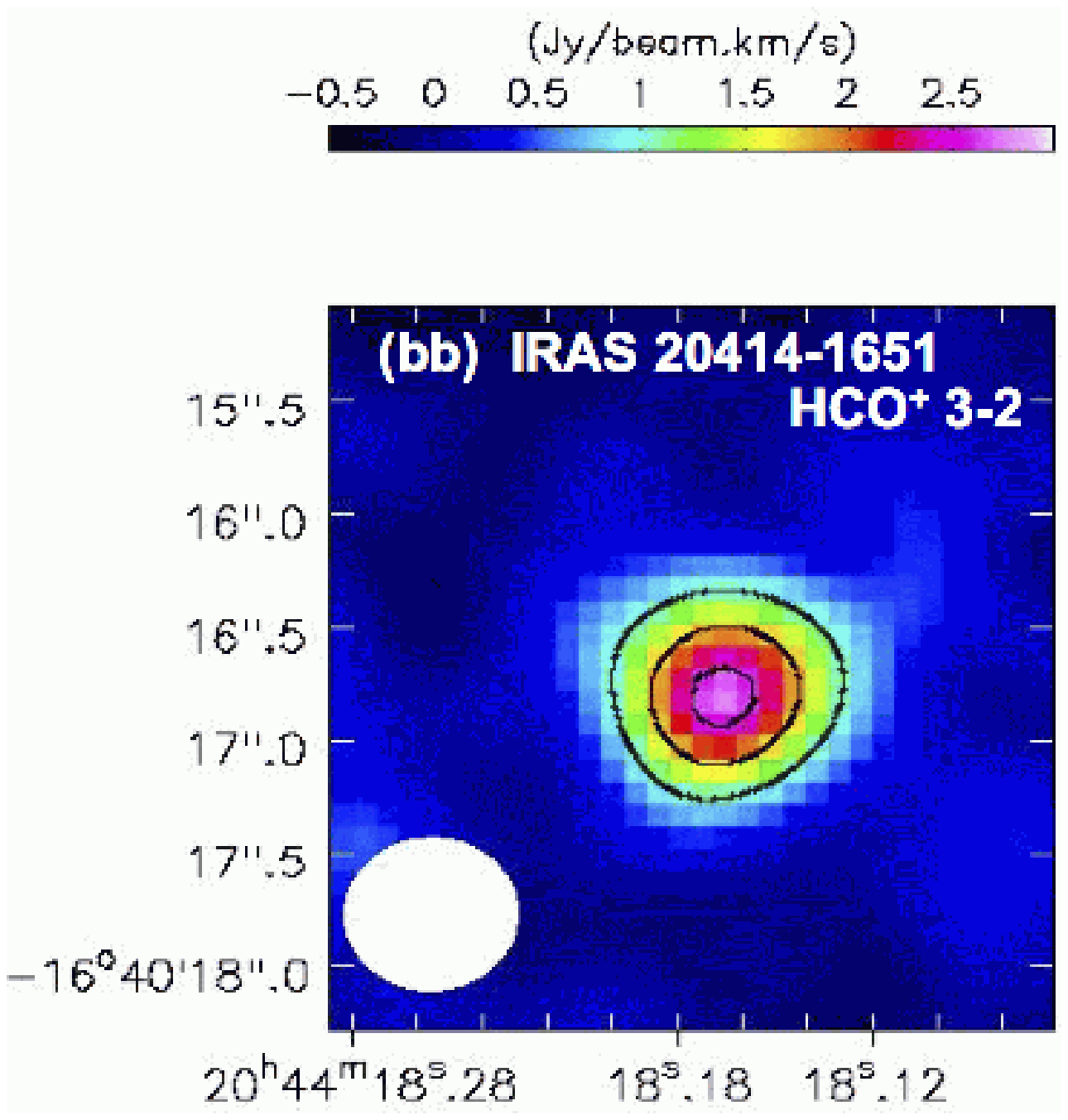}
\includegraphics[angle=0,scale=.314]{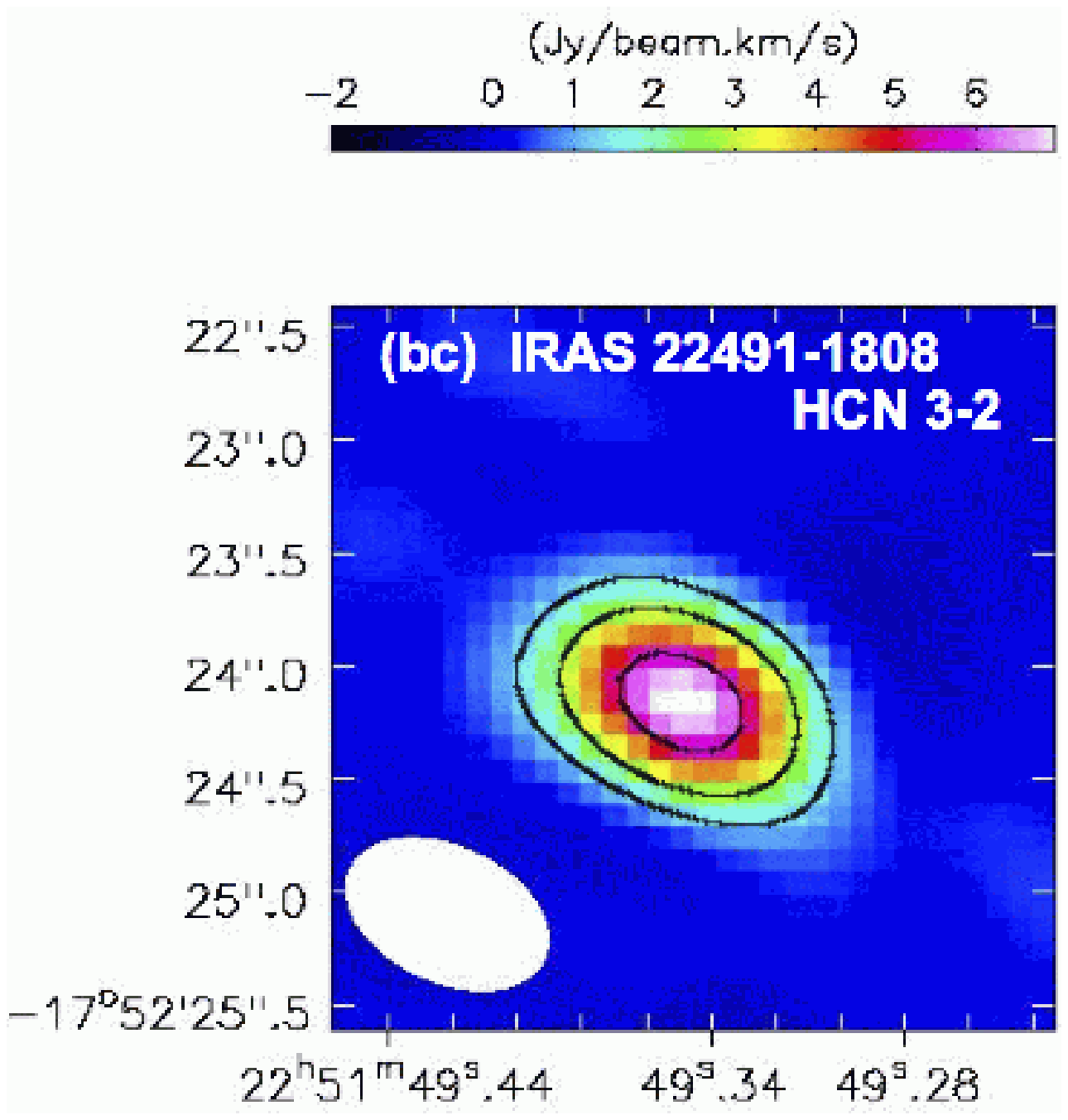} 
\includegraphics[angle=0,scale=.314]{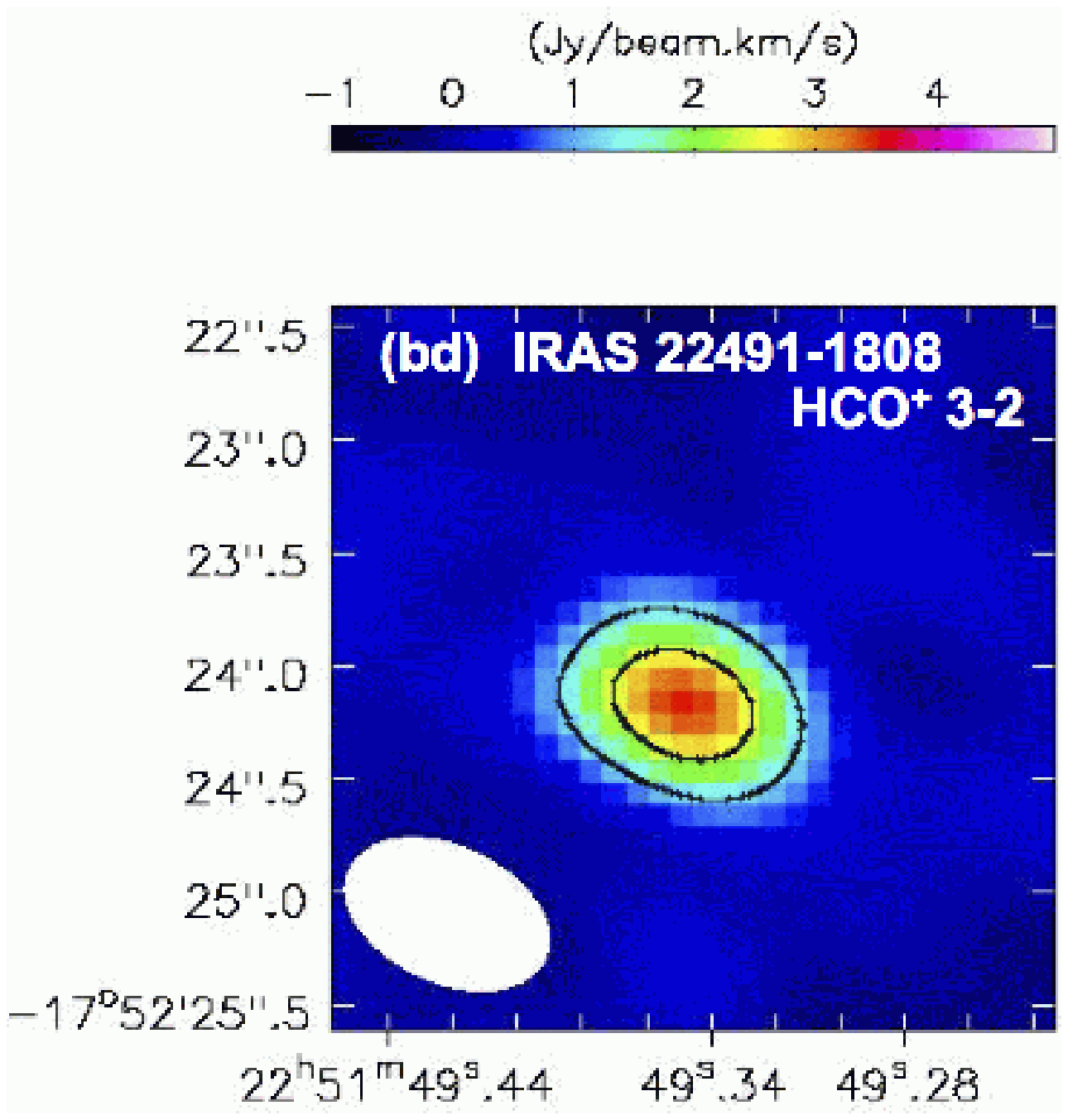} \\
\caption{
Integrated intensity (moment 0) maps of the HCN J=3--2 and HCO$^{+}$ J=3--2
emission lines of the observed ULIRGs with $>$3$\sigma$ detection.
The abscissa and ordinate are right ascension and declination in ICRS, 
respectively. 
The contours are 
5$\sigma$, 10$\sigma$, 20$\sigma$ for IRAS 00188$-$0856 HCN J=3--2,
5$\sigma$, 10$\sigma$, 15$\sigma$ for IRAS 00188$-$0856 HCO$^{+}$ J=3--2,
4$\sigma$, 9$\sigma$ for IRAS 03250+1606 HCN J=3--2,
5$\sigma$, 8$\sigma$ for IRAS 03250+1606 HCO$^{+}$ J=3--2,
3$\sigma$, 5$\sigma$, 7$\sigma$ for IRAS 04103$-$2838 HCN J=3--2,
3$\sigma$, 6$\sigma$ for IRAS 04103$-$2838 HCO$^{+}$ J=3--2,
4$\sigma$, 8$\sigma$, 12$\sigma$ for IRAS 09039+0503 HCN J=3--2,
4$\sigma$, 6$\sigma$, 8$\sigma$ for IRAS 09039+0503 HCO$^{+}$ J=3--2,
6$\sigma$, 12$\sigma$, 24$\sigma$ for IRAS 10378+1108 HCN J=3--2,
5$\sigma$, 10$\sigma$, 20$\sigma$ for IRAS 10378+1108 HCO$^{+}$ J=3--2,
4$\sigma$, 8$\sigma$, 12$\sigma$ for IRAS 11095$-$0238 HCN J=3--2,
4$\sigma$, 6$\sigma$, 9$\sigma$ for IRAS 11095$-$0238 HCO$^{+}$ J=3--2,
4$\sigma$, 8$\sigma$ for IRAS 13335$-$2612 N HCN J=3--2,
4$\sigma$, 7$\sigma$ for IRAS 13335$-$2612 N HCO$^{+}$ J=3--2,
3$\sigma$, 4$\sigma$ for IRAS 13335$-$2612 S HCN J=3--2,
3$\sigma$ for IRAS 13335$-$2612 S HCO$^{+}$ J=3--2,
6$\sigma$, 12$\sigma$, 24$\sigma$ for IRAS 14348$-$1447 SW HCN J=3--2,
5$\sigma$, 9$\sigma$, 13$\sigma$ for IRAS 14348$-$1447 SW HCO$^{+}$ J=3--2,
4$\sigma$, 8$\sigma$, 12$\sigma$ for IRAS 14348$-$1447 NE HCN J=3--2,
4$\sigma$, 6$\sigma$ for IRAS 14348$-$1447 NE HCO$^{+}$ J=3--2,
4$\sigma$, 8$\sigma$, 16$\sigma$, 24$\sigma$ for IRAS 16090$-$0139 HCN J=3--2,
4$\sigma$, 8$\sigma$, 16$\sigma$ for IRAS 16090$-$0139 HCO$^{+}$ J=3--2,
4$\sigma$, 9$\sigma$, 14$\sigma$ for IRAS 21329$-$2346 HCN J=3--2,
5$\sigma$, 10$\sigma$, 15$\sigma$ for IRAS 21329$-$2346 HCO$^{+}$ J=3--2,
5$\sigma$, 9$\sigma$ for IRAS 23234+0946 HCN J=3--2,
4$\sigma$, 9$\sigma$ for IRAS 23234+0946 HCO$^{+}$ J=3--2,
4$\sigma$, 10$\sigma$, 16$\sigma$ for IRAS 00091$-$0738 HCN J=3--2,
5$\sigma$, 9$\sigma$, 13$\sigma$ for IRAS 00091$-$0738 HCO$^{+}$ J=3--2,
4$\sigma$, 8$\sigma$, 16$\sigma$ for IRAS 00456$-$2904 HCN J=3--2,
4$\sigma$, 8$\sigma$, 12$\sigma$ for IRAS 00456$-$2904 HCO$^{+}$ J=3--2,
4$\sigma$, 9$\sigma$, 14$\sigma$ for IRAS 01004$-$2237 HCN J=3--2,
4$\sigma$, 9$\sigma$, 14$\sigma$ for IRAS 01004$-$2237 HCO$^{+}$ J=3--2,
5$\sigma$, 10$\sigma$, 15$\sigma$ for IRAS 01166$-$0844 SE HCN J=3--2,
5$\sigma$, 7$\sigma$, 9$\sigma$ for IRAS 01166$-$0844 SE HCO$^{+}$ J=3--2,
5$\sigma$, 11$\sigma$, 17$\sigma$ for IRAS 01298$-$0744 HCN J=3--2,
4$\sigma$, 8$\sigma$, 12$\sigma$ for IRAS 01298$-$0744 HCO$^{+}$ J=3--2,
5$\sigma$, 10$\sigma$, 15$\sigma$ for IRAS 01569$-$2939 HCN J=3--2,
5$\sigma$, 9$\sigma$, 13$\sigma$ for IRAS 01569$-$2939 HCO$^{+}$ J=3--2,
2.5$\sigma$, 3.3$\sigma$ for IRAS 02411+0353 HCN J=3--2,
2.8$\sigma$, 3.2$\sigma$ for IRAS 02411+0353 HCO$^{+}$ J=3--2,
4$\sigma$, 8$\sigma$, 12$\sigma$ for IRAS 10190+1322 HCN J=3--2,
4$\sigma$, 7$\sigma$, 10$\sigma$ for IRAS 10190+1322 HCO$^{+}$ J=3--2,
4$\sigma$, 6$\sigma$, 8$\sigma$ for IRAS 11506+1331 HCN J=3--2,
4$\sigma$, 7$\sigma$, 10$\sigma$ for IRAS 11506+1331 HCO$^{+}$ J=3--2,
3$\sigma$, 6$\sigma$ for IRAS 13509+0442 HCN J=3--2, 
3$\sigma$, 5$\sigma$, 7$\sigma$ for IRAS 13509+0442 HCO$^{+}$ J=3--2,
5$\sigma$, 12$\sigma$, 19$\sigma$ for IRAS 22206$-$2715 HCN J=3--2, 
4$\sigma$, 10$\sigma$, 16$\sigma$ for IRAS 22206$-$2715 HCO$^{+}$ J=3--2,
3$\sigma$, 5$\sigma$, 10$\sigma$, 20$\sigma$ for IRAS 12112$+$0305 NE 
HCN J=3--2, 
3$\sigma$, 5$\sigma$, 10$\sigma$, 15$\sigma$ for IRAS 12112$+$0305 NE
HCO$^{+}$ J=3--2,
3$\sigma$, 4$\sigma$ for IRAS 12112$+$0305 SW HCN J=3--2, 
4$\sigma$, 5$\sigma$, 6$\sigma$ for IRAS 12112$+$0305 SW HCO$^{+}$ J=3--2, 
5$\sigma$, 10$\sigma$, 20$\sigma$ for IRAS 20414$-$1651 HCN J=3--2, 
4$\sigma$, 8$\sigma$, 12$\sigma$ for IRAS 20414$-$1651 HCO$^{+}$ J=3--2,
5$\sigma$, 10$\sigma$, 20$\sigma$ for IRAS 22491$-$1808 HCN J=3--2, and
5$\sigma$, 10$\sigma$ for IRAS 22491$-$1808 HCO$^{+}$ J=3--2.
IRAS 10485$-$1447 is not shown because of no molecular line detection.
The 1$\sigma$ rms noise level in each source is tabulated 
in Tables 4 and 5 (column 3).
Beam sizes are shown as filled circles in the lower-left region.
No cut-off in signal to noise ratio is applied.
}
\end{center}
\end{figure}

\clearpage

\begin{figure}
\begin{center}
\includegraphics[angle=0,scale=.45]{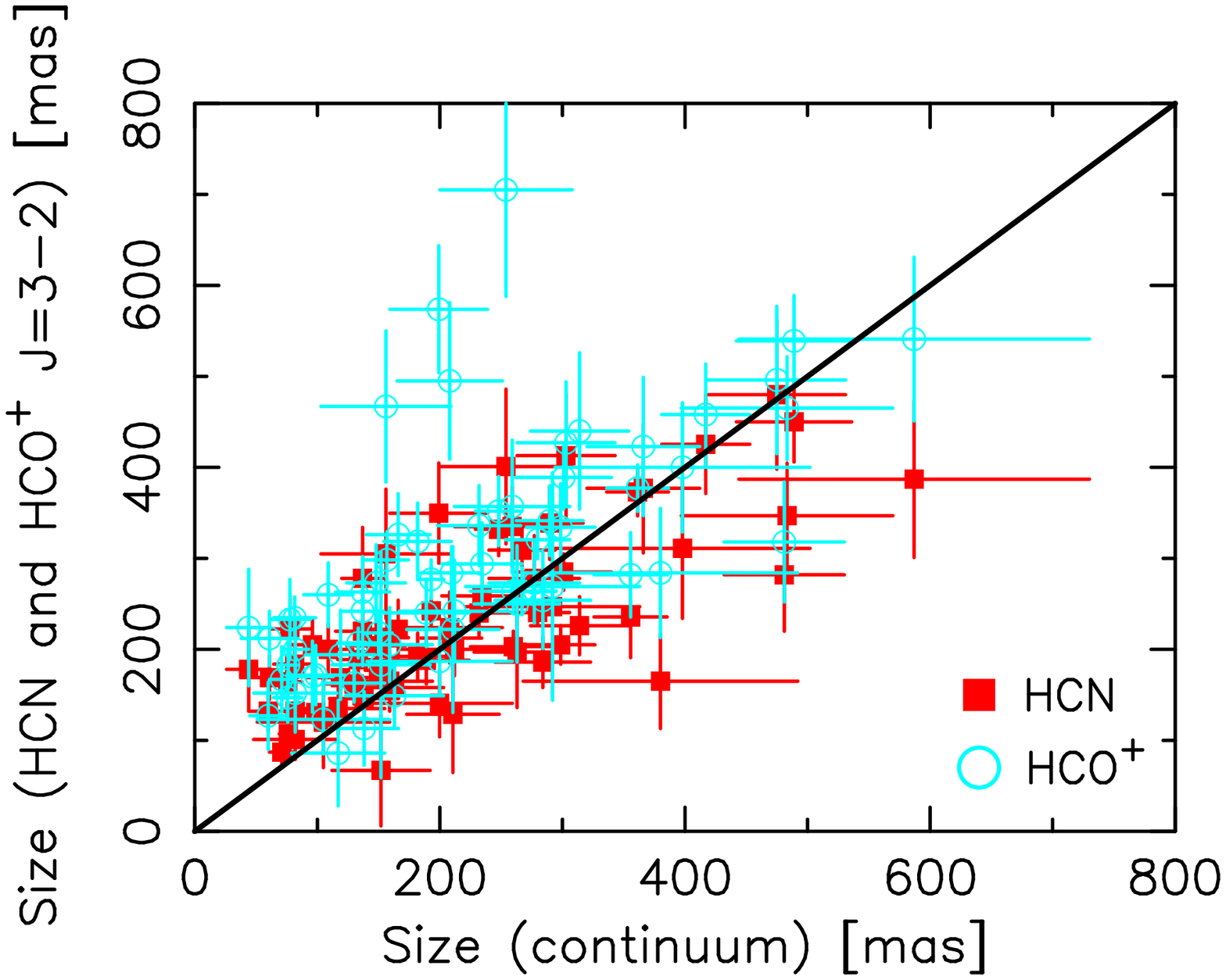}
\includegraphics[angle=0,scale=.45]{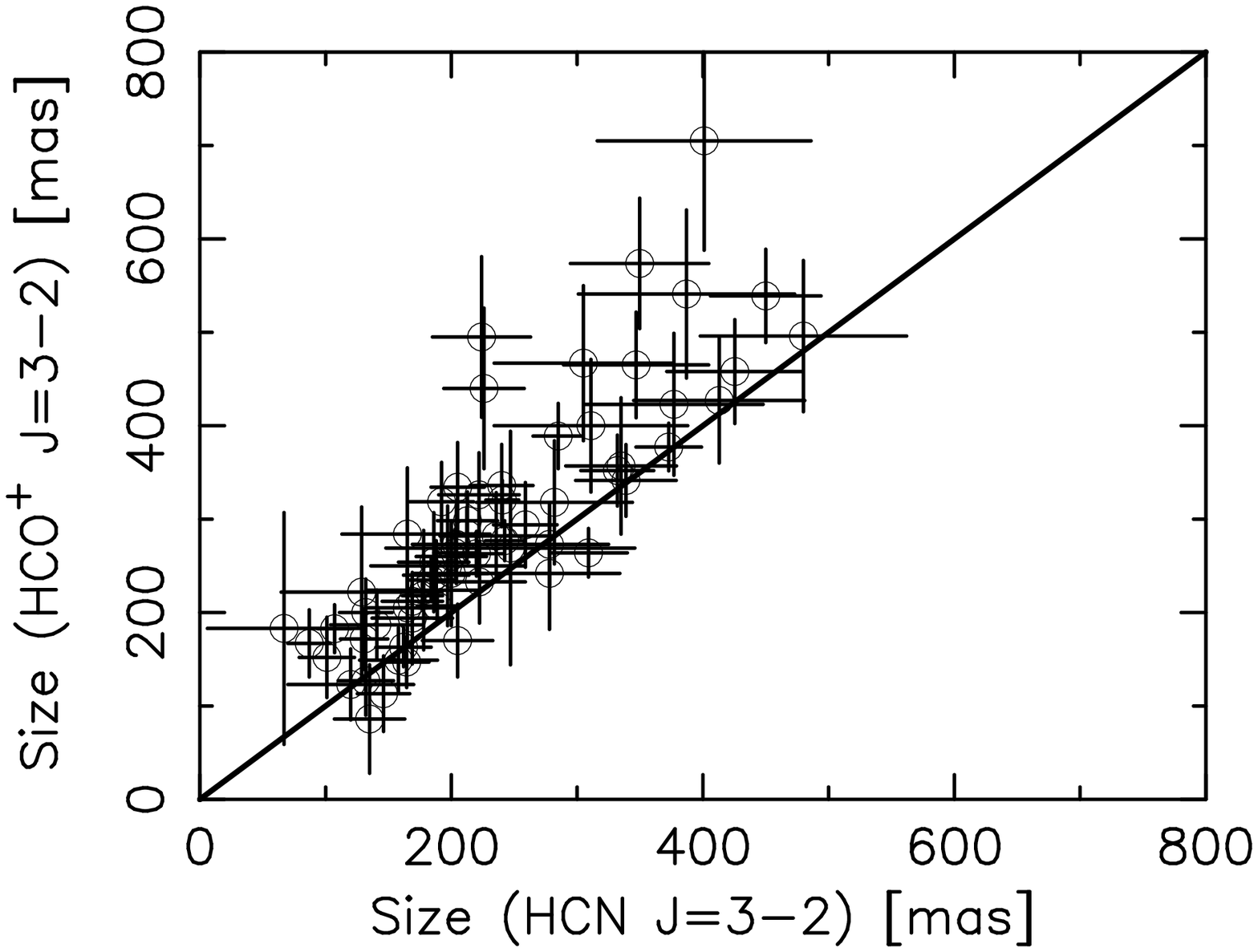}
\caption{
Comparison of the intrinsic emission size after deconvolution 
for the continuum, HCN J=3--2, and HCO$^{+}$ J=3--2
for ULIRGs observed with small beam sizes (0$\farcs$1--0$\farcs$2).
Both major and minor axis values as well as their geometric means 
are plotted.
The thick solid line indicates the same size between the abscissa 
and ordinate.
{\it (Left)}: The abscissa is the continuum emission size (in mas).
The ordinate is the size of the HCN J=3--2 (red squares) or 
HCO$^{+}$ J=3--2 (light blue circles) emission line (in mas).
{\it (Right)}: The size of the HCN J=3--2 (abscissa) and HCO$^{+}$ J=3--2 
(ordinate) emission lines (in mas).
}
\end{center}
\end{figure} 

\begin{figure}
\begin{center}
\includegraphics[angle=0,scale=.8]{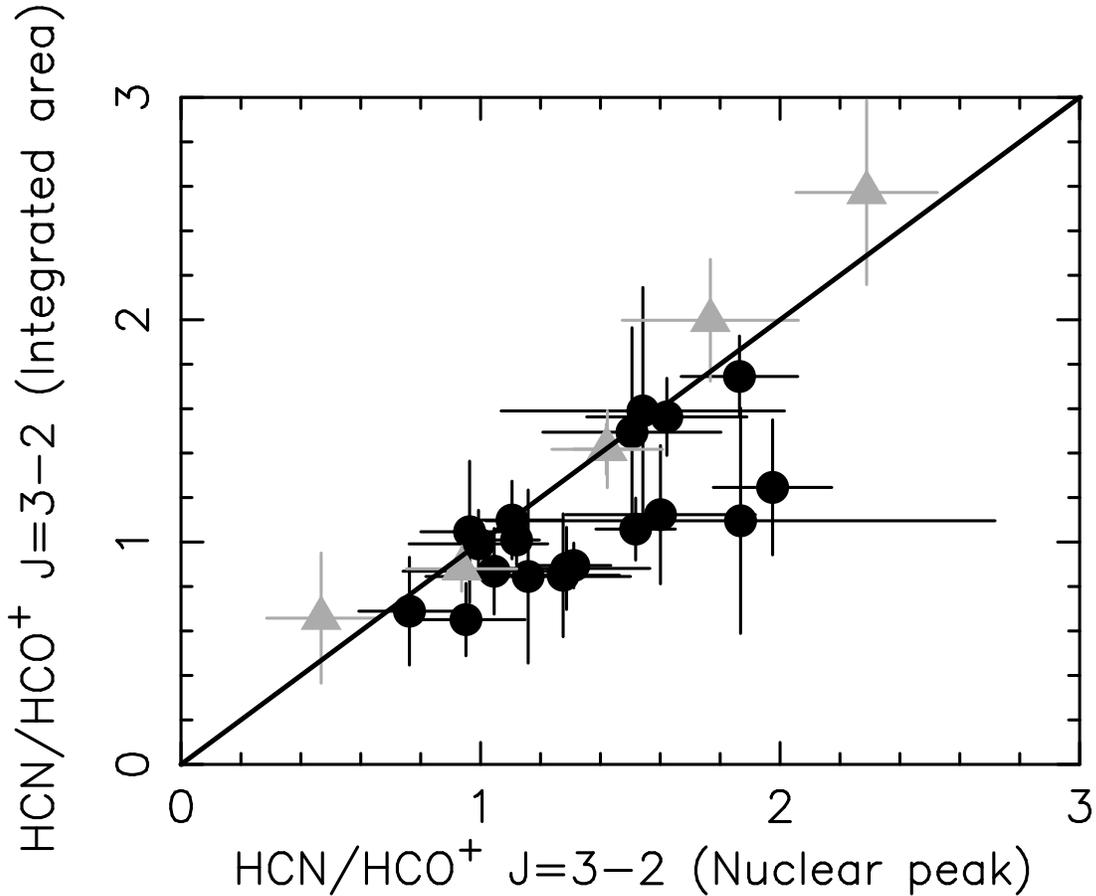}
\caption{
Comparison of the HCN-to-HCO$^{+}$ J=3--2 flux ratio at the nuclear 
continuum peak position within the beam size (abscissa) and within a 
1--2$''$ diameter circular aperture (ordinate).
The thick solid line indicates the same ratio between the abscissa 
and ordinate.
For ULIRGs observed in ALMA Cycle 5 with 0$\farcs$1--0$\farcs$2 
synthesized beam sizes, 1$''$ diameter circular apertures are employed 
for the ordinate. 
These ULIRGs are plotted as filled circles.
For ULIRGs observed with larger synthesized beam size 
(0$\farcs$5--0$\farcs$9), 
1$\farcs$5 (IRAS 21329$-$2346, IRAS 22206$-$2715, and IRAS 20414$-$1651) 
or 2$''$ (IRAS 12112$+$0305 NE and SW, and IRAS 22491$-$1808) 
diameter circular apertures are used, depending on the growth curve of 
HCN J=3--2 and HCO$^{+}$ J=3--2 molecular line fluxes. 
These ULIRGs are plotted as filled gray triangles, where IRAS 22206$-$2715 
and IRAS 20414$-$1651 are almost at the same location.
IRAS 09039$+$0503 and IRAS 11095$-$0238 display double nuclear molecular 
emission with $\sim$0$\farcs$5 separation (Figs. 3g, 3h, 3k, and 3l), 
for which measurements of molecular line fluxes with $\gtrsim$1$''$ diameter 
circular apertures suffer from contamination from the other nucleus in 
the abscissa. 
These two ULIRGs are not plotted.
}
\end{center}
\end{figure} 

\begin{figure}
\begin{center}
\hspace*{0.0cm}
\includegraphics[angle=0,scale=.4]{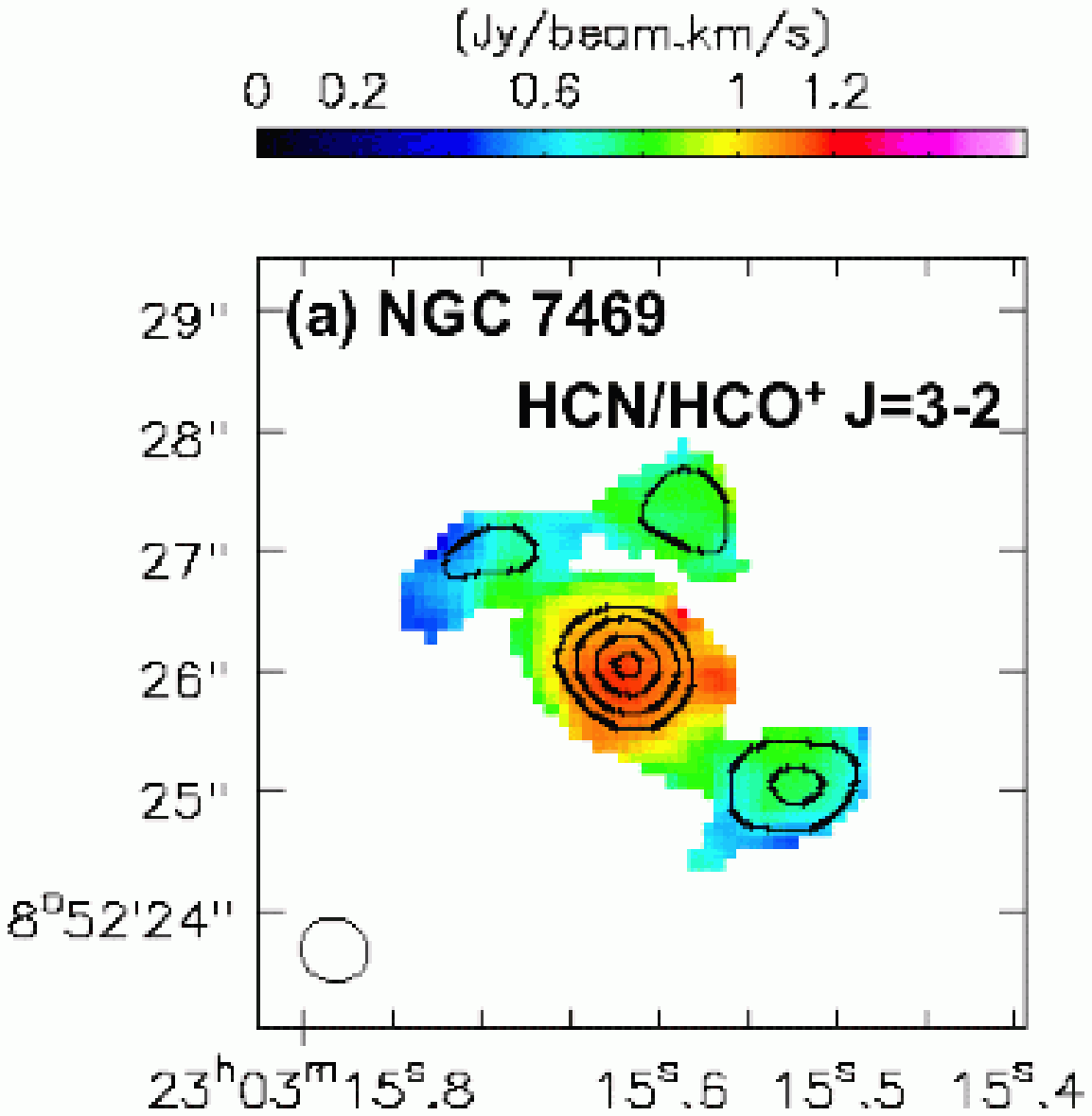}
\includegraphics[angle=0,scale=.4]{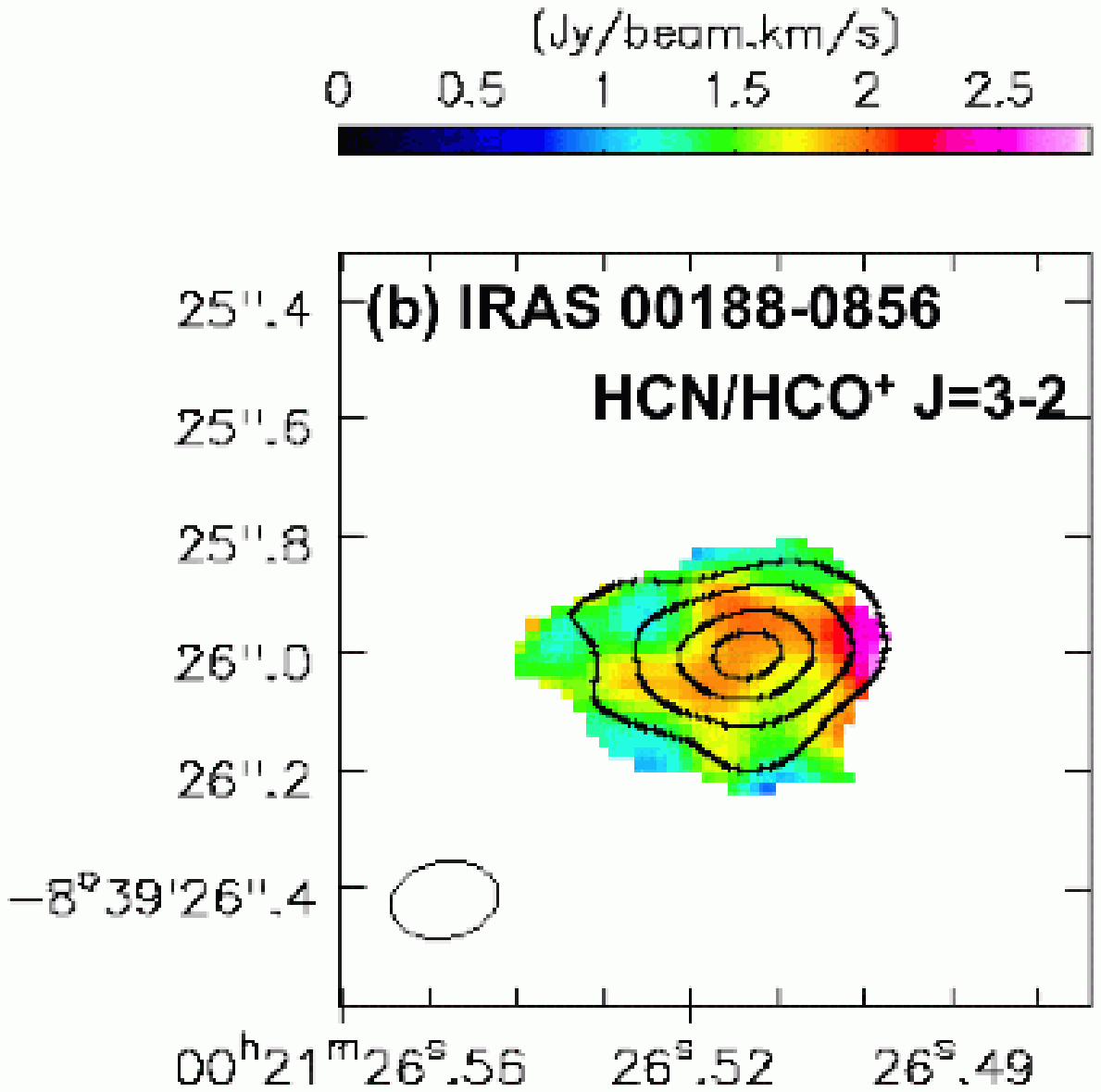}
\includegraphics[angle=0,scale=.4]{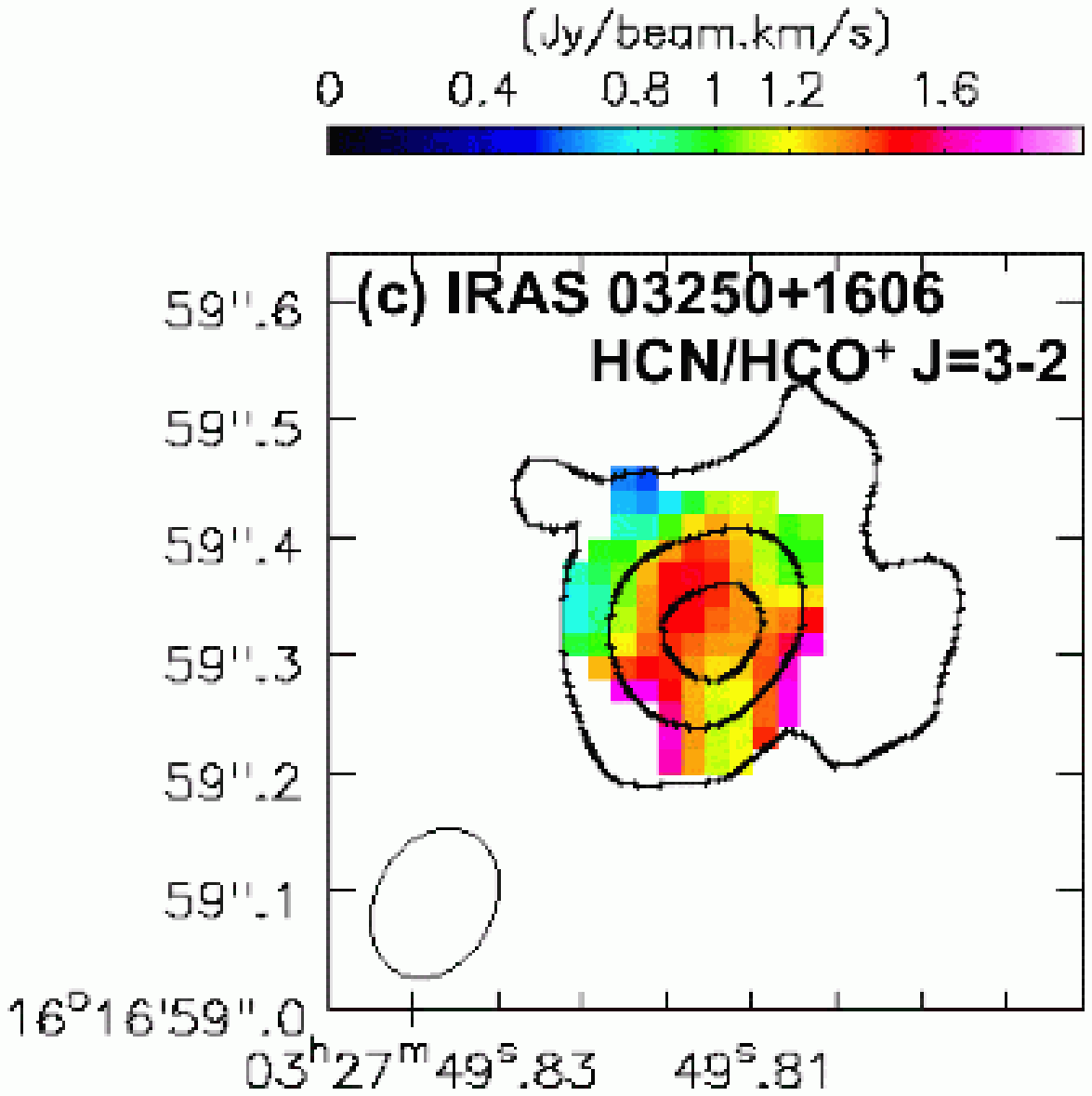} \\
\includegraphics[angle=0,scale=.4]{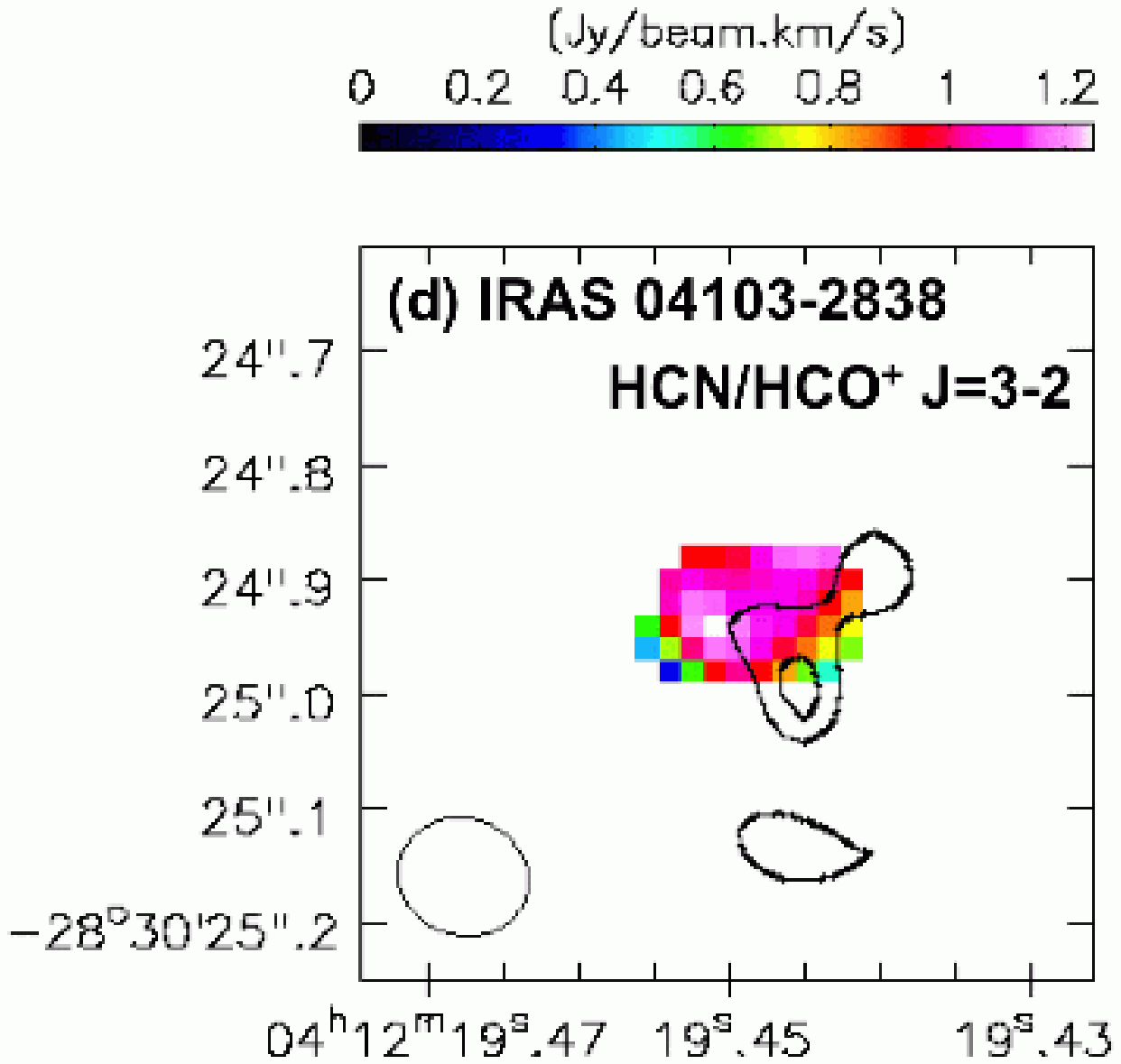} 
\includegraphics[angle=0,scale=.4]{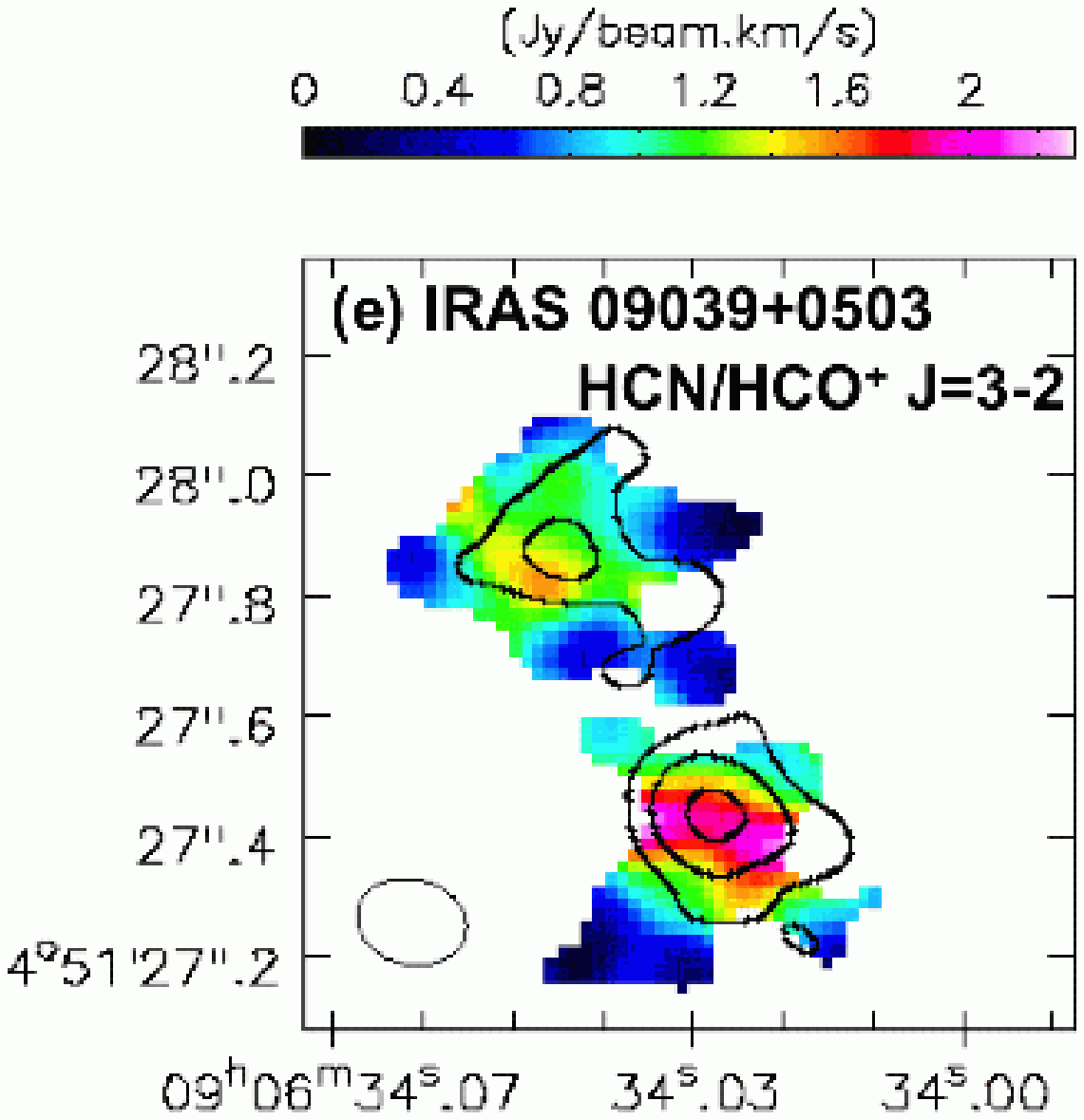} 
\includegraphics[angle=0,scale=.4]{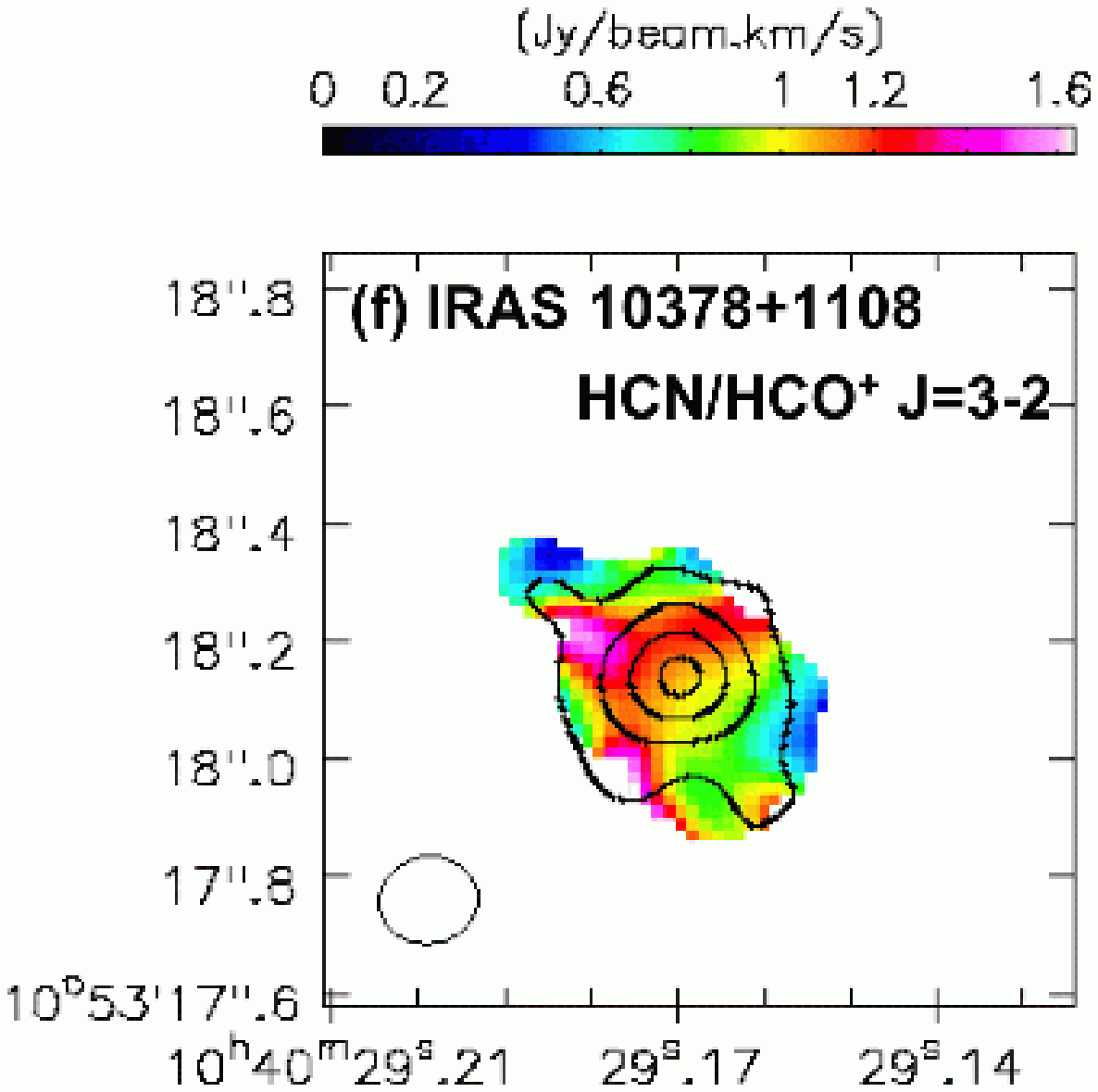} \\
\includegraphics[angle=0,scale=.4]{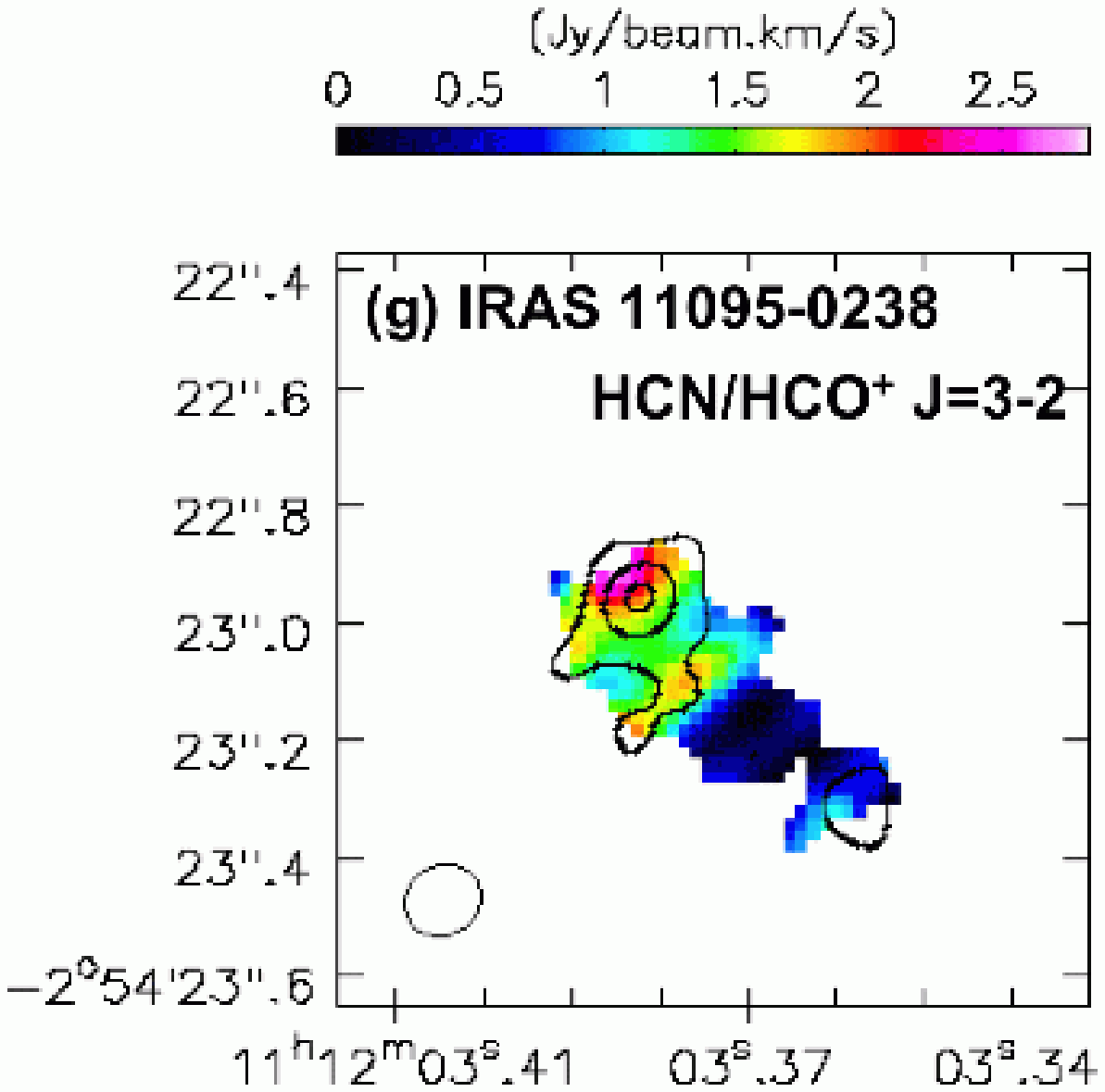} 
\includegraphics[angle=0,scale=.4]{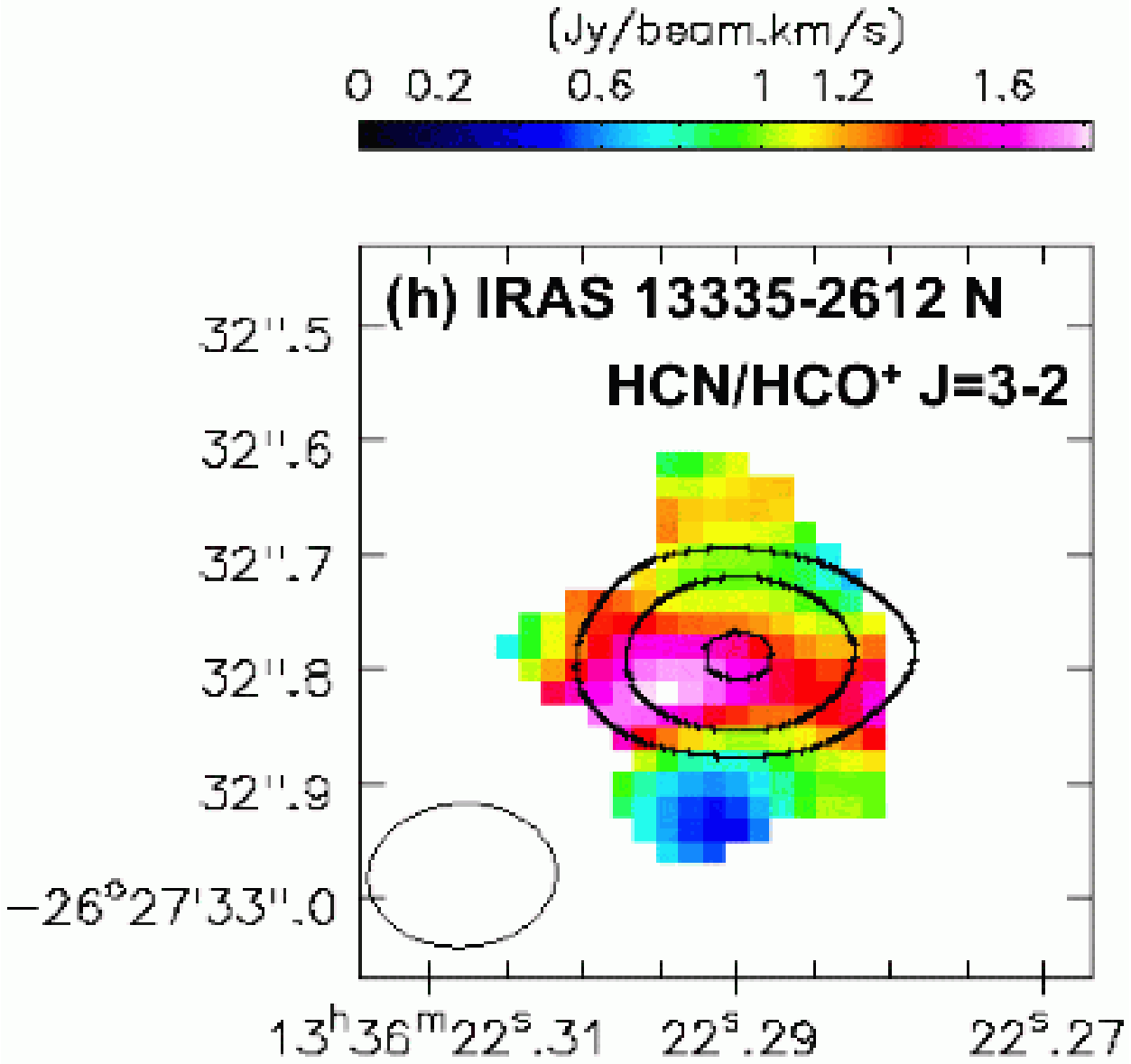}
\includegraphics[angle=0,scale=.4]{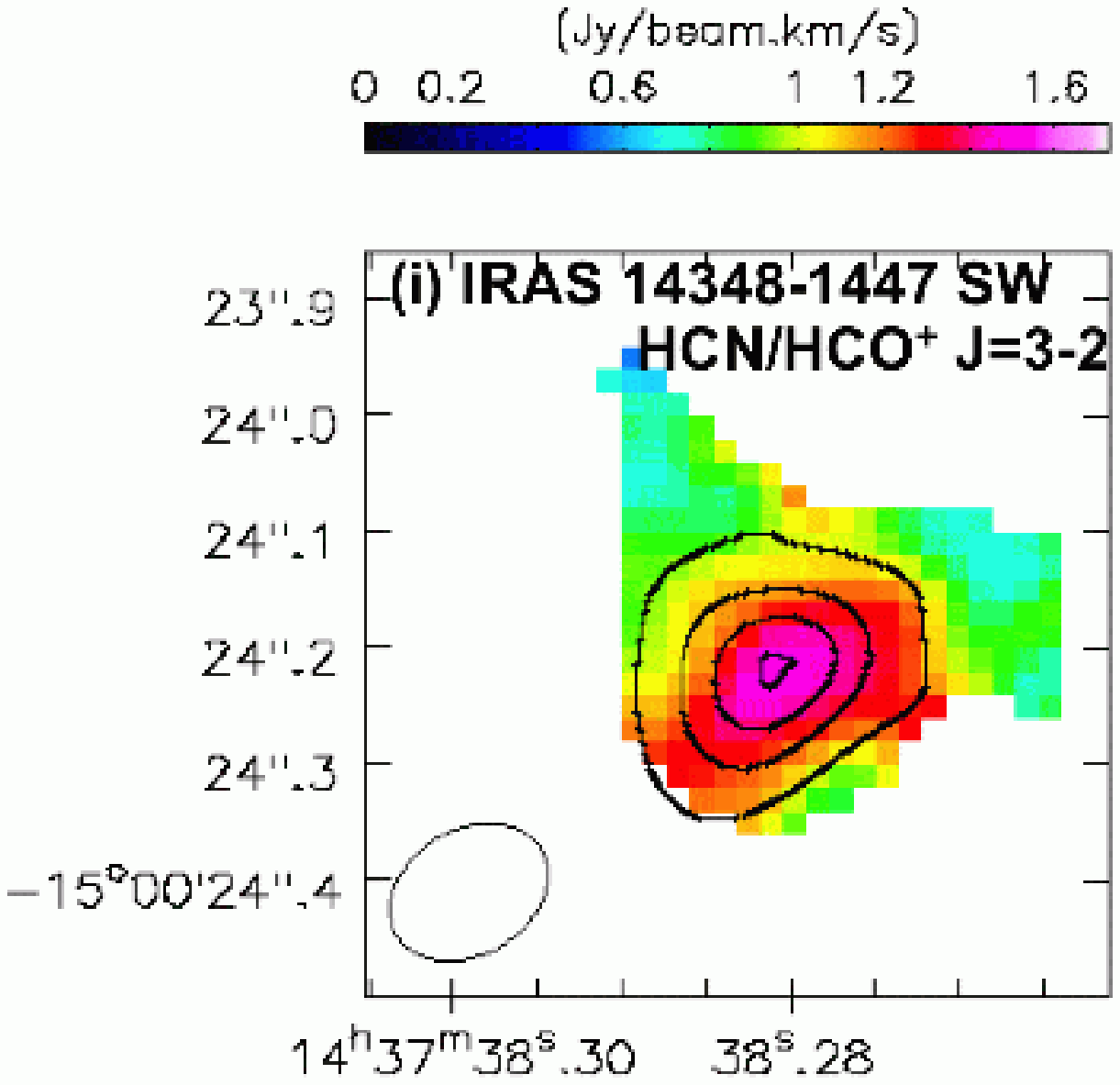} \\
\includegraphics[angle=0,scale=.4]{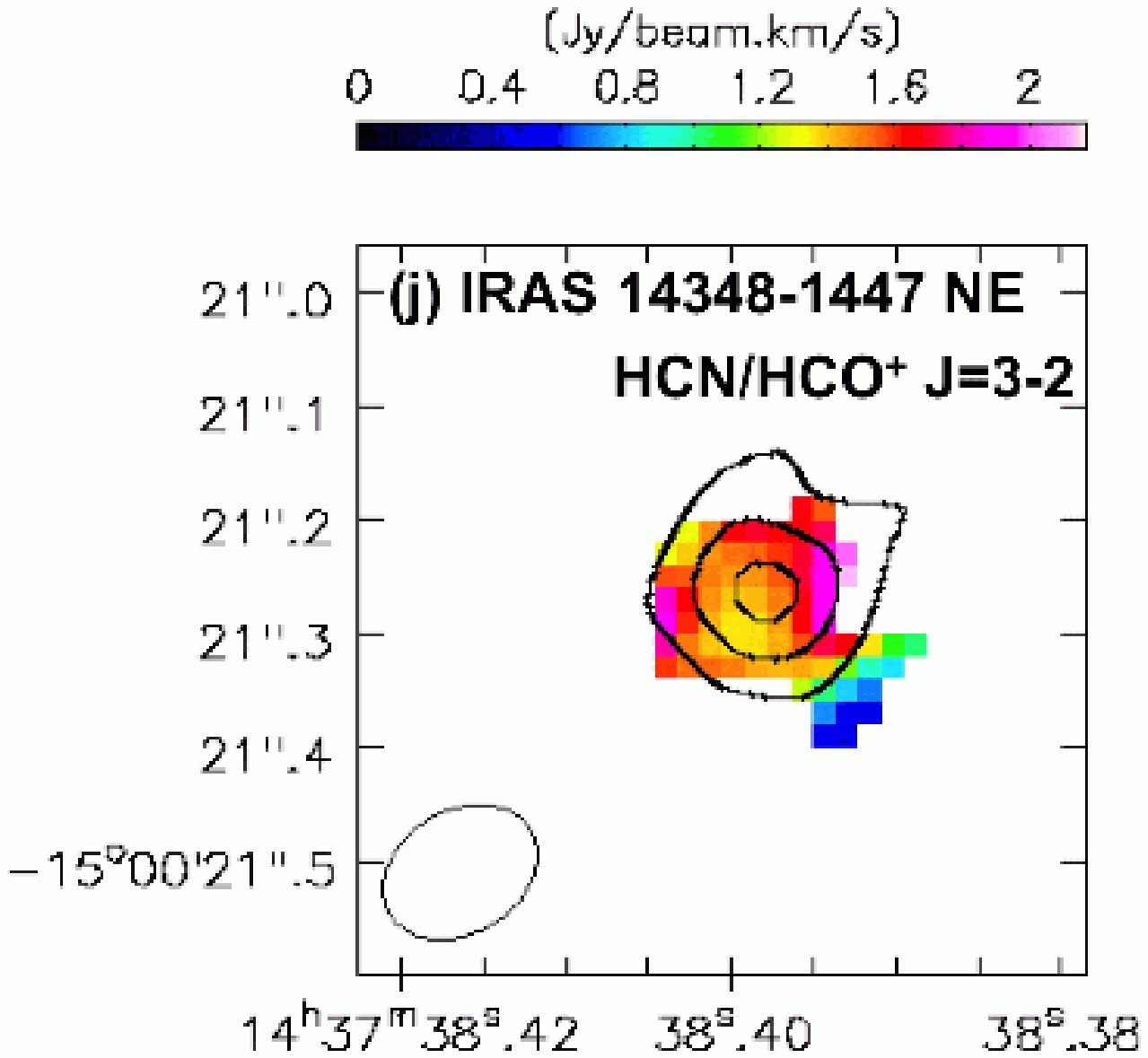} 
\includegraphics[angle=0,scale=.4]{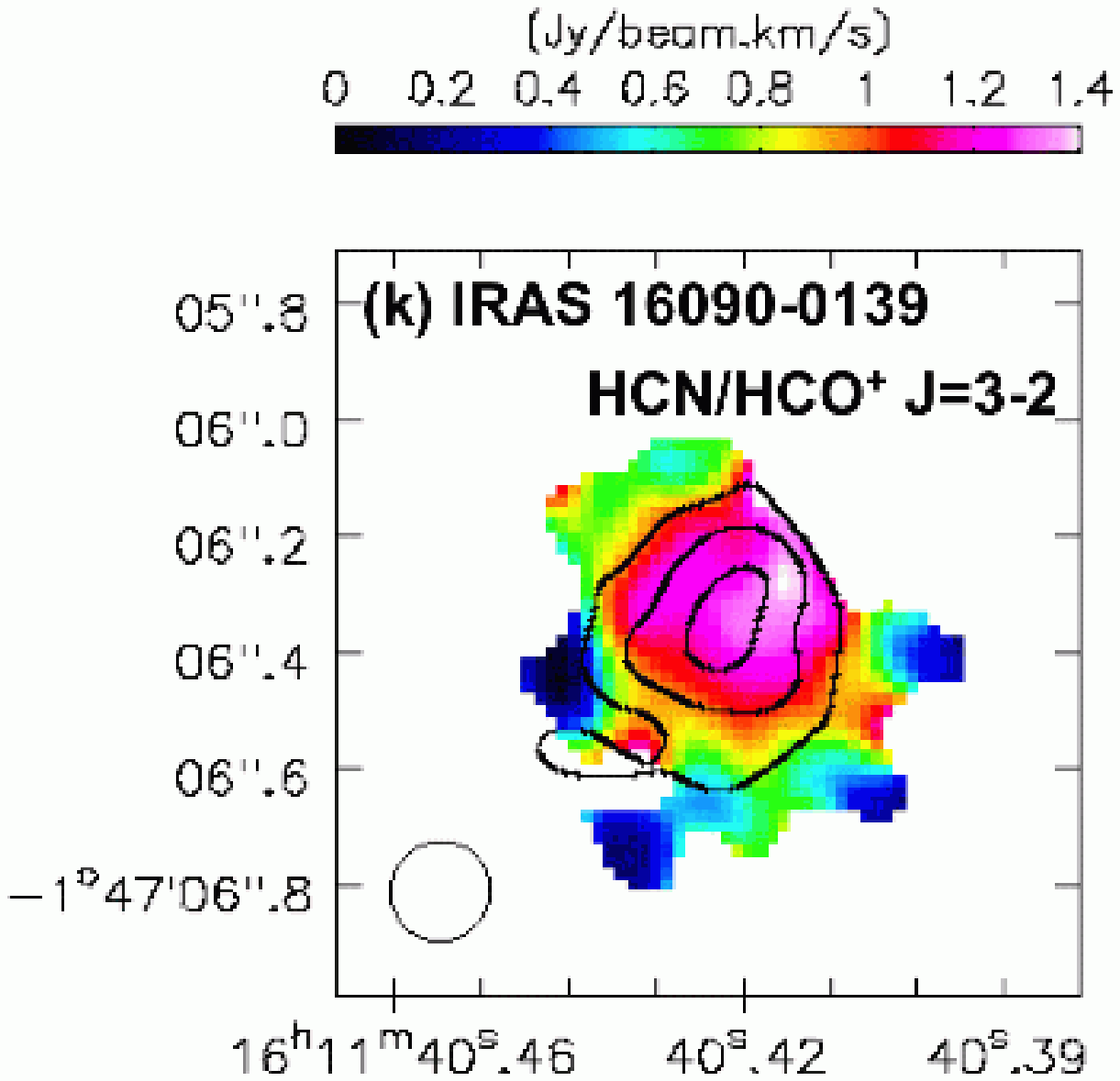}
\includegraphics[angle=0,scale=.4]{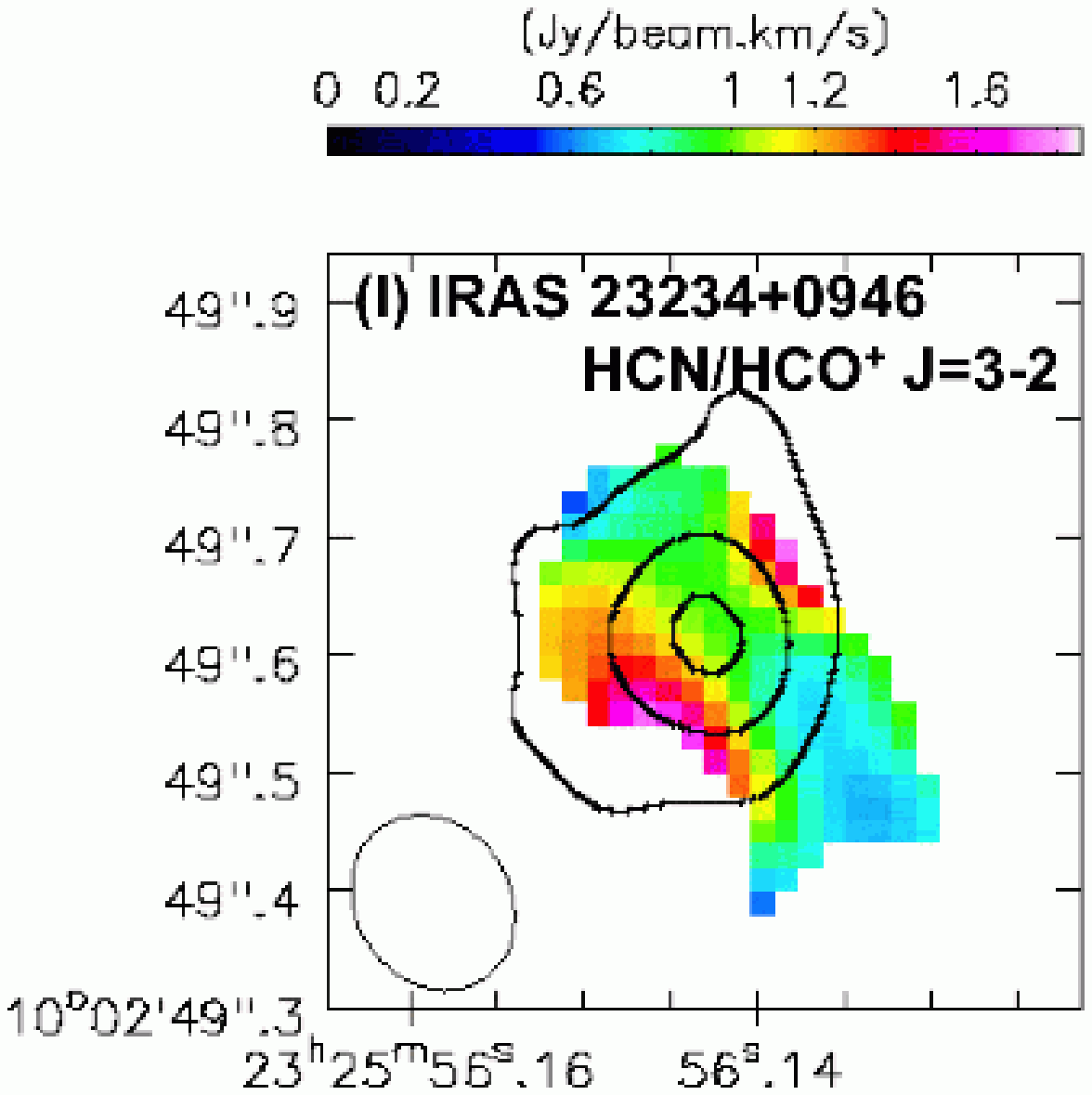} \\
\includegraphics[angle=0,scale=.4]{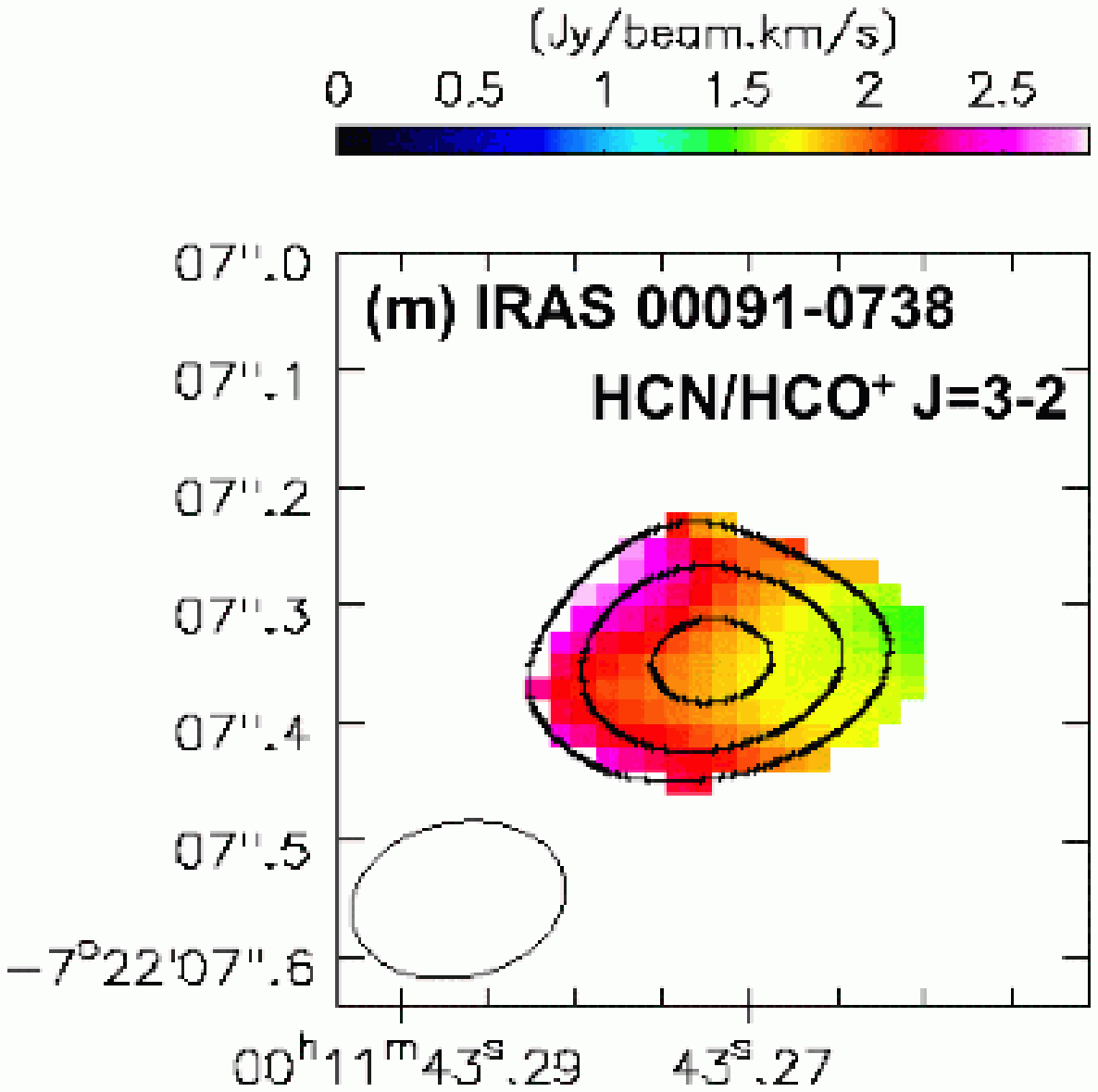} 
\includegraphics[angle=0,scale=.4]{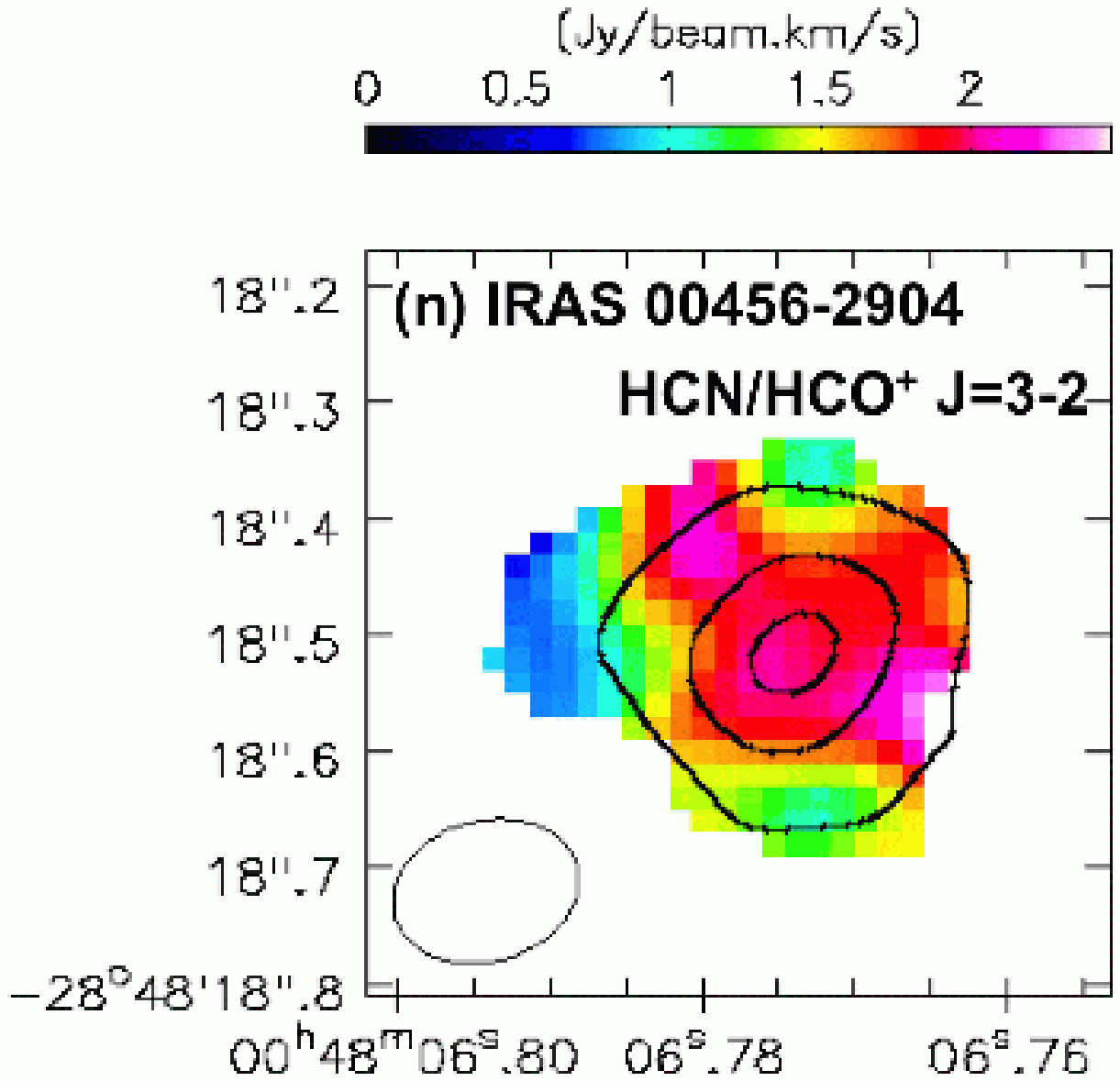}
\includegraphics[angle=0,scale=.4]{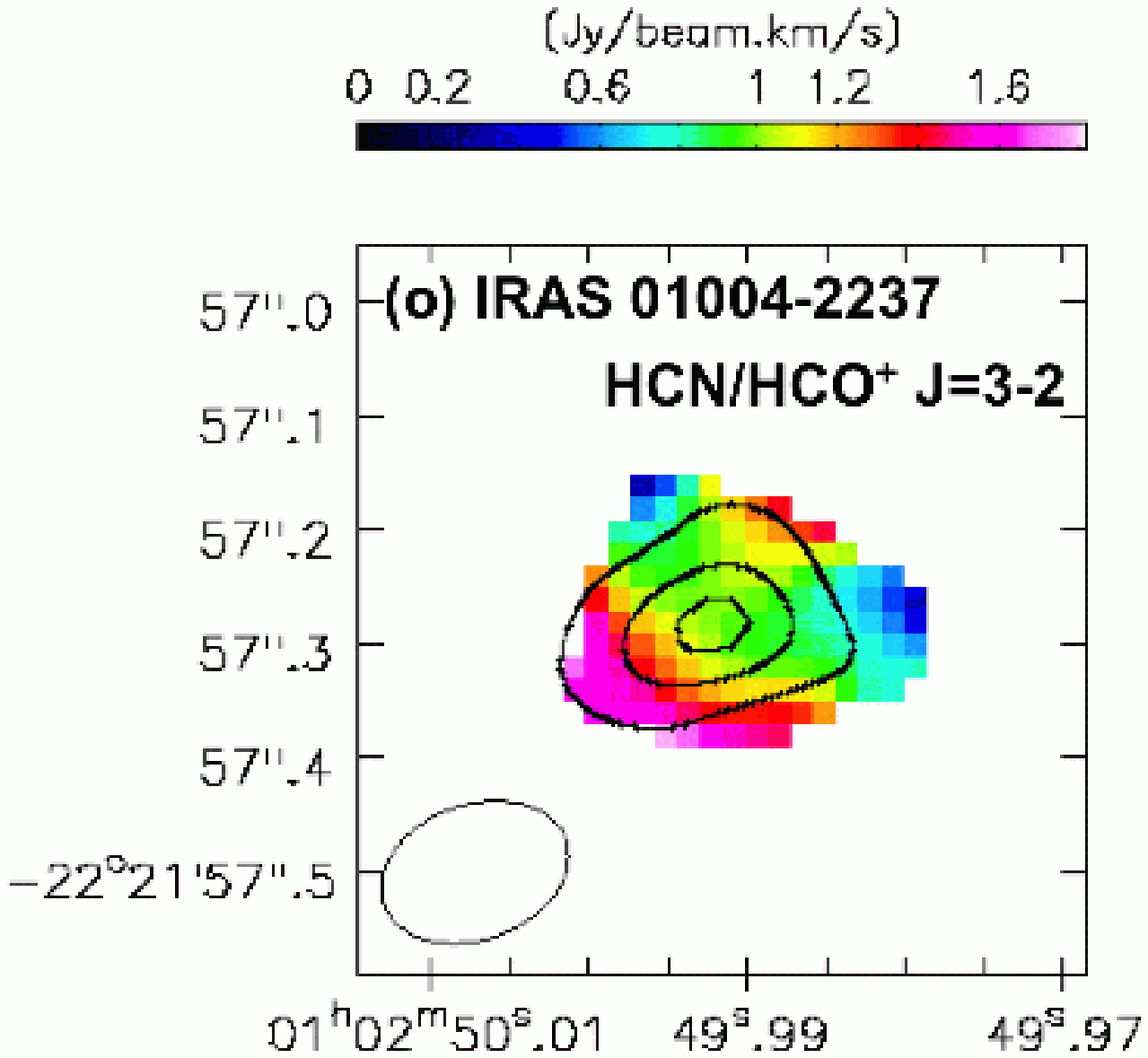} \\
\end{center}
\end{figure} 

\clearpage

\begin{figure}
\begin{center}
\includegraphics[angle=0,scale=.4]{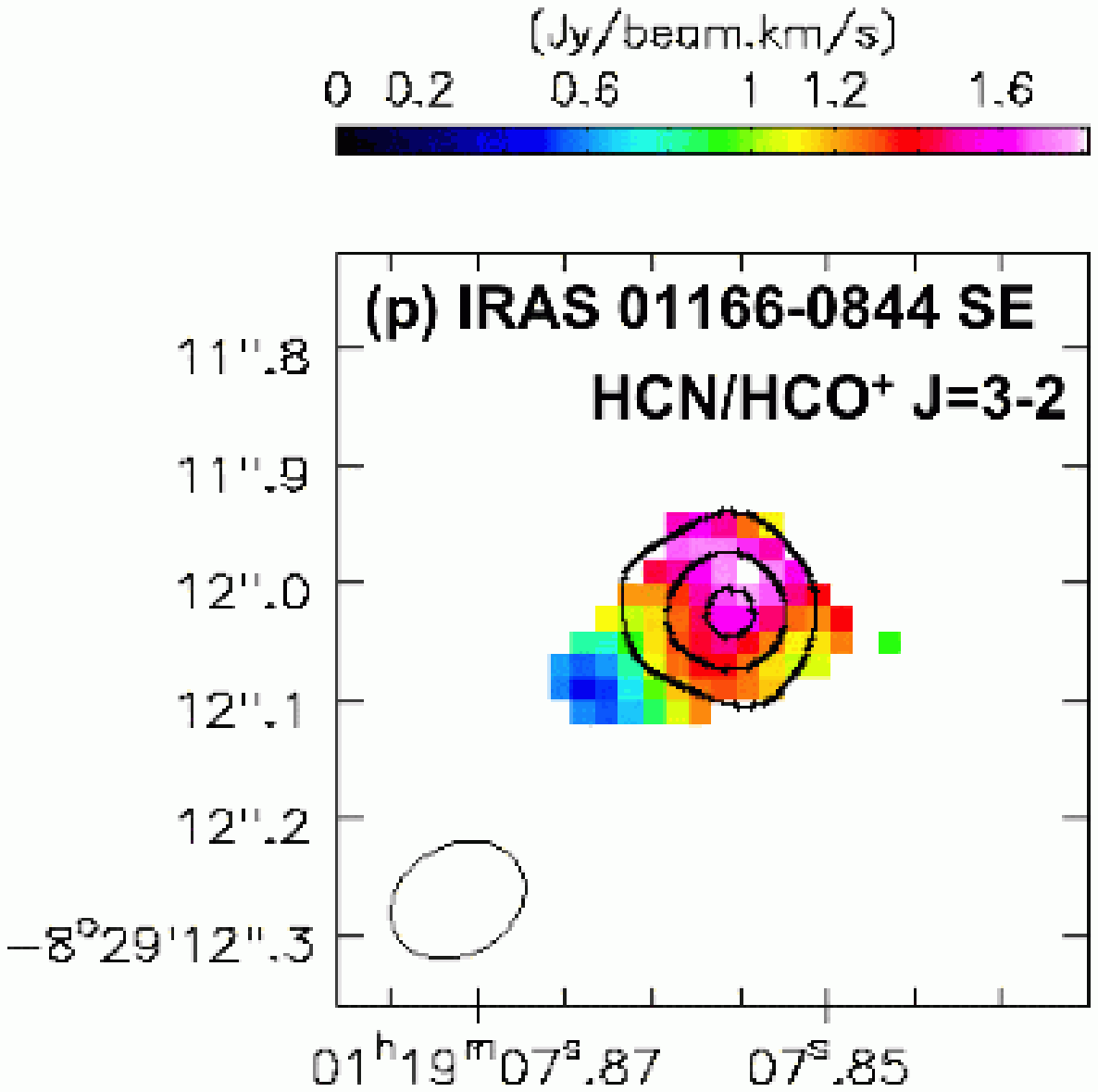} 
\includegraphics[angle=0,scale=.4]{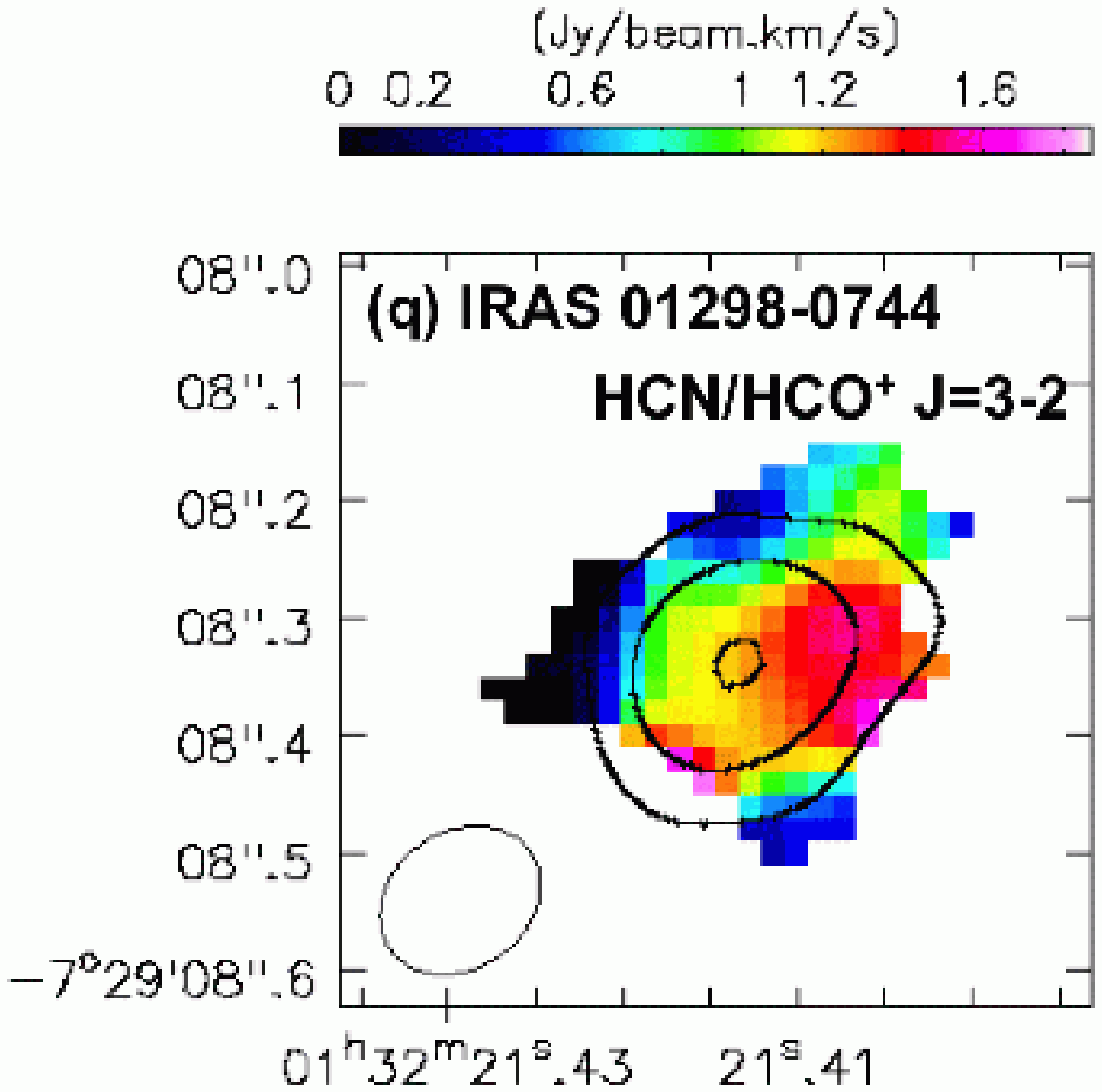}
\includegraphics[angle=0,scale=.4]{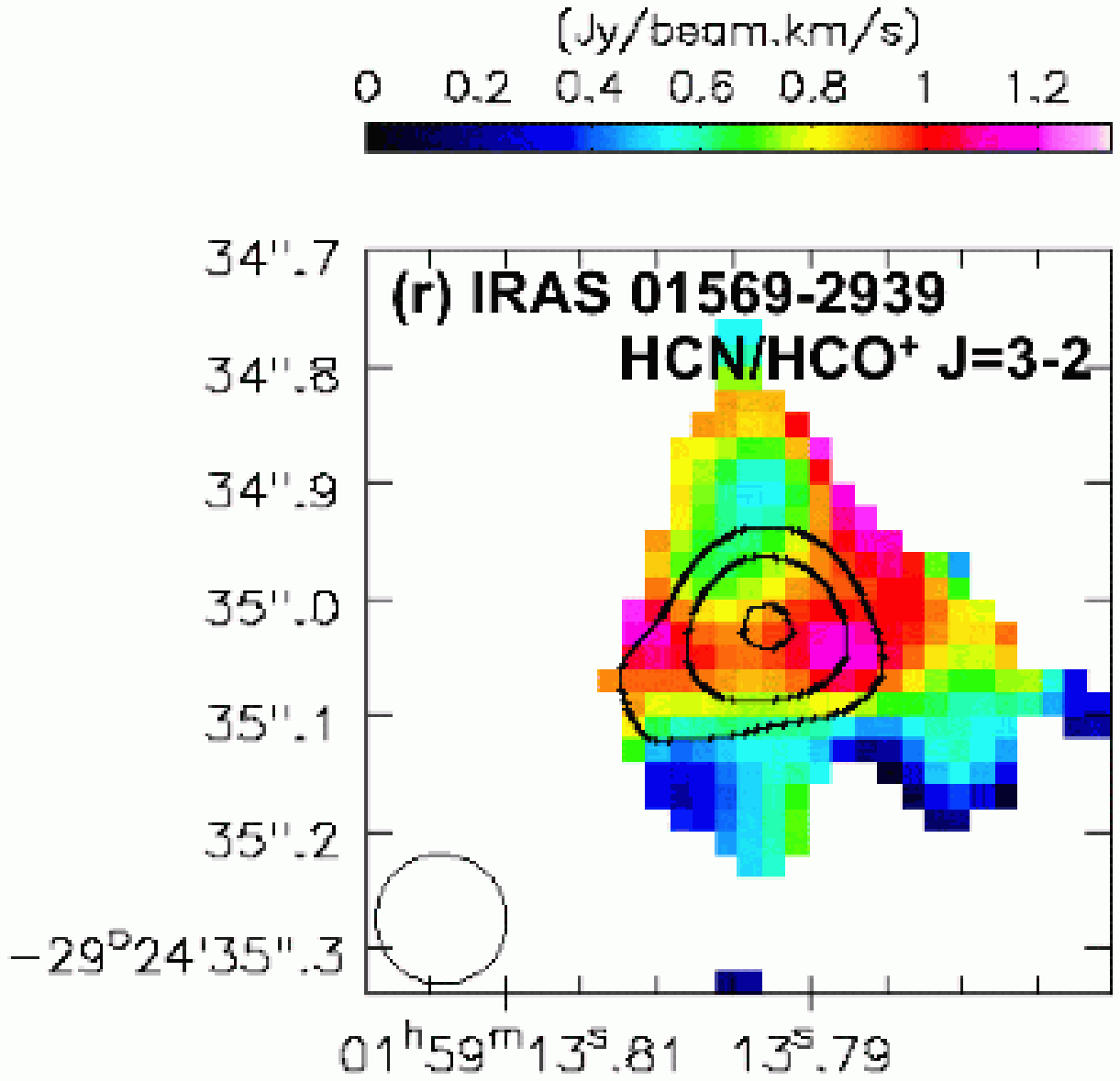} \\
\includegraphics[angle=0,scale=.4]{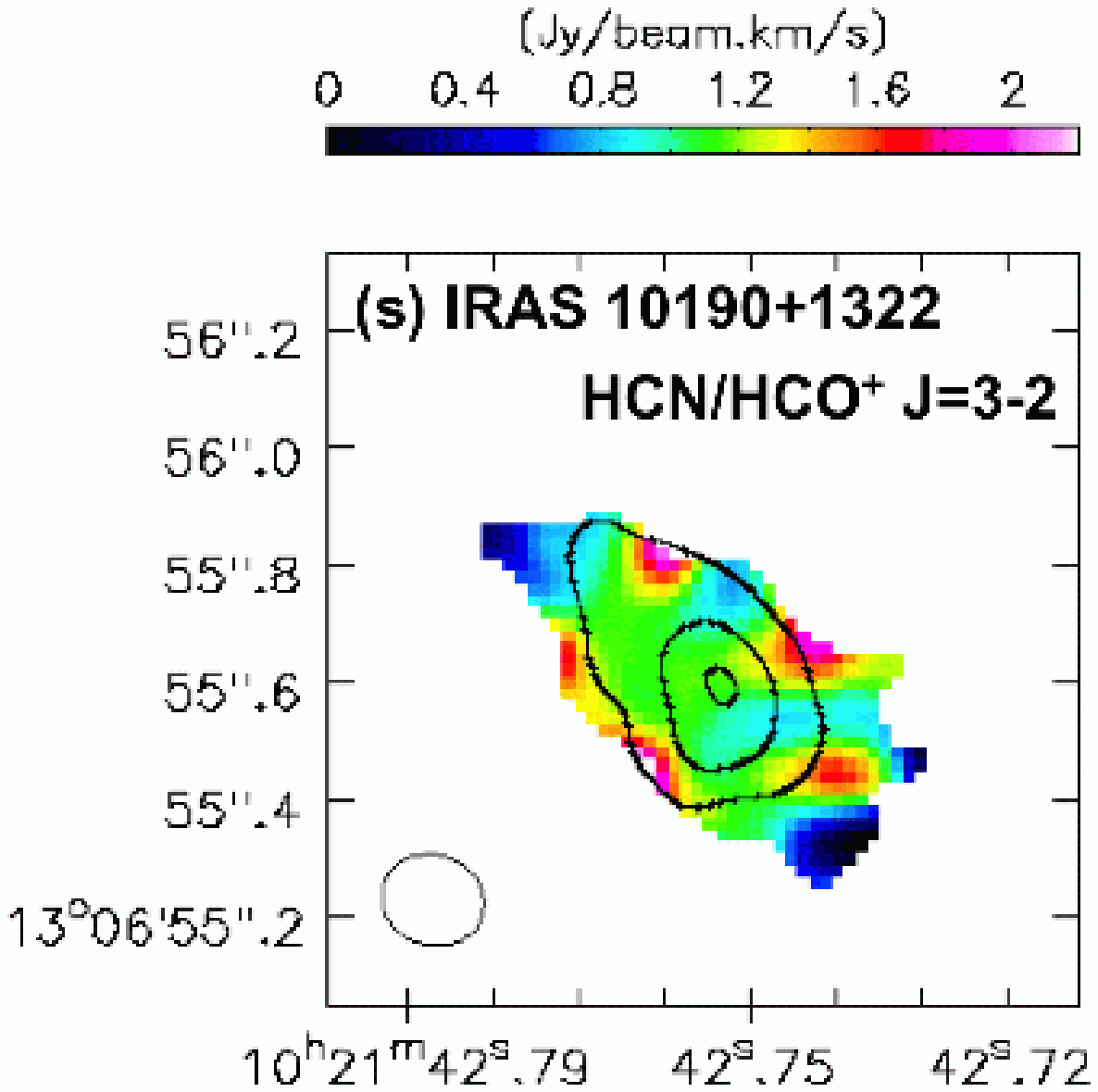} 
\includegraphics[angle=0,scale=.4]{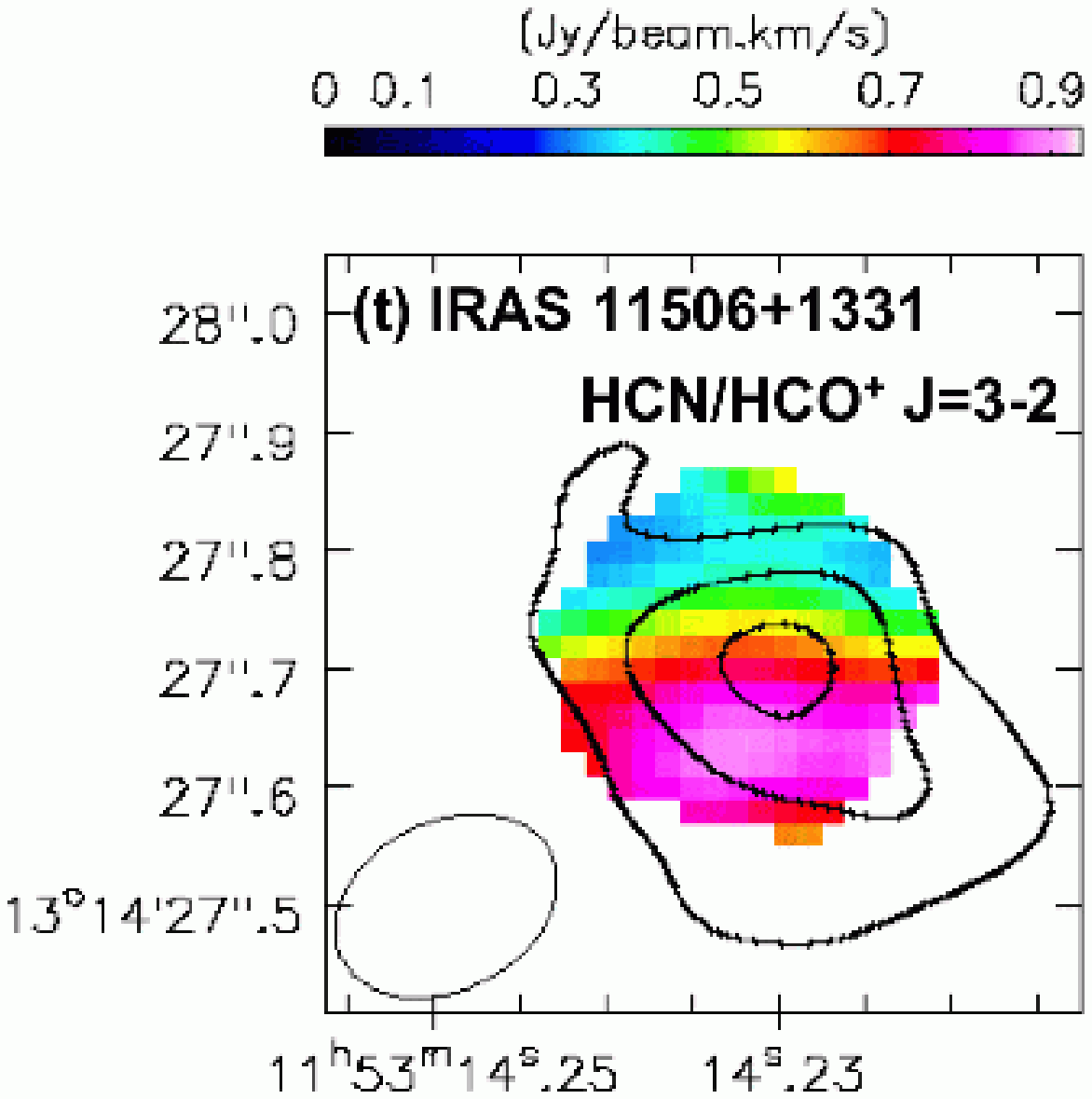}
\includegraphics[angle=0,scale=.4]{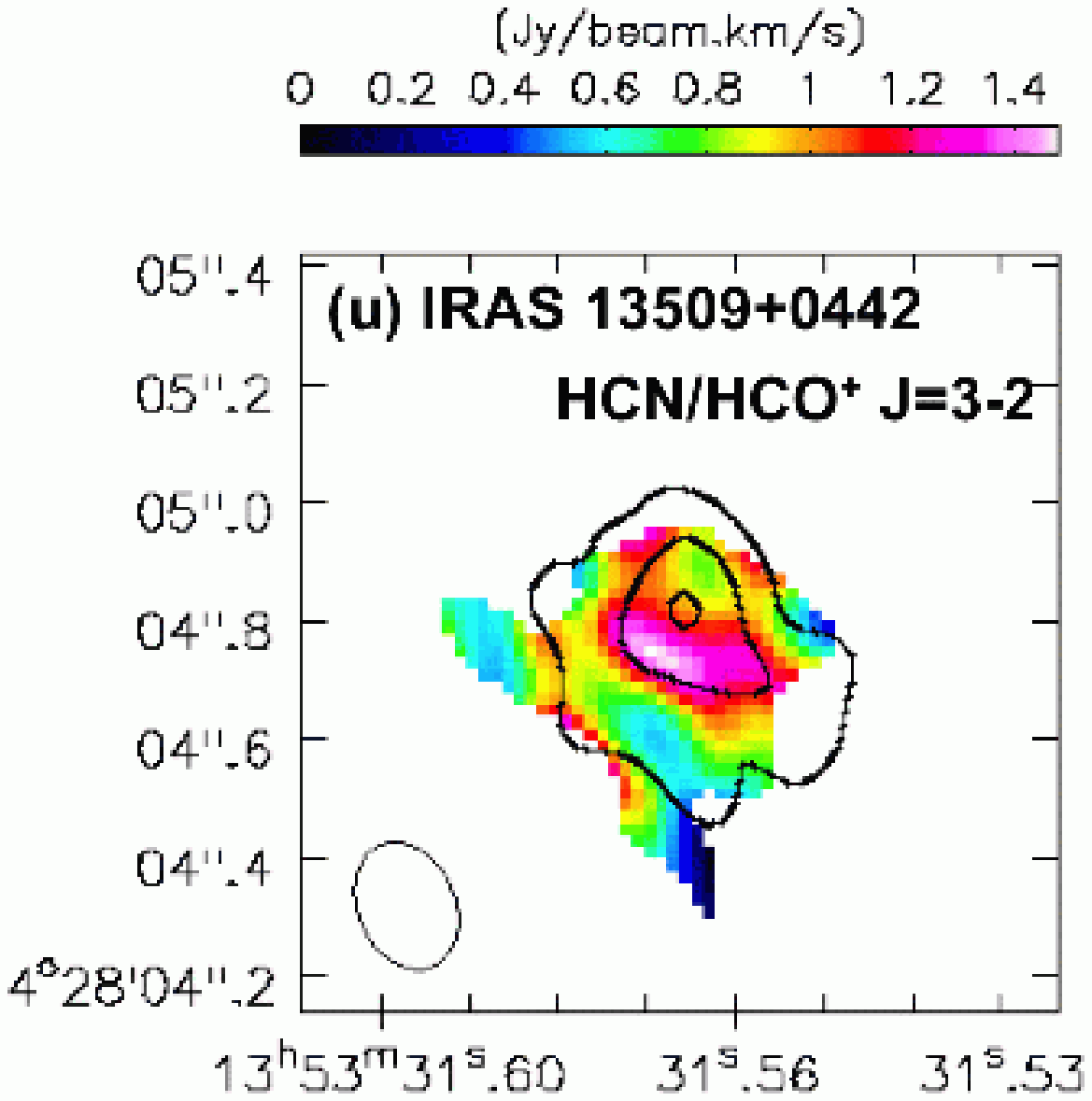}
\vspace{0.5cm} 
\caption{ 
Ratios of the HCN J=3--2 to HCO$^{+}$ J=3--2 flux (in Jy km s$^{-1}$) 
are shown as the images for ULIRGs observed with small (0$\farcs$1--0$\farcs$2) 
beam sizes.
Continuum emission is overplotted as the black contours to indicate the 
nuclear location.
The flux ratio map for the nearby well-studied optically identified 
luminous AGN, NGC 7469, is provided as a reference (Fig. 6a), 
as it clearly shows an elevated HCN-to-HCO$^{+}$ J=3--2 flux ratio 
at the AGN-dominated nucleus, 
compared to circumnuclear starburst-dominated regions \citep{ima16c}.
The continuum contours are 5$\sigma$, 10$\sigma$, 20$\sigma$, and 
30$\sigma$ \citep{ima16c}.
For ULIRGs, the contour levels are the same as those in Figure 1, except 
IRAS 00188$-$0856 (5$\sigma$, 10$\sigma$, 20$\sigma$, 30$\sigma$), 
IRAS 10378$+$1108 (3$\sigma$, 10$\sigma$, 20$\sigma$, 30$\sigma$), 
IRAS 13335$-$2612 N (6$\sigma$, 12$\sigma$, 24$\sigma$), 
IRAS 14348$-$1447 SW (5$\sigma$, 10$\sigma$, 15$\sigma$, 20$\sigma$), 
IRAS 14348$-$1447 NE (5$\sigma$, 9$\sigma$, 13$\sigma$), 
IRAS 23234$+$0946 (6$\sigma$, 12$\sigma$, 18$\sigma$), and 
IRAS 11506$+$1331 (5$\sigma$, 8$\sigma$, 11$\sigma$).
IRAS 10485$-$1447 and IRAS 02411$+$0353 are not shown because molecular 
emission lines are too faint to create meaningful flux ratio maps. 
Only pixels whose HCO$^{+}$ J=3--2 emission line fluxes (i.e., 
denominator) are above a certain threshold are displayed to prevent 
the resulting ratio map from being dominated by noise.
}
\end{center}
\end{figure} 

\begin{figure}
\begin{center}
\includegraphics[angle=0,scale=.72]{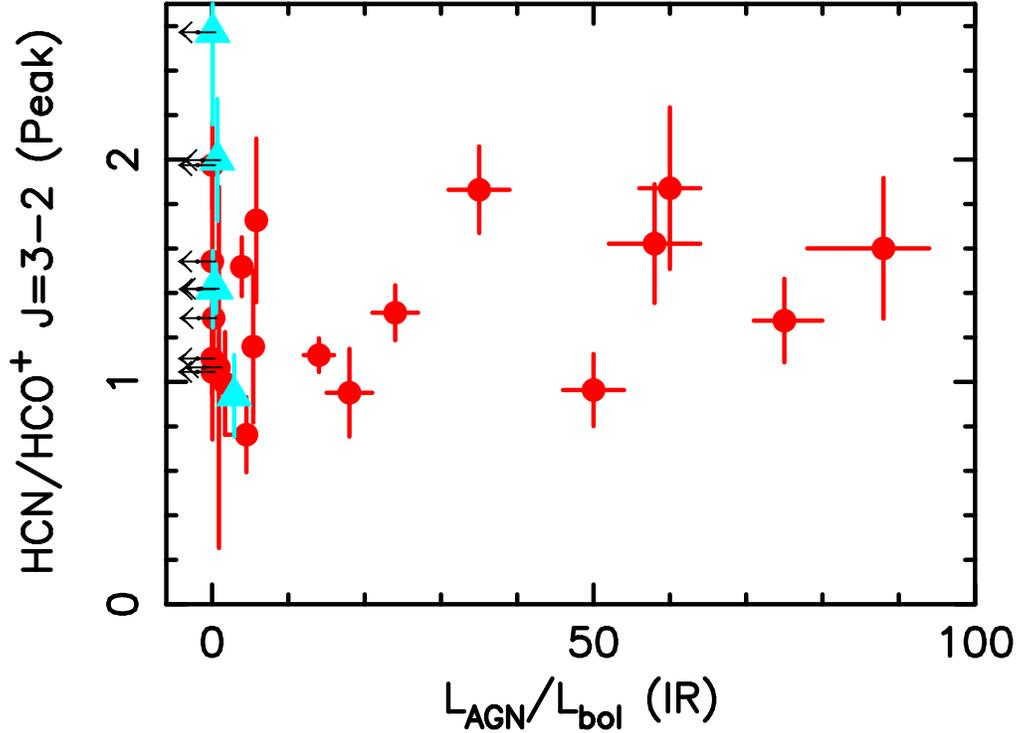} 
\caption{
Comparison of infrared spectroscopically estimated AGN bolometric 
contribution (in \%) by 
\citet{nar10} (abscissa) and HCN-to-HCO$^{+}$ J=3--2 flux ratios 
at the nuclear continuum peak position within the beam size (ordinate) 
in the observed ULIRGs.
ULIRGs observed with small (0$\farcs$1--0$\farcs$2) beam sizes are 
plotted as red circles.
Those observed with large (0$\farcs$5--0$\farcs$9) beam sizes are plotted 
as light blue triangles. 
For several small-separation ($<$4$''$) double-nuclei ULIRGs, 
while the bolometric contributions of the AGNs based on Spitzer IRS infrared 
spectra are for both nuclei combined \citep{nar10}, the derived 
HCN-to-HCO$^{+}$ flux ratios at J=3--2 shown in Table 8 (column 2) 
are for the individual nuclei separately. 
For IRAS 09039$+$0503, IRAS 11095$-$0238, IRAS 13335$-$2612, 
IRAS 14348$-$1447, and IRAS 12112$+$0305, we plot the 
HCN-to-HCO$^{+}$ J=3--2 flux ratios for the SW, NE, N, SW, and NE nuclei, 
respectively, because they are brighter than the other nuclei in the 
continuum (Fig. 1) and thus are judged to dominate the bolometric 
luminosities.
}
\end{center}
\end{figure}

\begin{figure}
\begin{center}
\includegraphics[angle=0,scale=.45]{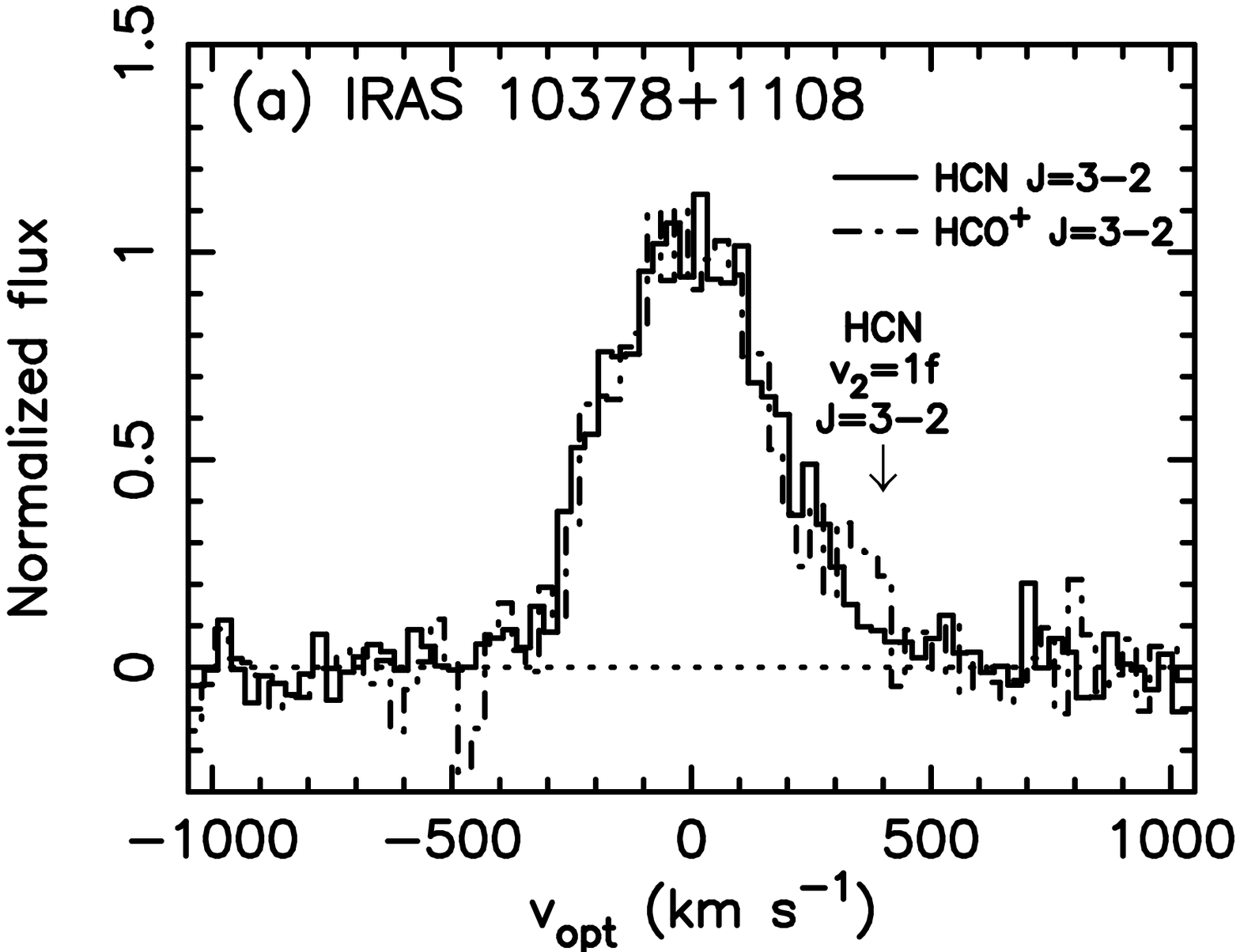} 
\includegraphics[angle=0,scale=.45]{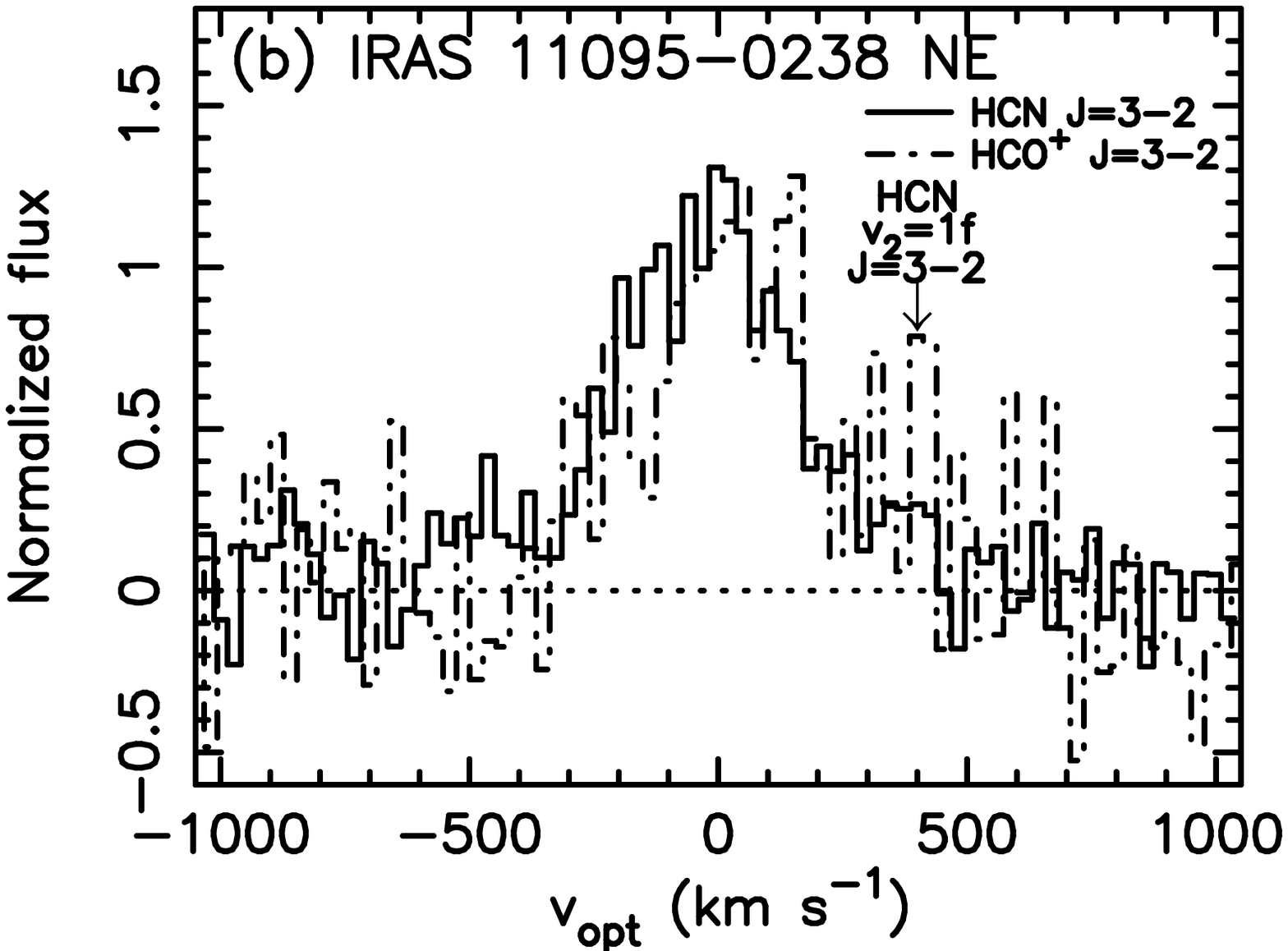} 
\caption{
Comparison of the emission line profile of HCN J=3--2 and HCO$^{+}$ 
J=3--2 in the nuclear beam-sized spectra of IRAS 10378$+$1108 and 
IRAS 11095$-$0238 NE, both of which show possible flux excess 
attributed to the vibrationally excited (v$_{2}$=1f) HCN J=3--2 
emission line.
The abscissa is velocity in (km s$^{-1}$) relative to the systemic 
value. The ordinate is normalized flux.
The expected velocity of the HCN v$_{2}$=1f J=3--2 line is indicated as 
a downward arrow.
}
\end{center}
\end{figure}

\begin{figure}
\begin{center}
\includegraphics[angle=0,scale=.45]{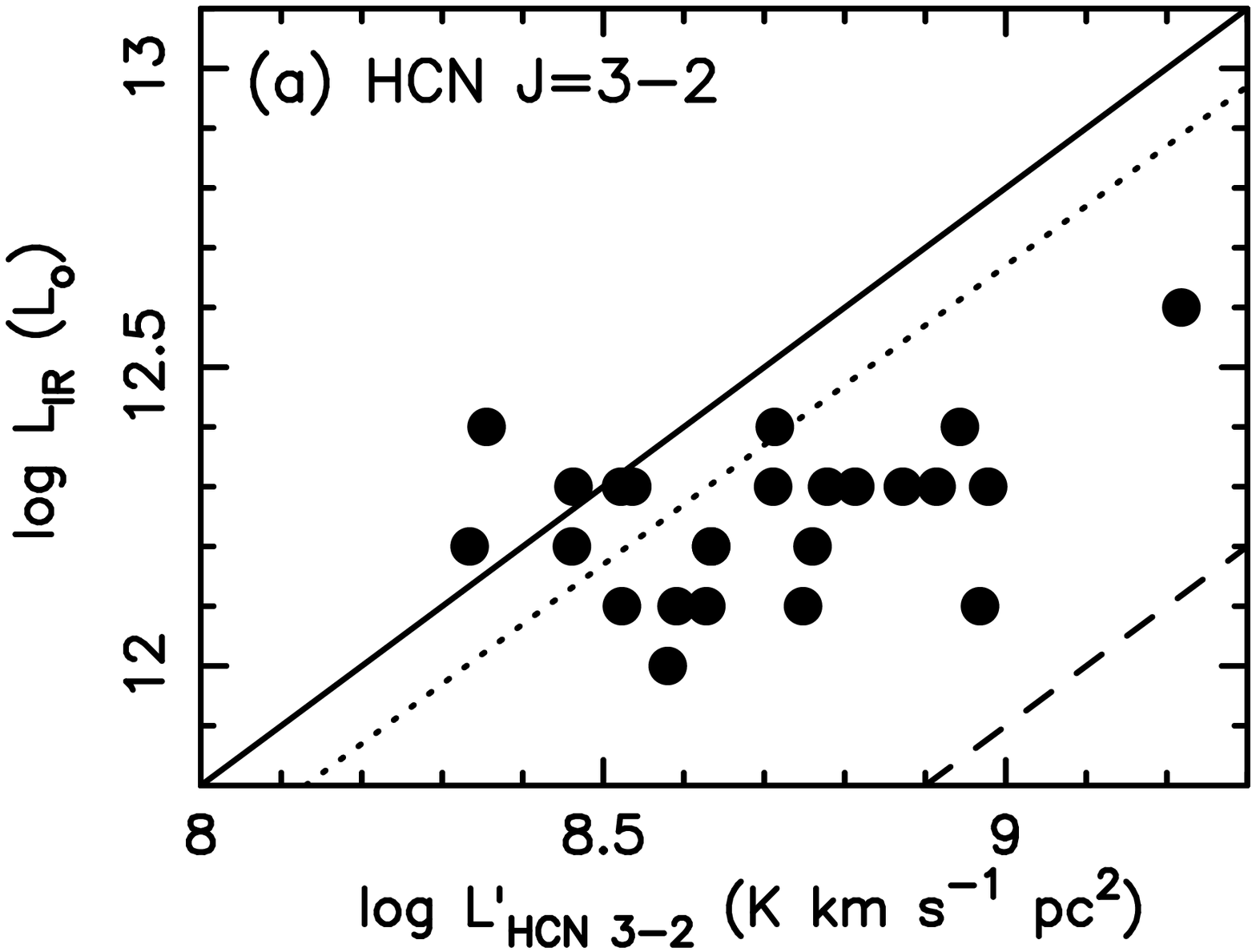} 
\includegraphics[angle=0,scale=.45]{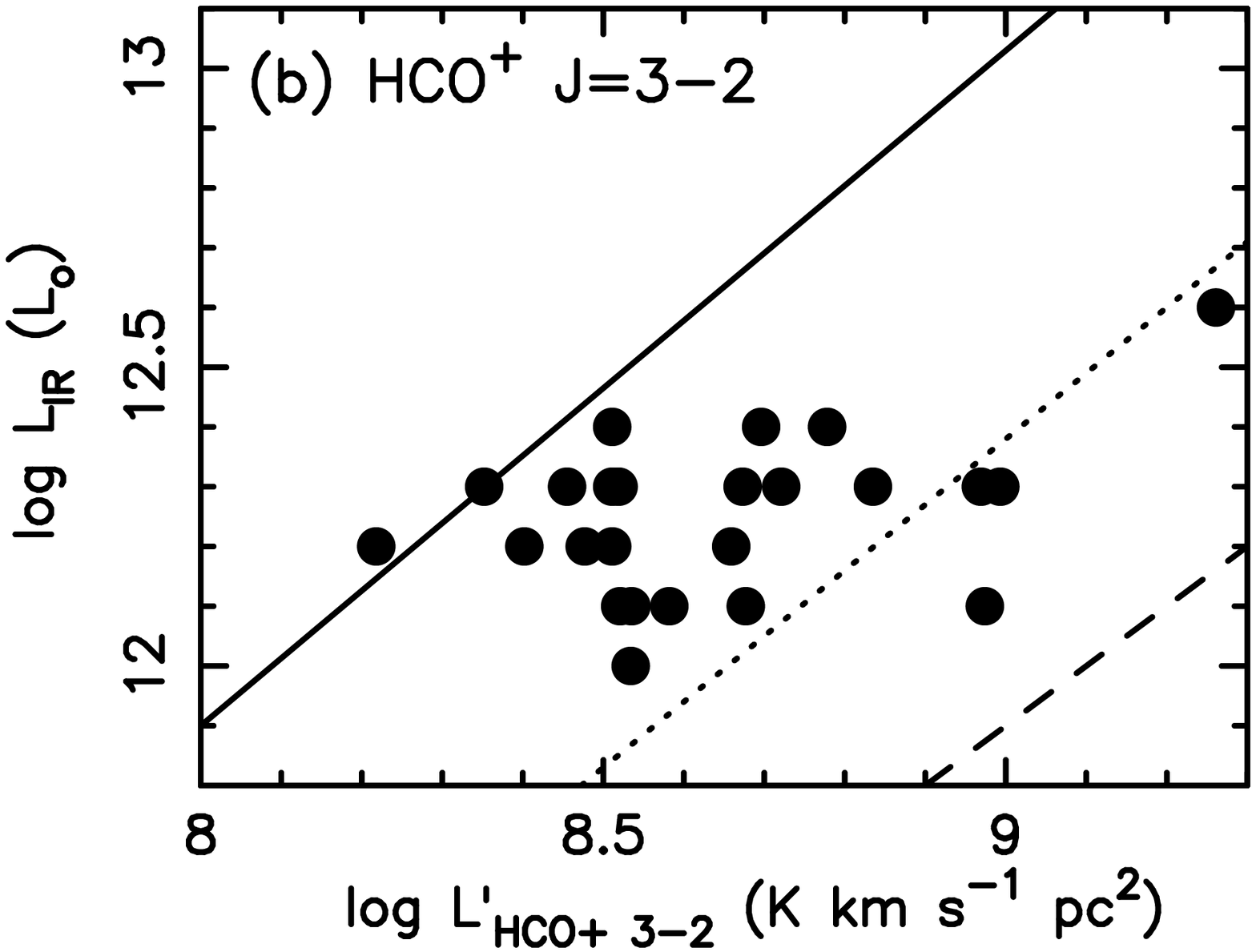} 
\caption{
Comparison of (a) HCN J=3--2 and (b) HCO$^{+}$ J=3--2 emission line 
luminosities (in K km s$^{-1}$ pc$^{2}$) (abscissa), with infrared 
luminosity (in L$_{\odot}$) (ordinate). 
The molecular line luminosities are estimated based on the Gaussian 
fits in spatially integrated spectra within the 1--2$''$ diameter circular 
apertures (Table 9).
The solid straight lines in panels (a) and (b) are the best-fit lines for HCN
J=4--3 (log$L$$_{\rm IR}$ = 1.00log$L$$'$$_{\rm HCN(4–3)}$ + 3.80) and
HCO$^{+}$ J=4--3 (log$L$$_{\rm IR}$ = 1.13log$L$$'$$_{\rm HCO+(4–3)}$ + 2.83) 
for various types of galaxies \citep{tan18}. 
The dotted straight lines in panels (a) and (b) are the best-fit lines for HCN
J=4--3 (log$L$$_{\rm IR}$ = 1.00log$L$$'$$_{\rm HCN(4–3)}$ + 3.67) and
HCO$^{+}$ J=4--3 (log$L$$_{\rm IR}$ = 1.10log$L$$'$$_{\rm HCO+(4–3)}$ + 2.48) 
by \citet{zha14}. 
The dashed straight line in panel (a) is the best-fit line for HCN
J=1--0 (log$L$$_{\rm IR}$ = 1.00log$L$$'$$_{\rm HCN(1–0)}$ + 2.90) 
\citep{gao04b}.
The same line is overplotted for HCO$^{+}$ J=1--0 in panel (b), because the 
conversion factor from molecular emission line luminosity to dense 
molecular gas mass is comparable for HCN J=1--0 and HCO$^{+}$ J=1--0 
\citep{ler15,ler17}.
}
\end{center}
\end{figure}

\begin{figure}
\begin{center}
\includegraphics[angle=0,scale=.3]{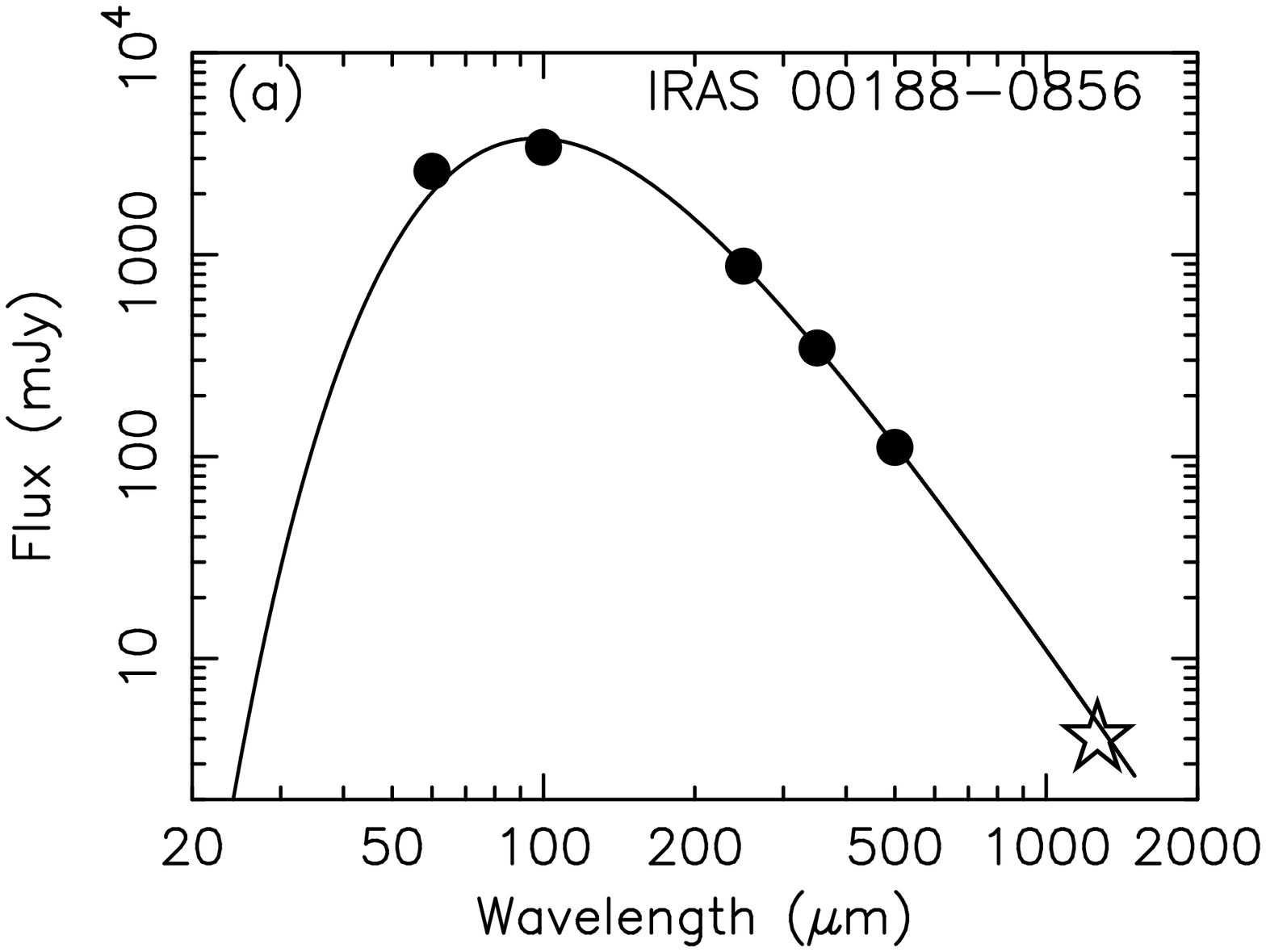}
\includegraphics[angle=0,scale=.3]{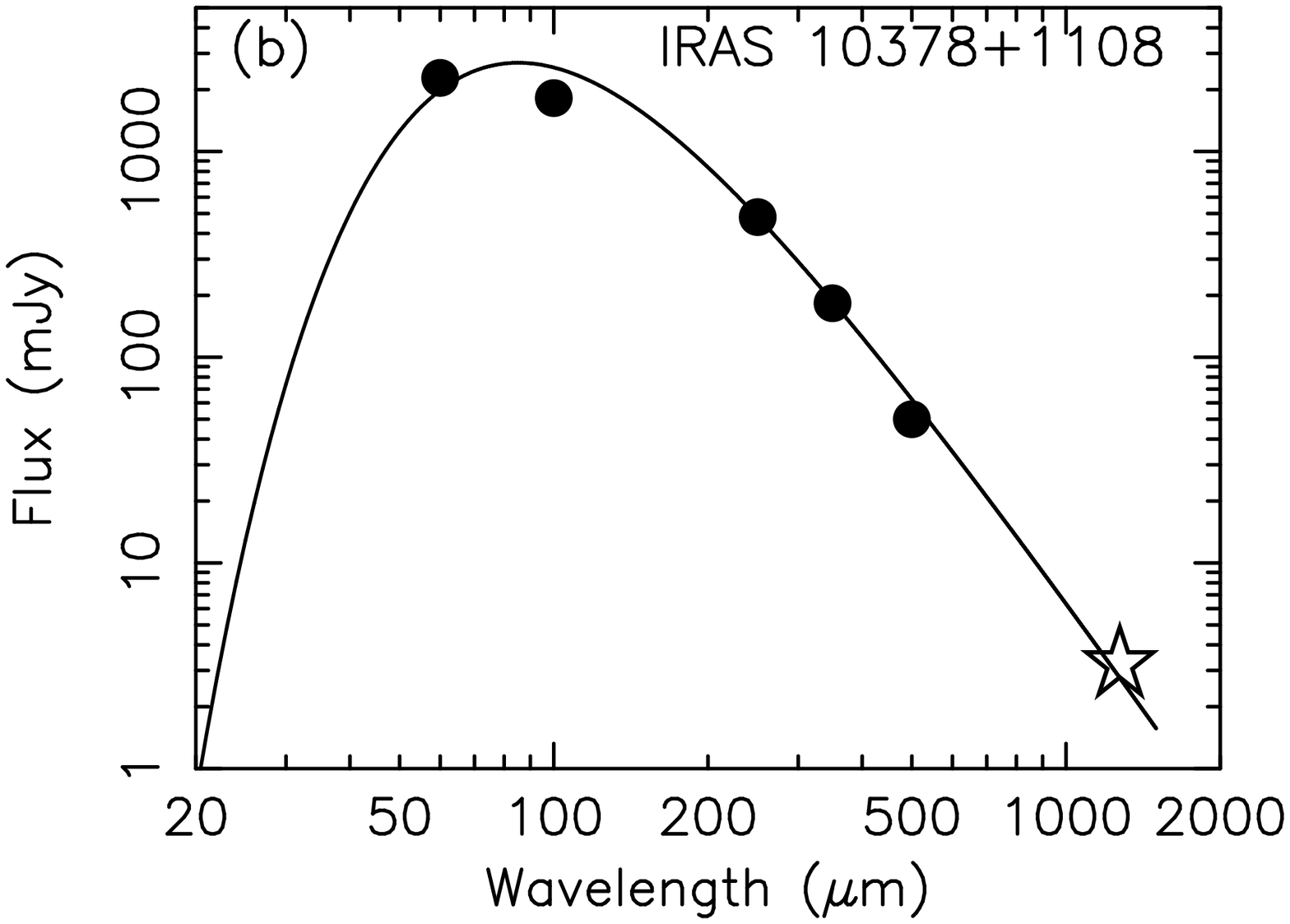}
\includegraphics[angle=0,scale=.3]{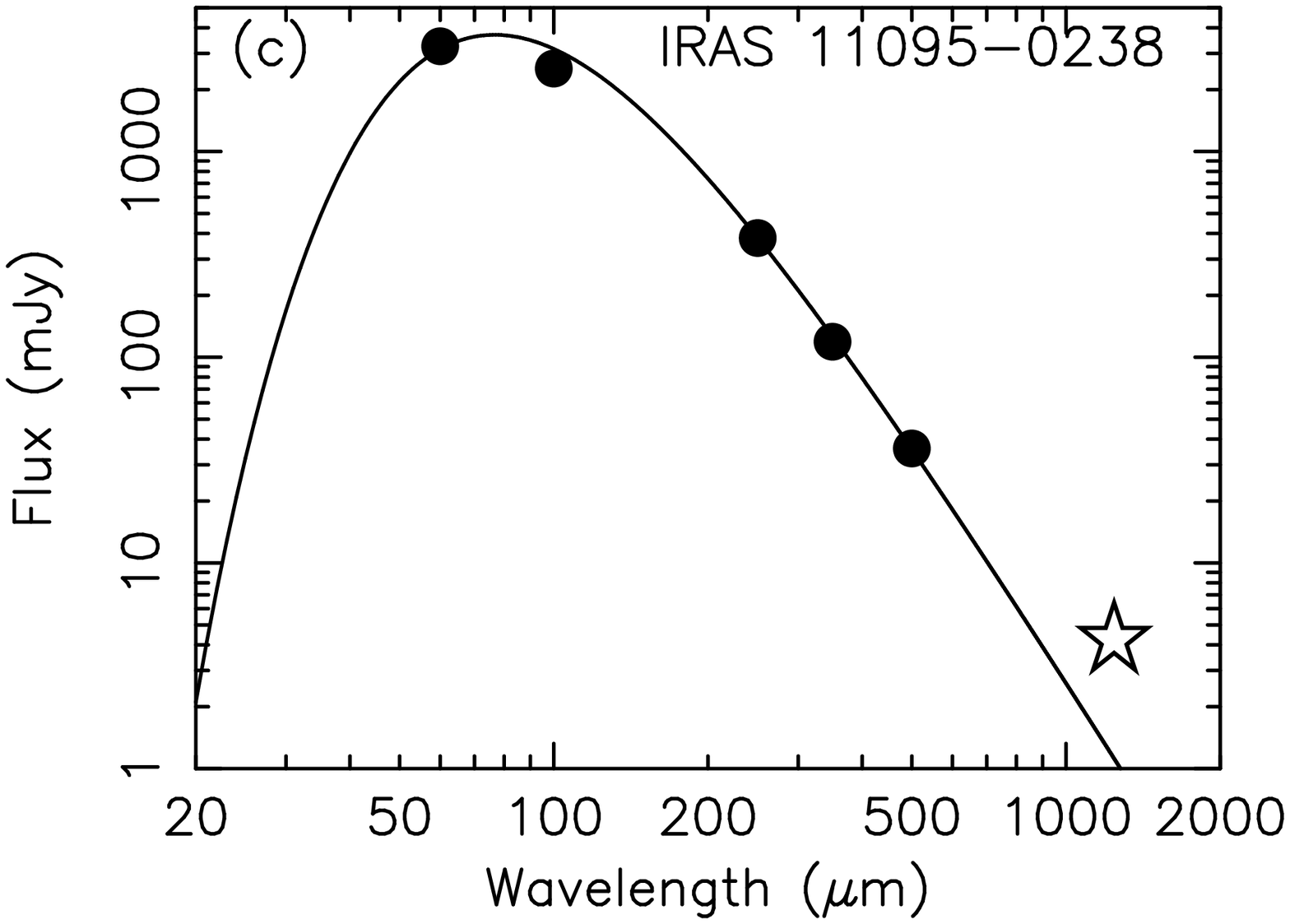} \\
\includegraphics[angle=0,scale=.3]{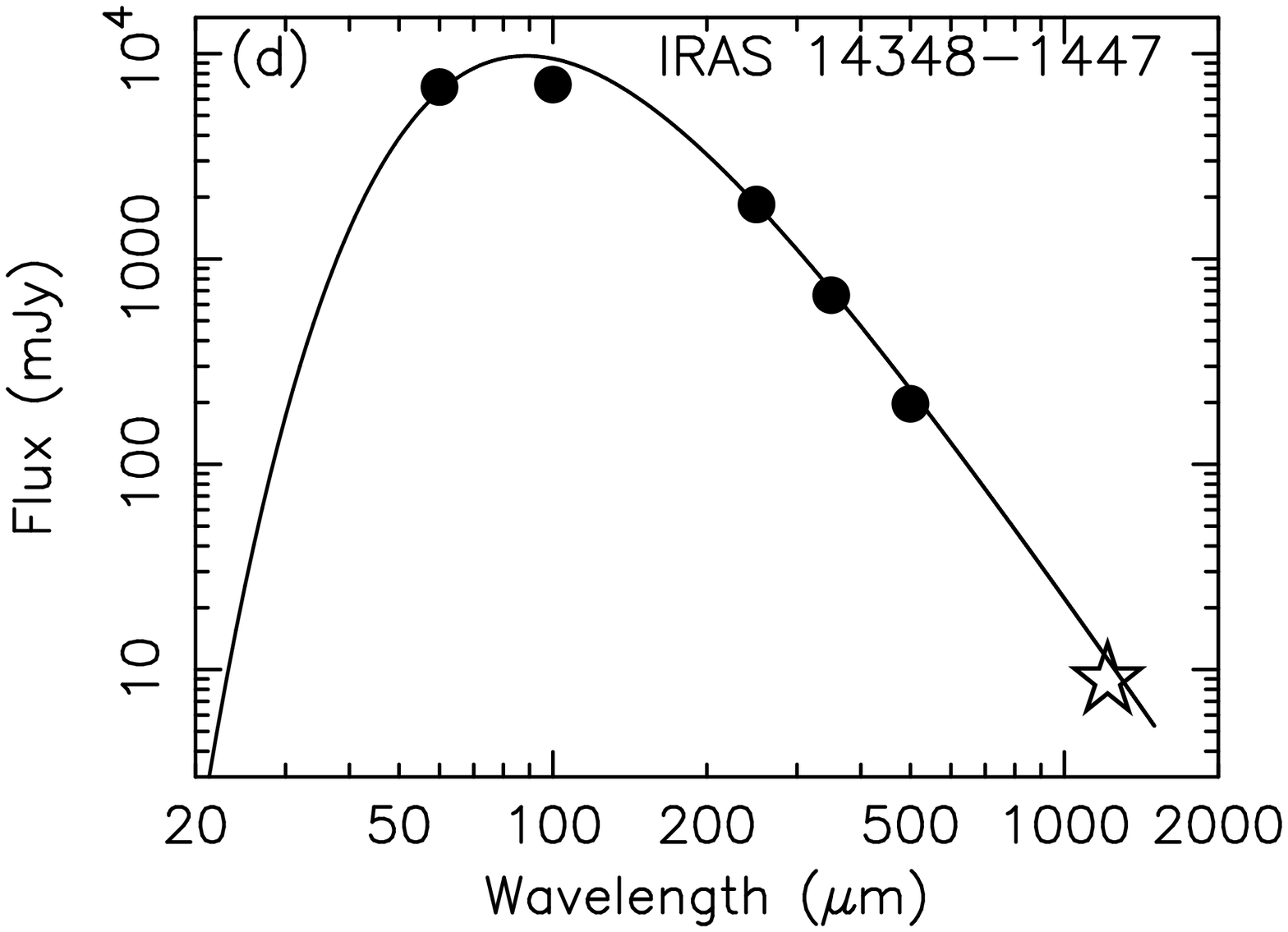}
\includegraphics[angle=0,scale=.3]{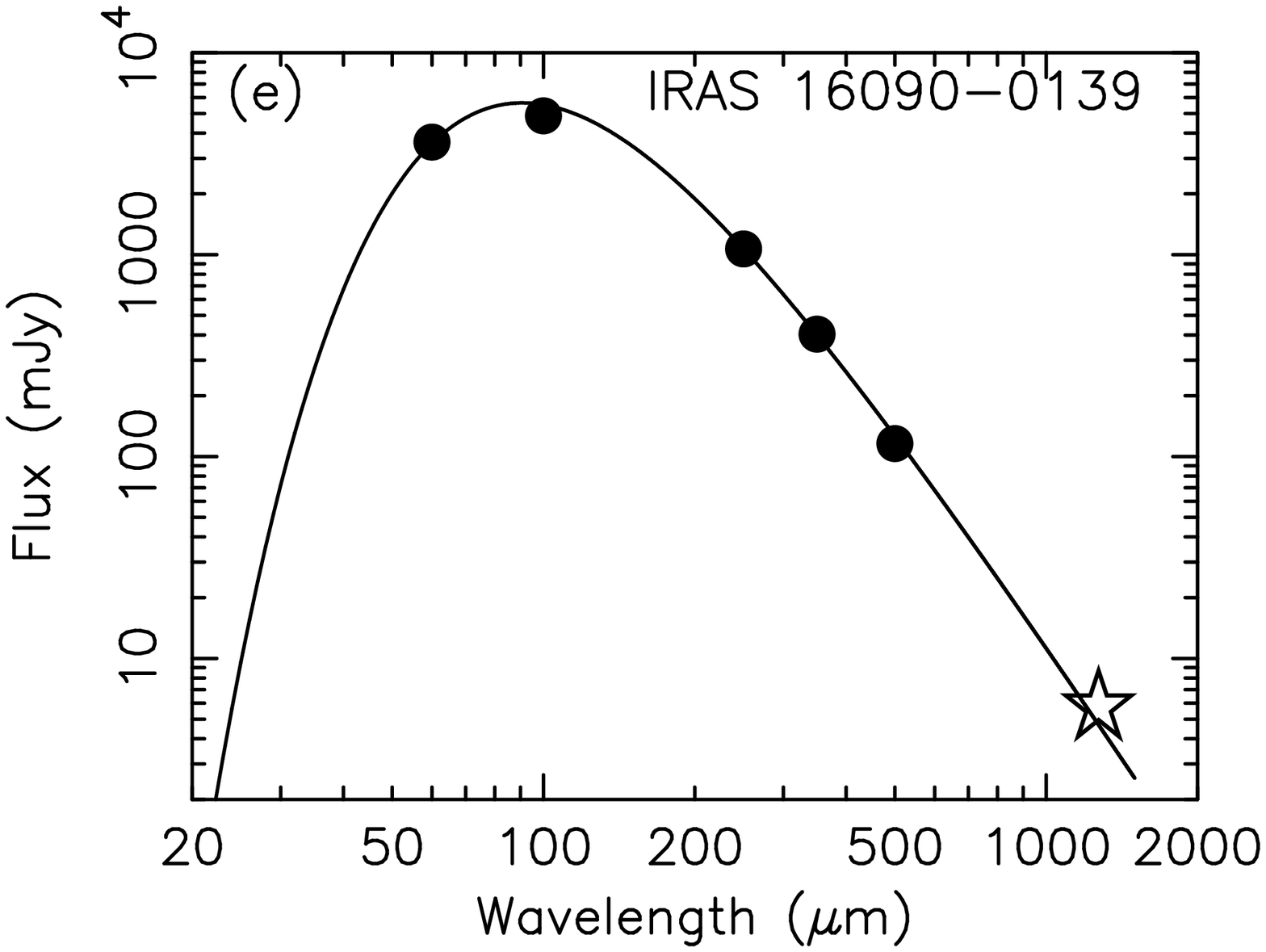}
\includegraphics[angle=0,scale=.3]{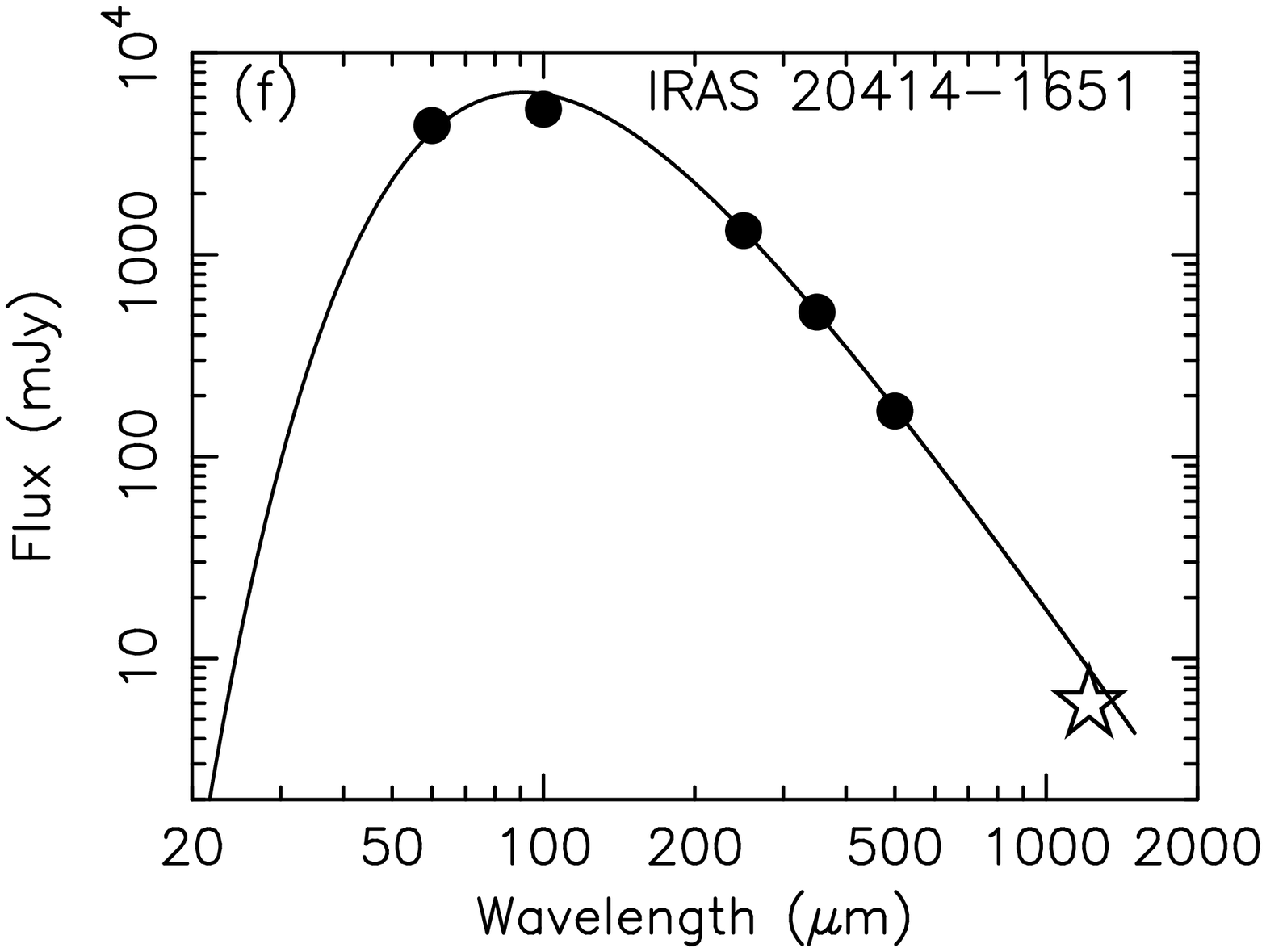} \\
\includegraphics[angle=0,scale=.3]{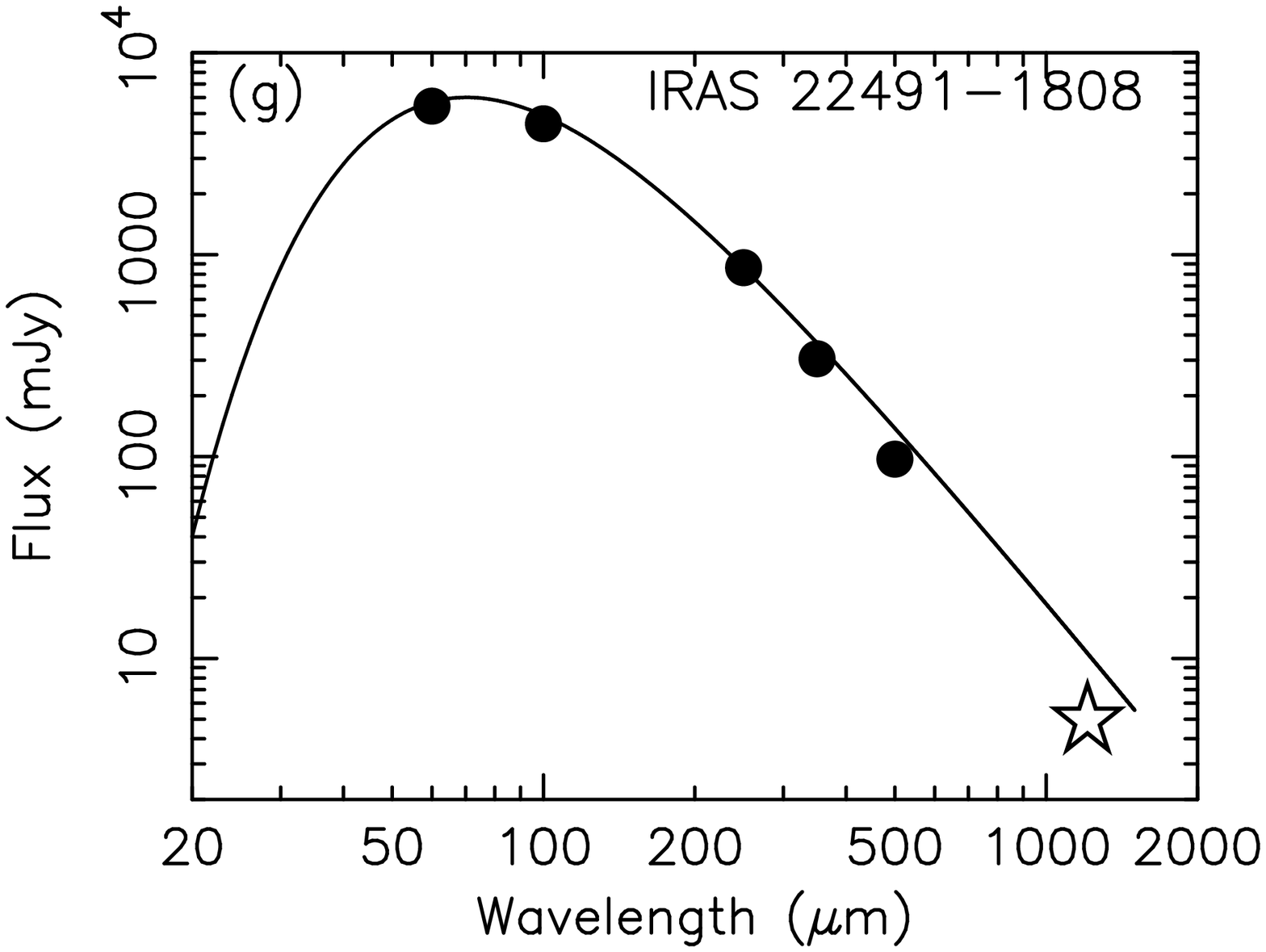}
\caption{
Spectral energy distribution (SED) at 20--2000 $\mu$m for selected 
ULIRGs for which Herschel photometry at 250 $\mu$m, 350 $\mu$m, and 
500 $\mu$m are available \citep{cle18}, in addition to the IRAS 
60 $\mu$m and 100 $\mu$m photometric measurements (Table 1).
These data are displayed as filled circles. 
The data points at $\sim$1250 $\mu$m (open stars) are from our ALMA 
measurements of spatially integrated continuum fluxes (Table 10).
The best-fit graybody curves by \citet{cle18} are overplotted as thick 
curved lines after normalization at the 250-$\mu$m flux.
}
\end{center}
\end{figure}

\begin{figure}
\begin{center}
\includegraphics[angle=0,scale=.223]{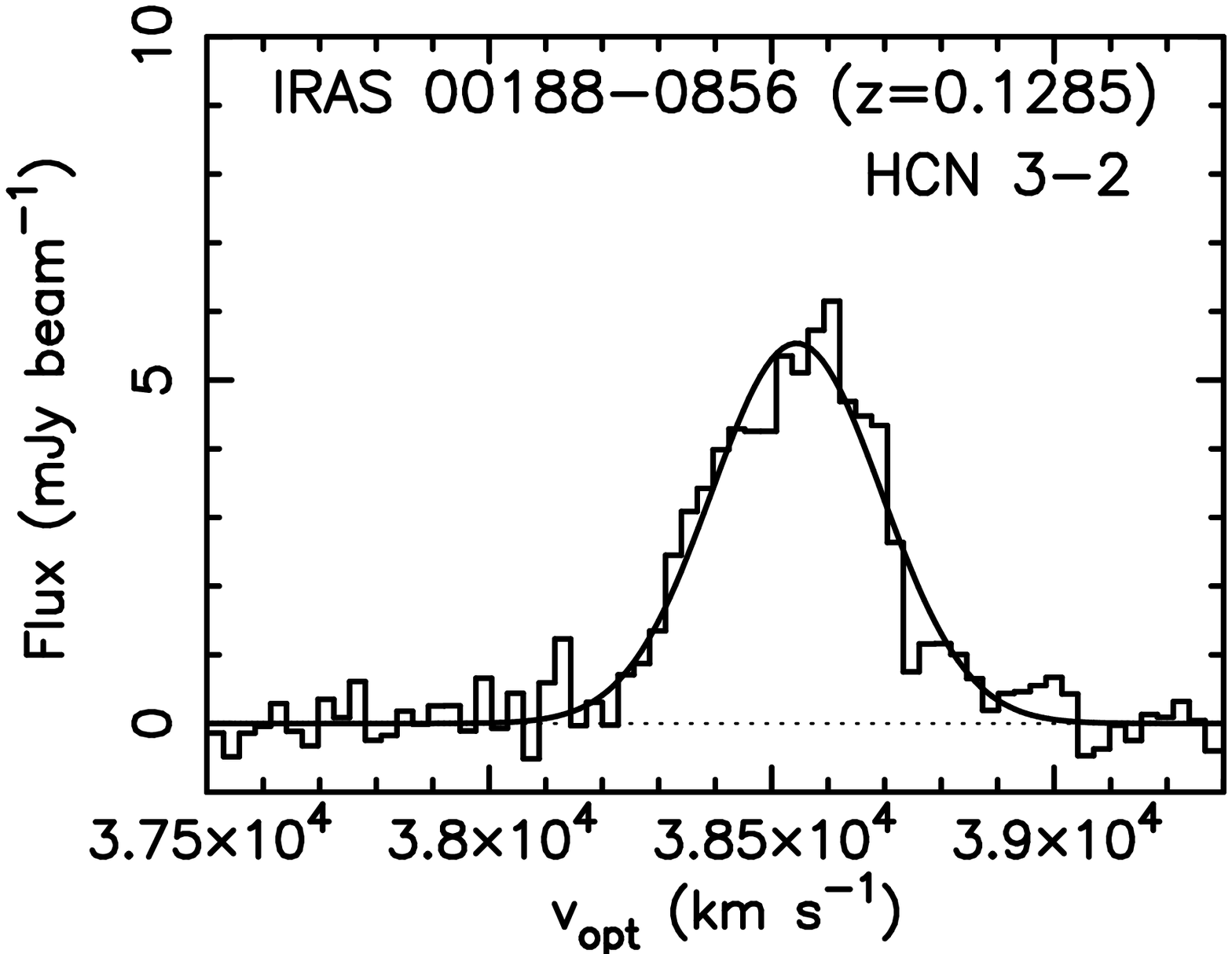} 
\includegraphics[angle=0,scale=.223]{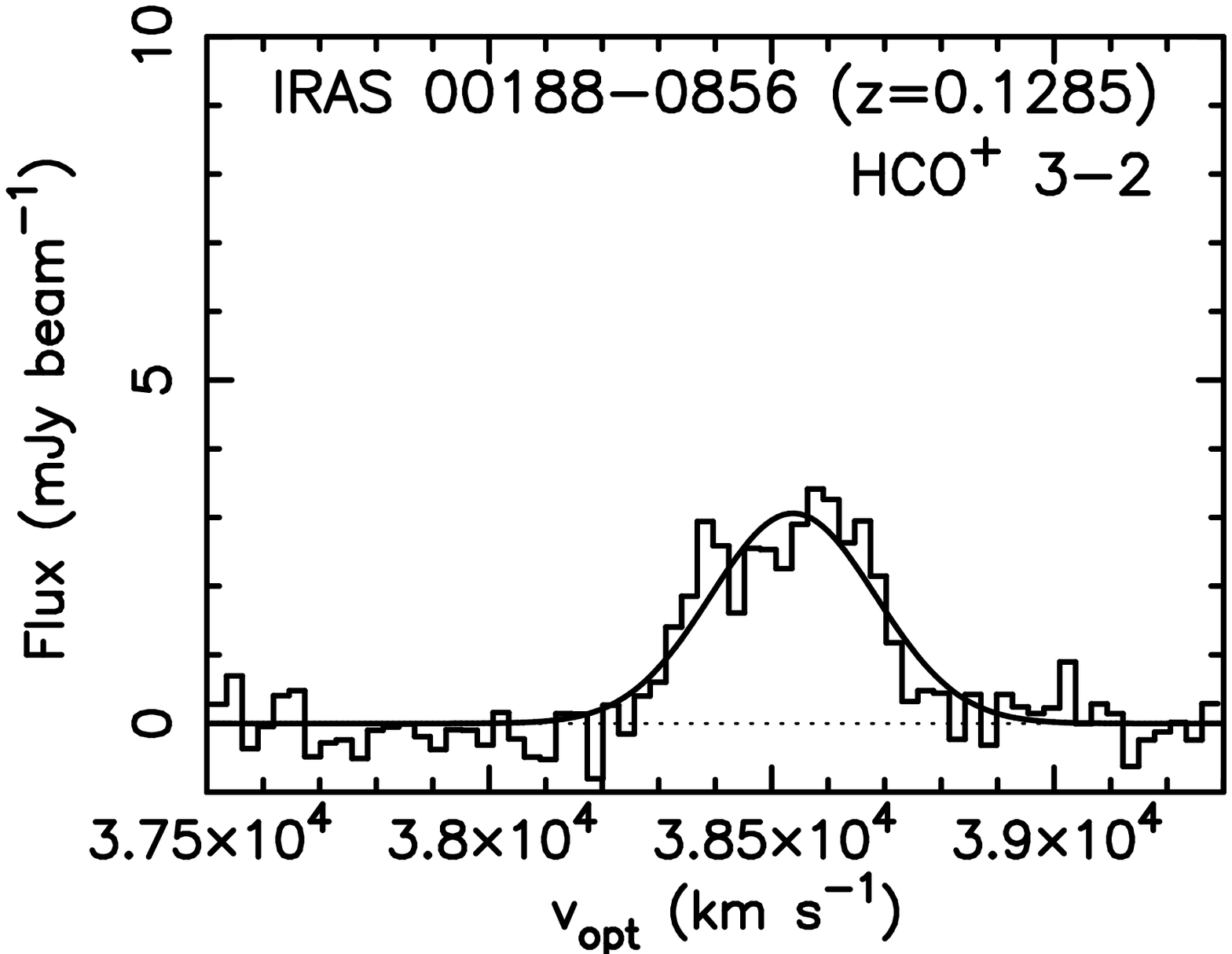} 
\includegraphics[angle=0,scale=.223]{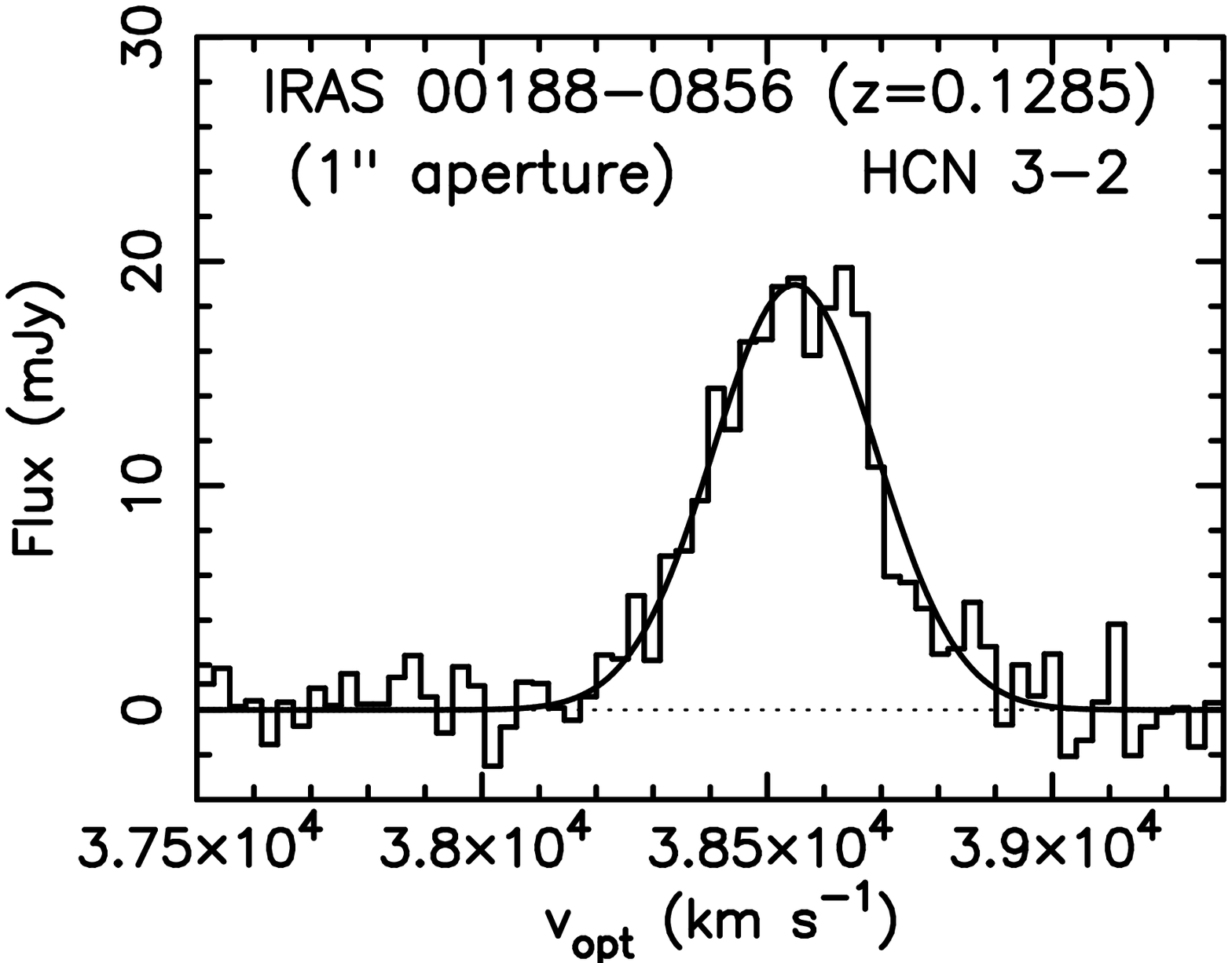} 
\includegraphics[angle=0,scale=.223]{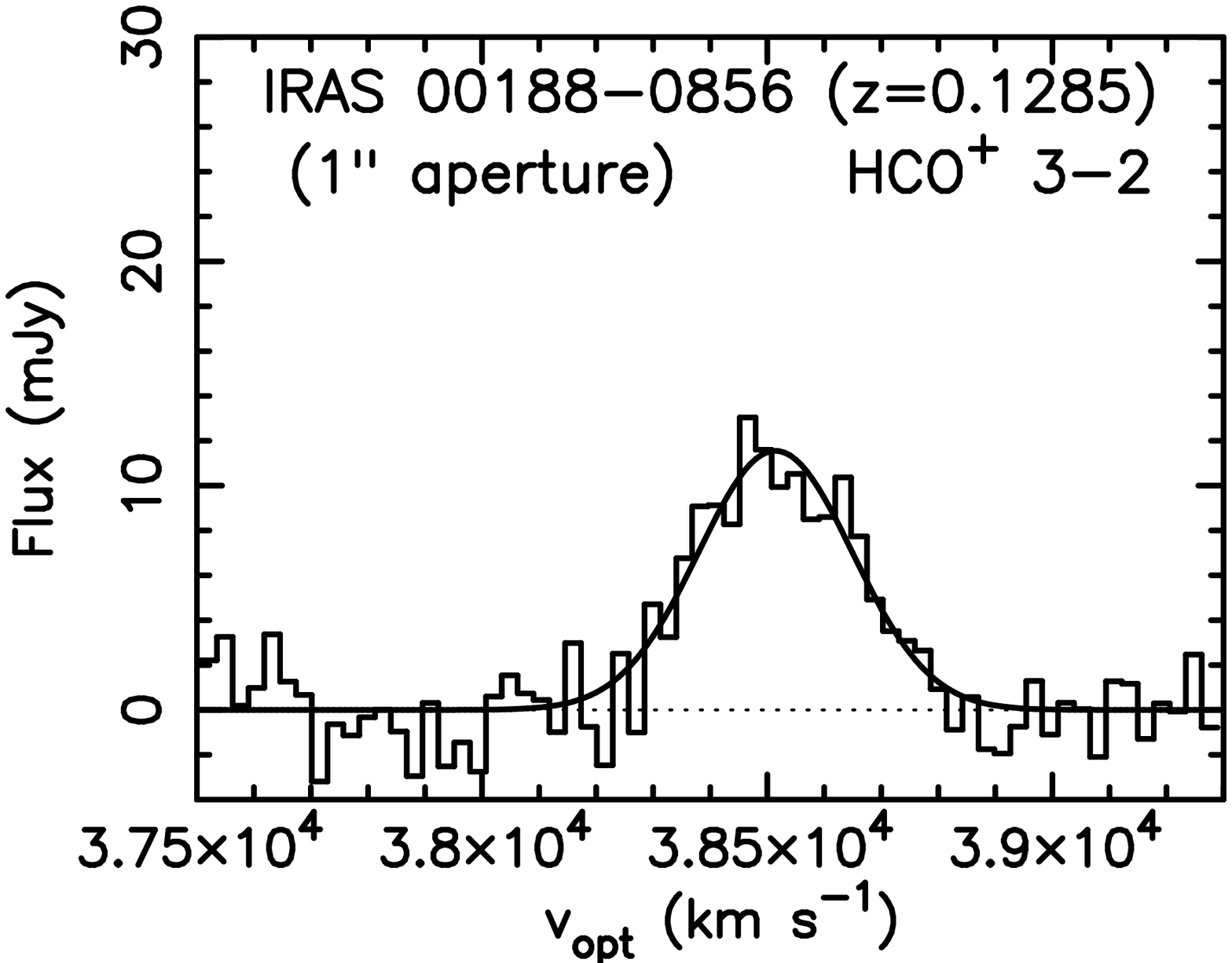} \\
\includegraphics[angle=0,scale=.223]{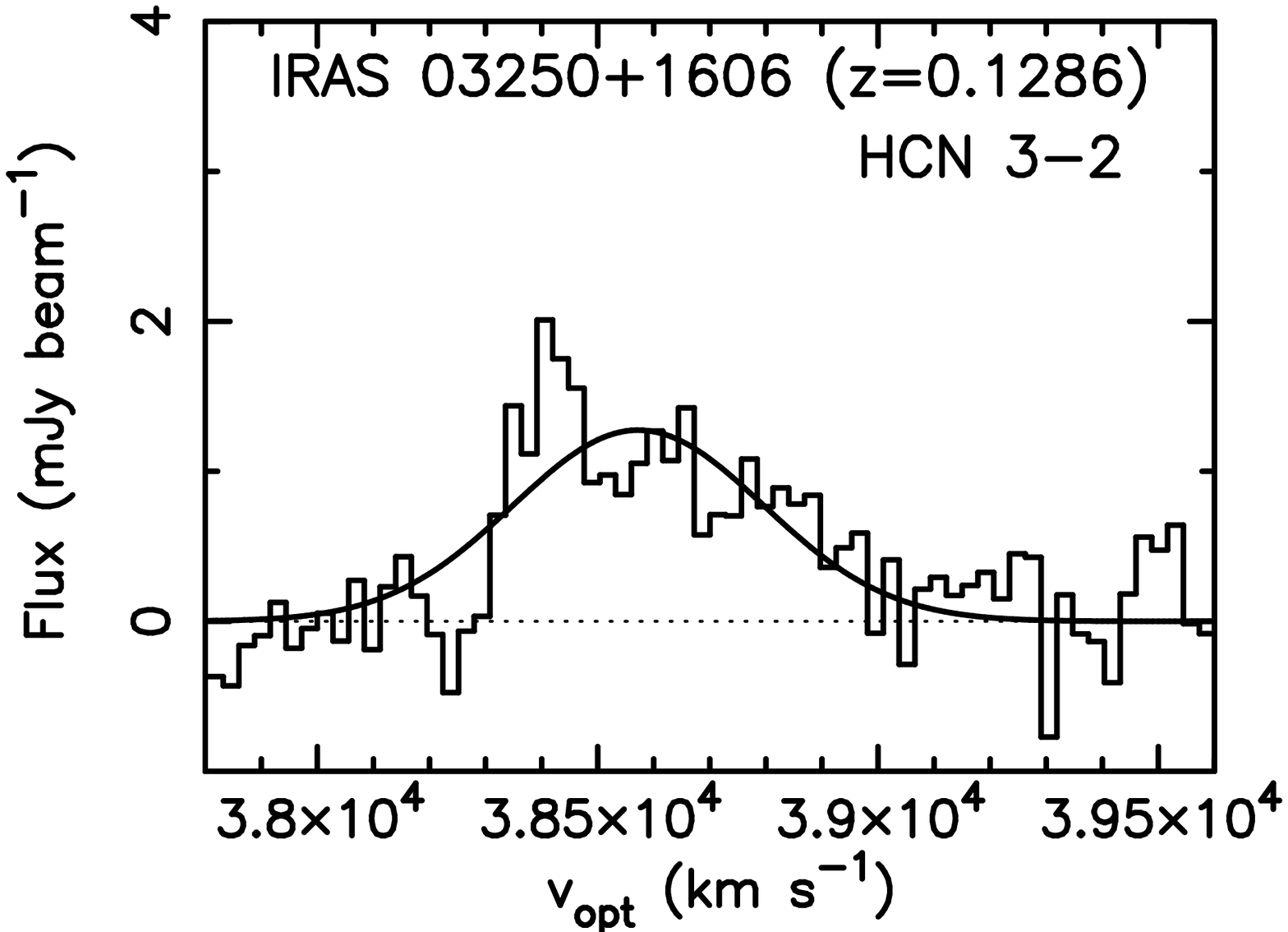} 
\includegraphics[angle=0,scale=.223]{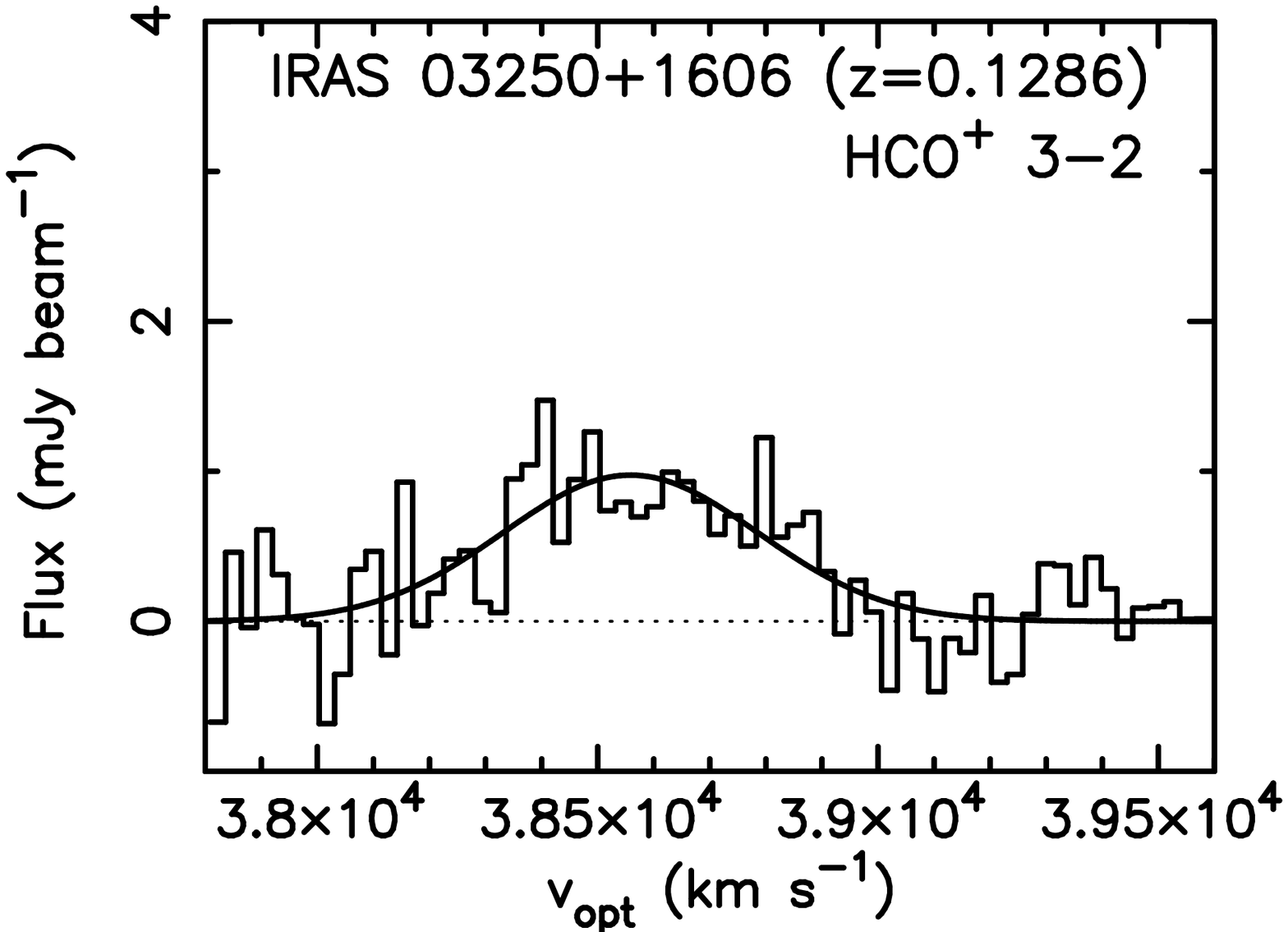} 
\includegraphics[angle=0,scale=.223]{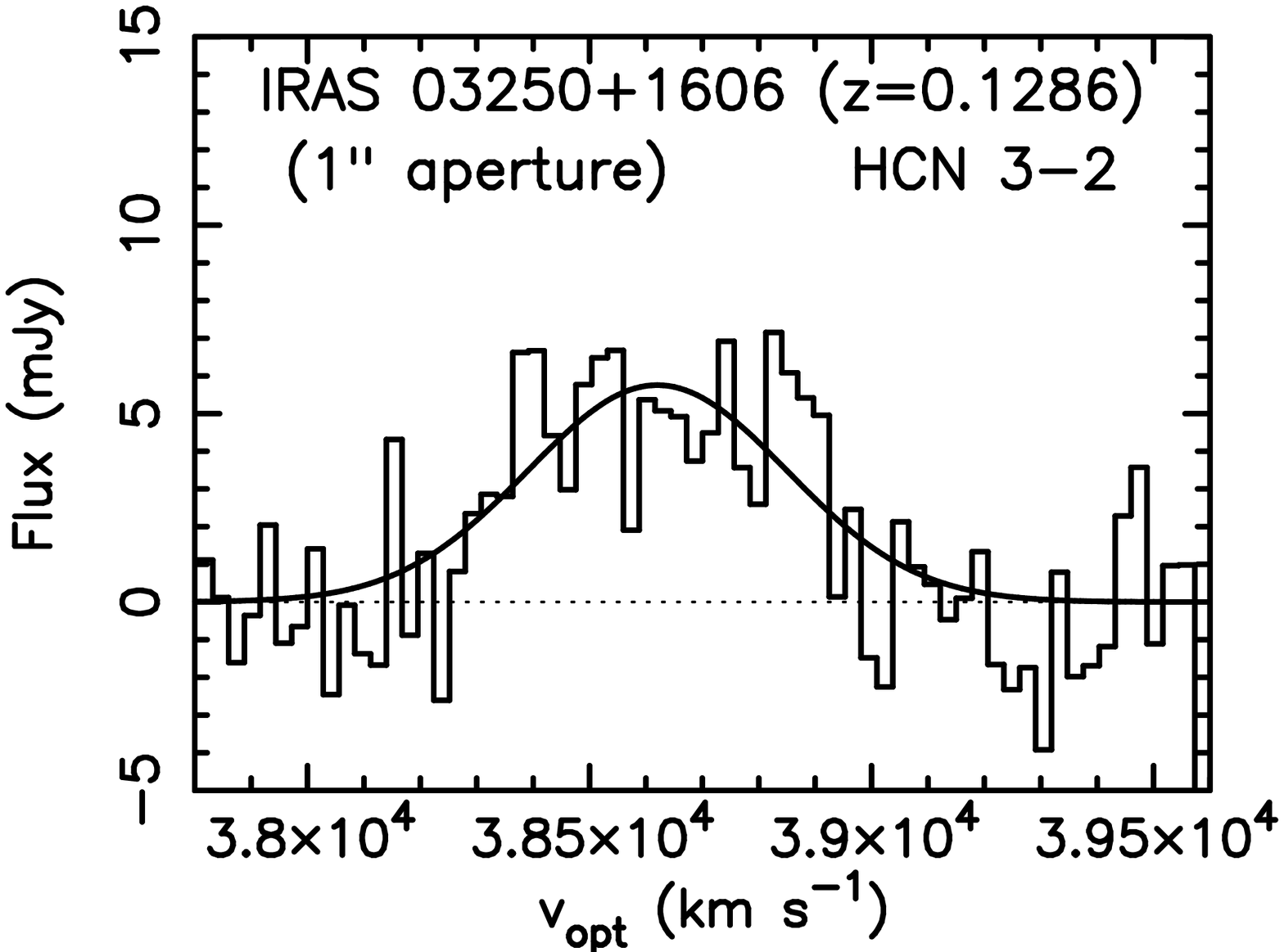} 
\includegraphics[angle=0,scale=.223]{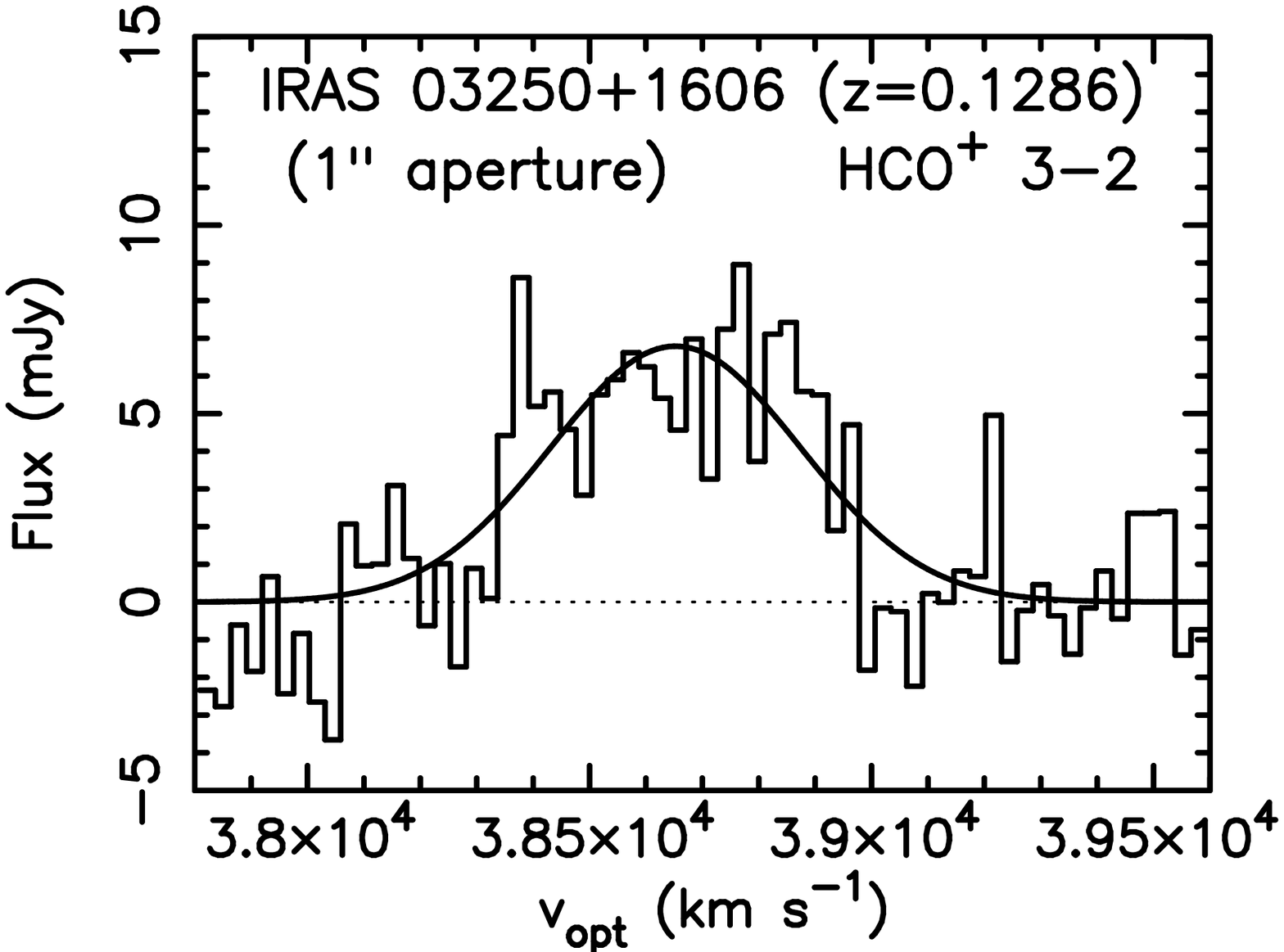} \\
\includegraphics[angle=0,scale=.223]{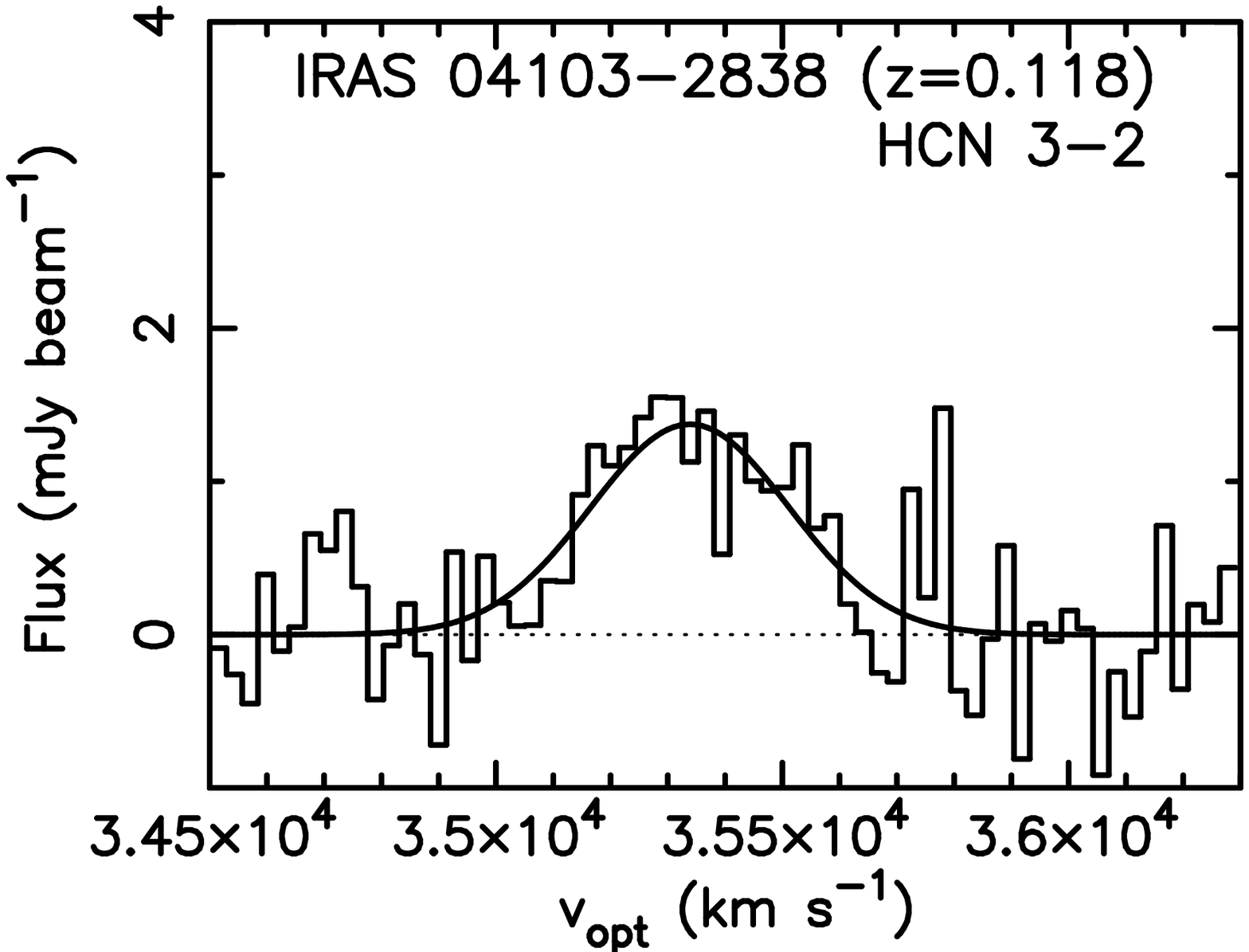} 
\includegraphics[angle=0,scale=.223]{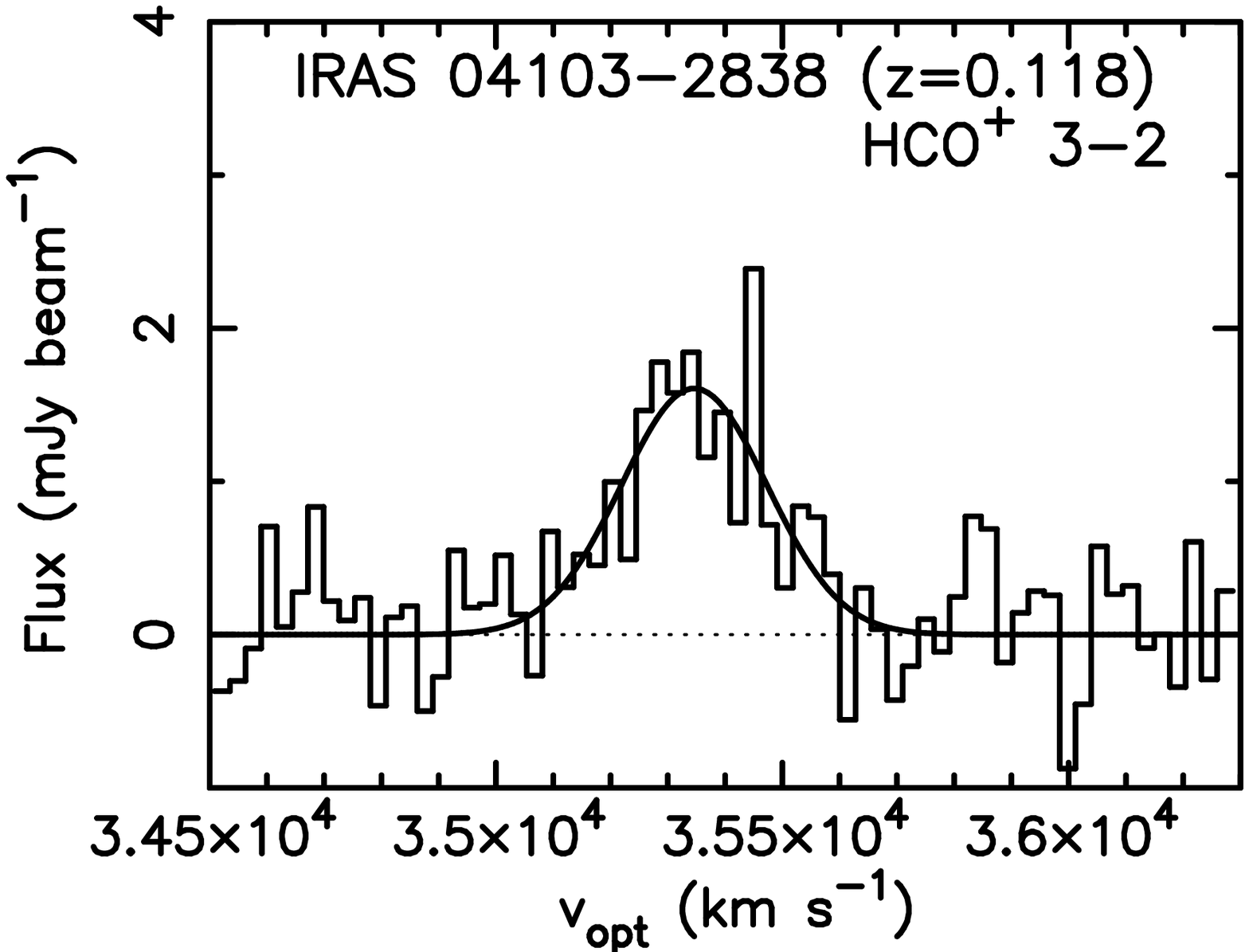} 
\includegraphics[angle=0,scale=.223]{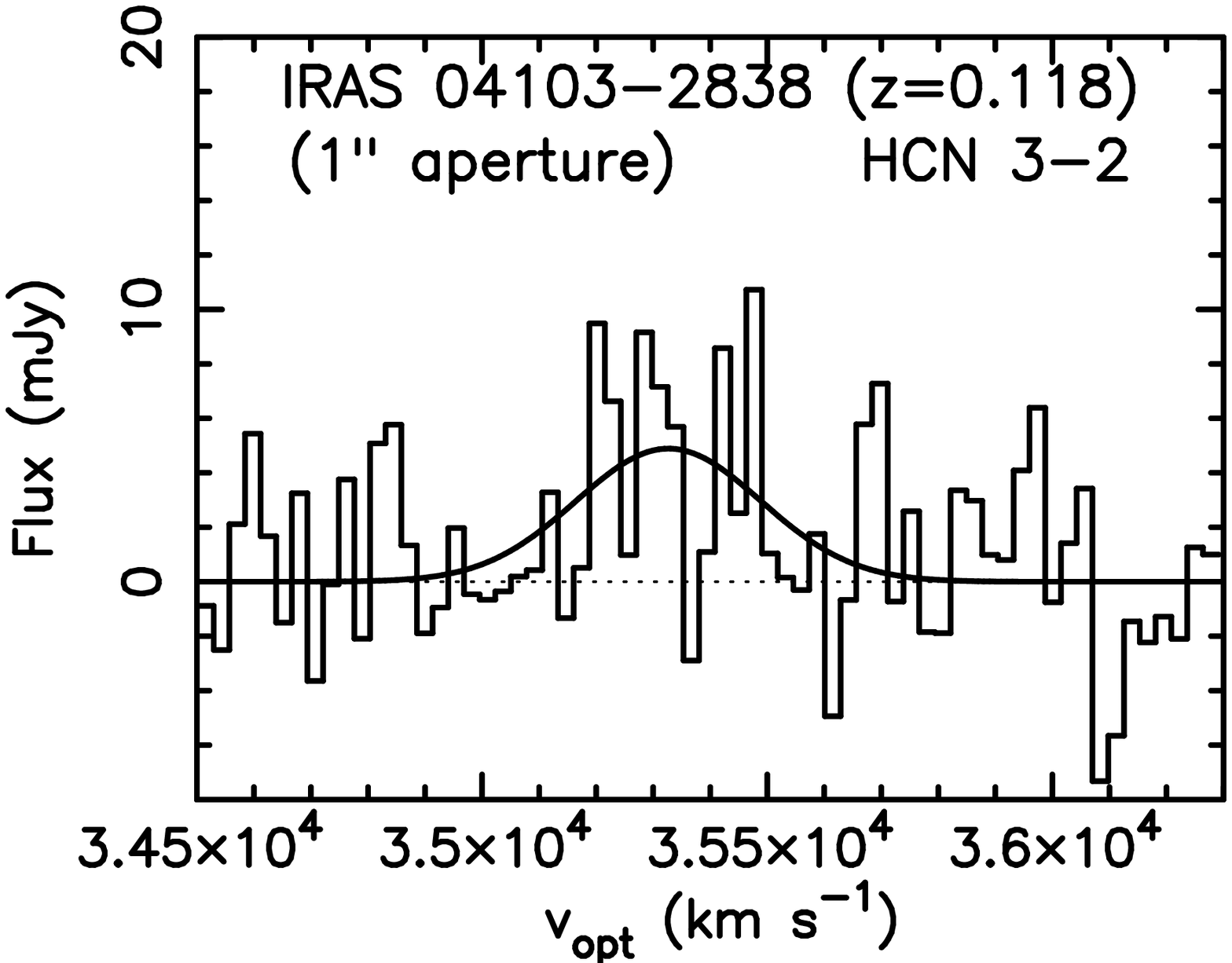} 
\includegraphics[angle=0,scale=.223]{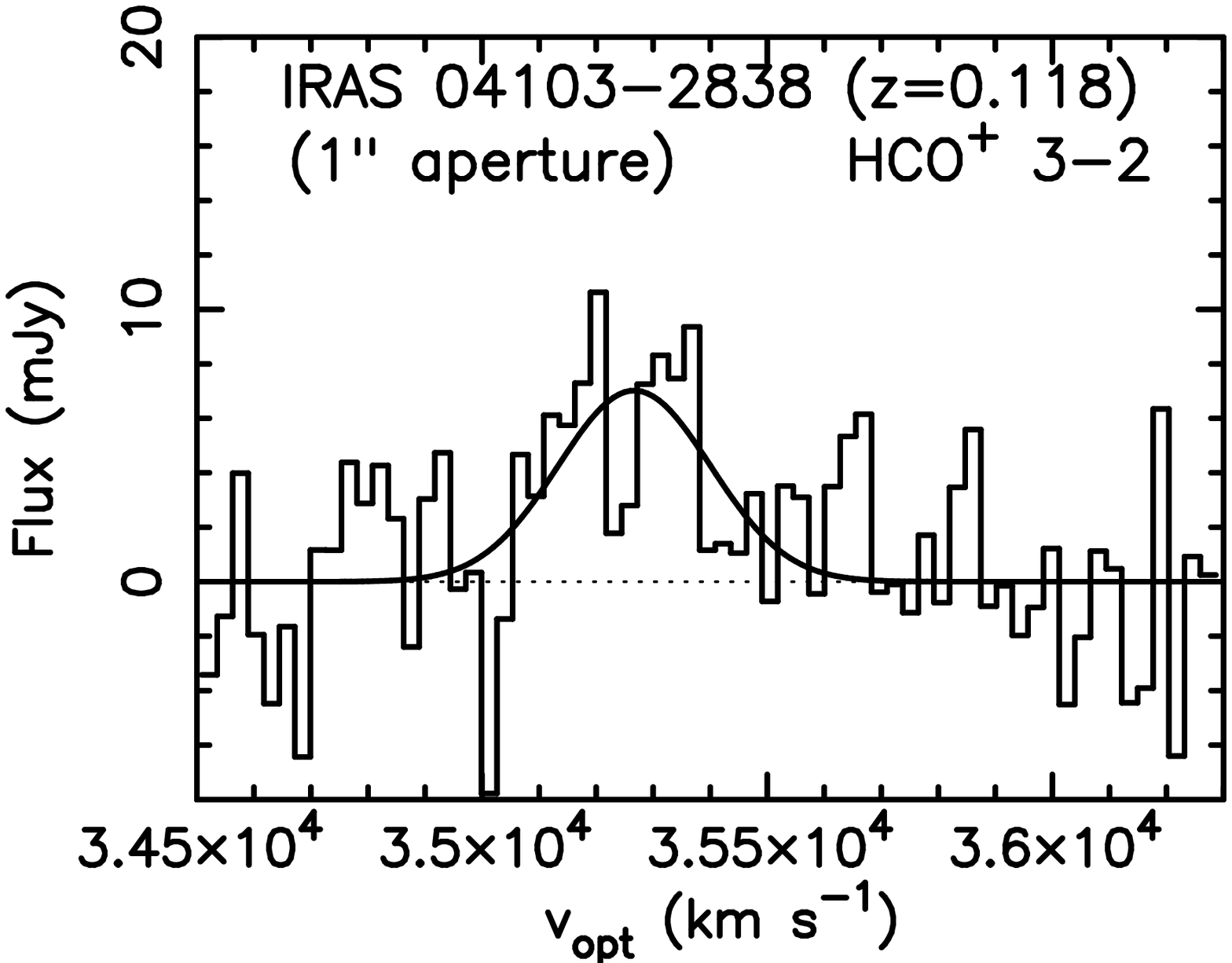} \\
\includegraphics[angle=0,scale=.223]{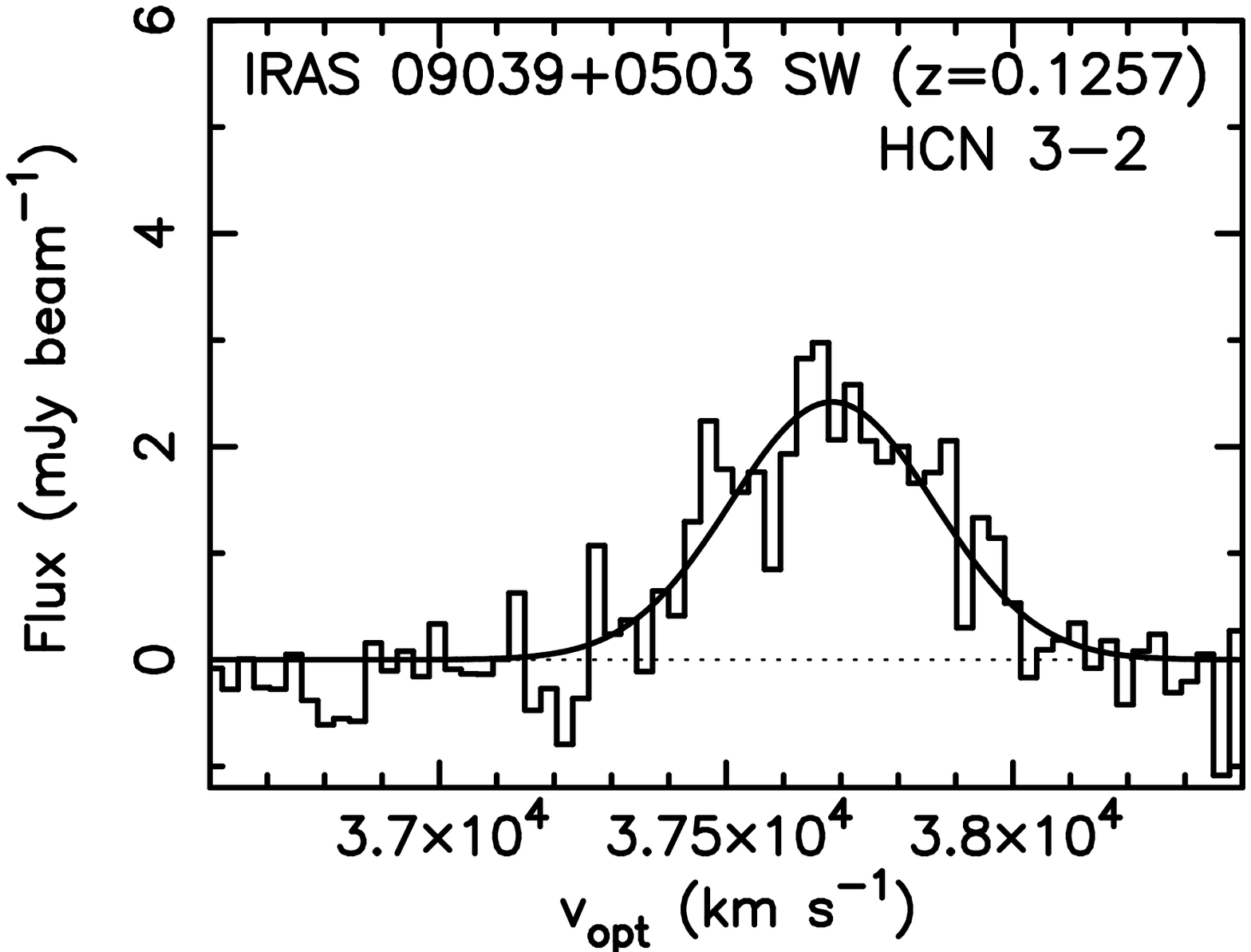} 
\includegraphics[angle=0,scale=.223]{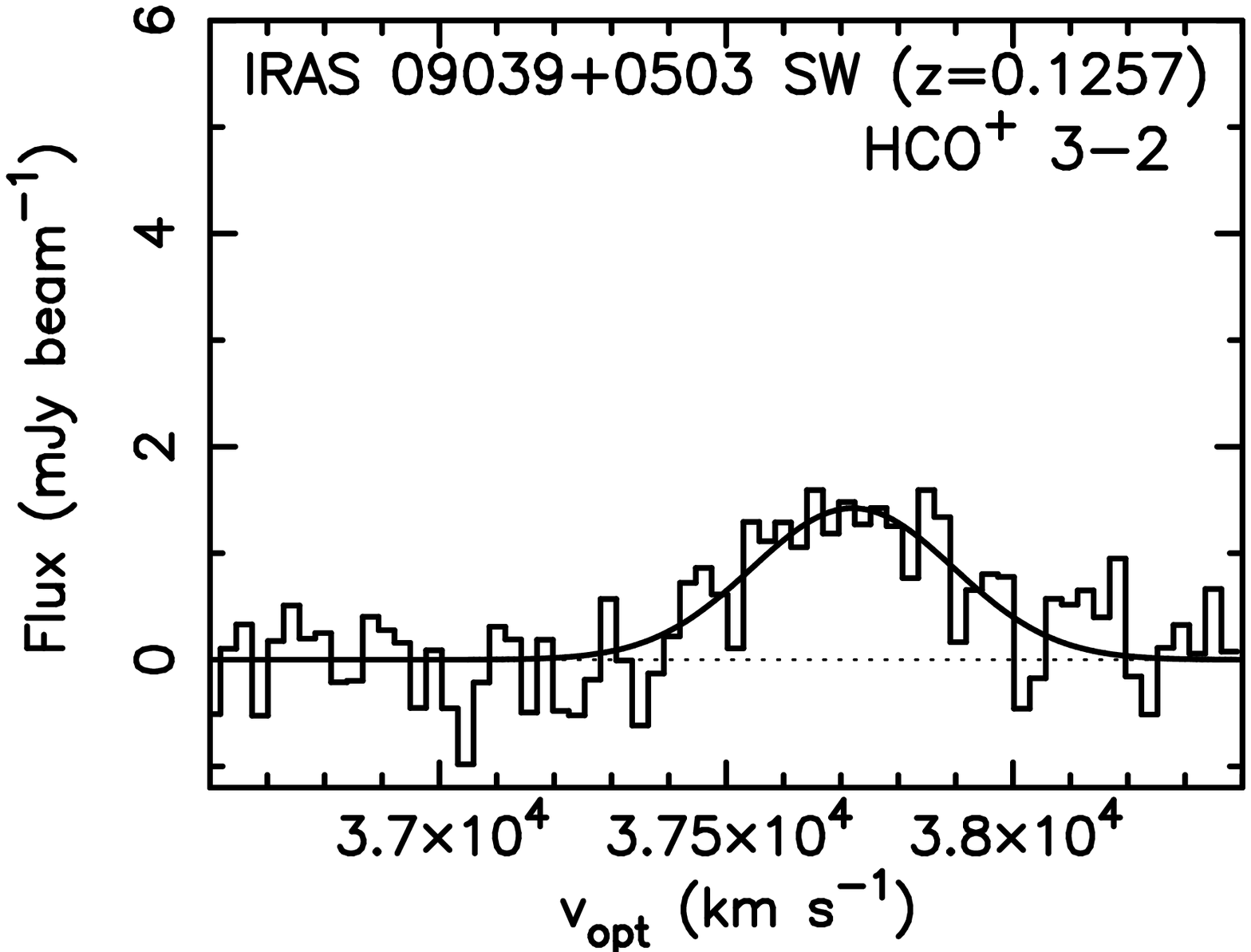} 
\includegraphics[angle=0,scale=.223]{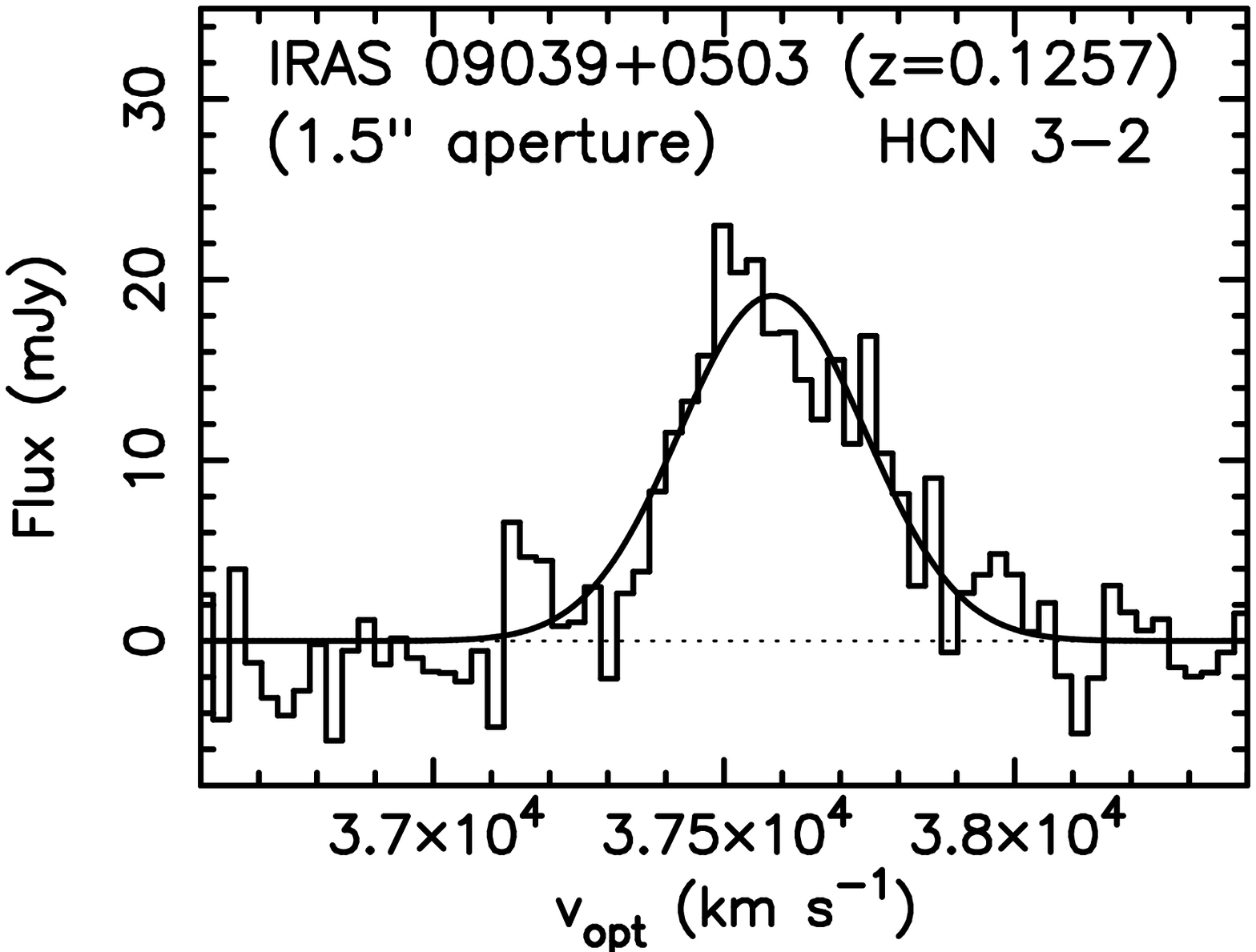} 
\includegraphics[angle=0,scale=.223]{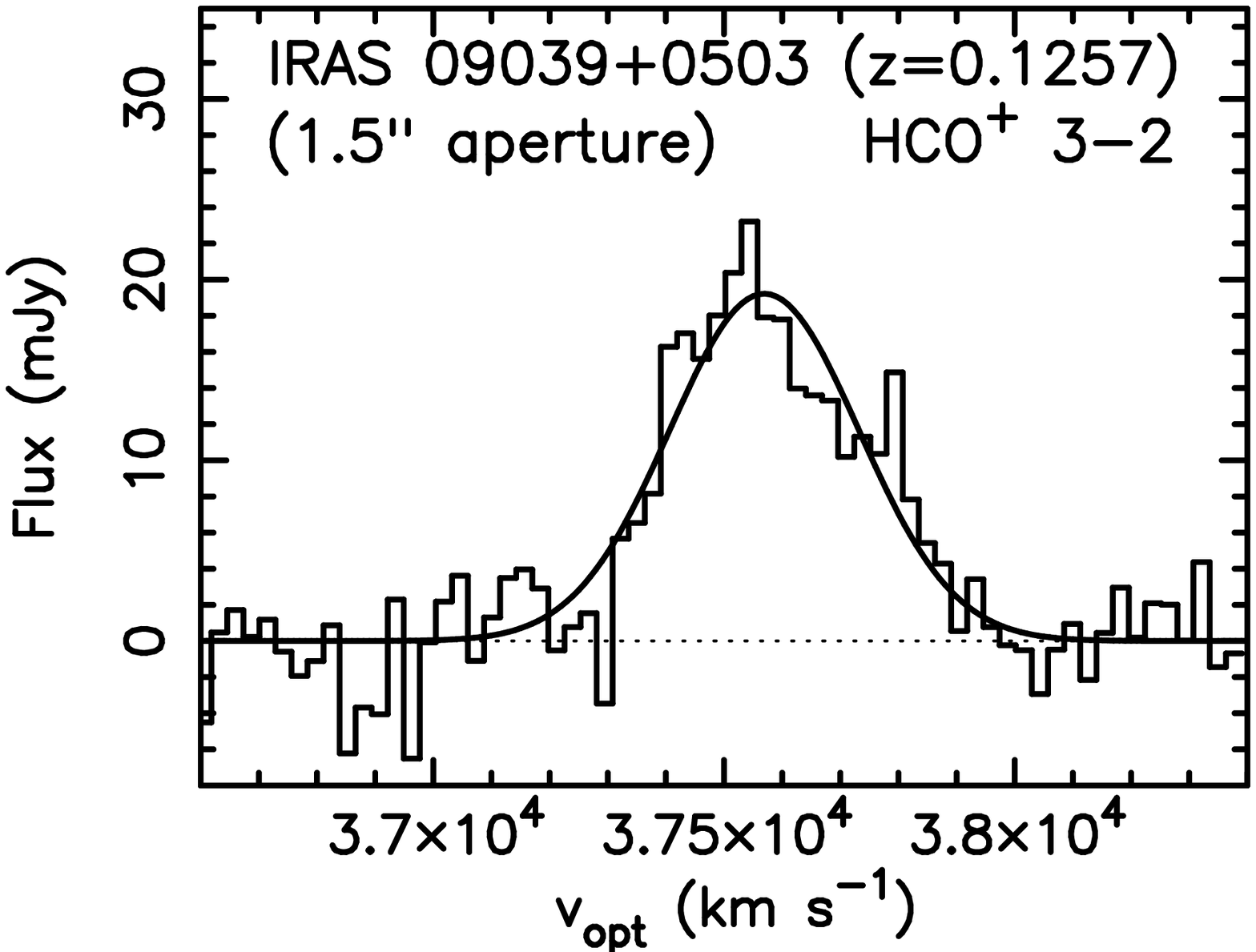} \\
\hspace*{-9.2cm}
\includegraphics[angle=0,scale=.223]{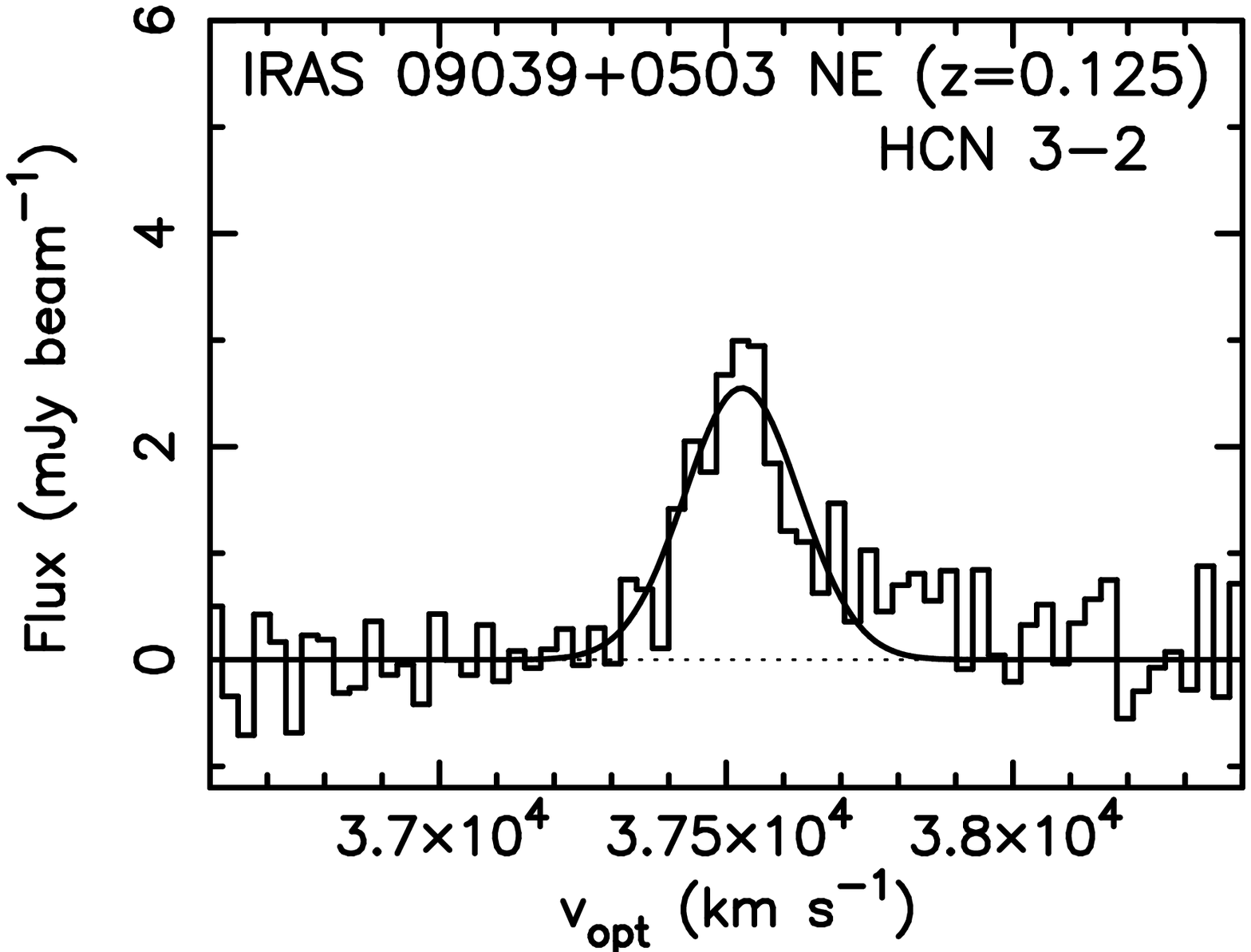} 
\includegraphics[angle=0,scale=.223]{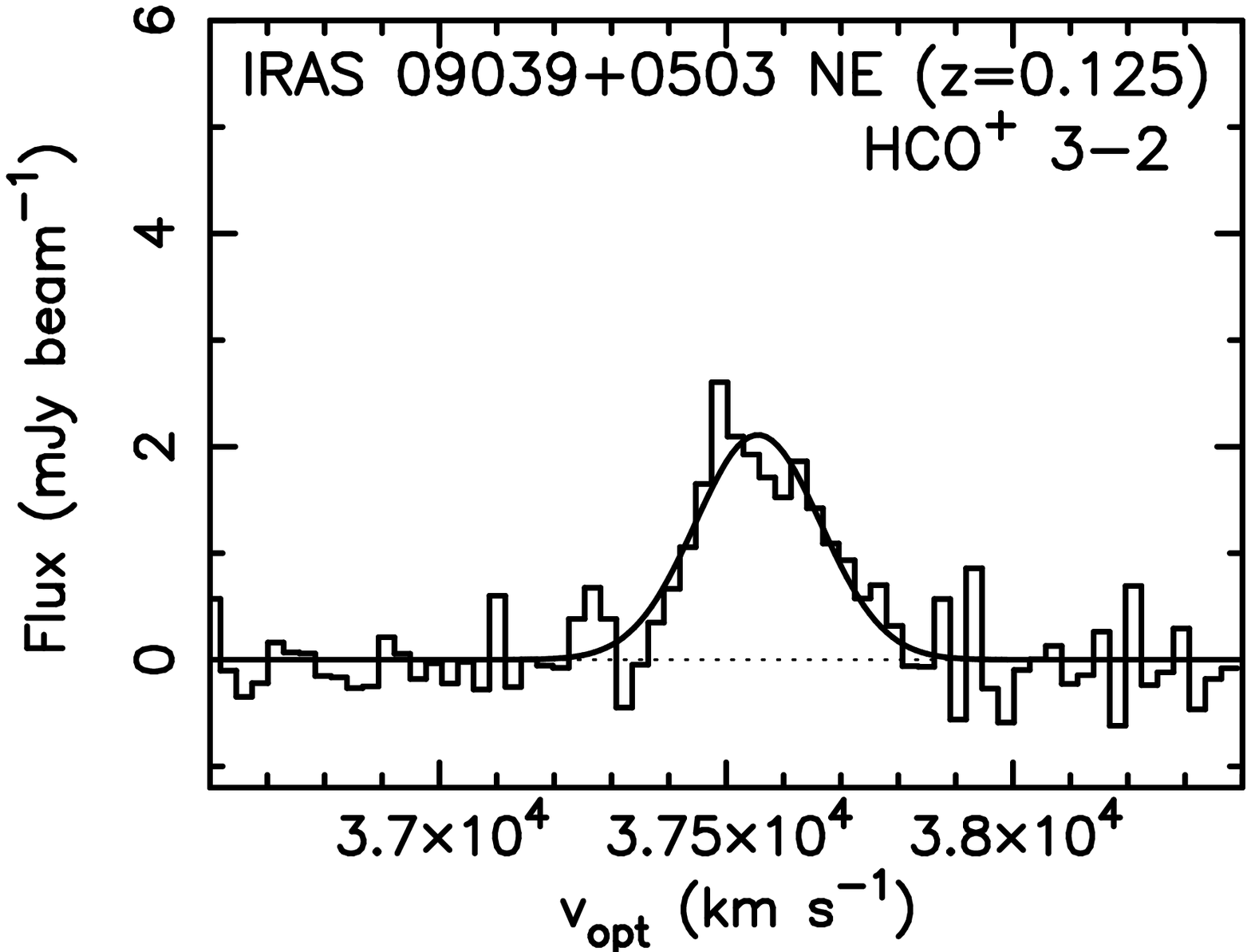} \\
\includegraphics[angle=0,scale=.223]{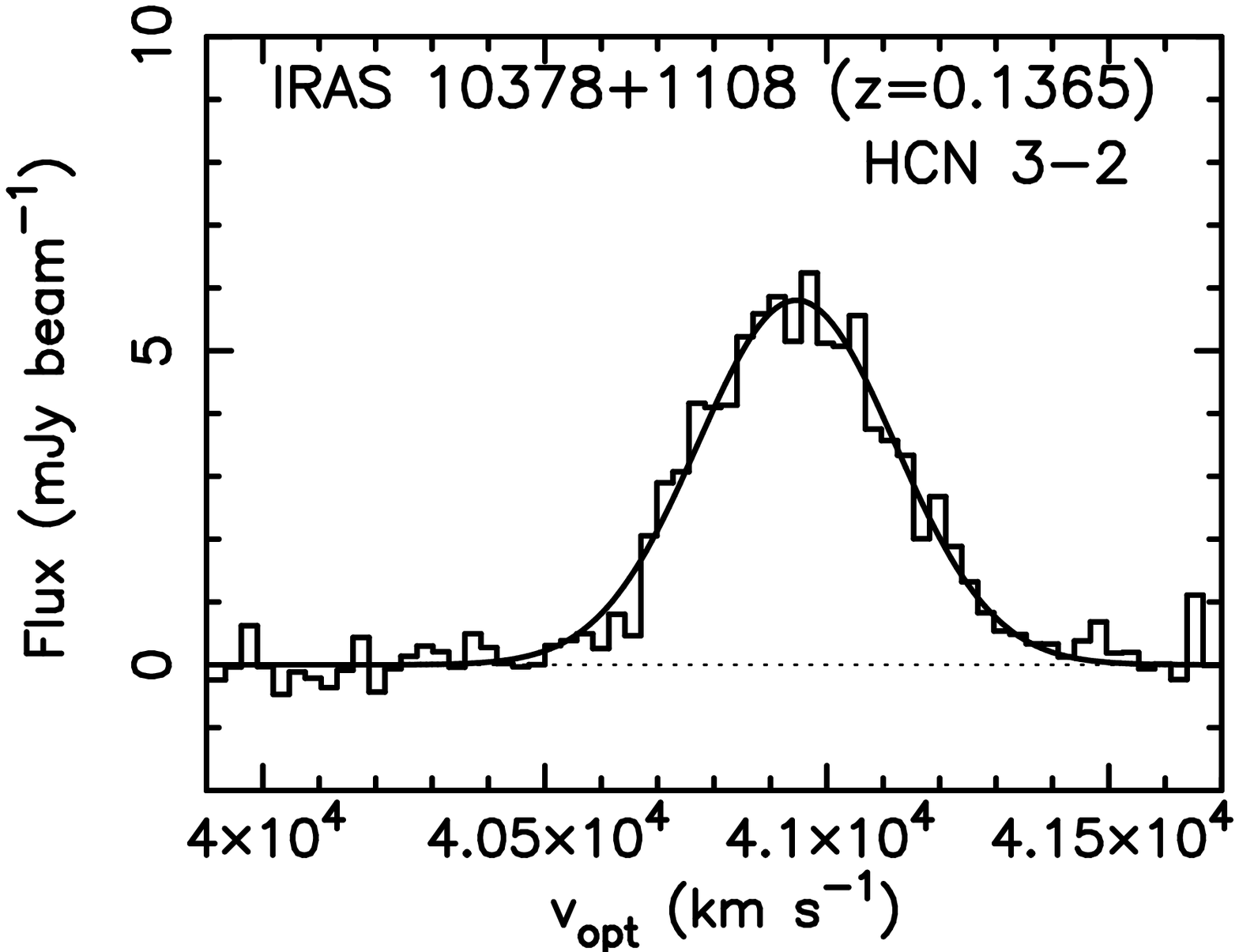} 
\includegraphics[angle=0,scale=.223]{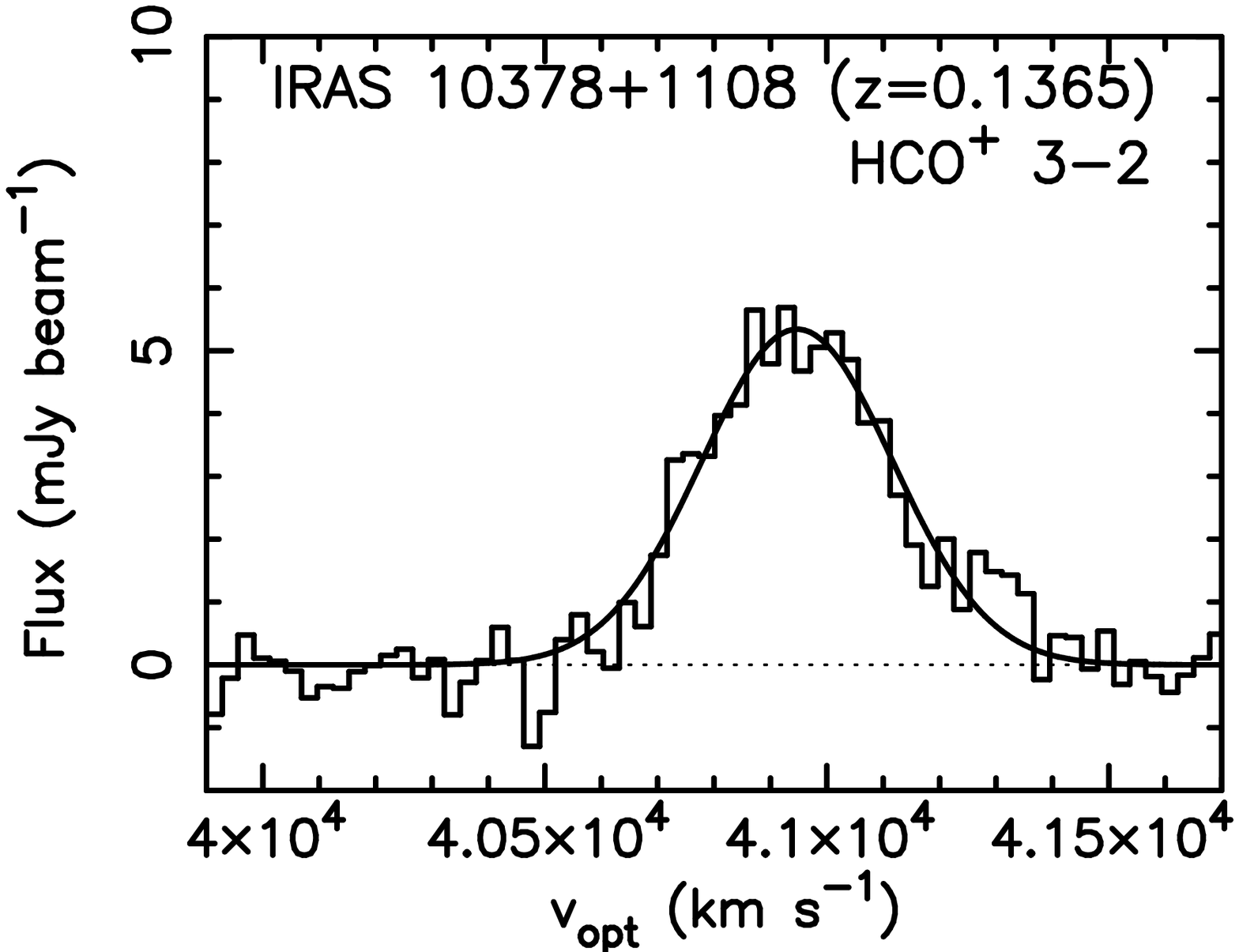} 
\includegraphics[angle=0,scale=.223]{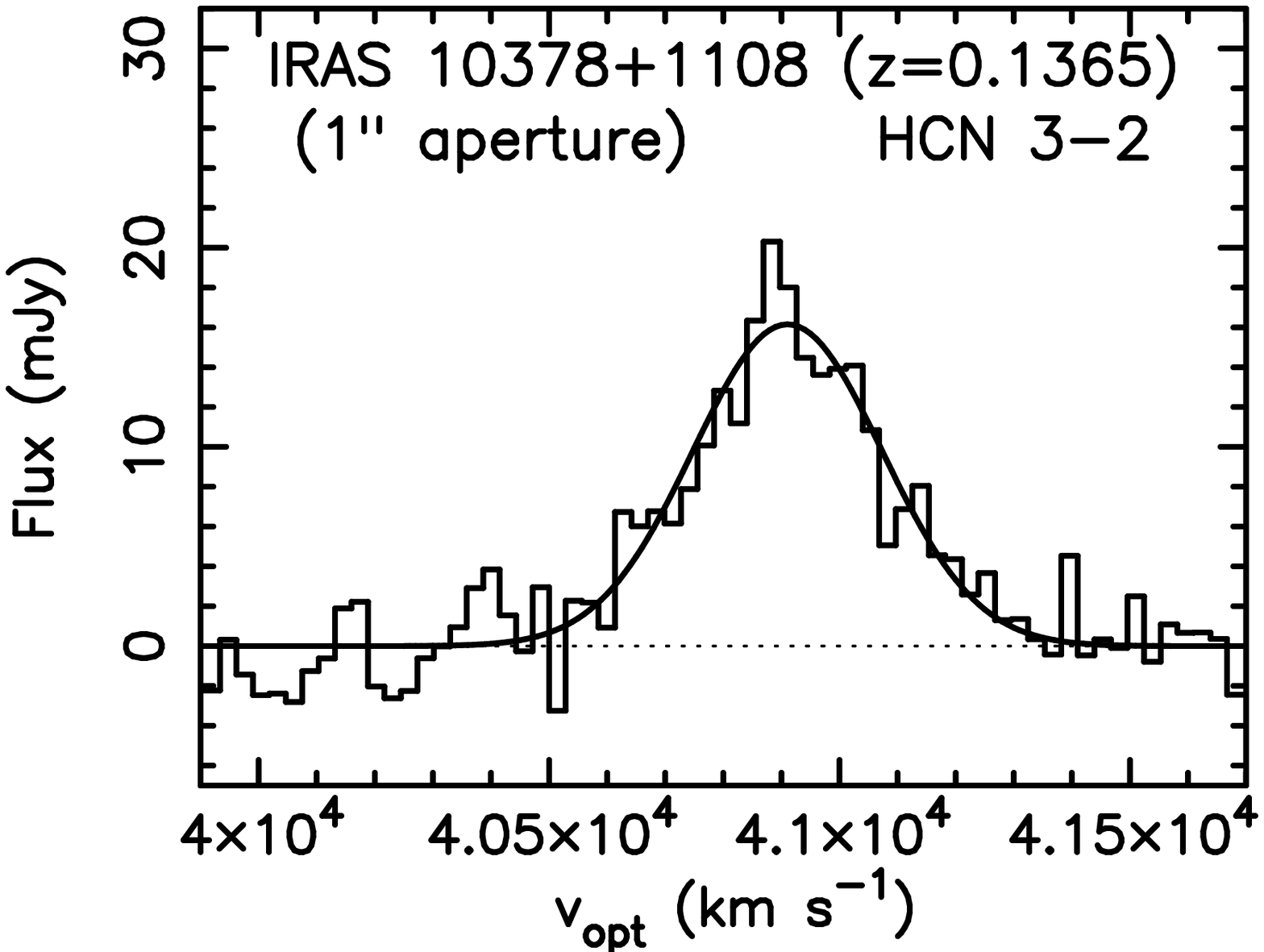} 
\includegraphics[angle=0,scale=.223]{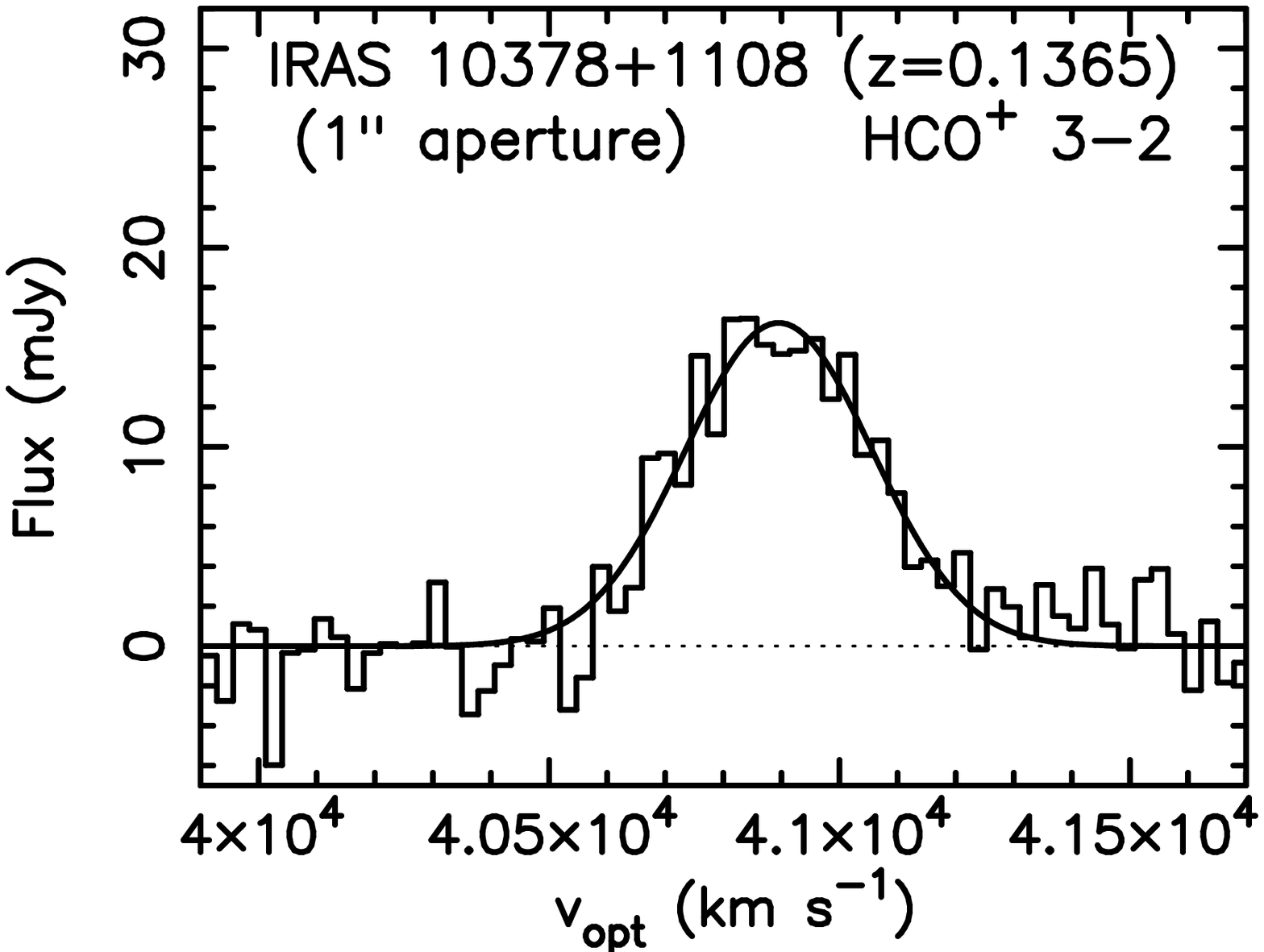} \\
\includegraphics[angle=0,scale=.223]{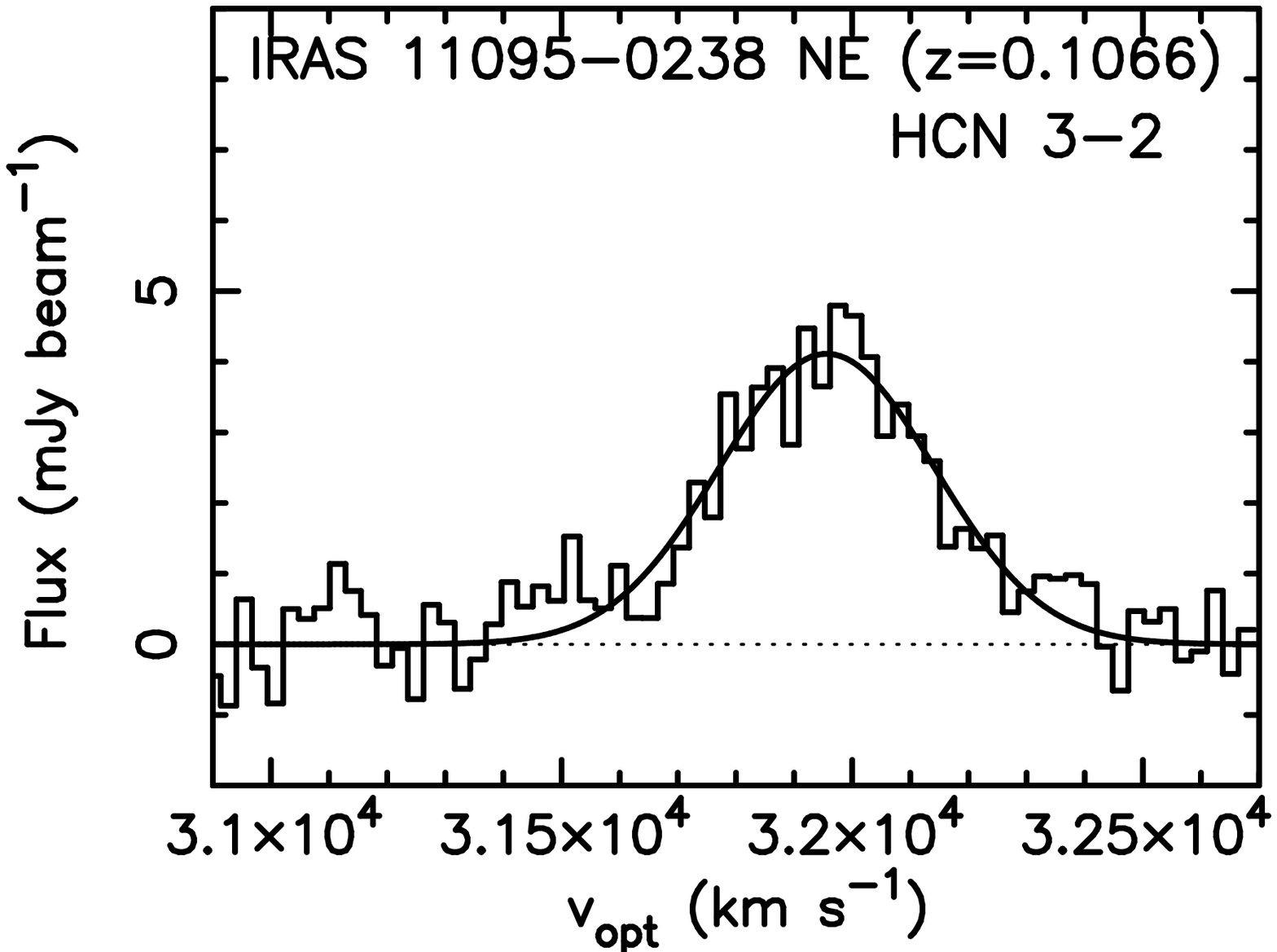} 
\includegraphics[angle=0,scale=.223]{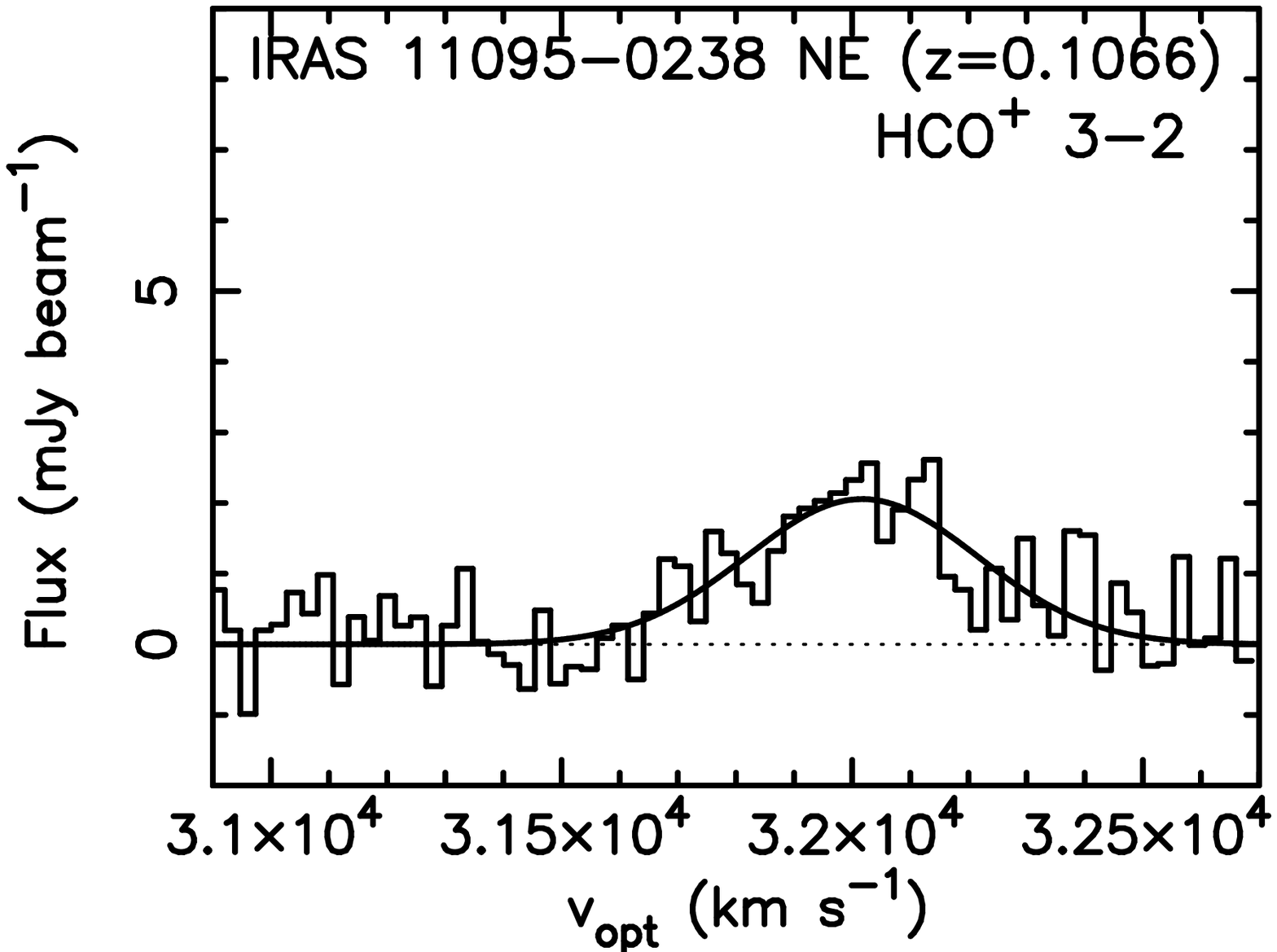} 
\includegraphics[angle=0,scale=.223]{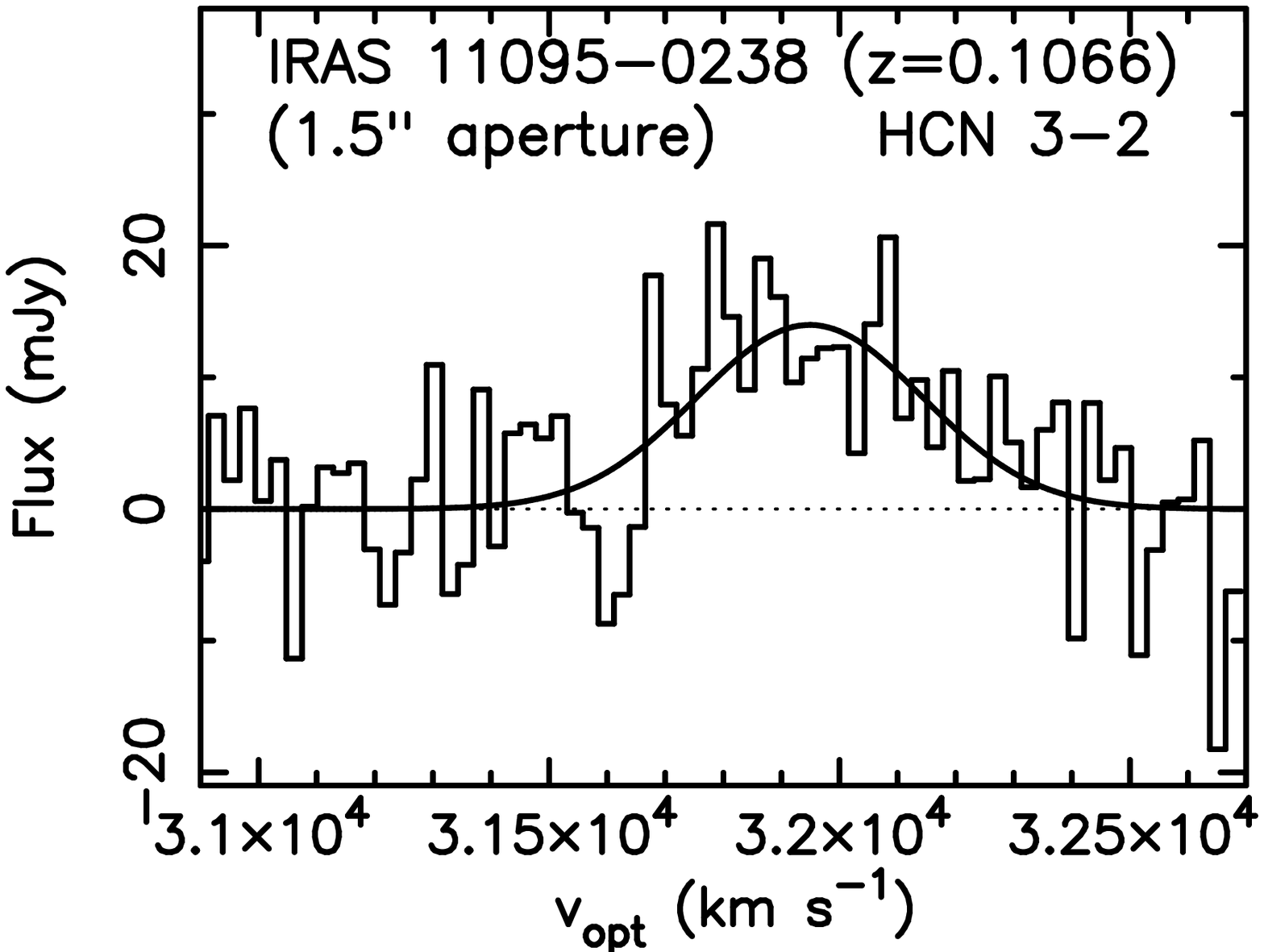} 
\includegraphics[angle=0,scale=.223]{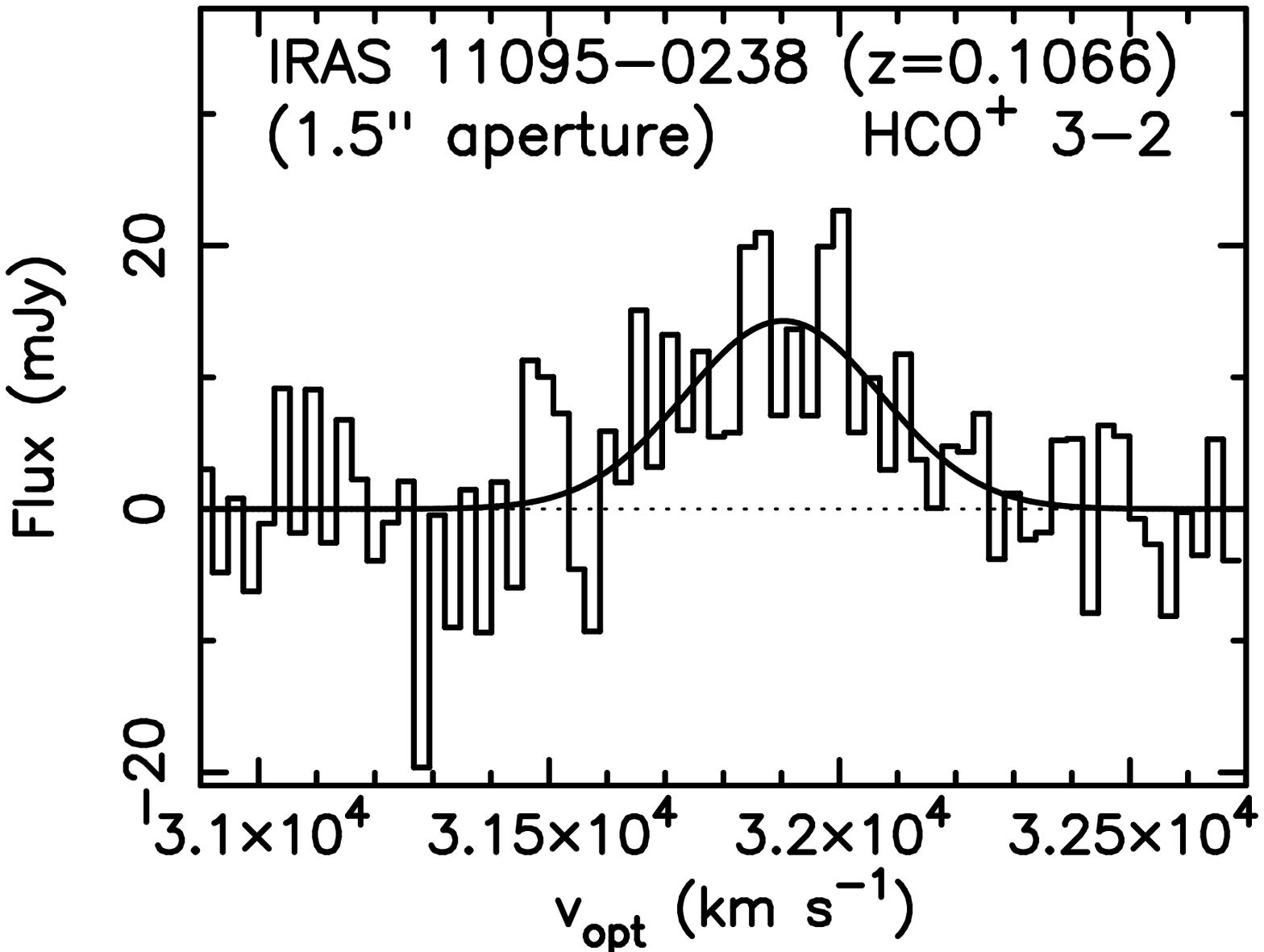} \\
\end{center}
\end{figure}

\clearpage

\begin{figure}
\begin{center}
\hspace*{-9.2cm}
\includegraphics[angle=0,scale=.223]{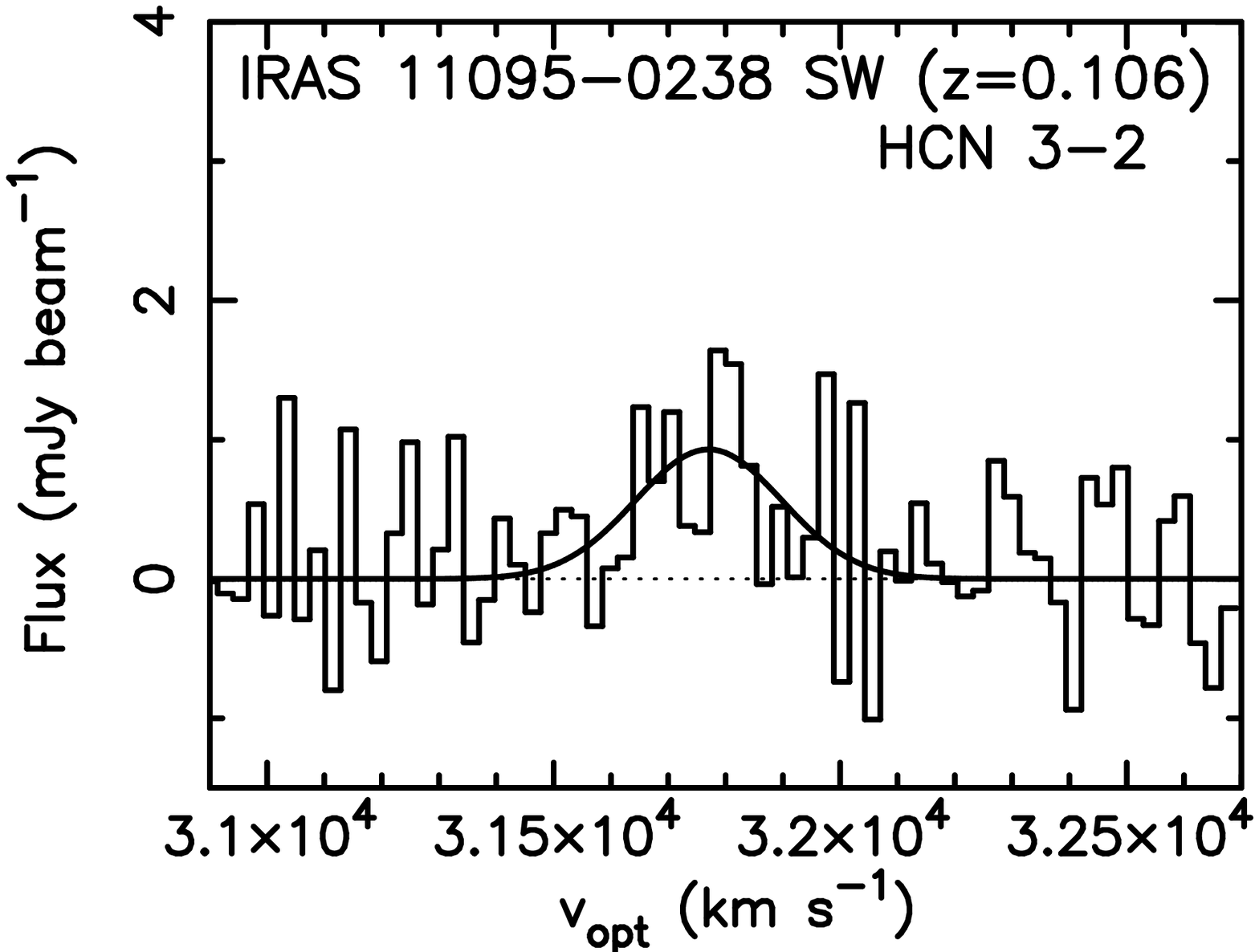} 
\includegraphics[angle=0,scale=.223]{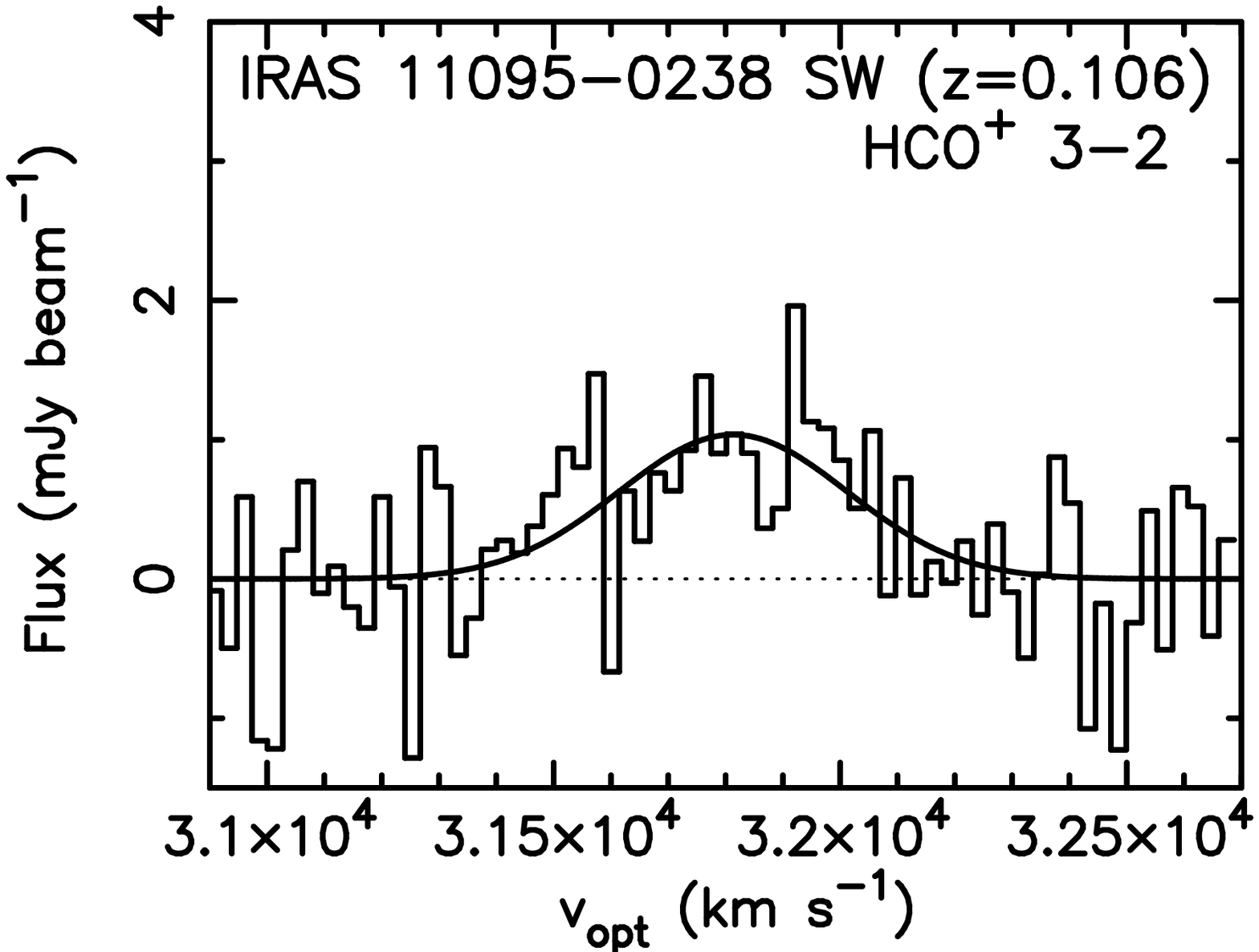} \\
\includegraphics[angle=0,scale=.223]{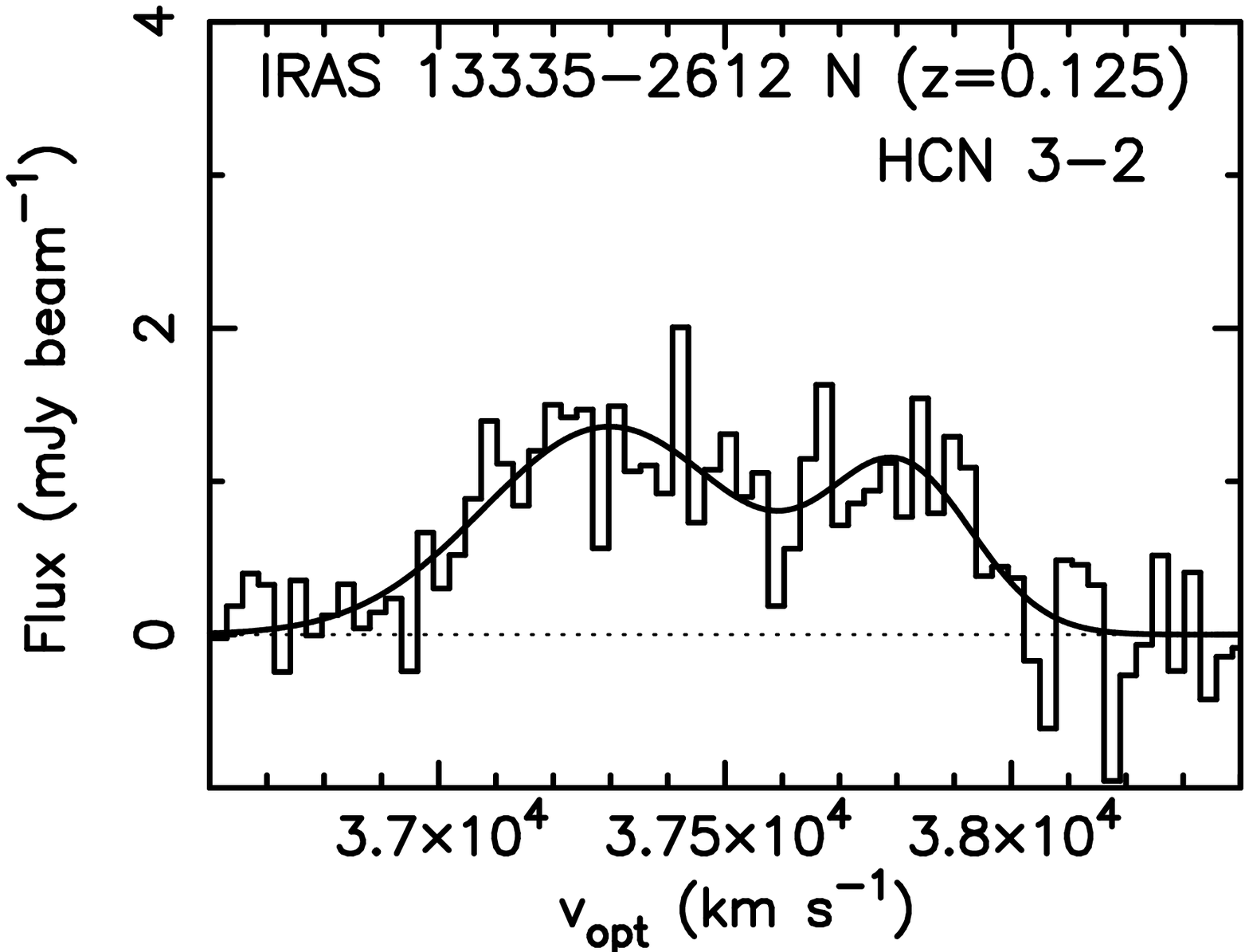} 
\includegraphics[angle=0,scale=.223]{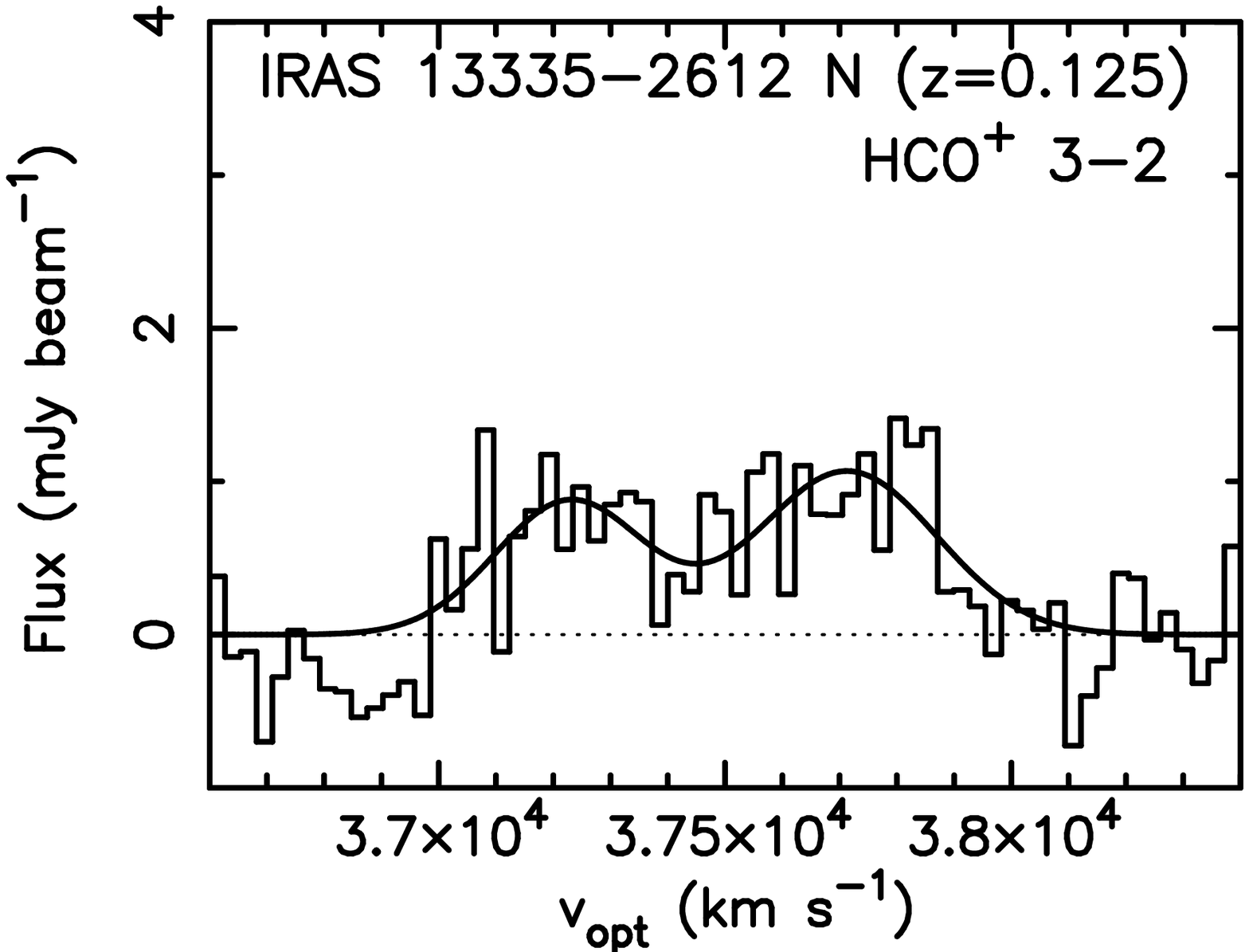} 
\includegraphics[angle=0,scale=.223]{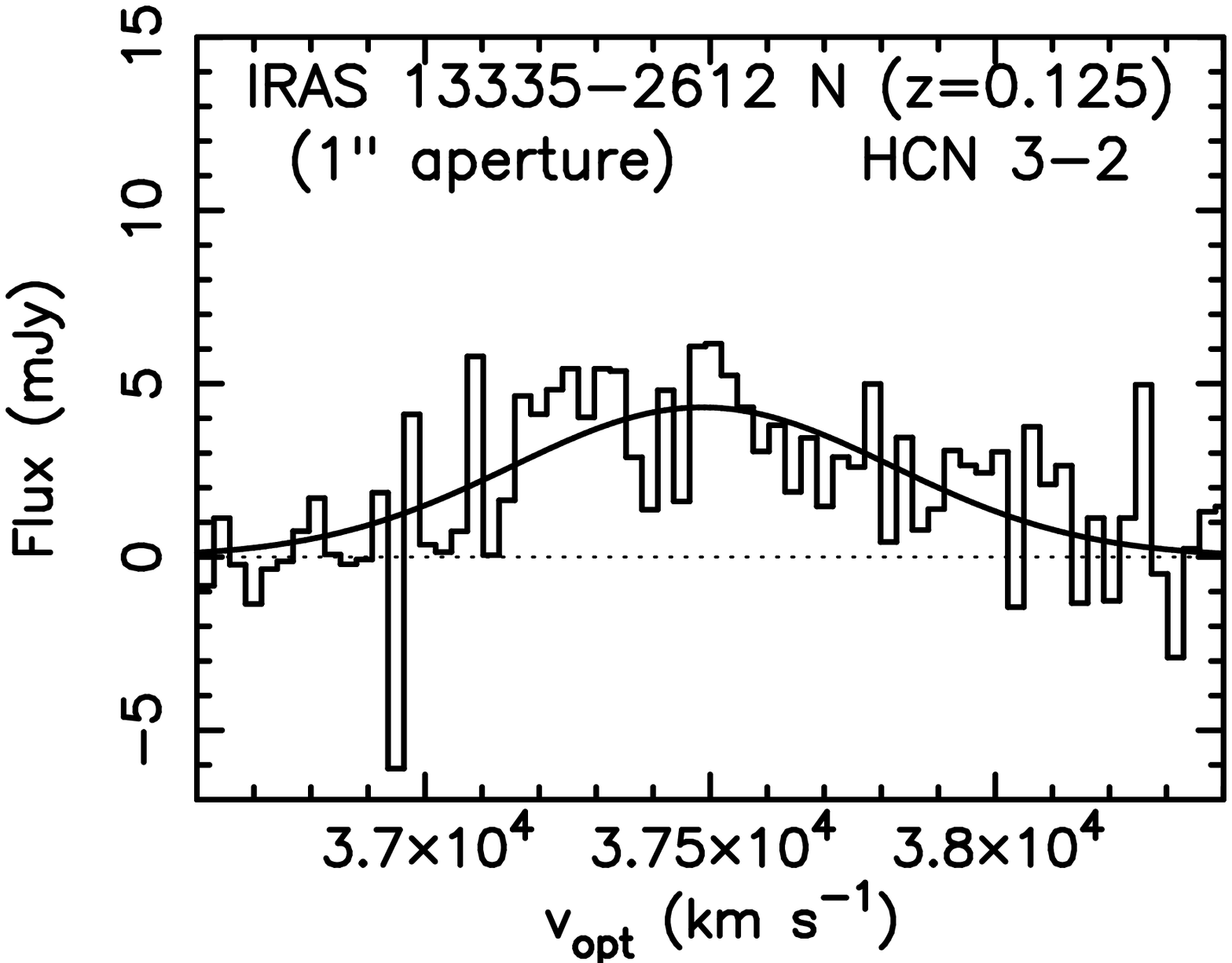} 
\includegraphics[angle=0,scale=.223]{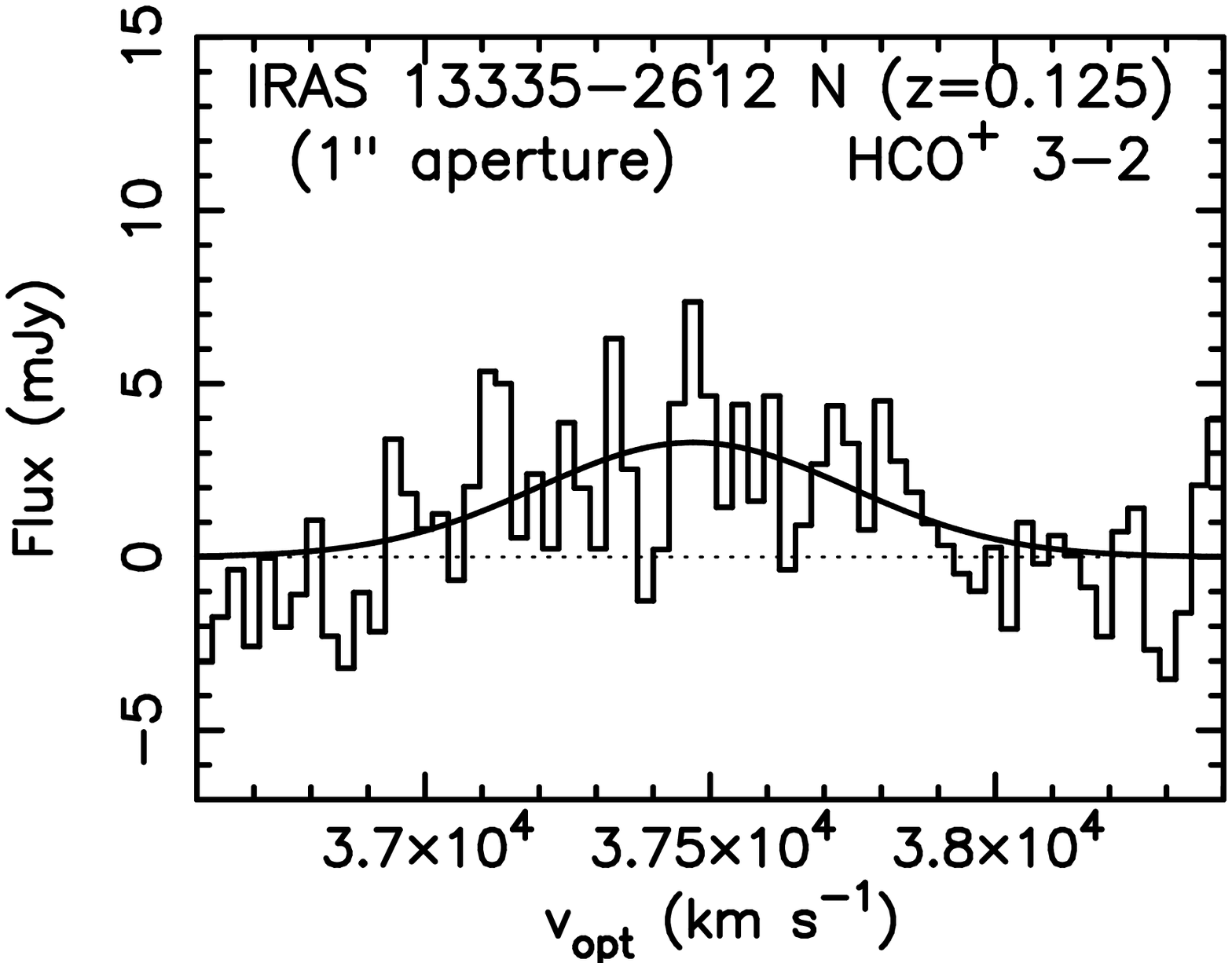} \\
\includegraphics[angle=0,scale=.223]{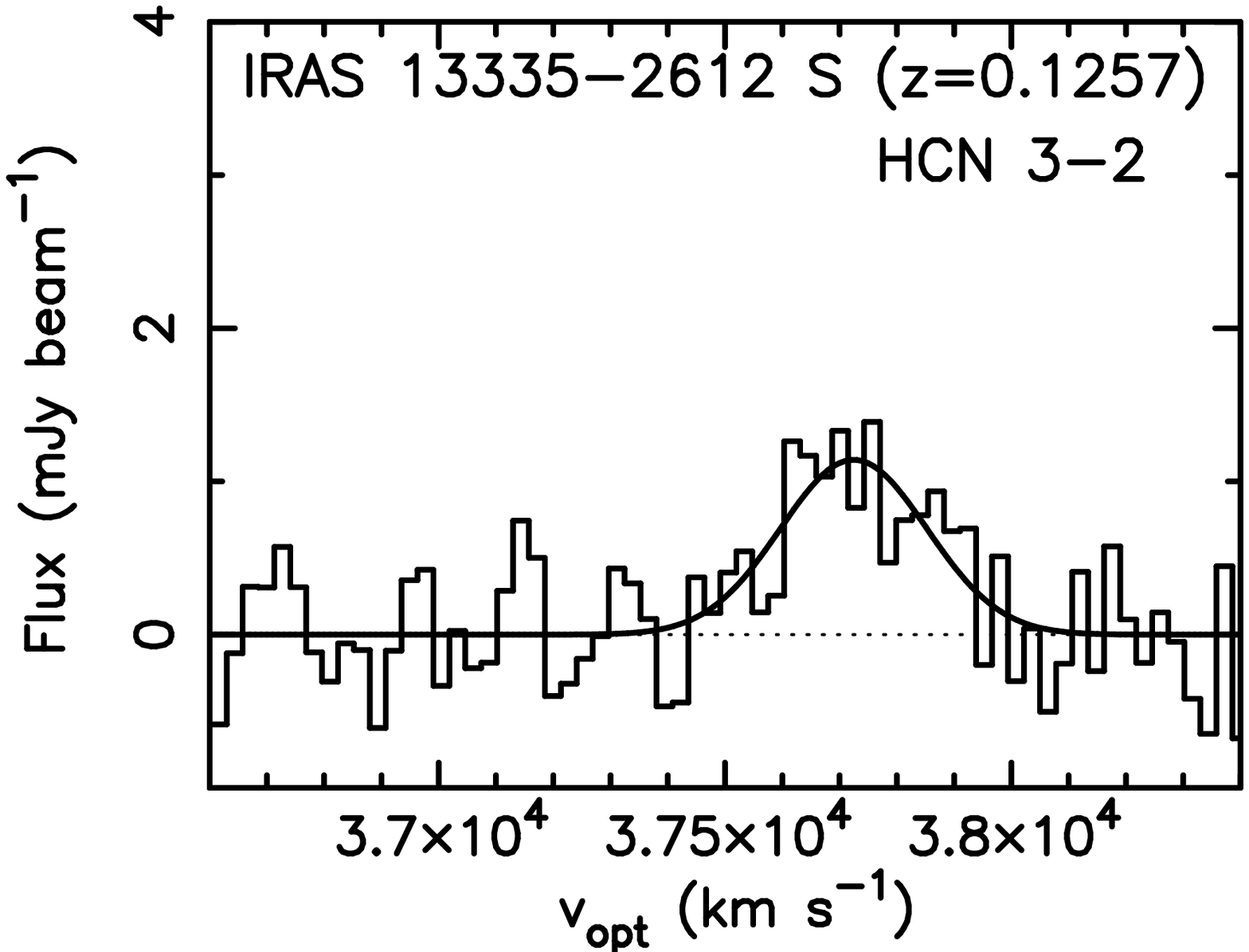} 
\includegraphics[angle=0,scale=.223]{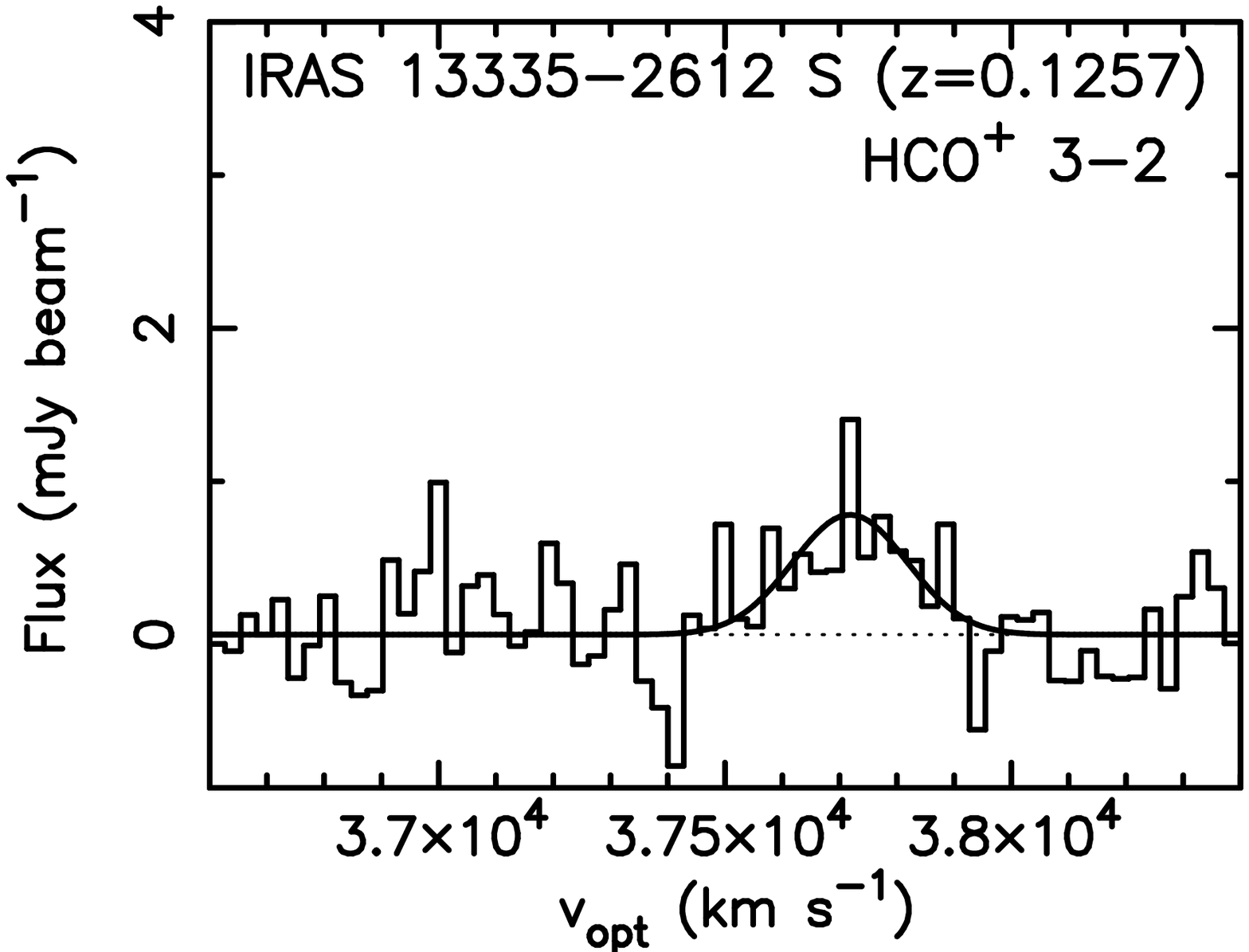} 
\includegraphics[angle=0,scale=.223]{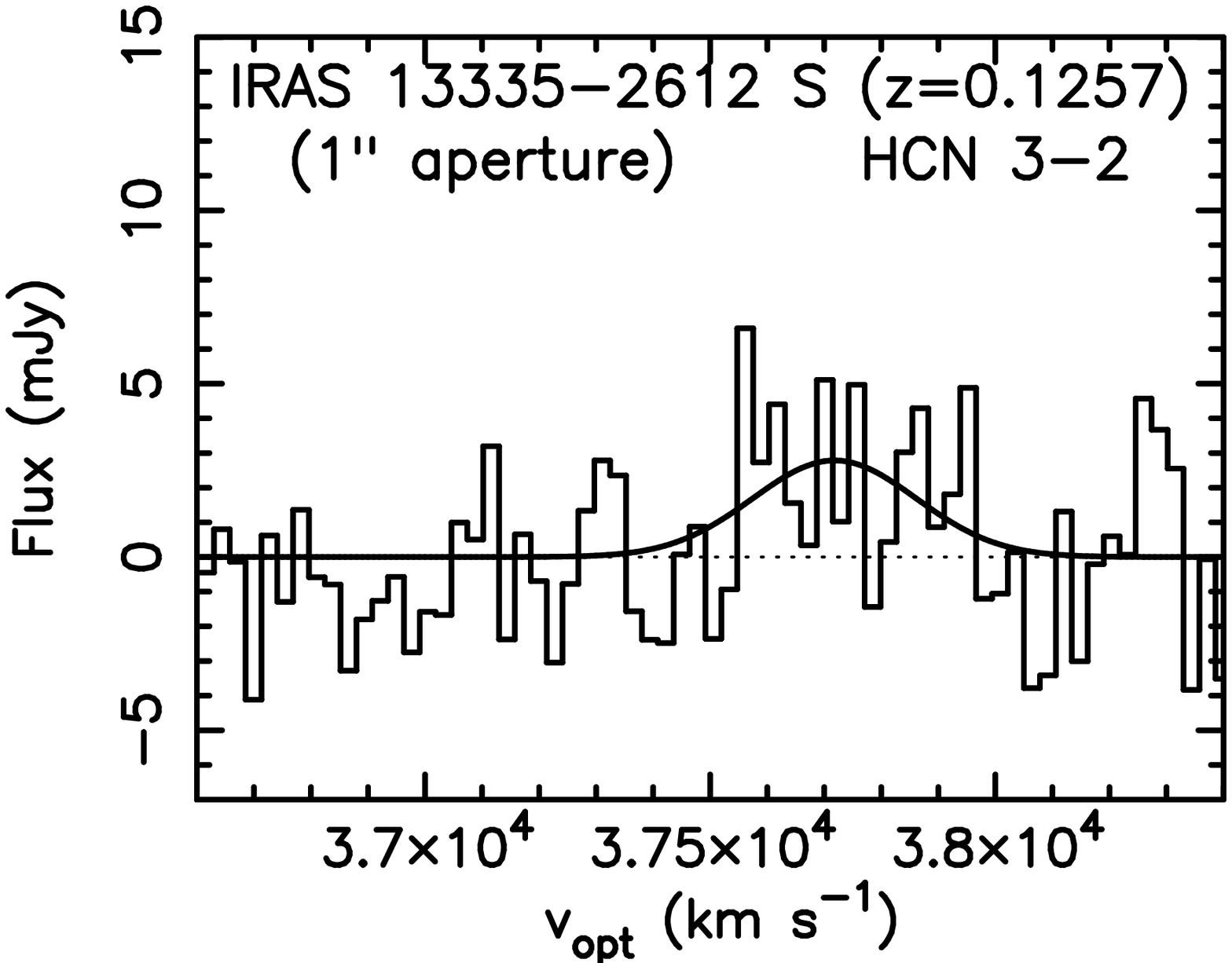} 
\includegraphics[angle=0,scale=.223]{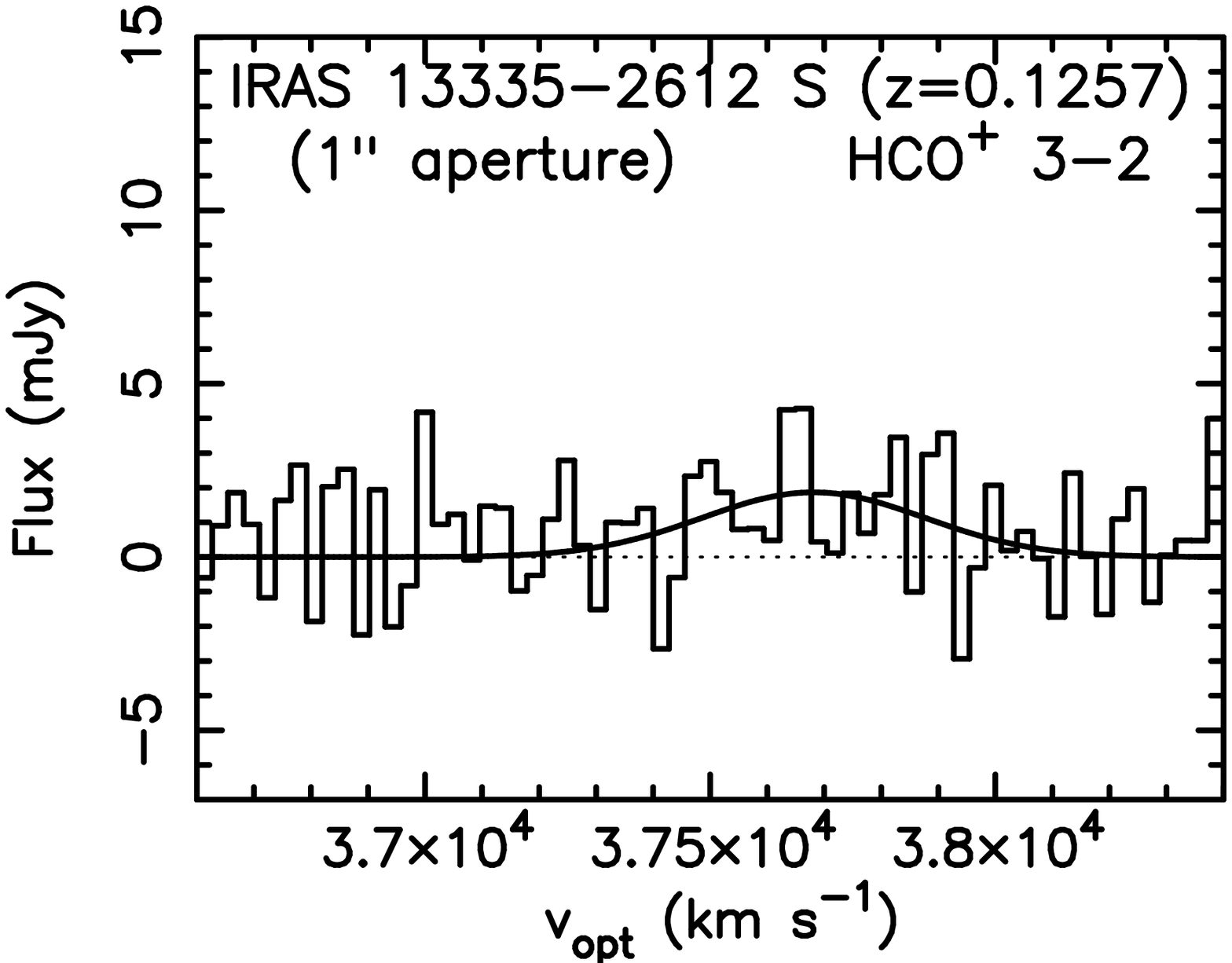} \\
\includegraphics[angle=0,scale=.223]{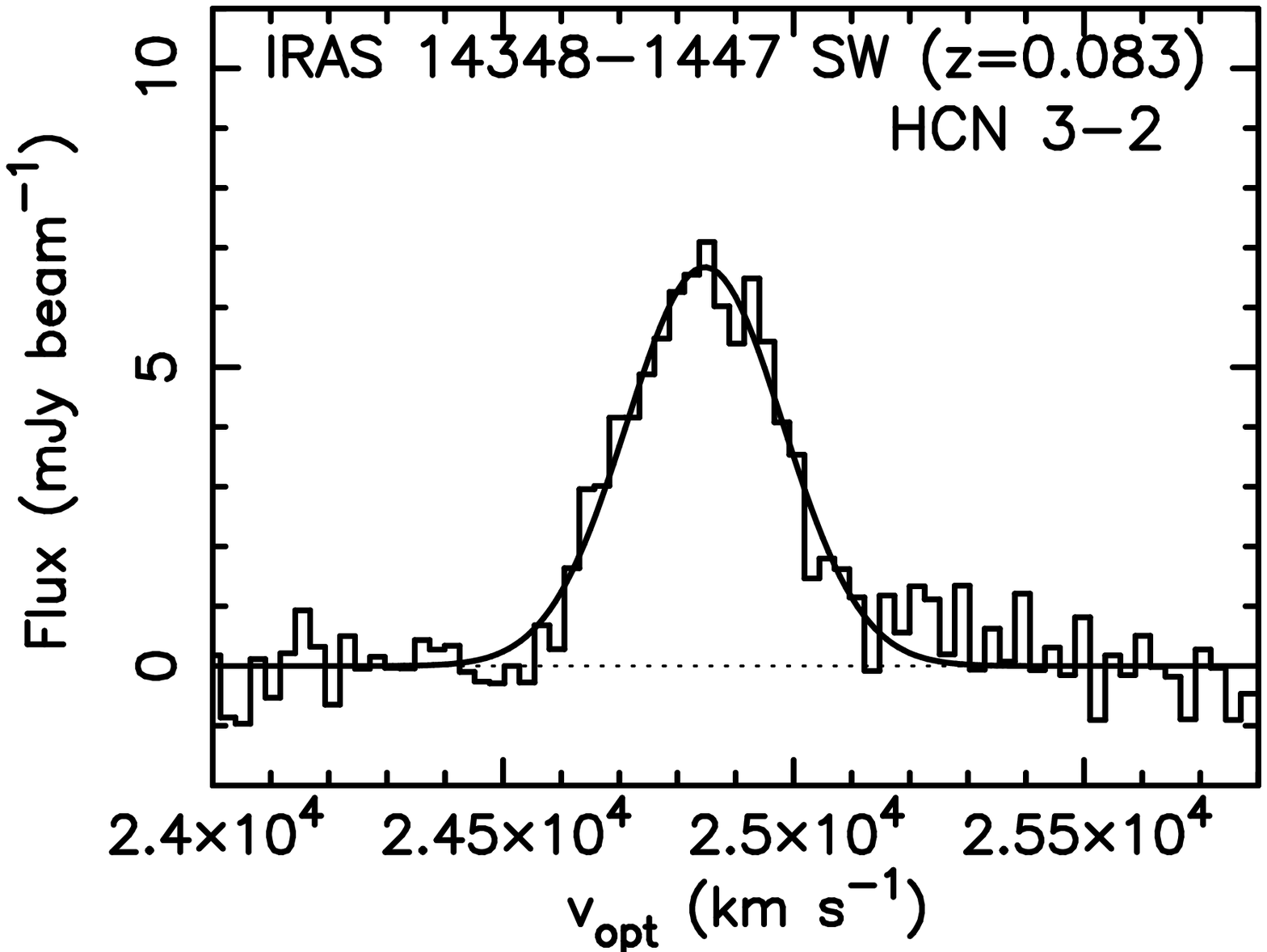} 
\includegraphics[angle=0,scale=.223]{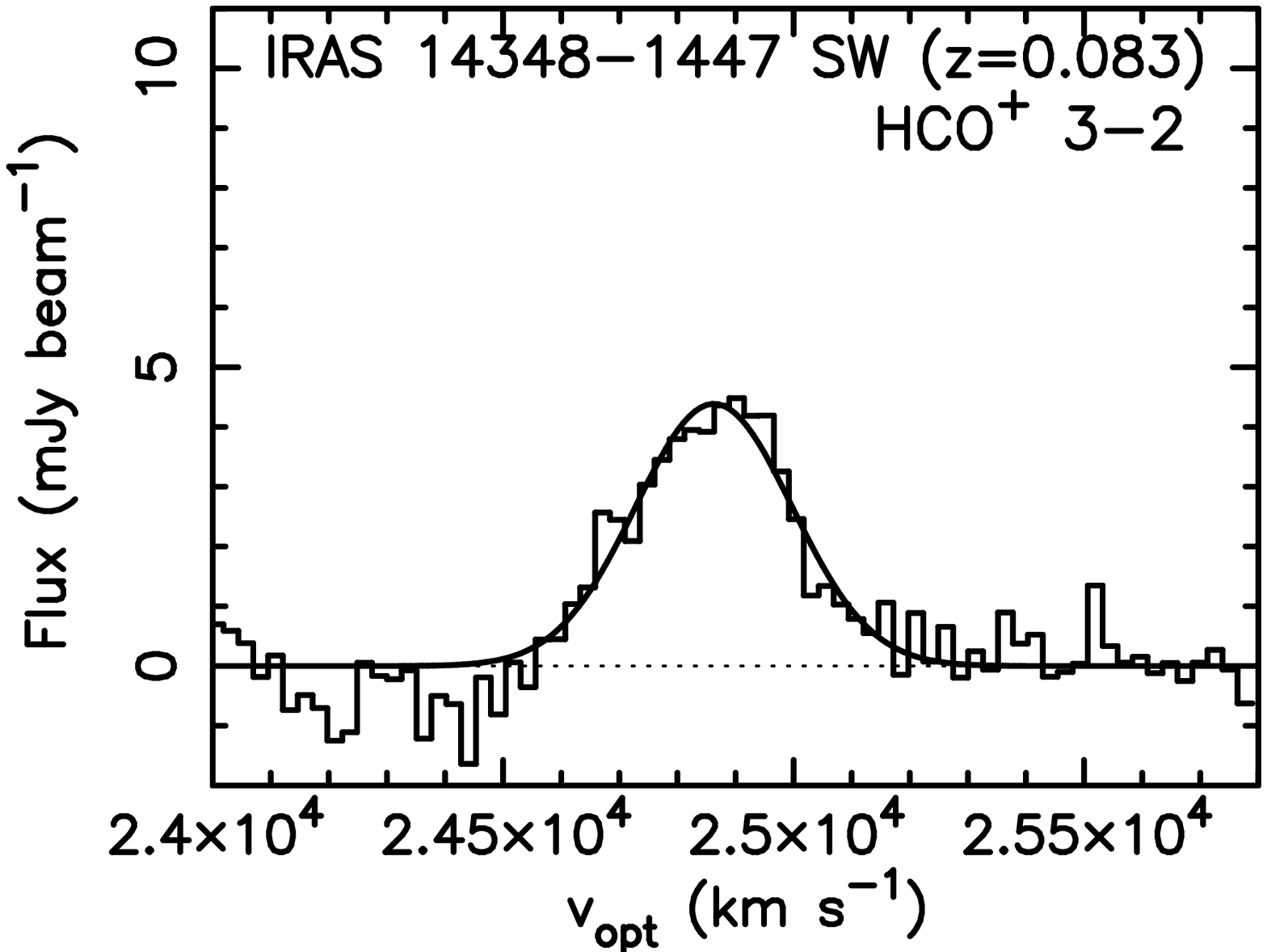} 
\includegraphics[angle=0,scale=.223]{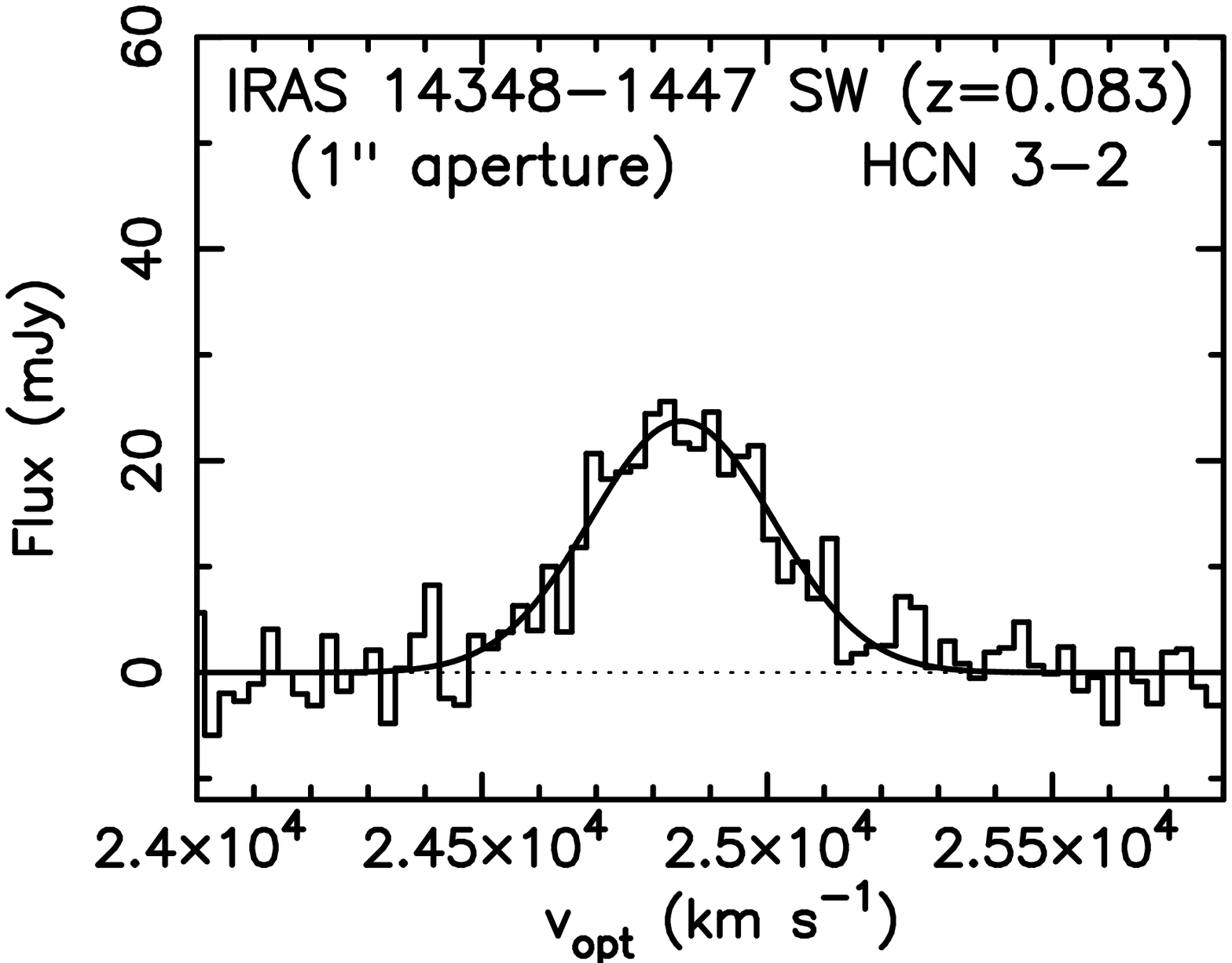} 
\includegraphics[angle=0,scale=.223]{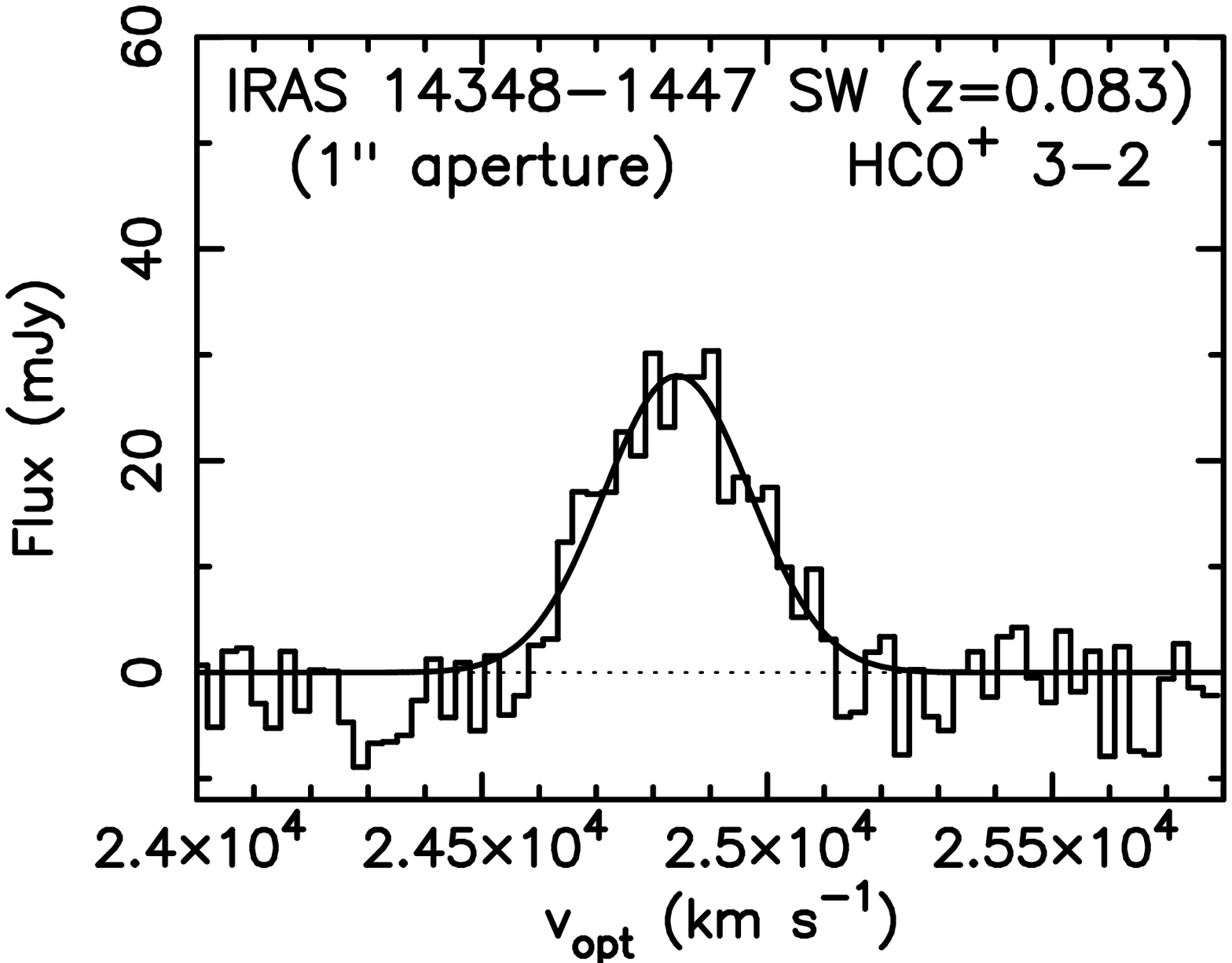} \\
\includegraphics[angle=0,scale=.223]{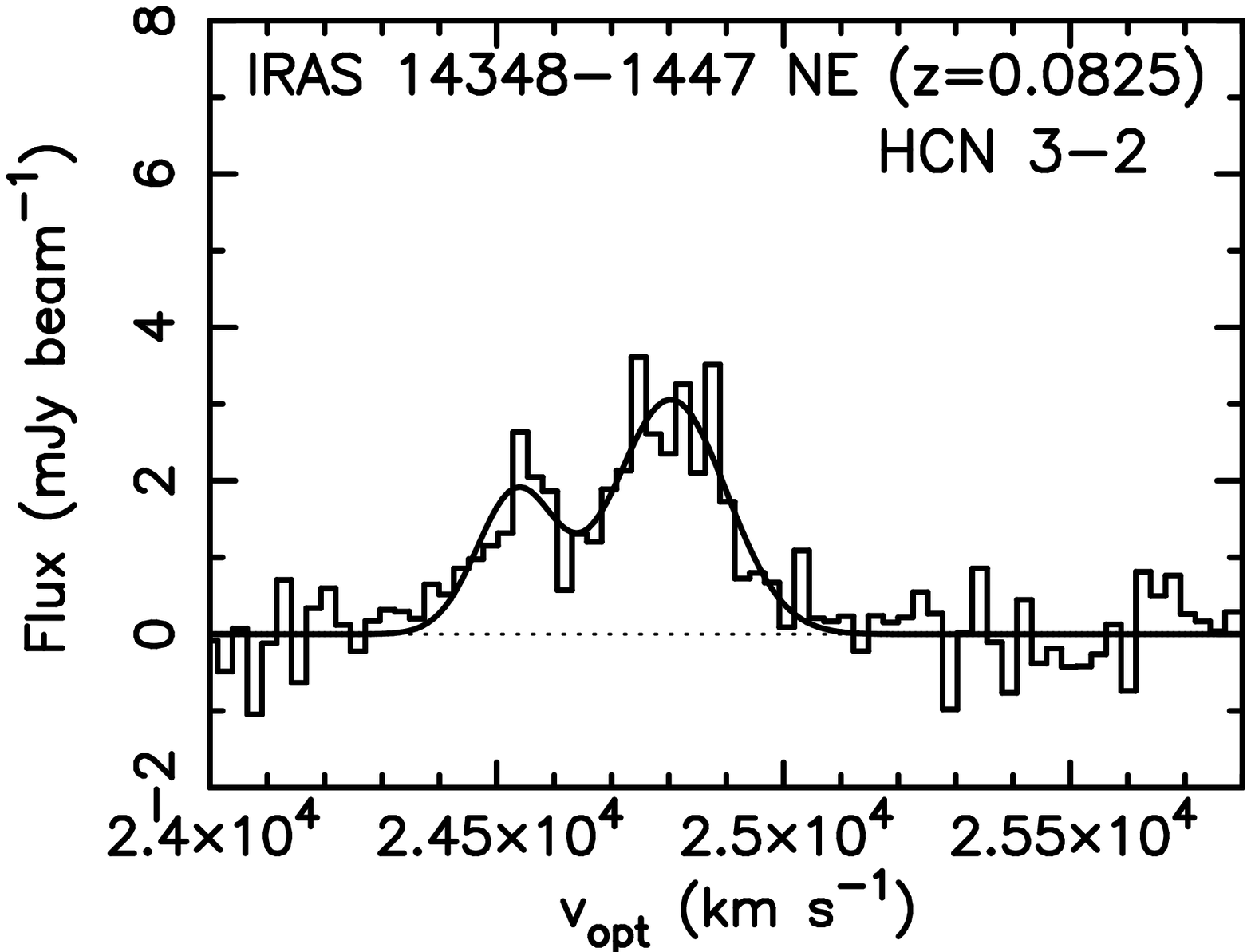} 
\includegraphics[angle=0,scale=.223]{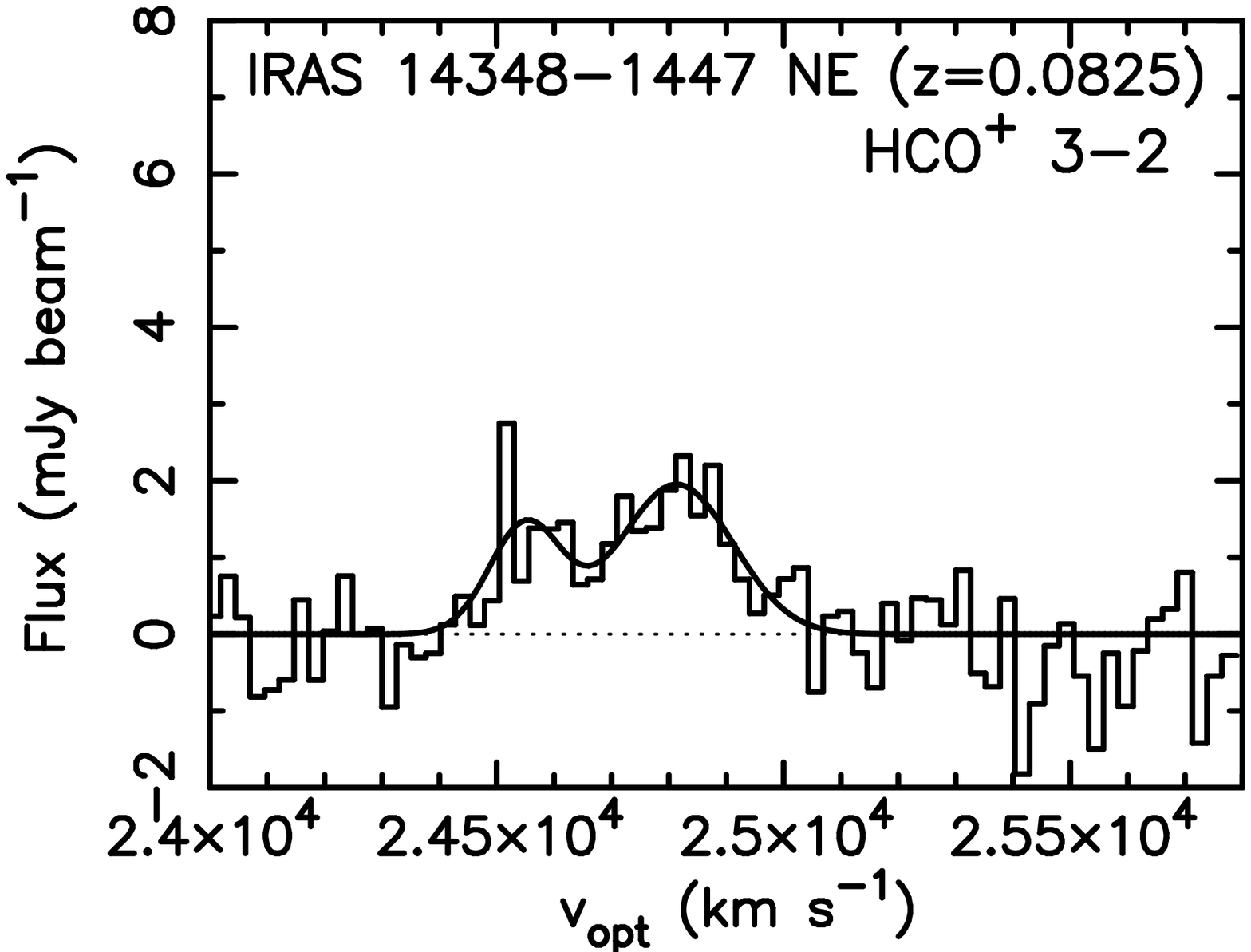} 
\includegraphics[angle=0,scale=.223]{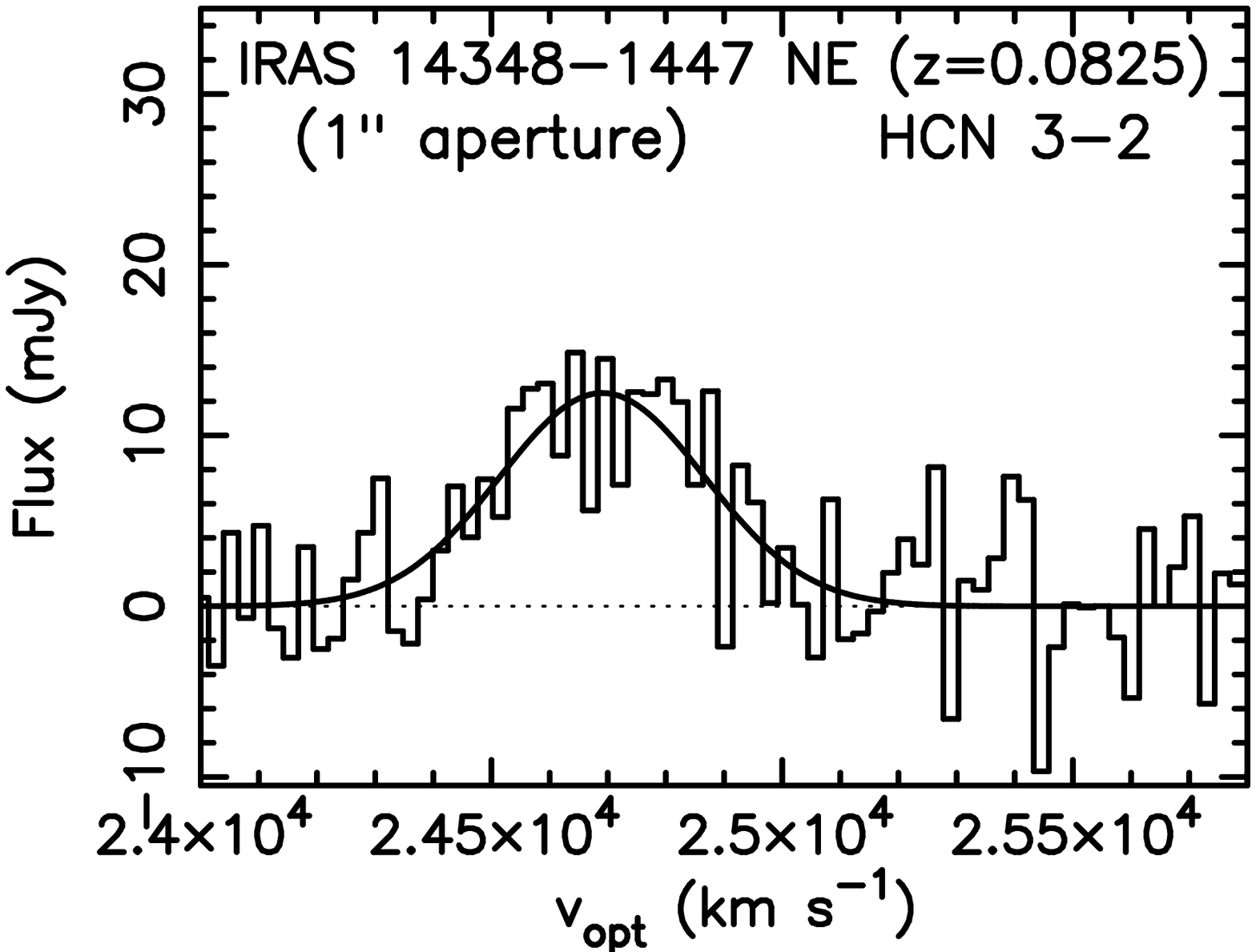} 
\includegraphics[angle=0,scale=.223]{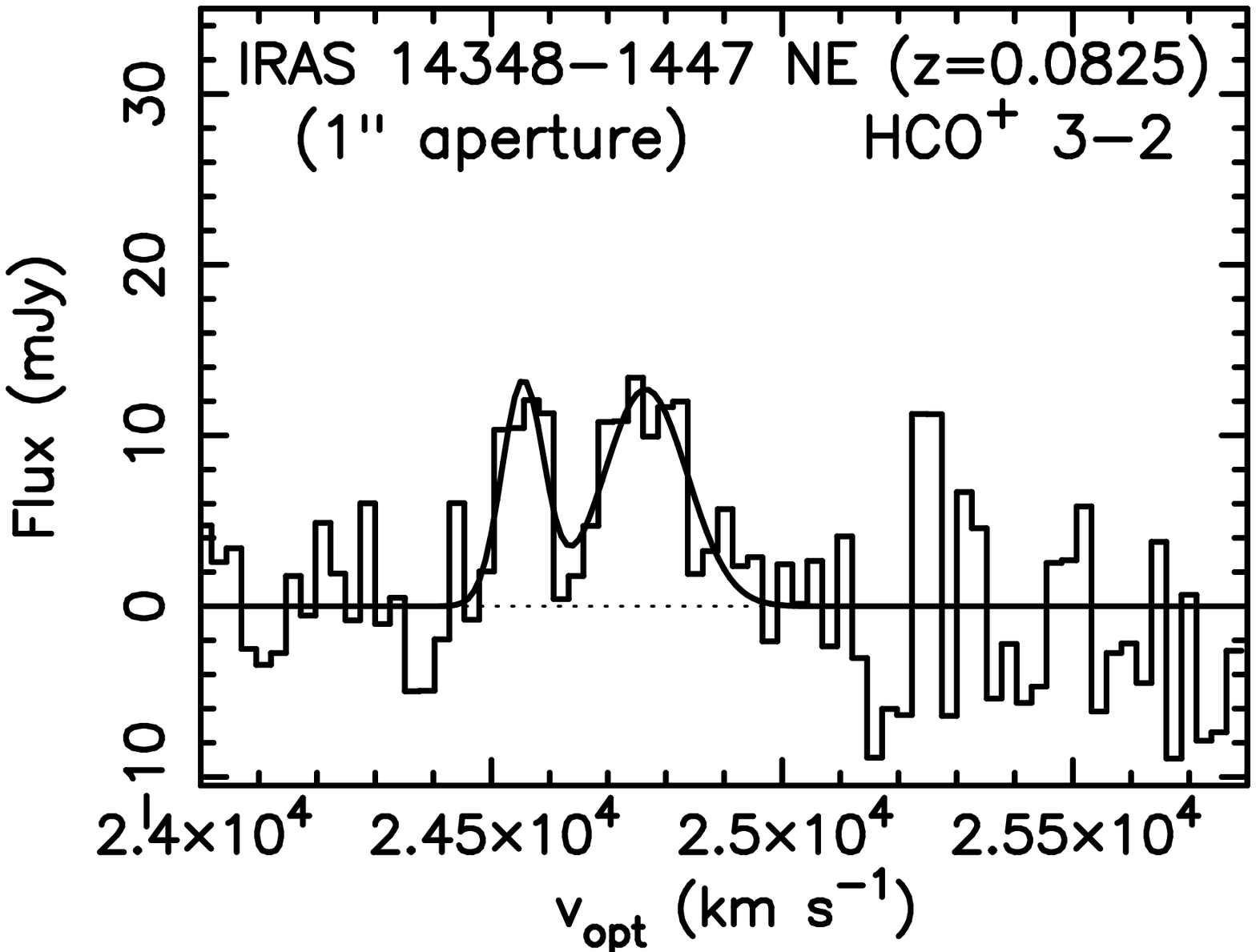} \\
\includegraphics[angle=0,scale=.223]{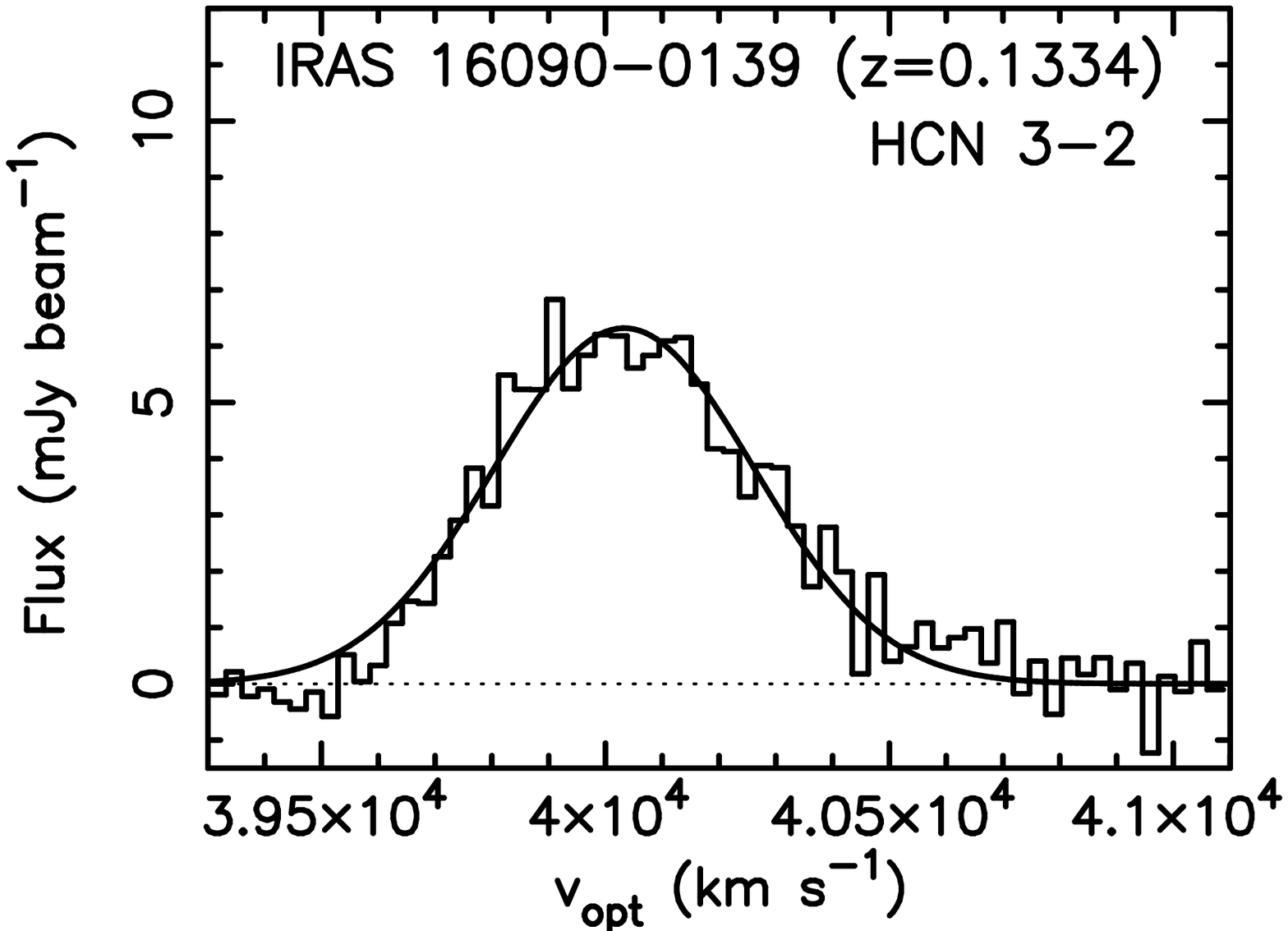} 
\includegraphics[angle=0,scale=.223]{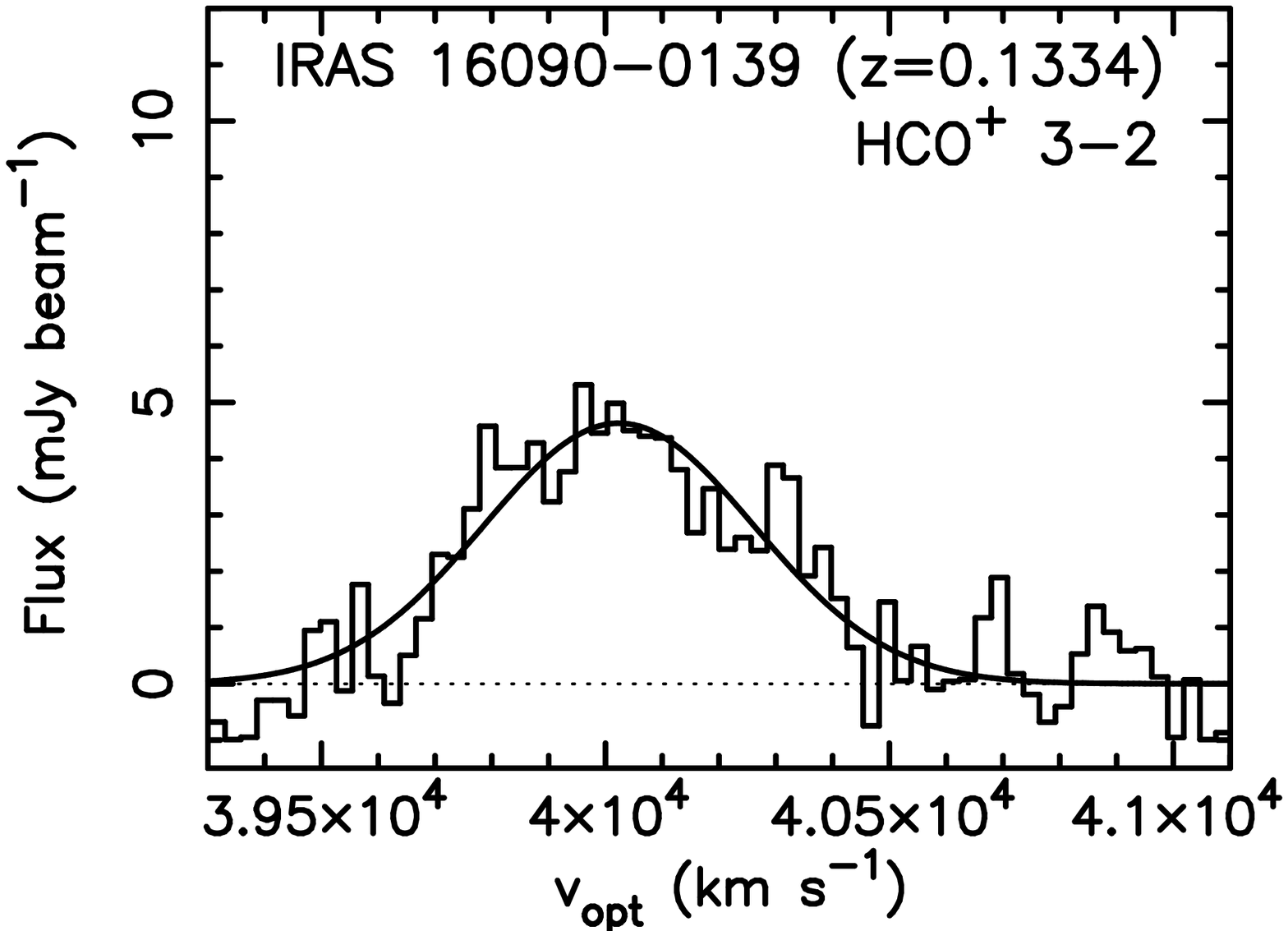} 
\includegraphics[angle=0,scale=.223]{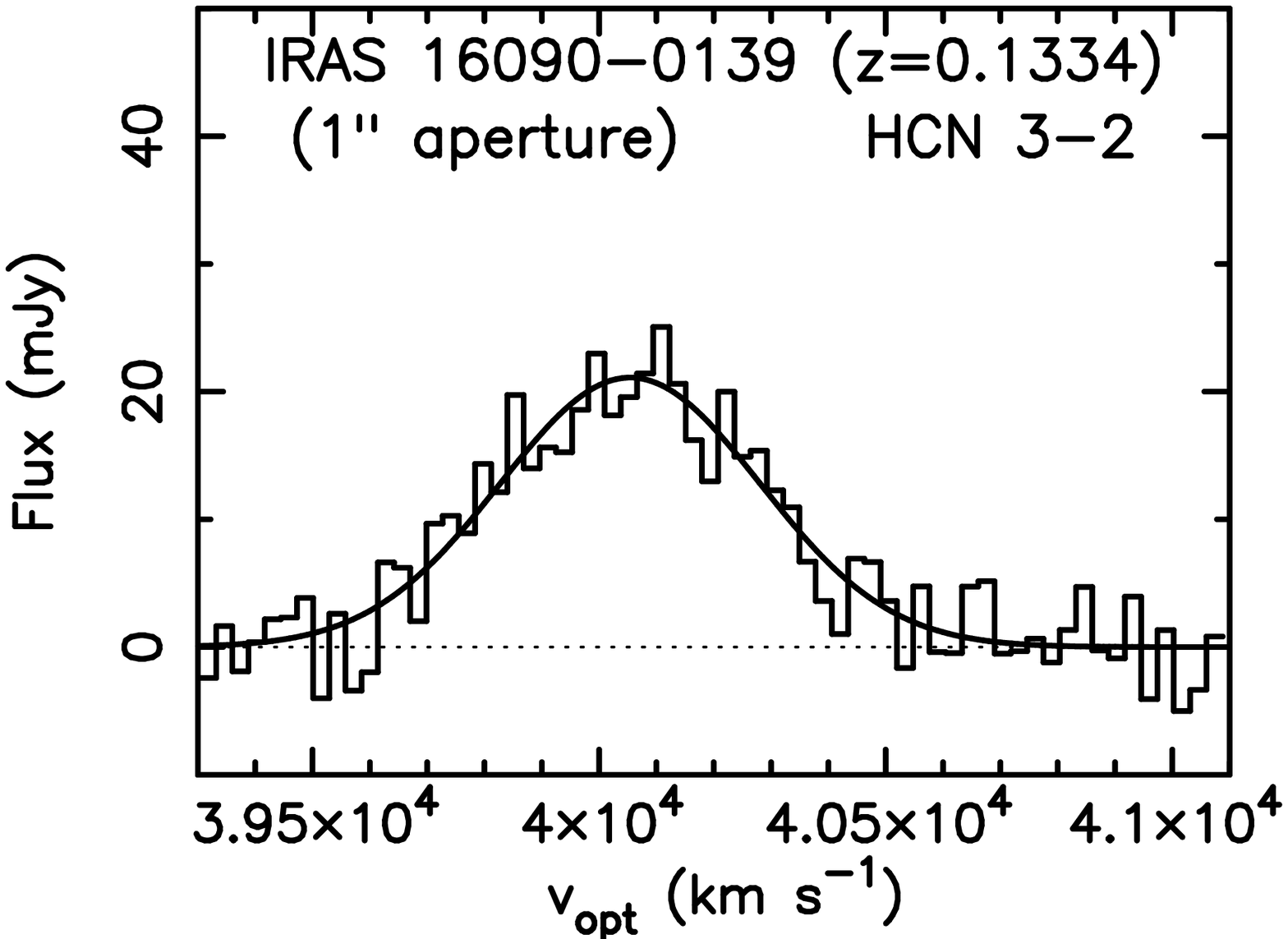} 
\includegraphics[angle=0,scale=.223]{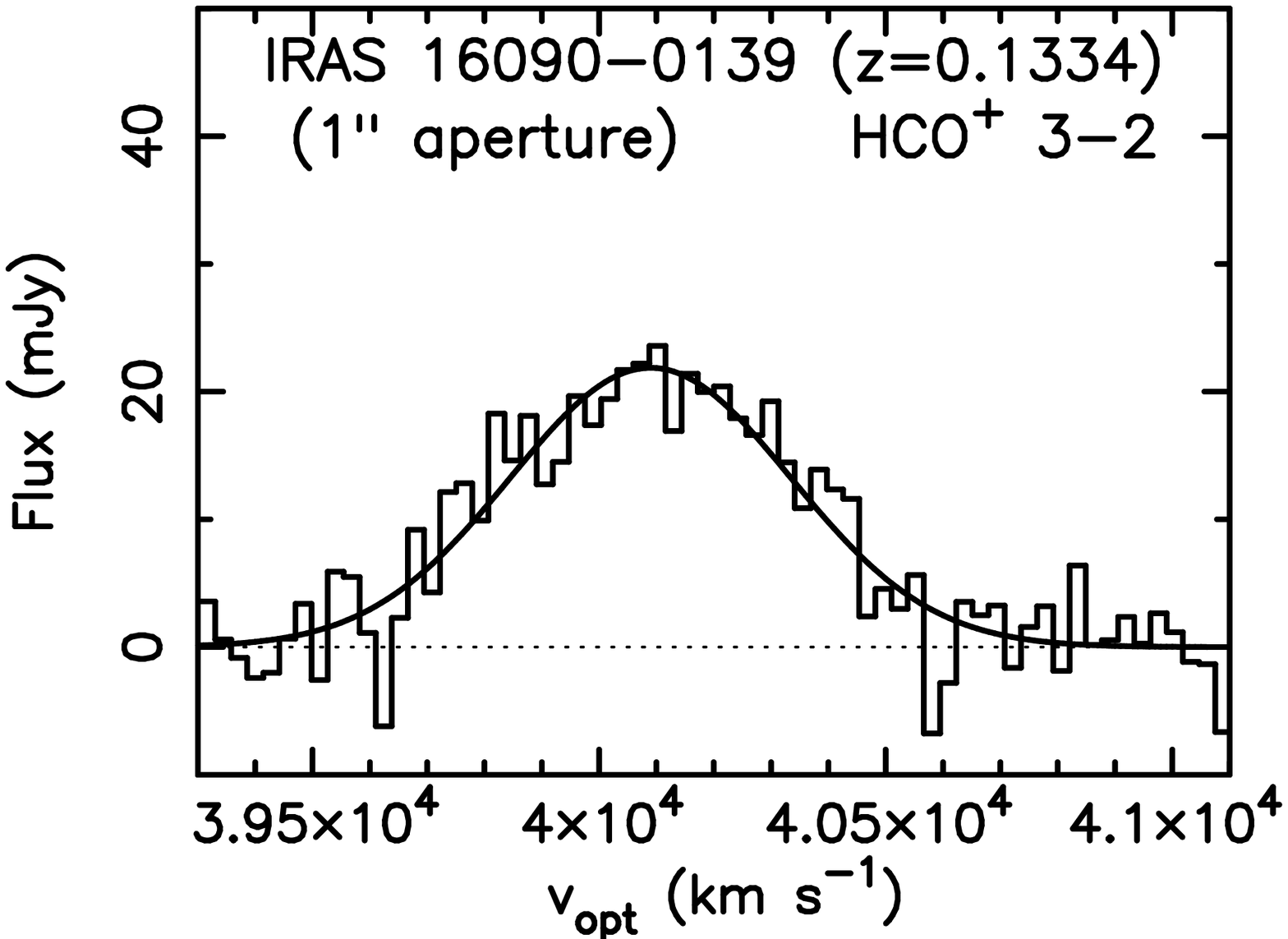} \\
\includegraphics[angle=0,scale=.223]{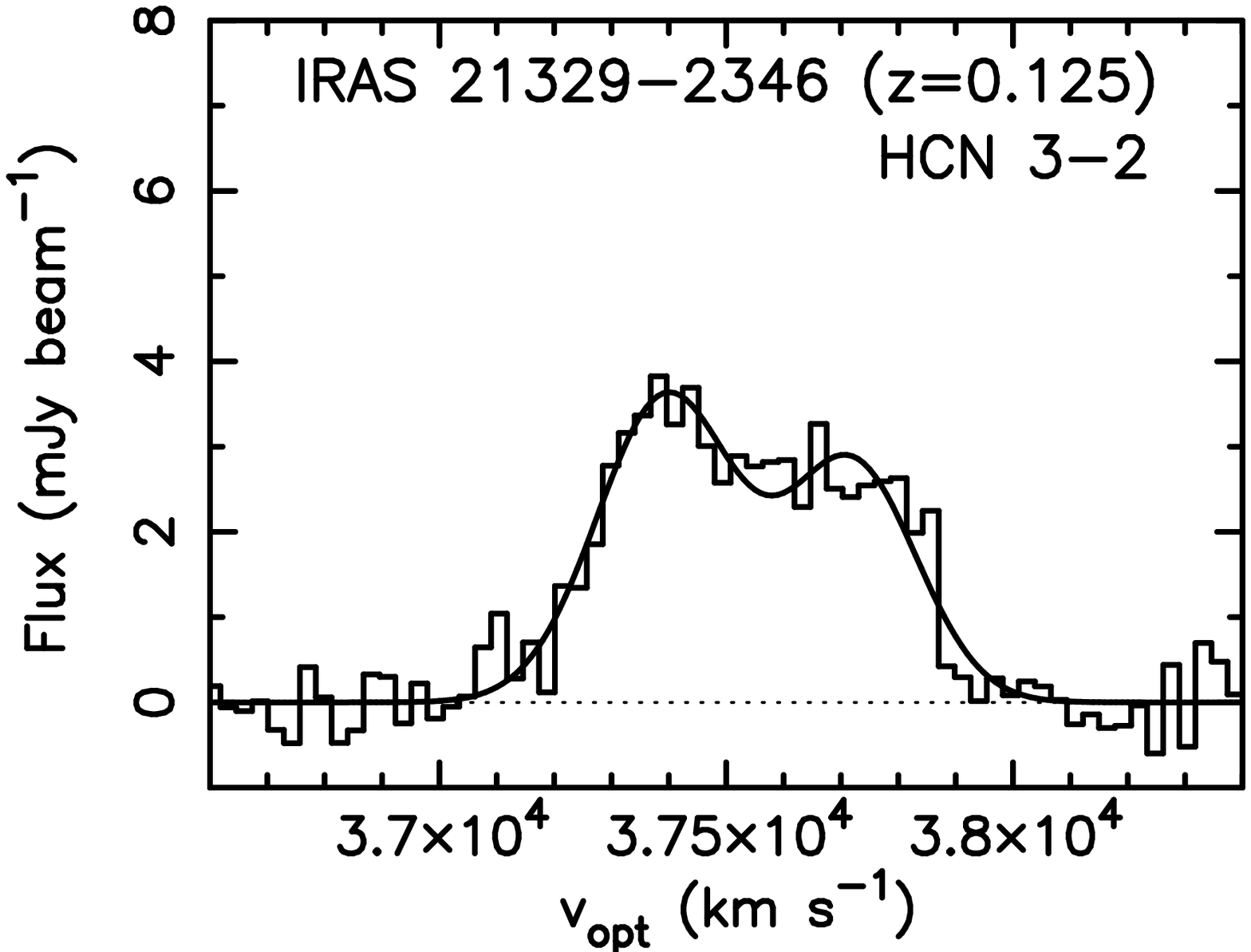} 
\includegraphics[angle=0,scale=.223]{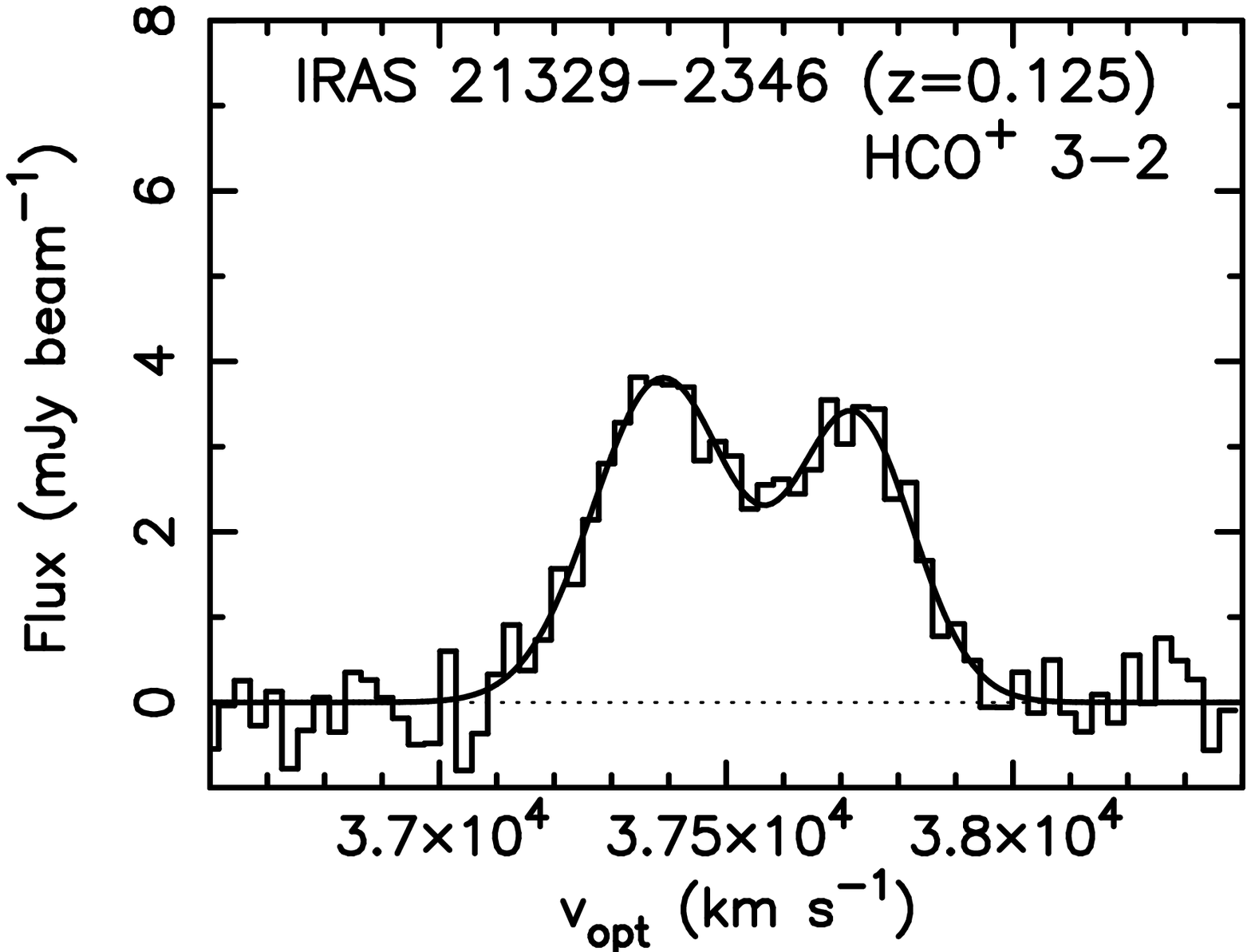} 
\includegraphics[angle=0,scale=.223]{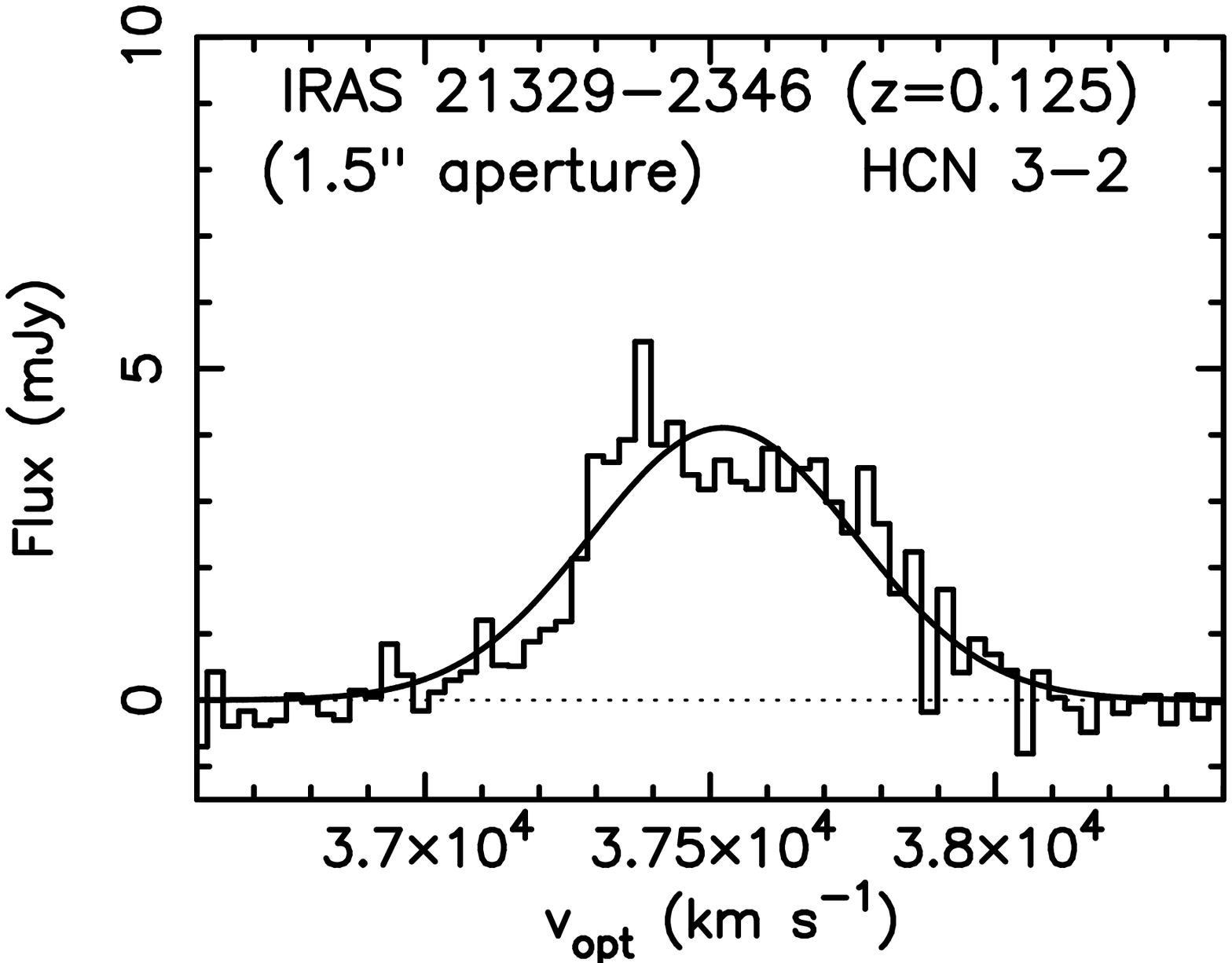} 
\includegraphics[angle=0,scale=.223]{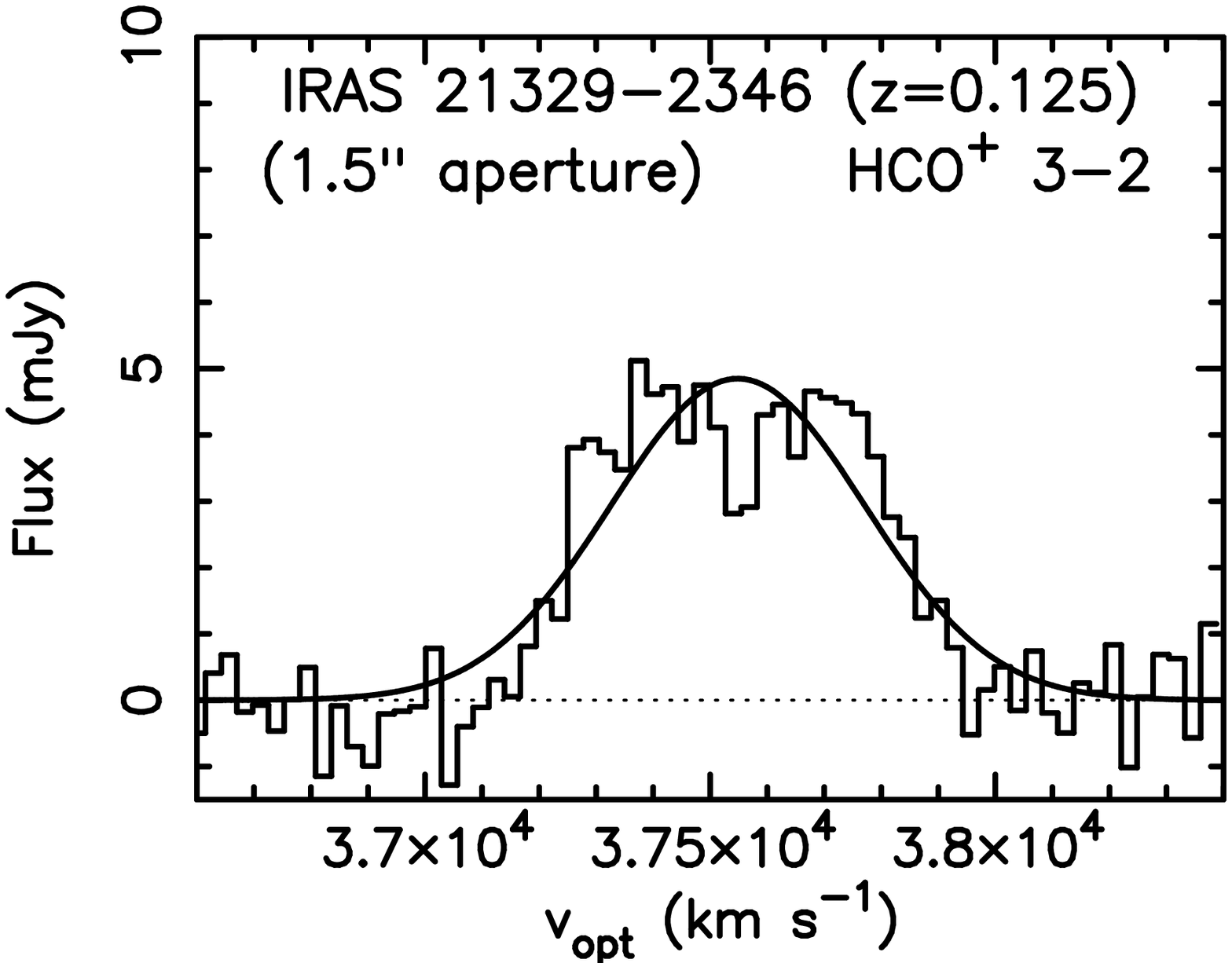} \\
\end{center}
\end{figure}

\clearpage

\begin{figure}
\begin{center}
\includegraphics[angle=0,scale=.223]{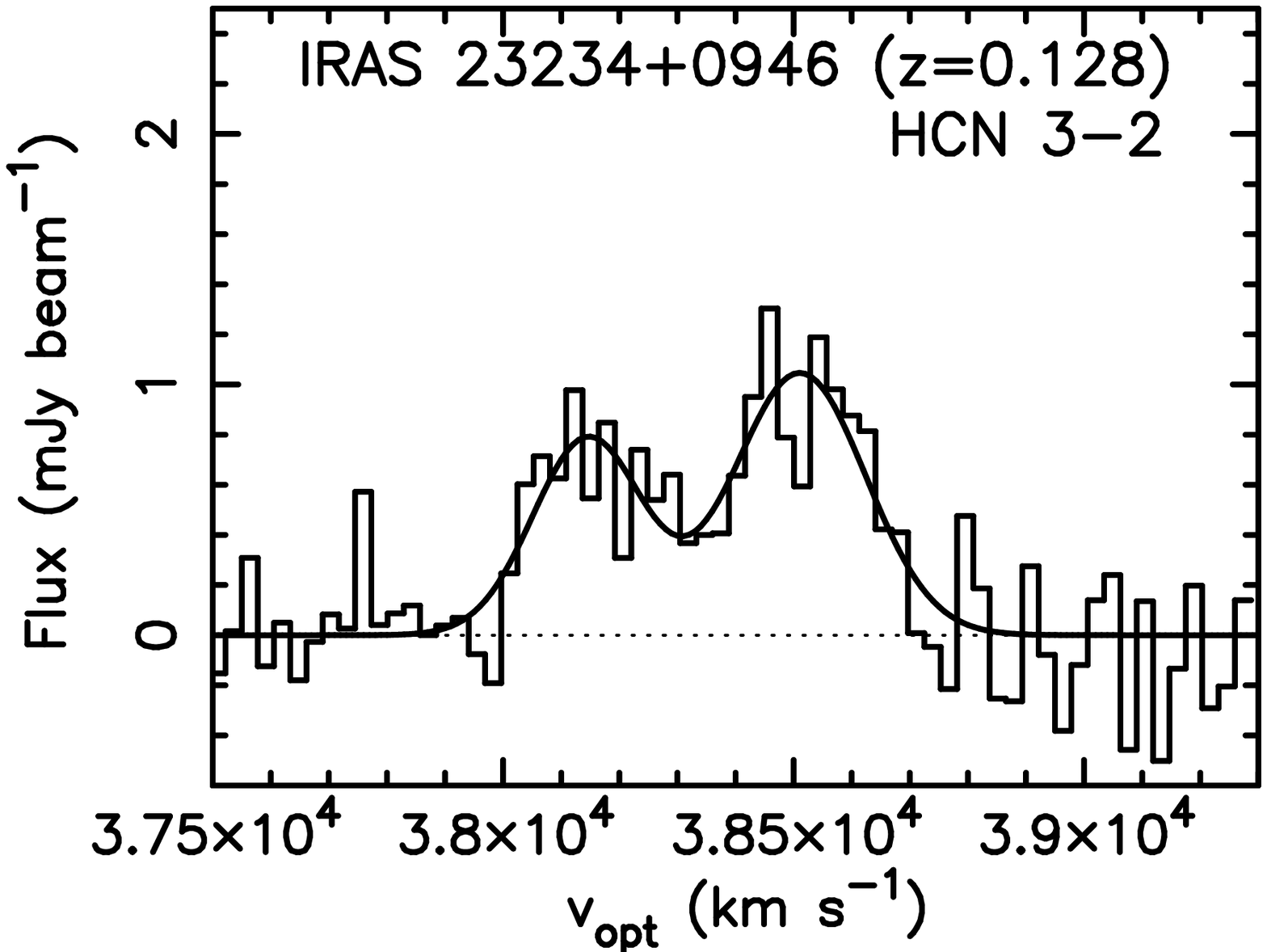} 
\includegraphics[angle=0,scale=.223]{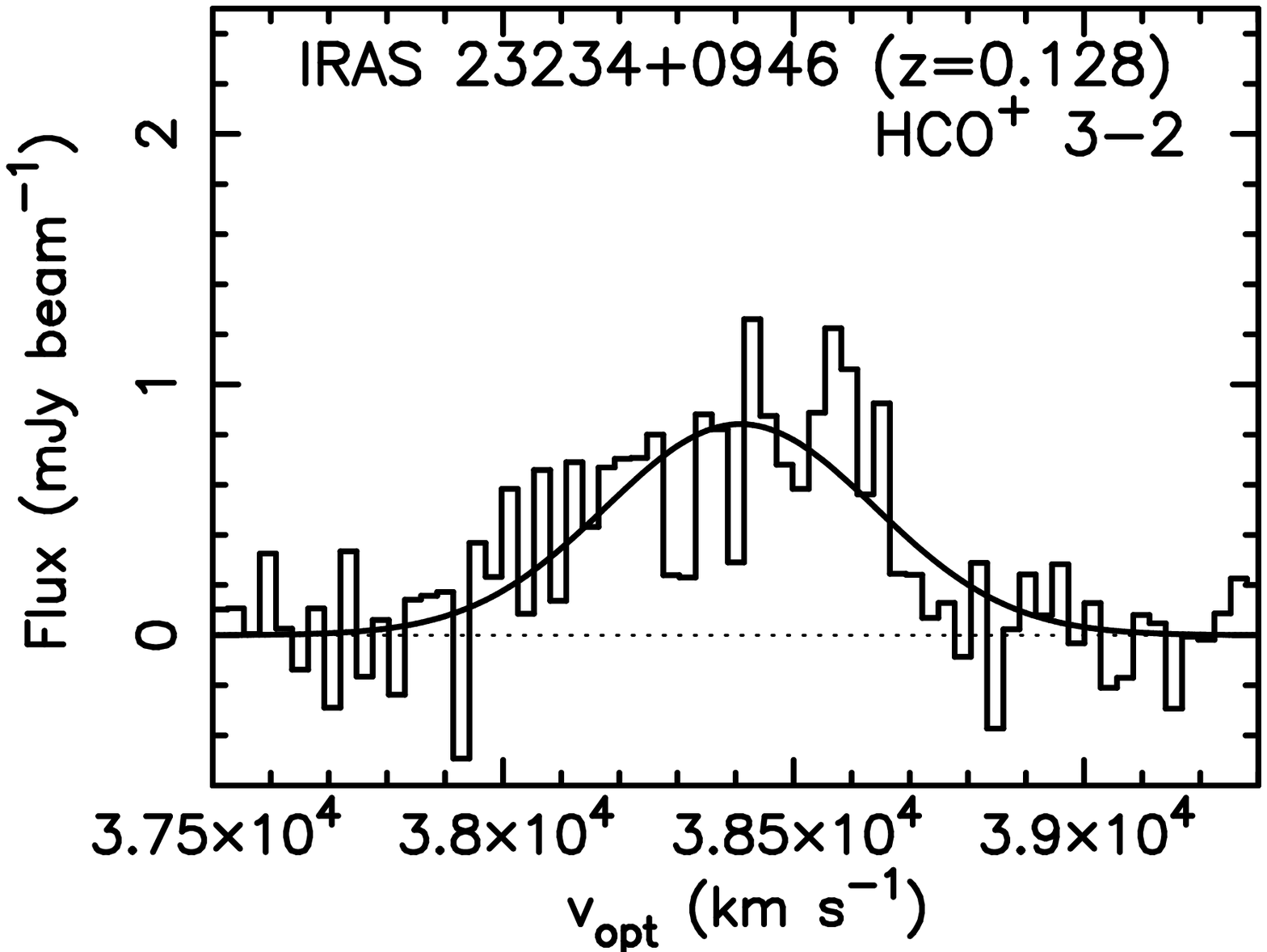} 
\includegraphics[angle=0,scale=.223]{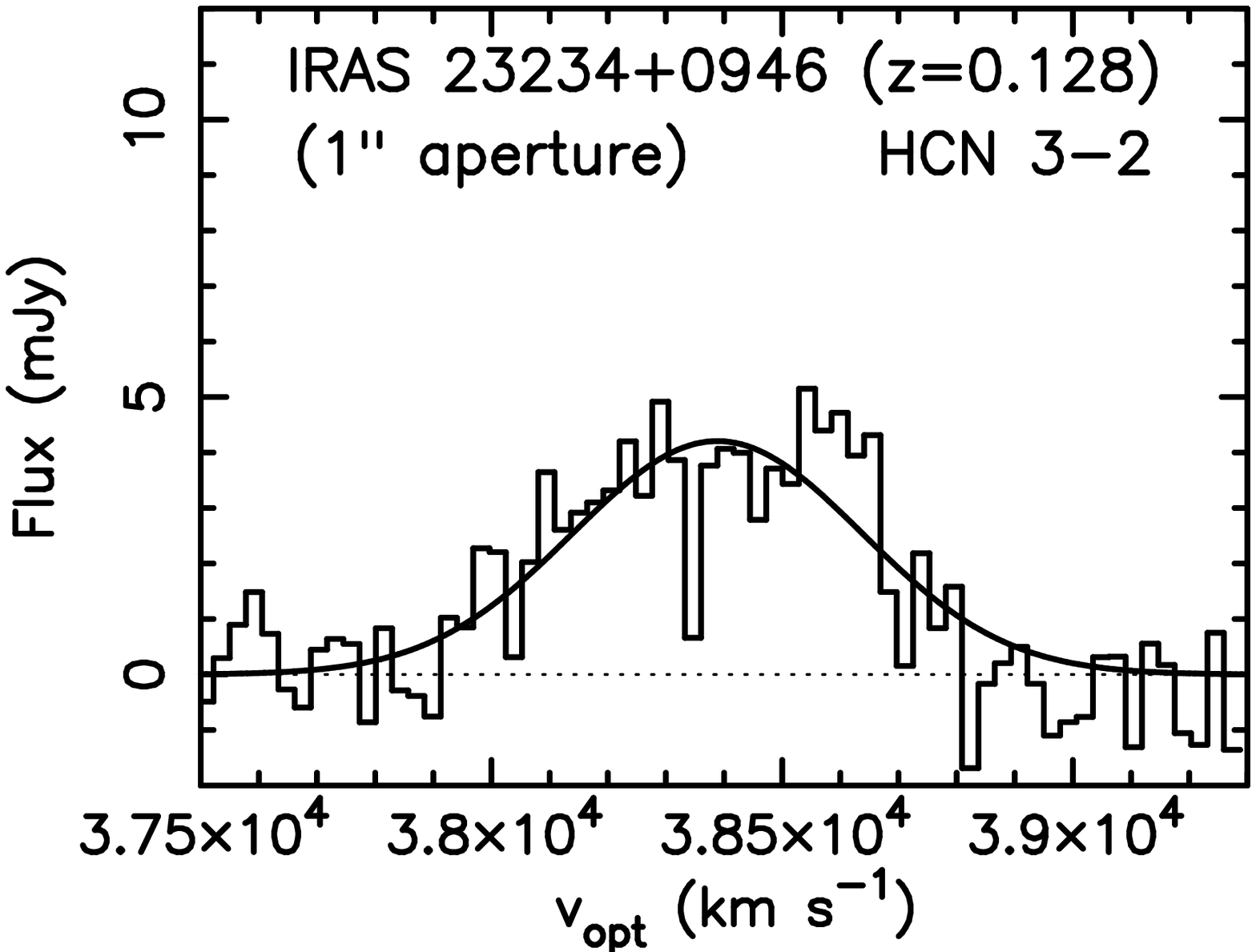} 
\includegraphics[angle=0,scale=.223]{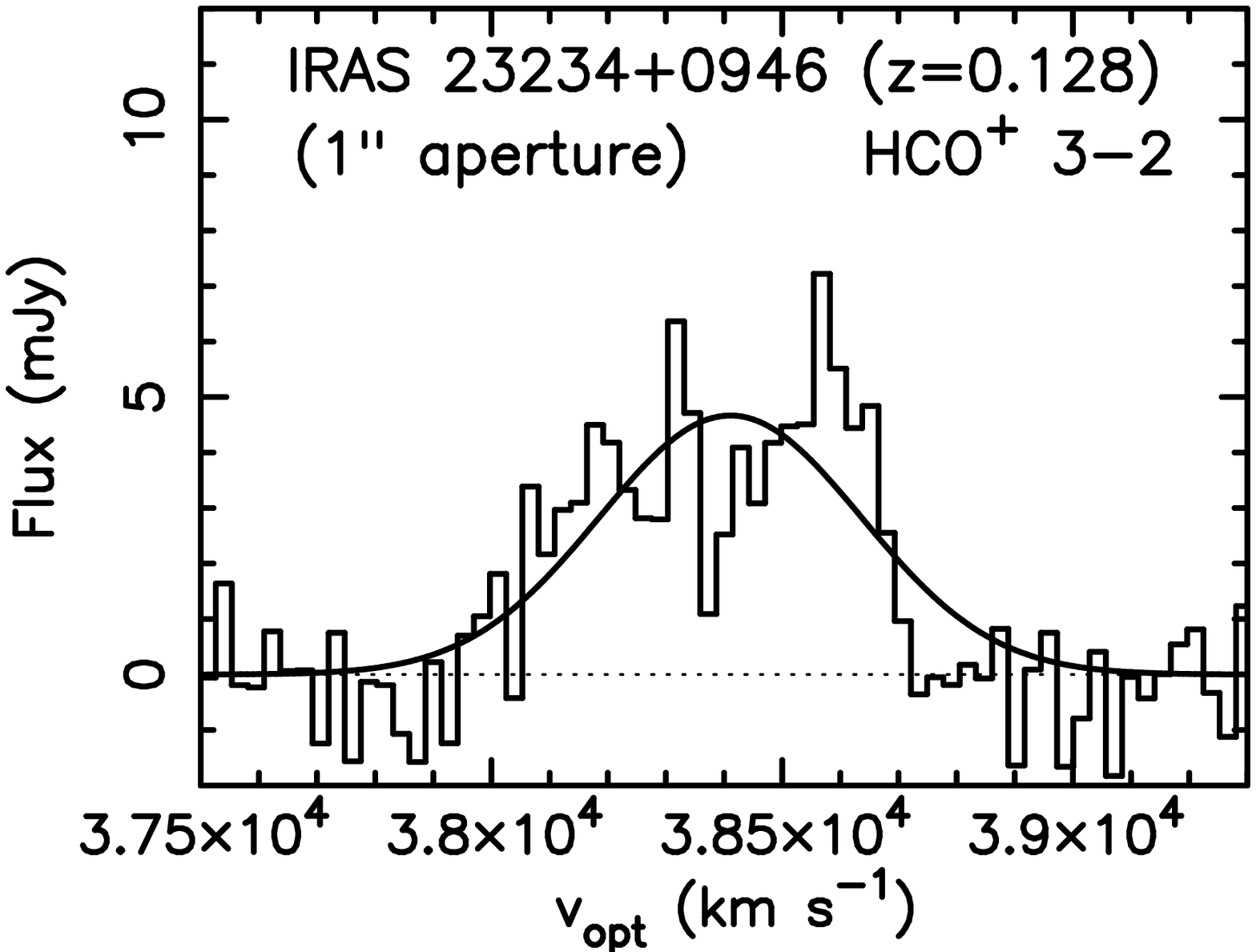} \\
\includegraphics[angle=0,scale=.223]{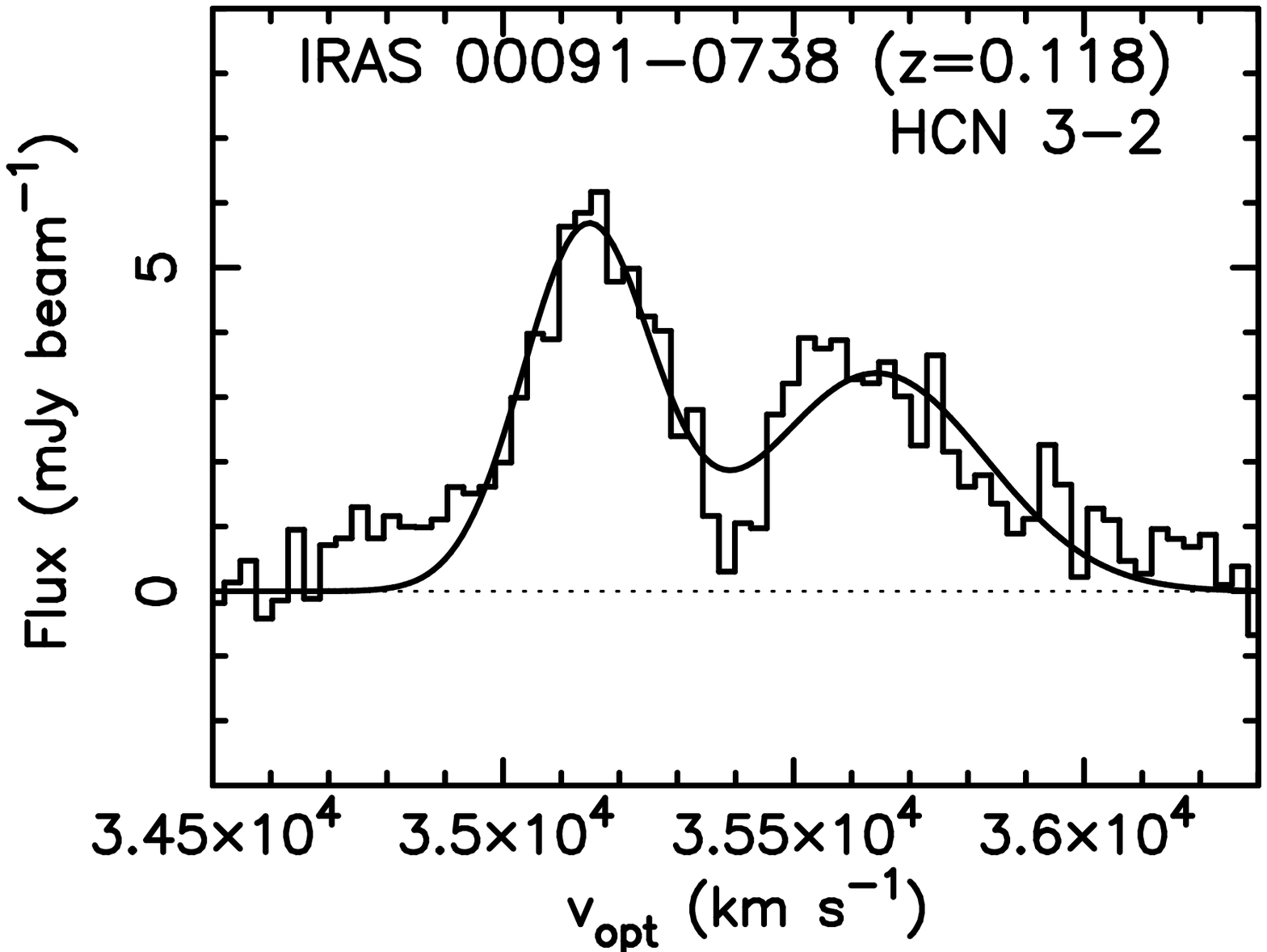} 
\includegraphics[angle=0,scale=.223]{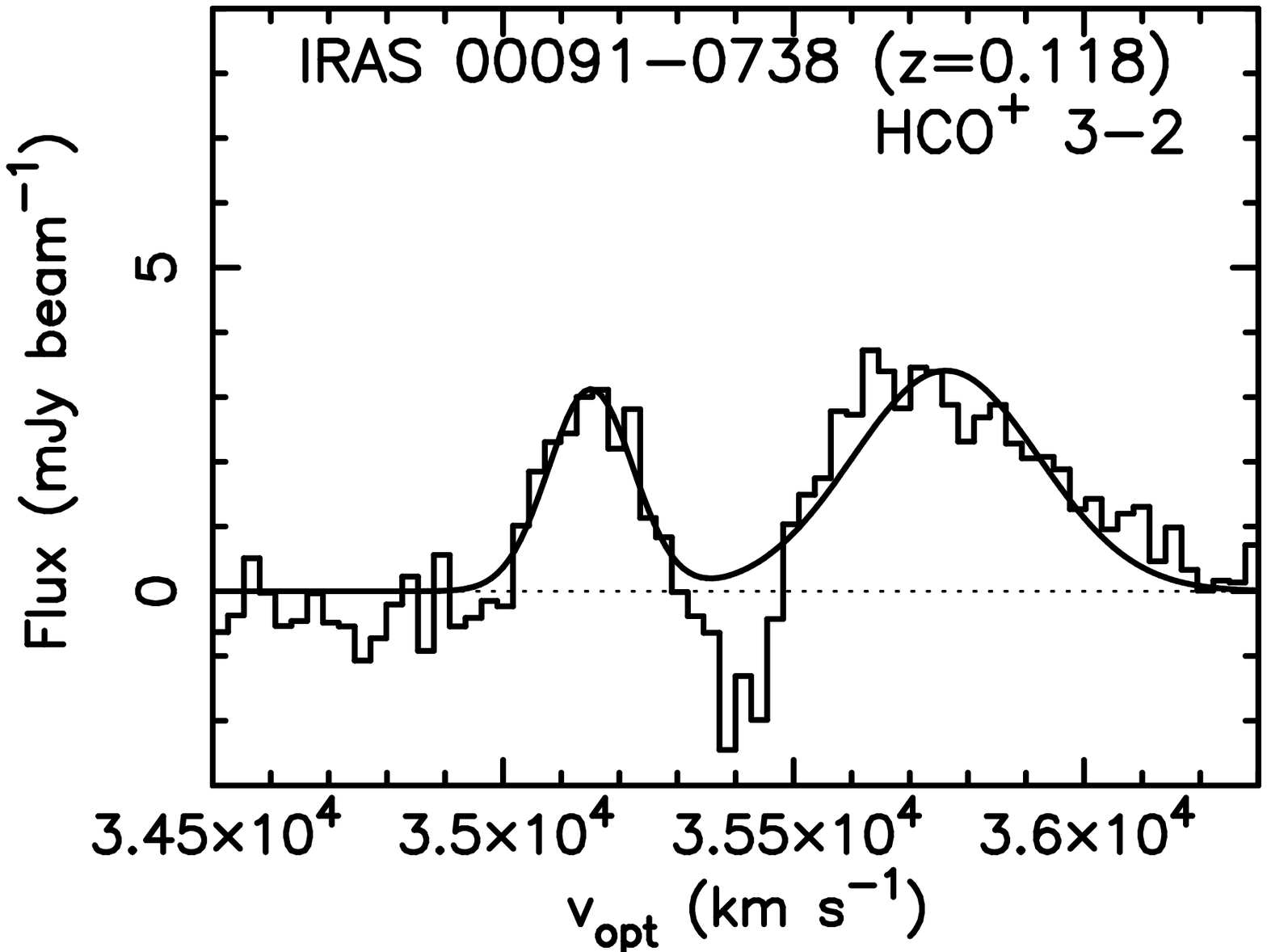} 
\includegraphics[angle=0,scale=.223]{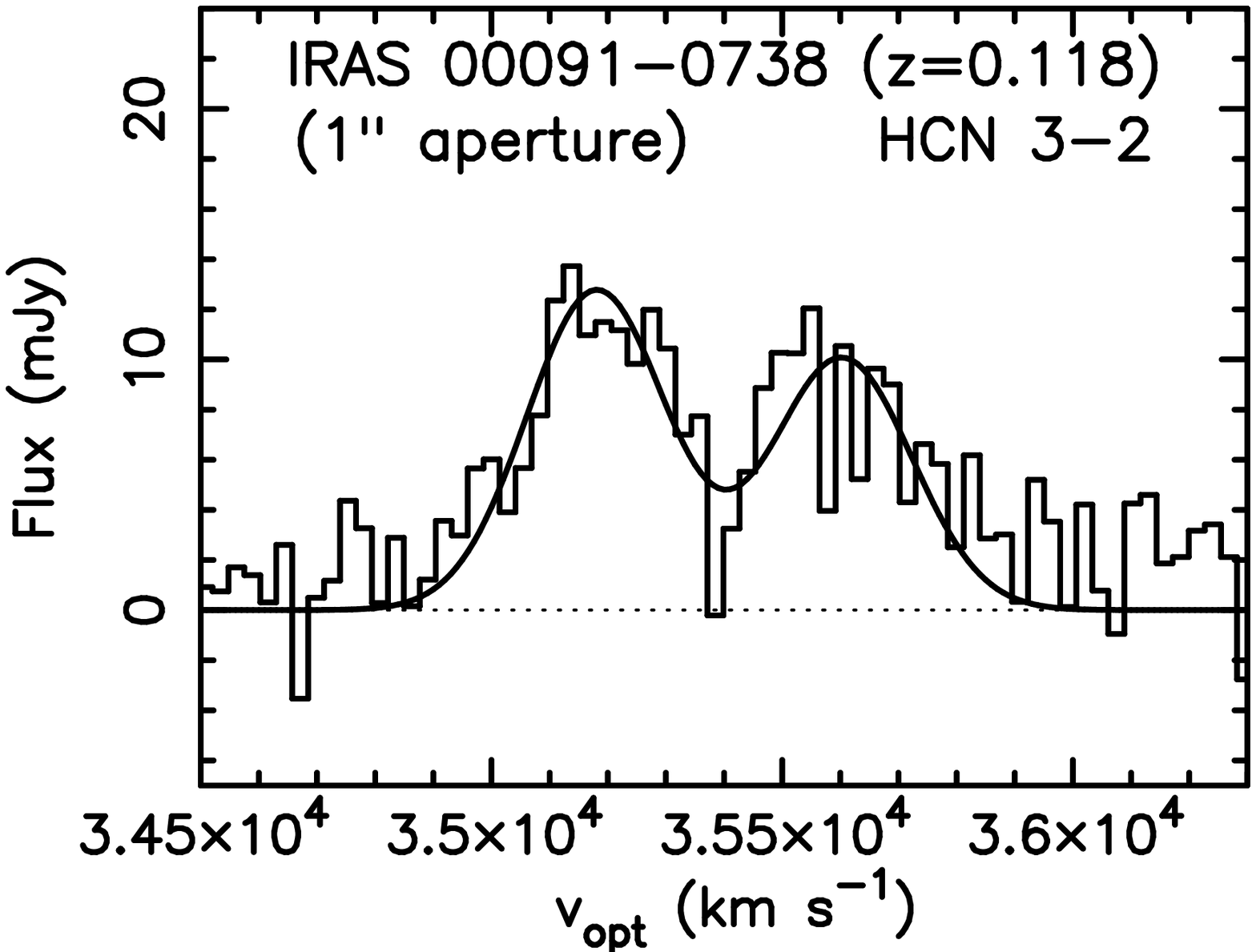} 
\includegraphics[angle=0,scale=.223]{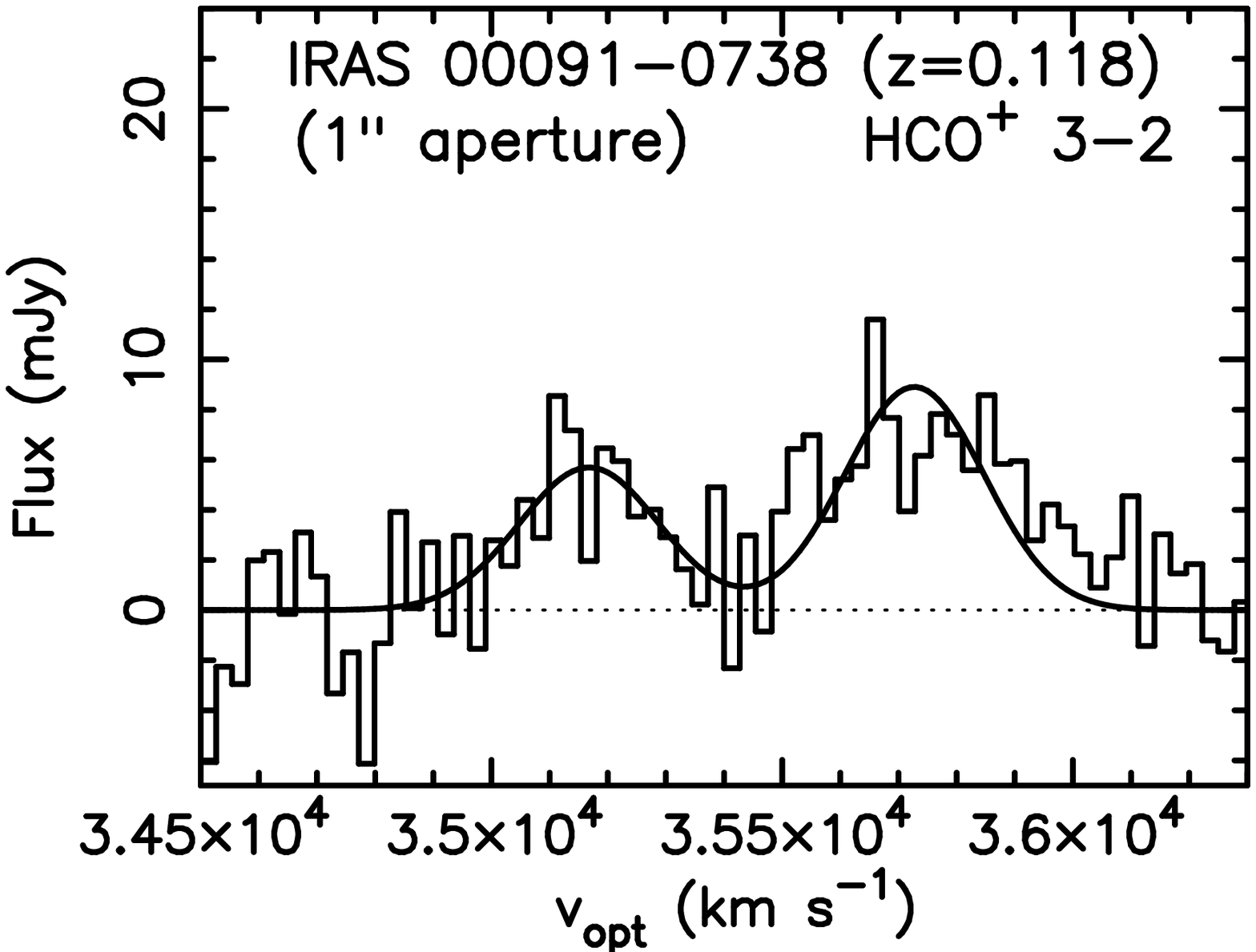} \\
\includegraphics[angle=0,scale=.223]{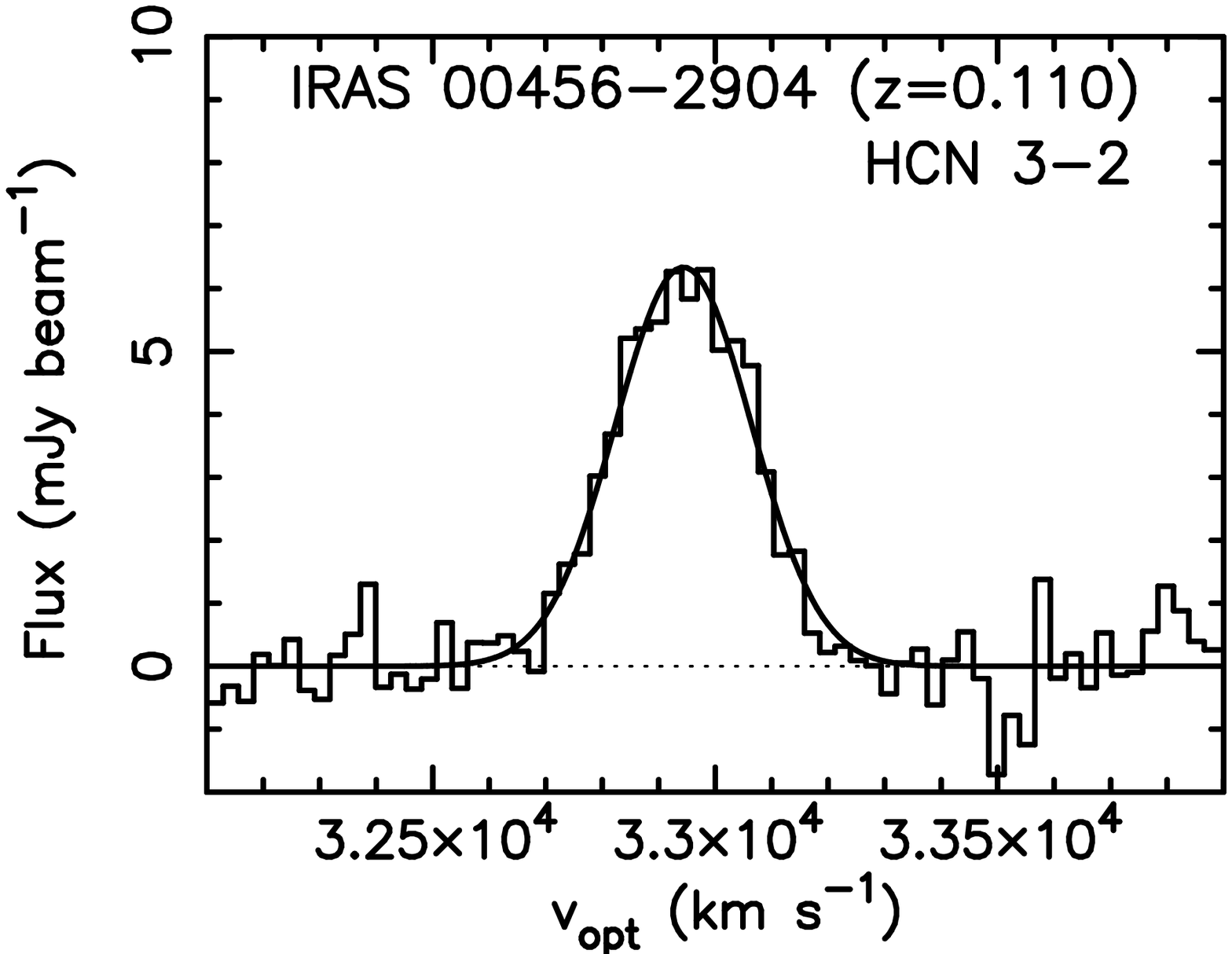} 
\includegraphics[angle=0,scale=.223]{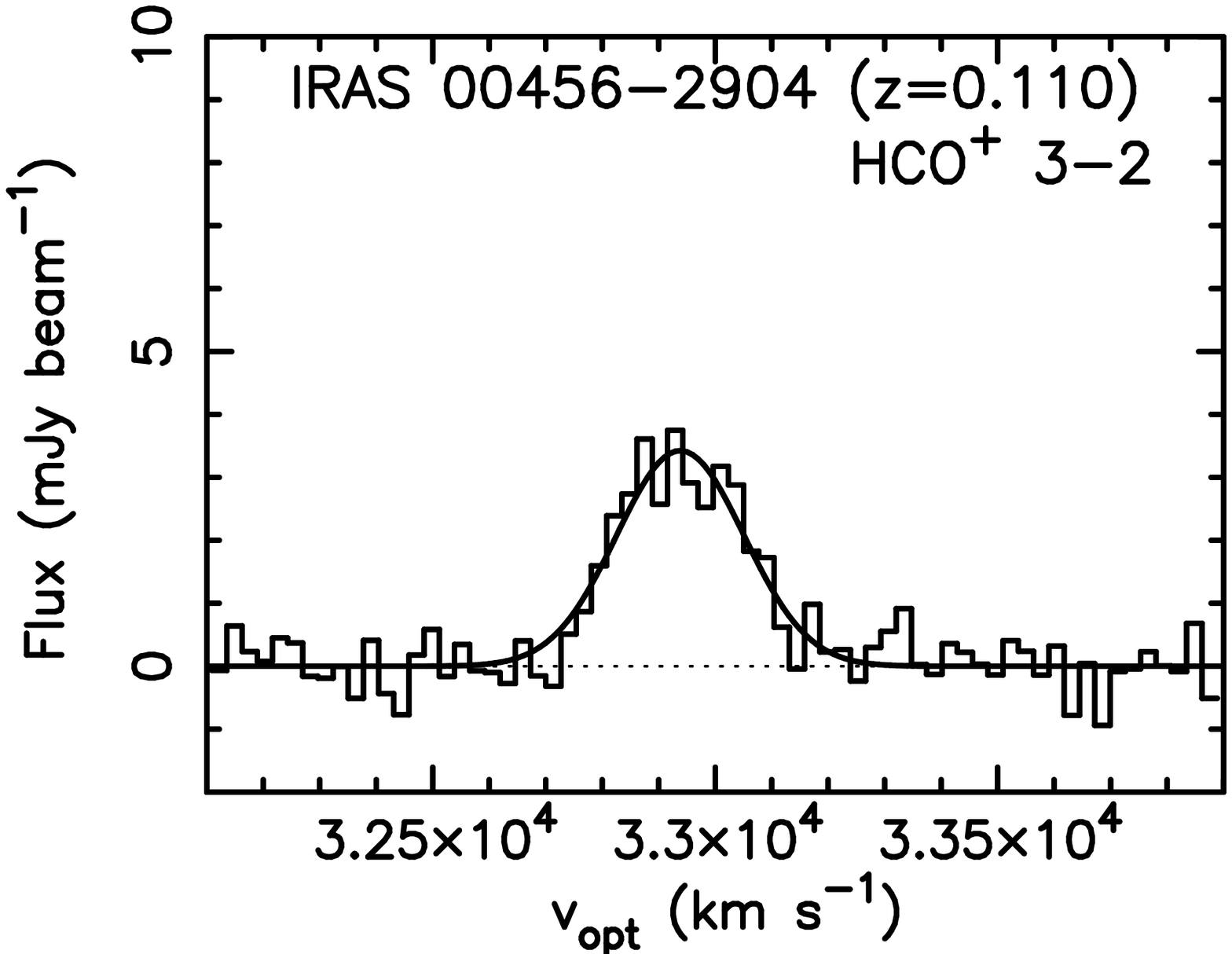} 
\includegraphics[angle=0,scale=.223]{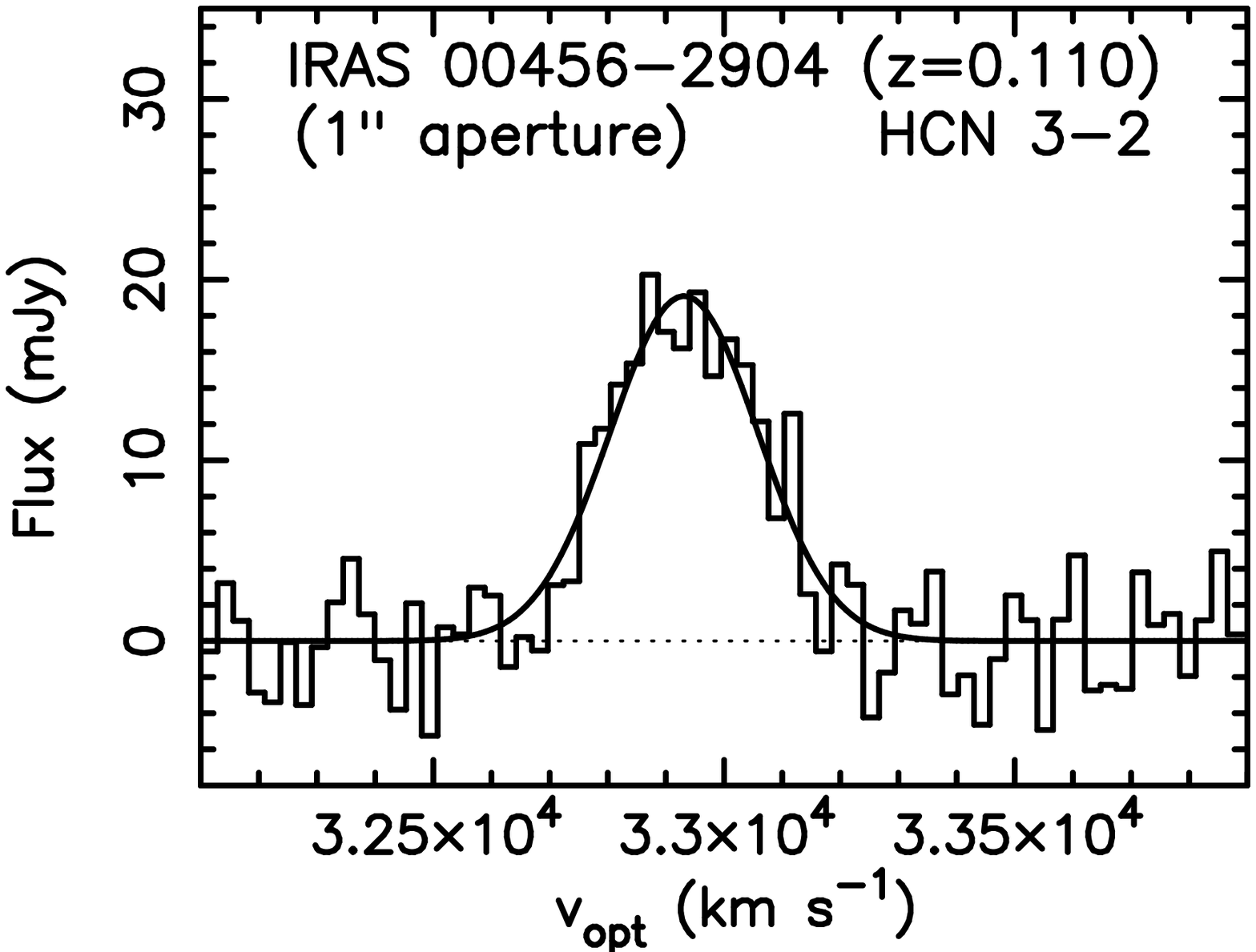} 
\includegraphics[angle=0,scale=.223]{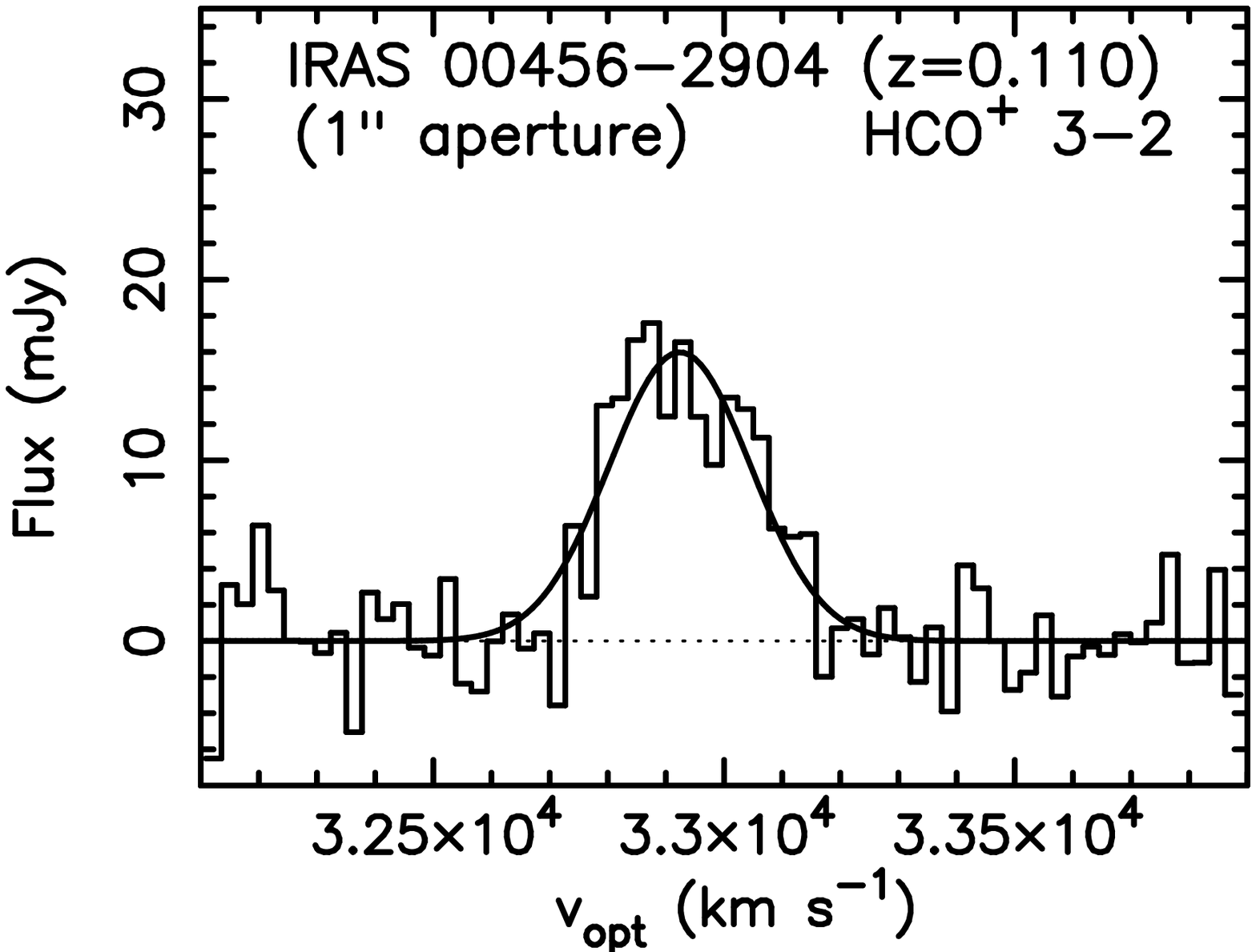} \\
\includegraphics[angle=0,scale=.223]{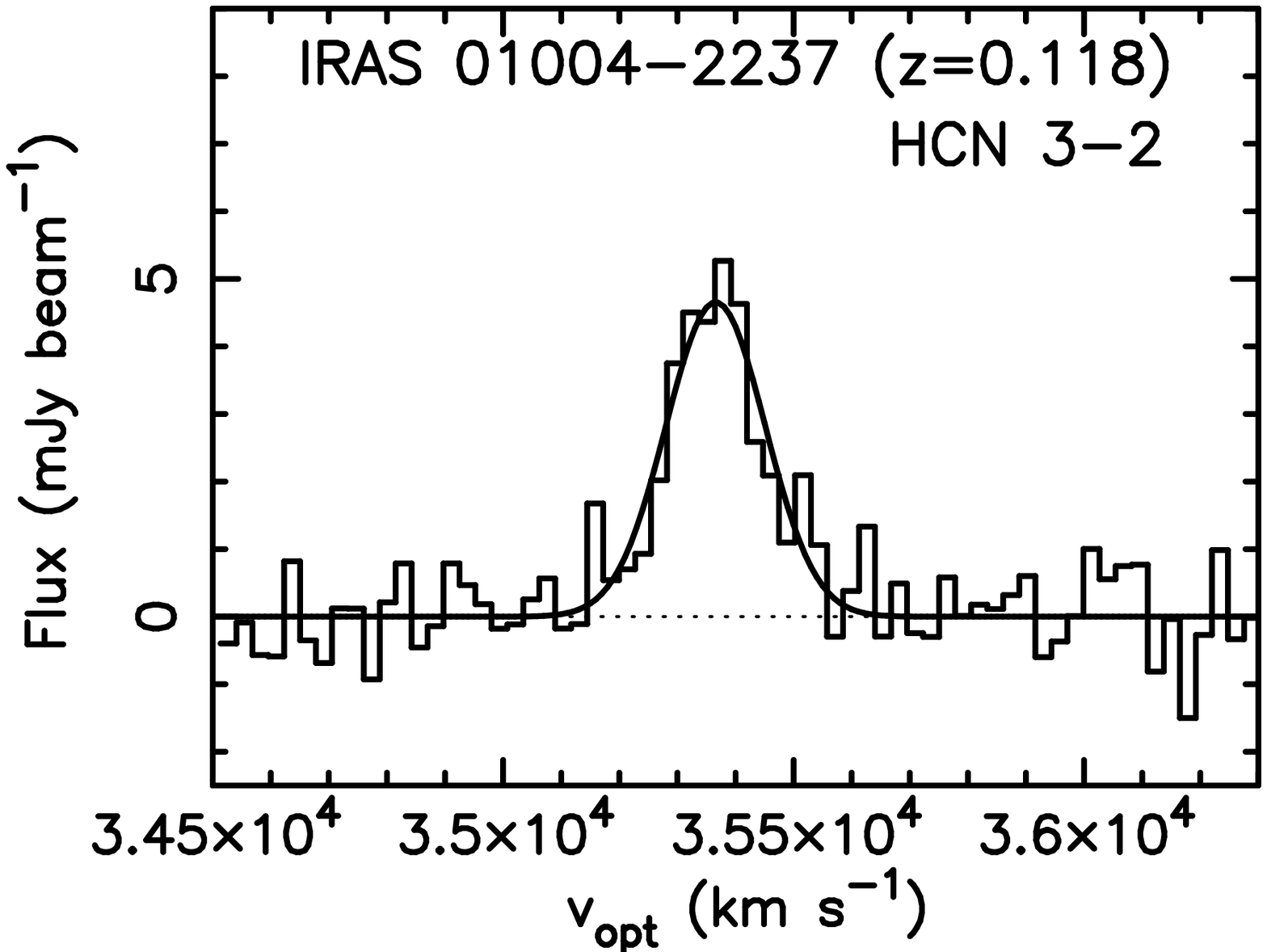} 
\includegraphics[angle=0,scale=.223]{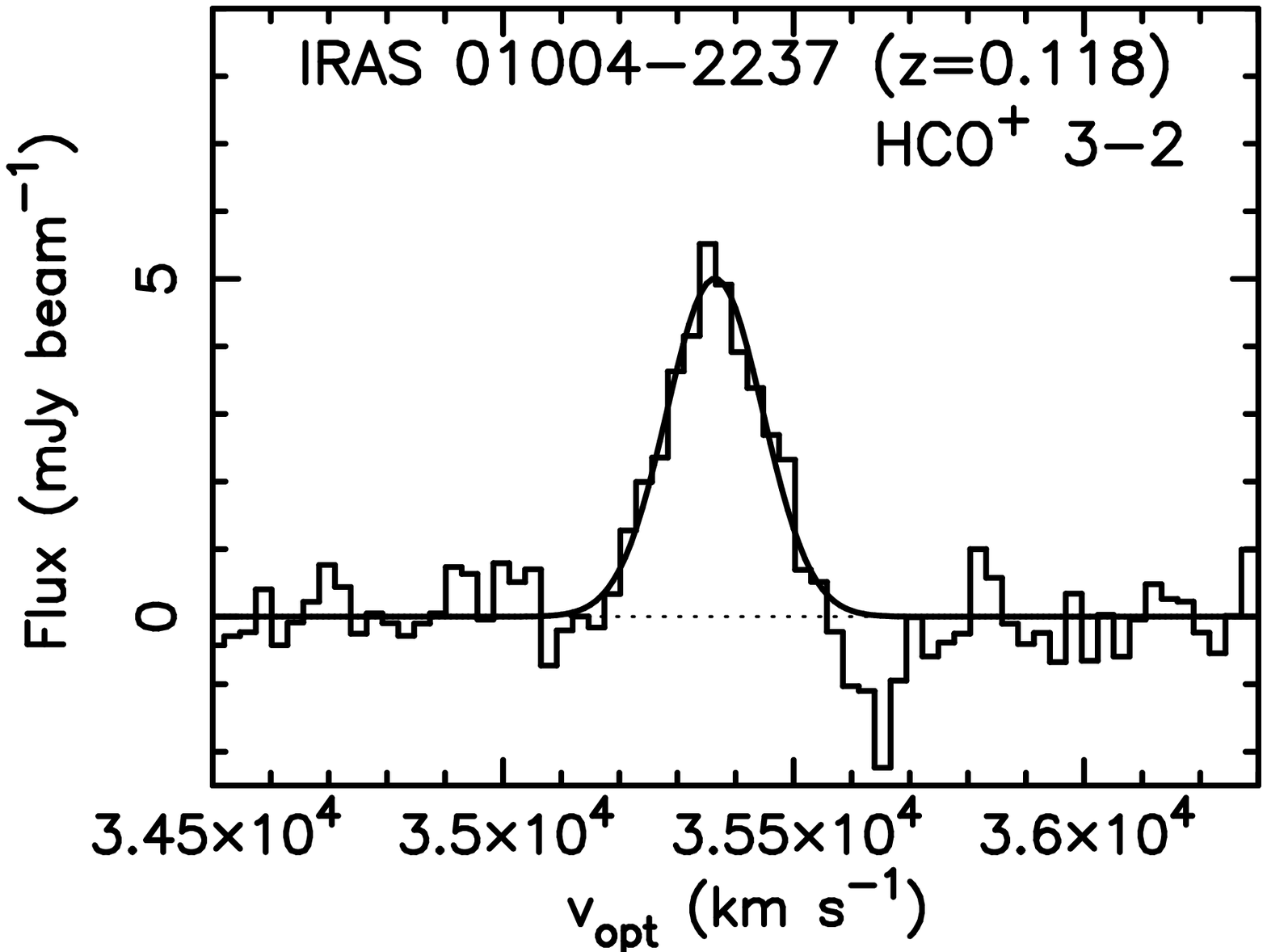} 
\includegraphics[angle=0,scale=.223]{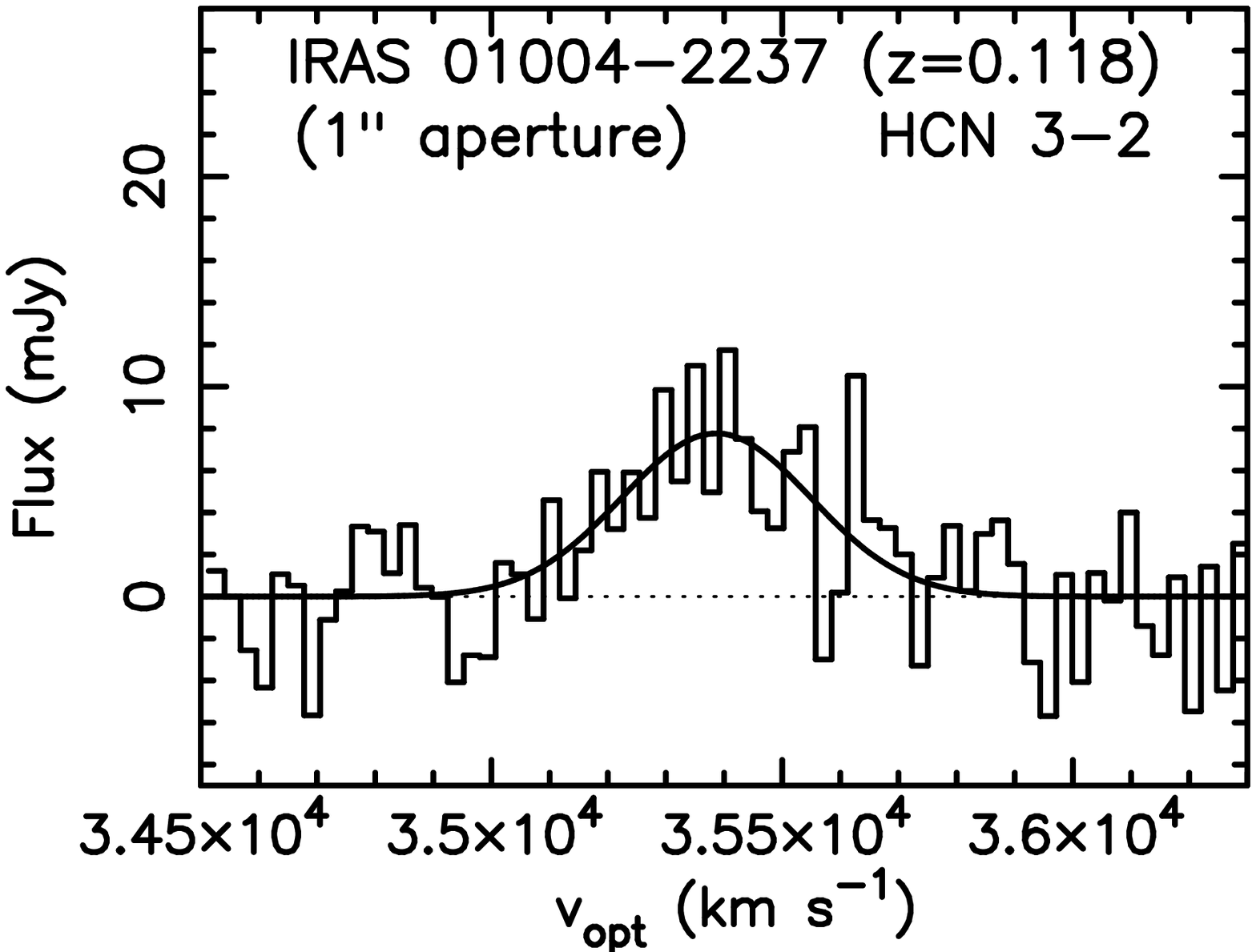} 
\includegraphics[angle=0,scale=.223]{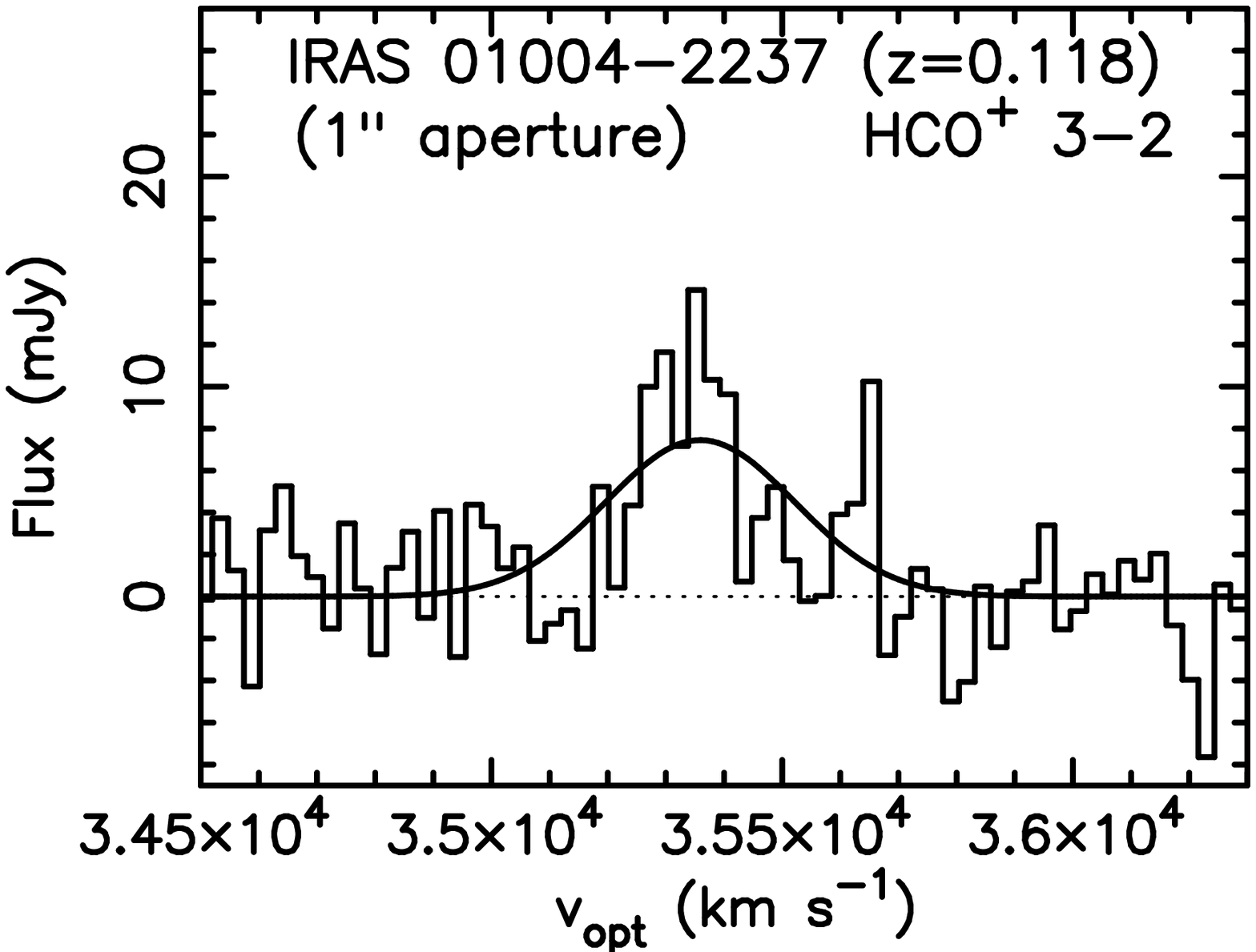} \\
\includegraphics[angle=0,scale=.223]{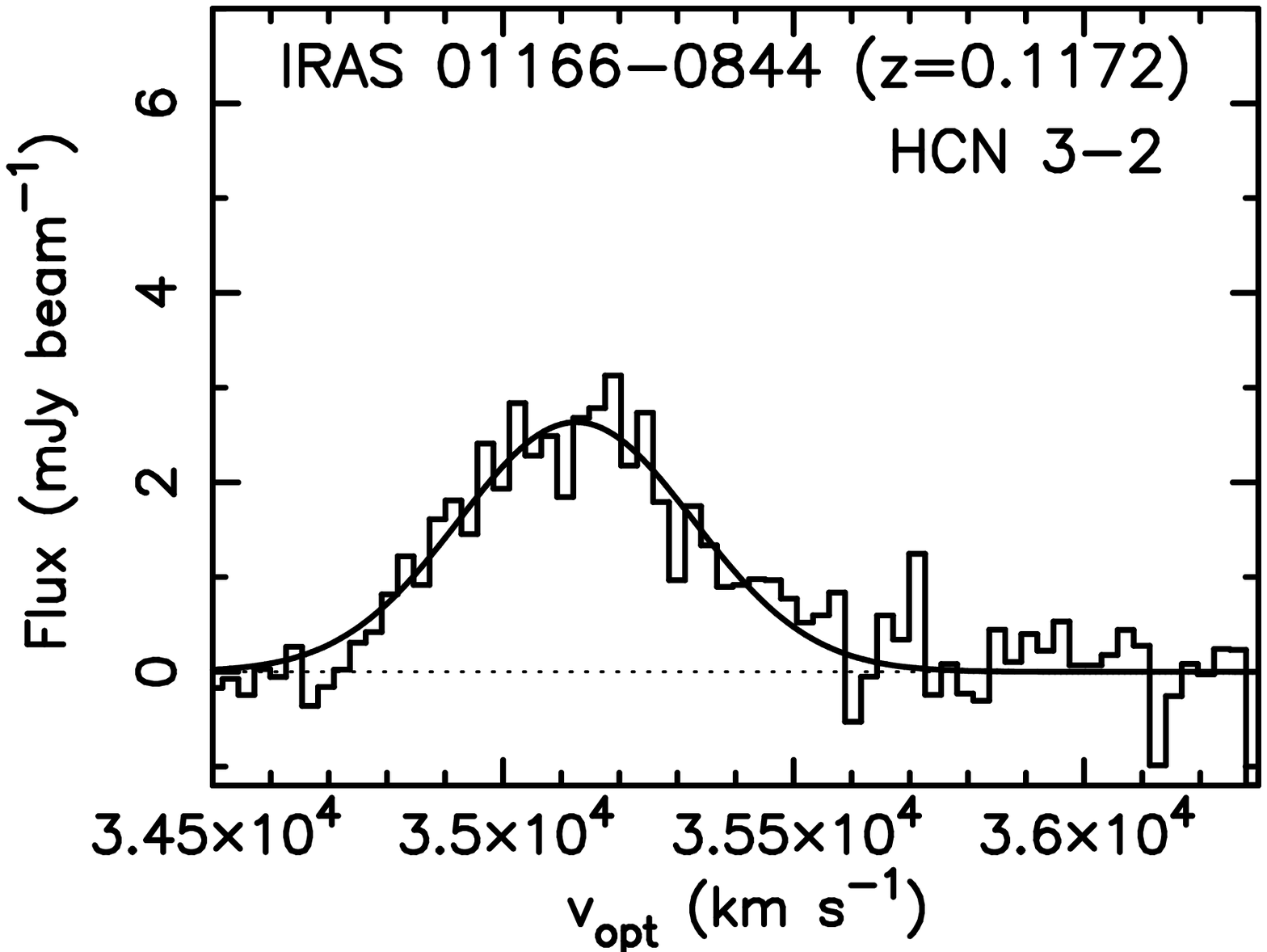} 
\includegraphics[angle=0,scale=.223]{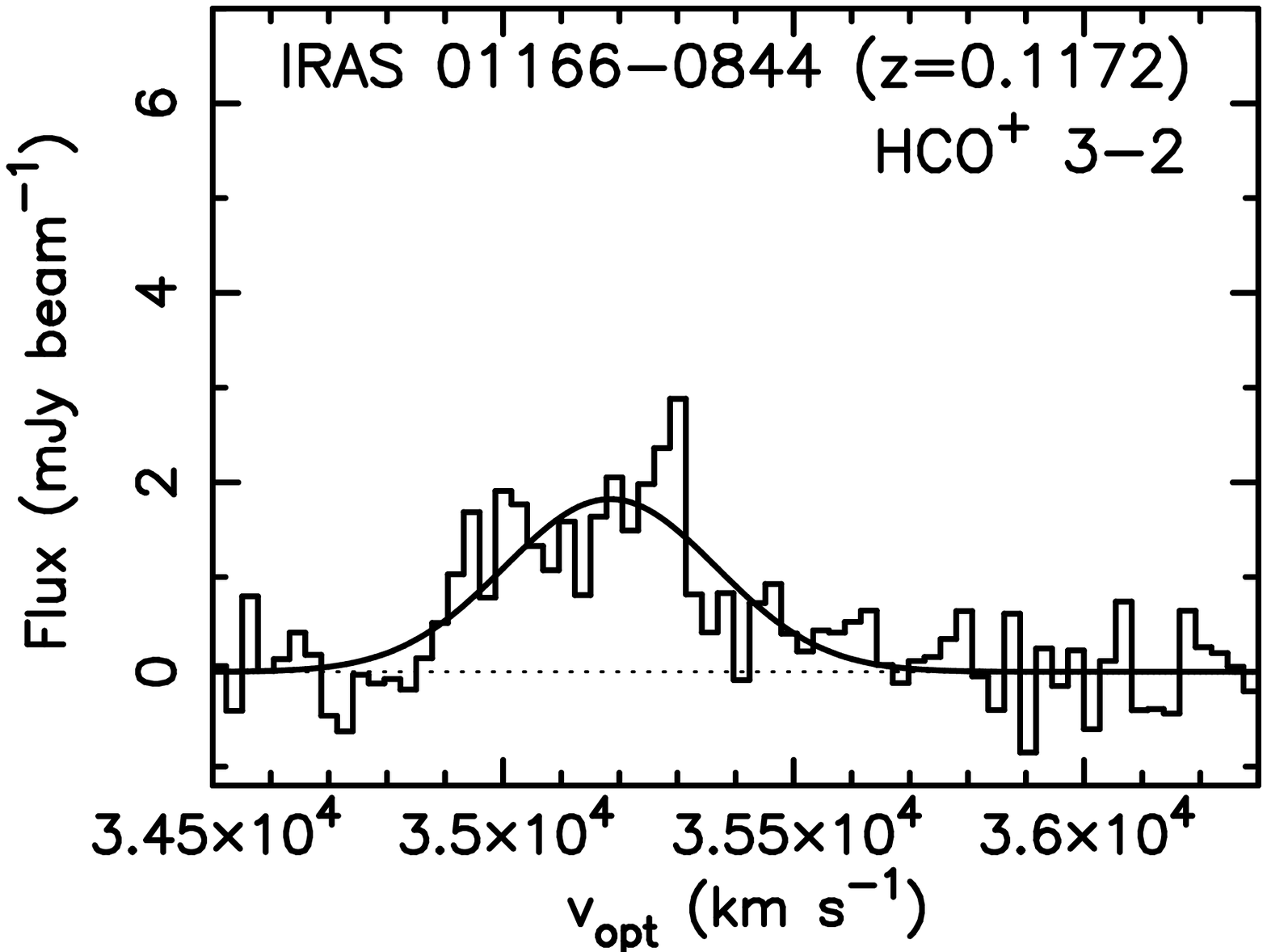} 
\includegraphics[angle=0,scale=.223]{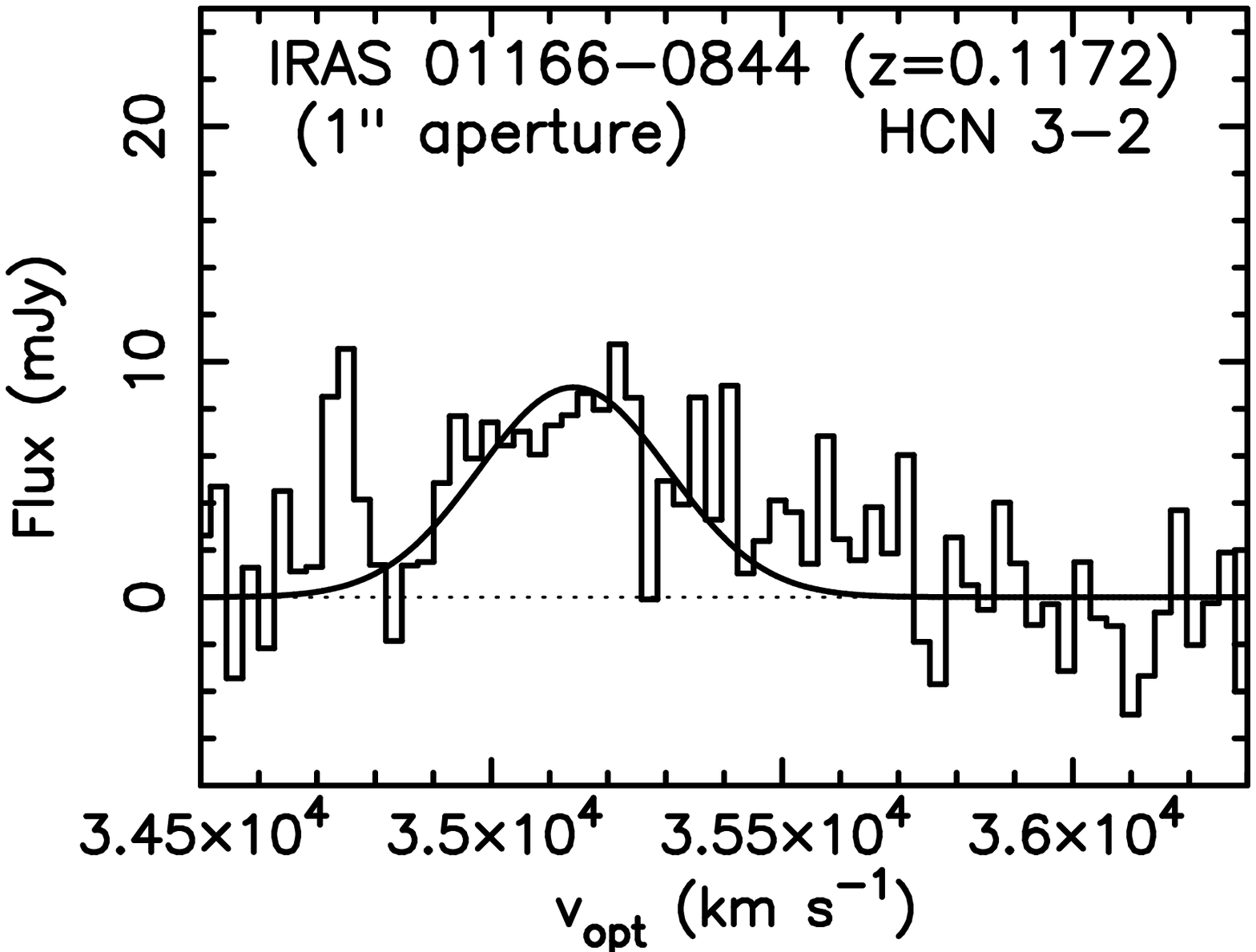} 
\includegraphics[angle=0,scale=.223]{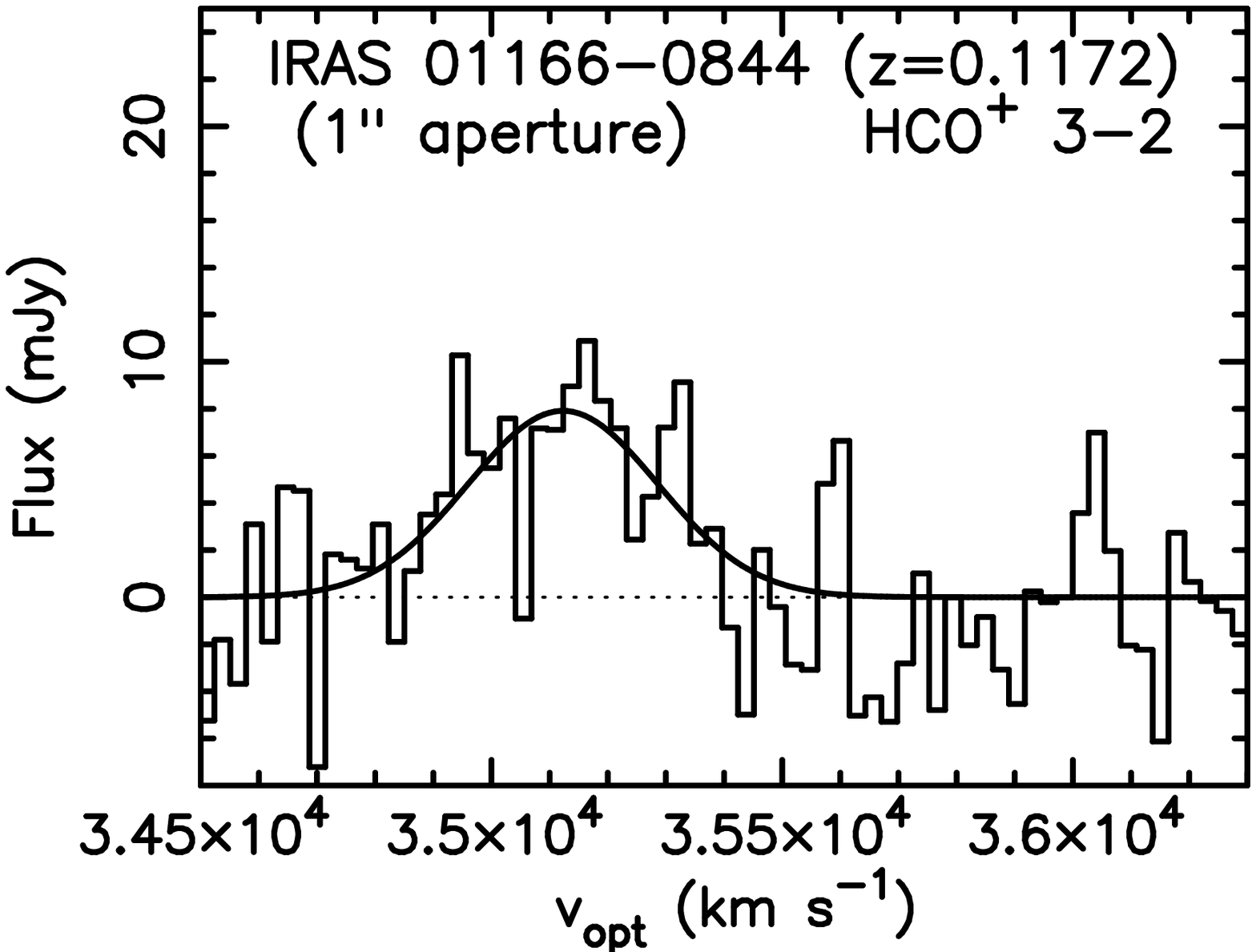} \\
\includegraphics[angle=0,scale=.223]{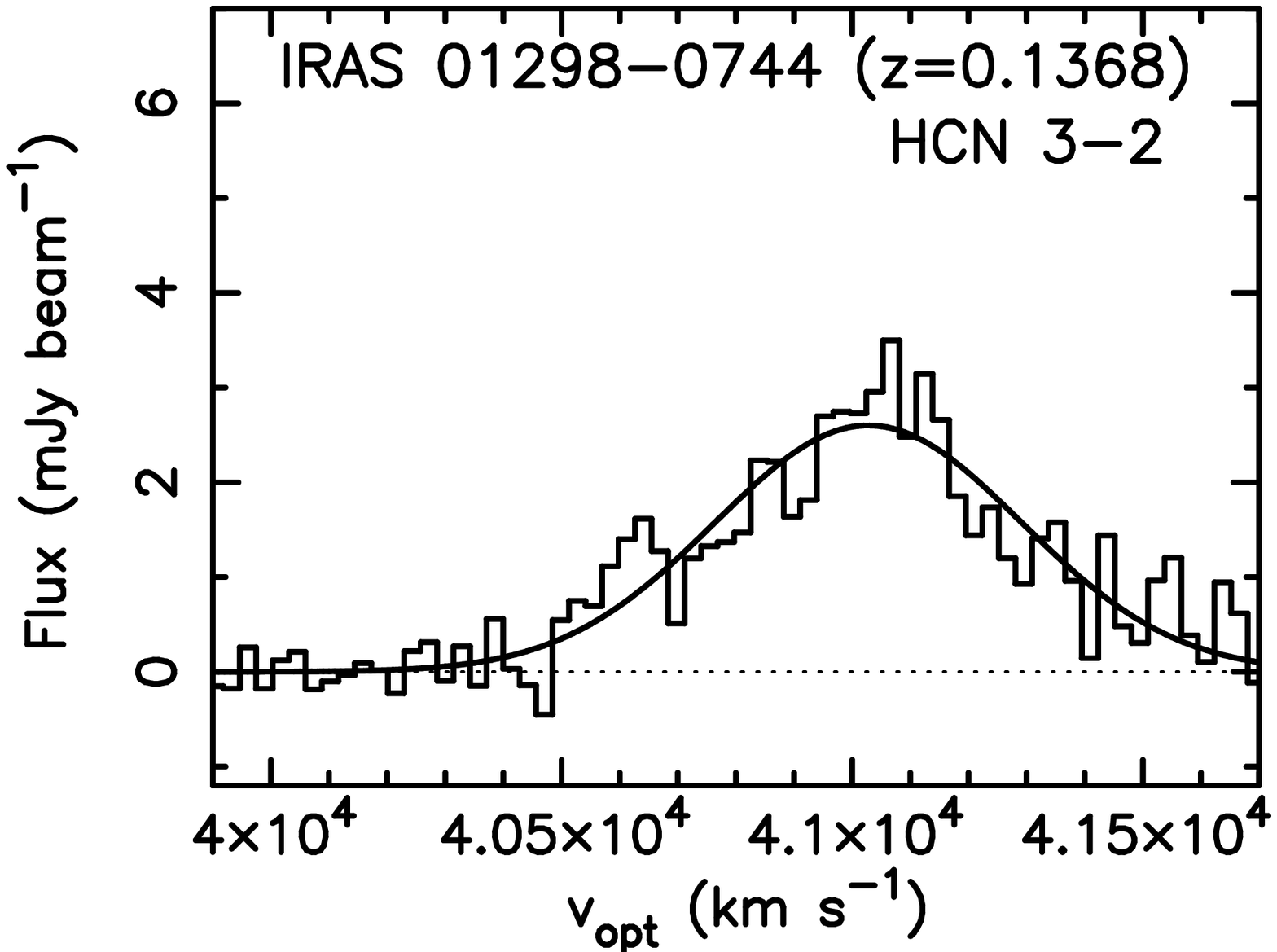} 
\includegraphics[angle=0,scale=.223]{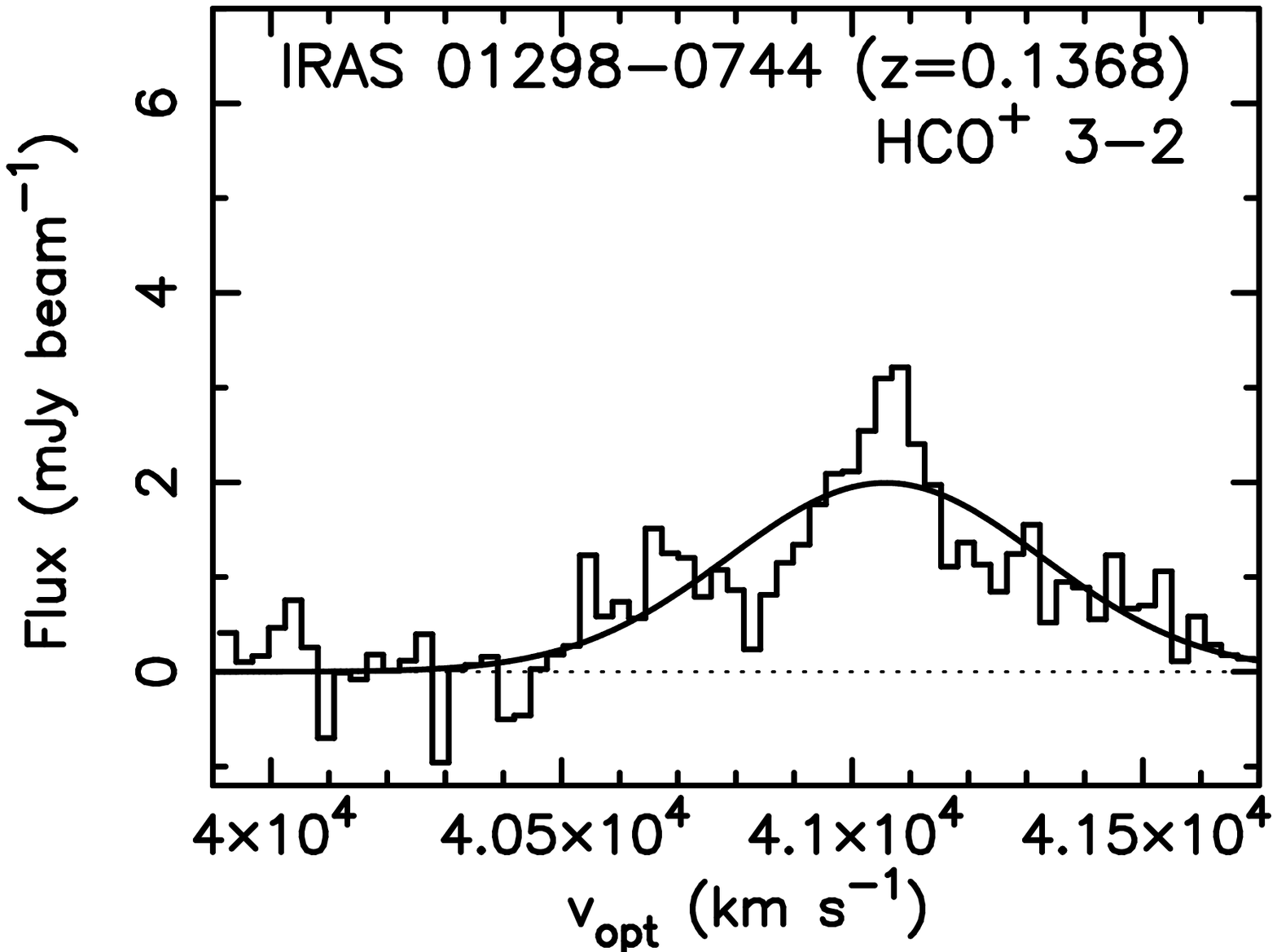} 
\includegraphics[angle=0,scale=.223]{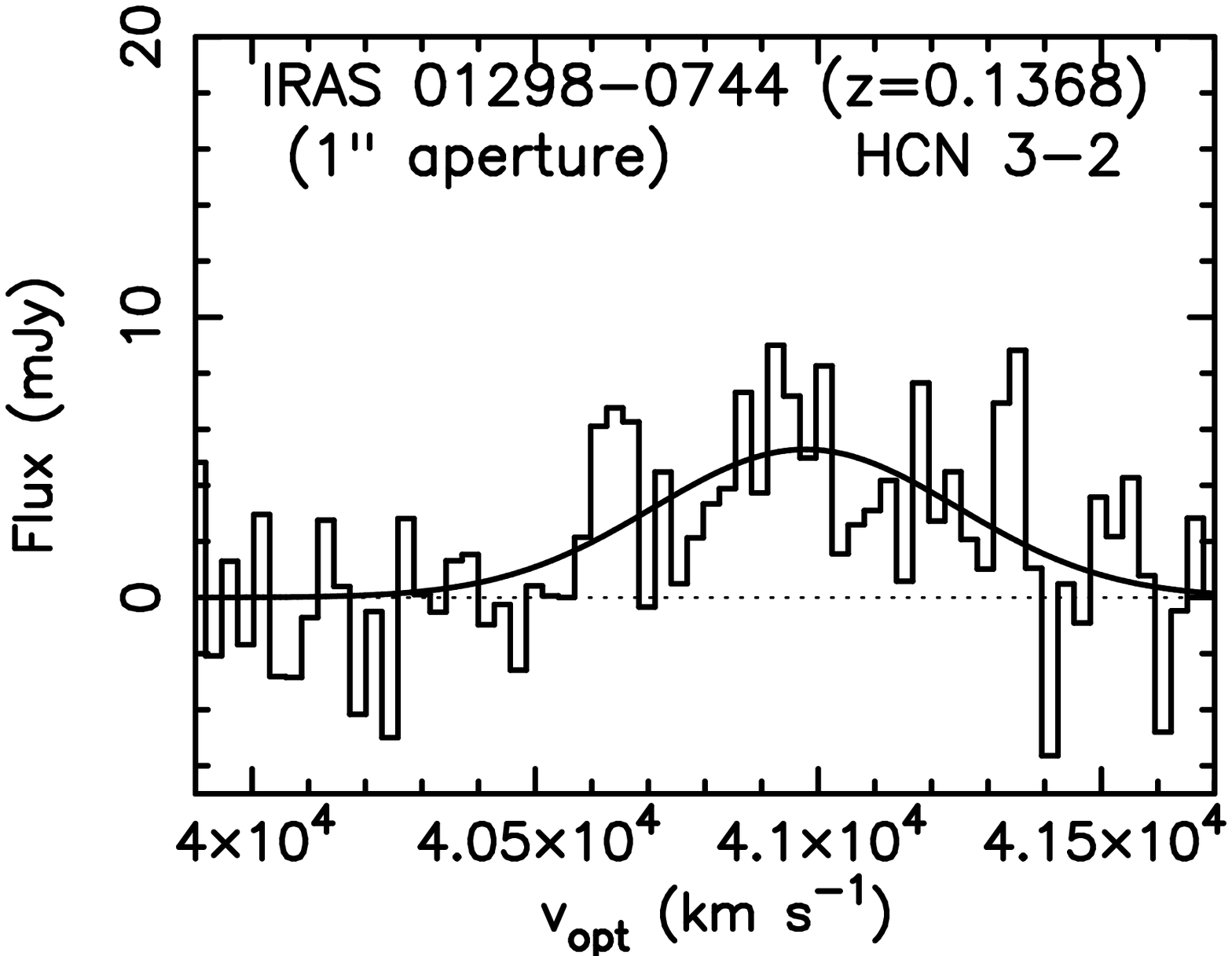} 
\includegraphics[angle=0,scale=.223]{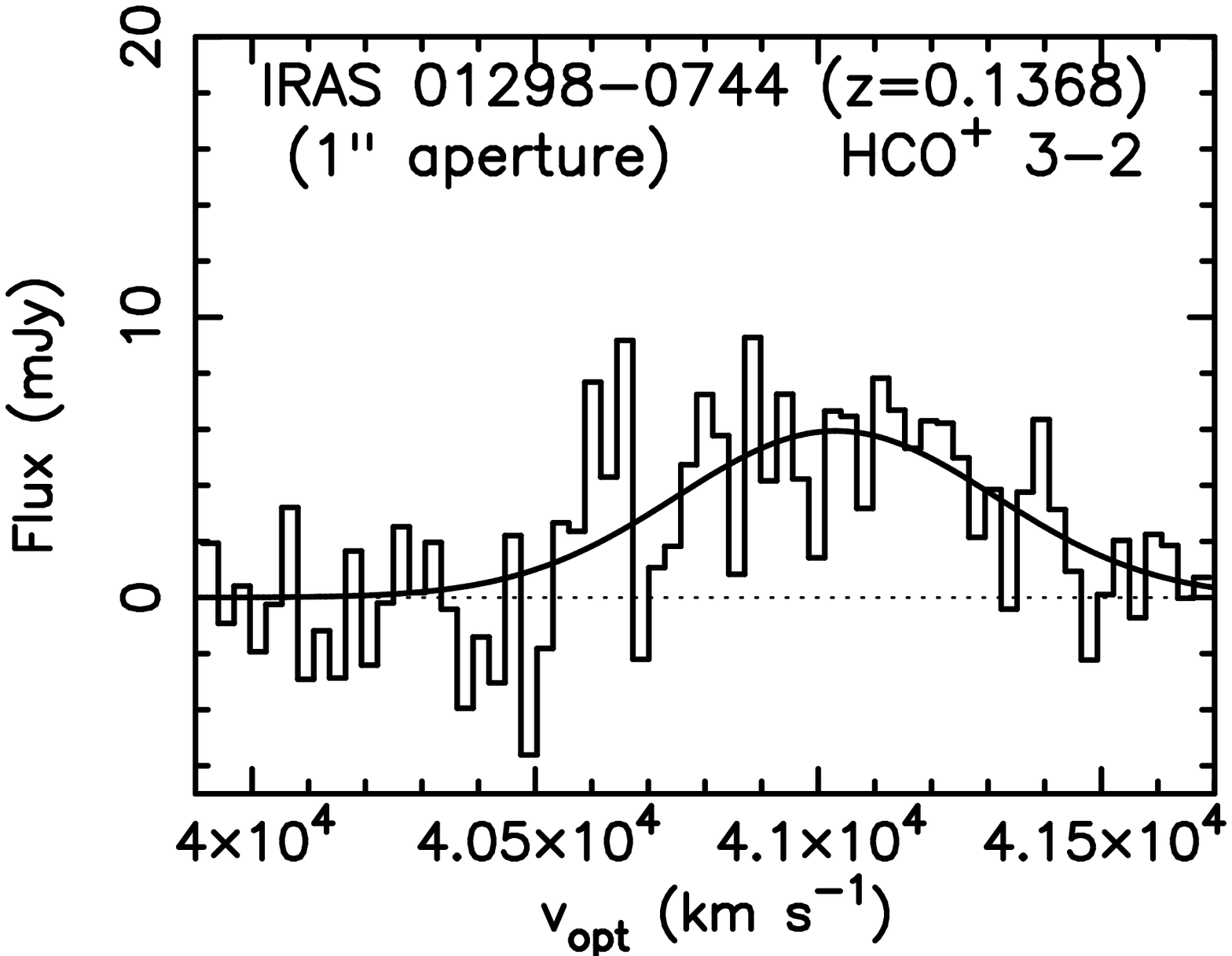} \\
\includegraphics[angle=0,scale=.223]{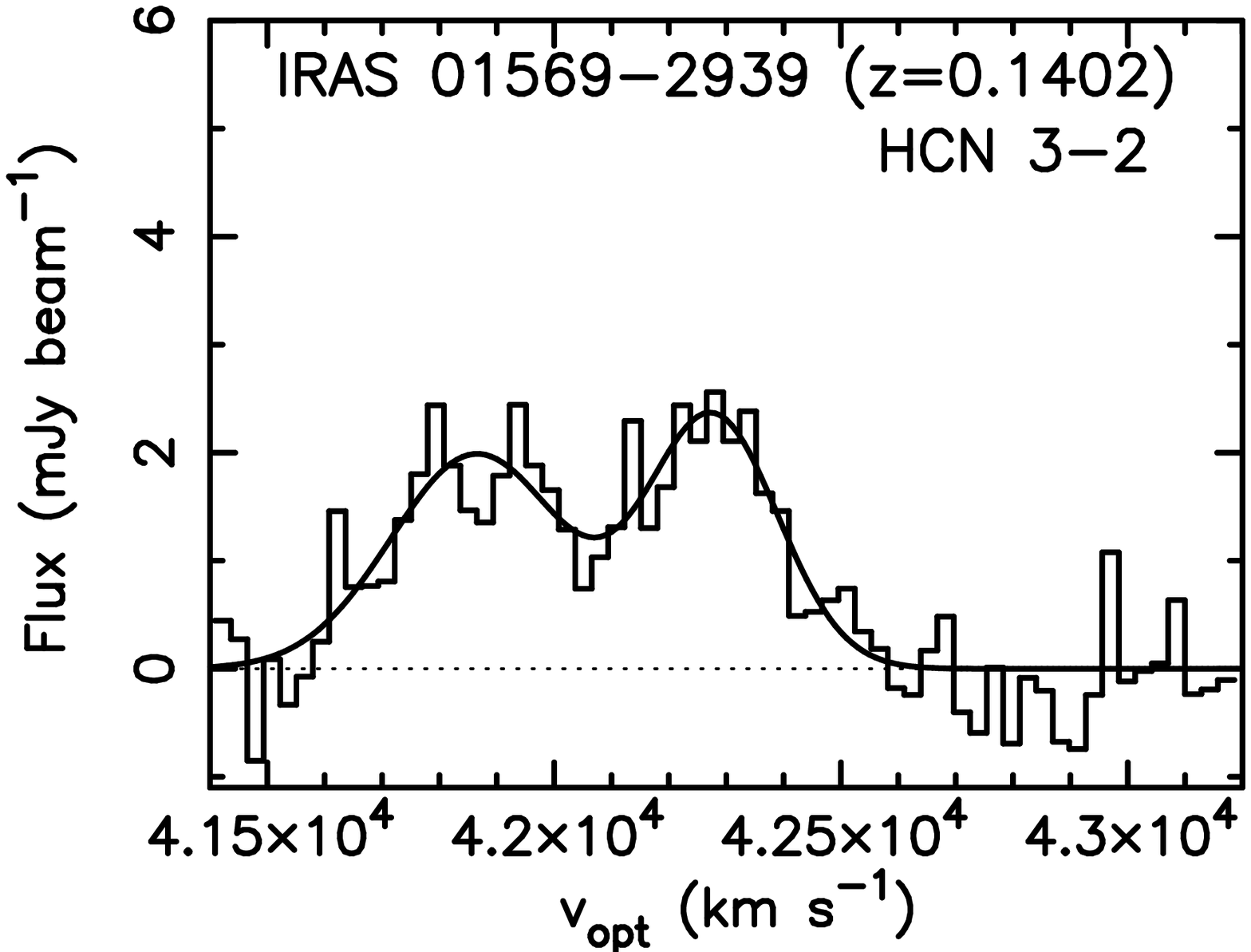} 
\includegraphics[angle=0,scale=.223]{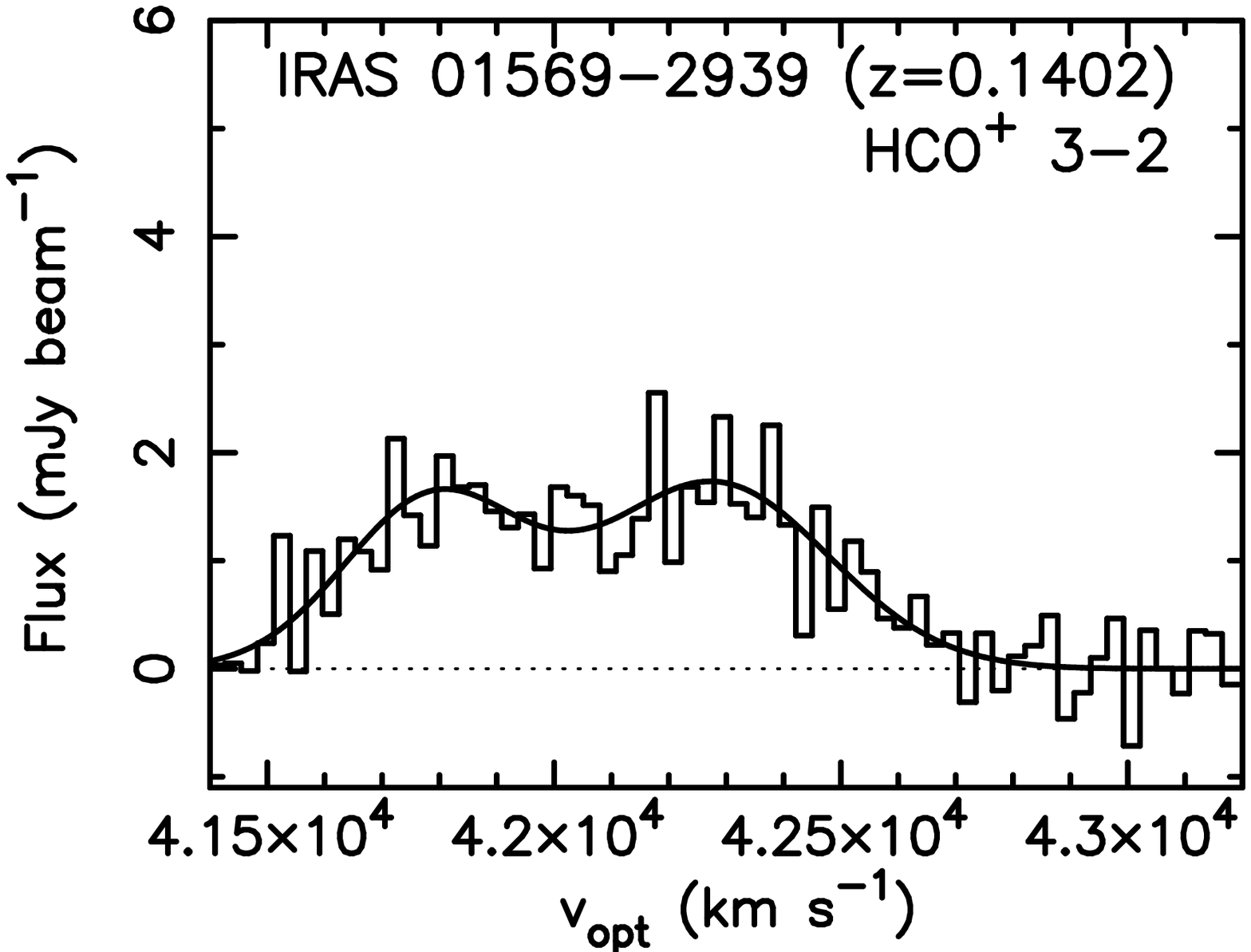} 
\includegraphics[angle=0,scale=.223]{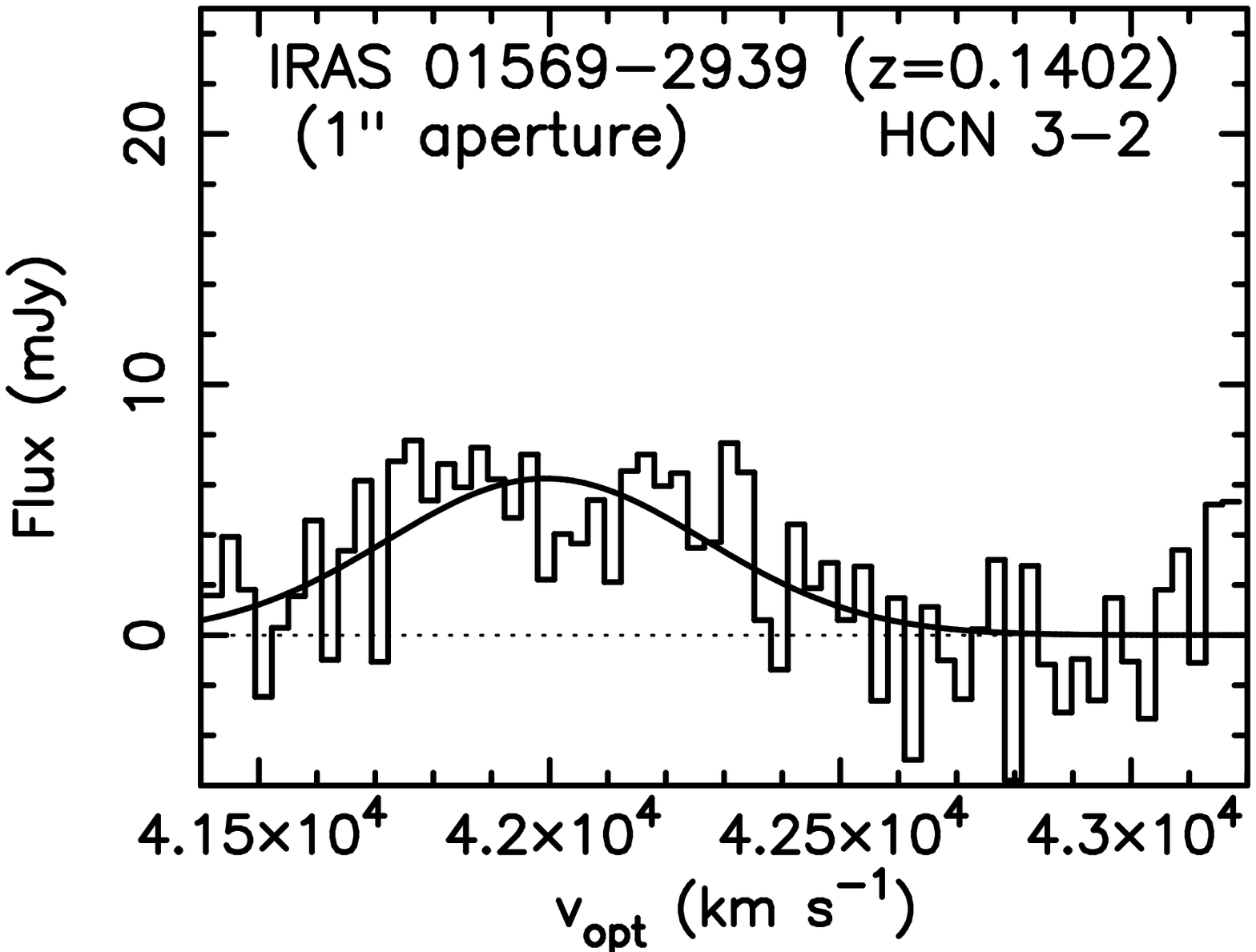} 
\includegraphics[angle=0,scale=.223]{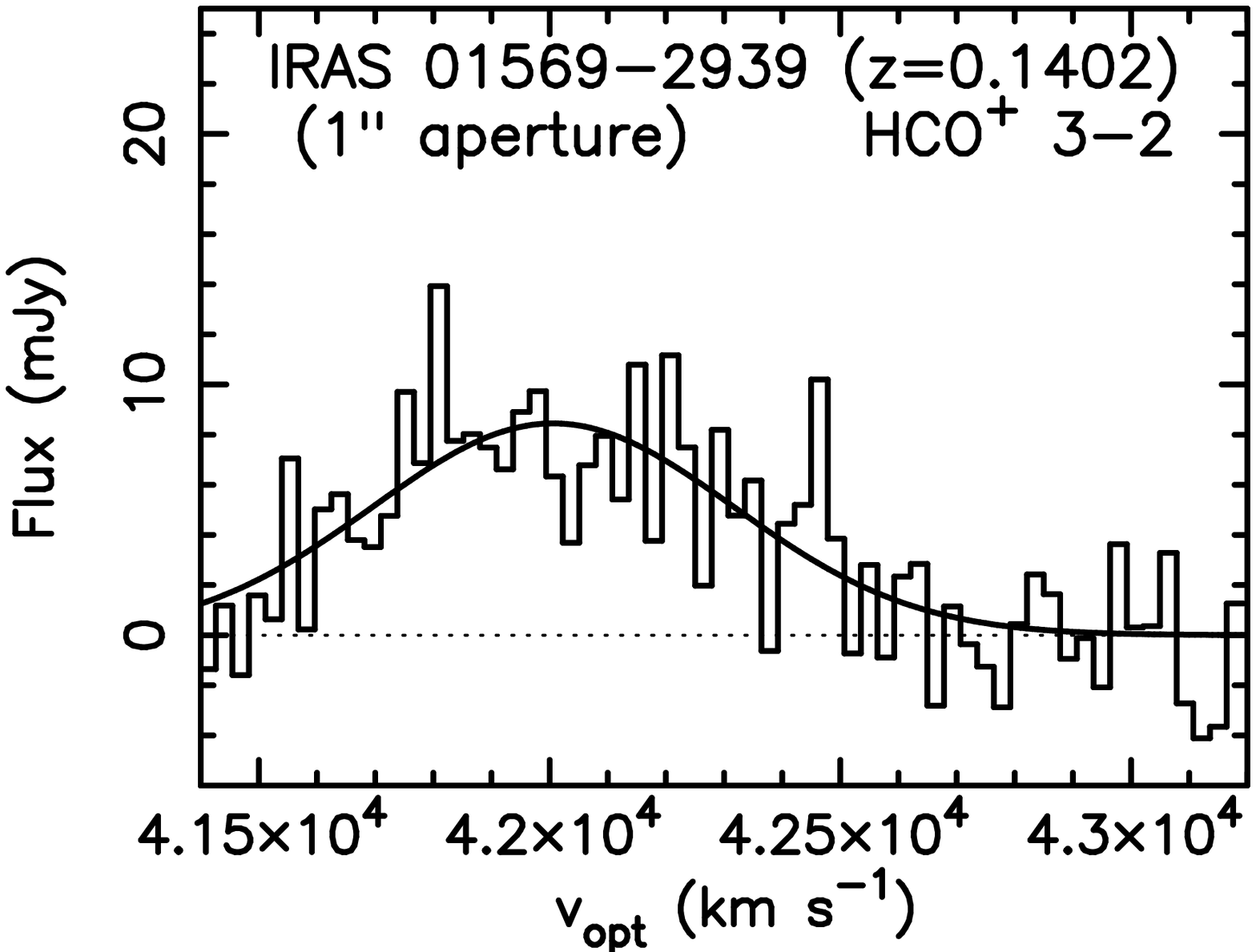} \\
\end{center}
\end{figure}

\clearpage

\begin{figure}
\begin{center}
\hspace*{-9.2cm}
\includegraphics[angle=0,scale=.223]{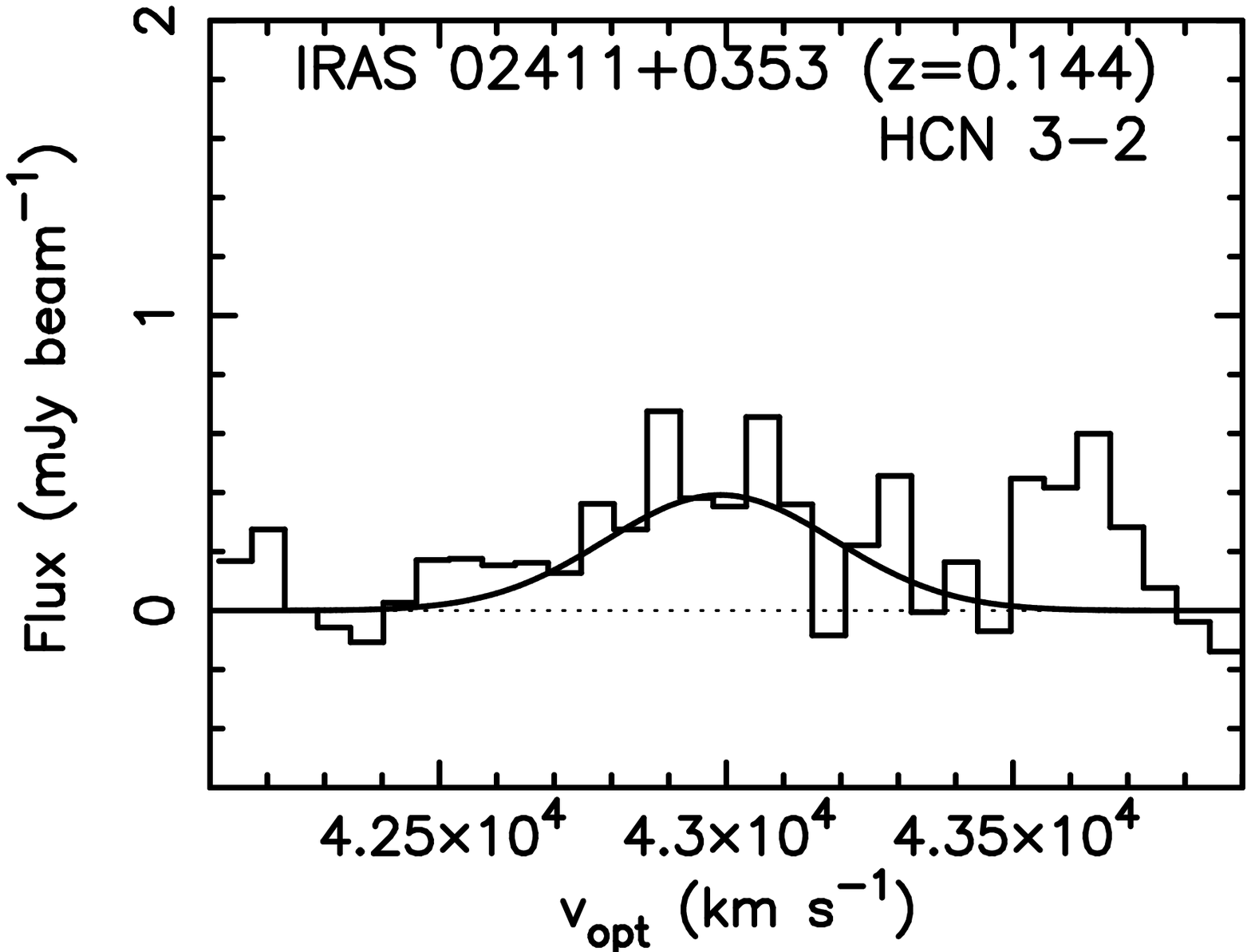} 
\includegraphics[angle=0,scale=.223]{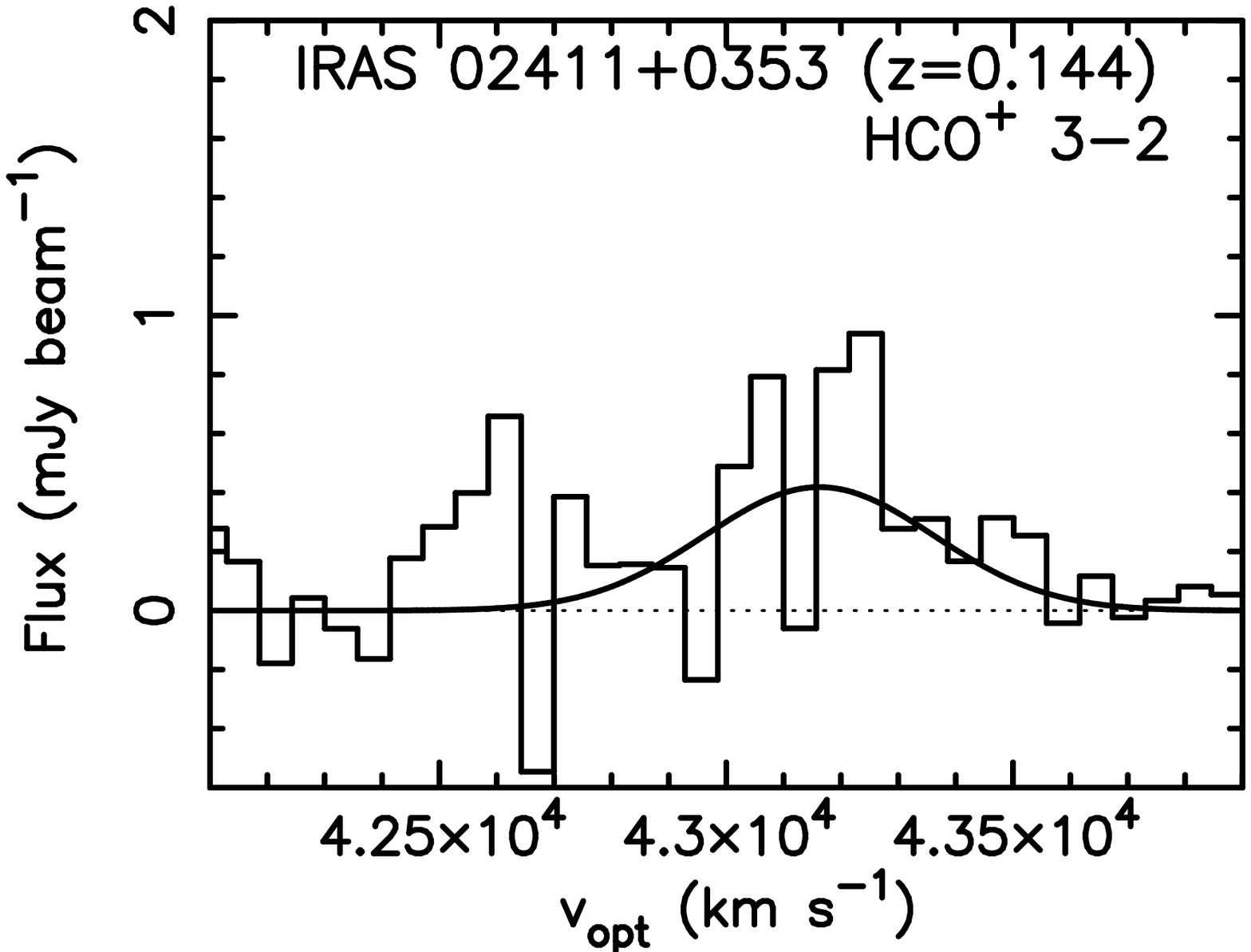} \\
\includegraphics[angle=0,scale=.223]{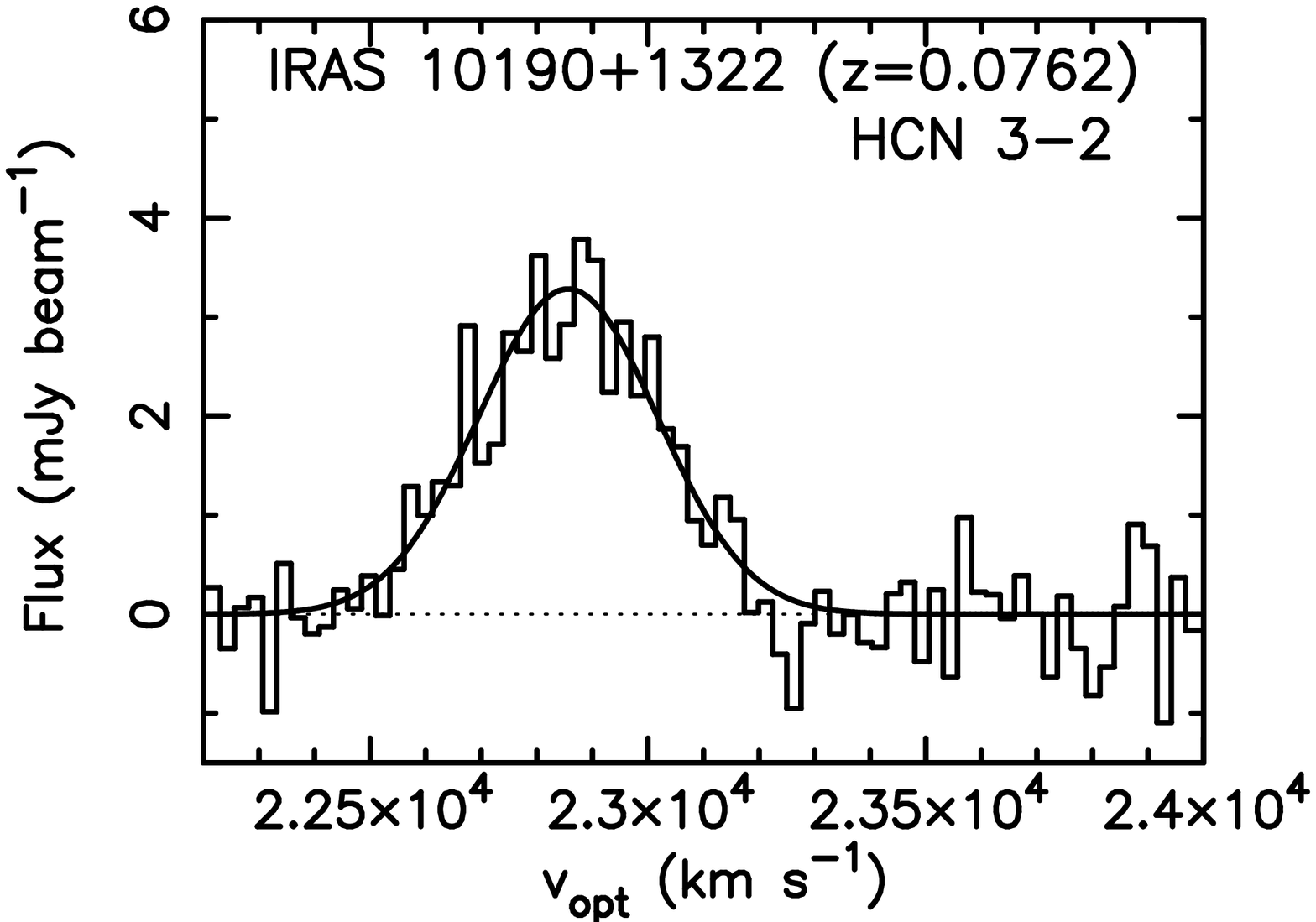} 
\includegraphics[angle=0,scale=.223]{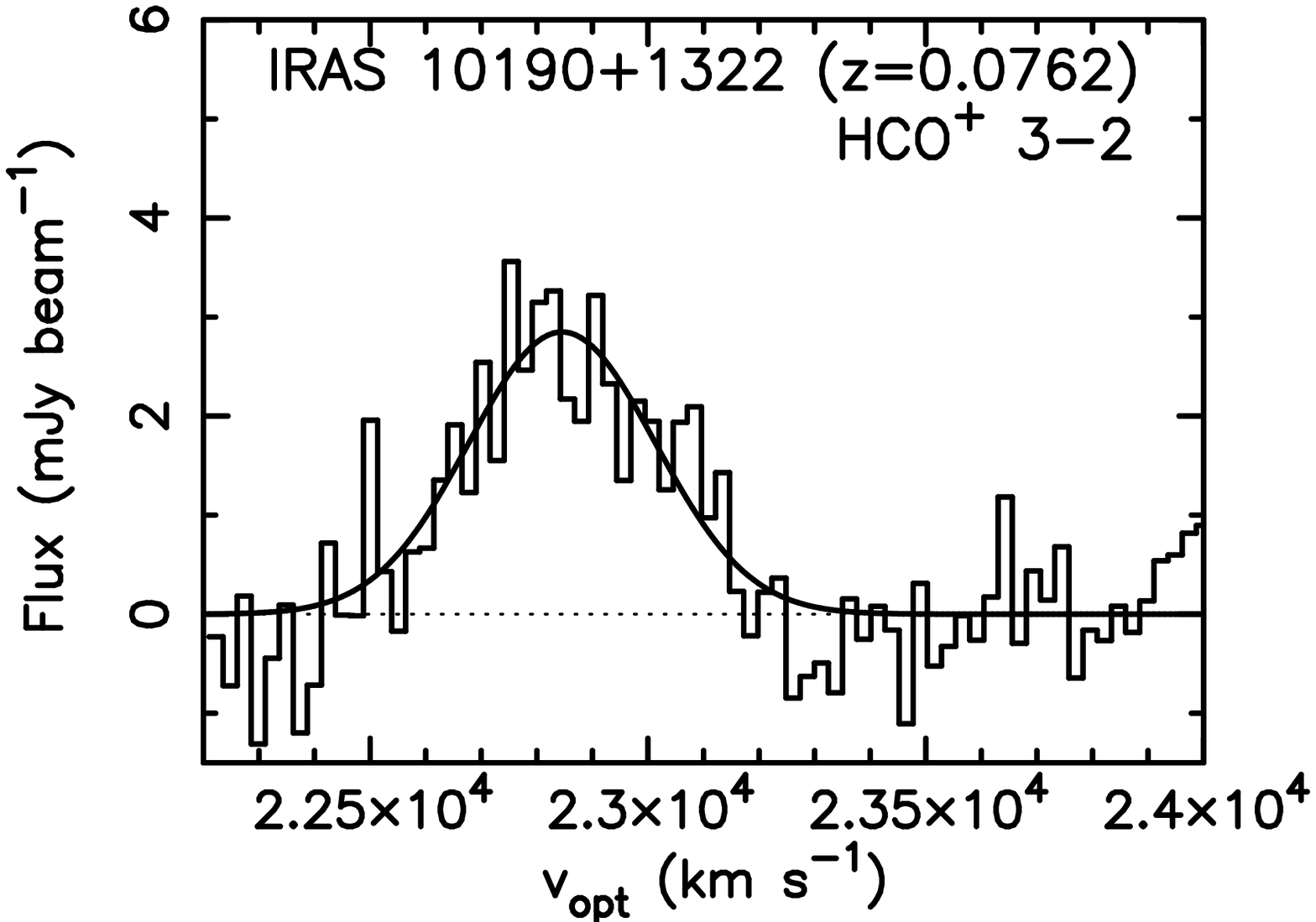} 
\includegraphics[angle=0,scale=.223]{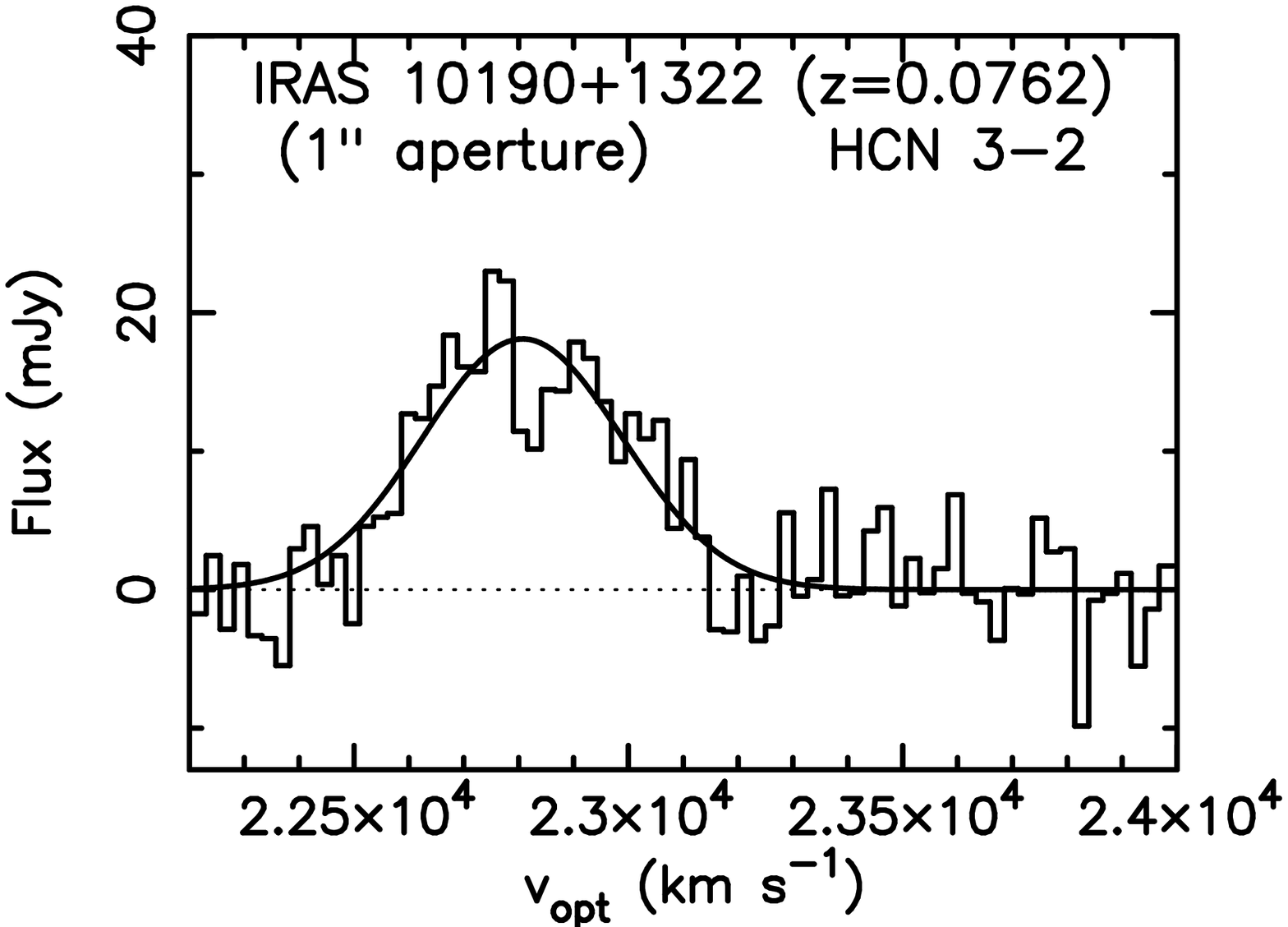} 
\includegraphics[angle=0,scale=.223]{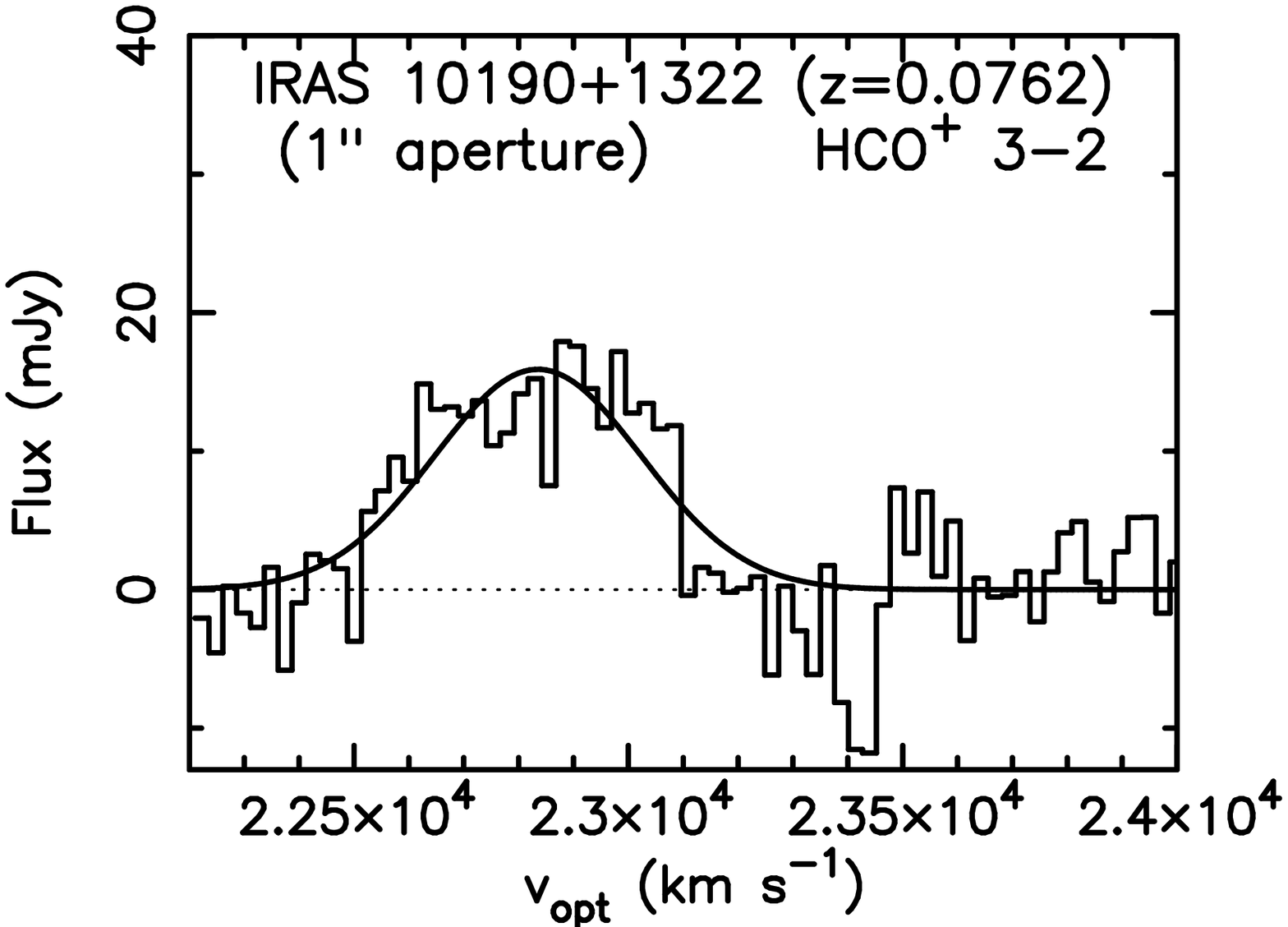} \\
\includegraphics[angle=0,scale=.223]{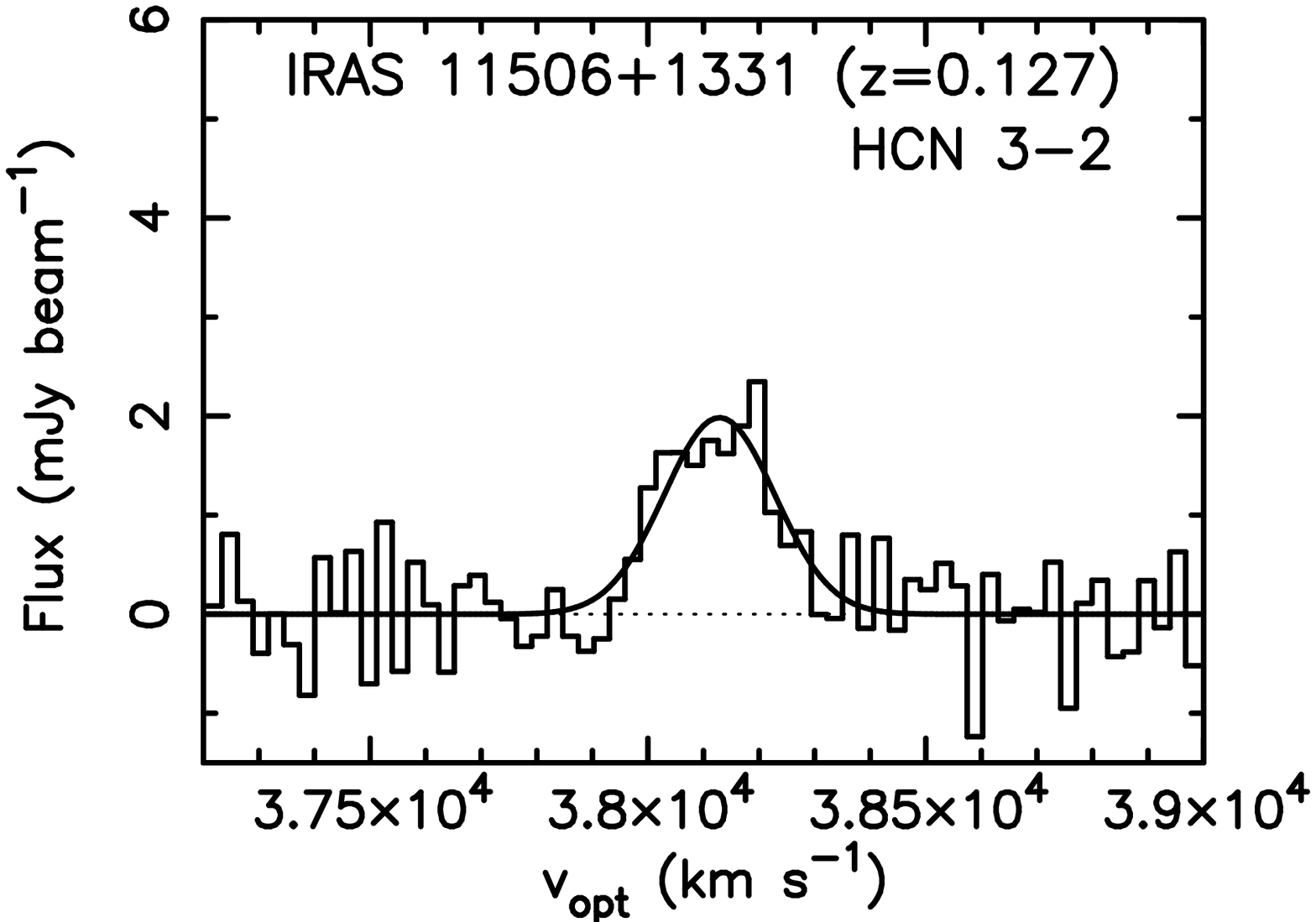} 
\includegraphics[angle=0,scale=.223]{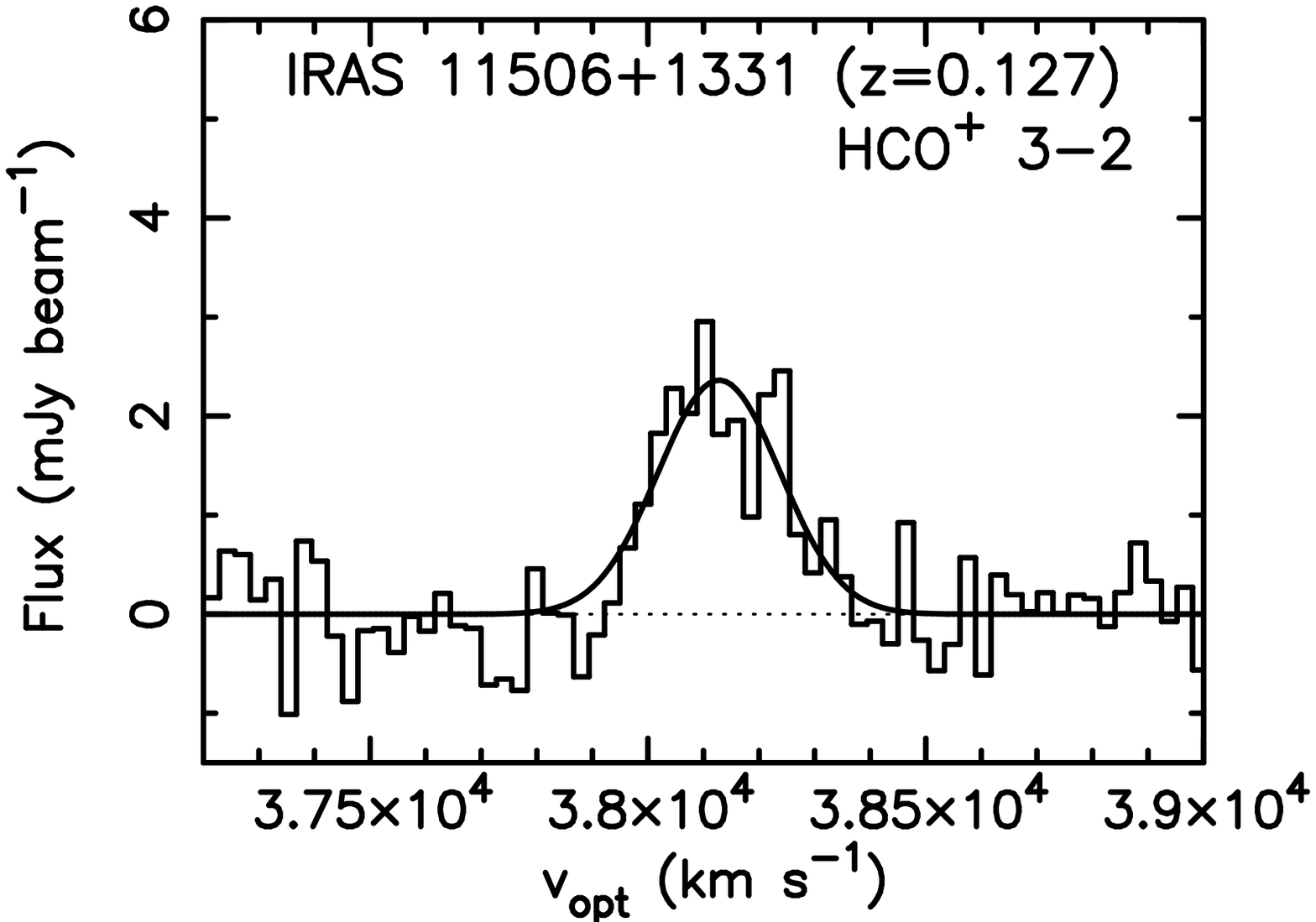} 
\includegraphics[angle=0,scale=.223]{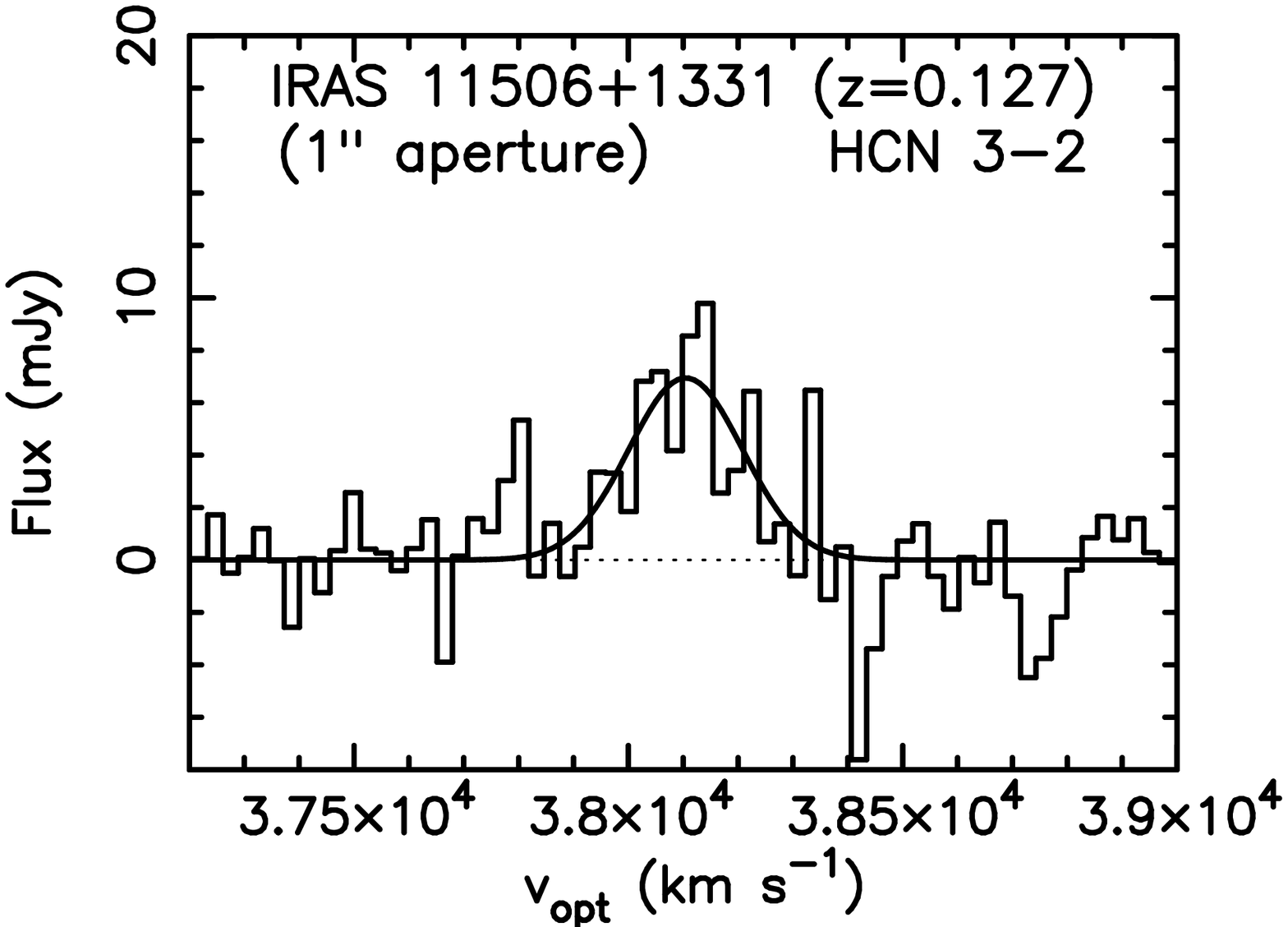} 
\includegraphics[angle=0,scale=.223]{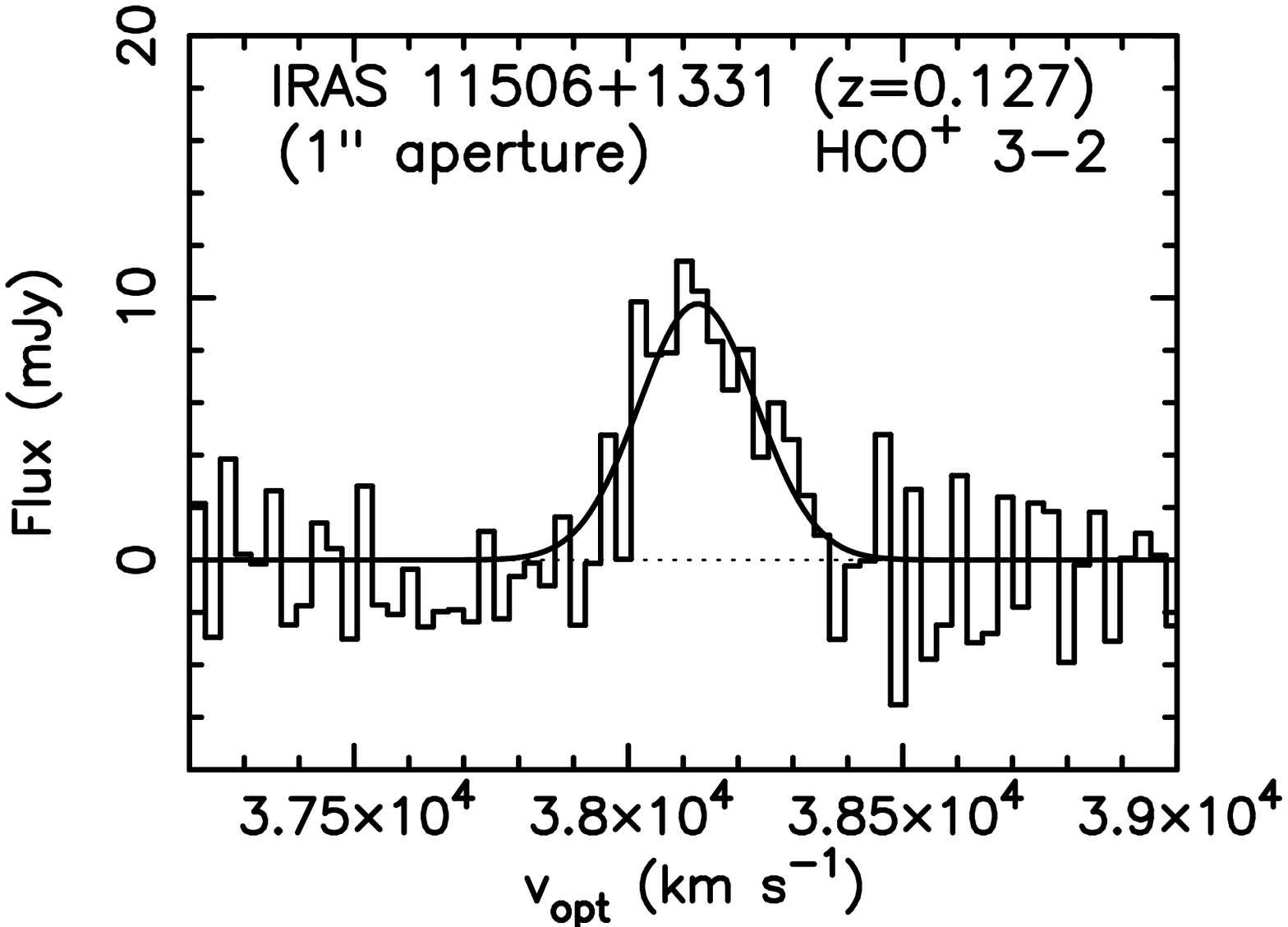} \\
\includegraphics[angle=0,scale=.223]{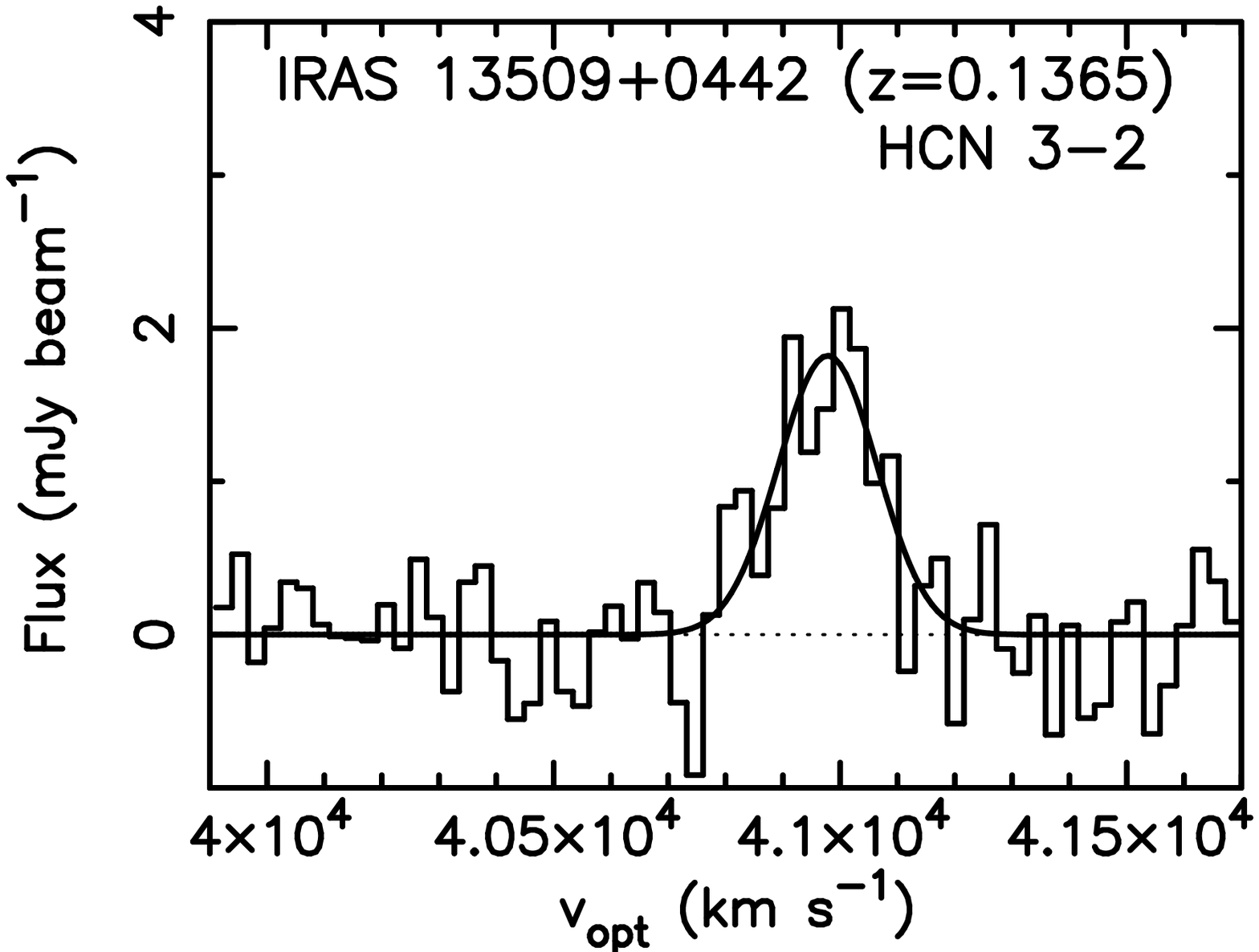} 
\includegraphics[angle=0,scale=.223]{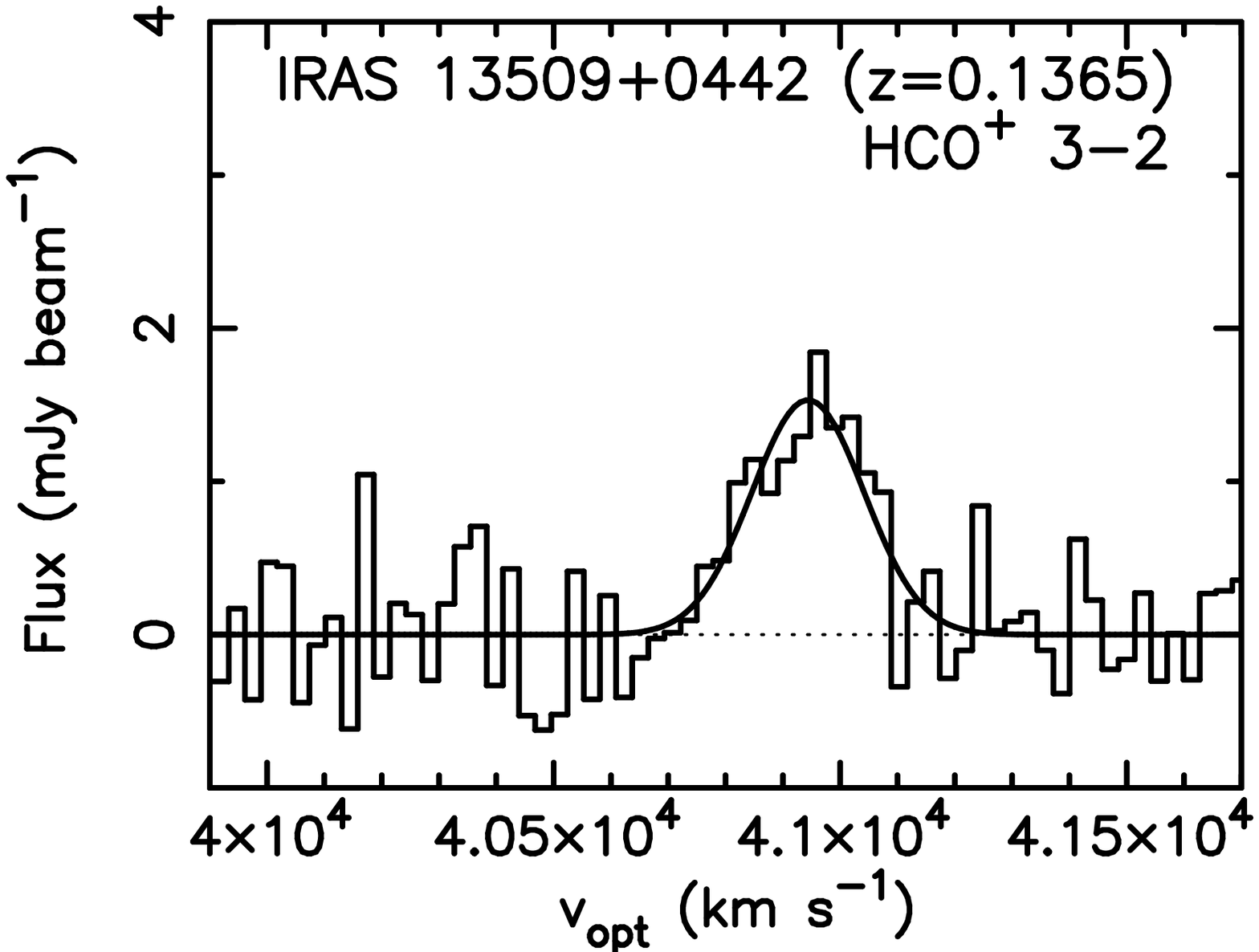} 
\includegraphics[angle=0,scale=.223]{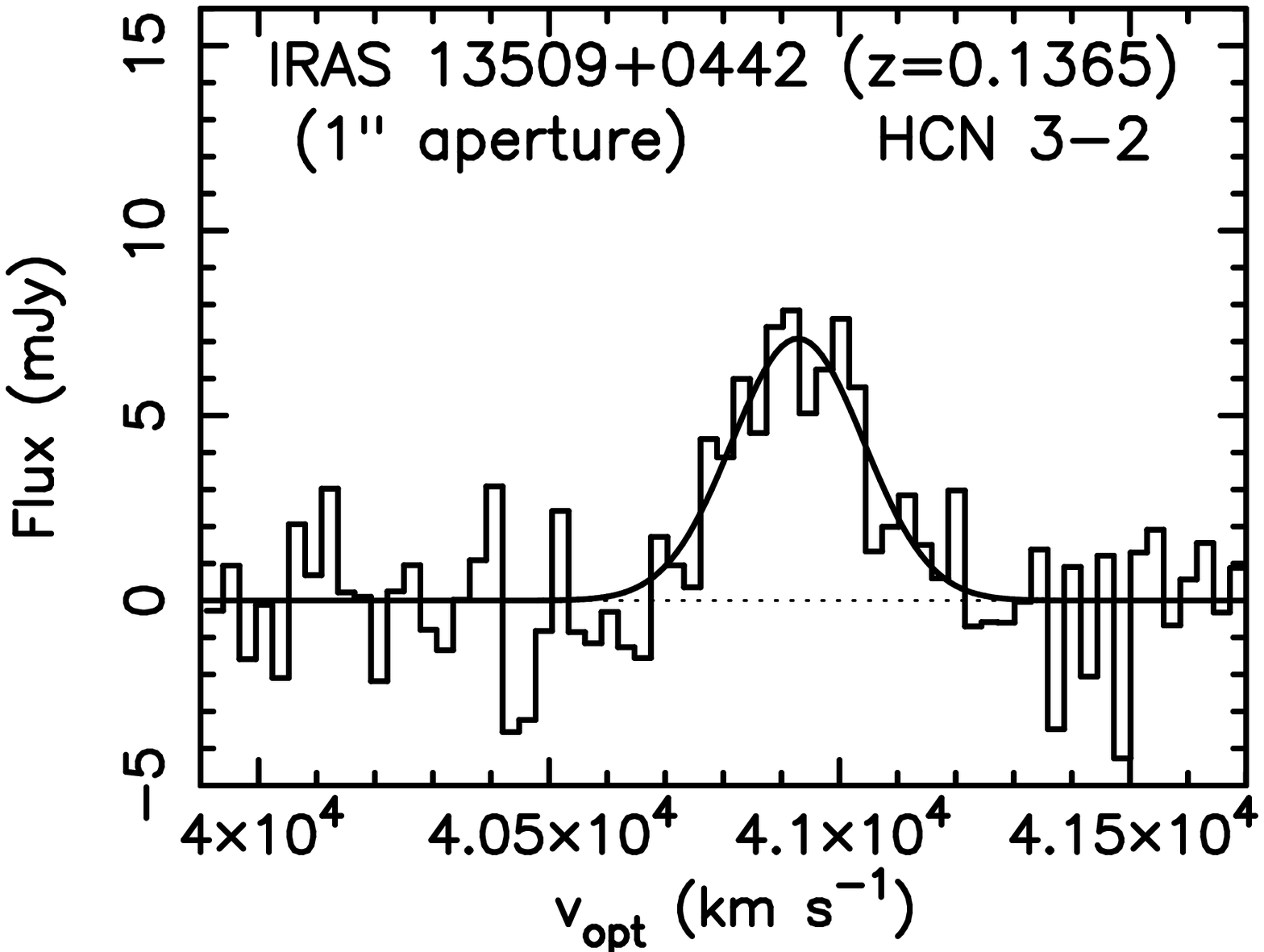} 
\includegraphics[angle=0,scale=.223]{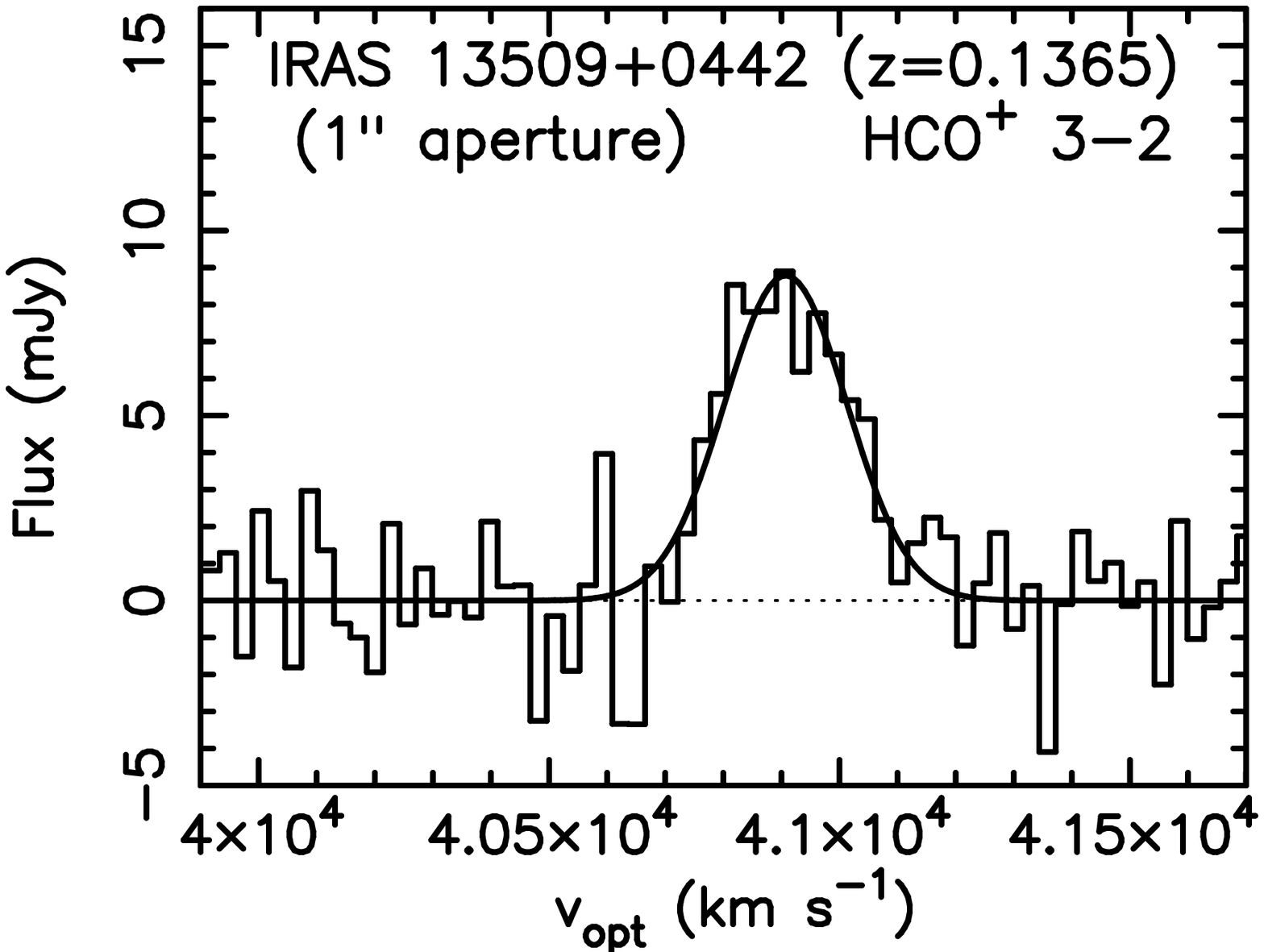} \\
\includegraphics[angle=0,scale=.223]{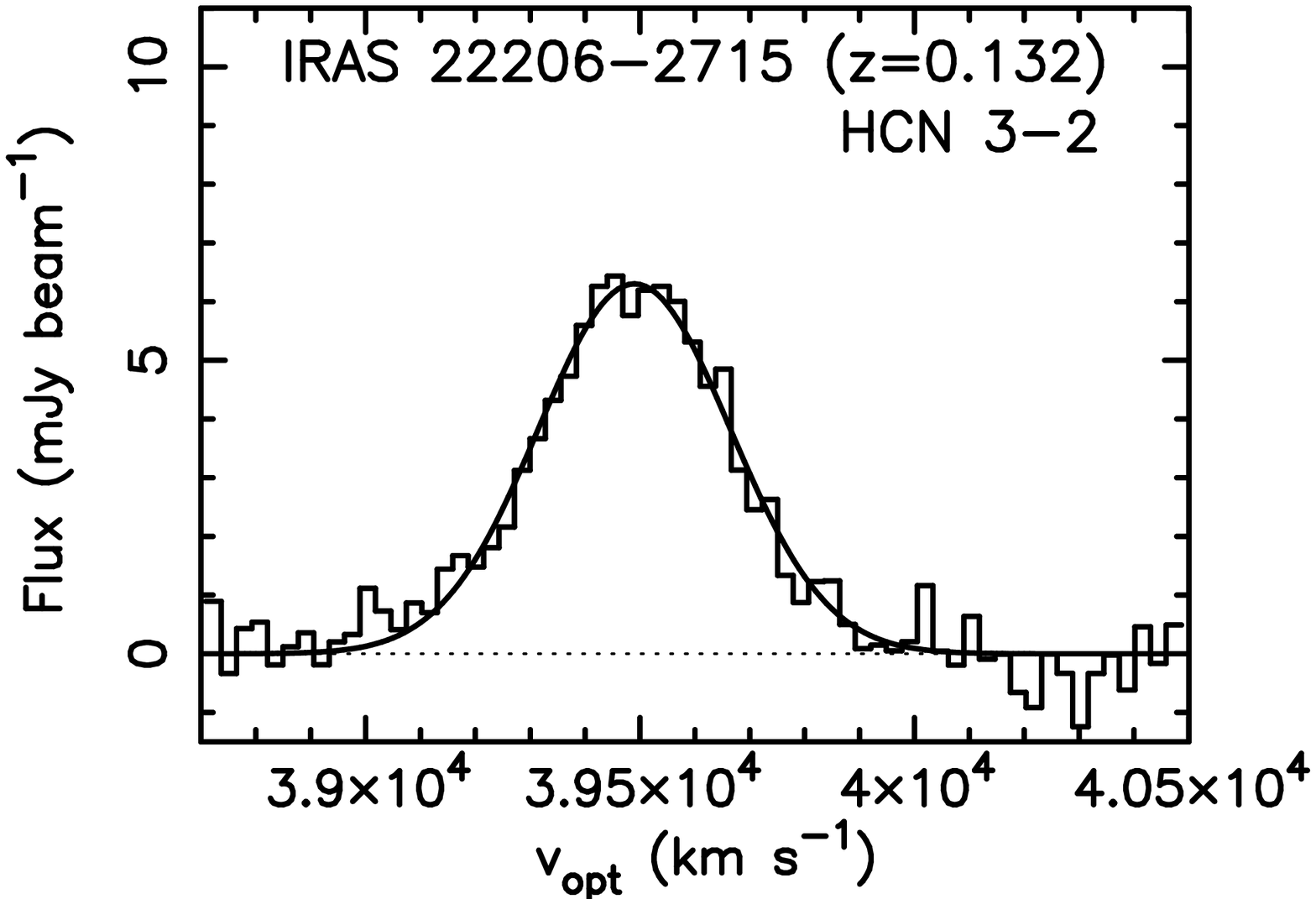} 
\includegraphics[angle=0,scale=.223]{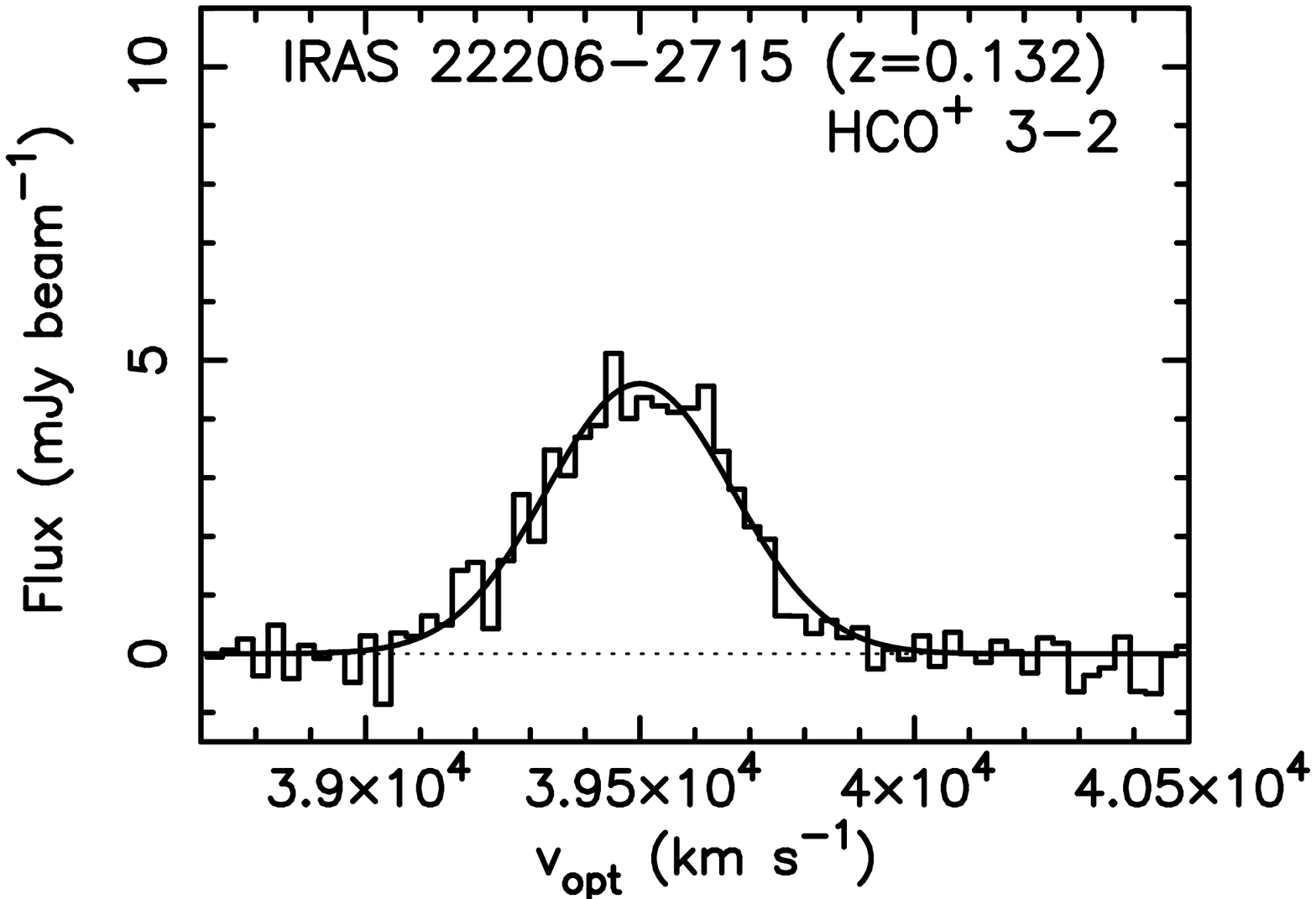} 
\includegraphics[angle=0,scale=.223]{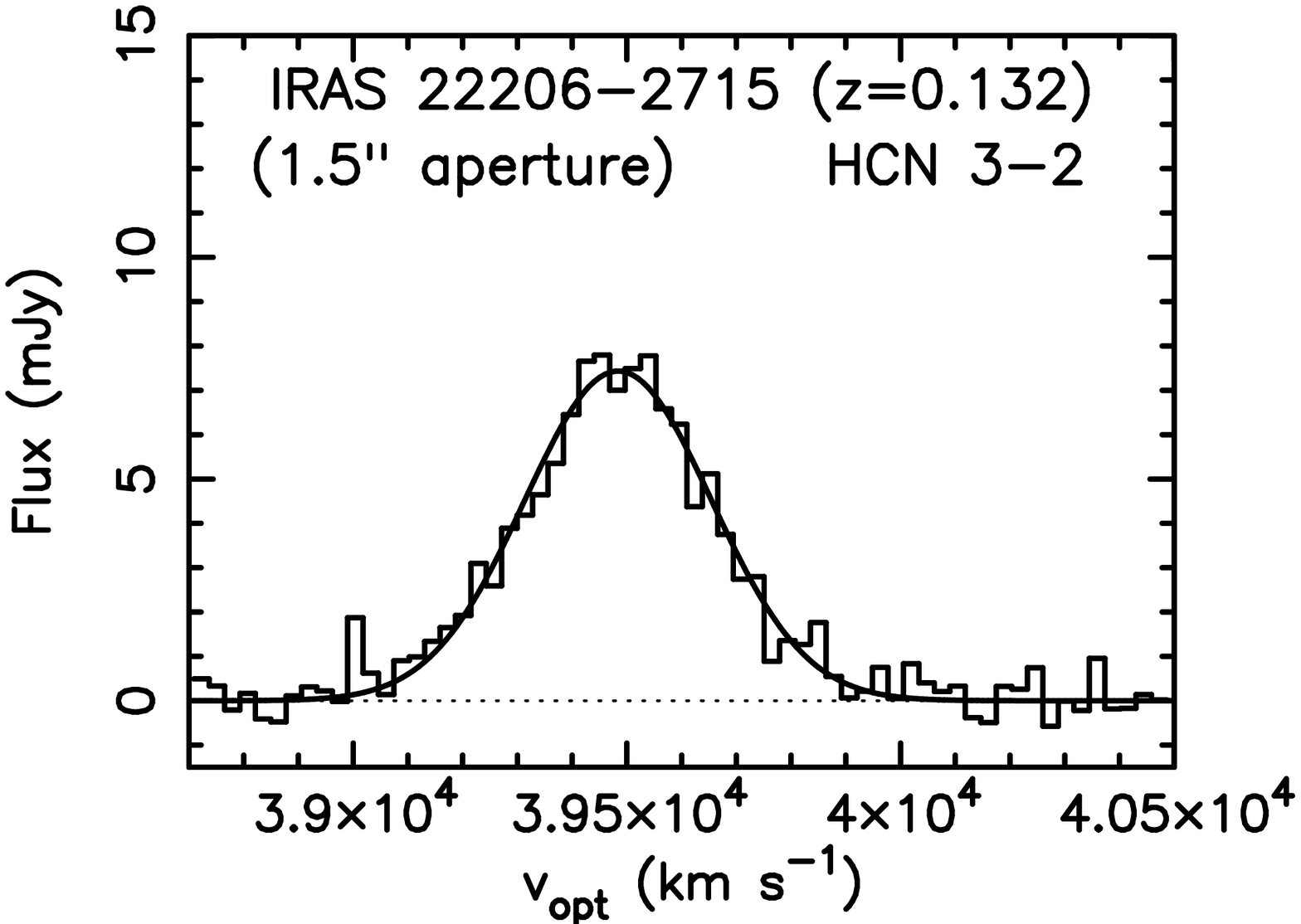} 
\includegraphics[angle=0,scale=.223]{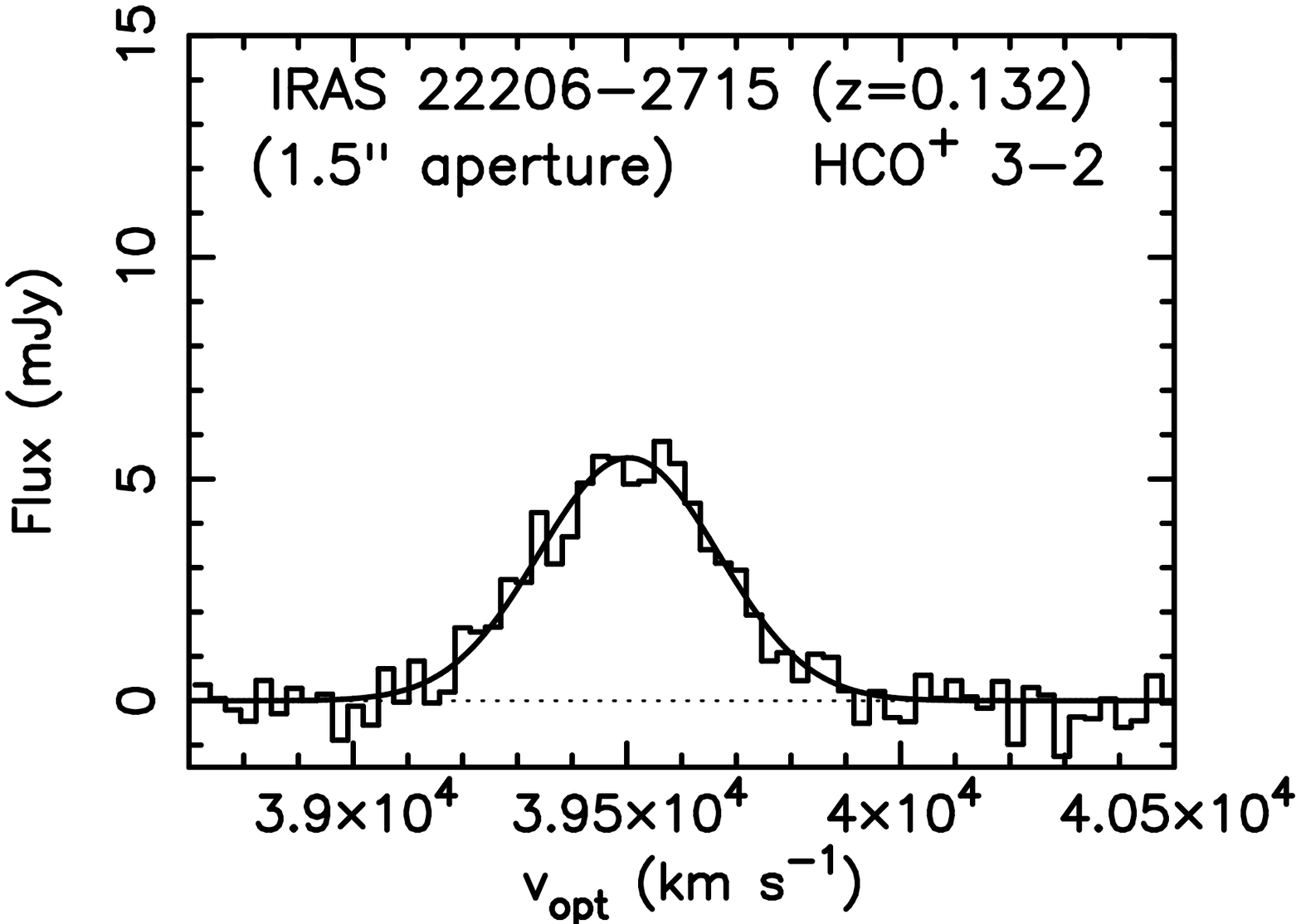} \\
\includegraphics[angle=0,scale=.223]{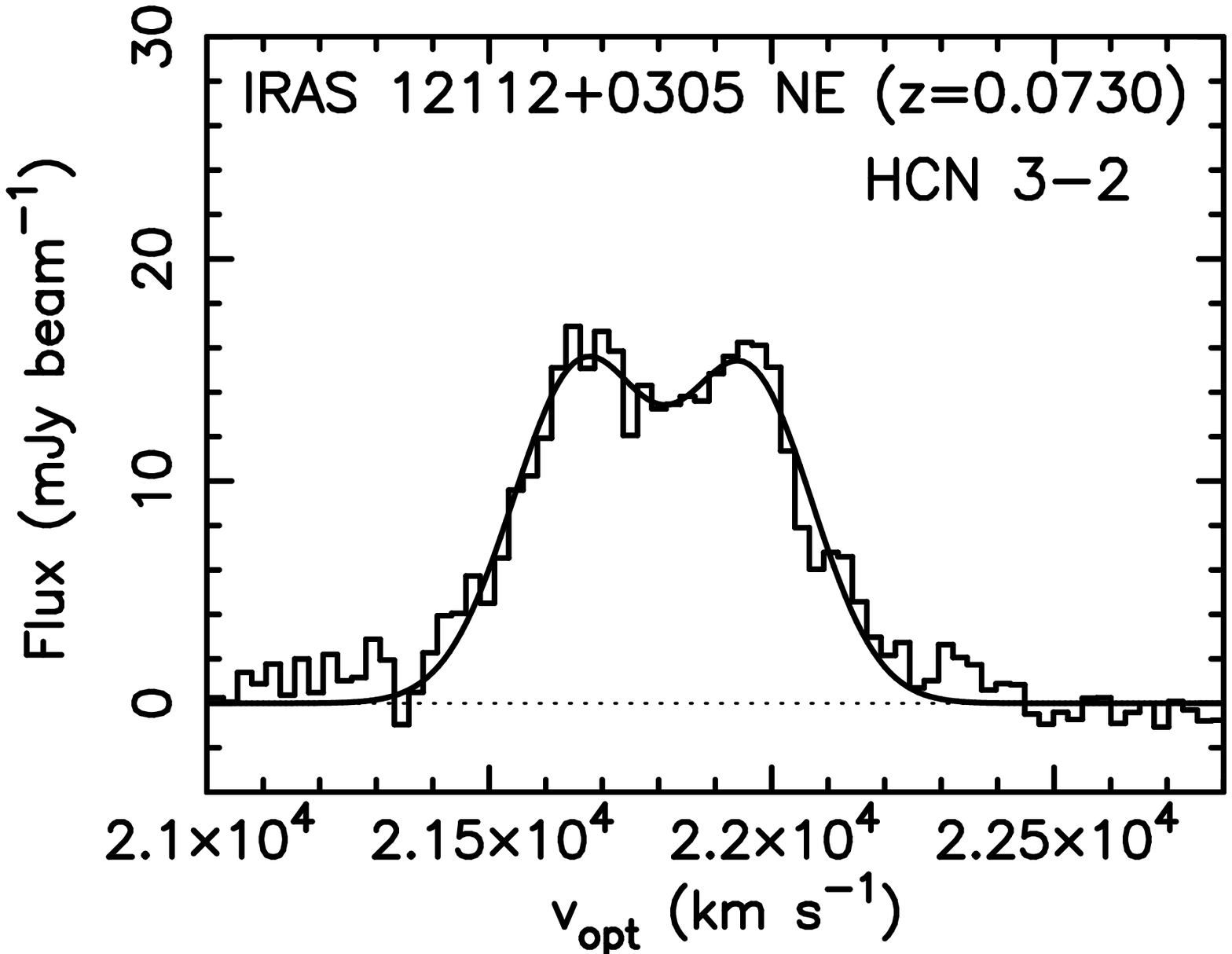} 
\includegraphics[angle=0,scale=.223]{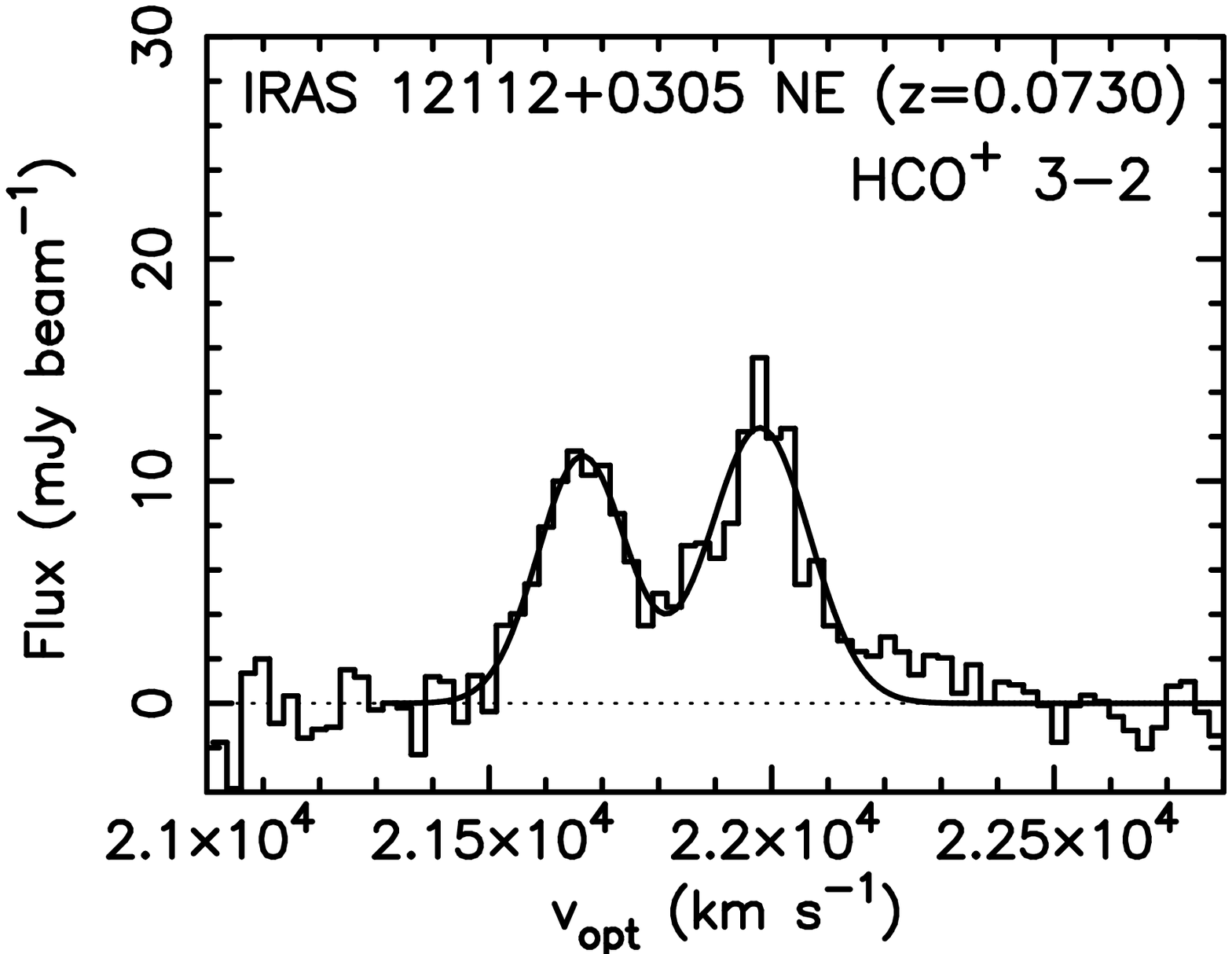} 
\includegraphics[angle=0,scale=.223]{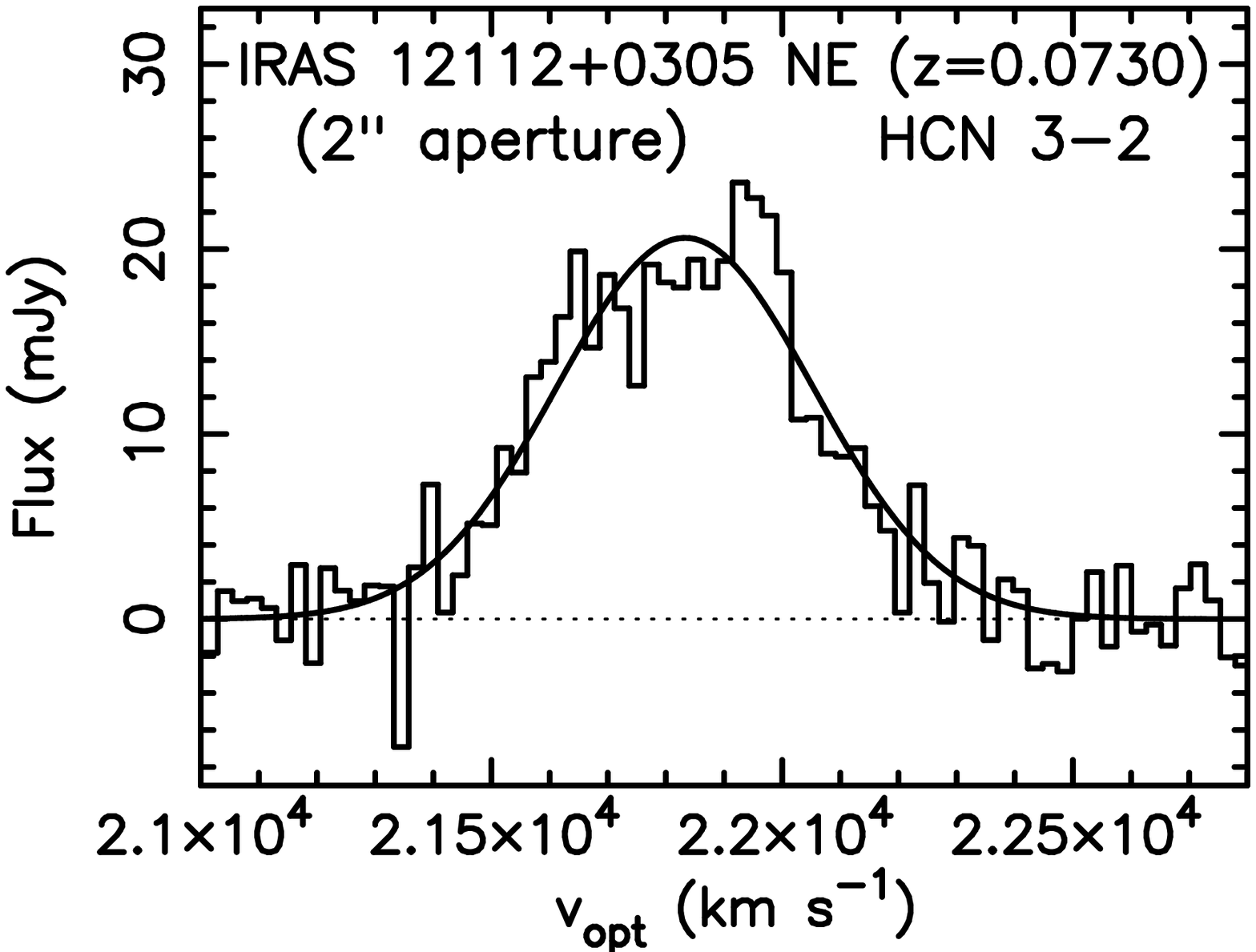} 
\includegraphics[angle=0,scale=.223]{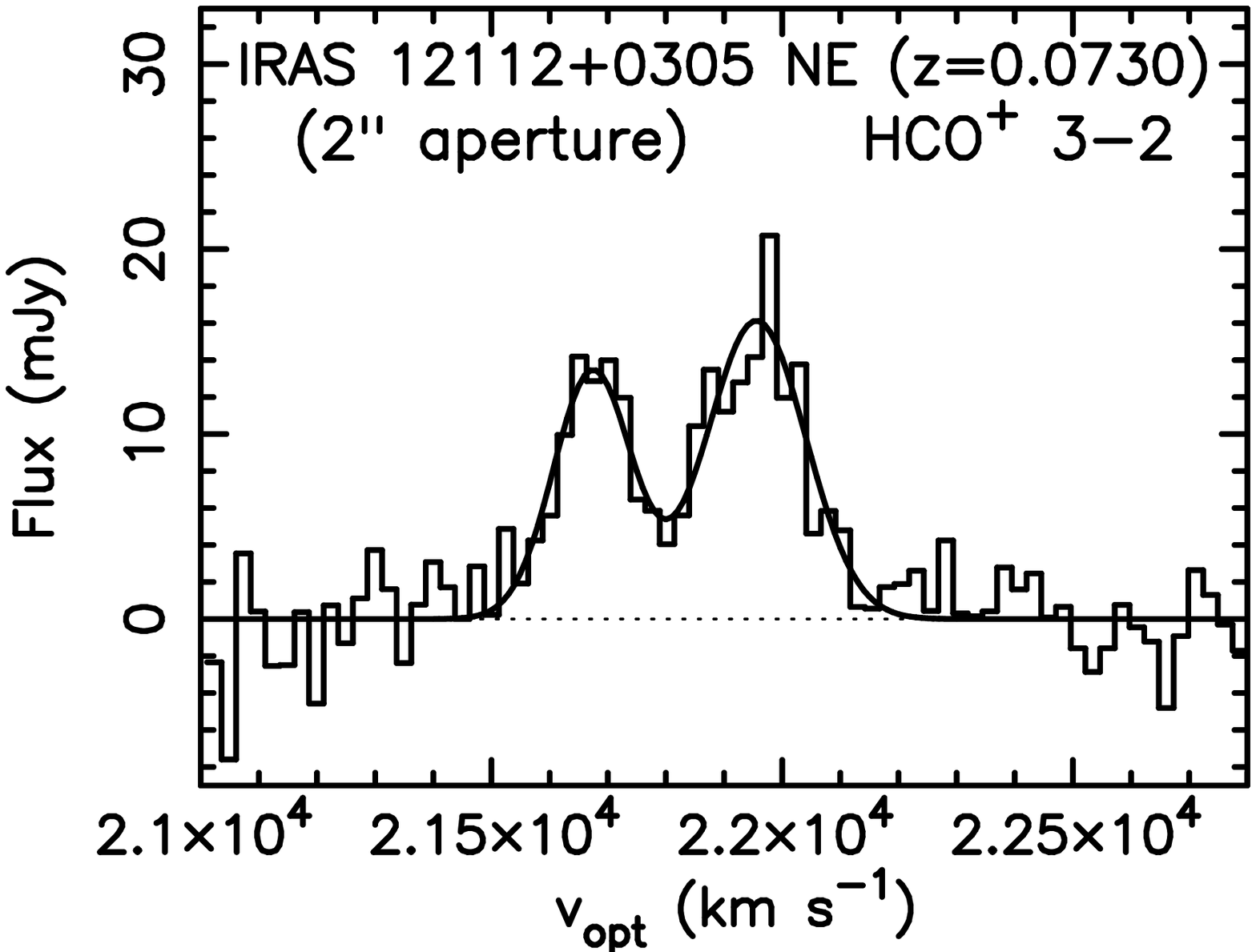} 
\includegraphics[angle=0,scale=.223]{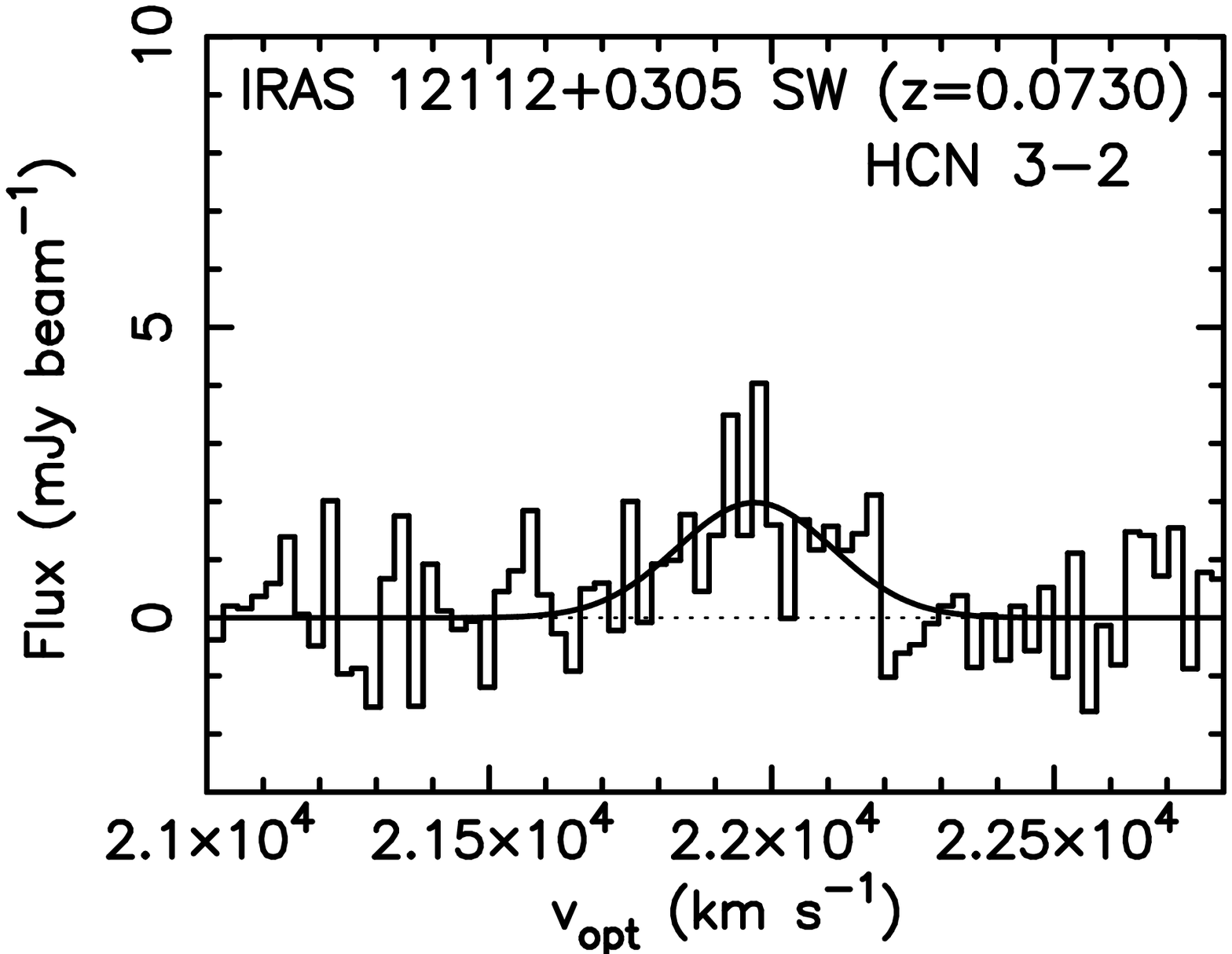} 
\includegraphics[angle=0,scale=.223]{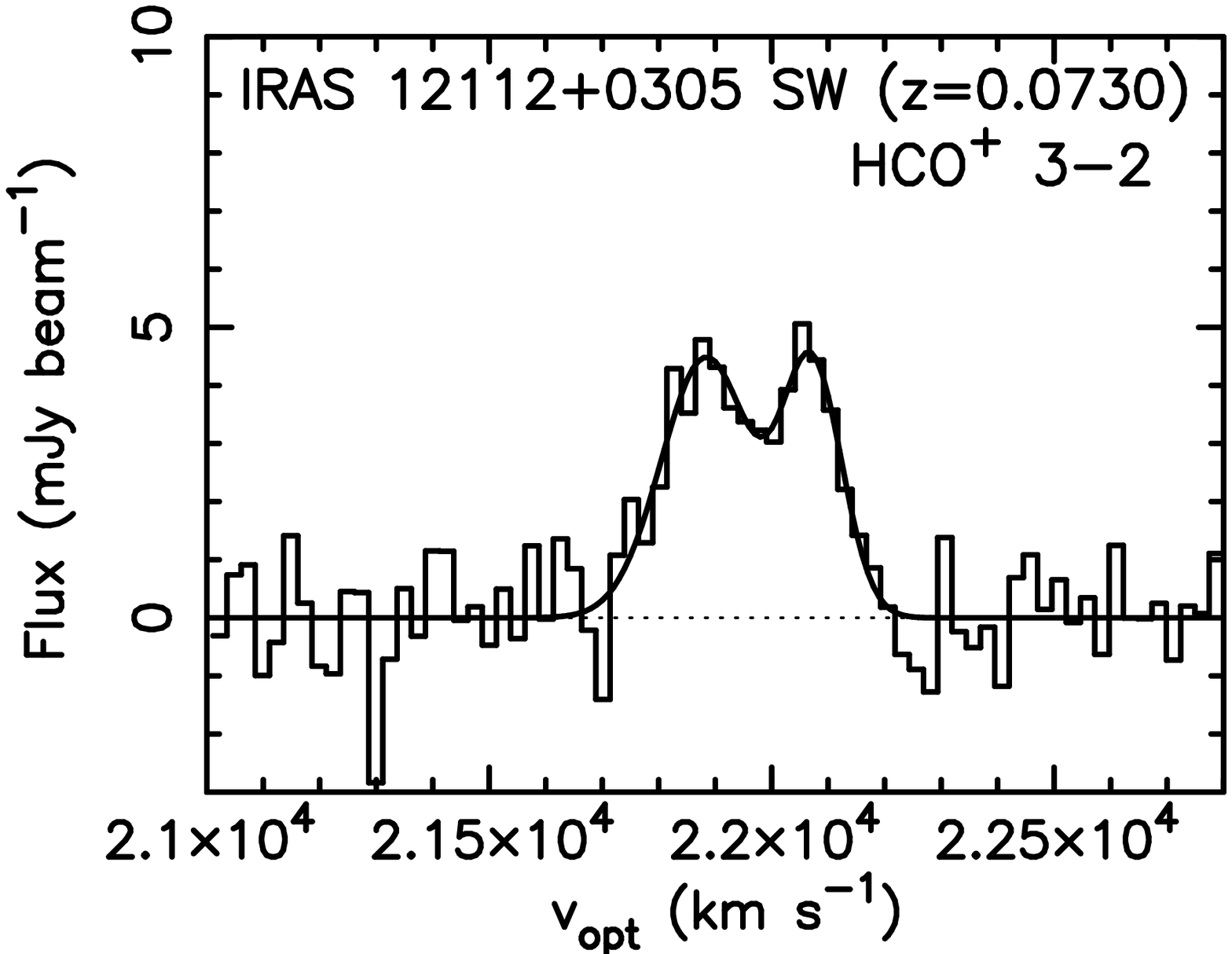} 
\includegraphics[angle=0,scale=.223]{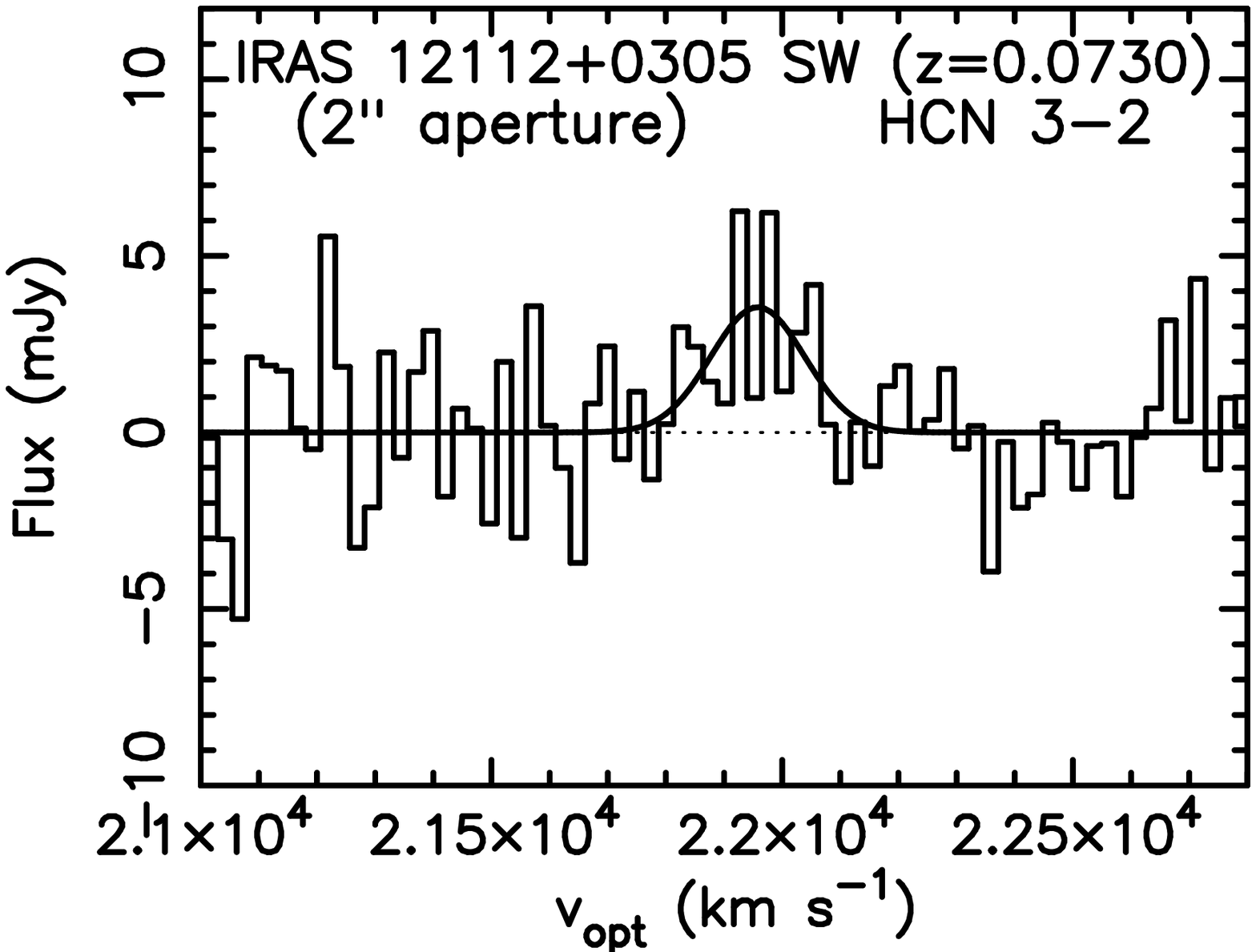} 
\includegraphics[angle=0,scale=.223]{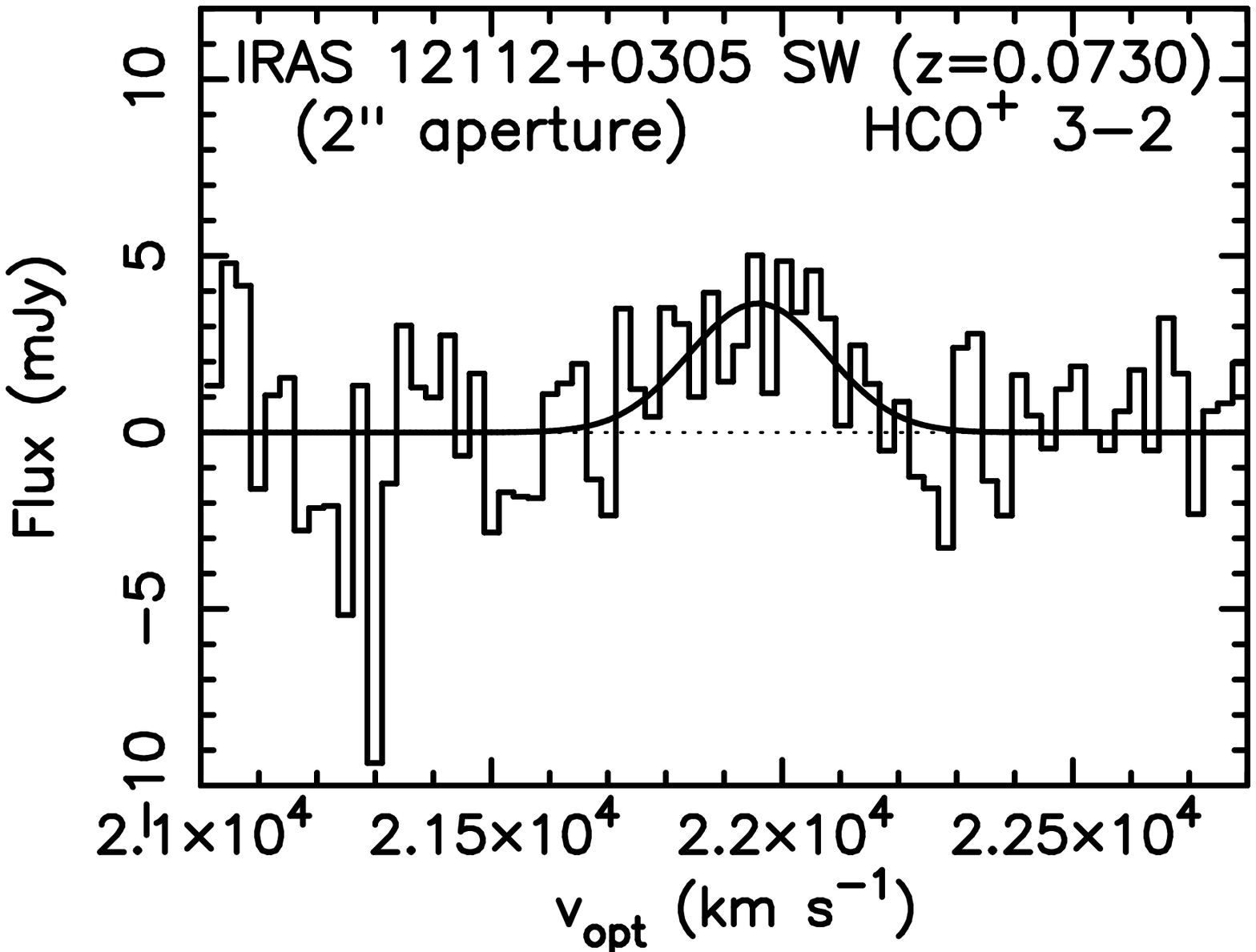} \\
\end{center}
\end{figure}

\clearpage

\begin{figure}
\begin{center}
\includegraphics[angle=0,scale=.223]{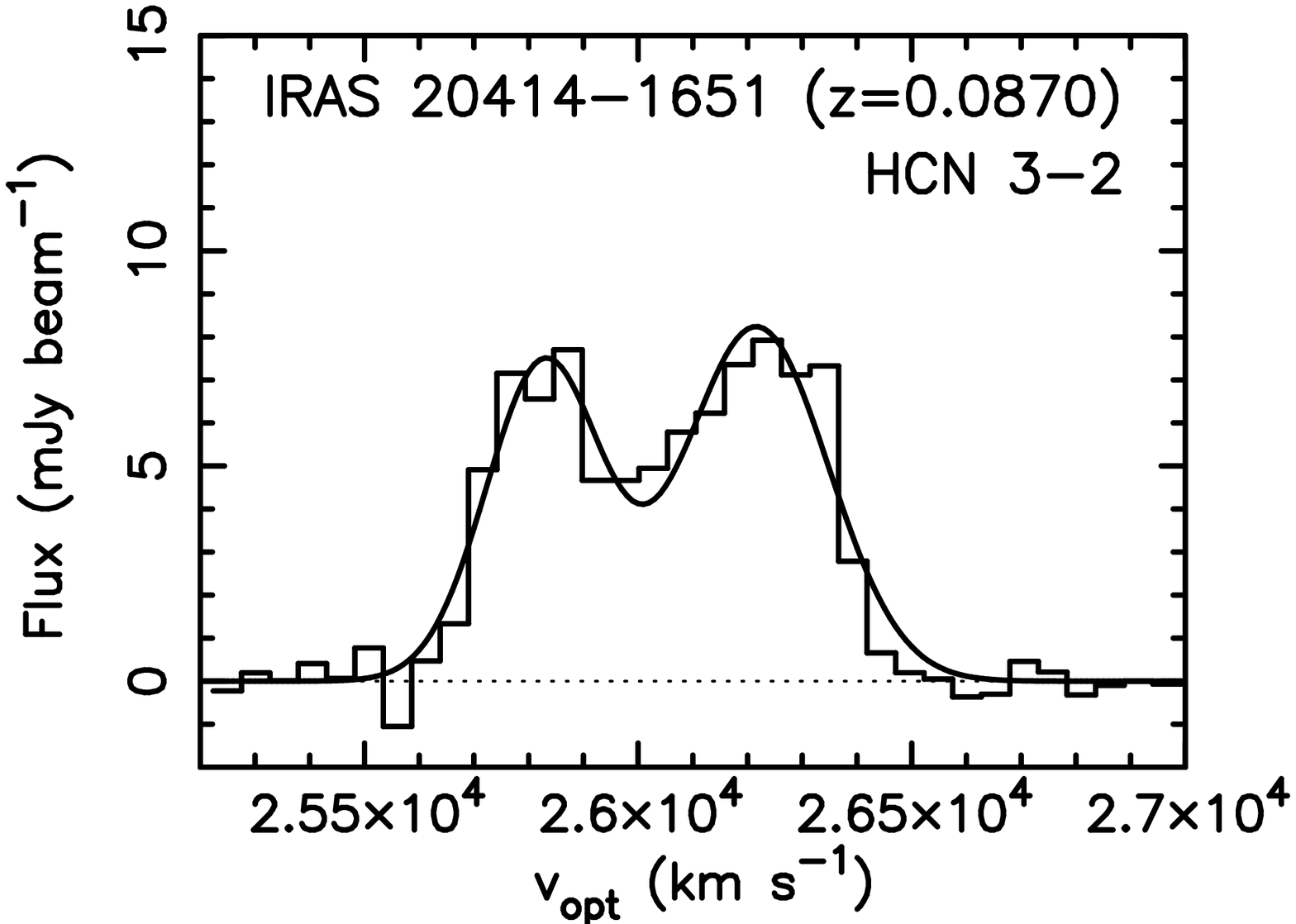} 
\includegraphics[angle=0,scale=.223]{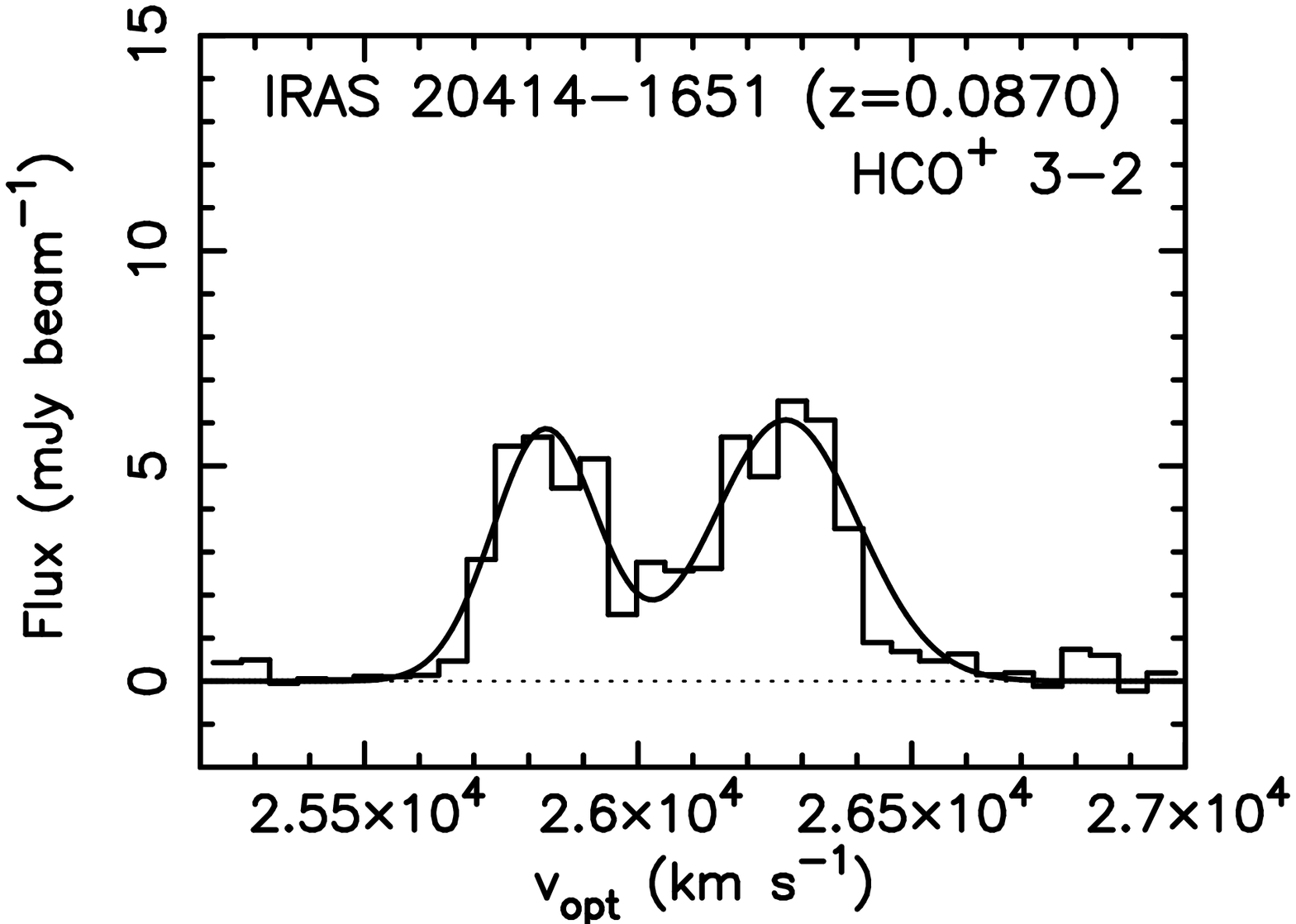} 
\includegraphics[angle=0,scale=.223]{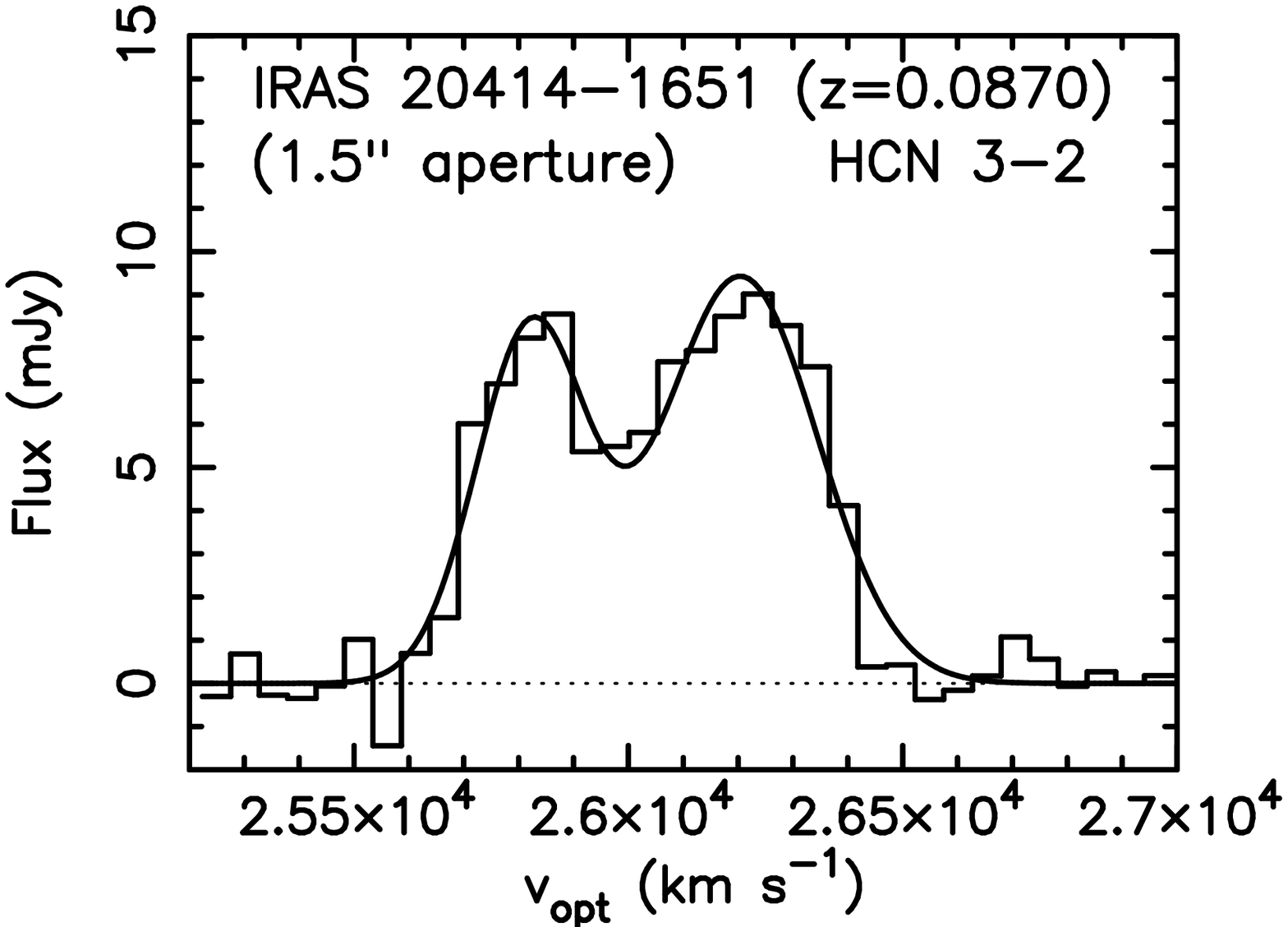} 
\includegraphics[angle=0,scale=.223]{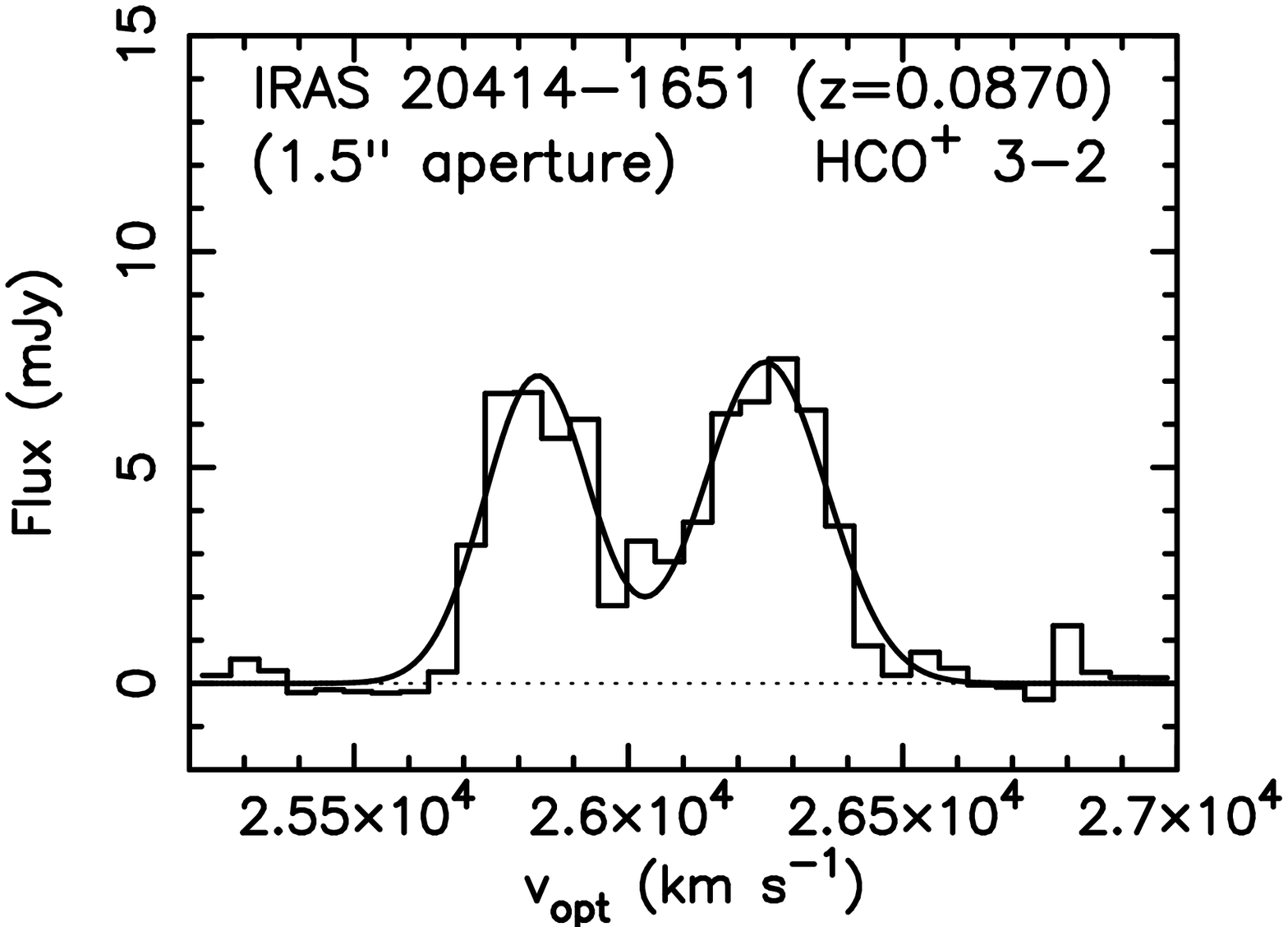} \\
\includegraphics[angle=0,scale=.223]{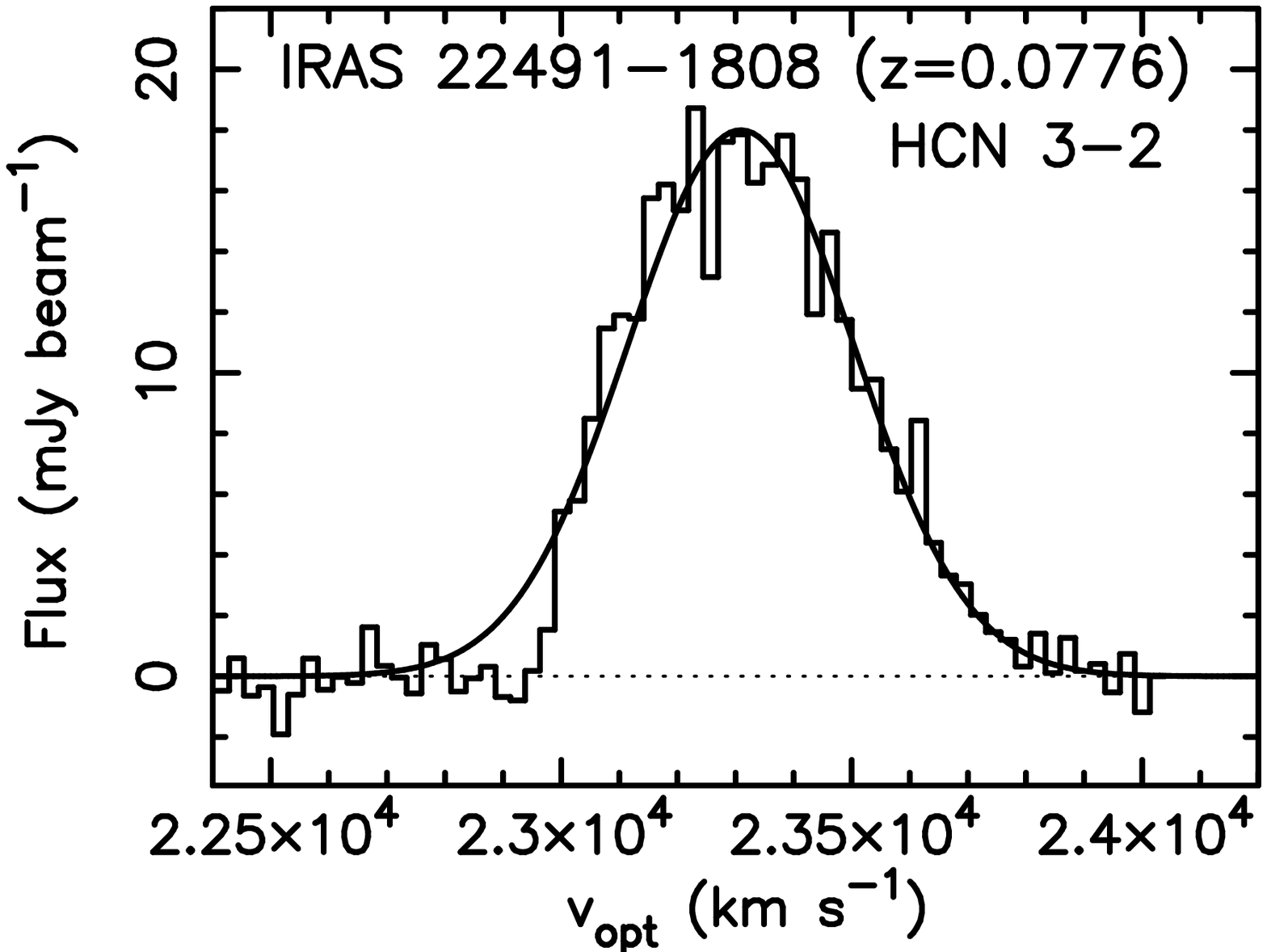} 
\includegraphics[angle=0,scale=.223]{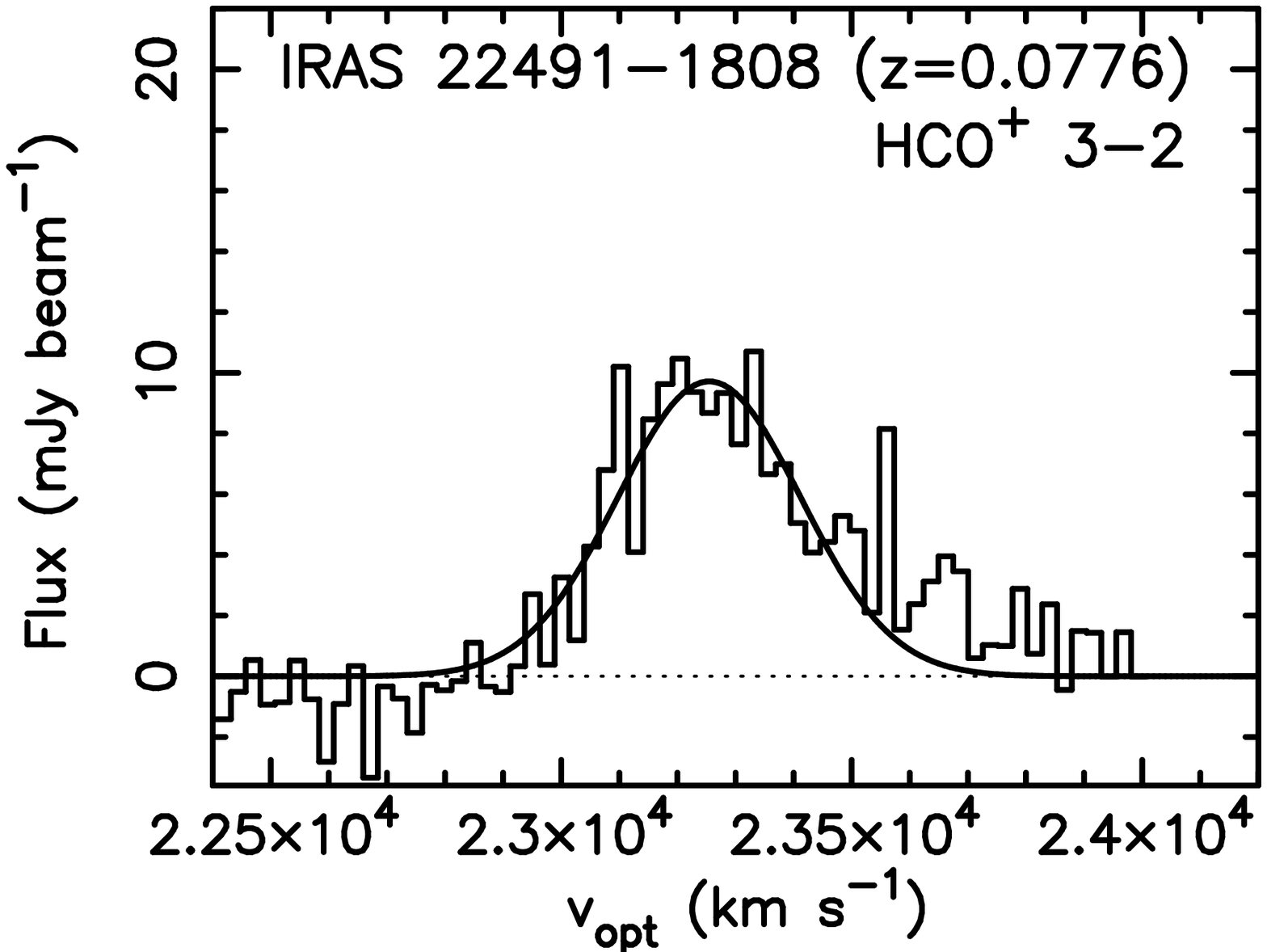} 
\includegraphics[angle=0,scale=.223]{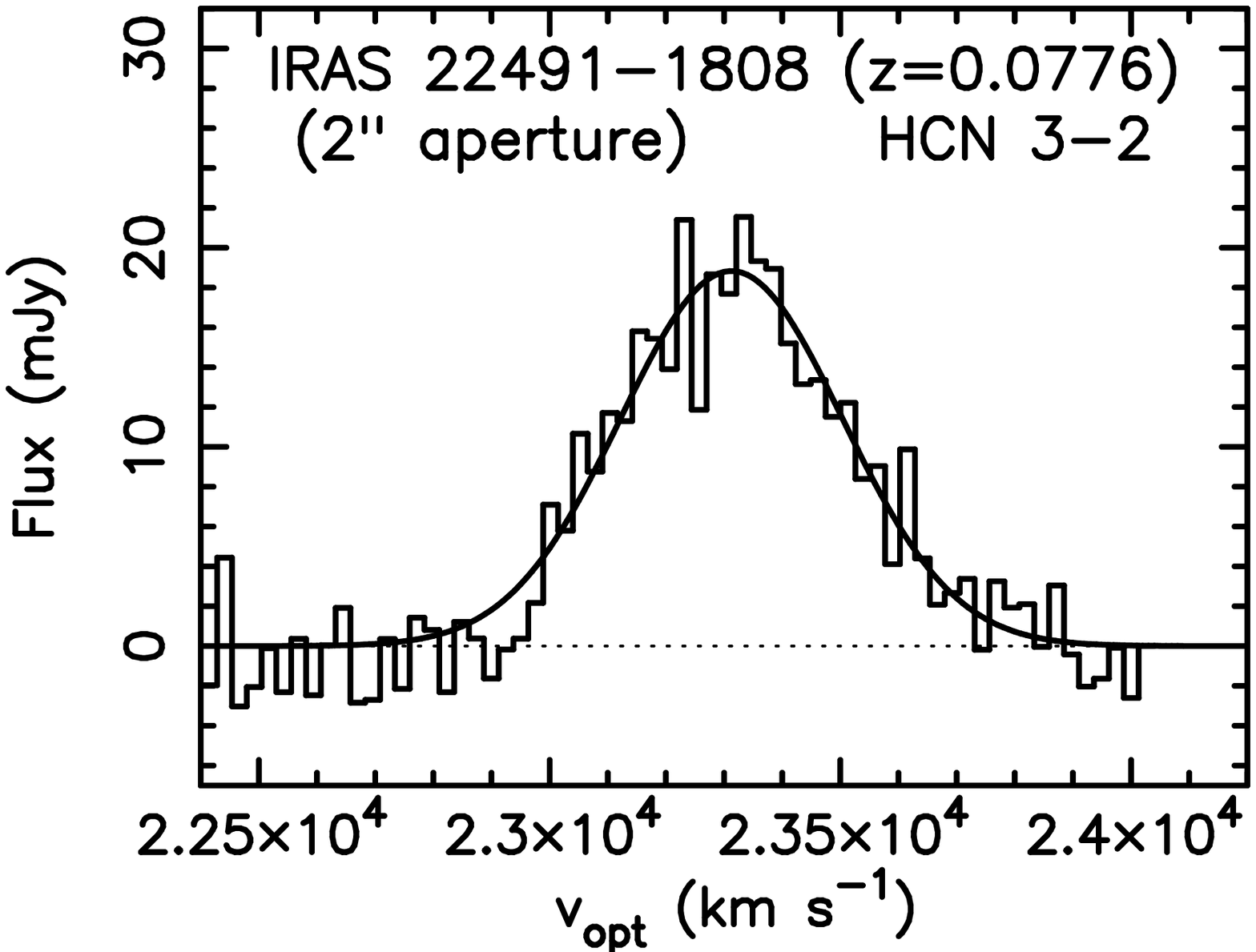} 
\includegraphics[angle=0,scale=.223]{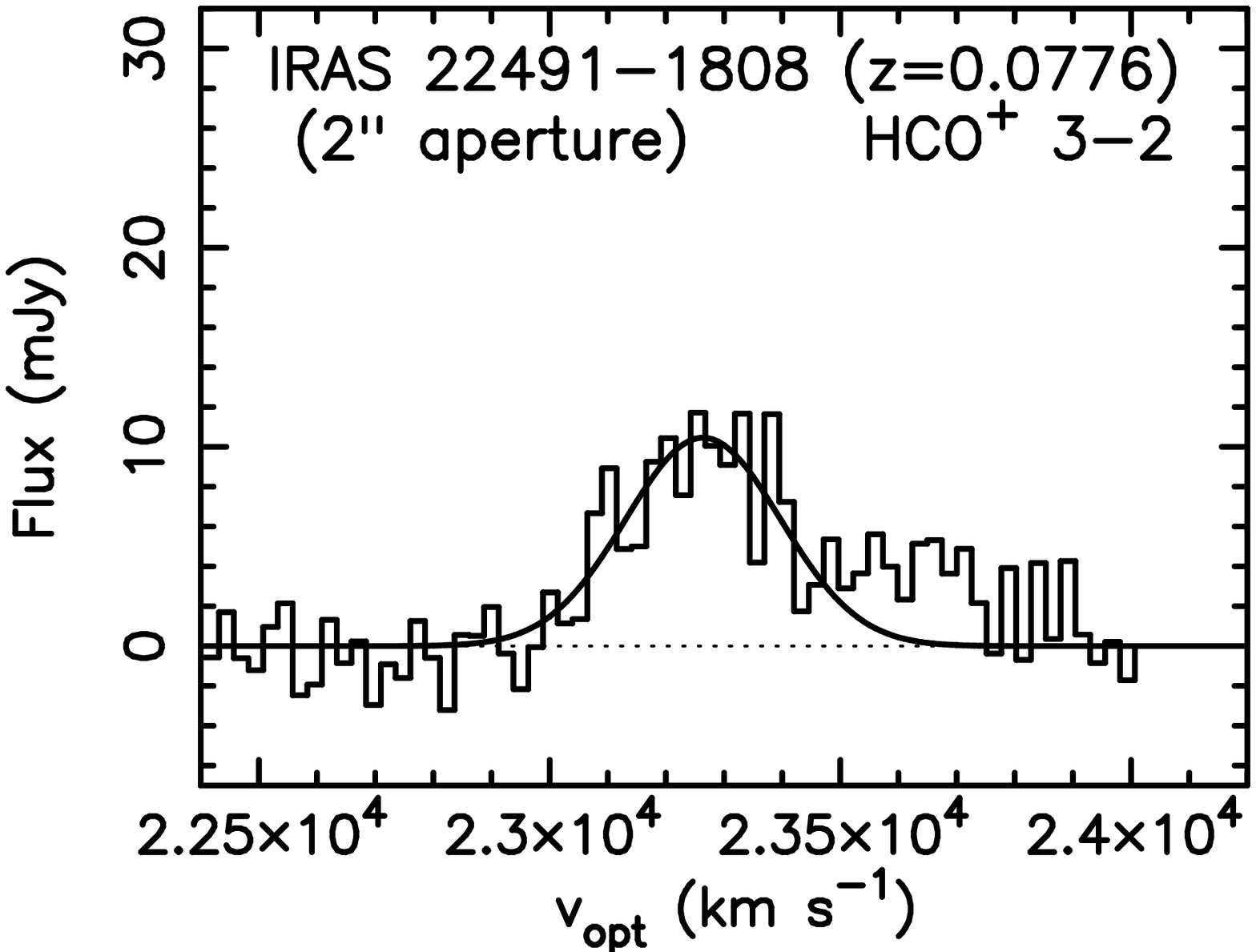} \\
\caption{Gaussian fit of the detected emission line in Figure 2.
The solid curved line is the best Gaussian fit and the dotted straight line 
is the zero flux level.}
\end{center}
\end{figure}

\begin{figure}
\begin{center}
\includegraphics[angle=0,scale=.314]{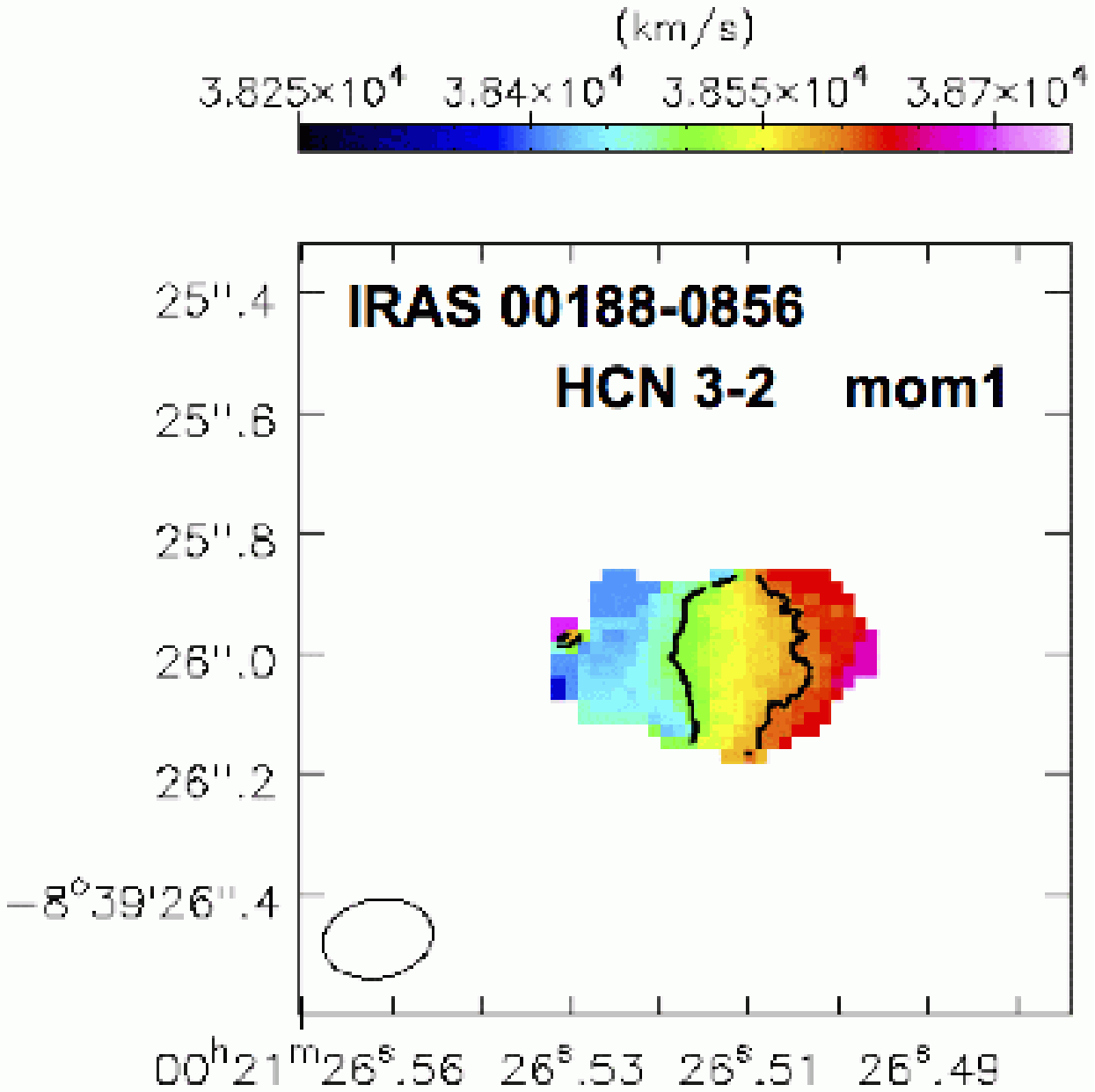} 
\includegraphics[angle=0,scale=.314]{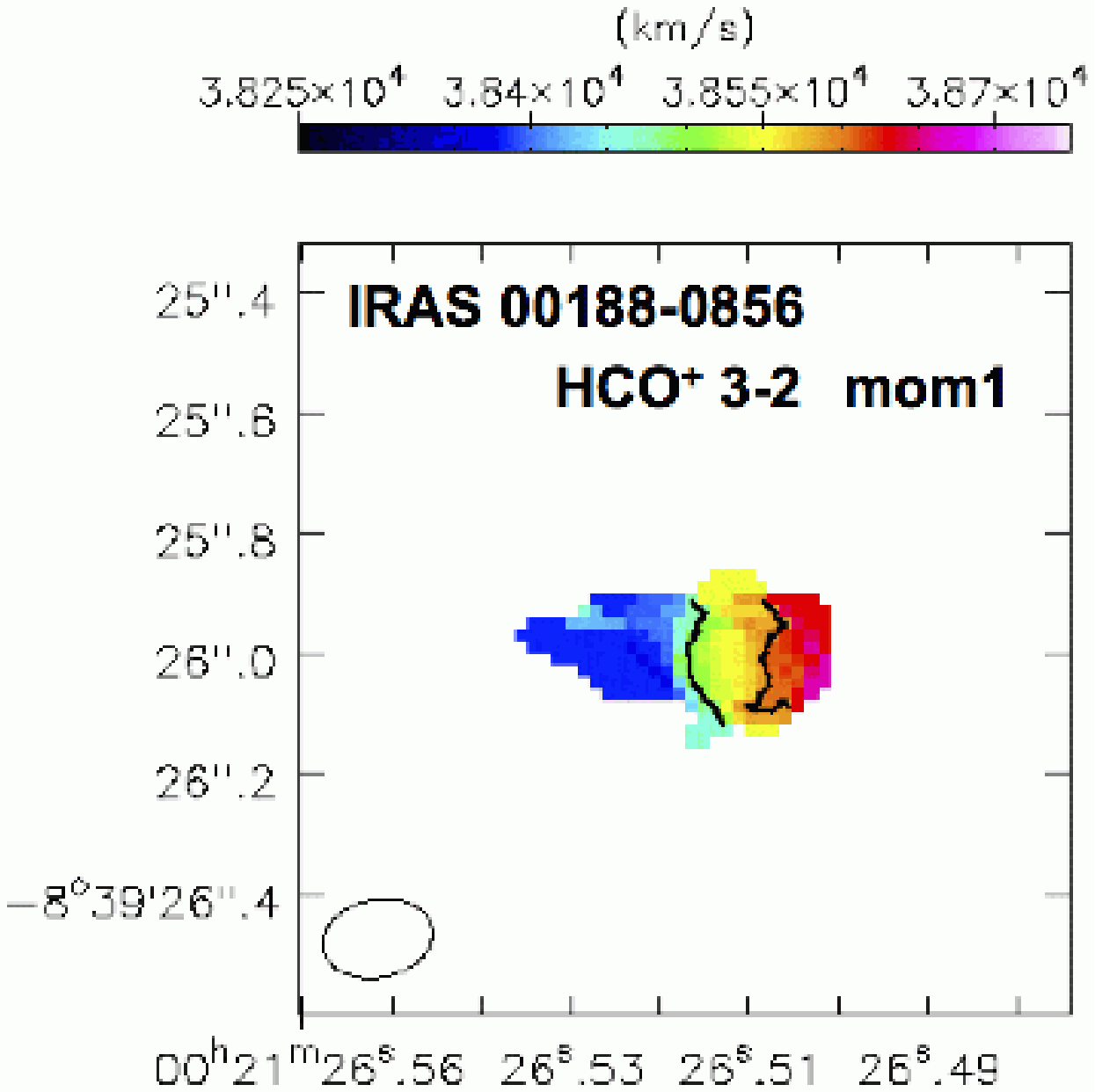} 
\includegraphics[angle=0,scale=.314]{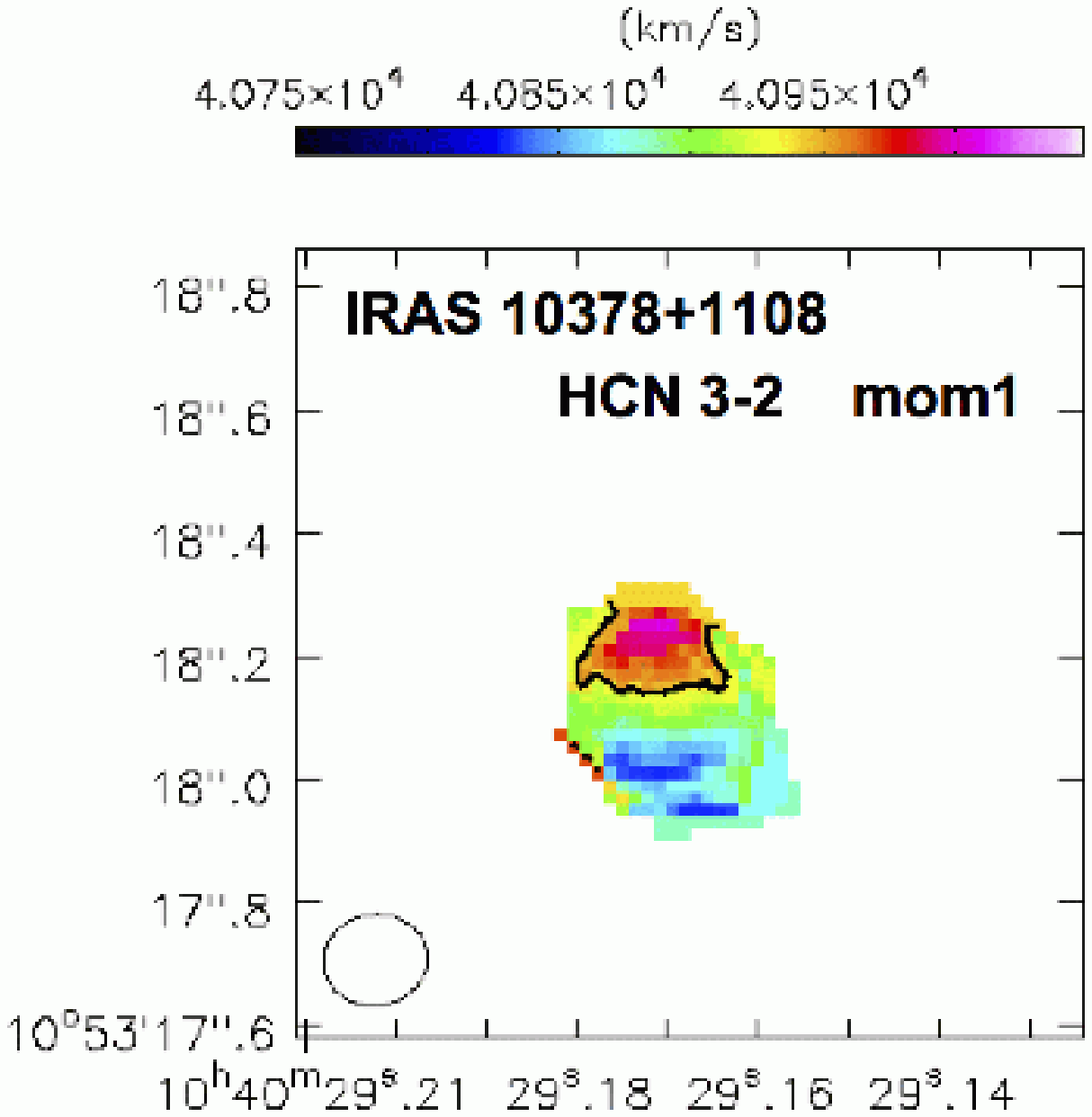} 
\includegraphics[angle=0,scale=.314]{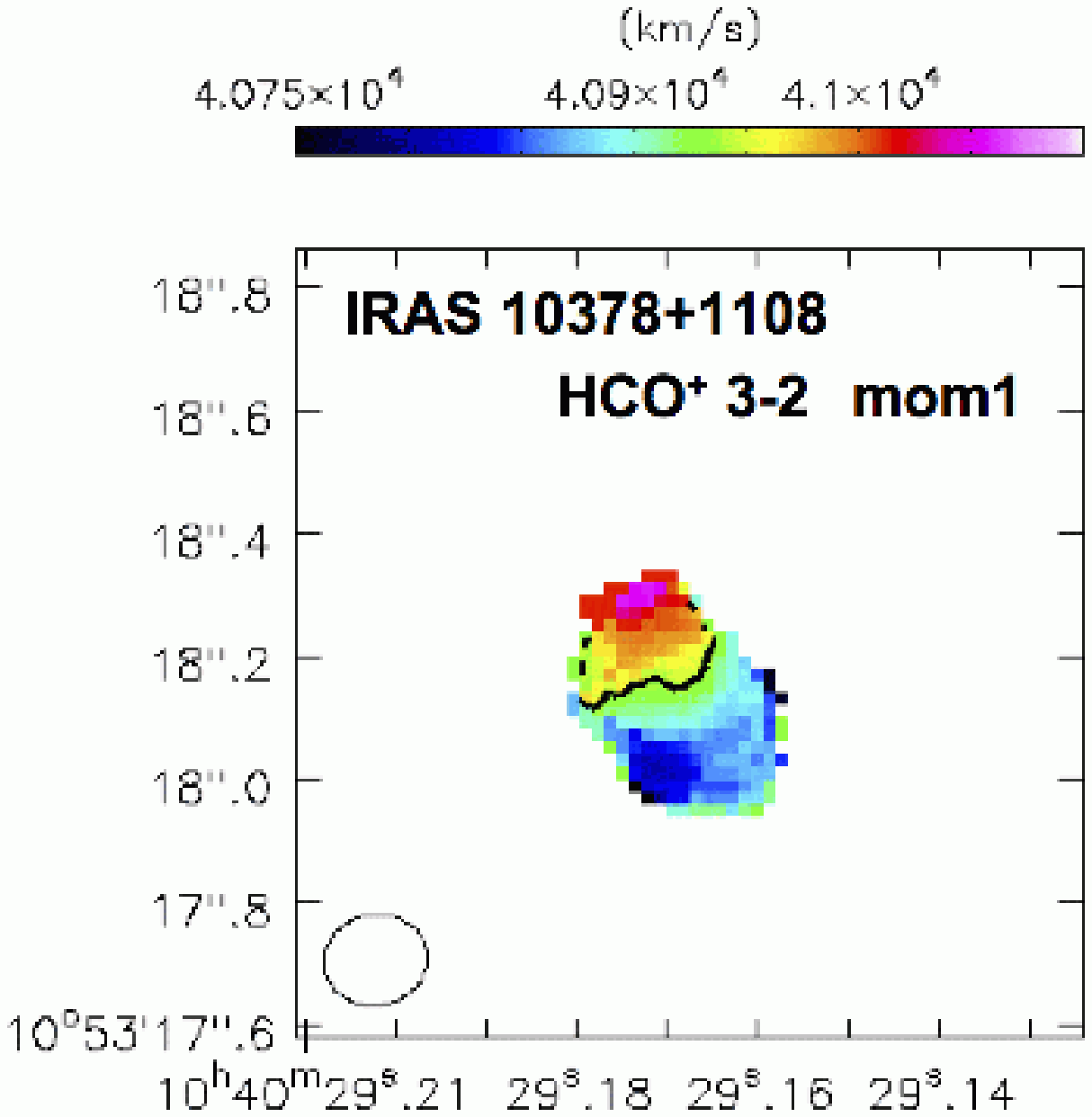} \\
\includegraphics[angle=0,scale=.314]{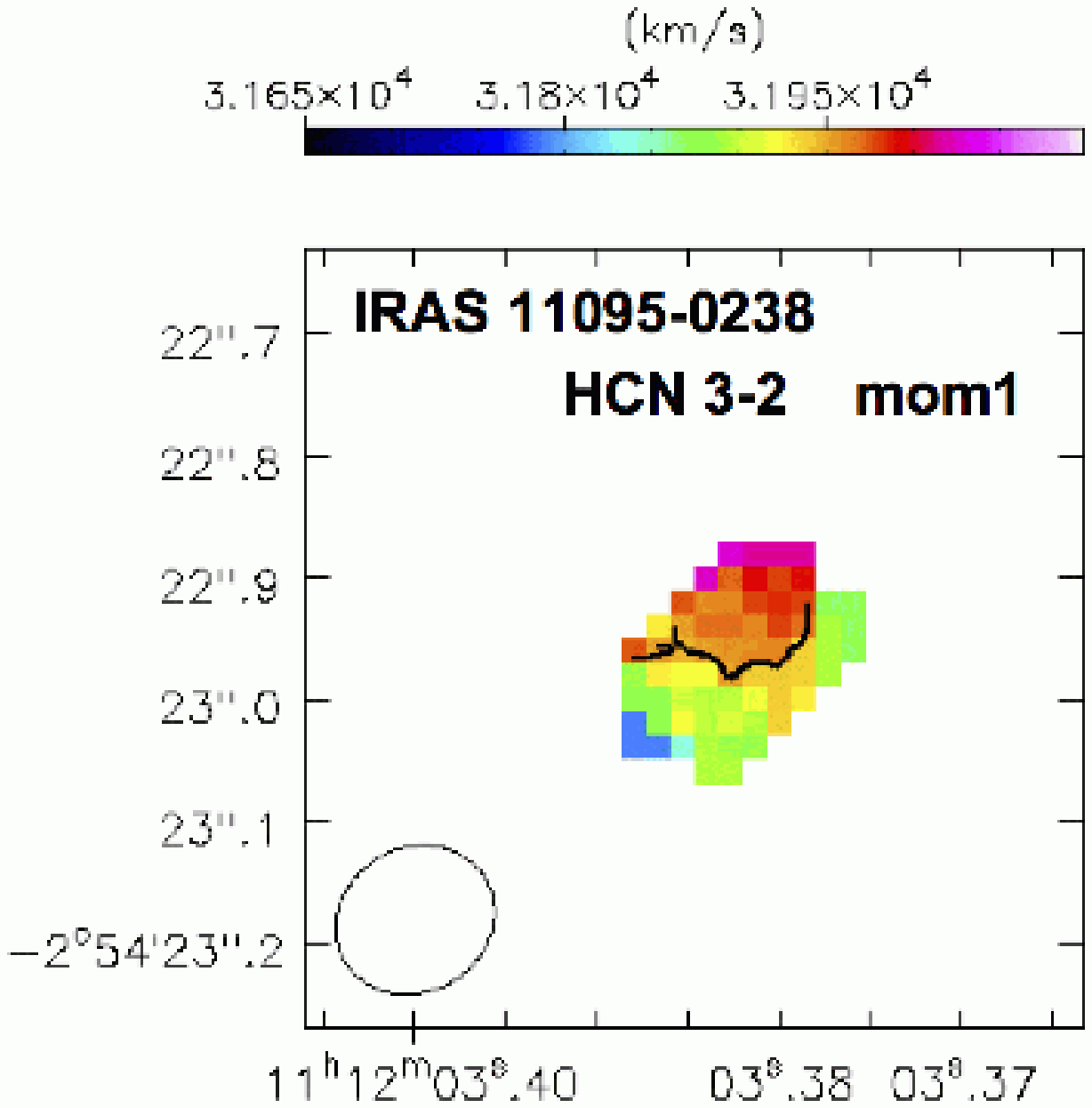} 
\includegraphics[angle=0,scale=.314]{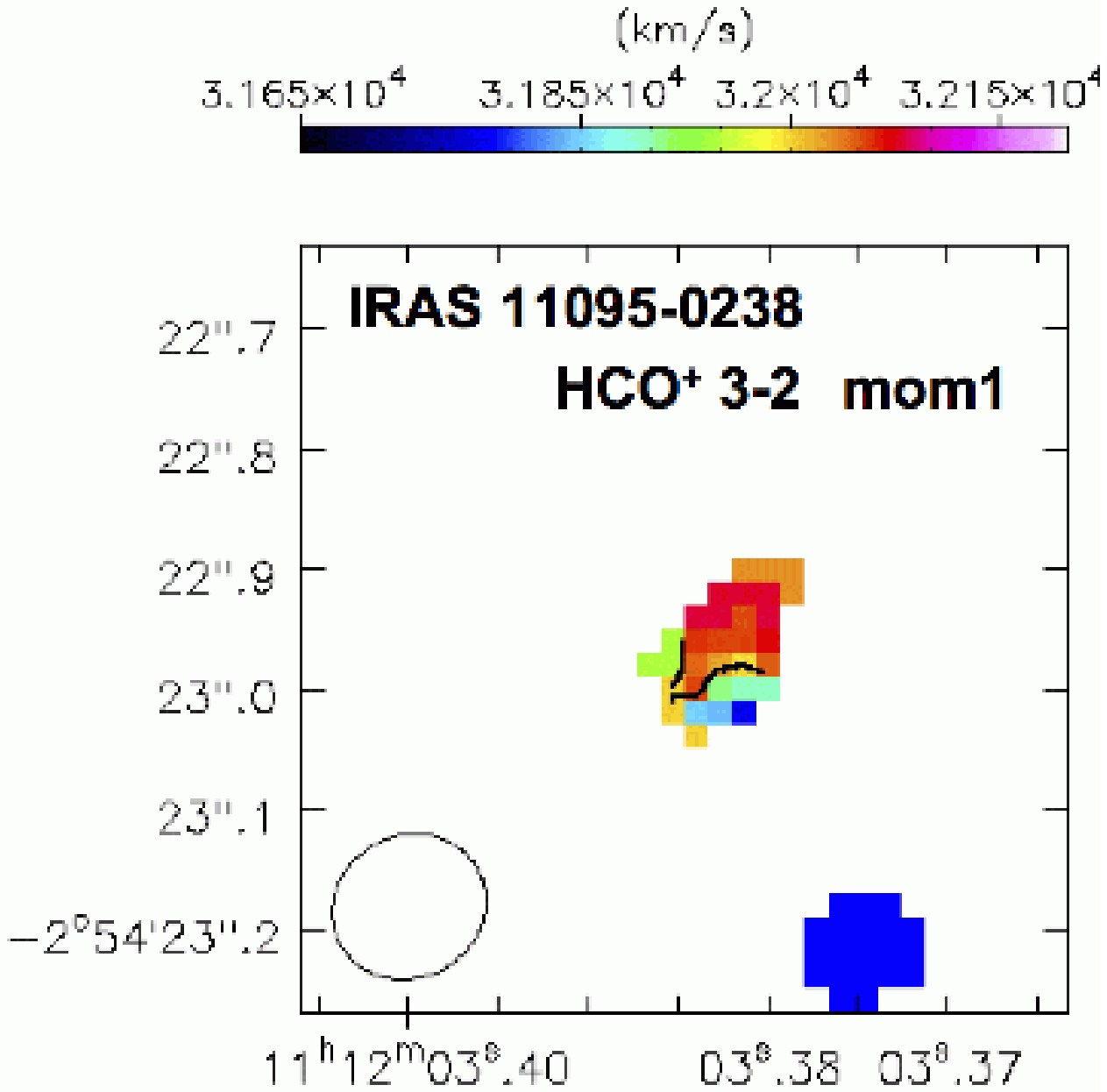} 
\includegraphics[angle=0,scale=.3104]{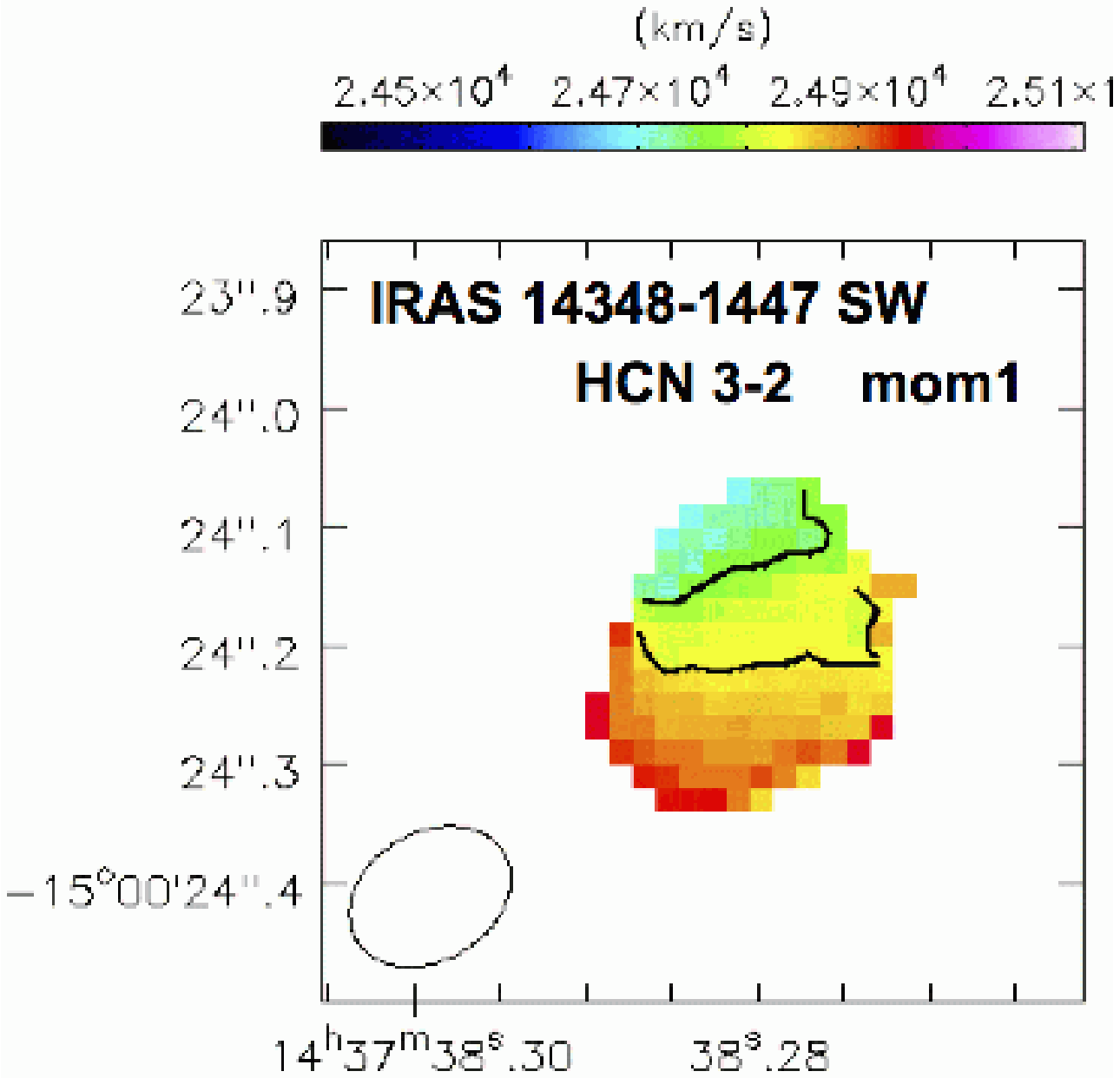} 
\includegraphics[angle=0,scale=.314]{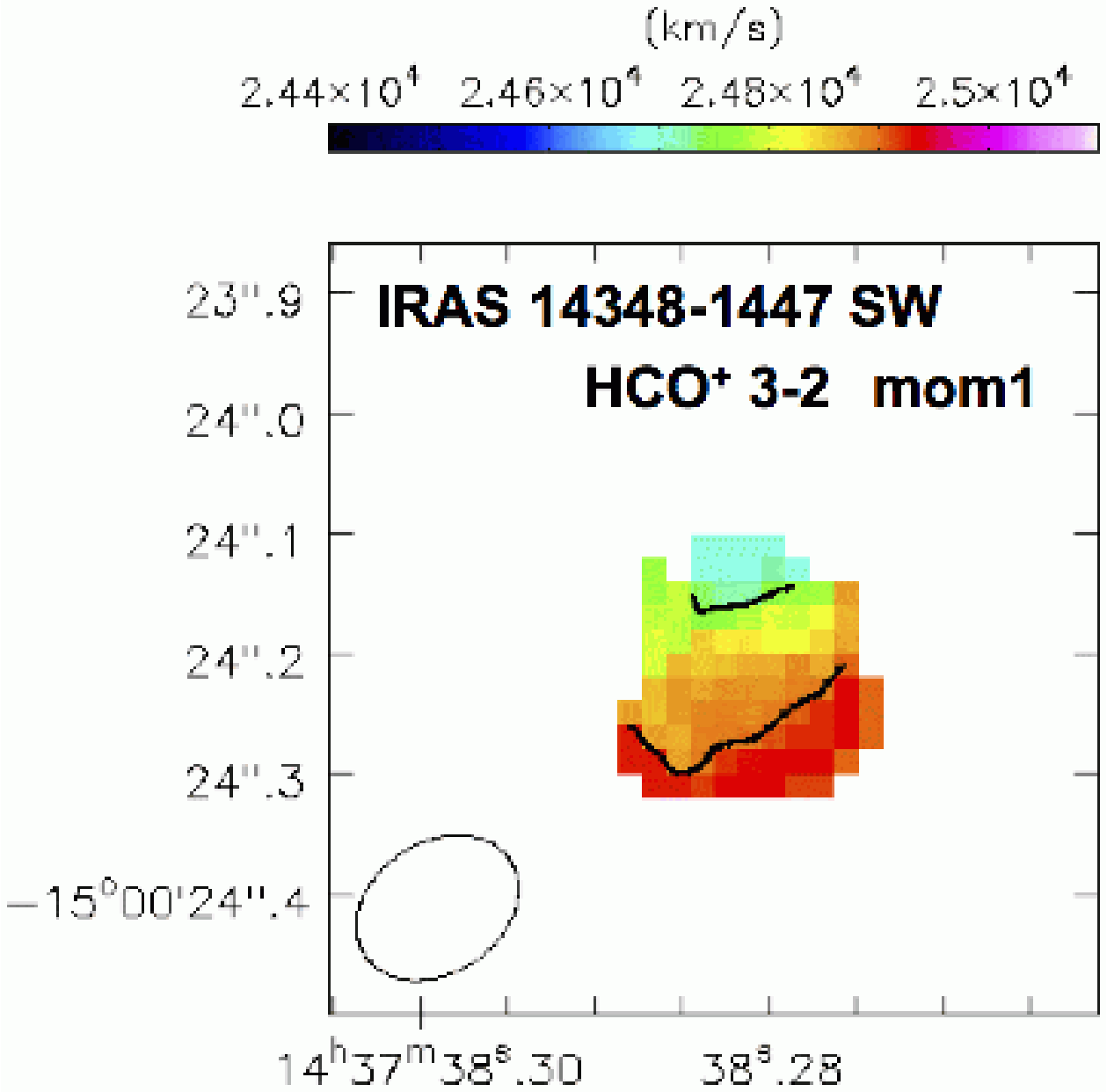} \\
\includegraphics[angle=0,scale=.3104]{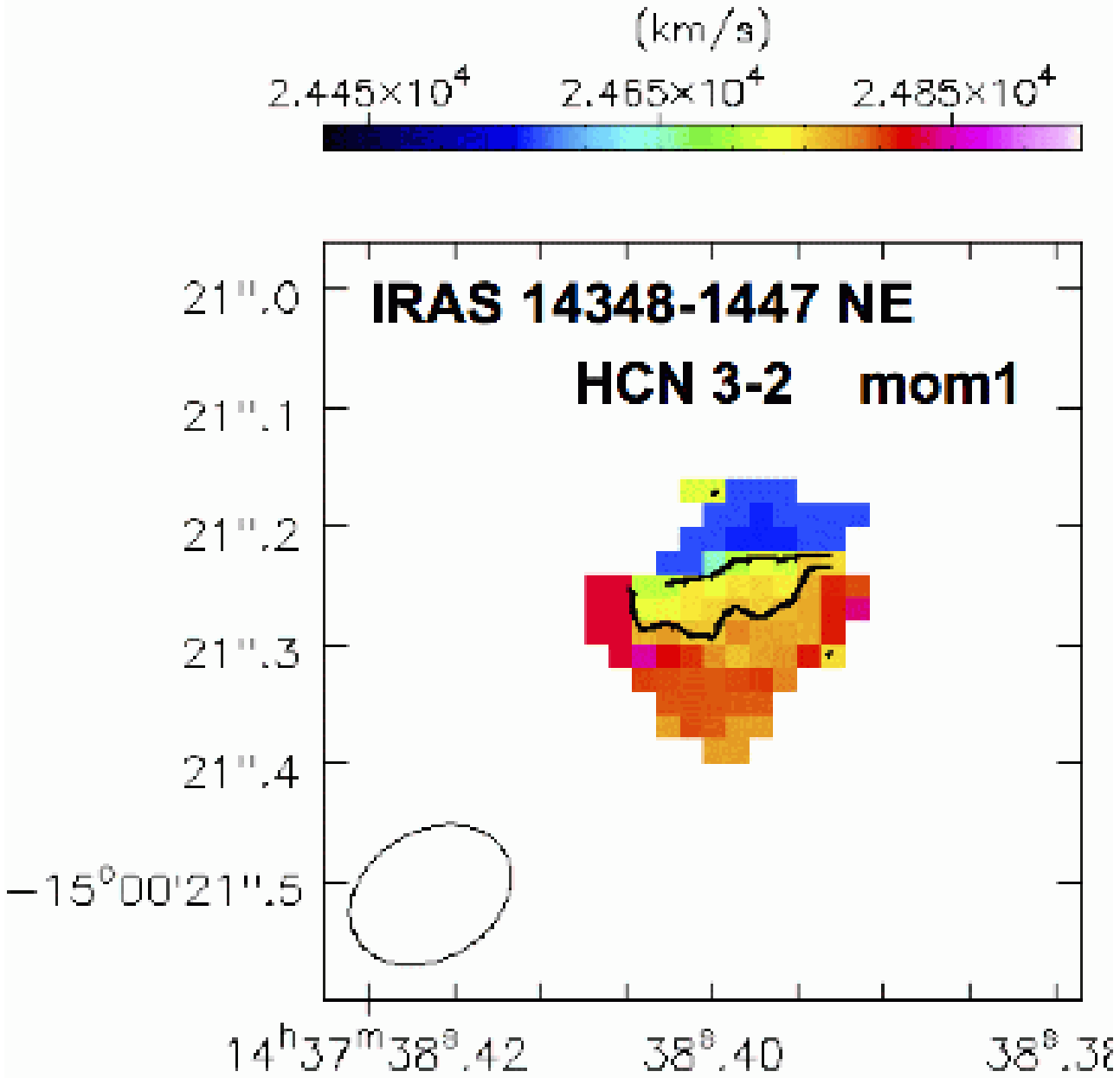} 
\hspace*{4cm}
\includegraphics[angle=0,scale=.314]{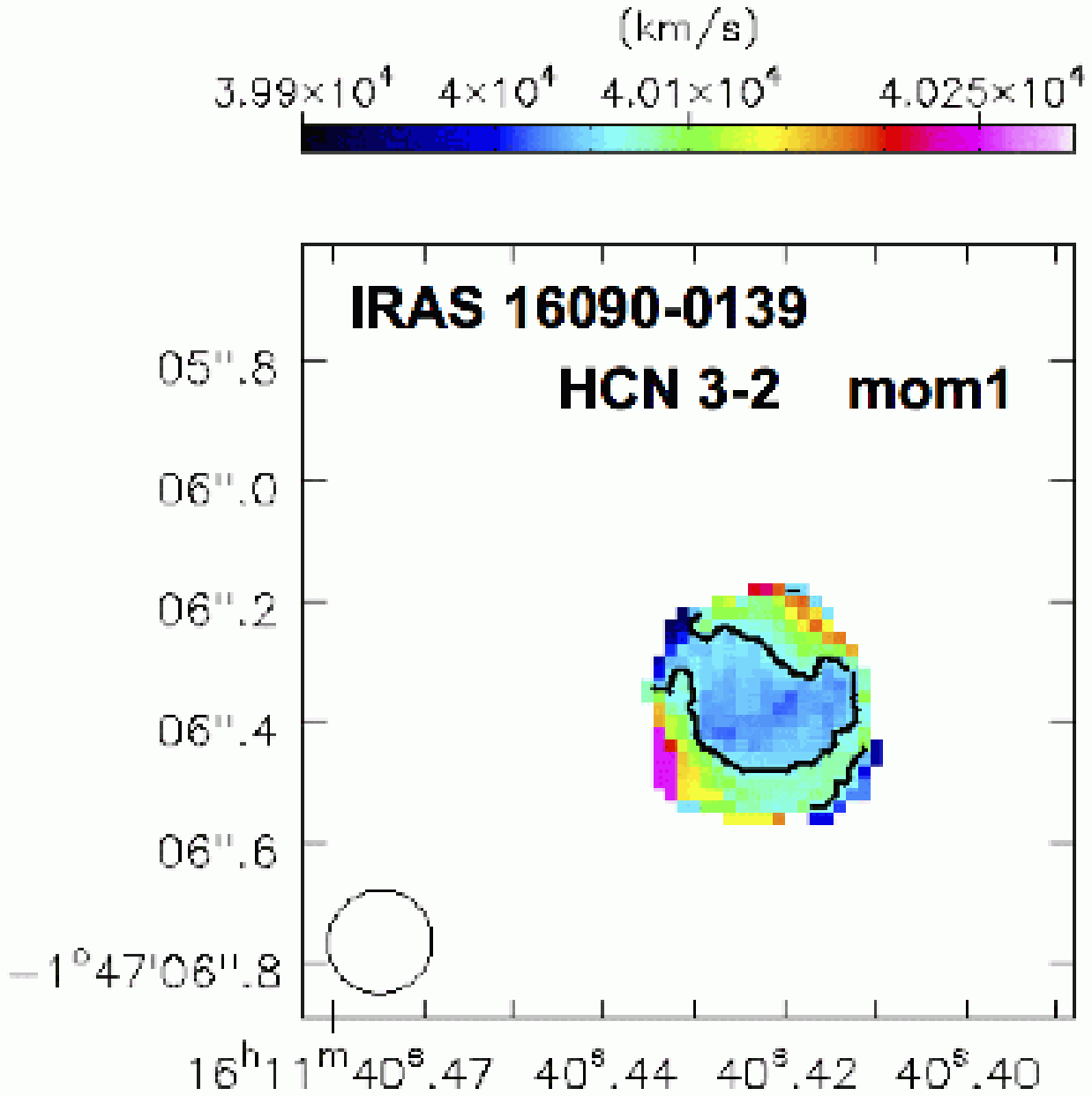} 
\includegraphics[angle=0,scale=.314]{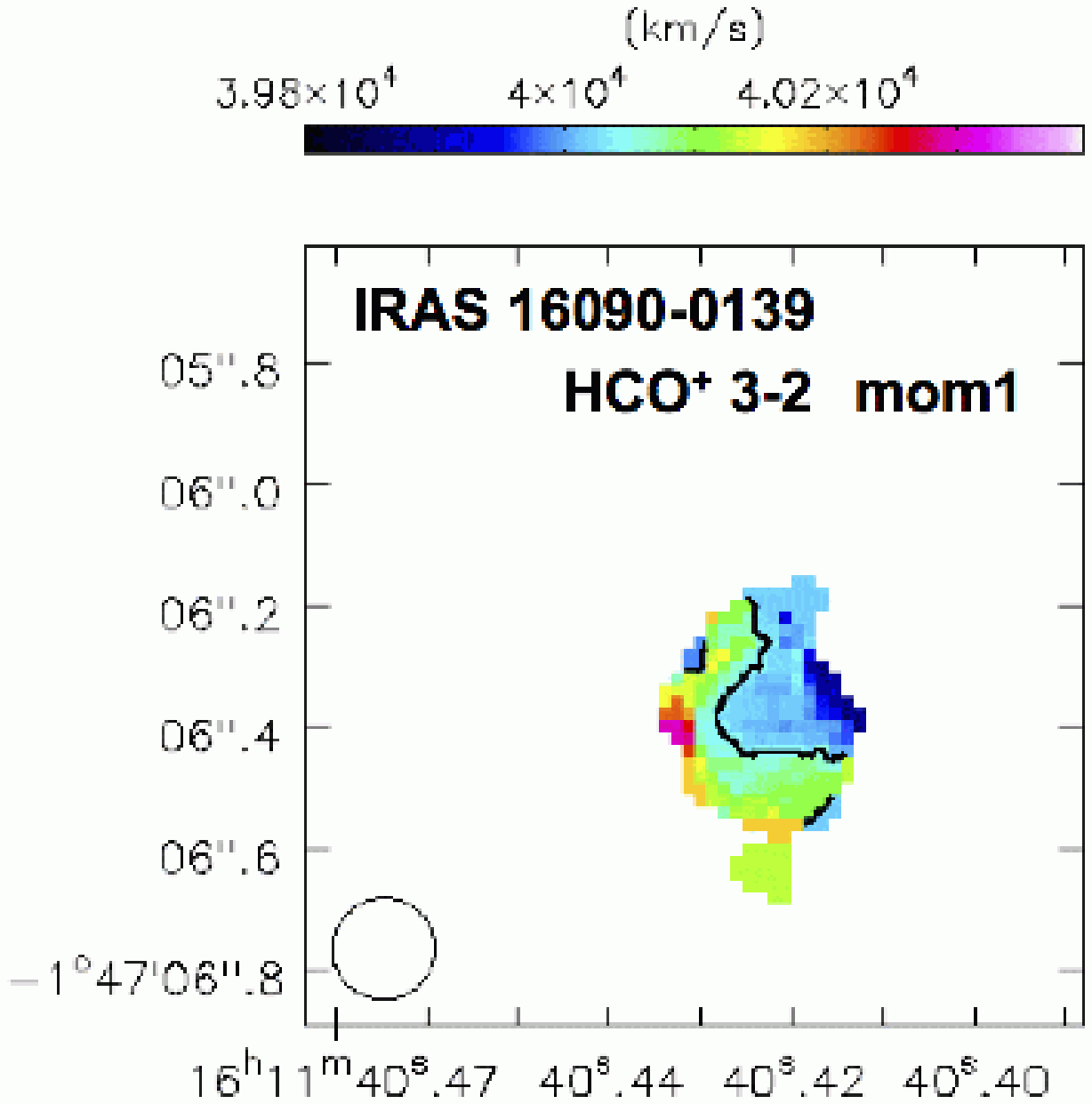} \\ 
\includegraphics[angle=0,scale=.314]{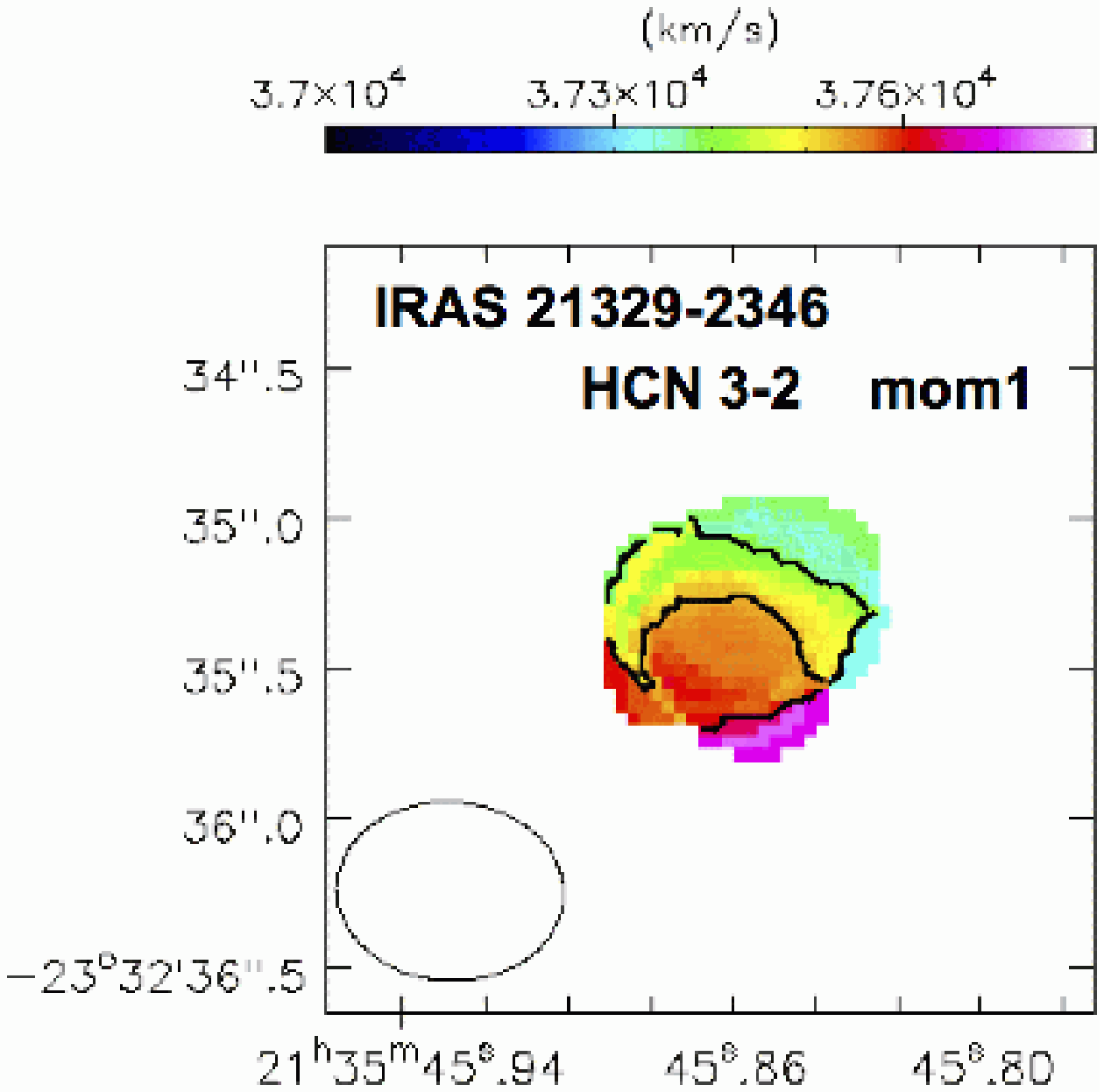} 
\includegraphics[angle=0,scale=.314]{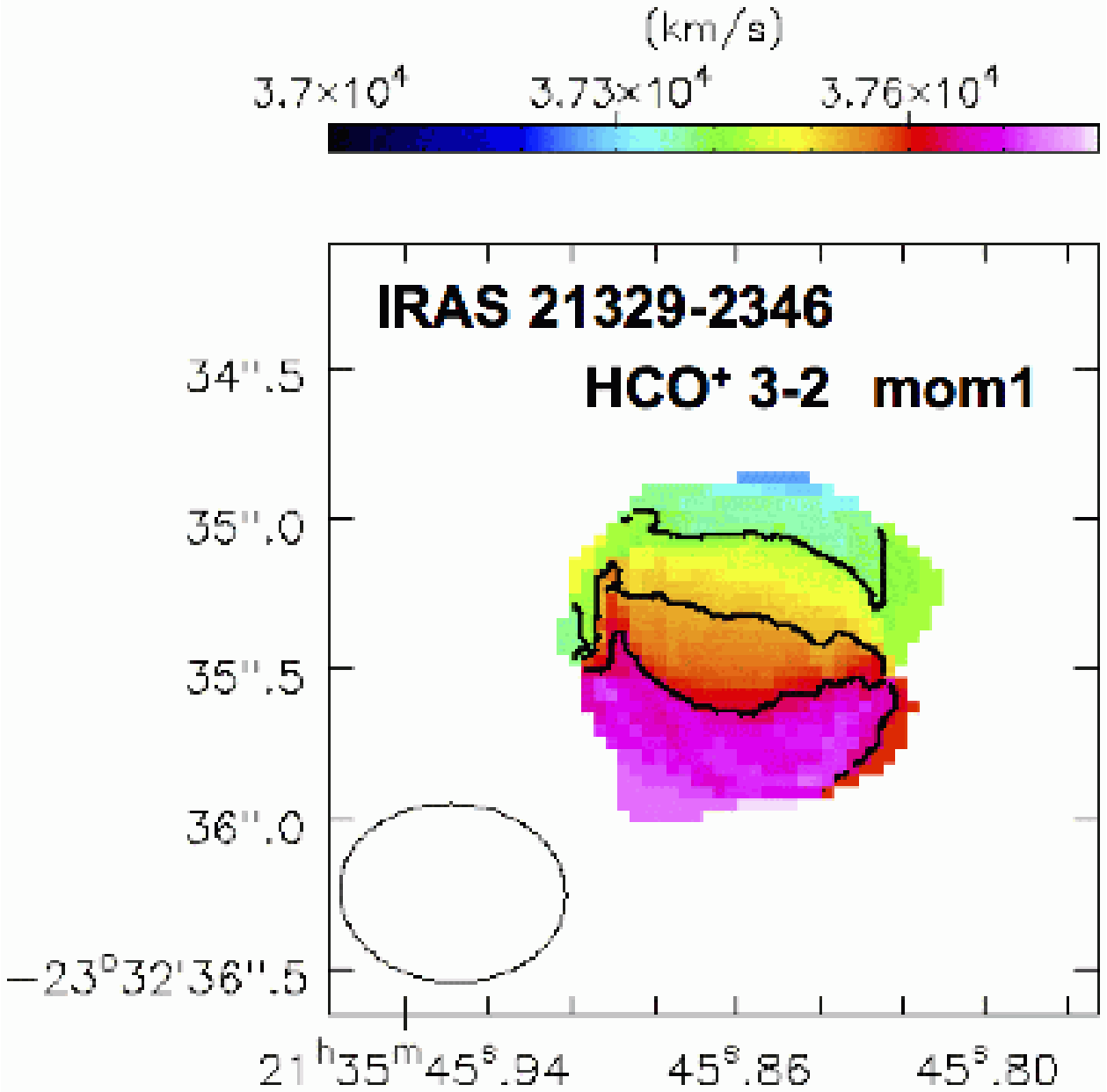} 
\includegraphics[angle=0,scale=.314]{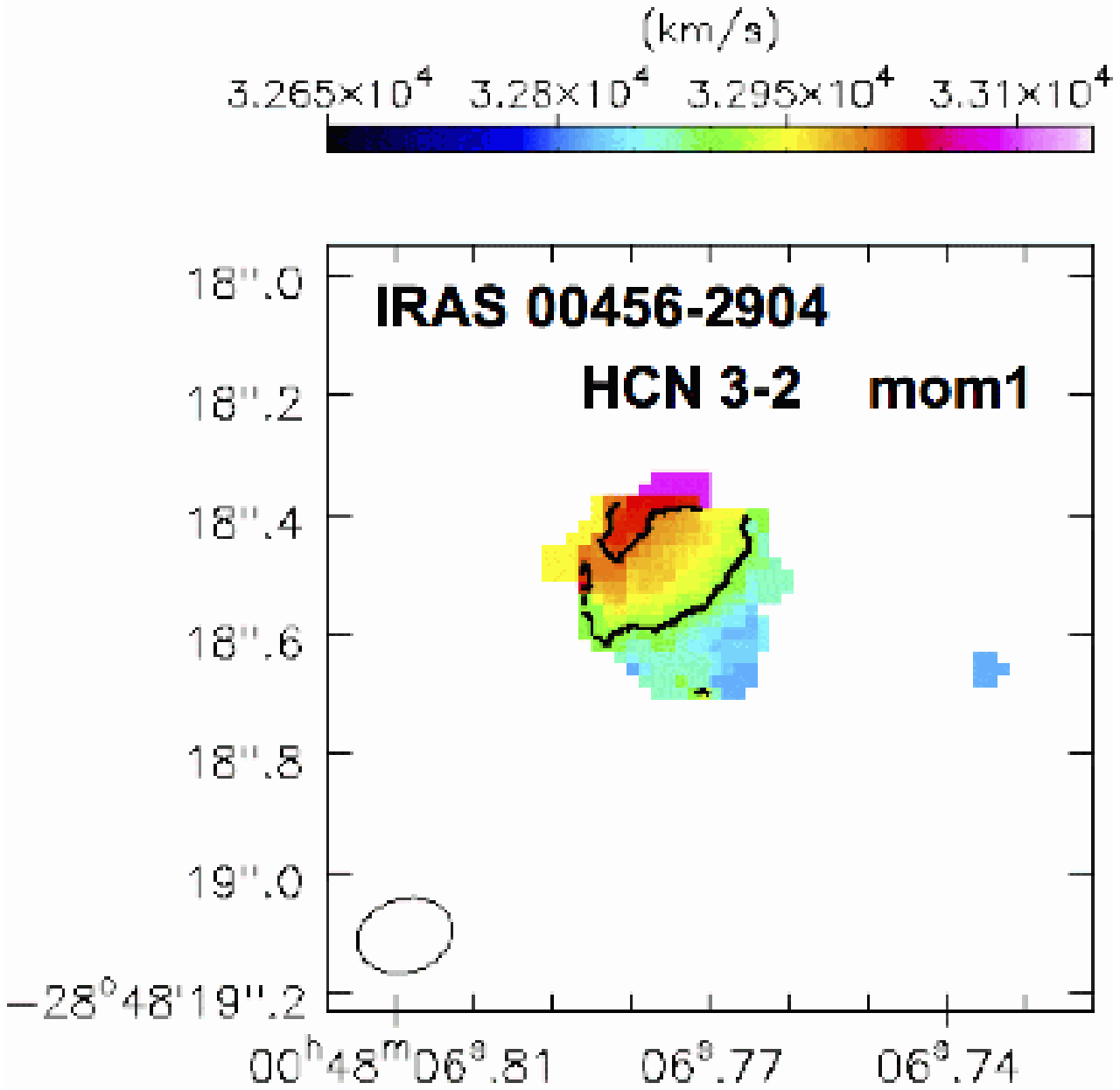} 
\includegraphics[angle=0,scale=.314]{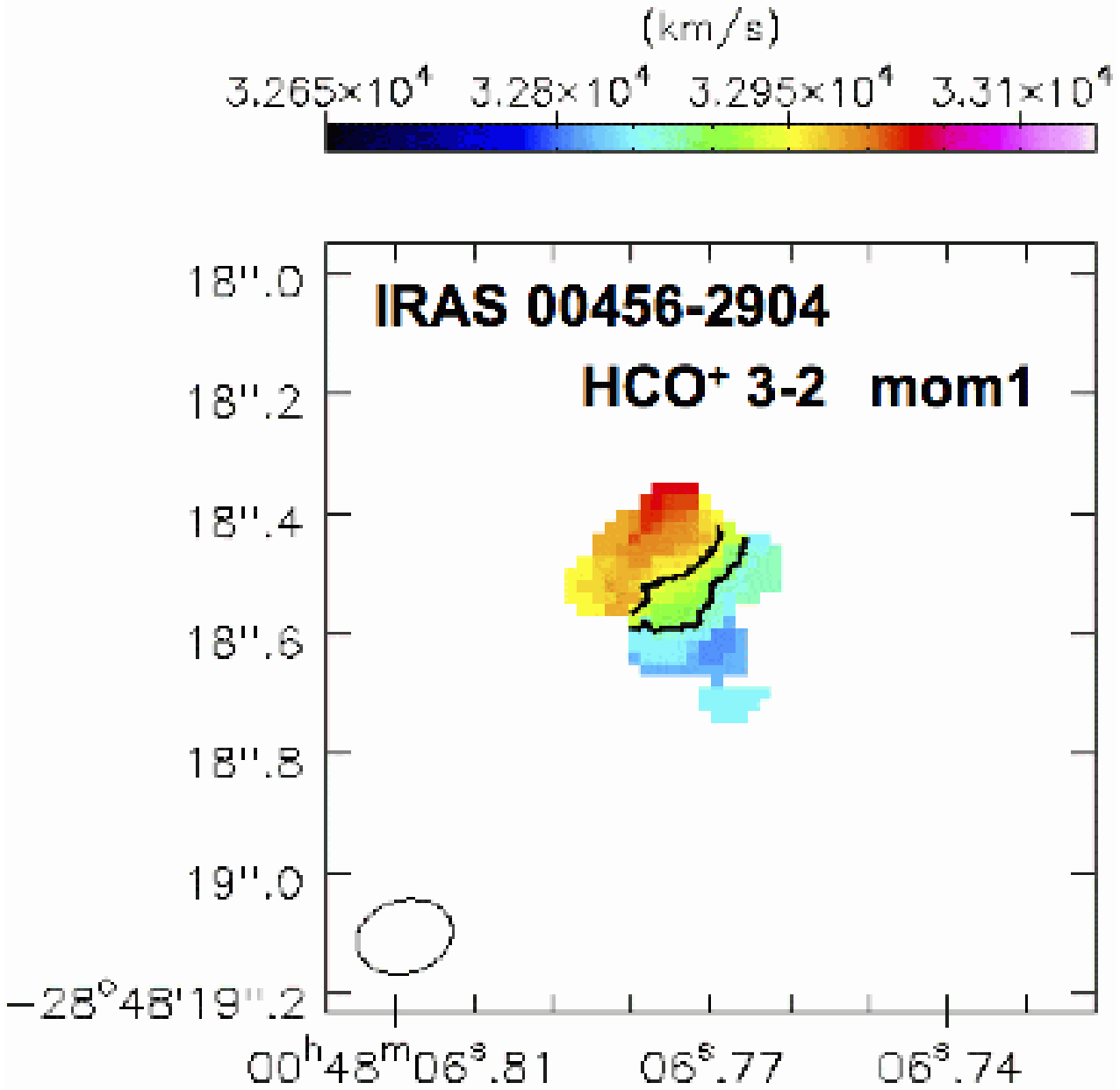} \\
\includegraphics[angle=0,scale=.314]{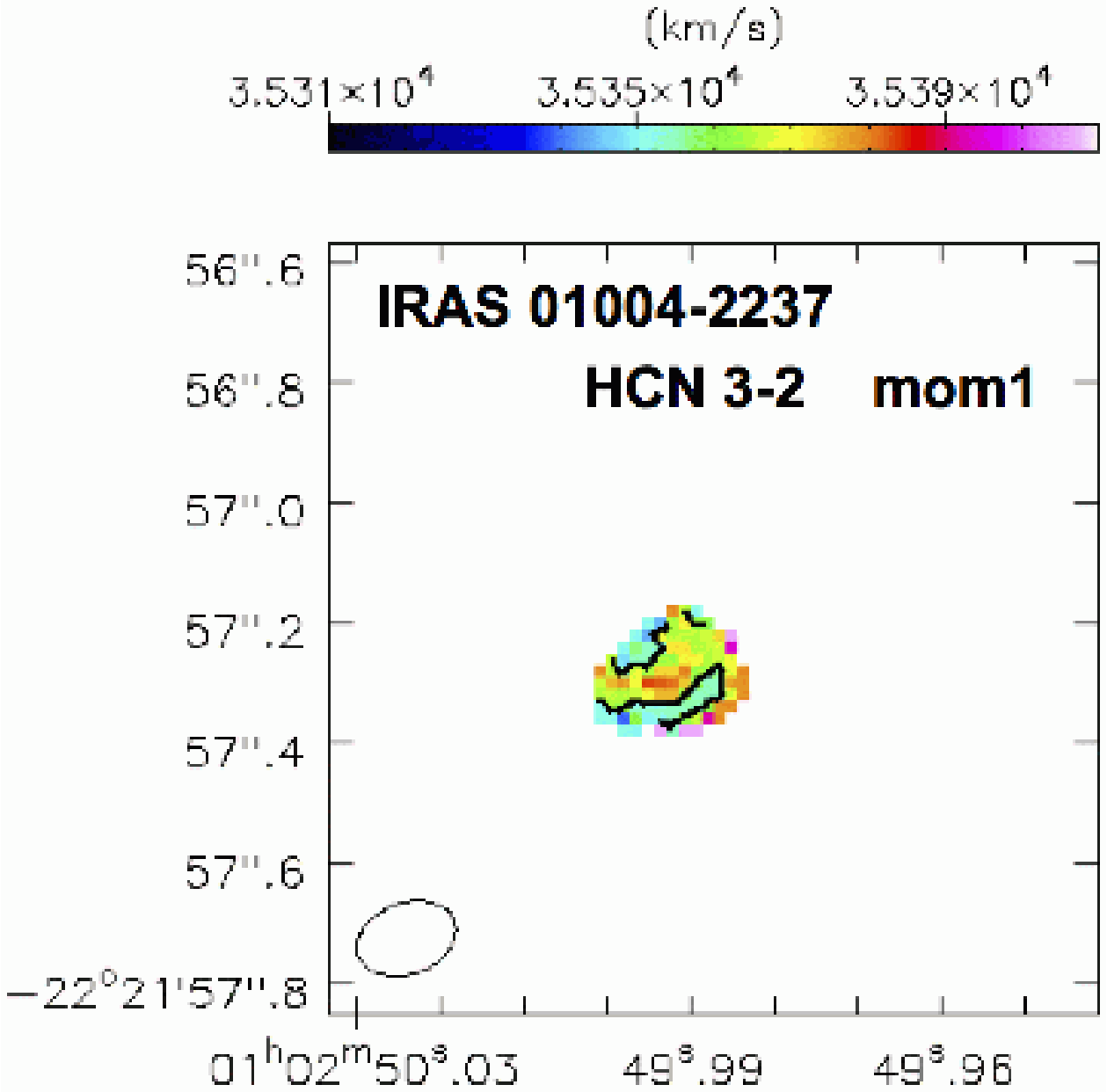} 
\includegraphics[angle=0,scale=.314]{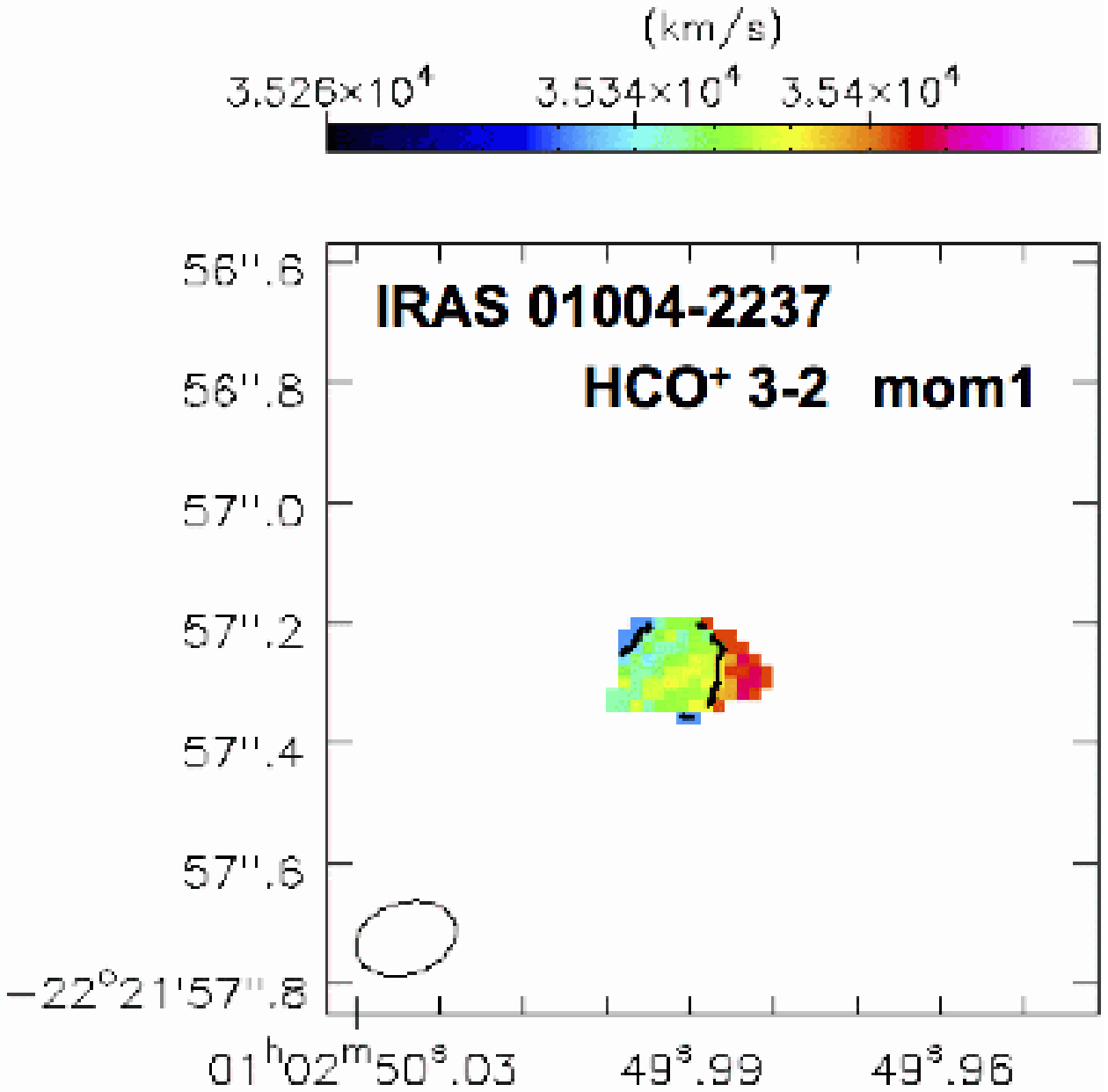} 
\includegraphics[angle=0,scale=.314]{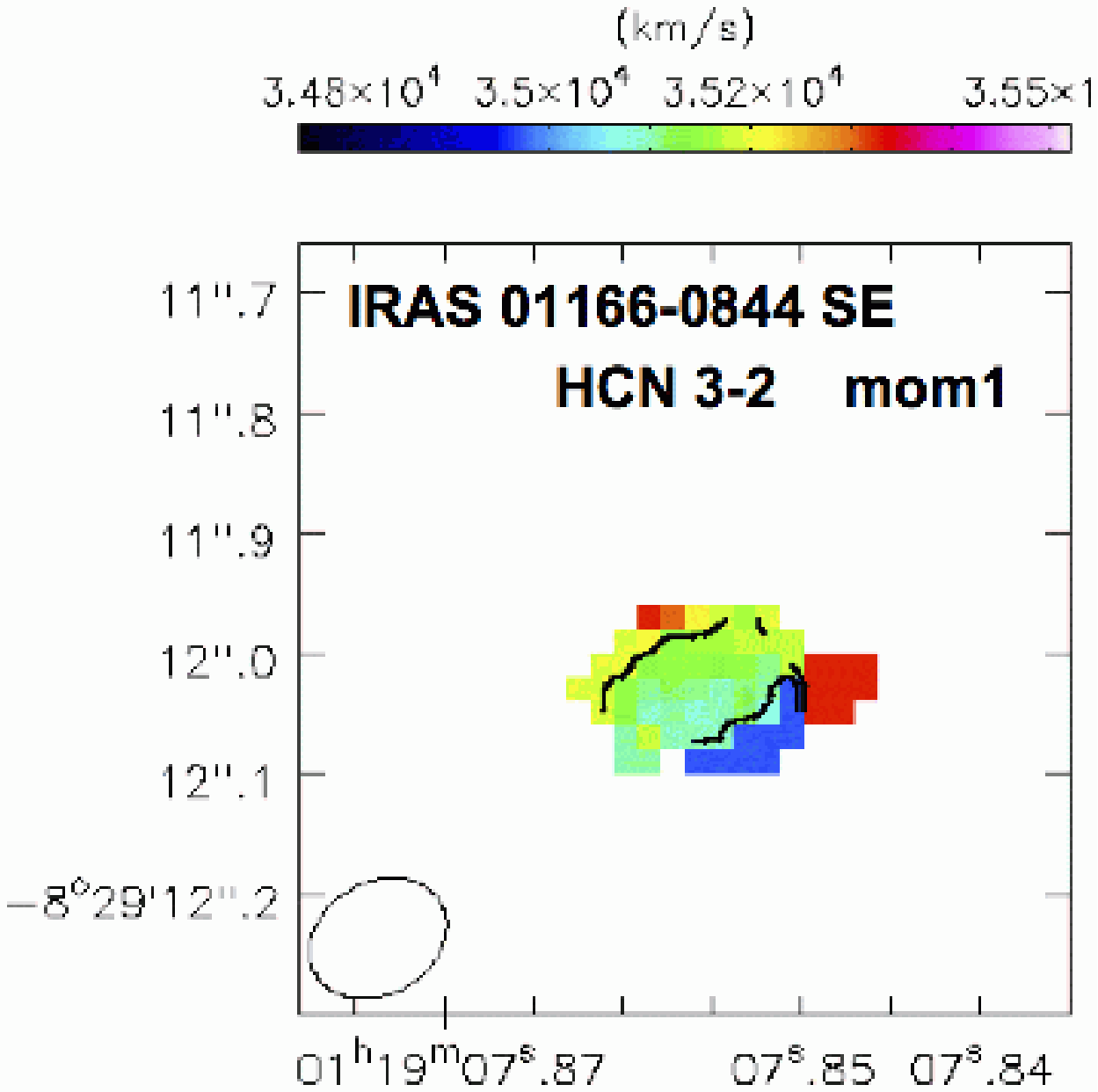} 
\includegraphics[angle=0,scale=.314]{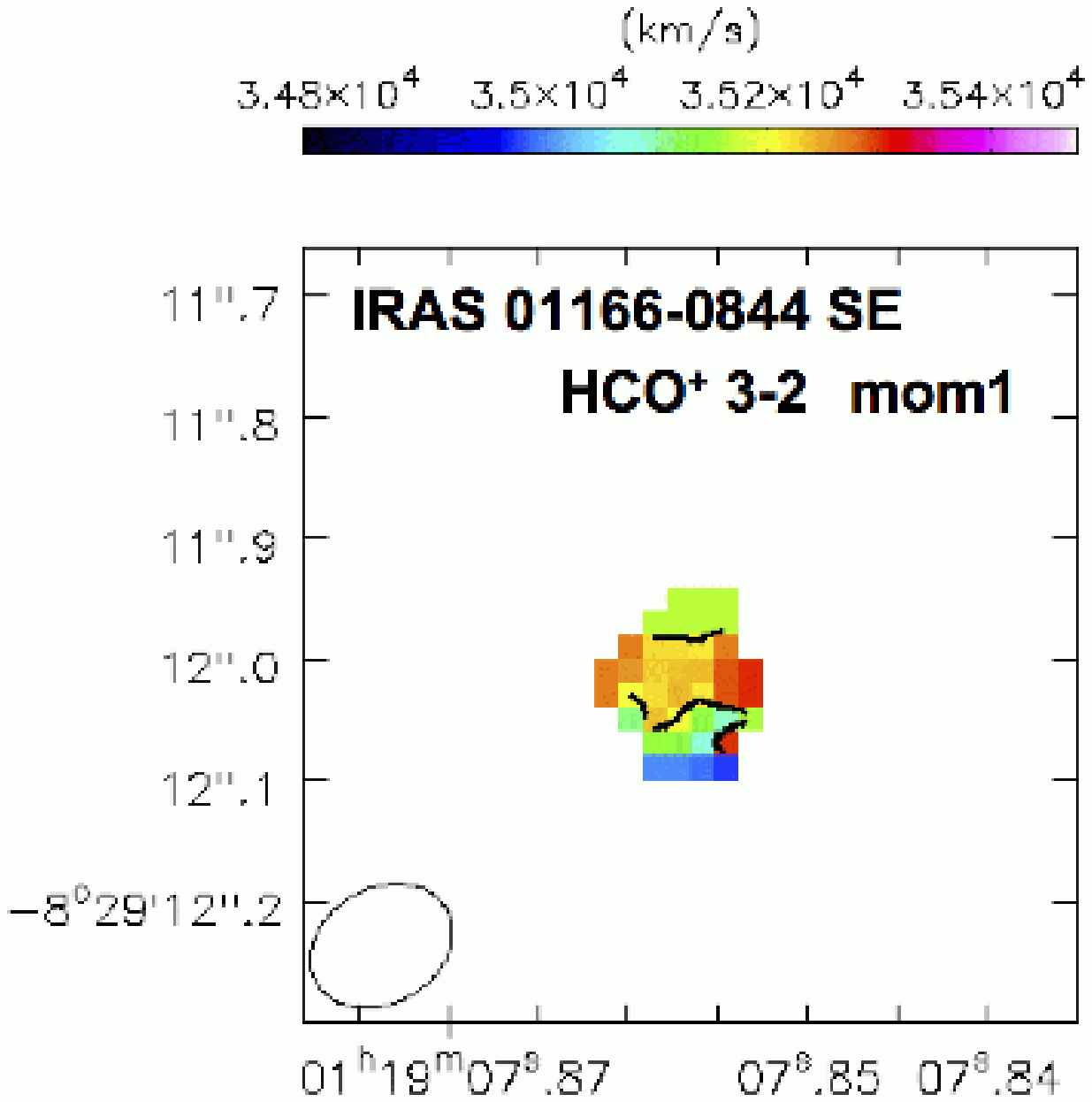} \\
\includegraphics[angle=0,scale=.314]{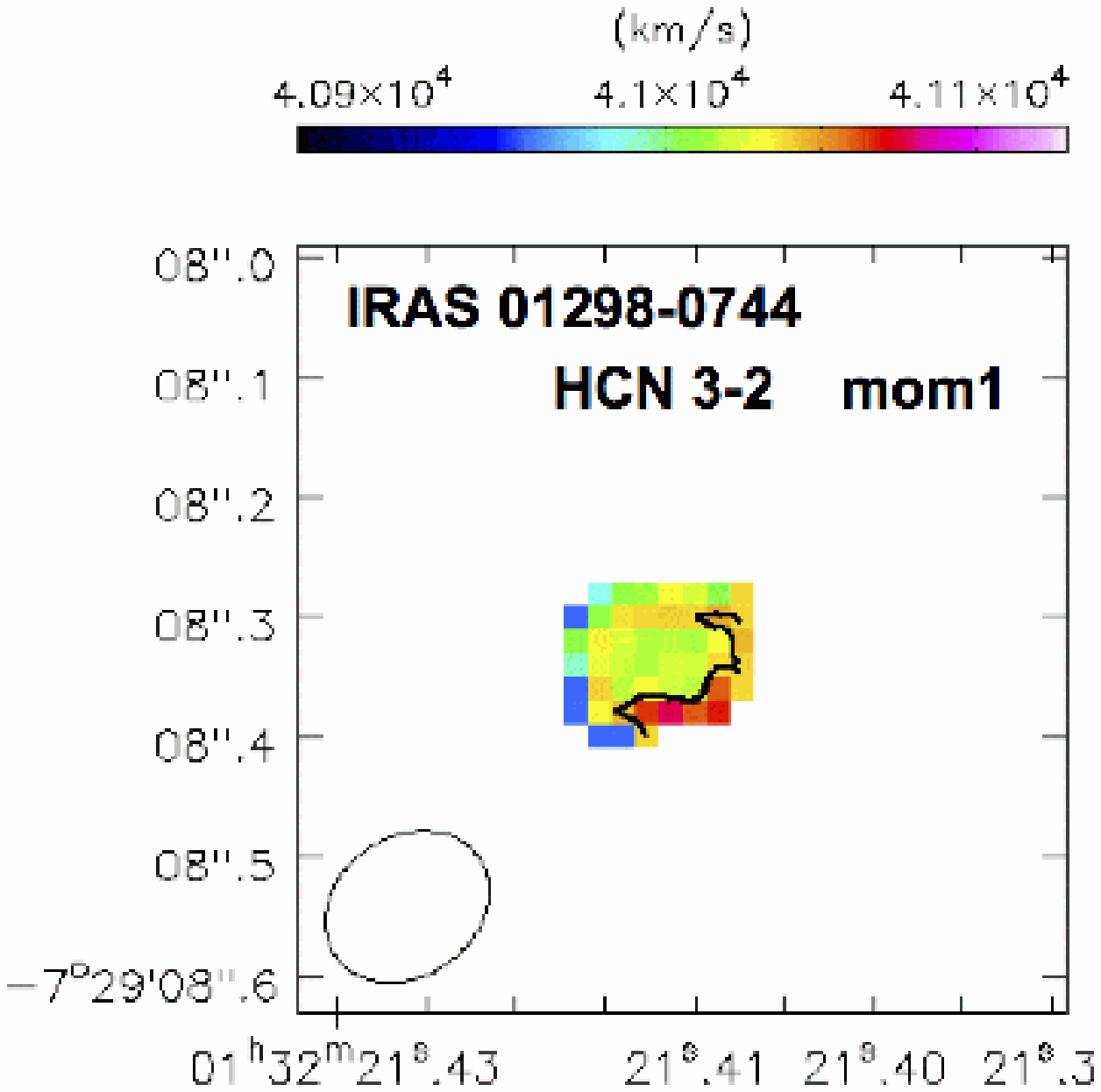} 
\includegraphics[angle=0,scale=.314]{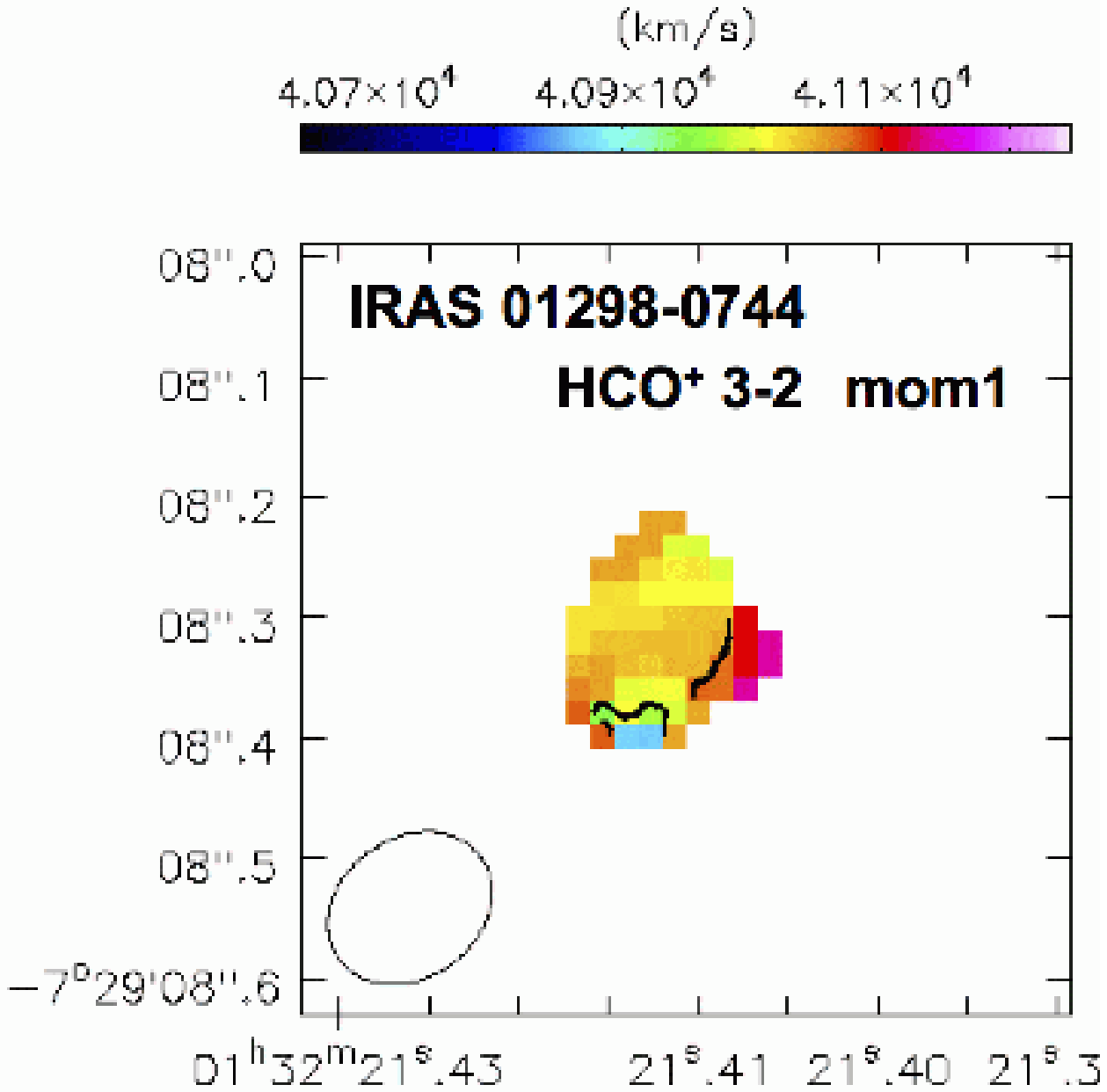} 
\includegraphics[angle=0,scale=.314]{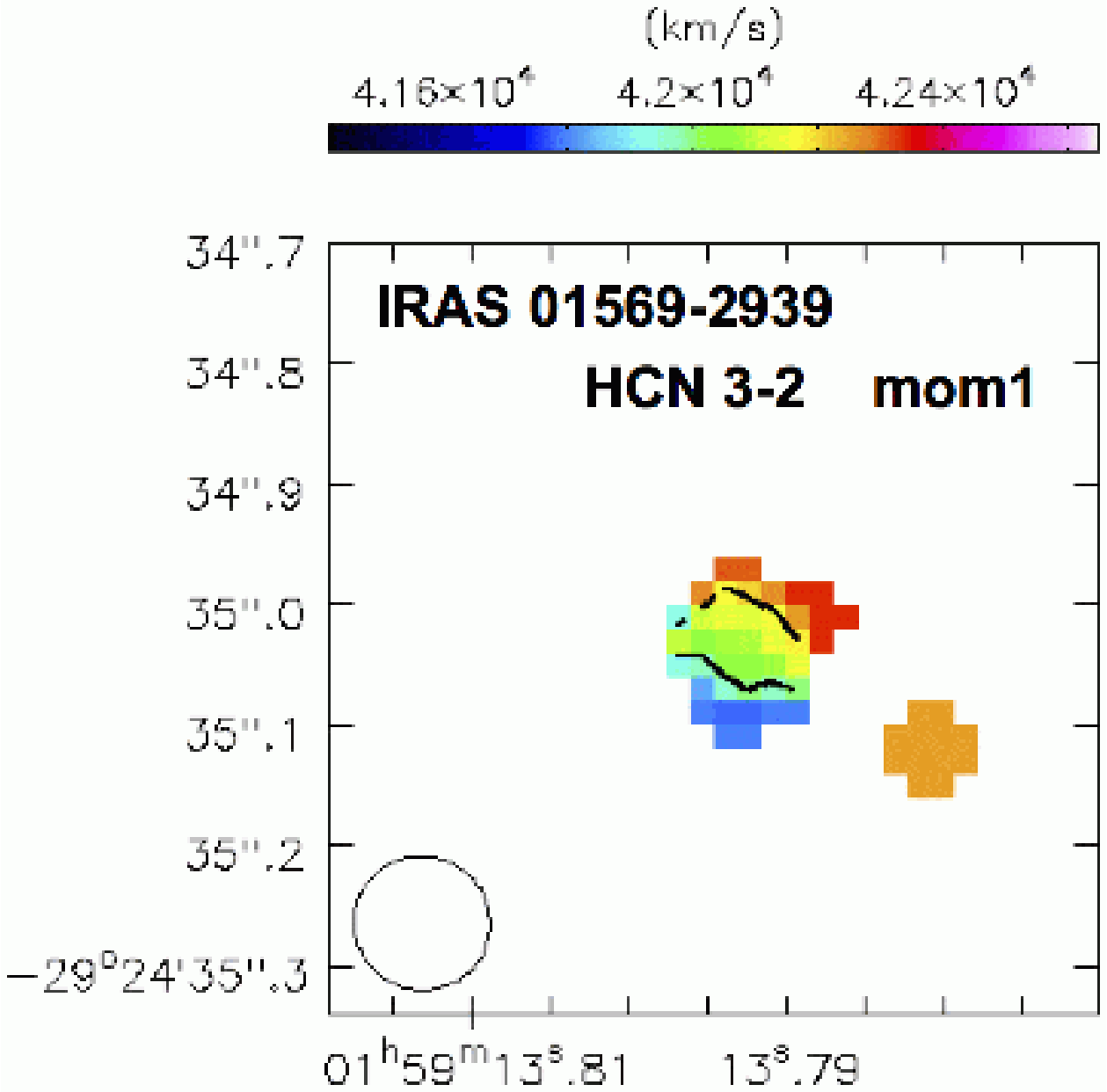} 
\includegraphics[angle=0,scale=.314]{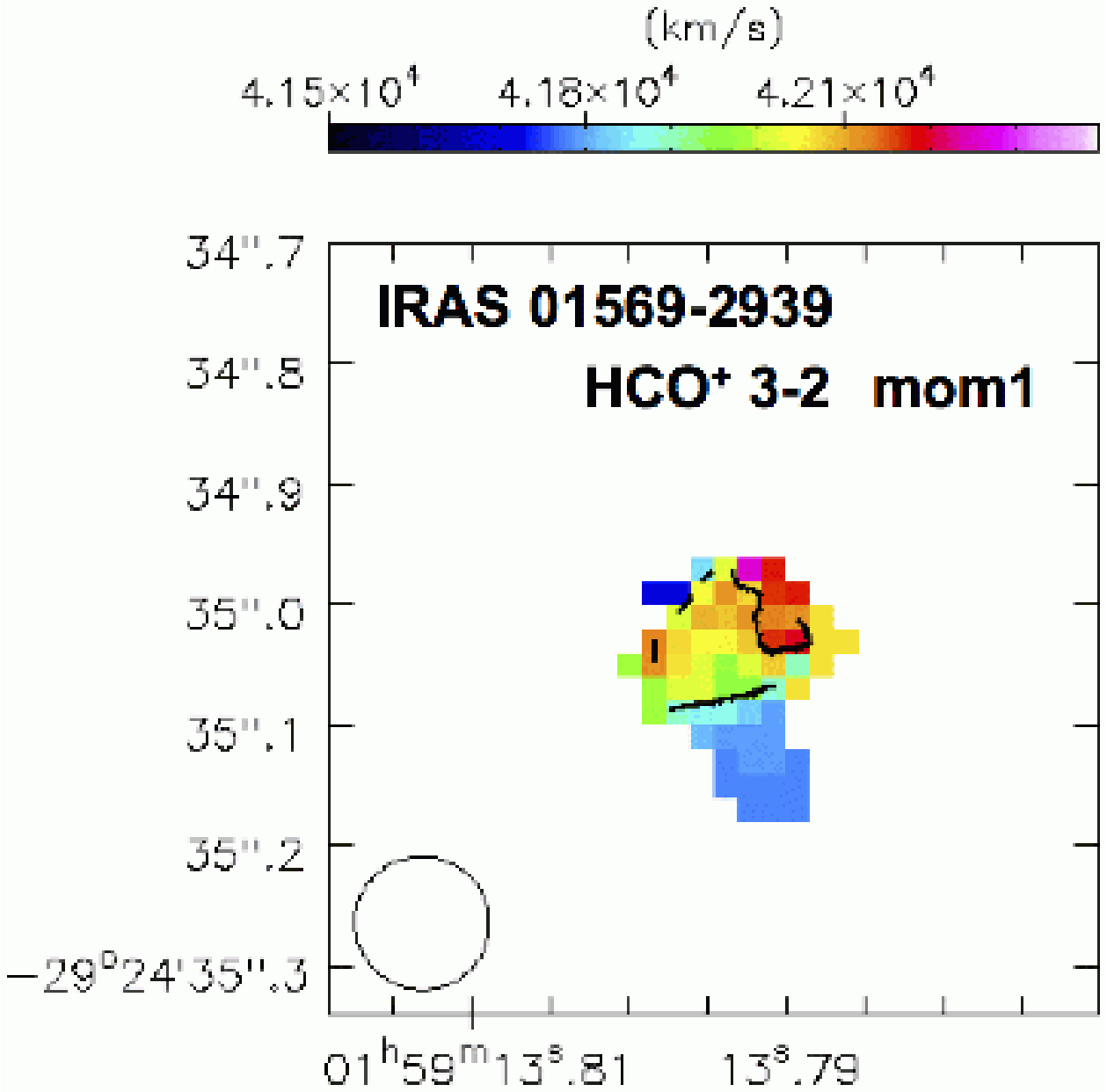} \\
\end{center}
\end{figure}

\clearpage

\begin{figure}
\begin{center}
\includegraphics[angle=0,scale=.314]{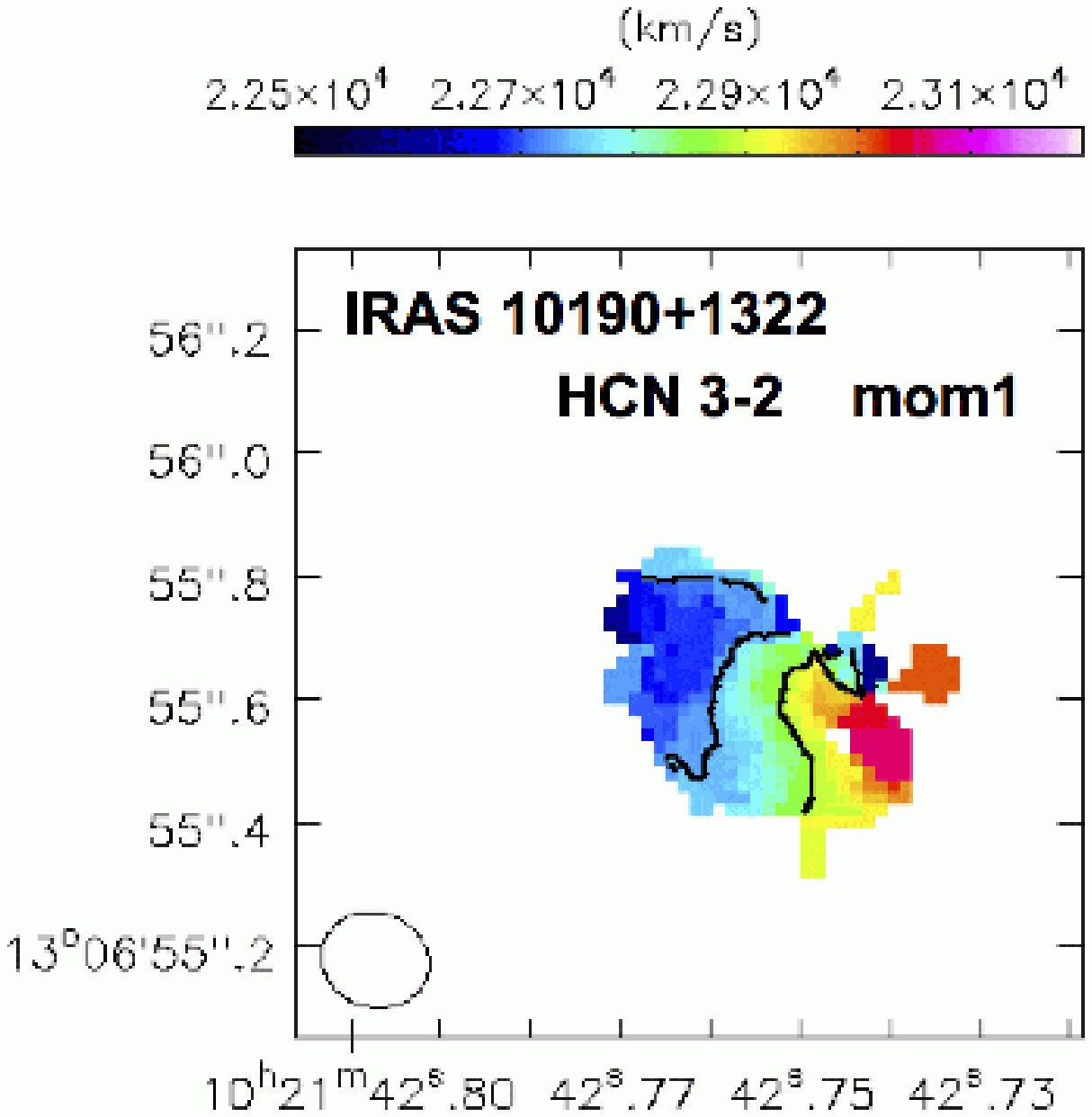} 
\includegraphics[angle=0,scale=.314]{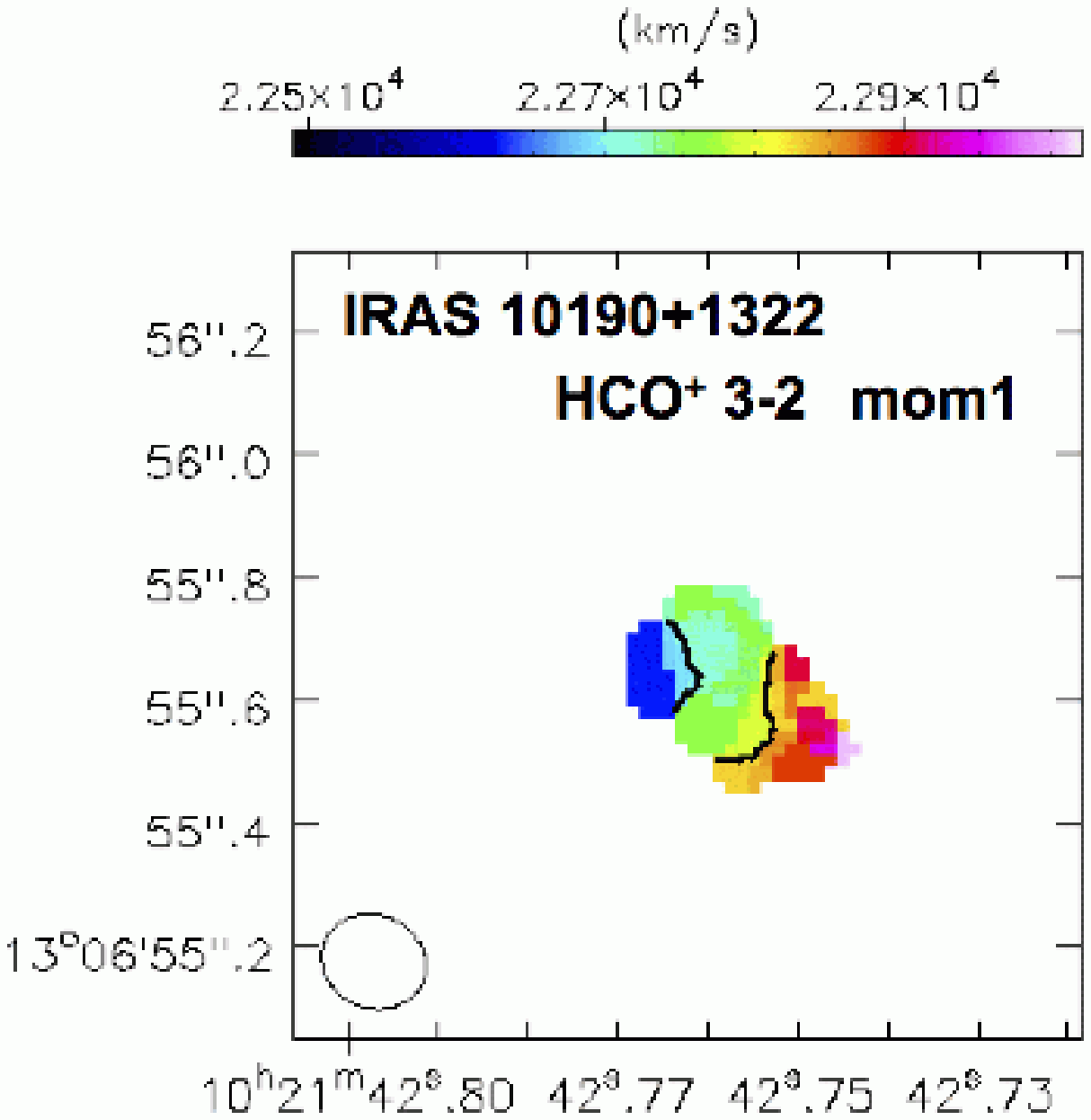} 
\includegraphics[angle=0,scale=.314]{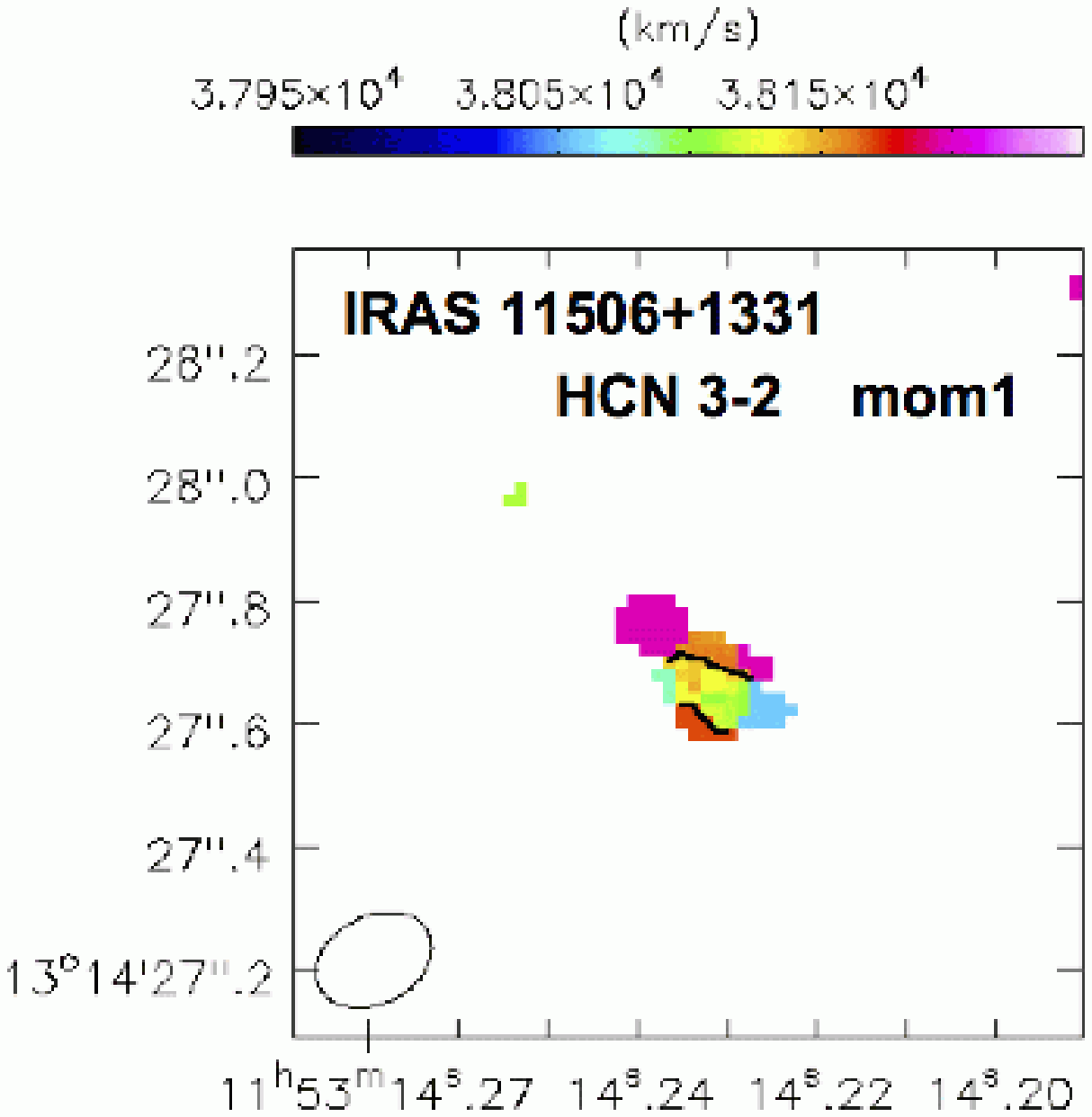} 
\includegraphics[angle=0,scale=.314]{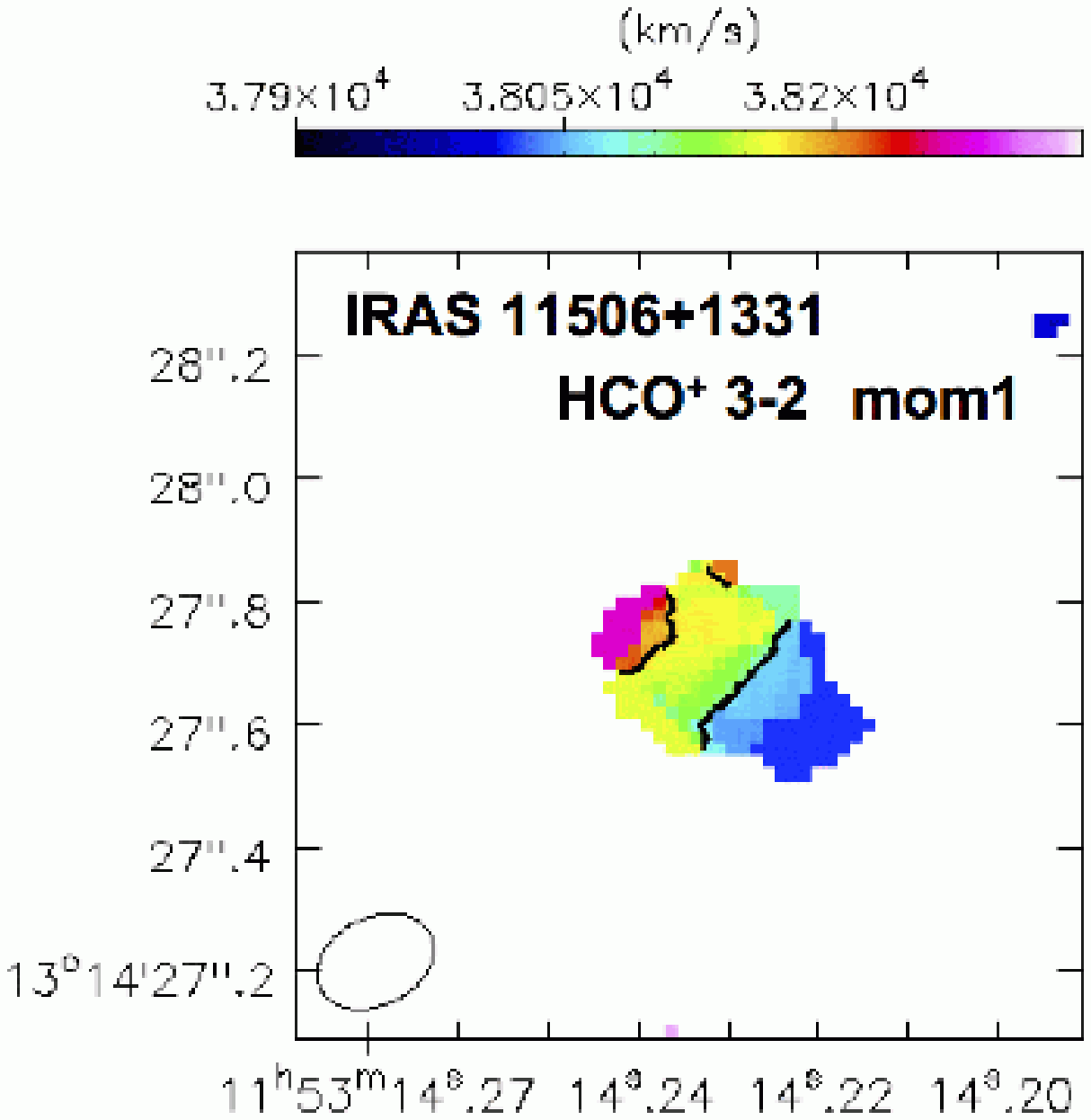} \\
\hspace*{-8.6cm}
\includegraphics[angle=0,scale=.314]{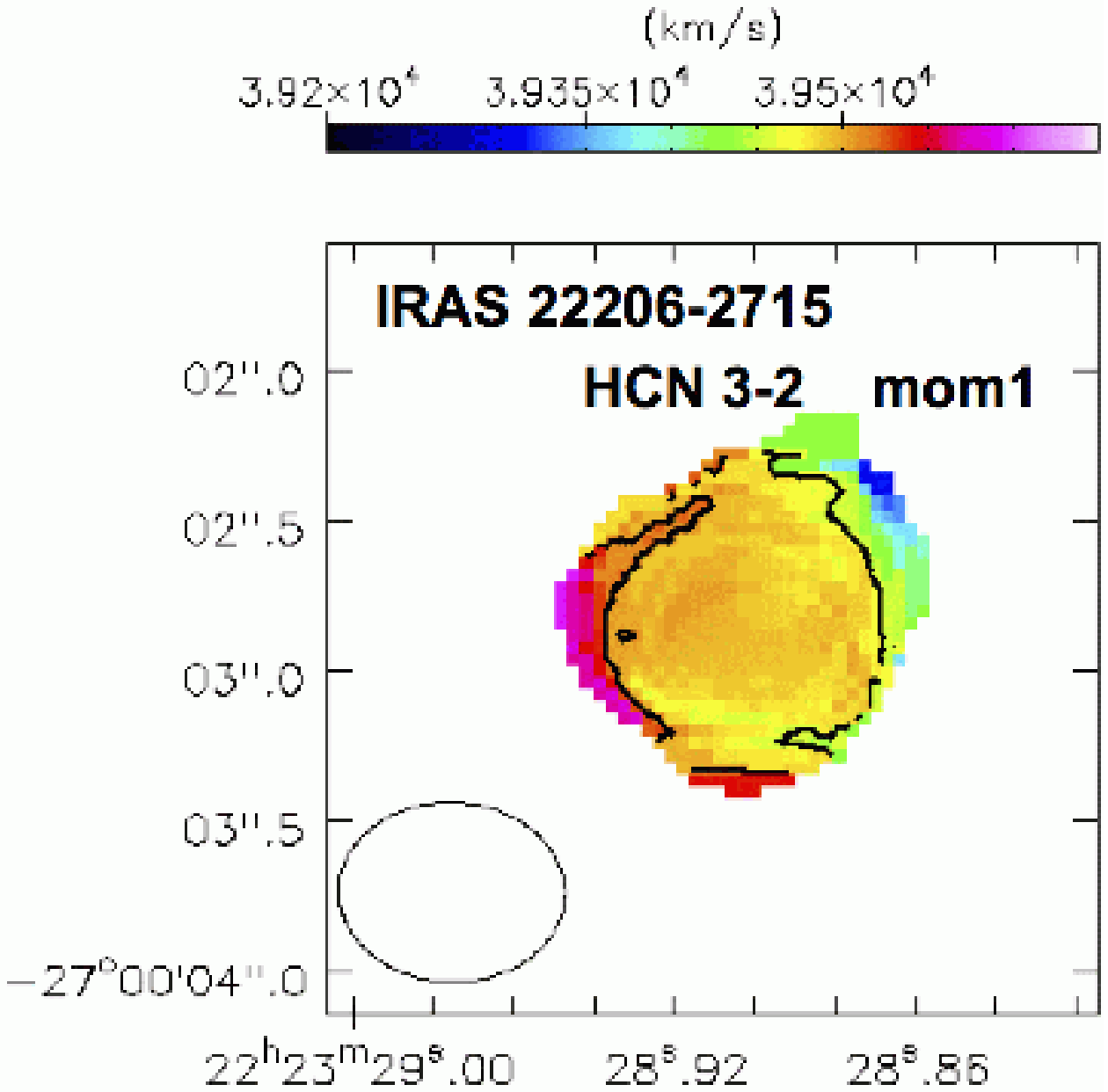} 
\includegraphics[angle=0,scale=.314]{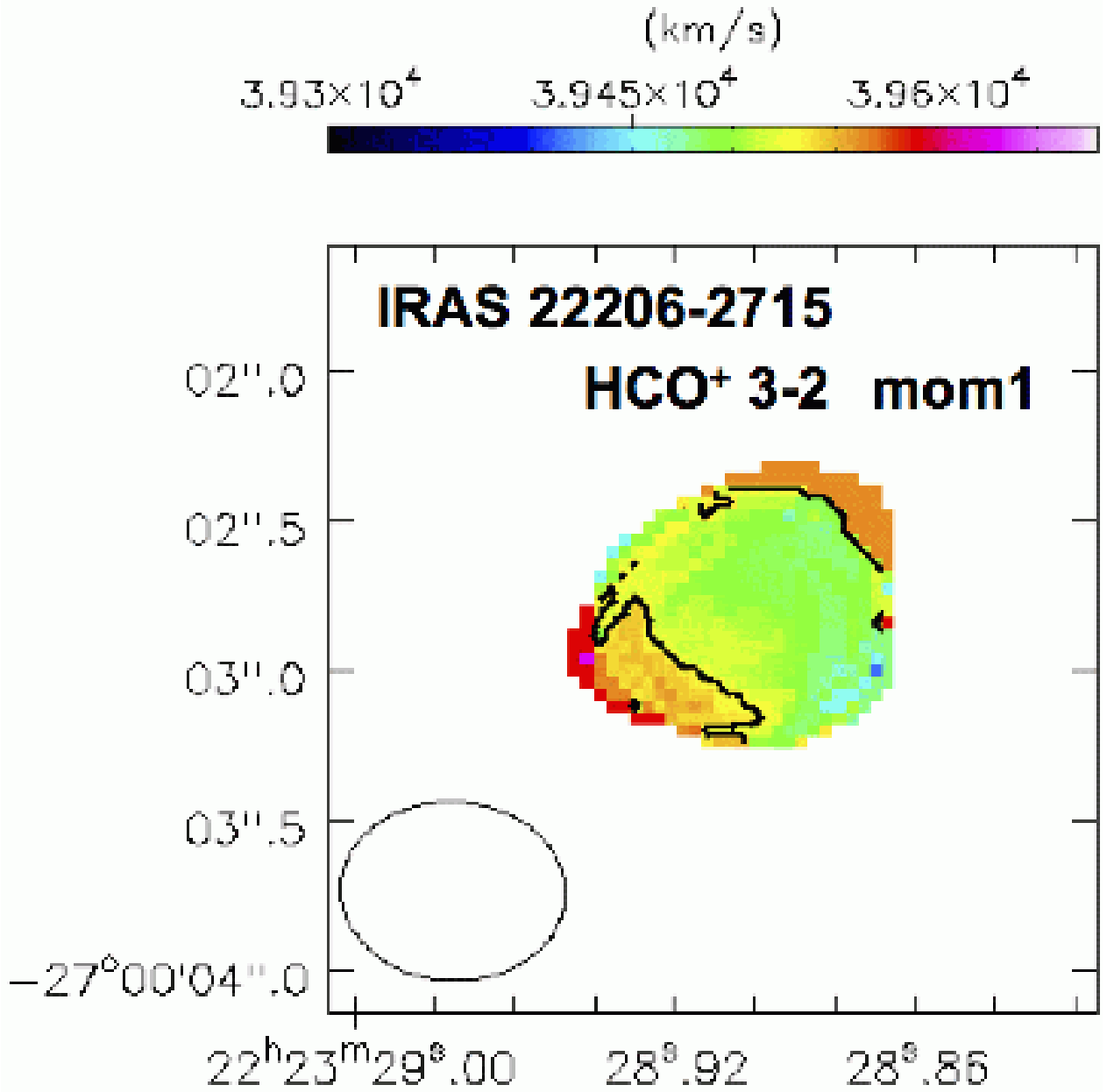} 
\caption{
Intensity-weighted mean velocity (moment 1) map of the HCN J=3--2 
and HCO$^{+}$ J=3--2 emission lines for selected ULIRGs with 
sufficiently high detection significance.
The abscissa and ordinate are right ascension and declination in ICRS,
respectively.
The contours are 
38500, 38600 km s$^{-1}$ for IRAS 00188$-$0856 HCN J=3--2 and HCO$^{+}$ J=3--2,
40940 km s$^{-1}$ for IRAS 10378+1108 HCN J=3--2 and HCO$^{+}$ J=3--2,
31950 km s$^{-1}$ for IRAS 11095$-$0238 HCN J=3--2 and HCO$^{+}$ J=3--2,
24770, 24840 km s$^{-1}$ for IRAS 14348$-$1447 SW HCN J=3--2,
24750, 24900 km s$^{-1}$ for IRAS 14348$-$1447 SW HCO$^{+}$ J=3--2,
24695, 24760 km s$^{-1}$ for IRAS 14348$-$1447 NE HCN J=3--2,
40060 km s$^{-1}$ for IRAS 16090$-$0139 HCN J=3--2,
40040 km s$^{-1}$ for IRAS 16090$-$0139 HCO$^{+}$ J=3--2,
37400, 37530, 37630 km s$^{-1}$ for IRAS 21329$-$2346 HCN J=3--2 and 
HCO$^{+}$ J=3--2,
32910, 33010 km s$^{-1}$ for IRAS 00456$-$2904 HCN J=3--2,
32880, 32950 km s$^{-1}$ for IRAS 00456$-$2904 HCO$^{+}$ J=3--2,
35360 km s$^{-1}$ for IRAS 01004$-$2237 HCN J=3--2,
35335, 35385 km s$^{-1}$ for IRAS 01004$-$2237 HCO$^{+}$ J=3--2,
35080, 35180 km s$^{-1}$ for IRAS 01166$-$0844 SE HCN J=3--2,
35200 km s$^{-1}$ for IRAS 01166$-$0844 SE HCO$^{+}$ J=3--2,
41040 km s$^{-1}$ for IRAS 01298$-$0744 HCN J=3--2,
40990, 41070 km s$^{-1}$ for IRAS 01298$-$0744 HCO$^{+}$ J=3--2,
42000, 42200 km s$^{-1}$ for IRAS 01569$-$2939 HCN J=3--2,
41920, 42120 km s$^{-1}$ for IRAS 01569$-$2939 HCO$^{+}$ J=3--2,
22740, 22880 km s$^{-1}$ for IRAS 10190+1322 HCN J=3--2,
22710, 22820 km s$^{-1}$ for IRAS 10190+1322 HCO$^{+}$ J=3--2,
38150 km s$^{-1}$ for IRAS 11506+1331 HCN J=3--2, 
38080, 38180 km s$^{-1}$ for IRAS 11506+1331 HCO$^{+}$ J=3--2,
39460, 39510 km s$^{-1}$ for IRAS 22206$-$2715 HCN J=3--2, and
39530 km s$^{-1}$ for IRAS 22206$-$2715 HCO$^{+}$ J=3--2.
For IRAS 14348$-$1447 NE, the map of the faint HCO$^{+}$ emission line 
is not shown.
The moment 1 maps of IRAS 12112$+$0305 NE, IRAS 20414$-$1651, and 
IRAS 22491$-$1808 were found in \citet{ima16c} and are not 
presented here.
Beam sizes are shown as open circles in the lower-left region.
An appropriate cut-off is applied to prevent the resulting maps
from being dominated by noise.
}
\label{ta1}
\end{center}
\end{figure}

\clearpage

\begin{figure}
\begin{center}
\includegraphics[angle=0,scale=.628]{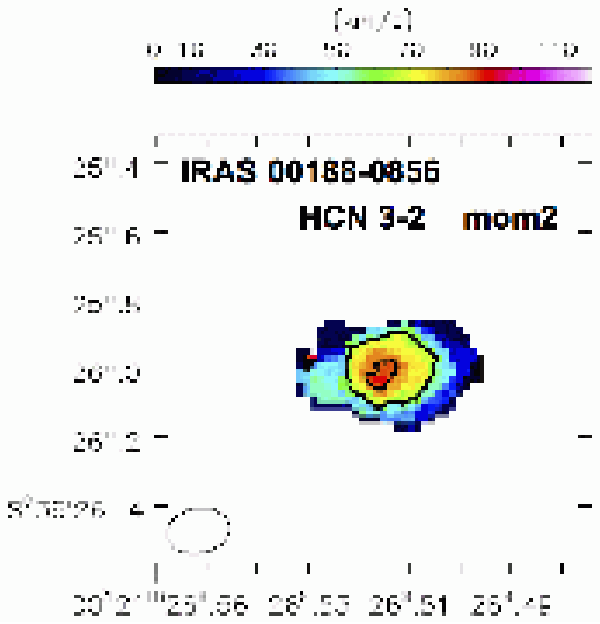} 
\includegraphics[angle=0,scale=.628]{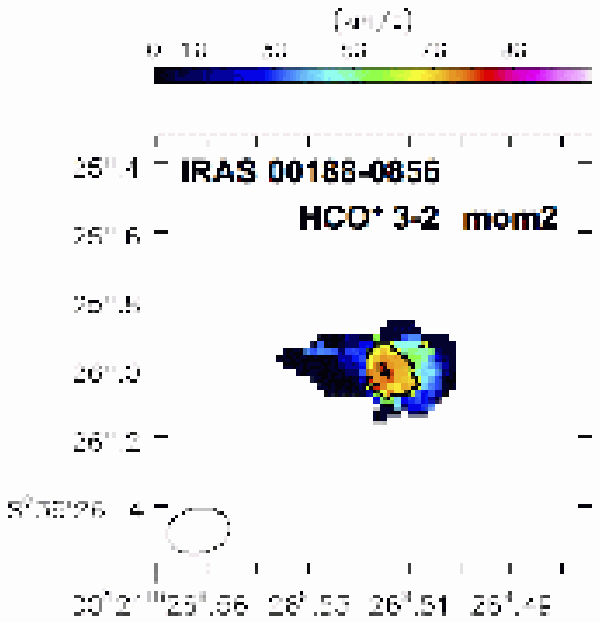}
\includegraphics[angle=0,scale=.628]{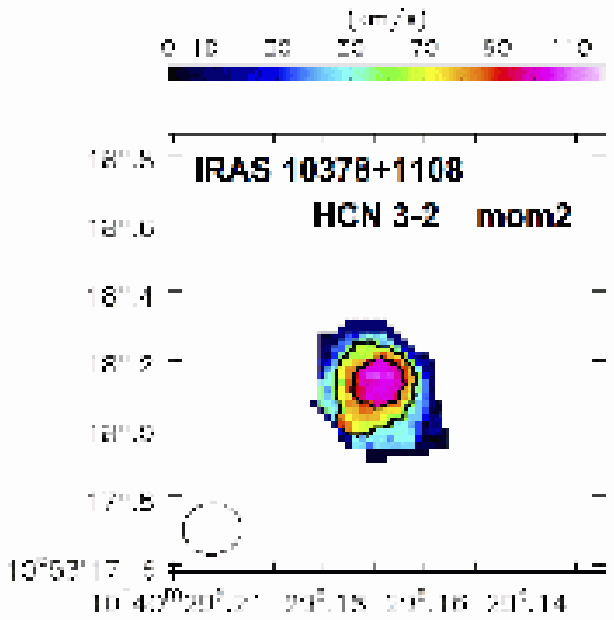} 
\includegraphics[angle=0,scale=.628]{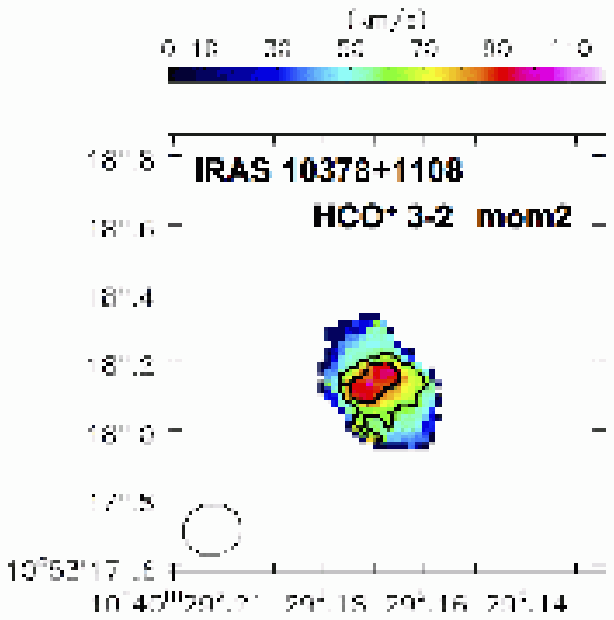}\\ 
\includegraphics[angle=0,scale=.628]{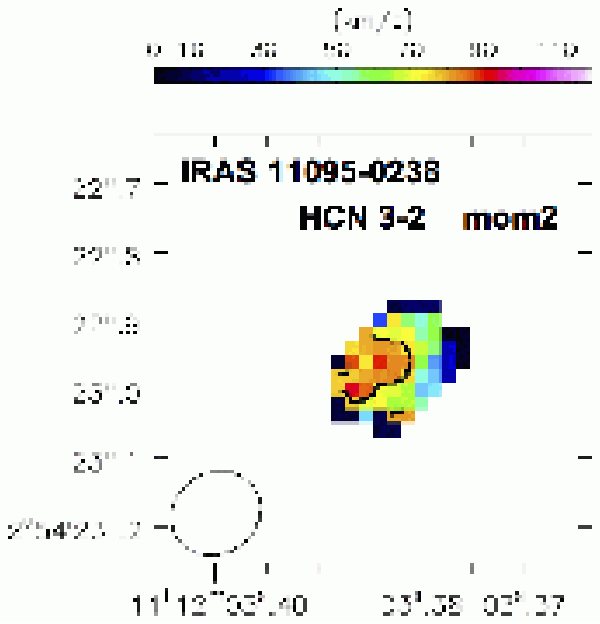} 
\includegraphics[angle=0,scale=.628]{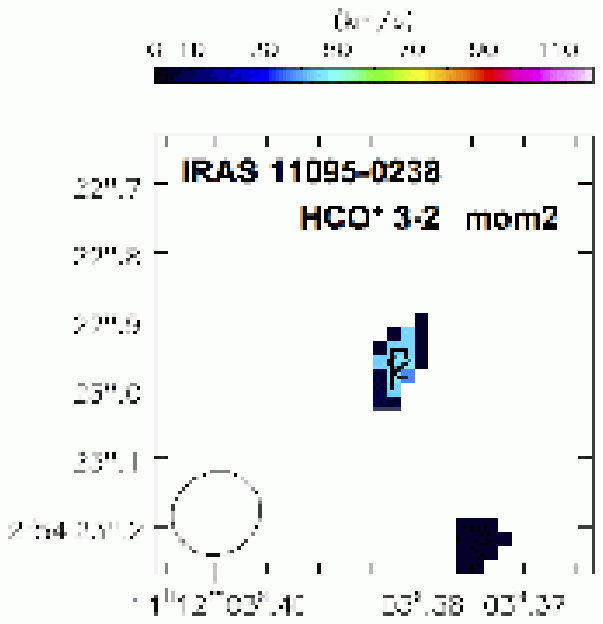} 
\includegraphics[angle=0,scale=.628]{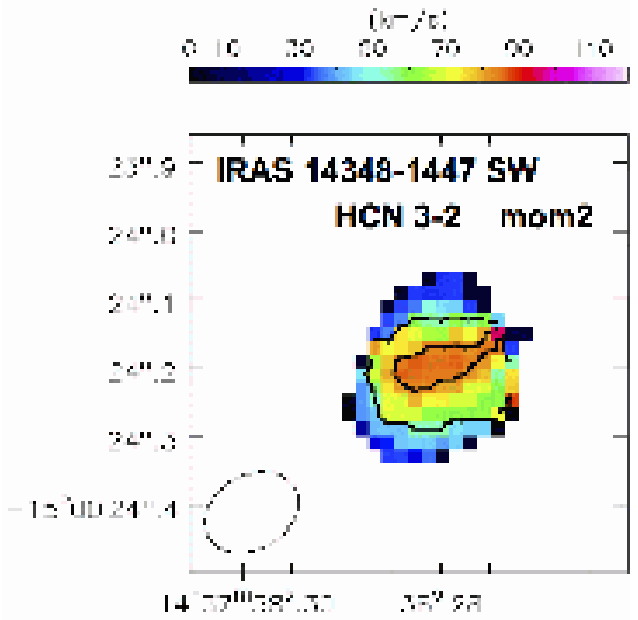} 
\includegraphics[angle=0,scale=.628]{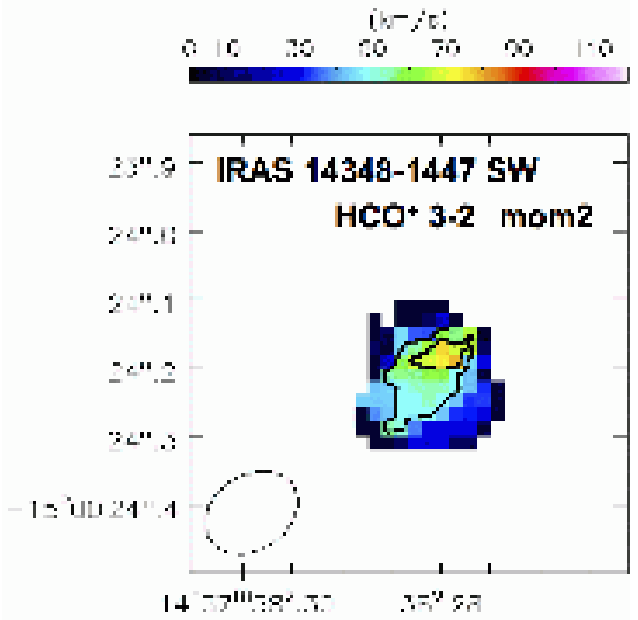} \\
\includegraphics[angle=0,scale=.628]{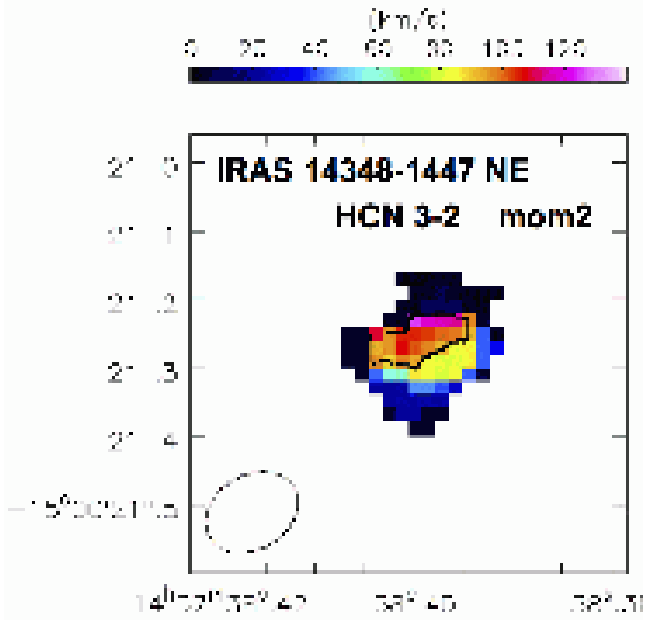} 
\hspace*{4cm}
\includegraphics[angle=0,scale=.628]{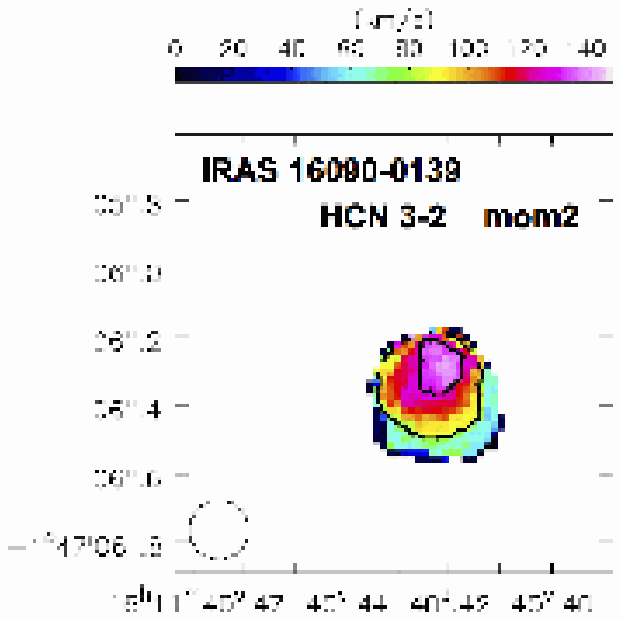}
\includegraphics[angle=0,scale=.628]{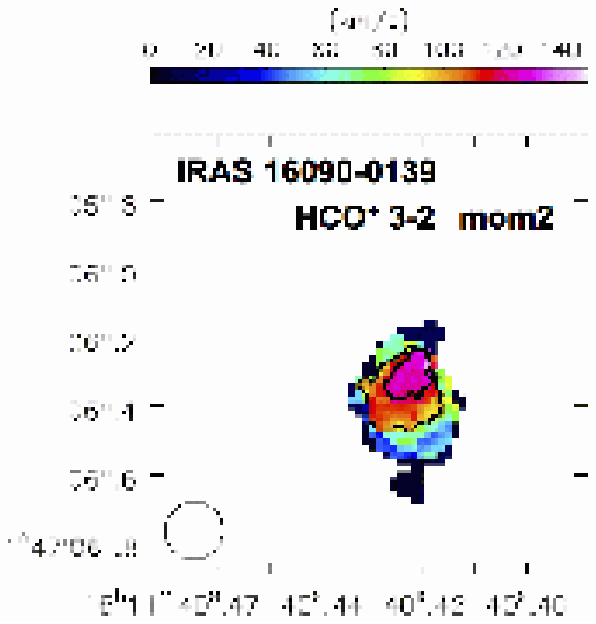}\\ 
\includegraphics[angle=0,scale=.628]{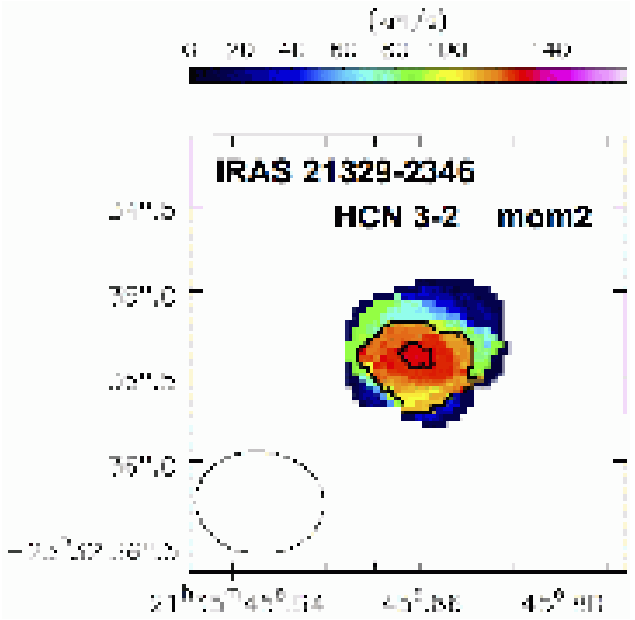}
\includegraphics[angle=0,scale=.628]{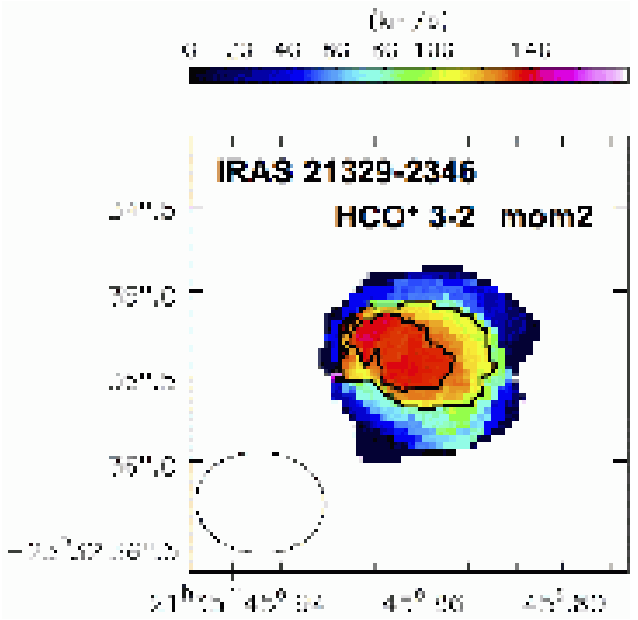}
\includegraphics[angle=0,scale=.628]{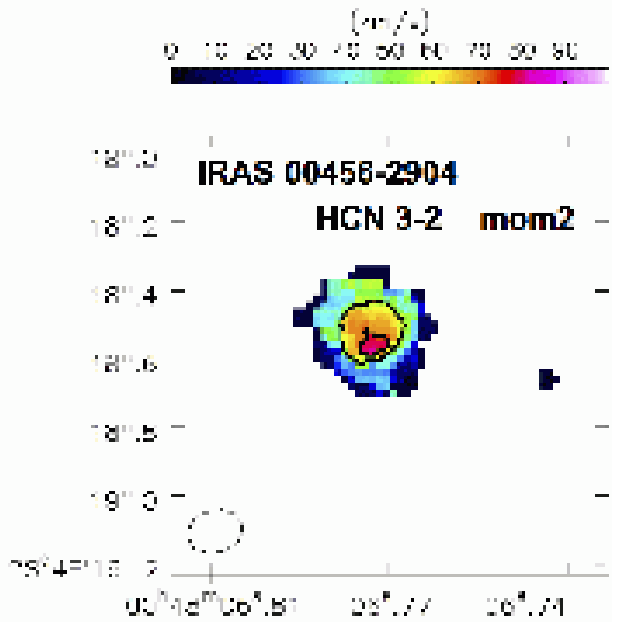} 
\includegraphics[angle=0,scale=.628]{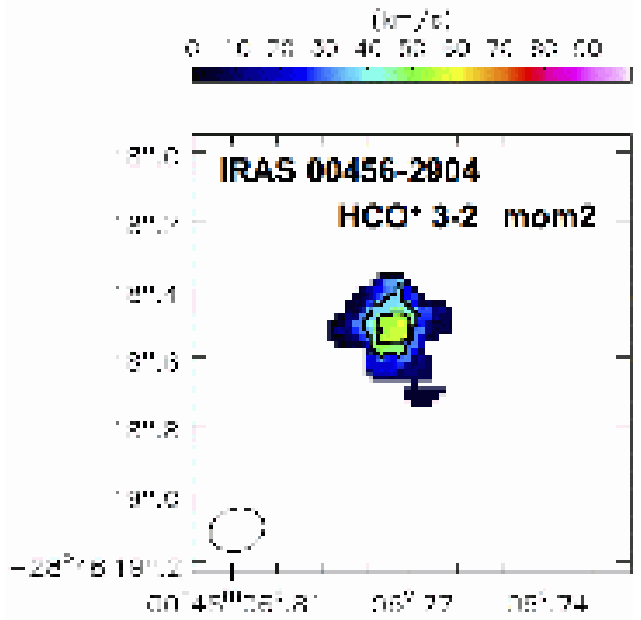}\\ 
\includegraphics[angle=0,scale=.628]{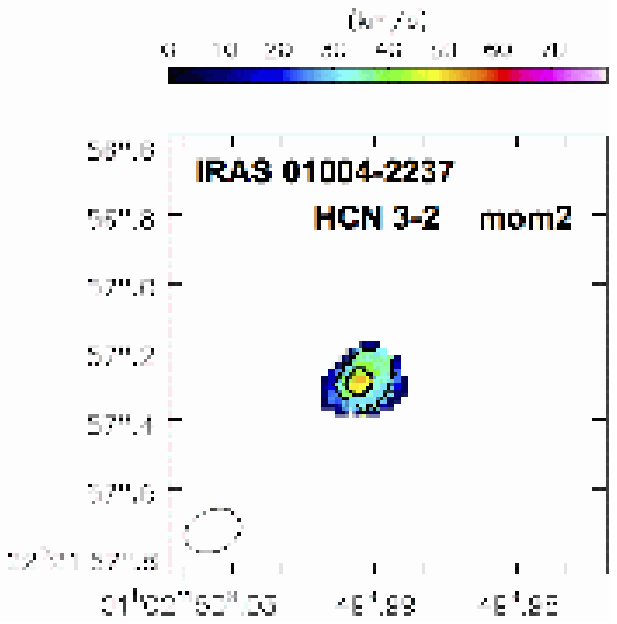} 
\includegraphics[angle=0,scale=.628]{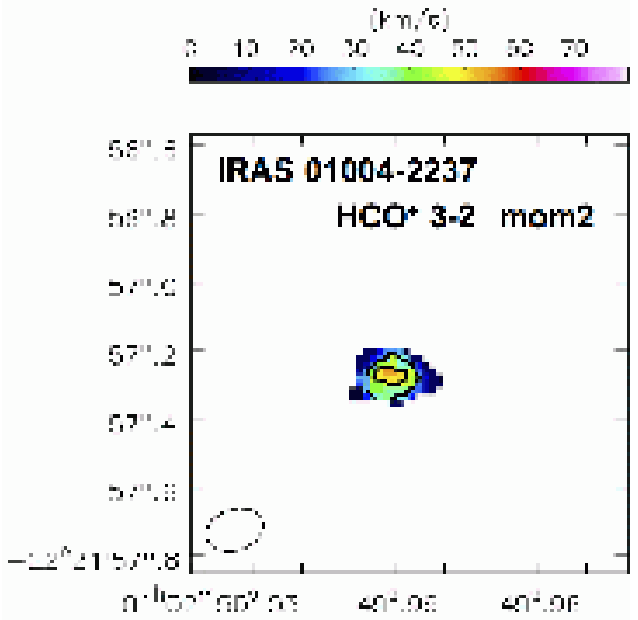} 
\includegraphics[angle=0,scale=.628]{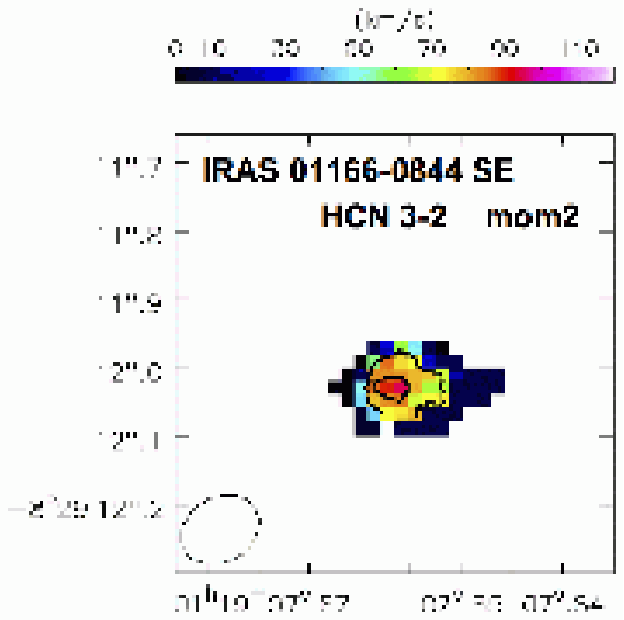} 
\includegraphics[angle=0,scale=.628]{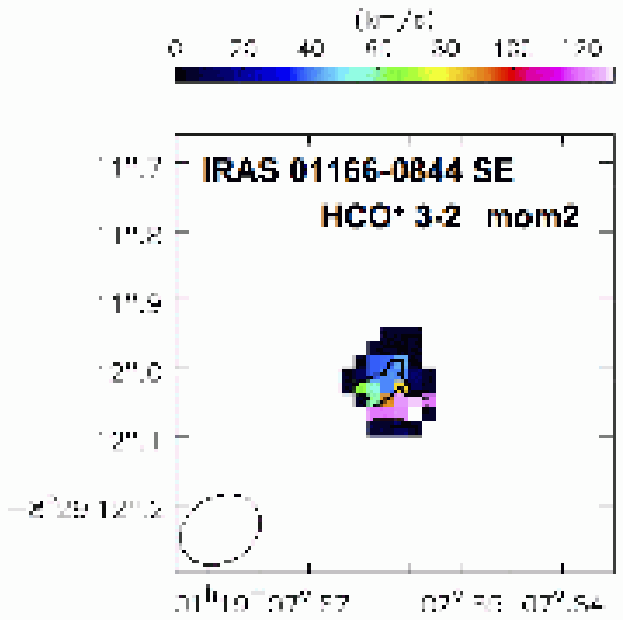}\\ 
\includegraphics[angle=0,scale=.628]{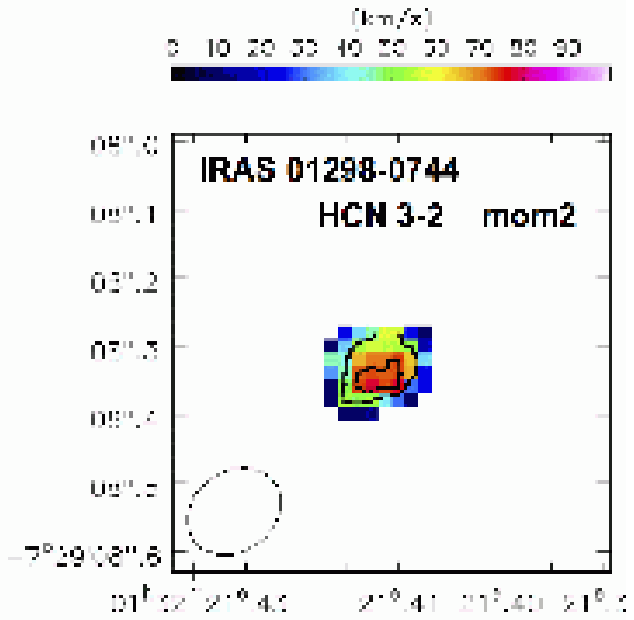} 
\includegraphics[angle=0,scale=.628]{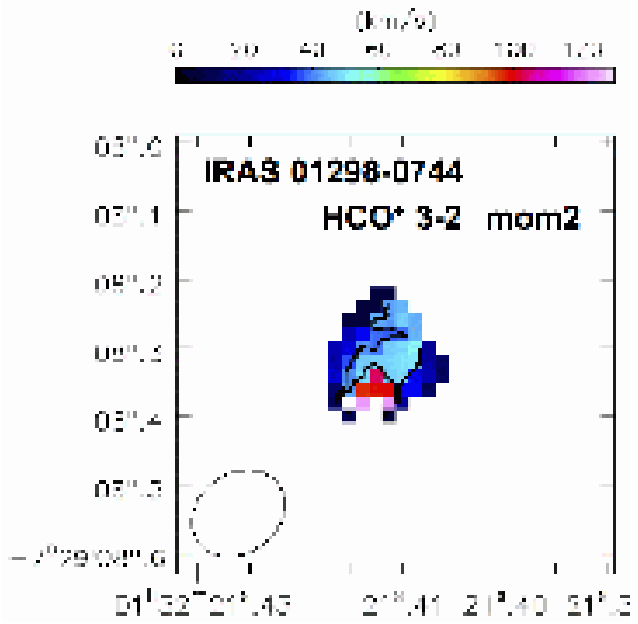} 
\includegraphics[angle=0,scale=.628]{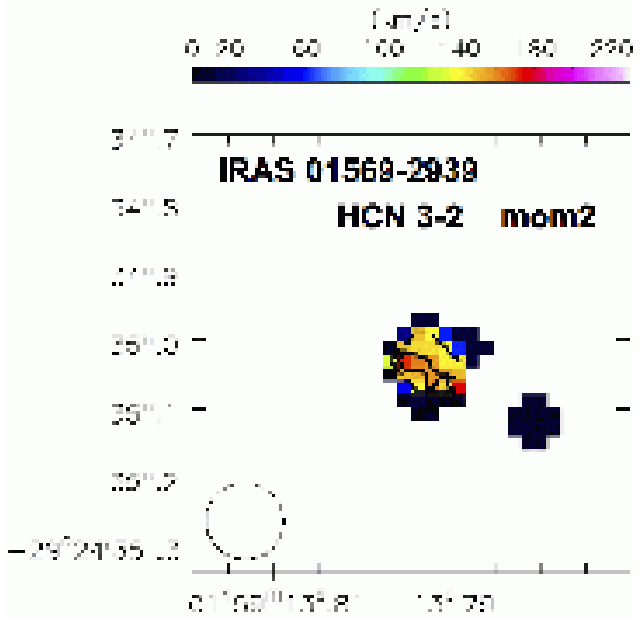} 
\includegraphics[angle=0,scale=.628]{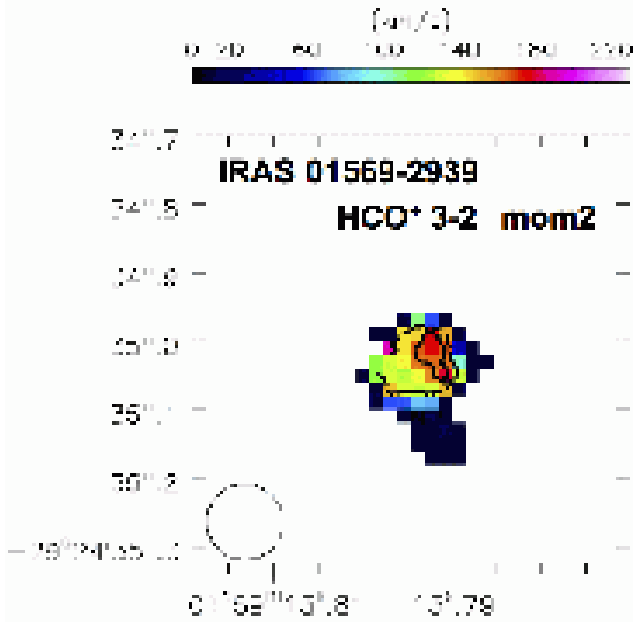}\\ 
\end{center}
\end{figure}

\clearpage

\begin{figure}
\begin{center}
\includegraphics[angle=0,scale=.628]{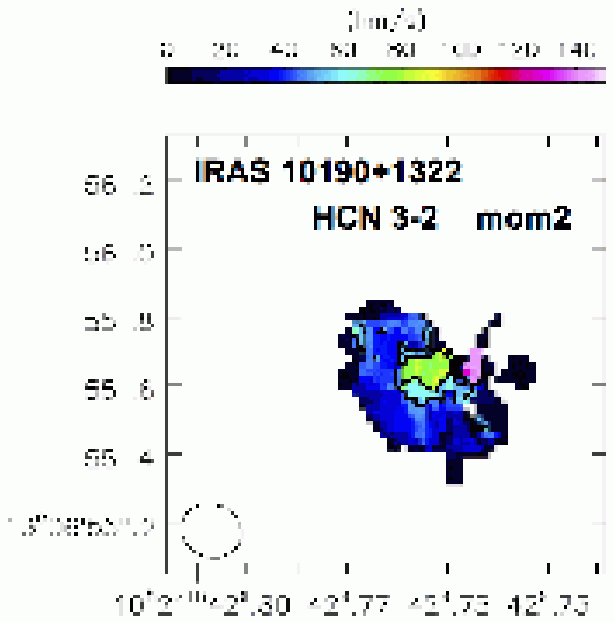} 
\includegraphics[angle=0,scale=.628]{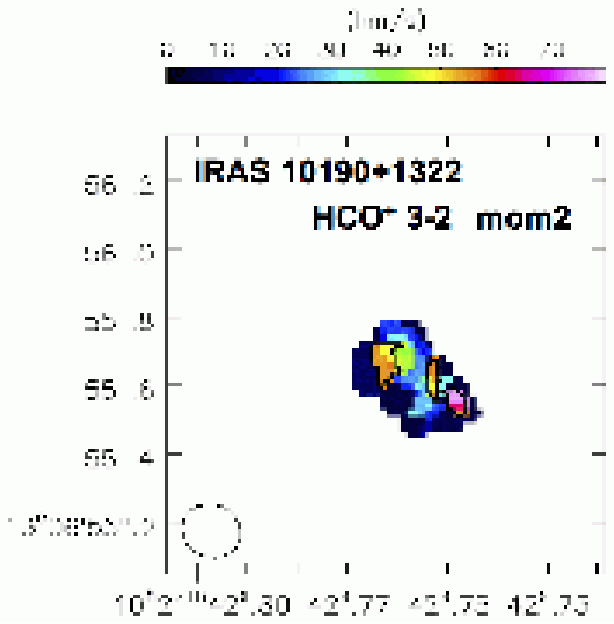} 
\includegraphics[angle=0,scale=.628]{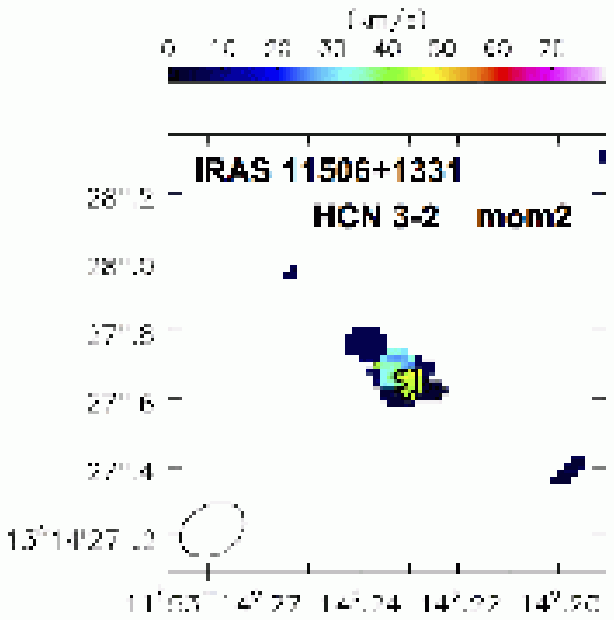}
\includegraphics[angle=0,scale=.628]{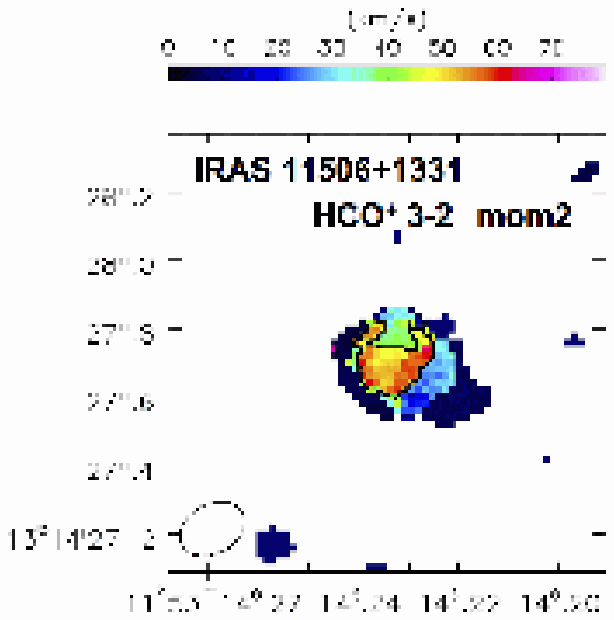}\\ 
\hspace*{-8.6cm}
\includegraphics[angle=0,scale=.628]{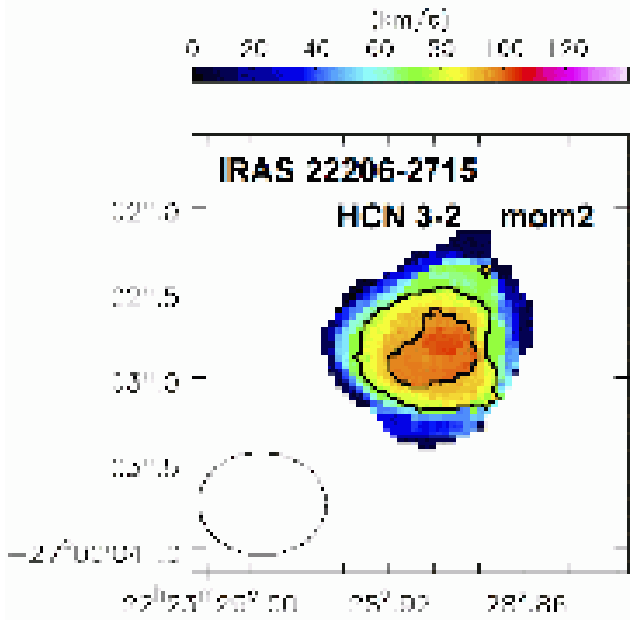}
\includegraphics[angle=0,scale=.628]{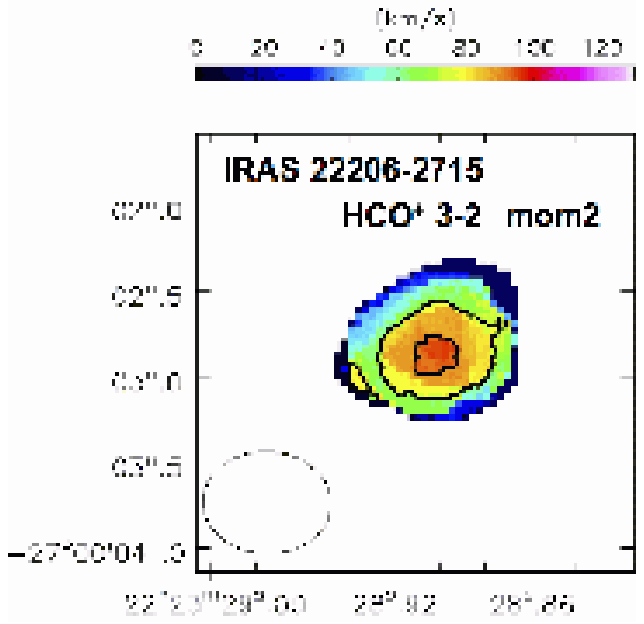}
\caption{
Intensity-weighted velocity dispersion (moment 2) map of the HCN 
J=3--2 and HCO$^{+}$ J=3--2 emission lines for selected ULIRGs 
with sufficiently high detection significance.
The abscissa and ordinate are right ascension and declination in ICRS,
respectively.
The contours are 
60, 85 km s$^{-1}$ for IRAS 00188$-$0856 HCN J=3--2,
60, 80 km s$^{-1}$ for IRAS 00188$-$0856 HCO$^{+}$ J=3--2,
60, 90 km s$^{-1}$ for IRAS 10378+1108 HCN J=3--2,
60, 85 km s$^{-1}$ for IRAS 10378+1108 HCO$^{+}$ J=3--2,
75 km s$^{-1}$ for IRAS 11095$-$0238 HCN J=3--2,
44 km s$^{-1}$ for IRAS 11095$-$0238 HCO$^{+}$ J=3--2,
55, 83 km s$^{-1}$ for IRAS 14348$-$1447 SW HCN J=3--2,
47, 65 km s$^{-1}$ for IRAS 14348$-$1447 SW HCO$^{+}$ J=3--2,
90 km s$^{-1}$ for IRAS 14348$-$1447 NE HCN J=3--2,
80, 130 km s$^{-1}$ for IRAS 16090$-$0139 HCN J=3--2,
90, 120 km s$^{-1}$ for IRAS 16090$-$0139 HCO$^{+}$ J=3--2,
102, 130 km s$^{-1}$ for IRAS 21329$-$2346 HCN J=3--2,
100, 130 km s$^{-1}$ for IRAS 21329$-$2346 HCO$^{+}$ J=3--2,
55, 72 km s$^{-1}$ for IRAS 00456$-$2904 HCN J=3--2,
36, 50 km s$^{-1}$ for IRAS 00456$-$2904 HCO$^{+}$ J=3--2,
30, 42 km s$^{-1}$ for IRAS 01004$-$2237 HCN J=3--2,
34, 48 km s$^{-1}$ for IRAS 01004$-$2237 HCO$^{+}$ J=3--2,
65, 85 km s$^{-1}$ for IRAS 01166$-$0844 SE HCN J=3--2,
40, 80 km s$^{-1}$ for IRAS 01166$-$0844 SE HCO$^{+}$ J=3--2,
50, 72 km s$^{-1}$ for IRAS 01298$-$0744 HCN J=3--2,
40, 60 km s$^{-1}$ for IRAS 01298$-$0744 HCO$^{+}$ J=3--2,
120, 156 km s$^{-1}$ for IRAS 01569$-$2939 HCN J=3--2,
120, 164 km s$^{-1}$ for IRAS 01569$-$2939 HCO$^{+}$ J=3--2,
49, 70 km s$^{-1}$ for IRAS 10190+1322 HCN J=3--2,
45 km s$^{-1}$ for IRAS 10190+1322 HCO$^{+}$ J=3--2,
42 km s$^{-1}$ for IRAS 11506+1331 HCN J=3--2, 
45 km s$^{-1}$ for IRAS 11506+1331 HCO$^{+}$ J=3--2,
75, 96 km s$^{-1}$ for IRAS 22206$-$2715 HCN J=3--2, and
74, 91 km s$^{-1}$ for IRAS 22206$-$2715 HCO$^{+}$ J=3--2.
For IRAS 14348$-$1447 NE, the map of the faint HCO$^{+}$ emission line 
is not shown.
The moment 2 maps of IRAS 12112$+$0305 NE, IRAS 20414$-$1651, and 
IRAS 22491$-$1808 were found in \citet{ima16c} and are not 
presented here.
Beam sizes are shown as open circles in the lower-left region.
An appropriate cut-off is applied.
}
\end{center}
\end{figure}

\begin{figure}
\begin{center}
\includegraphics[angle=0,scale=.6]{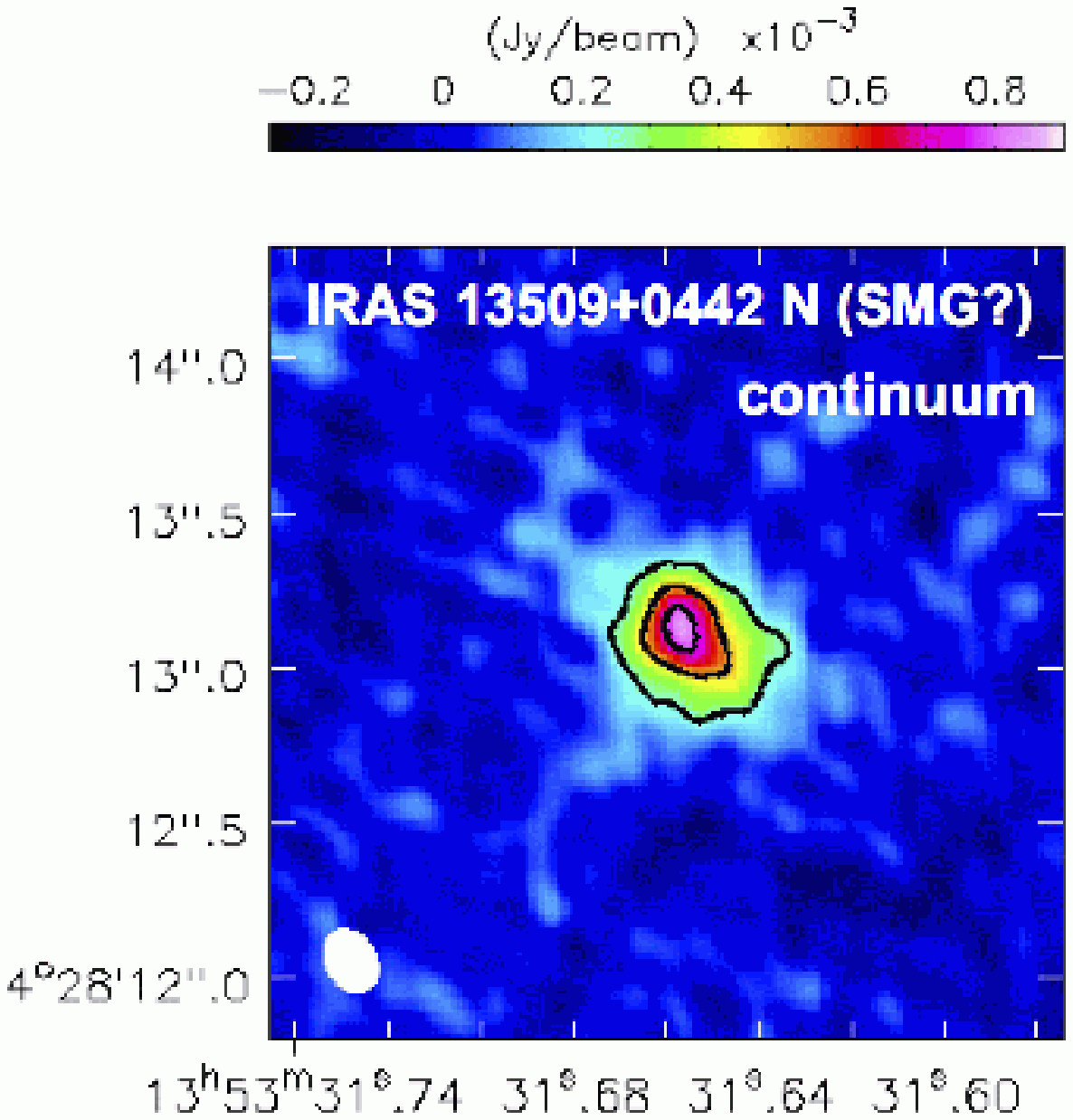}
\hspace{1cm}
\includegraphics[angle=0,scale=.45]{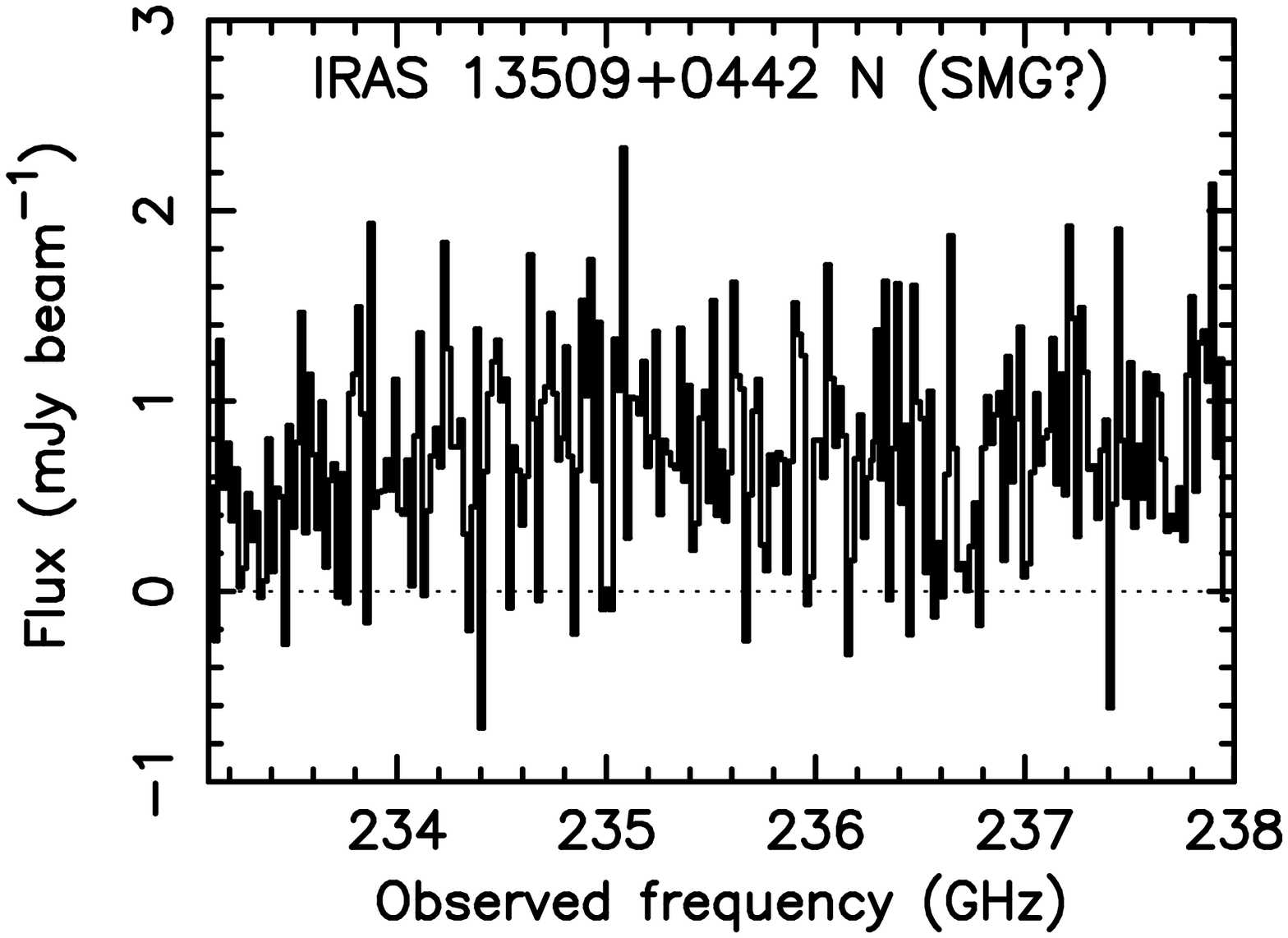}
\vspace{0.5cm}
\caption{
Continuum image (left) and spectrum without continuum subtraction 
at the nuclear peak position within the beam size (right) of the bright 
continuum-emitting source at $\sim$10$''$ northern side of 
IRAS 13509$+$0442 \citep{ima16c,ima18b}.
The contours represent 4$\sigma$, 10$\sigma$, 15$\sigma$ 
(1$\sigma$ = 0.054 mJy beam$^{-1}$) in the left panel.
}
\end{center}
\end{figure}

\end{document}